\documentclass[11pt]{cernrep}
 
\usepackage{epsfig, url, psfrag, color}
\usepackage{fancyhdr, multind}
\usepackage[square,comma,numbers]{natbib}
\usepackage{dcolumn, bm, mathrsfs, xspace, axodraw, graphicx,subfigure}
\usepackage{array,amsfonts,amssymb,amsmath,longtable}

\usepackage{tev4lhc}

\renewcommand{\theequation}{\thesubsection.\arabic{equation}}
\renewcommand{\thefigure}{\thesubsection.\arabic{figure}}
\renewcommand{\thetable}{\thesubsection.\arabic{table}}
\begin{document}



\date{\today}
\documentlabel{hep-ph/yymmnnn \\ FERMILAB-CONF-07-052}


\title{\bf Tevatron-for-LHC Report: Top and Electroweak Physics}



\author{\center (The TeV4LHC-Top and Electroweak Working Group) 
C.E.~Gerber,$^8$
P.~Murat,$^1$ 
T.M.P.~Tait,$^{10}$ 
D.~Wackeroth,$^{16}$
A.~Arbuzov,$^{24}$
D.~Bardin,$^{24}$
U.~Baur,$^{16}$
J.A.~Benitez,$^3$
S.~Berge,$^{26}$
S.~Bondarenko,$^{24}$
E.E~.Boos,$^{14}$ 
M.T.~Bowen,$^{2}$
R.~Brock,$^3$
V.E.~Bunichev,$^{14}$
J.~Campbell,$^{12}$
F.~Canelli,$^{1,5}$
{Q.-H.~Cao},$^{11}$
C.M.~Carloni Calame,$^{28,35}$
F.~Chevallier,$^{17}$
P.~Christova,$^{24}$
{C.~Ciobanu},$^9$ 
S.~Dittmaier,$^{25}$
L.~V.~Dudko,$^{14}$ 
S.D.~Ellis,$^{2}$
A.I.~Etienvre,$^{20}$ 
F.~Fiedler,$^6$
A.~Garcia-Bellido,$^{2}$
A.~Giammanco,$^{18}$
D.~Glenzinski,$^1$
P.~Golonka,$^{30,31}$
C.~Hays,$^{22,34}$
S.~Jadach,$^{30,31}$
S.~Jain,$^{21}$
L.~Kalinovskaya,$^{24}$
M.~Kr\"amer,$^{26}$
A.~Lleres,$^{17}$
J.~L\"uck,$^{15}$ 
A.~Lucotte,$^{17}$
{A.~Markina},$^{14}$
G.~Montagna,$^{33,28}$
P.M.~Nadolsky,$^{10}$
O.~Nicrosini,$^{28}$
F.I.~Olness,$^{27}$
W.~P{\l}aczek,$^{30,32}$
R.~Sadykov,$^{24}$
V.I.~Savrin,$^{14}$ 
{R.~Schwienhorst},$^3$
A.V.~Sherstnev,$^{14}$
S.~Slabospitsky,$^{19}$
B.~Stelzer,$^5$
M.J.~Strassler,$^{2}$
Z.~Sullivan,$^{10}$
F.~Tramontano,$^{13}$
A.~Vicini,$^{36,29}$
W.~Wagner,$^{15}$
Z.~Was,$^{30,31}$
G.~Watts,$^2$
M.~Weber,$^1$
S.~Willenbrock,$^9$
U.K.~Yang,$^7$
C.-P.~Yuan,$^3$
J.~Zhu$^{23}$
}

\institute{
\mbox{$^1$ Fermilab},
\mbox{$^2$ Univ. of Washington}, 
\mbox{$^3$ Michigan State Univ.}, 
\mbox{$^4$ Univ. of Rochester}, 
\mbox{$^5$ Univ. of California, Los Angeles}, 
\mbox{$^{6}$ Ludwig-Maximilians-Universit{\"a}t M{\"u}nchen},
\mbox{$^7$ Univ. of Manchester}, 
\mbox{$^8$ Univ. of Illinois at Chicago}, 
\mbox{$^9$ Univ. of Illinois at Urbana-Champaign}, 
\mbox{$^{10}$ Argonne National Laboratory}, 
\mbox{$^{11}$ Univ. of California, Riverside}, 
\mbox{$^{12}$ Univ. of Glasgow}, 
\mbox{$^{13}$ Univ. di Napoli},
\mbox{$^{14}$ Moscow State Univ.}, 
\mbox{$^{15}$ Univ. of Karlsruhe},
\mbox{$^{16}$ State Univ. of New York at Buffalo},
\mbox{$^{17}$ Laboratoire de Physique Subatomique \& Cosmologie, Grenoble},
\mbox{$^{18}$ Universit\'e Catholique de Louvain},
\mbox{$^{19}$ Institute for High Energy Physics, Protvino},
\mbox{$^{20}$ CEA-Saclay},
\mbox{$^{21}$ Univ. of Oklahoma},
\mbox{$^{22}$ Duke Univ.},
\mbox{$^{23}$ State Univ. of New York, Stony Brook},
\mbox{$^{24}$ JINR Dubna}, 
\mbox{$^{25}$ Max-Planck Institut f{\"u}r Physik (Werner-Heisenberg-Institut)}, 
\mbox{$^{26}$ RWTH Aachen} ,
\mbox{$^{27}$ Southern Methodist Univ.}, 
\mbox{$^{28}$ INFN - Sezione di Pavia}, 
\mbox{$^{29}$ INFN - Sezione di Milano}, 
\mbox{$^{30}$ CERN}, 
\mbox{$^{31}$ Institute of Nuclear Physics, Cracow}, 
\mbox{$^{32}$ Jagiellonian Univ.},
\mbox{$^{33}$ Univ. of Pavia},
\mbox{$^{34}$ Oxford Univ.},
\mbox{$^{35}$ Univ. of Southampton},
\mbox{$^{36}$ Univ. of Milano} 
}

\maketitle

\begin{abstract}
  The top quark and electroweak bosons ($W^\pm$ and $Z$) represent the
  most massive fundamental particles yet discovered, and as such refer
  directly to the Standard Model's greatest remaining mystery: the
  mechanism by which all particles gained mass.  This report
  summarizes the work done within the top-ew group of the
  Tevatron-for-LHC workshop. It represents a collection of both
  Tevatron results, and LHC predictions.  The hope is that by
  considering and comparing both machines, the LHC program can be
  improved and aided by knowledge from the Tevatron, and that particle
  physics as a whole can be enriched.  The report includes
  measurements of the top quark mass, searches for single top quark
  production, and physics of the electroweak bosons at hadron
  colliders.
\end{abstract}

\newpage
\pagestyle{plain}
\setcounter{page}{2}
\setcounter{tocdepth}{2}
\tableofcontents

\newpage
\def\pt{$p_T$}                          
\def\et{$E_T$}                          
\def\met{\mbox{${\hbox{$E$\kern-0.6em\lower-.1ex\hbox{/}}}_T$}} 
\def\htran{$H_T$}                       
\def\aplan{$\cal{A}$}                   
\def\htp{$H_{T2}'$}                     
\def\hp{$H_{\parallel}$}                  
\def\ktmp{$K_{Tmin}'$}
\def\mw{$M_W$}                          

\section{Introduction}

The top quark and electroweak bosons ($W^\pm$ and $Z$) represent the
most massive fundamental particles yet discovered.  Thus, they are not
only the newest additions to the Standard Model (SM) of particle
physics, but are also the most interesting because their large masses
refer directly to the SM's greatest remaining mystery: the mechanism
by which all particles gained mass.  The $SU(2) \times U(1)$ gauge
structure of SM is successful at describing all interactions, and is
an essential ingredient for the theoretical consistency of the theory,
but if the symmetry were exact it would require all particles to be
massless.  Thus, the symmetry must be spontaneously broken.  In the SM
itself this is defined by introducing a Higgs boson together with a
potential that insures it has a non-zero expectation value in empty
space.  However, even if this assumption is correct, the Higgs has as
yet eluded experimental observation, and it could be that the SM
description is incomplete or even simply incorrect.  Theoretical
arguments related to the hierarchy of scales or triviality of the
Higgs potential, further suggest that the SM description is at best a
stand-in for some more natural explanation.  Unravelling the details
of the true nature of electroweak symmetry-breaking (EWSB) is one of
the most pressing challenges awaiting particle physics.

The top and electroweak bosons, as the most massive objects in the SM,
are those which felt the symmetry breaking the most profoundly.  Thus,
they must couple the most strongly to the agent of EWSB (be it a SM
fundamental Higgs, or the result of some more plausible dynamics) and
a detailed study of their properties represents an excellent chance to
learn indirectly about EWSB itself.  They are interesting in their own
right and are produced copiously at both the Tevatron and the LHC.
Thus, it is natural as run II of the Tevatron draws to a close, and
the LHC era begins, to examine how well we can measure all of the
quantities needed to describe these particles, and how the two
machines may complement each other in our quest to explore EWSB
through study of massive objects.

In addition to the interest in top and the electroweak bosons in their
own right, they are also interesting ``standard candles'' that may
allow us to understand the SM predictions at the LHC in the light of
Tevatron data.  It may be that the resolution of the EWSB dynamics
involves new particles, and their observation as we probe the energy
frontier may be more striking than deviations in the properties of
top, $W$, or $Z$ from SM predictions.  If so, a key ingredient to
observing these new states is that we be able to infer very precisely
what the SM prediction for any given signature should be.  Top and the
electroweak bosons have signatures which can be extremely distinctive
at hadrons colliders, including charged leptons, missing energy, hard
jets, and massive resonances in distributions.  Understanding how to
predict signals involving these objects at the LHC can benefit from
Tevatron data, and the Tevatron can provide a laboratory to test out
analysis ideas in a better understood environment, before they become
essential at the LHC.

This report is collection of both Tevatron results, and LHC
predictions.  The hope is that by considering and comparing both
machines, the LHC program can be improved and aided by knowledge from
the Tevatron, and that particle physics as a whole can be enriched by
combining information from both machines.  Subsequent chapters deal
with measurement of the top quark mass, searches for single top quark
production, and understanding the physics of the electroweak bosons at
hadron colliders.


 
\section{Measurement of the top quark mass}
\label{sec:topmass}

\def\mrm{\mathrm}
\def\ra{\rightarrow}

\newcommand{\ljets}{\ensuremath{\ell}+\rm jets\xspace}
\newcommand{\psgn}{\ensuremath{P_{\rm sgn}}\xspace}
\newcommand{\pbkg}{\ensuremath{P_{\rm bkg}}\xspace}
\newcommand{\wjets}{\ensuremath{W}+\rm jets\xspace}
\newcommand{\ptcaljet} {p_T^{jet}}
\newcommand{\ptparton} {p_{T}^{parton}}
\newcommand{\ptpartjet} {p_{T}^{particle}}
\newcommand{\mev}{\ensuremath{\mathrm{Me\kern -0.1em V}}}
\newcommand{\gevc}{\ensuremath{\mathrm{Ge\kern -0.1em V/c}}}
\newcommand{\mevc}{\ensuremath{\mathrm{Me\kern -0.1em V/c}}}
\newcommand{\gevcs}{\ensuremath{\mathrm{Ge\kern -0.1em V/c^{2}}}}
\newcommand{\mevcs}{\ensuremath{\mathrm{Me\kern -0.1em V/c^{2}}}}
\newcommand{\roots}{\ensuremath{\sqrt{s}}}
\newcommand{\Mt}{\ensuremath{M_{t}}}
\newcommand{\at}{\symbol{64}}
\newcommand{\mreco}{\ensuremath{m_t^{reco}}}
\newcommand{\mjj}{\ensuremath{m_{\mathrm{jj}}}\xspace}
\newcommand{\jes}{\ensuremath{\mathrm{JES}}\xspace}
\newcommand{\chisq}{\ensuremath{\chi^{2}}\xspace}
\newcommand{\chisqmin}{\ensuremath{\chi^{2}_{min}}\xspace}
\newcommand{\gevcc}[1]{\ensuremath{#1~\mathrm{GeV}/c^{2}}}

\newcommand{\sye}[1]{\ensuremath{~\pm #1}}
\newcommand{\ase}[2]{\ensuremath{^{~+ #1}_{~- #2}}}
\newcommand{\asi}[2]{\ensuremath{^{~- #1}_{~+ #2}}}
\newcommand{\ptmissvec}{\ensuremath{\vec{p}_T}\hspace{-2.5ex}/\hspace{1.5ex}\xspace}

\sloppy

\subsection{Introduction}\label{sec:topew_mass_intro}
\textbf{Contributed by:~T.~Tait}

The top quark mass is one of the fundamental parameters of the standard model
(SM), related to the top's coupling to the Higgs by the tree level relationship
$m_t = y_t v$.  The top mass, like all SM fermion masses is a manifestation
of the breaking of the electroweak symmetry from $SU(2)_L \times U(1)_Y$ to
electromagnetism.  As the heaviest fermion, the top felt this symmetry-breaking
the most strongly, and thus is a natural laboratory to learn about the
dynamics of the breaking.  Thus, the hope is that precision measurements
of the top quark will either confirm the SM's picture of electroweak breaking,
or show deviations which will point the way to a more complete theory.

Even within the SM, the top's large mass implies that it is special.  The
large coupling of the Higgs boson to top inferred from the mass suggests
that rates for processes involving both Higgs and top can be large.  The
top essentially determines the Higgs coupling to two gluons (induced by
a loop of top quarks) and is significant in determing the coupling to two
photons (complementing a loop of $W$ bosons).  The top mass is an essential
input in determining the SM prediction for these processes.  In addition, the
top contribution to flavor-violating processes in the SM (such as
bottom- or strangeness-number violating processes, which occur at loop level
in the Standard Model)
is usually dominant,
because the large top mass disrupts the GIM mechanism and permits these
processes to take place.

\begin{figure}[htbp]
\centering 
\includegraphics[width = .75\textwidth]{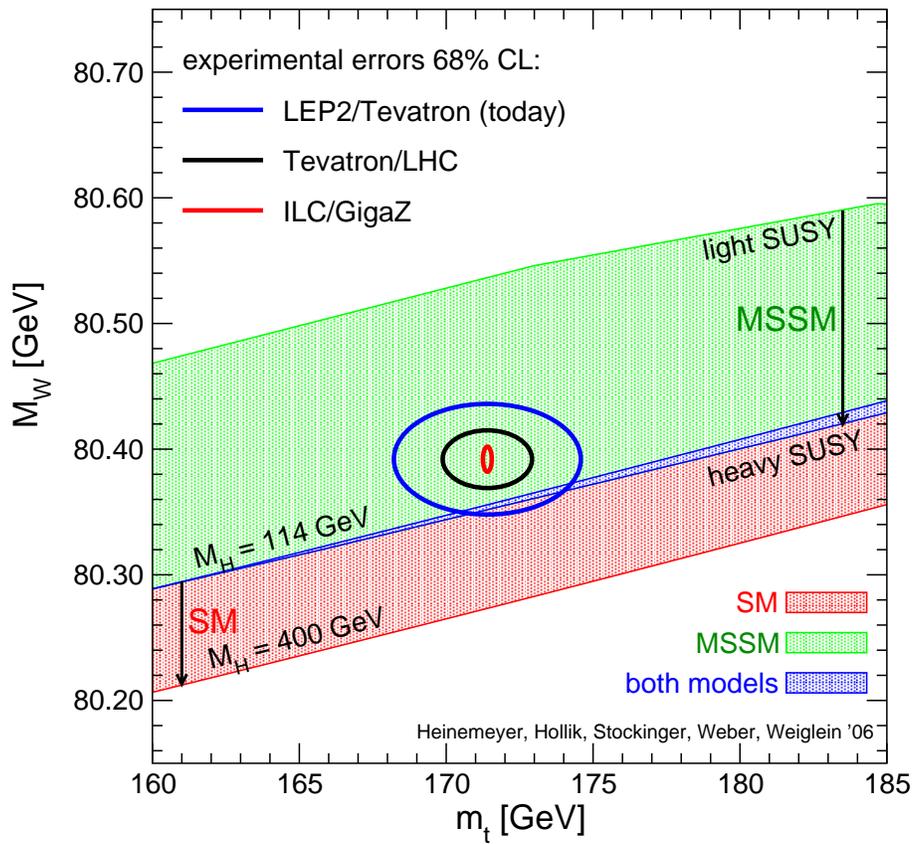}
\caption{Current experimental results for $m_W$ and $m_{top}$, and expected accuracies at the next generation of colliders, 
compared with predictions obtained within the Standard Model and the MSSM 
\cite{Heinemeyer:2004gx}.  }
\label{fig:mtmw}
\end{figure}

Perhaps the most famous role the top mass plays in the Standard Model is
through the corrections to electroweak precision observables at loop level.
The precision of experiments at LEP and SLAC is enough to be sensitive to
loop contributions of the Higgs and top, and thus given the top mass
measured at the Tevatron, the precision data can be used to predict the
as yet unknown mass of the Higgs boson.
The most important  
top mass dependence contribution to the Electroweak observables 
arises via the one-loop radiative correction term $\Delta r$
\cite{radiative_corr}, related to the W mass through the relation: 
$m_W^2 \, = \, \frac{\pi \alpha}{\sqrt{2} G_F\sin^2\Theta_W} (1 \, + \, \Delta r)$. 
$\Delta r$ depends on the top mass via terms proportional to 
$m_t^2/m_Z^2$, while the Higgs mass gives rise to terms proportional 
to $\log m_H /m_Z$~: therefore, the dependence on the Higgs mass is much 
weaker than the dependence on the top mass and without a precise measurement
of $m_t$, no information about $m_H$ can be extracted.
The current value of the top mass
($m_{top}$ = 172.7 $\pm$ 2.9 GeV/c$^2$)  \cite{tevatron}
results in the following constraints on the Higgs boson mass: 
$m_H \, = \, 91 ^{+45} _{-32}$ GeV/c$^2$, 
$m_H  \, \leq \,  186$ GeV/c$^2$ at 95$\%$ C.L. limit. 
The allowed region in the ($m_W$, $m_t$) plane is displayed in 
Fig.~\ref{fig:mtmw}, 
for different Higgs boson masses, in the SM and in the MSSM.   

Even in theories beyond the Standard Model, the large top mass can imply
a special role for top.  In the
minimal supersymmetric standard model (MSSM), the Higgs quartic
interaction is determined from gauge couplings, and requires 
$m_H \leq m_Z$ at tree level.  This would be 
largely ruled out by the LEP-II searches for the Higgs boson, if it were not
for the quantum corrections from loops of the top quark -- large because
the large top mass implies strong coupling to the Higgs, the MSSM would
be excluded by the null LEP search.  As it is, the precise value of the top
mass determines a prefered range of MSSM Higgs masses.  As a further test of
physics beyond the Standard Model, the top mass (along with the strong
coupling constant) determines the SM prediction for the rate of $t \bar{t}$
production, and correlated measurements of the top mass and cross section
thus test theories which contain new $t \bar{t}$ production mechanisms, or
objects which decay like the top quark and thus can be confused in the
top event sample.

The following sections describe the methods used at the Tevatron to
measure the top quark mass in the various channels, 
summarize the systematic uncertainties that
dominate the results, and explain the techniques for combination of results.
In addition, the expectation for the top mass measurement during Tevatron 
Run II, and the plans for the LHC are also included. 

\subsection{Top Mass Determination at the Tevatron}
\subsubsection*{Introduction}
\textbf{Contributed by:~C.~Gerber}

\begin{figure}[htbp]
\centering
\includegraphics[width = .75\textwidth]{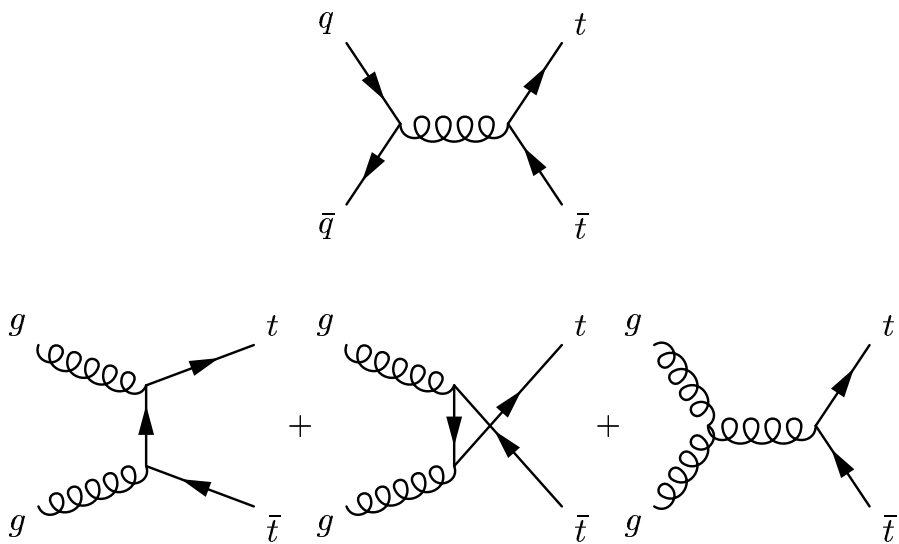}
\caption{Leading order Feynman diagrams for the production of \ttbar pairs
at the Tevatron.}
\label{fig:topproduction}
\end{figure}

The top quark is pair-produced in $p\overline{p}$
collisions through quark-antiquark
annihilation and gluon-gluon fusion.
The Feynman diagrams of the leading order (LO) subprocesses are shown in
Fig.~\ref{fig:topproduction}.
At Tevatron energies, the $q\bar q
\rightarrow t\bar t$ process dominates, contributing 85\% of the cross
section. The $gg \rightarrow t\bar t$ process contributes the remaining 15\%.

Within the SM, the top quark decays via the weak interaction to a $W$ boson and
a $b$ quark, with a branching fraction
$Br(t\rightarrow Wb) >$ 0.998.
The \ttbar pair decay channels are classified as follows:
the {\em dilepton channel}, where both $W$ bosons decay leptonically
into an electron or a muon ($ee$, $\mu\mu$, $e\mu$);
the {\em $\ell$+jets channel}, where one of the $W$
bosons decays leptonically and the other hadronically ($e$+jets,
$\mu$+jets); and the {\em all-jets channel}, where both $W$ bosons decay
hadronically. A fraction of the $\tau$ leptons decays
leptonically to an electron or a muon, and two neutrinos. These events
have the same signature as events in which the $W$ boson decays
directly to an electron or a muon and are treated as part of the
signal in the $\ell$+jets channel. In addition,
dilepton events in which one of the leptons is not identified
are also treated as part of the signal in the $\ell$+jets channel.
Two $b$ quarks are present in the final state of a \ttbar event
which distinguishes it from most of the background processes.
As a consequence, identifying
the bottom flavor of the corresponding jet can be used as a selection
criteria to isolate the \ttbar signal.

\subsubsection*{Template Method}
\textbf{Contributed by:~U.-K.~Yang}

The template method relies on 
reconstructed distributions of the top quark mass 
 from Monte Carlo
 for a wide range of mass values.
 The top quark mass is then extracted by comparing the reconstructed 
 top quark mass distribution 
 from data to the Monte Carlo mass templates using a likelihood fit.

 In this method, the reconstructed top quark mass ($\mreco$) 
 in each event is obtained
 by using kinematic constraints on the top quark decay products. 
 We require that both $t$ and $\bar{t}$ 
 have the same mass, and that two $W$ particles have mass
 equal to 80.42 GeV (PDG value).
 For the \ljets channel these constraints are sufficient 
 to construct $\mreco$, even though the 
 longitudinal momentum of the neutrino is not measured.
 For the dilepton channel these constraints are not sufficient 
due to the two missing neutrinos. 
 We therefore have to assume some kinematic distributions 
 based on the Standard Model when calculating the reconstructed top quark 
 mass for each event.

\paragraph{Lepton+jet channel} 
  The $\ttbar$ events in the \ljets channel are selected by requiring 
  a high-pt lepton (electron or muon), large transverse missing energy 
  ($\met$),  
  and at least four jets.
  Even though kinematic constraints on the top pair system are sufficient 
to define all four vectors of the top quark decay products, 
  we still need to figure out the correct jet-parton assignments.
  This task is very challenging, because the association between 
  partons  from the top quark decay and reconstructed jets is complicated 
  by many processes, for instance parton shower, hadronization, 
  and jet reconstruction. In addition, the observed jet energy is not 
  precisely measured and additional jets are present in the event from
  initial and final state gluon radiation. Only 50\% of the time
  the leading four jets contain four hard-scattered partons from the top 
  quark decays. In this analysis we perform a kinematic $\chisq$ 
  fit to choose the best assignment
  and to extract the reconstructed mass $\mreco$ for each event. 
  The $\chisq$ expression is given by:
 \begin{eqnarray}
 \label{eq_chi2}
 \chisq &=&
 \Sigma_{i = \ell, 4 jets} \frac{(p_T^{i,fit} - p_T^{i,meas})^2}{\sigma_i}
 + \Sigma_{j = x,y} \frac{(p_j^{UE,fit} - p_j^{UE,meas})^2}{ \sigma_j}
 \nonumber \\
 &+& \frac{(M_{jj} - M_W)^2 }{\Gamma_W^2}
 + \frac{(M_{\ell\nu} - M_W)^2 }{ \Gamma_W^2}
 + \frac{(M_{bjj} - M_t)^2 }{ \Gamma_t^2}
 + \frac{(M_{b\ell\nu} - M_t)^2}{\Gamma_t^2}.
 \nonumber
 \end{eqnarray}
 where $\sigma_\ell$ and $\sigma_{jet}$ are the resolutions
 of the lepton and four leading jets, and $p_{x,y}^{UE}$ and
 $\sigma_{x,y}$ are corresponds to the unclustered energy in the calorimeter.  
 The jet energies are corrected to the parton-level.
 In each event there are 12 combinations for jet-parton assignment.
 We pick the combination with the lowest $\chisq$ as the best assignment. 
 An additional requirement of $\chisqmin < 9$
 is found to give the best expected statistical uncertainty 
 on the top quark mass. This requirement effectively rejects 
 badly reconstructed $\ttbar$ events or background events). 

\begin{figure}[h]
\begin{center} 
 \includegraphics[height=.38\textheight]{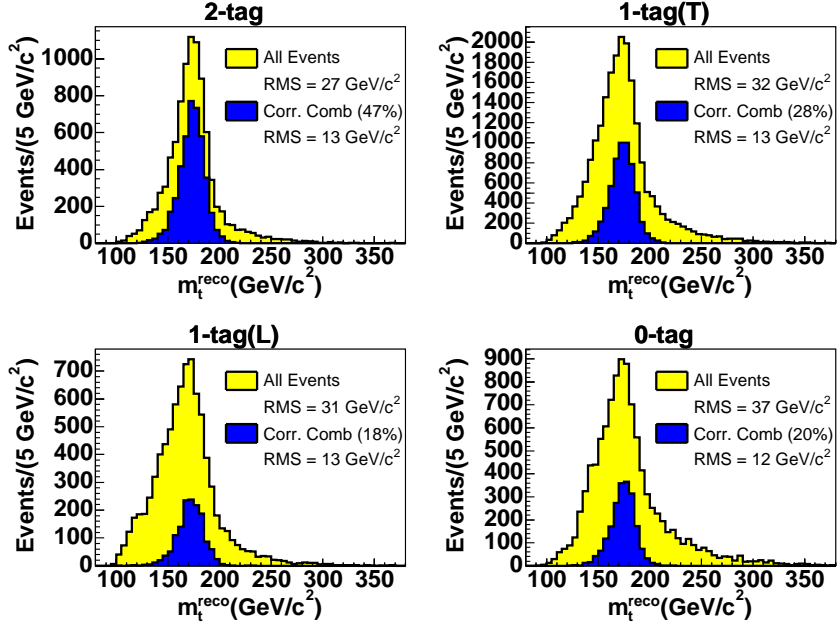}
 \caption[\gevcc{178} reconstructed mass templates.]
 {The light histograms show the reconstructed top quark mass
 distribution for the $\gevcc{178}$ \HW $t\bar t$ sample,  
 and the blue histogram for the correct jet-parton assignment.}
 \label{fig:tmp175}
  \end{center}
 \end{figure}

 Information from $b$ tagging is very powerful in finding the 
 correct combination.
 To improve the statistical power of the measurement, CDF divides the 
 sample based
 on the number of tagged $b$ jets (0, 1, and 2-tags) whereas D0 uses
 only events with tagged $b$ jets.
 A typical reconstructed top mass distribution for signal Monte 
 Carlo (178 GeV sample) is shown in Fig.~\ref{fig:tmp175}. 
 The blue histogram in the same figure shows the case
 for the correct jet-parton assignment. As can be observed, 
 the resolution of the reconstructed 
 mass is much better with more $b$-tagged jets.

\begin{figure}
\begin{center} 
\includegraphics[width=.48\columnwidth]{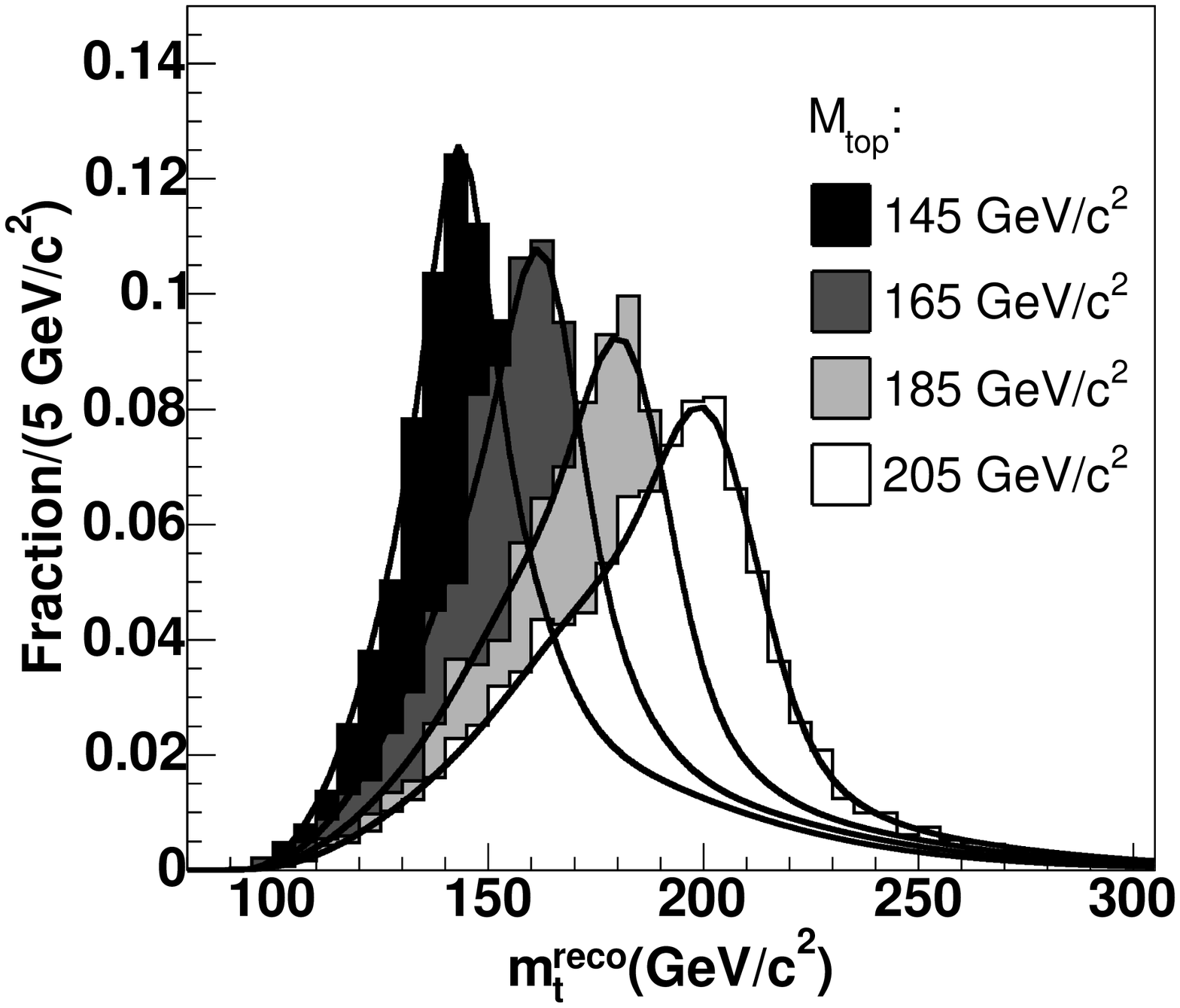}
\includegraphics[width=.48\columnwidth]{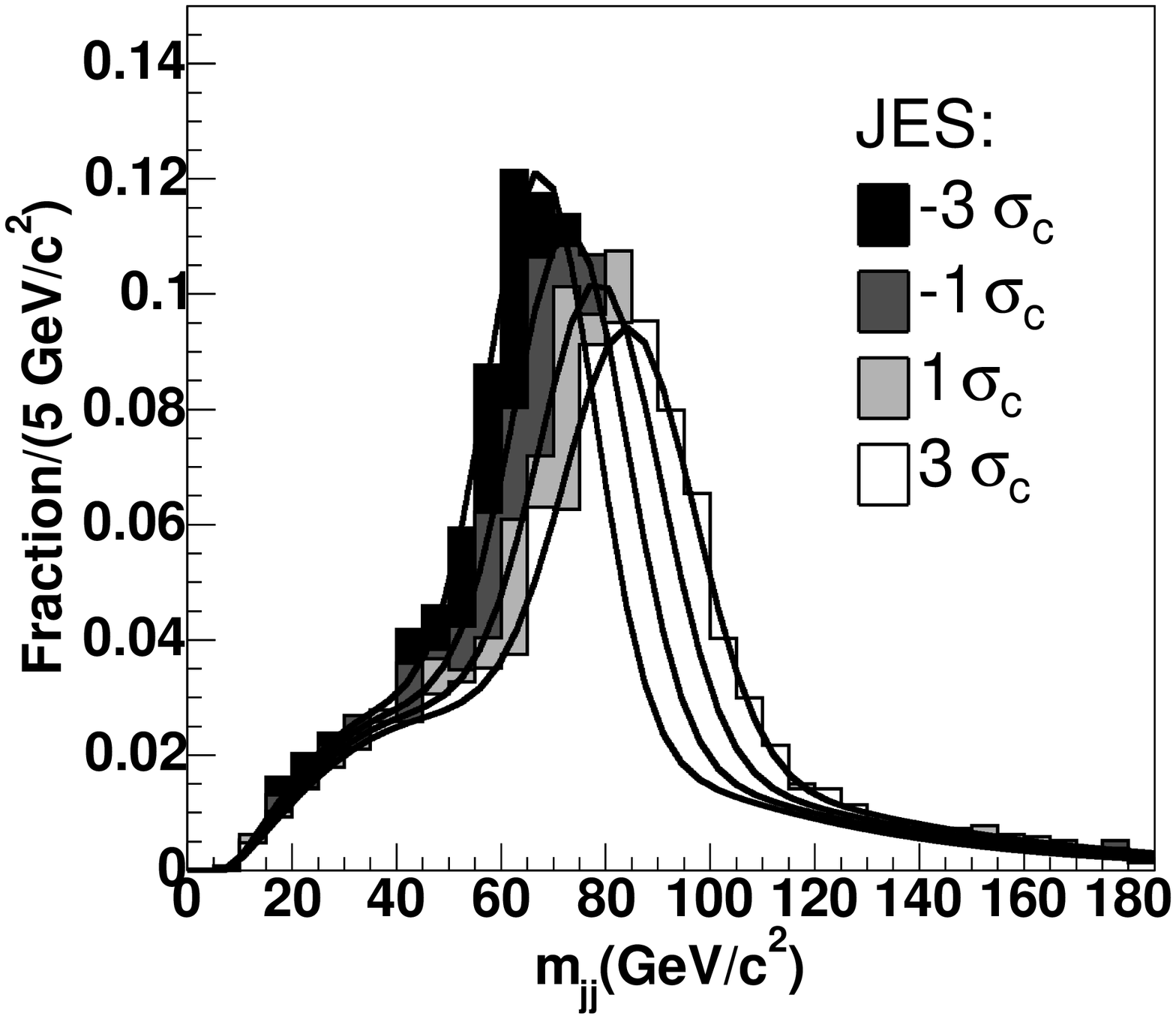}
\caption[Signal and Wjj templates.]
{[Left] Signal $\mreco$ templates for 1-btag(T) sample are shown with top quark masses ranging 
from $\gevcc{145}$ to $\gevcc{205}$ and with $\jes$ set to 0. [Right] Signal $\mjj$ templates
for the 2-btag sample are shown with different values of the $\jes$.}
\label{fig:sig_tmp}
 \end{center}
\end{figure}

 The uncertainty in the jet energy scale is the dominant systematic error on 
 the determination of the top quark mass. 
 We use the dijet mass $\mjj$ from hadronic $W$ boson decay 
to reduce this error. 
 The quantity $\mjj$ is sensitive to the jet energy scale but is 
 relatively insensitive 
 to the true top quark mass. Thus, we can calibrate the jet energy scale
 in situ while reconstructing the top quark mass. CDF has used
 both the $\mjj$ templates and the a priori determination of $\jes$ described 
 in Sec.~\ref{sec:CDFjes}. All pairs of untagged jets are used
 to get the best sensitivity to the jet energy scale. 
 Parameterized signal templates
 for the $\mreco$ and $\mjj$ are shown in Fig.~\ref{fig:sig_tmp}.
 In the tagged samples, the size of backgrounds is small. 
 Most of the background
 comes from $W$ boson production associated with real heavy 
 flavor jets, or associated jets 
 with a misidentified $b$-jet (mistags), 
 and QCD backgrounds due to fake leptons. 
 Background templates for the $W+$ jets with heavy flavor 
 production and mistags 
 are obtained from ALPGEN Monte Carlo samples. 
 The mistag template is also used for the QCD background, 
 because the non-isolated lepton data
 (QCD enriched sample) shows a very similar shape to the mistag sample.

 The reconstructed mass distribution from data is finally compared to parameterized signal templates 
 for different values of top quark mass and jet energy scale, and background templates
 using an unbinned likelihood fit. Gaussian constraints on the prior jet energy scale
 and expected background rate are used. Thus, the likelihood fit to the data returns 
 the number of signal events, the true top quark pole mass and the jet energy scale.
 This simultaneous fit to the top quark mass and the jet energy scale 
 results in significant reduction of the total uncertainty 
 as more data is added to the analysis because 
 the dominant systematic uncertainty, 
 the jet energy scale, is part of the statistical error.
 Currently, the template method used by both CDF and D0 treats all events 
 equally regardless of the different mass resolution in each event. 
 We might be able to improve 
 the resolution on the top quark mass by introducing a weight to each event. 
 
\paragraph{Dilepton channel} 
 Reconstruction of the top quark mass $\mreco$ in the dilepton channel 
 is difficult because much of the final state kinematic information is lost.
 The template method has to make kinematic assumptions on 
 unconstrained variables, and obtain the probability distribution of 
 the reconstructed top quark mass for each event.
 The most probable value of this distribution is taken as 
 $\mreco$ for each event.
 An unbinned likelihood fit is performed to parameterized 
 signal and background templates to extract a top quark mass from data, 
 like was done in the \ljets channel.
CDF has developed three template methods, 
  depending on the choice of the assumed kinematic 
  distributions. The neutrino-$\eta$ weighting method (NWA) 
  uses the $\eta$ distributions
   of the two neutrinos; the full kinematic method (KIN) 
   uses the $P_z$ of the $\ttbar$ system; 
  and the neutrino-$\phi$ weighting method (PHI) uses
 the $\phi$ of the two neutrinos.

  In the NWA method, we calculate $\mreco$ for possible solutions 
  for various 
   $\eta$ values of the neutrinos. 
  A probability for each solution ($p$) is given by the 
  measured missing energy ($\met$)
   and its resolution ($\sigma_x$, $\sigma_y$).
 \begin{eqnarray}
 p = exp \left(-\frac{(\met _x -p_x^\nu -p_x^{\bar{\nu}})^2}
                      {2\sigma_x^2}                             \right)
       exp \left(-\frac{(\met _y -p_y^\nu -p_y^{\bar{\nu}})^2}
                      {2\sigma_y^2}                             \right).
 \nonumber
 \end{eqnarray}
  The top quark mass that maximizes this probability is taken 
  as $\mreco$ for each event.
  The template method D0 has developed is similar, 
  but the probability for each solution
  is based on the prediction of the matrix element.

\subsubsection*{Kinematic Methods}
\textbf{Contributed by:~U.-K.~Yang}

 In the previous section we have shown that the 
 reconstructed top quark mass 
  $\mreco$ has a strong linear correlation with the true top quark mass. 
  However, the method relies heavily
  on the calibration of the jet energy scale. CDF has developed a 
  novel method which uses the transverse decay length of $b$-hadrons 
  from top decays to measure the top quark mass. 
  This method avoids the jet energy scale uncertainty as it relies
  on measurements by the tracking system. 

\begin{figure}[t]
\begin{center} 
 \includegraphics[height=.38\textheight]{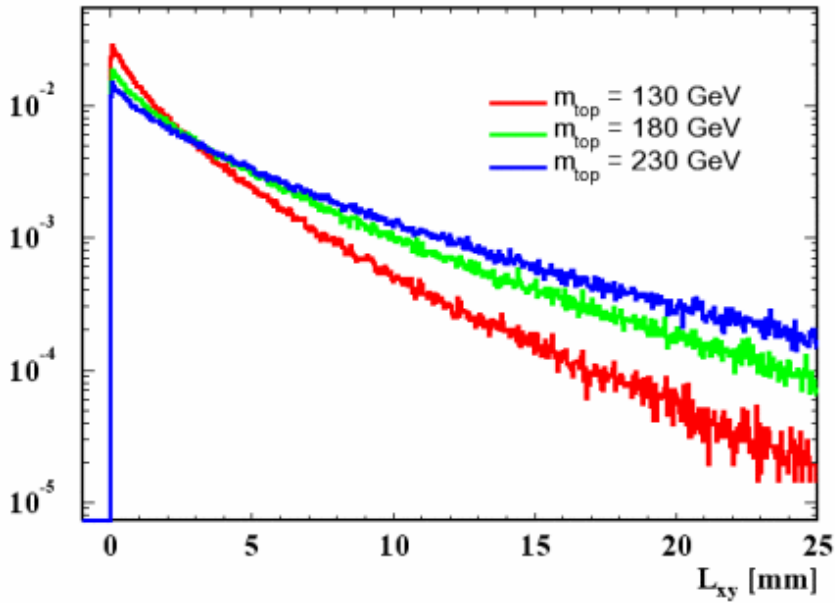}
 \caption[$L_{xy}$ distribution]
 {The $L_{xy}$ distributions for three different top mass values are shown.}
 \label{fig:lxy_mtop}
  \end{center}
 \end{figure}

  In the rest frame of the top quark, the boost factor ($\gamma$) to the $b$ quark 
  from the top decay can be written as
 \begin{eqnarray}
 \gamma = \frac{m_t^2+m_b^2-m_W^2}{2 m_t m_b} \sim 0.4 \frac{m_t}{m_b}.
 \nonumber
 \end{eqnarray}
  We can see that the top quark mass is strongly correlated with $\gamma$
  as long as the top quarks are produced at rest. At the Tevatron,
  top quarks are mostly produced nearly at rest 
  given that the transverse momentum of the top quark
  is small compared to its mass. 
  Thus, the average lifetime of the $b$ hadrons
  can be used to extract the top quark mass. 
  CDF used the transverse decay length
  of the $b$-hadrons ($L_{xy}$) as a measure of the lifetime of the $b$ hadrons. 
  Fig.~\ref{fig:lxy_mtop}
  shows $L_{xy}$ distributions for three different top quark masses. We can
  see that the $L_{xy}$ distribution has good sensitivity to the top quark mass.
 Because this method requires only a tagged $b$-jet from top quark decays,
 events with three jets in the \ljets are included.
 Dilepton events can be easily included in the method, which we plan to
 do in the near future.

 The transverse decay length $L_{xy}$ is obtained using 
 the secondary vertex 
 algorithm (SecVtx). Once SecVtx finds a secondary vertex, 
 $L_{xy}$ is calculated
 as the projection of the secondary vertex position to the jet axis. 
 This method requires an efficient SecVtx algorithm
 and an accurate simulation of the $L_{xy}$, which has been tested 
 using a heavy-flavor enriched data sample (mainly $\bbbar$ ). 
 CDF finds good agreement 
 in the average value of $L_{xy}$ within 1.4\% between data and 
 simulated events.
 The average values of the $L_{xy}$ distributions are calibrated
 for various true top quark mass values, including contributions 
 from backgrounds. The top quark mass is obtained by a simple fit to the 
 average value of $L_{xy}$ from the data.
 Currently the source of the largest systematic error at CDF 
 comes from inaccurate simulation of $L_{xy}$ including 
 imprecise knowledge of $b$-hadron lifetimes.



\subsubsection*{Matrix Element Method}
\textbf{Contributed by:~F.~Canelli and F.~Fiedler}

\noindent Both CDF and D0 have implemented methods to extract the maximum
possible information on the top quark mass from their limited
\ttbar event samples and thus minimize the (statistical and overall) error.
In these measurements, a probability density as a function of the
assumed top quark mass is calculated for each individual observed
event by evaluating the differential cross-sections for production of
top-antitop pairs of a given mass and for production of background
events~\cite{Kondo:1988yd,Abazov:2004cs}.
The probability densities from all events are combined into one
probability for the event sample, from which the value of the top
quark mass is extracted.
If the probability is calculated not only as a function of the assumed
top quark mass, but also of the jet energy scale, both parameters can
be measured simultaneously.
Both CDF and D0 have reported a measurement in the \ljets channel
using the matrix element (ME) method~\cite{bib:CDFme,bib:D0me}.
In addition, CDF have applied the ME method to the dilepton
channel~\cite{Abulencia:2005uq,bib:CDFmedil}, and CDF have measured the top quark mass
in the \ljets channel using the dynamical likelihood 
method (DLM)~\cite{Abulencia:2005ak}.

In general, the probability density $P_{evt}$ for one event to be
observed in the detector can be expressed as the sum of 
probability densities \psgn for signal and $\pbkg^i$ for $n$
background processes as
\begin{equation}
  \label{eq:MEpevt}
    P_{evt}
  = 
      f_{\rm sgn} \psgn
    + \sum_{i=1}^{n} f_{\rm bkg}^i \pbkg^{\, i}
  \ .
\end{equation}
Here, $f_{\rm sgn}$ is the signal fraction of the event sample, and
the $f_{\rm bkg}^i$ denote the fractions of events from the background
sources, where $f_{\rm sgn}+\sum_{i=1}^{n} f_{\rm bkg}^i = 1$.
The probability density for a given partonic final state to be
produced in the hard scattering process is proportional to the
differential cross-section ${\rm d}\sigma_{\rm hs}$ of the
corresponding process.
The differential cross-section for \ttbar production will depend on
the assumed top quark mass.
To obtain the differential cross-section in \ppbar collisions, the
differential cross-section for the hard scattering process has to be
convoluted with the parton density functions (PDF) of the proton and
antiproton.
The finite detector resolution is taken into account via a convolution
with transfer functions (TF) that describe the detector response.
These transfer functions are derived from Monte Carlo simulated
events.

For a measured event $x$, the signal probability density as a function
of assumed top quark mass $m_t$ becomes
\begin{equation}
  \label{eq:MEpsgn}
    \psgn(x;m_t)
  = 
    \frac{1}{N}
    \sum_{\rm comb}
    \int\limits_{q_1,q_2,y} 
    \sum_{\rm flav} 
    {\rm d}q_1 {\rm d}q_2 f_{\rm PDF}(q_{1}) f_{\rm PDF}(q_{2}) 
    {\rm d}\sigma_{\rm hs}(y;m_t)
    T\!F(x,y)
\end{equation}
(similarly for the backgrounds).
Here, ${\rm d}\sigma_{\rm hs}(y;m_t)$ denotes the differential hard
scattering cross-section for \ttbar production, and $T\!F(x,y)$ is the
probability to observe $x$ in the detector when $y$ was produced.
A sum over all flavors (flav) of colliding partons has to be
performed, including the relevant PDFs.
The integration is over the entire 6-particle phase space of all
possible partonic final states $y$ that could have led to the event
$x$, and over the momentum fractions $q_i$ of the colliding partons
inside the proton/antiproton.
The integration is performed numerically, and assumptions on the
detector response (e.g.\ good lepton momentum resolution compared to
the jet energy resolution) allow to reduce the dimension of the
integration space.
The quantity $N$ ensures that the probability is normalized.
The sum over jet-parton assignments (comb) is discussed below.

The event selection for the \ljets analyses (ME and DL) requires an
energetic isolated charged lepton (electron or muon), missing
transverse energy, and exactly four hadronic jets.
The reconstructed jets in the detector cannot be assigned
unambiguously to the partons described by the differential
cross-section.
Without the identification of $b$ jets, there are 24 possible
assignments of jets to partons.
In events with identified $b$ jets, this number (and also the 
fraction of background events) is reduced.
For the ME measurement in the dilepton channel, events with two
energetic charged leptons, missing transverse energy, and two hadronic jets
are selected, which amounts to 2 possible jet-parton assignments
per event.
All relevant possibilities for assignment of jets to partons are taken
into account as indicated in Eq.~(\ref{eq:MEpsgn}).

In the D0 and CDF ME measurement in the \ljets channel, \ttbar
production is described with the leading order matrix element, and
\wjets background is described using matrix-elements from subroutines 
of the Vecbos Monte Carlo generator, while QCD multijet background is
not handled explicitly in the probability calculation.
Jet and charged lepton angles as well as electron energies are assumed
to be well-measured in the probability calculation.
A likelihood function is determined for the event sample as a function
of top quark mass, jet energy scale, and of the parameter $f_{\rm
sgn}$ defined in Eq.~(\ref{eq:MEpevt}).
The event selection and jet energy scale are taken into account in the
normalization of the signal probability, and the background
probability normalization is determined such that the parameter
$f_{\rm sgn}$ reproduces the \ttbar fraction in the event sample.
The top quark mass and jet energy scale are then determined in a fit
to the likelihood.

For the CDF ME measurement in the dilepton channel, also the leading
order matrix element is used.
The background considered for this measurement are Drell-Yan
production with extra jets, \W pairs with jets, and single \W
production with jets one of which is misidentified as a lepton.
So far, the jet energy scale uncertainty is treated as an external
error.

In the dynamical likelihood (DL) technique used by the CDF
collaboration in the \ljets channel, the integration over all possible
partonic final states is performed with a Monte Carlo technique, where
the mass of the leptonically decaying \W boson is generated according
to the Breit-Wigner form, and parton energies according to a transfer
function.
Backgrounds are then not treated explicitly in the likelihood calculation;
instead, the measured top quark mass is corrected for the effect 
of presence of background in the event sample.

\subsubsection*{Ideogram Method}
\textbf{Contributed by:~M.~Weber}
\newcommand{\mmm}{\ensuremath{x_{\rm fit}}\xspace}
\newcommand{\barq}{\ensuremath{\bar{q}}\xspace}
\newcommand{\barb}{\ensuremath{\bar{b}}\xspace}
\newcommand{\mtop}{\ensuremath{m_{\rm top}}\xspace}
\newcommand{\fsgn}{\ensuremath{f_{\rm sgn}}\xspace}
%
\def\pt{$p_T$}                          
\def\et{$E_T$}                          
\def\met{\mbox{${\hbox{$E$\kern-0.6em\lower-.1ex\hbox{/}}}_T$}} 
\def\htran{$H_T$}                       
\def\aplan{$\cal{A}$}                   
\def\htp{$H_{T2}'$}                     
\def\hp{$H_{\parallel}$}                  
\def\ktmp{$K_{Tmin}'$} 
\def\mw{$M_W$}                          
As in the Matrix Element analyses a likelihood is calculated for
each event as a function of the assumed top quark mass taking into account all possible jet assignments and the
probability that the event was signal or background. 
The approach is very similar to a technique, which was used by the
DELPHI experiment~\cite{Eur.Phys.J.C2.581,Phys.Lett.B462.410,Phys.Lett.B511.159,Mulders:2001aa,Mulders:2001mp} to extract the mass of the W boson at
LEP.
As in the Matrix Element method the likelihood is described as a convolution of a
physics function and the detector resolution. The difference, however,
is that in the ideogram method a kinematic constrained fit is used to 
describe the detector resolution, and the physics function is
simplified to a relativistic Breit-Wigner describing the average of
the invariant masses of the supposed top and anti-top quark that were
produced in the event.
The ME methods are based on matrix element integrations which require significant computing resources.
The approximations of the signal and background probability functions used in the ideogram method result in approximately a factor 1000 faster processing times.
This is a major technical advantage of the ideogram method especially considering running an analysis multiple times for systematics evaluation and parameter optimization.
The probability $P_{evt}$ is the same as for the Matrix Element methods in equation \ref{eq:MEpevt}.
$P_{sgn}$ and $P_{bkg}$ are functions of the full set of observables that characterize the event $x$.
The event observables $x$ can be divided in two groups.
One set was chosen to provide good separation between signal and background events while minimizing the correlation with the mass information in the event.
These topological variables are used to construct a discriminant $D$.
The other event information used is the mass information \mmm from 
the constrained kinematic fit, which will give the sensitivity to the top mass.
The variables used in the low-bias discriminant $D$ are the same as developed in Run I \cite{run1prdmass}.
The first variable $x_1 \equiv \met$ it the missing transverse energy.
The second variable  $x_2 \equiv \cal{A}$ is the aplanarity, which is the least eigenvalue of the laboratory normalized momentum tensor of the jets and the $W$ boson. 
$x_3 \equiv \frac{H_{T2}}{H_{\parallel}}$ measures the event centrality, where $H_{\parallel}$ is the scalar sum of $\left|p_{z}\right|$ of the jets, isolated lepton, and the neutrino. $H_{T2}$ is the sum of the $\left|p_{T}\right|$ of the jets
excluding the leading jet.
$x_4  \equiv \frac{\Delta R^{min}_{jj} \cdot E_{T}^{lesser\ j}}{E_{W}^{T}}$ is a measure of the jet separation folded together with the transverse energy of the reconstructed W. 
$\Delta R_{ij}$ is the least distance in $\eta-\phi$ space between any two of the four leading jets.
$E_{T}^{lesser j}$ is the smaller of the two jets \et\ 's.
The transverse energy of the W is defined as the sum of $\left|p_{T}^{l}\right|$ and $\left|p_{T}^{\nu}\right|$.
For each variable $x_i$ we determine the probability density functions $s_i$ for \ttbar signal and $b_i$ for \wjets background from MC.
We assume these to be nearly uncorrelated and we write $s = \prod_{i}{s^{w_i}_i}$ and $b = \prod_{i}{b^{w_i}_i}$.
With the weights $w_i$ slightly adjusted away from unity for $x_{1,2,3,4}$ the correlation to the top quark mass was nullified.
A discriminant $D$ is built from $s({\boldmath x})$ and $b({\boldmath x})$ as:
$$
D({\boldmath x}) = \frac{s({\boldmath x})}{s({\boldmath x})+ b({\boldmath x})}
$$
We do use a parametrized form for $D$ where the ratios $s(x_i)/b(x_i)$ were parametrized with 
polynomial fits~\footnote{Additional 
transformations to the $x_i$s before the fit were done for the 
functions to be better approximated by polynomials: 
$x_1'=exp[-(max(0,\sqrt{\frac{3(x_1/(1 \mathrm GeV)-5)}{2})]}$, 
$x_2'=\exp(-11x_2)$, $x_3' = \ln(x_3)$, $x_4'=\sqrt{x_4}$.}.
%
The fitted mass information \mmm is a set of kinematic variables, calculated from a constrained kinematic fit to the reconstructed jets, lepton and missing transverse energy.
The procedure explained in section~\ref{sec:detdescr} 
corrects the measured jets for the portion of the showers which spread outside the jet cone, but not for any radiation outside the cone.
We do further correct the jet energies to that of the fragmented partons in the MC.
To derive this correction we use MC events where the jets could be
matched to the partons of the \ttbar decay and compare the jet energy
to the parton energy information from the MC.
The constrained fit technique is the same as used in the D0 Run I template mass analysis \cite{run1prdmass}.
The fit is performed by minimizing a  $\chi^2$ defined as:
\begin{equation}
\chi^{2} = (\vec{x}-\vec{x}_{M})G(\vec{x}-\vec{x}_{M})^{T}
\end{equation}
where $\vec{x}_{M}$ is a vector of measured variables, $\vec{x}$ is a vector
of fitted variables, and $G$ is the inverse error matrix of the measured
quantities. $G^{-1}$ is taken to be diagonal.
The $\chi^2$ is minimized subject to the kinematic constraints
$m(t\rightarrow l\nu b) = m(\bar t\rightarrow q\barq\barb)$, $m(l\nu)=M_W$ and $m(q\barq) = M_W$.
The minimization algorithm uses the method of Lagrange Multipliers; the nonlinear constraint equations are solved using an iterative technique.
The fitted mass $m_{\mathrm fit}$ and its uncertainty $\sigma_{\mathrm fit}$ are taken at the minimum of the $\chi^2$.
For every event we run the kinematic fitter for each of the 12
possible permutations $i$ of assigning the 4-momenta of the
reconstructed jets to the partons $(\bbbar \qqbar)$.
The $W$-boson mass constraint on the leptonic side can result in a twofold ambiguity on the neutrino longitudinal momentum $p_\mathrm{z}$.
Both cases are considered and the fit is repeated for each initial guess.
To good approximation $D$ and \mmm are uncorrelated, and the \psgn and \pbkg probabilities can be written as the product of a probability to observe a value $D$ and a probability to observe \mmm:
%
\begin{equation}
  \label{eq:idsigfactorised}
    \psgn\left(x;\,\mtop\right) \equiv \psgn\left(D\right)\psgn\left(\mmm;\,\mtop\right)
\end{equation}
and
\begin{equation}
  \label{eq:idbgfactorised}
    \pbkg(x) \equiv \pbkg\left(D\right)\pbkg(\mmm)
\end{equation}

The normalized probability distributions of the $D$ discriminant for signal
$\psgn\left(D\right)$ and
background $\pbkg\left(D\right)$ are obtained from Monte Carlo simulation.
%
The permutations are weighted using weights $w_i$, which estimate the relative probability for a
certain jet permutation to be the correct one. 
The relative probability of each jet assignment $w_i$ purely depends on the
$\chi^2_i$ for the corresponding fit and is calculated as $ w_i = {\rm exp}(-\frac{1}{2} \chi^2_i)$.
The signal term in Eq.~\ref{eq:idsigfactorised} is calculated as 
\begin{equation}
\psgn\left(\mmm;\,\mtop \right) \equiv \sum^{24}_{i=1} w_i 
\left[\int^{300}_{100} {\bf G}(m_i,m',\sigma_i) \cdot {\bf BW}(m',m_{\rm t}) dm' \right]
\end{equation}
and the background term:
\begin{equation}
\pbkg (\mmm) \equiv \sum^{24}_{i=1} w_i \cdot BG(m_i)
\end{equation}

The signal term consists of the
compatibility of the solution with a certain value of
the top mass, taking
into account the estimated mass resolution $\sigma_i$ for each jet
permutation.
This is given by a convolution of a 
Gaussian resolution function ${\bf G}(m_i,m',\sigma_i)$ describing the 
experimental resolution with a relativistic Breit-Wigner 
${\bf BW}(m',m_{\rm t})$, representing the expected distribution of
the average of the two invariant masses of the top and anti-top quark
in the event, for a top mass $m_{\rm t}$.

For the background term a weighted sum 
${\rm BG}(m_i)$ is used, where BG(m) is the shape of the mass spectrum 
obtained from W+jets in MC simulation with all entries weighted
according to the permutation weight $w_i$ assigned to
each solution.
The Breit-Wigner and other permutation signal shape are normalized to unity on 
the integration interval: 100 to 300 GeV. This 
interval was chosen large enough not to bias the mass in the region of
interest. 
Since each event is independent the combined likelihood for the whole sample is calculated as the product of the single event likelihood curves:
$$
{\cal L}_{\rm samp}(m_{\rm t},\fsgn) = \prod_j  {\cal L}_{{\rm evt} j}(m_{\rm t},\fsgn)
$$
This likelihood is maximized with respect to the top mass $m_{\rm t}$ and the estimated fraction of signal in the sample \fsgn. 

%
%





\subsection{Systematic Uncertainties}
\label{sec:MtSyst}
\textbf{Contributed by: F.~Canelli, F.~Fiedler, M.~Weber, and U.-K.~Yang}

Systematic uncertainties arise from the modeling of physics processes
and from the simulation of the detector.  These two sources are 
described in the two following sections.

\subsubsection*{Physics Modeling}
\label{sec:physmod}

\paragraph{Signal Modeling:}
When $\ttbar$ events are produced in
association with a jet, the additional jet can be misinterpreted as
a product of the \ttbar decay.  
Such events are present in the simulated events used for the 
calibration of the method. We tuned the initial and final state
gluon radiations in \PY  by using the transverse momentum of Drell-Yan events
and extrapolated to the $Q^2$ region of the $\ttbar$ production. 
Uncertainties on the extra jets are estimated based on this tuning.
The difference between \ttbar cross-sections calculated
at leading and next-to-leading order is also used to estimate 
abundance of such events.
To assess the uncertainty in the modeling of these effects, 
their fraction is varied in the 
simulation.  
Also, the relative cross-section of the processes
$gg\to\ttbar$ and $\qqbar\to\ttbar$ is varied.

\paragraph{Background Modeling:} 
The main background in the lepton+jets channel is due to the
production of jets in association with a leptonically decaying $W$.
In order to study the sensitivity of the
measurement to the choice of background model, the 
factorization scale of $Q^2 = m_{\W}^{2} + \sum_{j} p_{T,j}^{2}$ 
used in the modeling of \wjets events is 
replaced by ${Q'}^2 = \left<p_{T,j}\right>^{2}$. 

\paragraph{PDF Uncertainty:} 
To study the systematic uncertainty on the top mass due to the choice
of PDF used to simulate signal and background events, the
variations provided with the next-to-leading-order PDF set 
CTEQ6M~\cite{Pumplin:2002vw} are used.  The 
result obtained with each of these variations is compared with
the result using the default CTEQ6M parametrization. 
The difference between the results obtained with the CTEQ
and MRST PDF sets is taken as another uncertainty.
Finally, the value of $\alpha_s$ is varied.
All errors are added in quadrature.

\paragraph{Bottom Fragmentation and Semileptonic Decays:} 

The estimate of the jet energy scale 
from a priori information and from $W \rightarrow j j$ decays do not give
direct information on the $b$-jets energy scale. The $b$-jets can behave 
differently from gluon and light quark jets because of their different fragmentation models, 
more abundant semi leptonic decays and different color flow in \ttbar events than $W$-daughter jets.
However, we find that a major uncertainty on the $b$-jet energy scale comes from common features
of the generic jets.
We study simulated $\ttbar$ events with different fragmentation models
for $b$-jets due to the choice of the model.

The reconstructed energy of $b$-jets containing a semileptonic
bottom or charm decay is in general lower than that of jets 
containing only hadronic decays.  This can only be taken into account for
jets in which a soft muon is reconstructed.  Thus, the fitted top
quark mass still depends on the semileptonic $b$ and $c$ decay
branching ratios.  They are varied within the bounds given 
by the LEP results~\cite{bib:Zbible}.

\subsubsection*{Jet Energy Scale}
\label{sec:detdescr}

Since the measurement of the top quark mass requires the determination
of the four-momenta of quarks which relies on the reconstruction of
hadronic jets resulting from fragmentation, the dominant systematic
uncertainty comes from our measurements of the jet energies.

At CDF and D0, jets are observed as clustered energy depositions in
the calorimeters. Both experiments use a cone algorithm defined with a
radius of $R_{jet}$=0.4 and $R_{jet}$=0.5 for CDF and D0,
respectively.  Measured jet energies are corrected to best describe
particle jets or partons energies. The accurate modeling of the
detector response as well as a good understanding of the fragmentation
process is an essential requirement for these corrections.

In the following, we describe the corrections to the measured jet
energy and the determination of the overall jet energy scale in CDF
and D0.  A more detailed explanation can be found
in~\cite{Abulencia:2005aj,Bhatti:2005ai} and~\cite{Abbott:1998xw}.

The overall jet energy scale is the dominant systematic uncertainty on
top quark mass measurements in the lepton+jets channel unless it is
determined simultaneously (``in situ'') with the top quark mass from
the same event sample.  Both CDF and D0 have shown such analyses
with a simultaneous measurement of the top quark mass and overall jet
energy scale.  But even for such an in situ calibration, systematic
errors still arise from the possible dependence of the jet energy
scale on the energy itself or on the position in the calorimeter.

\paragraph{CDF Jet Energy Scale}
\label{sec:CDFjes}

CDF uses the Monte Carlo simulation to determine the jet energy scale
allowing to correct an energy range from 8 GeV to 600 GeV. Therefore,
the major task involved in the determination of the jet energy scale
is the tuning and validation of the detector simulation as well as of
the physics modeling used in the simulation.

Before corrections are derived, the energy scale for the
electromagnetic calorimeter is set using electrons from the decay
$Z\rightarrow e^+e^-$ and the energy scale for the hadronic
calorimeter is set to the test-beam scale of $50$~GeV/$c$ charged
pions.

The corrections are divided in different levels to allow many
different analyses in different groups to use them and to create an
experiment-wide definition of jet energies. Firstly, measured jets are
corrected for all instrumental effects to a particle-level jet which
corresponds to the sum of the momenta of the hadrons, leptons, and
photons within the jet cone. Particle-level jets are then corrected to
parton level energies.

Since the simulation is used to correlate a particle jet to a
calorimeter jet a detailed understanding of the detector is needed.
The simulation is tuned to model the response of the calorimeter to
single particles by comparing the calorimeter energy measurement, $E$,
to the particle momentum, $p$, measured in tracking detectors.  Here,
measurements based on both test beam and CDF data taken during Run II
are used. The calorimeter simulation is most reliable in the central
part of the calorimeters since the tracking coverage in the forward
regions is limited. Therefore, the forward calorimeter jet response is
calibrated with respect to the central, to flatten out the jet
response versus the jet polar angle. This procedure also corrects for
the lower response in poorly instrumented regions of the
calorimeters. After tuning the simulation to the individual particles
response and achieving a jet response independent of the polar angle,
calorimeter jets are corrected to a particle jet, i.e. they are
corrected for the central calorimeter response. Since the correction
is derived from simulation, it is also important to ensure that the
multiplicity and momentum spectrum of particles in the data is well
reproduced by the simulation.

A further correction is made for pile-up of additional $p\bar{p}$
interactions. This pile-up can lead to an overestimate of the jet
energy if particles produced in the additional interactions happen to
overlap with those produced in the hard scattering process. Similarly,
the jet energy is also corrected for particles from the underlying
event, i.e.  interactions from spectator quarks and initial state QCD
radiation.

Since the jet cone is of finite size some particles originating from
the initial parton may escape from the jet cone either in the
fragmentation process or due to parton radiation. The out-of-cone
energy is measured in MC events. Depending on the analysis different
corrections are used. For matrix element based analyses these
corrections correspond to the transfer functions. Here, the full shape
of the mapping between particle jets and parton energies is used. The
template analysis uses an average correction of this mapping also
obtained from $t\bar t$ \HW MC.

The original parton transverse energy is estimated by correcting the
jet for all the above effects:
\begin{equation}
\ptparton=(\ptcaljet \times C_{\eta} - C_{MI}) \times C_{Abs} - C_{UE} + C_{OOC}=\ptpartjet - C_{UE} + C_{OOC}
\end{equation}
where $\ptparton$ is the transverse momentum of the parent parton the
procedure is aimed at, $\ptcaljet$ is the transverse momentum measured
in the calorimeter jet, $\ptpartjet$ is the transverse momentum of the
particle jet. The different factors in the corrections are:
$C_{\eta}$, ``$\eta$-dependent'' correction, ensures homogeneous
response over the entire angular range; $C_{MI}$, ``Multiple
Interaction'' correction, is the energy to subtract from the jet due
to pile-up of multiple $p\bar{p}$ interactions in the same bunch
crossing; $C_{Abs}$, ``Absolute'' correction, is the correction of the
calorimeter response to the momentum of the particle jet. Particle
jets can be compared directly to data from other experiments or
theoretical predictions which include parton radiation and
hadronization. $C_{UE}$ and $C_{OOC}$, the ``Underlying Event'' and
``Out-Of-Cone'' corrections, correct for parton radiation and
hadronization effects due to the finite size of the jet cone algorithm
that is used. Note that the $C_{UE}$ and $C_{OOC}$ corrections are
independent of the experimental setup, i.e. the CDF detector
environment. All the correction factors are determined as a function
of the jet transverse momentum but they apply to all components of the
four-momentum of the jet.

Various cross-checks using different physics processes ($\gamma$+jet,
$Z$+jet, $W$+jet) are done to validate the universality of the
procedure and verify the systematic uncertainties.

\begin{figure}[h]
 \begin{center}
 \includegraphics[width=0.6\linewidth,clip=]{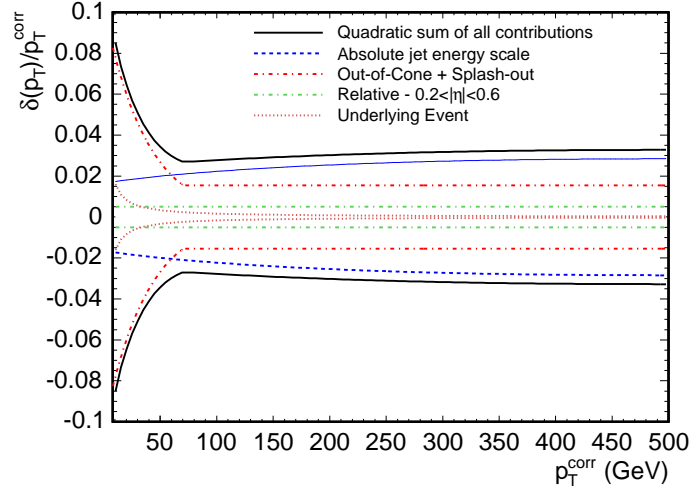}
 \caption{CDF jet energy scale systematic uncertainties as a
 function of the corrected jet $p_{T}$ in
 0.2$<|\eta|<$0.6. \label{fig:cdfjes}} \end{center}
\end{figure}

The systematic uncertainties take into account any differences
observed between the data and the simulation and possible systematic
biases in the procedure used to determine the corrections. Data and
Monte Carlo are compared in every step of the correction procedure,
and the uncertainties are added in quadrature. The final systematic
error on the jet energy scale is shown in Fig.~\ref{fig:cdfjes}. The
total systematic uncertainty on the jet energy scale varies between
8\% at low jet $p_T$ and 3\% at high jet $p_T$.  The systematic
uncertainties are largely independent of the corrections applied and
mostly arise from the modeling of jets by MC simulation and from
uncertainties in the calorimeter response to single particles.

For $p_T>60$ GeV/$c$ the largest contribution arises from the absolute
jet energy scale which is limited by the uncertainty of the
calorimeter response to charged hadrons. A further reduction of the
systematic uncertainties can be achieved by improving the tuning of
the simulation, and by including {\it in situ} single track data which
recently became available, replacing test beam data used so far in the
momentum region 7-20 GeV/$c$ and probably beyond.

At low $p_T$ the largest uncertainty arises from the out-of-cone
energy which can be improved by further studying differences between
the data and the predictions of \PY and \HW, and by
optimizing the fragmentation and underlying event model of both
generators.

\paragraph{D0 Jet Energy Scale}

\label{sec:D0jes}
The measured energy of a reconstructed jet is given by the sum of energies
deposited in the calorimeter cells associated with the jet by a cone
algorithm.
Several mechanisms cause this energy estimate to deviate from the
energy of the particle level jet:
\begin{itemize}
\item
  {\bf Energy Offset:} Energy in the clustered cells which is
  due to noise, underlying event, multiple interactions, energy
  pile-up, and uranium noise lead to an offset $E_{O}(R, \eta, {\cal L})$ 
  of jet
  energies. $E_{O}$ is determined from energy densities in minimum bias
  events.
\item 
  {\bf Calorimeter Response:} Jets consist of different
  particles (mostly photons, pions, kaons, protons and
  neutrons), for which the calorimeter response is
  different. Furthermore, the calorimeter responds slightly
  non-linearly to particle energies. The response 
  $R_{jet}(E_{jet}^{meas}, \eta)$ is determined with
  $\gamma$+jets events requiring transverse momentum balance. The
  photon scale is measured independently with high precision in $Z\rightarrow ee$ events.
\item 
  {\bf Showering Corrections:} Not all particles deposit their 
  energy within the jet cone.  The fraction $R_{cone}(R, E_{jet}^{meas}, \eta)$ 
  deposited inside the 
  cone of radius $R=\sqrt{(\Delta\eta)^2 + (\Delta\phi)^2}$ 
  is obtained from jet energy density profiles.
\end{itemize}
Consequently, the corrected particle level jet energy $E_{jet}^{corr}$ 
is obtained from the measured reconstructed
jet energy $E_{jet}^{meas}$ as
\begin{equation}
\label{eq:jetcorr}
E_{jet}^{corr} = \frac{E_{jet}^{meas} - E_O}{R_{jet} R_{cone}} \ .
\end{equation}

The offset energy $E_O$ is defined as the energy contribution to a jet that 
is not associated with the hard scattering process.
Contributions to the offset come from electronic noise, uranium noise,
pile-up, and energy from additional interactions underneath the interesting
physics process.
The shaping time of the D0 calorimeter readout electronics is longer
than the bunch crossing time of 396\,ns, so the signal from an earlier 
bunch crossing may contribute to the energy of the jet under consideration.
The offset energy is measured from minimum bias events, which are defined
as events triggered by the condition that the luminosity counters on both
sides of the interaction point are hit.
As a cross-check, the contribution from noise and pile-up is also measured
from events without a hard interaction.

As in Run\,I, D0 uses the missing $E_T$ projection fraction method to 
measure the calorimeter response from the $p_T$ imbalance in back-to-back
$\gamma$+jet events~\cite{Abbott:1998xw}.
For an ideal detector, the photon transverse momentum $p_T^\gamma$ and
the transverse momentum of the hadronic recoil $p_T^{had}$ are balanced.
However, because the calorimeter response to photons, $R^\gamma$, and 
hadronic jets, $R^{had}$, is different, an overall transverse momentum 
imbalance is observed:
\begin{equation}
  R^\gamma \vec{p}_T^{\,\gamma} + R^{had} \vec{p}_T^{\,had} = -\ptmissvec \ .
\end{equation}
The missing transverse momentum $\ptmissvec$ is corrected for
the electromagnetic calorimeter response $R^\gamma$, which is determined
from the position of the mass peak in $Z^0 \to e^+e^-$ events.
After that, the hadronic response is obtained as
\begin{equation}
  R^{had} = 1 + \frac{ \ptmissvec^{corr} \cdot \vec{p}_T^{\,\gamma} }
                     { (\vec{p}_T^{\,\gamma})^2 }
  \ .
\end{equation}
In events with one photon and exactly one jet, the jet response can be 
identified with the hadronic response.
The jet response is determined as a function of jet energy 
and pseudorapidity, and an additional correction is applied for jets
in the region between the central and endcap calorimeter cryostats.

Part of the jet energy may be deposited outside the jet cone because
of the finite lateral shower width and because charged particles may
be bent outside the cone by the magnetic field.
This effect is measured from energy density profiles of jets.
Because gluon emission and fragmentation processes also contribute to 
the energy density profile measurement, these effects are corrected for 
using Monte Carlo simulation.

Additional corrections are needed to reconstruct the energy of a jet
containing an identified semimuonic decay of a bottom or charm hadron.
The expected energy that the muon deposited in the calorimeter is
subtracted from the jet energy, and the muon momentum and the average
neutrino momentum added.
The average neutrino momentum has been obtained from simulated events
and is calculated as a function of the momentum of the muon and its transverse
momentum relative to the jet axis.

In each event, the missing transverse momentum is adjusted according
to the jet energy scale factors applied to all jets in the event.

The measurement technique for the top quark mass is calibrated using
Monte Carlo simulated events.
Consequently, the ratio between data and Monte Carlo of the jet energy
scale and the associated uncertainty are the relevant quantities for 
the top quark mass measurement.
Because the dependence of the jet energy scale on jet energy and 
pseudorapidity is not necessarily the same in data and Monte Carlo,
this ratio will depend on jet energy and pseudorapidity as well.
In measurements of the top quark mass at D0, these dependences are 
taken from the $\gamma$+jet measurement.
The overall data/Monte Carlo scale factor, $JES$, is either taken
from the $\gamma$+jet measurement as well, or determined
``in situ'' simultaneously with the top quark mass in measurements
using lepton+jets \ttbar events.

The jet energy scale for $b$ jets may be different from that for light
quark jets.
If the relative $b$/light quark jet energy scale is different in data 
from that in the simulation, then the measurement of the top quark 
mass is affected.
The uncertainty on this double ratio is estimated by varying the 
ratio of the calorimeter response to hadrons and electrons, and by 
varying the $b$ quark fragmentation model.
In addition, the double ratio is cross-checked with $\gamma$+jet
events where the jet is tagged by the presence of a secondary vertex.

If the overall $JES$ factor is determined simultaneously with the top
quark mass, the statistical error on the latter increases by a factor of
about 1.5.
The uncertainty on the energy dependence of the jet energy
scale measurement from $\gamma$+jet events contributes a systematic
error to the top quark mass measurement of 250 MeV~\cite{bib:D0me}.
Currently, the largest systematic uncertainty on the top quark mass
measurement with in situ $JES$ calibration comes from the knowledge
of the $b$/light quark jet energy scale ratio.
This systematic error is about 1 GeV.

\subsection{Top Mass Combination}
\label{sec:MtCombo}
\textbf{Contributed by:~D.~Glenzinski }

A world average top quark mass, \Mt, is obtained by combining the various 
Tevatron measurements.  The average is performed by the Tevatron Electroweak 
Working Group (TevEWWG) with working members from both the CDF and D0 
Collaborations. The most recent combination is described in detail in 
reference~\cite{MtSum05} and includes preliminary CDF and D0 measurements 
using about $350\:\pb$ of Run~II data.  A summary of the methodology is given 
in this section.

The combination takes into account all statistical and systematic correlations.
Measurements of \Mt\ in the lepton+jets (l+j), di-lepton (dil), and all-jet 
(all-j) channels from both CDF and D0 are combined using the analytic BLUE 
method~\cite{Lyons:1988rp,Valassi:2003mu} and cross-checked using a numerical $\chi^{2}$
minimization.  The experiments supply the inputs and the TevEWWG, in
collaboration with the experts from the experiment, specifies
the error categories.  The definition of error categories is driven by
the categories of uncertainties considered and their correlations.  For 
example, in an effort to more accurately account for the JES uncertainties 
correlated between the experiments, the JES is broken into several 
sub-categories.  The error categories are discussed in detail below.

At present, each experiment evaluates the associated systematic uncertainties
independently, often times using different techniques.  These differences can
effect the weight a particular input carries, and thus the world average \Mt\ 
as a result.   While at the present time these differences affect the average
\Mt\ at the level of $100$~\mevcs or less, the TevEWWG will focus on 
more accurately determining the intra-experiment correlations as the 
precision of the combination continues to improve.  For 
example, by specifying the methodology to use when quantifying particular 
classes of systematic uncertainties (e.g. the Signal Modeling, Background
Modeling, and JES uncertainties).  These discussions have already begun, 
although there is nothing concrete to report at this time.  Once LHC results 
become available, the precision on the world average \Mt\ may be such that 
these same specifications may also be important when including the new LHC 
results.  Thus it will be important to document any common methodologies used.

The following error categories are used when performing the \Mt\ combination:
\begin{description}
  \item{\bf Statistical:} The statistical uncertainty, calibrated to 
    correspond to $68\%$ coverage using pseudo-experiments to study the r.m.s.
    of the resulting pull distribution.

  \item{\bf Signal Modeling:} This includes modeling uncertainties related to
    ISR, FSR, PDF, and $\Lambda_{QCD}$ variations in \ttbar\ events.

  \item{\bf Background Modeling:} This includes modeling uncertainties related
    to fragmentation, $Q^{2}$, and normalization variations in background 
    events.

  \item{\bf Monte Carlo Generator:} This includes comparisons of fit biases
    introduced when using different Monte Carlo generators to simulate \ttbar\
    events.  This arguably double-counts some of the uncertainty in the
    ''Signal Modeling'' category and may be revised as we gain further
    confidence in the methodologies and variations used to quantify those
    modeling uncertainties.

  \item{\bf Fit:} This includes uncertainties from limited Monte Carlo 
    statistics, and other possible (small) residual biases related to
    the specific techniques used to determine \Mt\ for a given input.

  \item{\bf Uranium Noise:} Includes uncertainties specific to D0 Run I
    results which account for effects of noise in the Uranium
    calorimeter on the jet energy determination.

  \item{\bf In-Situ JES:} This is the uncertainty from the JES as determined
    using the mass of {\it{in situ}} $W\ra qq^{\prime}$ decays.  At this
    time this determination is completely statistics dominated and is thus
    treated as uncorrelated between CDF and D0.

  \item{\bf JES Modeling:} This includes modeling uncertainties from 
    fragmentation and out-of-cone showering variations which affect the 
    determination of corrections necessary to estimate the original parton
    energy from the measured jet energy.

  \item{\bf JES B-jet Modeling:} This includes modeling uncertainties specific
    to B-jets and includes fragmentation, color flow, and b-decay branching
    fraction variations.

  \item{\bf JES B-jet Response:} This includes uncertainties arising from 
    differences in the e/h ratio between light-quark-jets and B-jets and
    is specific to D0 Run~II.

  \item{\bf JES Relative Response:} This includes uncertainties arising from
    uncertainties associated with the $\eta$-dependent corrections made to 
    flatten the calorimeter response as a function of pseudo-rapidity.

  \item{\bf JES Calibration:} This includes uncertainties arising from the
    limited statistics of the calibration and control samples used to 
    determine several components of the JES corrections.
\end{description}
The techniques used to quantify these uncertainties are described in detail
in Section~\ref{sec:MtSyst} above.
The eight \Mt\ measurements presently included in the combination are
summarized in Table~\ref{tab:MtInputs}\footnote{The inputs listed in
Table~\ref{tab:MtInputs} are the same as those used in 
reference~\cite{MtSum05}.  Since then the two CDF Run~II measurements have
been finalized and published~\cite{CdfLjR2}\cite{CdfDlR2} with small
improvements to some of systematic uncertainties.  However, these improvements
have not yet been included in a new Tevatron combined \Mt.}.
Note that the CDF Run~II determination
in the lepton+jets channel uses both the {\it{in situ}} $W\ra qq^{\prime}$
mass and the external calibrations to determine the JES.  In order to
accurately account for the correlations with other inputs that measurement is 
recorded as two separate inputs with the JES components of the uncertainty 
appropriately divided while the remaining statistical and systematic 
uncertainties are taken to be $100\%$ correlated.  The combination of these 
two inputs yields the same statistical, systematic, and total uncertainty as 
the original measurement.

\begin{figure}[tbh]
  \centering
  \includegraphics[width=4.0in]{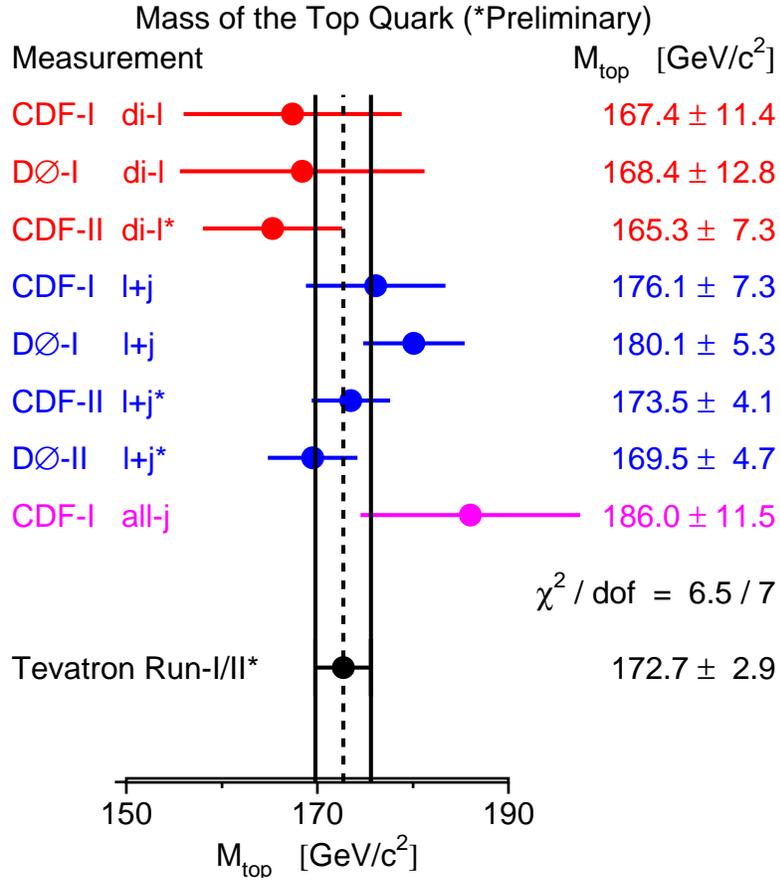}
  \caption{\label{fig:MtCombo}  The inputs for the Tevatron combined 
    \Mt\ combination and the resulting world average top quark mass, 
    obtained as described in Section~\ref{sec:MtCombo}. }
\end{figure}

In the combination, the categories of uncertainty discussed above are
assumed to have the following correlations among the various inputs:
\begin{itemize}
  \item The Statistical, Fit, and {\it{in situ}} JES uncertainties 
    are taken to be uncorrelated among all inputs.
  \item The Uranium Noise and JES Relative Response uncertainties are taken
    to be $100\%$ correlated among all inputs from the same experiment, 
    but uncorrelated between the experiments.
  \item The JES uncertainties from B-jet Response and Calibration are taken
    to be $100\%$ correlated among all inputs from the same experiement and
    data-taking period (ie. Run~I or Run~II) and uncorrelated otherwise.
  \item The Background uncertainties are taken to be $100\%$ correlated
    across all inputs in the same final-state (ie. all-j, l+j, or dil),
    regardless of experiment or data-taking period, and uncorrelated otherwise.
  \item The Signal, Monte Carlo, JES Modeling, and JES B-Jet Modeling 
    uncertainties are taken to be $100\%$ correlated across all inputs. 
\end{itemize}
The resulting global correlation coefficients are given in 
Table~\ref{tab:MtCorrelations} and yield a world average
\begin{equation}
  \Mt = 172.7 \pm 2.9\:\:\gevcs
\label{eq:MtAvg}
\end{equation}
with a $\chi^{2}/dof = 6.5/7$, corresponding to a $\chi^{2}$ probability
of $49\%$. 
The total uncertainty of $\pm2.9$~\gevcs\ is the quadrature sum of a 
Statistical
uncertainty of $\pm1.7$~\gevcs, a total JES uncertainty of $\pm2.0$~\gevcs, 
a Signal uncertainty of $0.9$~\gevcs, a Background uncertainty of $0.9$~\gevcs,
a Uranium Noise uncertainty of $0.3$~\gevcs, a Fit uncertainty of
$0.3$~\gevcs, and a Monte Carlo uncertainty of $0.2$~\gevcs.  The inputs and
the combined \Mt\ are all shown together in Fig.~\ref{fig:MtCombo} while
the pulls and weights of each input are given in Table~\ref{tab:MtPandW}.
The issue of negative weights, as observed in this case for one of the inputs,
is discussed in detail in reference~\cite{Lyons:1988rp} and arises when the
correlation coefficient is comparable to the ratio of the total uncertainties
between two measurements.

\begin{table}[tbh]
\begin{center}
\begin{tabular}{|l||rrr|rr||rrr|r|} \hline
             & \multicolumn{5}{c||}{Run~I Published} 
             & \multicolumn{4}{c|}{Run~II Preliminary} \\ 
             & \multicolumn{3}{c|}{CDF}
             & \multicolumn{2}{c||}{D0}
             & \multicolumn{3}{c|}{CDF}
             & \multicolumn{1}{c|}{D0}                 \\ \cline{2-10}
             & all-j & l+j  & dil
             & l+j   & dil  
             & (l+j)$_{i}$  & (l+j)$_{e}$ & dil
             & l+j                                     \\ \hline\hline
     Result  & 186.0 & 176.1 & 176.4
             & 180.1 & 168.4
             & \multicolumn{2}{c}{173.5} & 165.3
             & 169.5                                   \\ \hline
     Signal  & 1.8 & 2.6 & 2.8
             & 1.1 & 1.8
             & \multicolumn{2}{c}{1.1} & 1.5
             & 0.3                                     \\
 Background  & 1.7 & 1.3 & 0.3
             & 1.0 & 1.1
             & \multicolumn{2}{c}{1.2} & 1.6
             & 0.7                                     \\
  Generator  & 0.8 & 0.1 & 0.6
             & 0.0 & 0.0
             & \multicolumn{2}{c}{0.2} & 0.8
             & 0.0                                     \\
        Fit  & 0.6 & 0.0 & 0.7
             & 0.6 & 1.1
             & \multicolumn{2}{c}{0.6} & 0.6
             & 0.6                                     \\
 Ur.\ Noise  & 0.0 & 0.0 & 0.0
             & 1.3 & 1.3
             & \multicolumn{2}{c}{0.0} & 0.0
             & 0.0                                     \\ \hline
 Sub-total   & 2.7 & 2.9 & 3.0
             & 2.1 & 2.7
             & \multicolumn{2}{c}{1.7} & 2.4
             & 1.0                                     \\ \hline\hline
 JES {\it{in situ}}  & 0.0 & 0.0 & 0.0
             & 0.0 & 0.0
             & 4.2 & 0.0 & 0.0
             & 3.3                                     \\
$\:\:\:$Model  & 3.0 & 2.7 & 2.6
               & 2.0 & 2.0
               & 0.0 & 2.0 & 2.2
               & 0.0                                   \\
$\:\:\:$B-Model  & 0.6 & 0.6 & 0.8
                 & 0.7 & 0.7
                 & 0.6 & 0.6 & 0.8
                 & 0.7                                   \\
$\:\:\:$B-Resp.  & 0.0 & 0.0 & 0.0
                 & 0.0 & 0.0
                 & 0.0 & 0.0 & 0.0
                 & 0.9                                   \\
$\:\:\:$Rel-Resp.& 4.0 & 3.4 & 2.7
                 & 2.5 & 1.1
                 & 0.0 & 2.3 & 1.4
                 & 0.0                                   \\
$\:\:\:$ Calib.  & 0.3 & 0.7 & 0.6
                 & 0.0 & 0.0
                 & 0.0 & 0.0 & 0.0
                 & 0.0                                   \\ \hline
 JES Total   & 5.0 & 4.4 & 3.9
             & 3.3 & 2.4
             & 4.3 & 3.1 & 2.7
             & 3.5                                     \\ \hline\hline
 Syst Total  & 5.7 & 5.3 & 4.9
             & 3.9 & 3.6
             & 4.6 & 3.5 & 3.6
             & 3.6                                     \\ \hline
Statistical  &10.0 & 5.1 &10.3
             & 3.6 &12.3
             & \multicolumn{2}{c}{2.7} & 6.3
             & 3.0                                     \\ \hline\hline
      Total  &11.5 & 7.3 &11.4
             & 5.3 &12.8
             & \multicolumn{2}{c}{4.1} & 7.3
             & 4.7                                     \\ \hline
\end{tabular}
\caption{\label{tab:MtInputs} The inputs for the most recent world 
  average \Mt\ combination~\cite{MtSum05}.  All values are in \gevcs.  The
  CDF Run~II measurement in the l+j channel is specially treated as
  described in the text.  The total uncertainties are the quadrature
  sum of the individual uncertainties listed.}
\end{center}
\end{table}

\begin{table}[tbh]
\begin{center}
\begin{tabular}{|ll||rrr|rr||rrr|r|} \hline
           & & \multicolumn{5}{c||}{Run~I Published} 
             & \multicolumn{4}{c|}{Run~II Preliminary} \\ 
           & & \multicolumn{3}{c|}{CDF}
             & \multicolumn{2}{c||}{D0}
             & \multicolumn{3}{c|}{CDF}
             & \multicolumn{1}{c|}{D0}                 \\ \cline{3-11}
           & & all-j & l+j  & dil
             & l+j   & dil  
             & (l+j)$_{i}$  & (l+j)$_{e}$ & dil
             & l+j                                     \\ \hline\hline
 CDF I & all-j & 1.00 &  &  &  &  &  &  &  & \\
 CDF I &   l+j & 0.32 & 1.00 &  &  &  &  &  &  &  \\
 CDF I &   dil & 0.19 & 0.29 & 1.00 &  &  &  &  &  &  \\ \hline
 D0 I  &   l+j & 0.14 & 0.26 & 0.15 & 1.00 &  &  &  &  &  \\
 D0 I  &   dil & 0.07 & 0.11 & 0.08 & 0.16 & 1.00 &  &  &  &  \\ \hline
 CDF II &(l+j)$_i$ & 0.04 & 0.12 & 0.06 & 0.10 & 0.03 & 1.00 &  &  &  \\
 CDF II &(l+j)$_e$ & 0.35 & 0.54 & 0.29 & 0.29 & 0.11 & 0.45 & 1.00 &  & \\
 CDF II &  dil & 0.19 & 0.28 & 0.18 & 0.17 & 0.10 & 0.06 & 0.30 & 1.00 &   
 \\ \hline
 D0 II &   l+j & 0.02 & 0.07 & 0.03 & 0.07 & 0.02 & 0.07 & 0.08 & 0.03 & 1.00 
 \\ \hline
\end{tabular}
\caption{\label{tab:MtCorrelations} The matrix of global correlation
  coefficients between the \Mt\ measurements of Table~\ref{tab:MtInputs} 
  using the error categories and assuming the correlations described 
  in the text.}
\end{center}
\end{table}

\begin{table}[htb]
\begin{center}
\begin{tabular}{|l||rrr|rr||rrr|r|} \hline
             & \multicolumn{5}{c||}{Run~I Published} 
             & \multicolumn{4}{c|}{Run~II Preliminary} \\ 
             & \multicolumn{3}{c|}{CDF}
             & \multicolumn{2}{c||}{D0}
             & \multicolumn{3}{c|}{CDF}
             & \multicolumn{1}{c|}{D0}                 \\ \cline{2-10}
             & all-j & l+j  & dil
             & l+j   & dil  
             & (l+j)$_{i}$  & (l+j)$_{e}$ & dil
             & l+j                                     \\ \hline\hline
  Pull      & +1.19   & +0.51   & -0.48
            & +1.67   & -0.34
            & \multicolumn{2}{c}{+0.28}  & -1.11
            & -0.86                                    \\ \hline
Weight [\%] & +1.0    & -0.2    & +1.1
            & +18.8   & +2.1
            & \multicolumn{2}{c}{+36.0} & +8.0
            & +33.3                                    \\ \hline
\end{tabular}
\caption{\label{tab:MtPandW} The pull and weight of each input from 
  Table~\ref{tab:MtInputs} in the Tevatron combined \Mt\ determination 
  using the global correlation coefficients given in 
  Table~\ref{tab:MtCorrelations}.}
\end{center}
\end{table}

\subsection{Top Mass Expectations}
\label{sec:MtExpect}
\textbf{Contributed by:~D.~Glenzinski }

\begin{figure}[htb]
  \centering
  \includegraphics[width=4.0in]{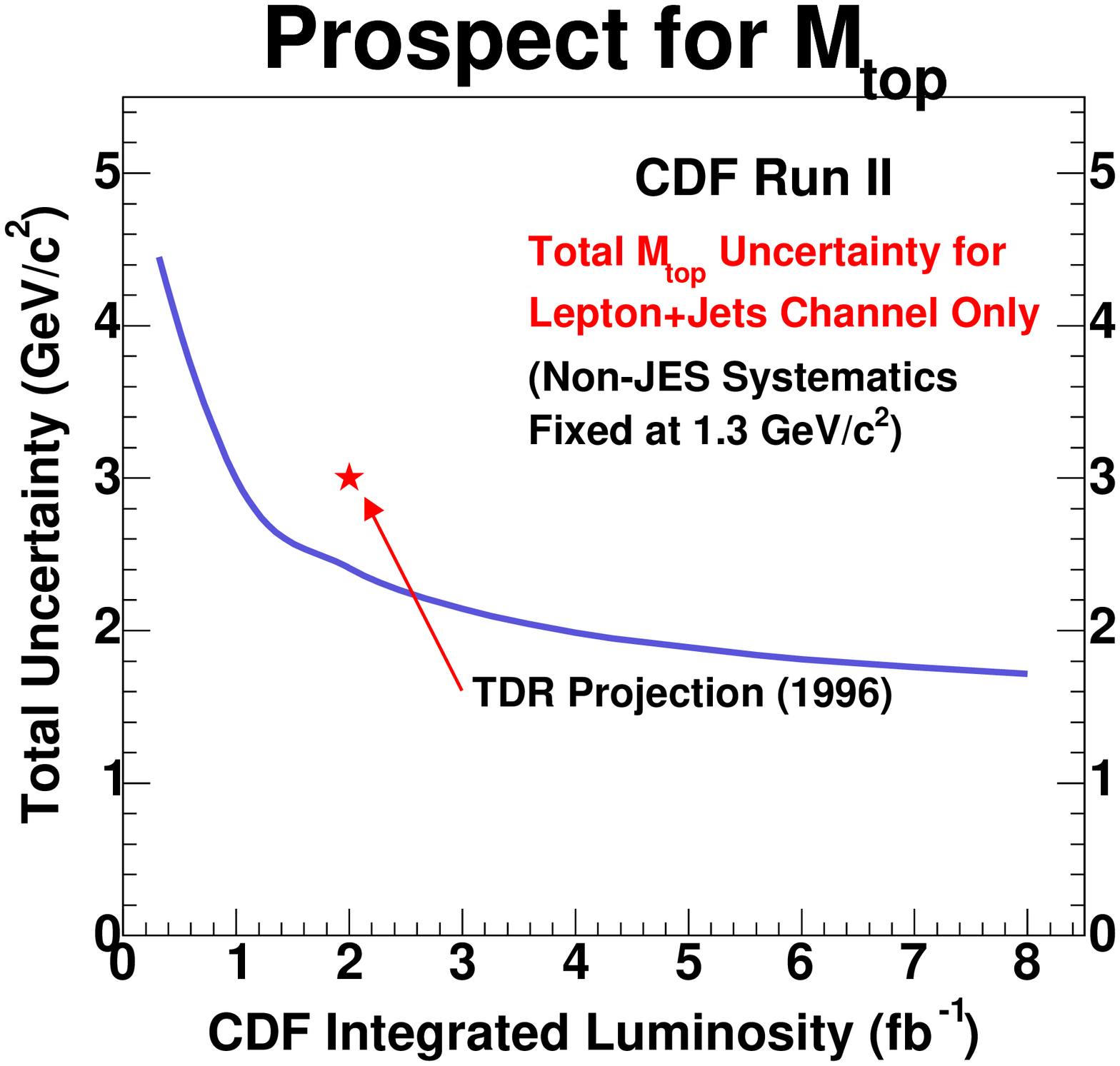}
  \caption{\label{fig:DMtCdfLj}  The total uncertainty on the top quark mass
    for the CDF lepton+jets channel, extrapolated to larger data-sets
    using the assumptions described in Section~\ref{sec:MtExpect} and
    based on the methodology described in reference~\cite{CdfLjR2}. }
\end{figure}

Using the new Run~II measurements as a basis, some simple extrapolations
have been performed in order to roughly estimate what the future sensitivity
of the Tevatron combined \Mt\ might be.  The present Run~II results each use
approximately $350\:\pb$, a factor of 15-20 less than the expected data set at
the end of Run~II.  It is important
to note that at present the JES uncertainty is effectively the weighted 
average of the {\it{in situ}} JES with the quadrature sum of the remaining 
JES uncertainties so that as the data sets increase the {\it{in situ}} 
determination will improve and will eventually come to dominate the JES
uncertainty.  Initial studies indicate that the total JES uncertainty will
fall to approximately $1.5\:\gevcs$ per experiment with $1\:\fb$ data-sets
(each) and to $\leq 1.0 \:\gevcs$ per experiment for data-sets exceeding
$4\:\fb$ each.
Even under the conservative assumption that {\it{only}} the 
Statistical and {\it{in situ}} JES uncertainties improve (proportional to 
$1/\sqrt{N}$) with increasing data-sets while all other uncertainties are 
fixed, the ultimate Tevatron combined sensitivity should readily fall below 
$\Delta\Mt<\pm 2\:\gevcs$.
Figure~\ref{fig:DMtCdfLj} shows how the total uncertainty in the l+j
channel for CDF is expected to evolve.  For this extrapolation the 
expected statistical uncertainties have been estimated by performing
Monte Carlo pseudo-experiments at several luminosity points, assuming
the Standard Model \ttbar\ production cross-section, and assuming the 
signal and background acceptances do not change with the increasing 
instantaneous luminosity necessary to meet the Run~II delivered luminosity 
goals.  With these conservative assumptions, CDF alone with this single
channel alone is expected to do better than the original TDR estimates for the
CDF combined sensitivity~\cite{CDFTDR}.  Using extrapolations for the other
inputs with the same conservative assumptions and repeating the combination
using the same error categories and correlations discussed above predict
an ultimate Tevatron combined sensitivity of $\Delta\Mt\leq\pm 1.5\:\gevcs$
for data-sets of $\geq4\:\fb$ collected per experiment.  It should be noted
that at present, as discussed above in Section~\ref{sec:MtSyst}, several
systematic uncertainties are limited by the statistics of the samples used
to quantify them.  Thus it is reasonable to expect that these, too, will 
improve with time to yield an even better Tevatron combined sensitivity.

\subsection{Top Mass Determination at the LHC}
\textbf{Contributed by:~A.I.~Etienvre, A.~Giammanco}

\subsubsection*{Introduction}

At the LHC, the top quark will be produced mainly in pairs through the hard process
$gg \, \rightarrow \, {\ttbar} $ (90$\%$) and $q\bar{q} \, \rightarrow \,
{\ttbar}$ (10$\%$)~;
the corresponding cross-section, at the next-to-leading order, is equal to 796
$^{+94} _{-91}$ pb~: therefore, we expect roughly 8 million {\ttbar} pairs to be produced 
 with 100 days at low luminosity (corresponding to an integrated luminosity of 10 fb$^{-1}$).\\
In order to ensure a similar contribution to the indirect measurement of the Higgs mass, 
the precision on $m_W$ and $m_t$ must fulfill the following relation~: 
$\Delta m_t \, \simeq \, 0.7 \, 10^{-2} \Delta m_W$. At LHC, we expect to 
reach an accuracy of 15 MeV/$c^2$ on $m_W$ and 1 GeV/$c^2$ on $m_t$. 
With these precision measurements, the relative precision on a Higgs 
boson mass of 115 GeV/c$^2$ would be of the order of 18$\%$ \cite{Snowmass}.
The various methods developed to measure the top mass at the LHC are
explained, together with their advantages, their disadvantages, 
and their corresponding systematic errors.

\subsubsection*{Systematic Uncertainties}
For the top mass analyses presented here, performed within ATLAS or CMS, several systematic uncertainties have been estimated. The main sources of errors, 
common to several analyses, are briefly described below.

\paragraph{Jet energy scale}

When the top quark is reconstructed via its hadronic decay (t $\rightarrow$ Wb $\rightarrow$ jjb), the accuracy of the measurement of its mass relies
 on a precise knowledge of the energy calibration for both light jets and b-jets. The energy of the two light jets can be calibrated
 precisely event by event using an in-situ calibration based on the W mass constraint \cite{PhDRoy}, while the b-jet energy scale has to be 
 calibrated independently~: therefore, their contributions to systematic errors are always estimated separately.\\
  \hspace{0.5cm}A jet energy scale calibration at the level of 1$\%$, for both light jets and b-jets, should be reached at LHC~: the corresponding errors on the top mass
measurement given below correspond to this level of precision. The estimation of an absolute jet energy scale uncertainty has been carried out applying 
 different miscalibration coefficients to the reconstructed jet energies~; a linear dependence has been observed.

\paragraph{Initial and final state radiation}

The presence of initial state radiation (ISR) of incoming partons and final state radiation (FSR) from the top decay products
has an impact on the top mass measurement. In order to estimate the uncertainty due to these radiations, 
 the top mass has been determined with ISR (FSR) switched on, at the generator level, and ISR (FSR) switched off. The systematic
 uncertainty on the top mass is taken to be 20 $\%$ of the corresponding mass shifts~: this should be a conservative estimate, 
assuming that ISR and FSR are known at a level of order of 10 $\%$ \cite{Beneke:2000hk}. 
 
\paragraph{b-quark fragmentation}

The systematic error due to an imperfect knowledge of the b-quark fragmentation has been estimated by varying the 
 Peterson parameter of the fragmentation function (equal to -0.006) within its experimental uncertainty (0.0025)~: the consecutive shift 
 on the top mass is taken as the systematic error on the top mass.
 
\paragraph{Background} 

The background of the top quark reconstruction is dominated by wrong
combinations in {\ttbar} events themselves (FSR, wrong association of the W to the 
corresponding b-jet,..). Varying the background shape and size in the fitting procedure of the top mass distribution gives access to 
 the resulting uncertainty on the top mass measurement. 

\subsubsection*{Top mass measurement in the lepton + jets channel}

The lepton plus jets channel will provide a large and clean sample of {\ttbar} events and is probably the most promising channel
 for an accurate measurement of the top mass. The main backgrounds are summarized in Table \ref{bdf}, with their corresponding cross sections 
and expected number of events at 10 fb$^{-1}$. Before any selection, 
 the signal over background ratio is of the order of $10^{-4}$. Events are selected by requiring one isolated lepton (electron or muon) with $p_T \, \geq \, 20$ GeV/c and 
$|\eta| \,  \leq \, 2.5$, $E_T^{miss} \, \geq \,  20$ GeV/${c^2}$, and at least 4 jets with $p_T \, \geq \, 40$ GeV/c and 
$|\eta| \, \leq \, 2.5$, of which two of them are required to be tagged as b-jets. Jets used for these analysis are reconstructed with a $\Delta R \, = \, 0.4$ \footnote{$\Delta R \, = \, \sqrt{\Delta \Phi^2 \, + \, \Delta \eta^2 }$} cone algorithm. After these cuts, S/B becomes much more favorable~: S/B $\simeq$ 30.  

\begin{table}
\begin{footnotesize}
\begin{center}
\begin{tabular}{||c|c|c||} \hline
Process & Cross section (pb) & Number of events @ 10 fb$^{-1}$ (millions)\\
\hline \hline
Signal & 250 & 2.5 millions \\
\hline \hline
$b\bar{b} \, \rightarrow \, l\nu \, + \, jets$ & 2.2 $10^6$ & 22 $10^3$  \\ \hline
$W \, + \, jets \, \rightarrow \, l\nu \, + \, jets$ & 7.8 $10^3$ & 78  \\ \hline
$Z \, + \, jets \, \rightarrow \, l^+l^- \, + \, jets$ & 1.2 $10^3$ & 12  \\ \hline
$WW \, \rightarrow \, l\nu \, + \, jets$ & 17.1 & 0.17  \\ \hline
$WZ \, \rightarrow \, l\nu \, + \, jets$ & 3.4 & 0.034 \\ \hline
$ZZ \, \rightarrow \, l^+l^- \, + \, jets$ & 9.2 & 0.092 \\ \hline \hline
\end{tabular}
\end{center}
\end{footnotesize}
\caption{\label{bdf} Main backgrounds to the lepton (l = e,$\mu$) + 
jets {\ttbar} signal.}
\end{table}

\paragraph{Top mass measurement using the hadronic top decay (\cite{etienvre}, \cite{top_atlas}, \cite{top_cms}, \cite{top_cms2})}

The top mass is estimated here from the reconstruction of the invariant mass of a three-jet system~: the two light jets from the W and one of the 
 two b-jets. The determination of this combination of three jets proceeds in two steps~: the choice of the two light jets, and the choice of the b-jet associated to the reconstructed hadronic W.

Events kept after the selection described above have at least two light jets above a given threshold on their 
 transverse momentum. In a first step, we select the hadronic W candidates in a mass window of $\pm \, 5 \, \sigma_{mjj}$ around the peak value of the 
 distribution of the invariant mass of the light jet pairs, made with events with only two light jets ($\sigma_{mjj}$ is the width of this distribution). \\ 
In order to reduce the incidence of a light-jet energy mis-measurement (due to the energy lost out of cone) on the precision of the top mass measurement, an in-situ calibration of these jets is performed, through a $\chi^2$ minimization procedure (\cite{PhDRoy}, \cite{top_atlas}). This minimization is applied event by event, for each light-jet pair combination. The expression of $\chi^2$, given by equation (\ref{insitu}), is the sum of three terms~: the first (and leading) one corresponds to the constrain of the jet pair invariant mass $m_{jj}$ to the PDG W mass ($m_W$)~; the others correspond to the jet energy correction factors, $\alpha_i \, (i\, = \, 1, 2)$, to be determined by this minimization ($\sigma_i \, (i\, = \, 1,2)$ is the resolution on the light jet energy).
  
\begin{equation} 
\label{insitu}
\chi^2 \, = \, \frac{(m_{jj}(\alpha_1,\alpha_2) \, - \, m_W)^2}{\Gamma_W^2} \, + \, \frac{(E_{j1}(1\, - \, \alpha_1))^2}{\sigma_1^2} \, + \, \frac{(E_{j2}(1\, - \, \alpha_2))^2}{\sigma_2^2}
\end{equation} 

The $\chi^2$ is minimized, event by event, for each light jet pair~; the light jet pair $j_1,j_2$ corresponding to the minimal $\chi^2$ is kept as the hadronic W candidate. This minimization procedure also leads to the corresponding energy correction factors $\alpha_1, \alpha_2$. The hadronic W is then reconstructed with the light jets chosen by this $\chi^2$ minimization.

\begin{figure}[h]
\begin{center}
\includegraphics[width= .6\textwidth]{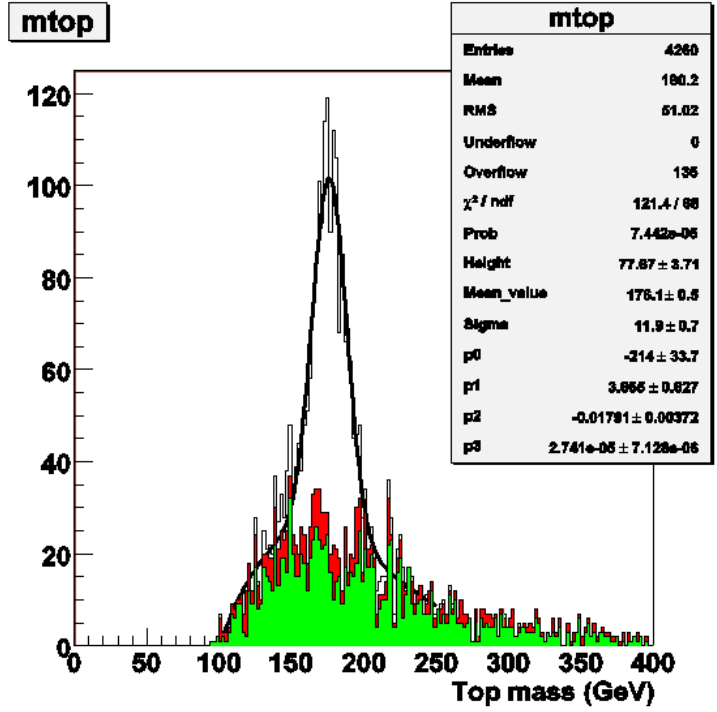}
\caption{\label{topmass}Top mass distribution, with the contribution from 
wrong W combinations, in green, and, in red, from wrong b-jet associations. 
This analysis has been performed using the MC@NLO generator and the 
full simulation of the ATLAS detector.}
\end{center}
\end{figure}

Several methods have been investigated to choose the b-jet among the two candidates, and the one giving the highest purity has been kept~: 
the b-jet associated to the hadronic W is the one leading to the highest $p_T$ for the top.\\
\hspace{0.5cm} The reconstructed three jets invariant mass is shown in 
Fig.~\ref{topmass}: the mass peak (176.1 $\pm$ 0.6 GeV/$c^2$) is in reasonable agreement
with the generated value (175 GeV/$c^2$); the width is equal to 
11.9 $\pm$ 0.7 GeV/$c^2$. The overall efficiencies and purities, 
with respect to lepton + jets events, are summarized in 
Table~\ref{efficiency}: we expect with this method 64,000 
events at 10 fb$^{-1}$, corresponding to a statistical error 
equal to 0.05 GeV/$c^2$.

\begin{table}[h]
\begin{center}
\begin{footnotesize}
\begin{tabular}{||c|c|c|c|c|c|c|c|c||} \hline

&  Efficiency ($\%$) & b purity ($\%$) & W purity ($\%$) & Top purity ($\%$)  \\
\hline
full mass window & 2.70 $\pm$ 0.005 &  56.0 $\pm$ 0.9 & 63.2 $\pm$ 0.9 & 40.5 $\pm$ 0.9          \\
mass window within $\pm 3 \, \sigma_{m_{top}}$ & 1.82 $\pm$ 0.04 &  69.1 $\pm$ 0.8 & 75.8 $\pm$ 0.8 & 58.6 $\pm$ 0.8  \\ 
\hline
\end{tabular}
\caption{\label{efficiency} Total efficiency and $W$, $b$ and 
top purity of the final selected events 
(MC@NLO, full simulation of the ATLAS detector), 
with respect to lepton (electron, muon) + jets events}
\end{footnotesize}
\end{center}
\end{table}

The dominant remaining background to lepton + jets {\ttbar} events comes from W + jets events. The contribution 
 to the top mass measurement is negligible~: the values of the mass peak (176.1 $\pm$ 0.6 GeV/$c^2$ for signal only, 176.2 $\pm$ 0.6 GeV/$c^2$ for signal plus background) and of the width (11.9 $\pm$ 0.7 GeV/$c^2$ for signal, 12.1 $\pm$ 0.7 GeV/$c^2$ for signal plus background )  are identical.

\paragraph{Top mass measurement using a kinematic fit \cite{top_atlas}}

An alternative method for the top mass measurement in the lepton plus jets channel consists in reconstructing 
 the entire {\ttbar} final state, in order to reduce the systematic error due to FSR.
 The hadronic part is reconstructed in a similar way to the previous section. The leptonic side
 can not be directly reconstructed due to the presence of the undetected neutrino, but can be estimated in three steps~:

\begin{itemize}

\item $p_T(\nu) \, = \, E_T^{miss}$

\item $p_z(\nu)$ is obtained by constraining the invariant mass of the lepton-neutrino system to the PDG W mass value~: this kinematic equation leads to two $p_z(\nu)$ solutions
 
\item the remaining b-jet is associated to the reconstructed W 
 
\end{itemize}   

The top mass determination is performed through a kinematic fit, relying on a $\chi^2$ based on mass constraints ($m_{jj} \, = \, m_W^{PDG} \, = \, m_{l\nu}$~; $ m_{jjb} \, = \, m_{l\nu b} $) and kinematic constraints (energy and direction of leptons and jets can vary within their resolutions). The minimization of this $\chi^2$ is performed event by event, for the two $p_z(\nu)$ solutions~: the one giving the lower $\chi^2$ is kept. The top mass is determined as the linear extrapolation of $m_{top}(\chi^2)$ for $\chi^2$ = 0.\\
With an efficiency equal to 1.1 $\%$, we expect with this method 26 000 events at 10 fb$^{-1}$, corresponding to a statistical error equal to 0.1 GeV/$c^2$. This analysis has been performed using a fast simulation of the ATLAS detector, and will be checked with a full simulation. 

\paragraph{Top mass measurement using large $p_T$ top events (\cite{top_atlas} \cite{top_pt_atlas})}

Thanks to the large amount of {\ttbar} events produced at LHC, a subsample of
lepton + jets {\ttbar} events, where the top quarks have a $p_T$ greater than 200 GeV/c, can be studied. The interest of such events is that the top and the anti-top are produced back-to-back in the laboratory frame, 
 so that their daughters will appear in distinct hemispheres of the detector~: therefore, the combinatorial background should be strongly reduced. \\
\hspace{0.5cm}Because of the high $p_T$(top), the three jets in one hemisphere tend to overlap. To overcome this problem, the top quark is reconstructed in a large calorimeter cone ($\Delta R$ in [0.8 - 1.8]), around the top quark direction. \\

\begin{figure}[htbp]
\begin{center}
\includegraphics[width= .8\textwidth]{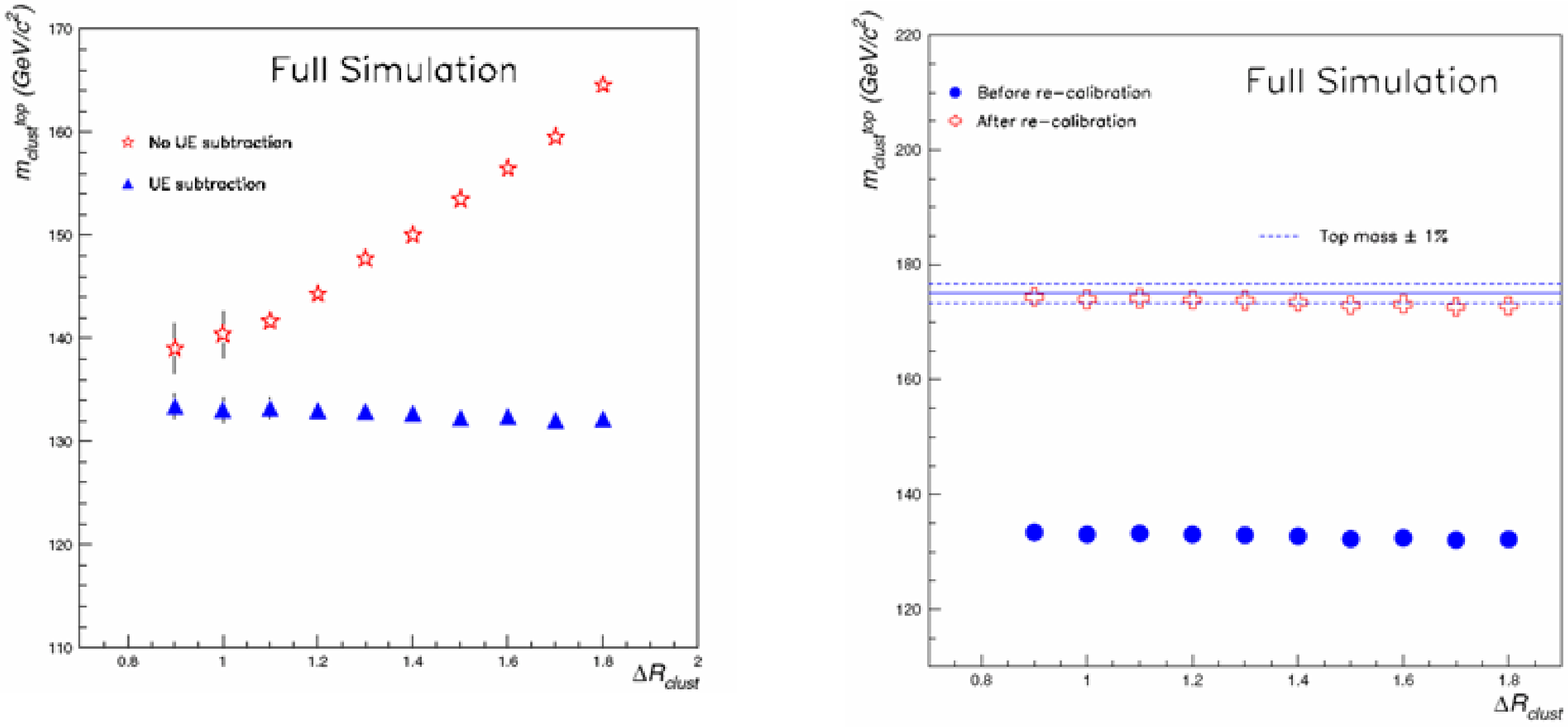}
\caption{ \label{UE} Fitted top mass reconstructed in a large calorimeter cluster as a function of the cluster size, for a subsample of events with $p_T(top) \, \geq  \, 200$ GeV/c, before and after UE subtraction, on the left. The plot on the right shows the effect of the mass scale recalibration. This analysis has been performed using the \PY generator for signal, and the full simulation of the ATLAS detector.}
\end{center}
\end{figure}

A strong dependence of the reconstructed top mass with the cone size has been observed and can be attributed to the Underlying Events (UE) contribution, evaluated to 45 MeV in a 0.1 X 0.1 calorimeter tower with the full simulation of the ATLAS detector. After UE subtraction, the top mass is independent of the cone size, but lower than the generated top mass by 25 $\%$ , as can be seen in Fig.~\ref{UE}. A mass scale recalibration, based on the hadronic W, is then applied and leads to an average top mass value consistent with the generated value (see Fig.~\ref{UE}).\\ 
With an efficiency equal to 2 $\%$ with respect to this subsample, we expect with this method 3600 events at 10 fb$^{-1}$, corresponding to a statistical error equal to 0.2 GeV/$c^2$.

\paragraph{Systematic uncertainties on the top mass measurement in the lepton + jets channel}

The systematic uncertainties on the top mass measurement are summarized in Table \ref{syste}, for the three methods explained above. It is possible to get rid of the error due to the light jet energy scale thanks to the in-situ calibration~; the dominant contribution comes from the FSR and the b-jet energy scale.
 
\begin{table}[h]
\begin{footnotesize}
\begin{center}
\begin{tabular}{||c|c|c|c||} \hline

Source of uncertainty &  Hadronic top  & Kinematic fit & High $p_T$ top sample   \\
                      &  $\delta m_{top}$ (GeV/c$^2$) &$\delta m_{top}$ (GeV/c$^2$)&$\delta m_{top}$ (GeV/c$^2$) \\ 
\hline
Light jet energy scale (1 $\%$)& 0.2 & 0.2 & \\
b-jet energy scale (1 $\%$)& 0.7 & 0.7 & \\
b-quark fragmentation & 0.1 & 0.1 & 0.3 \\
ISR& 0.1 & 0.1 & 0.1 \\
FSR& 1. & 0.5 & 0.1 \\
Combinatorial background& 0.1 & 0.1 & \\
Mass rescaling& & & 0.9 \\
UE estimate ($\pm$ 10 $\%$)& & & 1.3 \\
\hline
Total& 1.3 & 0.9 & 1.6 \\
\hline
Statistical error& 0.05 & 0.1 & 0.2\\
\hline
\end{tabular}
\end{center}
\end{footnotesize}
\caption{\label{syste} Systematic errors on the top mass measurements, 
in the lepton + jets channel, for the three methods described above.}
\end{table}

\subsubsection*{Top mass measurement in leptonic final states with J/$\psi$ (\cite{top_jpsi}, \cite{top_jpsi2})}

A last top mass determination can be carried out in the lepton+jets channel where a J/$\psi$ arises from the b-quark associated to the leptonic decaying W (Fig.~\ref{Jpsi}). The large mass of the J/$\Psi$ induces a strong correlation with the top mass, as will be shown below.
Although the overall branching ratio (5.5 $10^{-4}$) is low,
this analysis starts to be competitive with more traditional mass measurements already with the first 20 fb$^{-1}$.
This measure is expected to have an excellent resolution because of the very clean experimental reconstruction of the lepton three-vectors.
In the analysis presented in \cite{top_jpsi2}, in order to increase the available statistics, no attempt is made to correctly pair the J/$\psi$ to the lepton, when two isolated leptons are present: the top mass is extracted from the full distribution containing the combinatorial background.

\begin{figure}[h]
\begin{center}
\includegraphics[width= .55\textwidth]{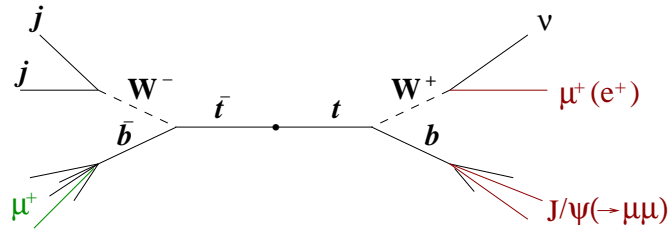}
\caption{ \label{Jpsi} Diagram of the {\ttbar} decay to semi-leptonic final state with J/$\Psi$.}
\end{center}
\end{figure}

Events are triggered using the inclusive lepton trigger. In events passing the trigger thresholds a J/$\psi$ is searched for by looking for same-flavour, opposite-sign leptons with invariant mass in the range [2.8,3.2] GeV/c$^2$ and forming an angle greater than 2 and lower than 35 degrees.
If a J/$\psi$ is found in an event, the isolated lepton with the highest $p_T$ and higher than 20 GeV/c is considered as the lepton candidate from the $W$ decay.
To reduce the background from non-top processes, the total scalar sum of the transverse jet momenta is required to be greater than 100 GeV/c. This cut is not applied if two isolated leptons are found, in order to preserve dileptonic $t\bar t$ events. If the flavour of the two leptons is the same, an explicit $Z$ veto is applied (removing events where the pair has invariant mass within 6 GeV/c$^2$ of the $Z$ mass).
To further reduce soft background and make the analysis less sensitive to systematic effects involving soft QCD, the cut on the transverse momentum of the isolated lepton is brought to 40 GeV/c.

\begin{figure}[h]
\begin{center}
\includegraphics[width= .95\textwidth]{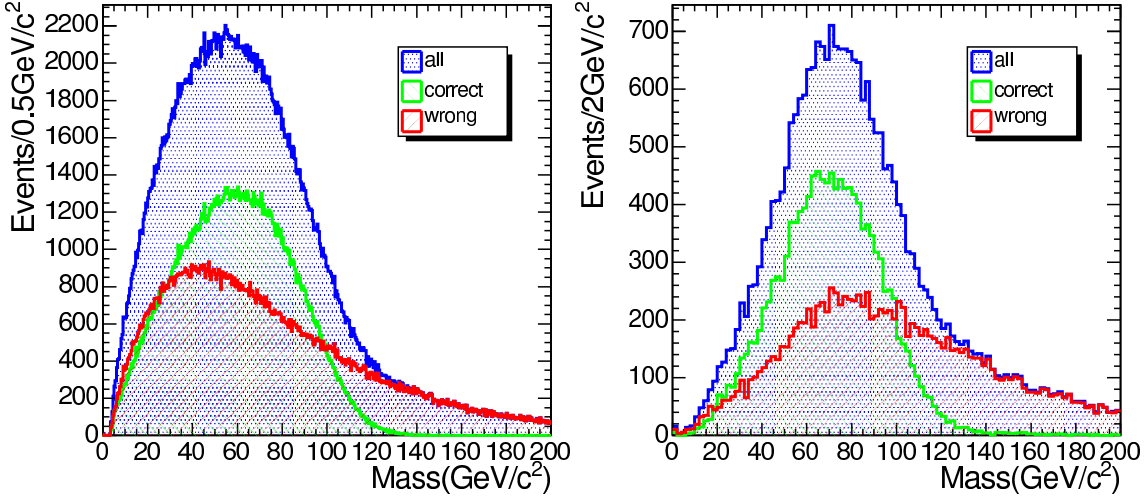}
\caption{ \label{Jpsi2} Lepton-J/$\psi$ invariant mass for $m_t = 175$ GeV/c$^2$ with 1 fb$^{-1}$ integrated luminosity, at generator level (left) and after full detector simulation and reconstruction (right).}
\end{center}
\end{figure}

\begin{figure}[h]
\begin{center}
\subfigure[]{\includegraphics[width=.45\textwidth]{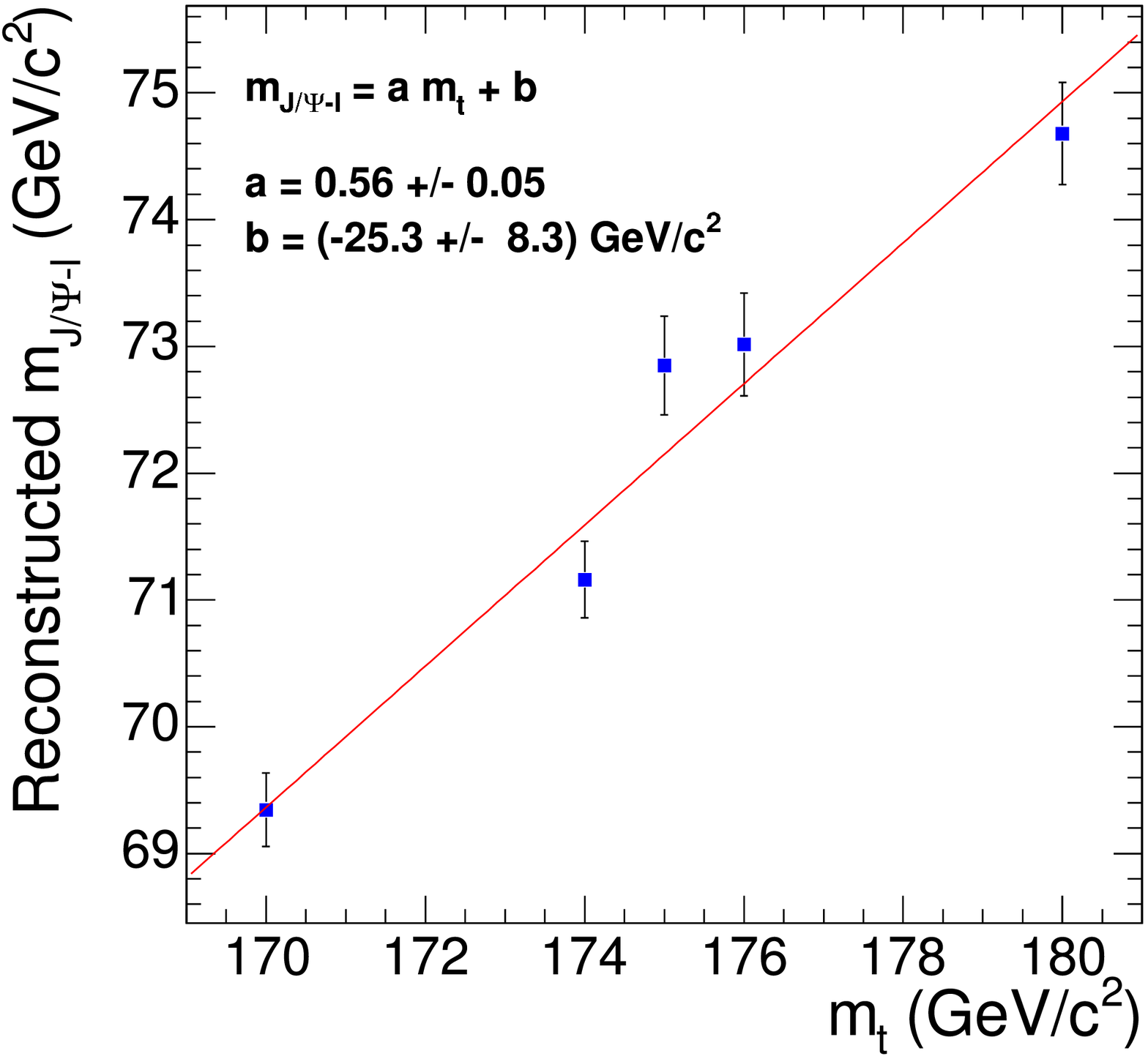}}
\subfigure[]{\includegraphics[width=.45\textwidth]{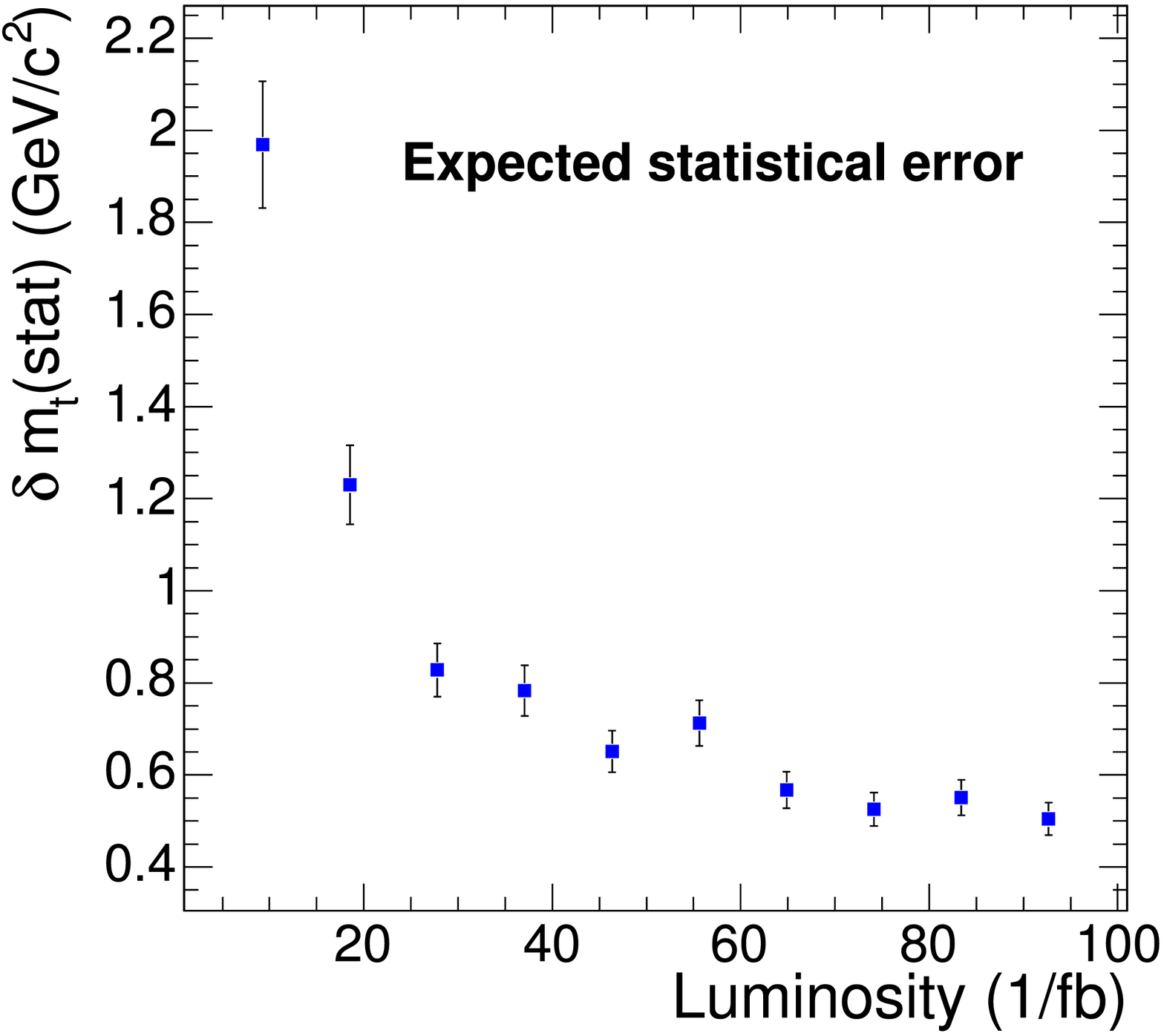}}
\caption{ \label{Jpsi3} Correlation between the reconstructed lJ/$\psi$ mass and the generated top quark mass (left), and expected statistical error as a function of integrated luminosity (right). This study has been performed with a fast simulation of the CMS detector \cite{top_cms}.}
\end{center}
\end{figure}

The observable most sensitive to the top mass is the position of the maximum of the three-lepton mass distribution, shown in Fig.~\ref{Jpsi2}. Its correlation to the top mass and the statistical error are shown in Fig.~\ref{Jpsi3}.

A statistical error of around 1.2 GeV/c$^2$ is expected after the first 20 fb$^{-1}$, and the systematic error, dominated by theory, is lower than 1.5 GeV/c$^2$ (only 0.5 GeV/c$^2$ of which come from instrumental uncertainties).
This analysis reduces to a minimum those systematics which are expected to dominate in more traditional estimations of the top mass, especially the ones from direct reconstruction, like the jet energy scale and the b-tagging efficiency. Therefore a reduction of the uncertainty on $m_t$ is expected when combining this to the direct measurements.

\subsubsection*{Top mass measurement in the dilepton channel \cite{top_atlas}, \cite{top_cms3}}

The dilepton channel is very clean, with a lower contribution of combinatorial background, but it can only provide an indirect top mass measurement, because of the presence of two undetected neutrinos in the final state. Events are selected requiring two leptons of opposite charge, with $p_T \,  \geq \, 20 $ GeV/c and $|\eta| \, \leq \,  2.5$, an $E^{miss}_T \, \geq \, 40$ GeV and 2 b-jets with $p_T \, \geq \, 25 $GeV/c and $|\eta| \, \leq \, 2.5$. After this selection, the ratio of signal over background is around 10.\\
The final state reconstruction relies on a set of six equations for the six unknown components of momenta of neutrino and antineutrino, based on kinematic conservation laws and assuming a given top mass value. This set of equations can provide more than one solution; then, weights are computed from kinematic Monte Carlo distributions of three variables (cos$\theta^*_{top}$, $E_{\nu}$ and $E_{\bar{\nu}}$), and the solution corresponding to the highest weight is kept. This weight is computed for several input top masses, and the top mass estimator corresponds to the maximum mean weight.

With an efficiency of 6.5 $\%$, 20 000 events are expected at 10 fb$^{-1}$. The statistical error on the top mass measurement is negligible (0.04 GeV/c$^2$). The systematic error, equal to 1.7 GeV/c$^2$, is dominated by the uncertainty on the parton distribution function (1.2 GeV/c$^2$). 

\subsubsection*{Top mass measurement in the all hadronic channel (\cite{top_atlas}, \cite{top_cms3})}

\begin{figure}[!h]
\begin{center}
\includegraphics[width= .6\textwidth]{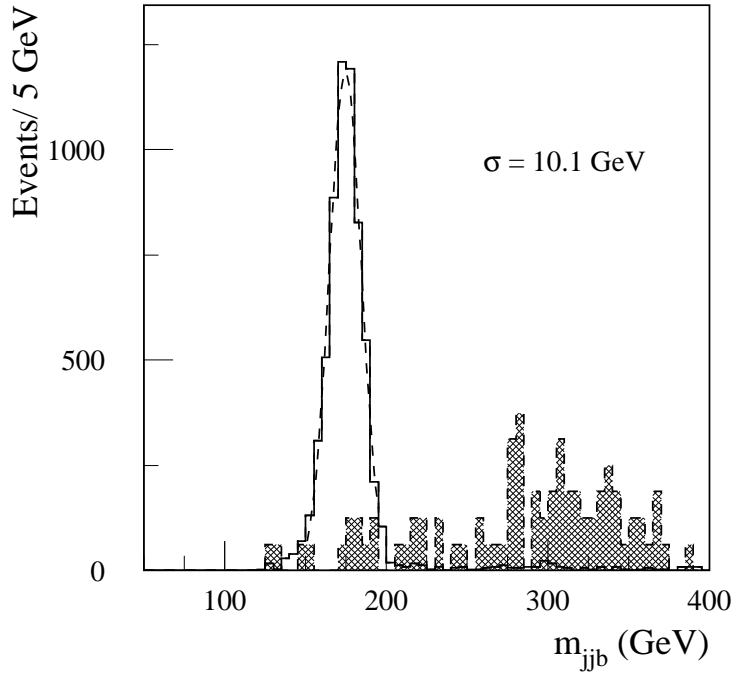}
\caption{ \label{had} Top mass distribution in the all hadronic channel, for the high $p_T$ top sample. The shaded area corresponds to the remaining QCD background. This study has been performed with a fast simulation of the ATLAS detector \cite{top_atlas}.}
\end{center}
\end{figure}

The main advantage of this channel is a full kinematic reconstruction of both sides, and its main disadvantage is the huge QCD multijet background~: before any selection, the ratio of signal over background is very low (10$^{-8}$). Events are selected requiring at least six jets with $p_T \,  \geq \,  40 $ GeV/c, and $|\eta| \, \leq \,  3$, and at least two b-jets with $p_T \, \geq \, 40 $ GeV/c, and $|\eta| \, \leq \, 2.5$. The final state reconstruction proceeds in two steps~: first, the choice of the two light jets pairs to form the two W bosons is performed through the minimization of a $\chi^2$ based on the W mass constraint. Both W candidates are then associated to the right b-jet minimizing a $\chi^2$ based on the equality of the top masses on both sides.
 In order to improve the signal over background ratio, the analysis can be restricted to a sample of high $p_T$ ($\geq$ 200 GeV/c) top and anti-top~: this ratio is finally favorable (S/B $\simeq$ 18).\\
\hspace{0.5cm}The top mass distribution is displayed in Fig.~\ref{had}. The overall efficiency, within the 130-200 GeV/c$^2$ top mass window, is equal to 0.08$\%$, corresponding to 3300 events at 10 fb $^{-1}$, and a statistical error equal to 0.18 GeV/c$^2$. The systematic error, of the order of 3 GeV/c$^2$, is dominated by the contribution of FSR (2.8 GeV/c$^2$).

\subsection{Conclusions}
\textbf{Contributed by:~F.~Canelli, A.I.~Etienvre, and D.~Glenzinski}

Impressive improvements have been achieved in the latest top-quark mass
measurements at the Tevatron. 
All the decay channels have explored new techniques
to address their major uncertainties and as a consequence all
measurements in all channels are currently systematic dominated.
There are still some improvements which are believed will be important. In
the all-jets channel it is possible to make an in-situ measurement of  the JES.
This could result in a measurement with the same precision as the  those in the
lepton+jets channel.
Once these channels have an in-situ JES measurement the remaining
uncertainty on the jet energy scale in all the channels will  predominantly
arising from the uncertainty on b-jets. We expect to reduce the  uncertainty on this
jet energy scale using $Z\to b\bar b$ events. Currently there has been some
progress on extracting an uncertainty from this sample but the  understanding
of the overwhelming background has been difficult. We expect to have  this
done in the next year. This will be more important for the dilepton  channel, where
an in-situ determination of the JES is not possible.
In the future we plan to combine different methods of analysis in the  same
channel. We have done this previously in the dilepton channel and  obtained
a significant improvement in the sensitivity since each method uses  different
information from the same dataset. We would like to do this in all  the channels.
The remaining systematic uncertainties should be revisited by both
experiments.  These will soon be the uncertainties dominating the top- quark
mass measurements. Currently the list of these uncertainties used by  D0 and
CDF is different and we don't  have a common way of applying them.
In the near future we should agree upon the best way to classify and  calculate
these uncertainties. Finally, there needs to be a quantitative study  of the effects of
Color Reconnection and other final state interactions.  Monte Carlo
generators which include these effects for $p\bar p\to t\bar t$ interactions  are only
recently becoming available.
All these improvements will get us to a precision of less than 1.5 GeV.

At the LHC, various top mass measurement methods have been investigated, 
in all decay channels of the top quark. The very large sample of {\ttbar} 
events that will be accumulated will allow a precision measurement 
after only one year of data taking at low luminosity (10 fb$^{-1}$)~: 
the statistical error on the top mass is negligible in all these methods 
except the method involving leptonic final states with $J/\Psi$. 
These analyses are differently sensitive to the various sources of 
systematic uncertainties~: therefore, this will allow reliable 
cross-checks between the various methods. The top quark mass should 
be measured at LHC with a precision of the order of 1 GeV/c$^2$, 
in the lepton plus jets channel.  

In all cases we need to be aware of physics limitations from 
Monte Carlo or  analysis approaches which would prevent us from 
reaching the levels of expected precision as soon
as possible so that we can mitigate their effects.


\clearpage

\section{Single Top Quark Physics}
\label{sec:singletop}

%
%
%
%
\subsection{Introduction}
\textbf{Contributed by:~C.~Ciobanu and R.~Schwienhorst}

The existence of the top quark was established in top quark pair events
produced via the strong interaction \cite{Abe:1995hr,Abachi:1995iq}, 
where quark-antiquark annihilation or gluon-gluon fusion leads to top-antitop pairs. 
The Standard Model (SM)
also allows for the top quark to be produced singly rather than in pairs via the electroweak 
charged current interaction, a mode typically referred to as single-top quark production.
At the time of this report, the single-top production mode 
is yet to be observed experimentally. Current searches at the Tevatron CDF and 
D\O~experiments are nearing in on this production mode as datasets in excess of 
1~fb$^{-1}$ are being accumulated. At the LHC, it is expected that the three different
production modes of single-top quark production can be observed individually.

Studying single-top quark production at hadron colliders is important for a number of
reasons. First, a measurement of the production cross section provides the only
direct measurement of the total top quark decay width and the CKM matrix element 
$|V_{tb}|^{2}$, without having to assume three quark generations or CKM matrix unitarity.
Second, measuring the spin polarization of single-top quarks and can be used to 
test the V-A structure of the top quark electroweak charged current interaction. 
Third, the presence of various new SM 
and non-SM phenomena may be inferred by observing deviations from the predicted rate 
of the single-top signal and by comparing different production modes.
Fourth, the single-top quark final state presents an irreducible background to several 
searches for SM or non-SM signals, for example Higgs boson searches in the associated 
production channel. 

This report is 
intended as a guide to the current issues in single-top quark physics at hadron colliders. 
Section~\ref{sec:singletoptheory} presents a theoretical perspective on single-top quark production. Studies
of single-top quark production at 
next-to-leading-order (NLO) are presented, followed by 
discussions of Monte Carlo modeling and its agreement with NLO results as well as strategies for
choosing event variables to optimize the signal-background separation.
Section~\ref{sec:singletoptev} presents the experimental challenges faced by single-top quark searches at 
the Tevatron. Recent studies from the CDF and D0 Collaborations are described, along with
sensitivity projections for the remainder of Run~II at the Tevatron. Section~\ref{sec:singletoplhc}
presents the experimental perspective from the LHC point of view. 
The connection between LHC and Tevatron single top searches are discussed in Section~\ref{sec:singletoptevlhc}.

%
%
%
%
\subsection {Theory}
\label{sec:singletoptheory}

%
%
%
\subsubsection*{Overview}
\label{sec:singletoptheoryoverview}

\textbf{Contributed by:~T.~Tait and S.~Willenbrock}

\begin{figure}[tb]
\begin{center}
\includegraphics[width=4.5in,clip]{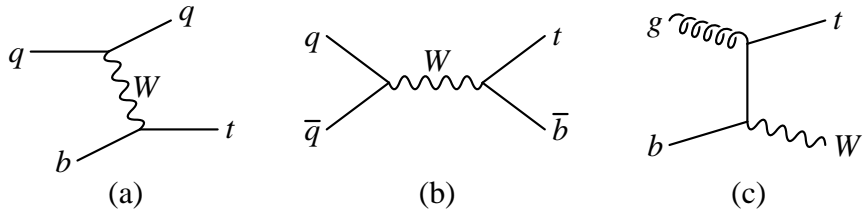}
\end{center}
\caption{Feynman diagrams for single-top-quark production in
hadron collisions: (a) $t$-channel process; (b) $s$-channel process;
(c) associated production (only one of the two diagrams for this process
is shown).}
\label{fig:singletop}
\end{figure}

At the Tevatron and the LHC, top quarks are mostly produced in
pairs, via the strong processes $q\bar q\to t\bar t$ (dominant at
the Tevatron) and $gg\to t\bar t$ (dominant at the LHC).  However,
there are also a significant number of top quarks that are produced
singly, via the weak interaction.  There are three separate
single-top-quark production processes, which may be characterized by
the virtuality of the $W$ boson (of four-momentum $q$) in the
process:
\begin{itemize}
\item $t$-channel: The dominant process involves a spacelike
$W$ boson ($q^2 \le 0$), as shown in Fig.~\ref{fig:singletop}(a)
\cite{Willenbrock:1986cr,Yuan:1989tc,Ellis:1992yw}. The virtual $W$
boson strikes a $b$ quark in the proton sea, promoting it to a top
quark. This process is also referred to as $W$-gluon fusion, because
the $b$ quark arises from a gluon splitting to $b\bar b$.
\item $s$-channel: If one rotates the $t$-channel diagram
such that the virtual $W$ boson becomes timelike, as shown in
Fig.~\ref{fig:singletop}(b),
one has another process that produces a single top quark
\cite{Cortese:1991fw,Stelzer:1995mi}. The virtuality of the $W$ boson
is $q^2 \ge (m_t+m_b)^2$.
\item Associated production: A single top quark may also be produced via
the weak interaction in association with a real $W$ boson ($q^2 =
M_W^2$), as shown in Fig.~\ref{fig:singletop}(c)
\cite{Heinson:1996zm,Tait:1999cf}. One of the initial partons is a
$b$ quark in the proton sea, as in the $t$-channel process.
\end{itemize}

The total cross sections for these three single-top-quark production
processes, calculated at next-to-leading-order in QCD, are listed in
Table~\ref{tab:sigma}, along with the cross section for the strong
production of top-quark pairs. Of the single-top processes, the
$t$-channel process has the largest cross section; it is nearly one
third as large as the cross section for top-quark pairs at both the
Tevatron and the LHC. The cross section for the $s$-channel process
is less than half that of the $t$-channel process at the Tevatron,
and is more than an order of magnitude less than the $t$-channel
process at the LHC. The $Wt$ process is negligible at the Tevatron,
but is significant at the LHC, with a cross section intermediate
between the $t$-channel and $s$-channel cross sections.

The cross sections for single-top production are all known at
next-to-leading-order in QCD, and have been calculated with
increasing sophistication over the years, such that they are now all
available as differential cross sections.  The $s$-channel process
has very little theoretical uncertainty
\cite{Smith:1996ij,Harris:2002md,Cao:2004ky,Cao:2004ap,Sullivan:2004ie,
Campbell:2004ch,Frixione:2005vw}, and the total cross section is
even known to next-to-next-to-leading order (in the large $N$ limit)
\cite{Chetyrkin:2000mq}.  The theoretical uncertainty is larger for
the dominant $t$-channel process
\cite{Bordes:1994ki,Stelzer:1997ns,Harris:2002md,Cao:2004ky,
Sullivan:2004ie,Campbell:2004ch,Cao:2005pq,Frixione:2005vw}.  The
$Wt$ process is also known at next-to-leading order
\cite{Tait:1999cf,Zhu:2002uj,Campbell:2004ch,Campbell:2005bb}, and
requires some care to separate out the large contribution from
$t\bar t\to tW\bar b$
\cite{Belyaev:2000me,Tait:1999cf,Campbell:2005bb}. Phenomenological
studies of single-top production have also been carried out with
increasing sophistication
\cite{Stelzer:1998ni,Belyaev:1998dn,Sullivan:2004ie,Sullivan:2005ar,
Bowen:2004my,Bowen:2005xq}.

\begin{table}
\begin{center}
\begin{tabular}{|c|c|c|} \hline
$\sigma$(pb) & Tevatron & LHC \\ \hline $t$-channel & 1.98 & 246 \\
\hline $s$-channel & 0.88 & 10.6 \\ \hline $Wt$  &   0.14 & 68
\\ \hline $t\bar t$ & 6.7 & 860 \\ \hline \hline
\end{tabular}
\caption[fake]{Total cross sections (pb) for single-top-quark
production and top-quark pair production at the LHC, for $m_t= 175$
GeV.  The next-to-leading-order $t$-channel and $s$-channel cross
sections are from Ref.~\cite{Sullivan:2004ie}.  The
next-to-leading-order cross section for the $Wt$ process is from
Ref.~\cite{Campbell:2005bb} (adjusted for $m_t=175$ GeV).  The
next-to-leading-order cross section for $t\bar t$ production is from
Ref.~\cite{Cacciari:2003fi} (Tevatron) and Ref.~\cite{Beneke:2000hk}
(LHC).} \label{tab:sigma}
\end{center}
\end{table}

Within the standard model, there are several reasons for studying
the production of single top quarks at the Tevatron and the LHC.
First, the cross sections for single-top-quark processes are
proportional to $|V_{tb}|^2$. These processes provide the only known
way to directly measure $V_{tb}$. In contrast, the observed fact
that $BR(t\to Wb)\approx 1$ \cite{Acosta:2005hr} only tells us that
$V_{tb}\gg V_{ts},V_{td}$. If there are just three generations of
quarks, as favored by precision electroweak data, then we already
know $V_{tb}=0.9990-0.9992$ at 90\% CL \cite{Eidelman:2004wy}. In
this case single-top production may be regarded as a test of the
standard model, including the generation of the $b$-quark sea from
gluon splitting.

Another reason for studying single-top production is that it these
processes are backgrounds to other signals.  For example,
single-top-quark events are backgrounds to some signals for the
Higgs boson
\cite{Campbell:2005bb,Ladinsky:1990ut,Stange:1994bb,Moretti:1997ng}.
Thus it is important to have a good understanding of single top both
theoretically and experimentally.  Single top will also serve a testing
ground for important theoretical tools needed to correctly model
Higgs physics.  For example, if no signal of physics beyond the Standard
Model is manifest in single top production, the $t$-channel production mode
will server to constrain the bottom quark parton distribution function, 
important for Higgs production from initial states including heavy quarks.
Just as in the weak boson fusion mode of Higgs production, the 
$t$-channel mode also contains a $t$-channel $W$ exchange and the associated
forward tagging jets, and thus single top represents an experimental
insight into a key characteristic of the Higgs signal.

A third reason is that single top quarks are produced with nearly
$100\%$ polarization, due to the weak interaction
\cite{Heinson:1996zm,Carlson:1993dt,Mahlon:1996pn,Mahlon:1999gz}.
This polarization serves as a test of the $V-A$ structure of the
top-quark charged-current weak interaction.

Single top is also interesting beyond the standard model.  New
physics can influence single-top-quark production by inducing
non-standard weak interactions
\cite{Carlson:1993dt,Carlson:1994bg,Datta:1996gg,Tait:1997fe,Hikasa:1998wx,Boos:1999dd,Tait:2000sh,Chen:2005vr}, via loop effects
\cite{Atwood:1996pd,Simmons:1996ws,Li:1996bh,Li:1996ir,Li:1997qf,Bar-Shalom:1997si}, or by providing new sources of single-top-quark events \cite{Tait:1997fe,Tait:2000sh,Simmons:1996ws,Malkawi:1995dm,Datta:1997us,Oakes:1997zg,Han:1998tp,Datta:2000gm}.
The three modes of single top production each respond quite differently to
different realizations of physics beyond the Standard Model
\cite{Tait:2000sh}.  The $s$-channel
mode is very sensitive to an exotic charged boson which couples to top and 
bottom.  Because the exchanged particle is time-like, there is the possibility
(if it is heavier than the top) that it can be produced on-shell, resulting
in a large enhancement of the cross section.  On the other hand, while a
FCNC interaction (such as $Z$-$t$-$c$) would allow new $s$-channel processes
such as $q \overline{q} \rightarrow Z^* \rightarrow t \overline{c}$, these
are difficult to extract from backgrounds, because there is no longer a final
state $b$ quark that can be tagged.  So the experimentally measured $s$-channel
cross section would not include the FCNC events.  Specific theories which
predict an enhancement of the $s$-channel rate are theories with a $W^\prime$
\cite{Chivukula:1995gu,Muller:1996dj,Malkawi:1996fs,He:1999vp,Batra:2003nj,Batra:2004vc}
or charged Higgs,
both of which can result in $s$-channel rates different from the SM by
factors of few at either Tevatron or LHC
\cite{Tait:2000sh,Sullivan:2003xy,Sullivan:2002jt}.

The $t$-channel mode is insensitive to heavy charged bosons.  The reason for
this is that the $t$-channel exchange results in a space-like momentum, which
never can go on-shell, and thus the amplitude for the heavy particle is always
suppressed by the mass of the heavy boson, $1/M_B^2$.  However, the FCNC
processes can have a drastic effect on the $t$-channel mode.  Because they 
involve new interactions between the top quark, a boson ($\gamma$, $Z$, 
$g$, or $H$), and one of the light quarks, ($c$ or $u$), the $t$-channel mode
can be enhanced.  For example, in the case of a $Z$-$t$-$c$ interaction there
is the process $q c \rightarrow q t$ with a $Z$ exchanged.  The fact that
high energy proton collisions contain more $c$ quarks than $b$ quarks
further enhances the new physics contribution compared to the SM piece.

\begin{figure}
  \includegraphics[height=.4\textheight]{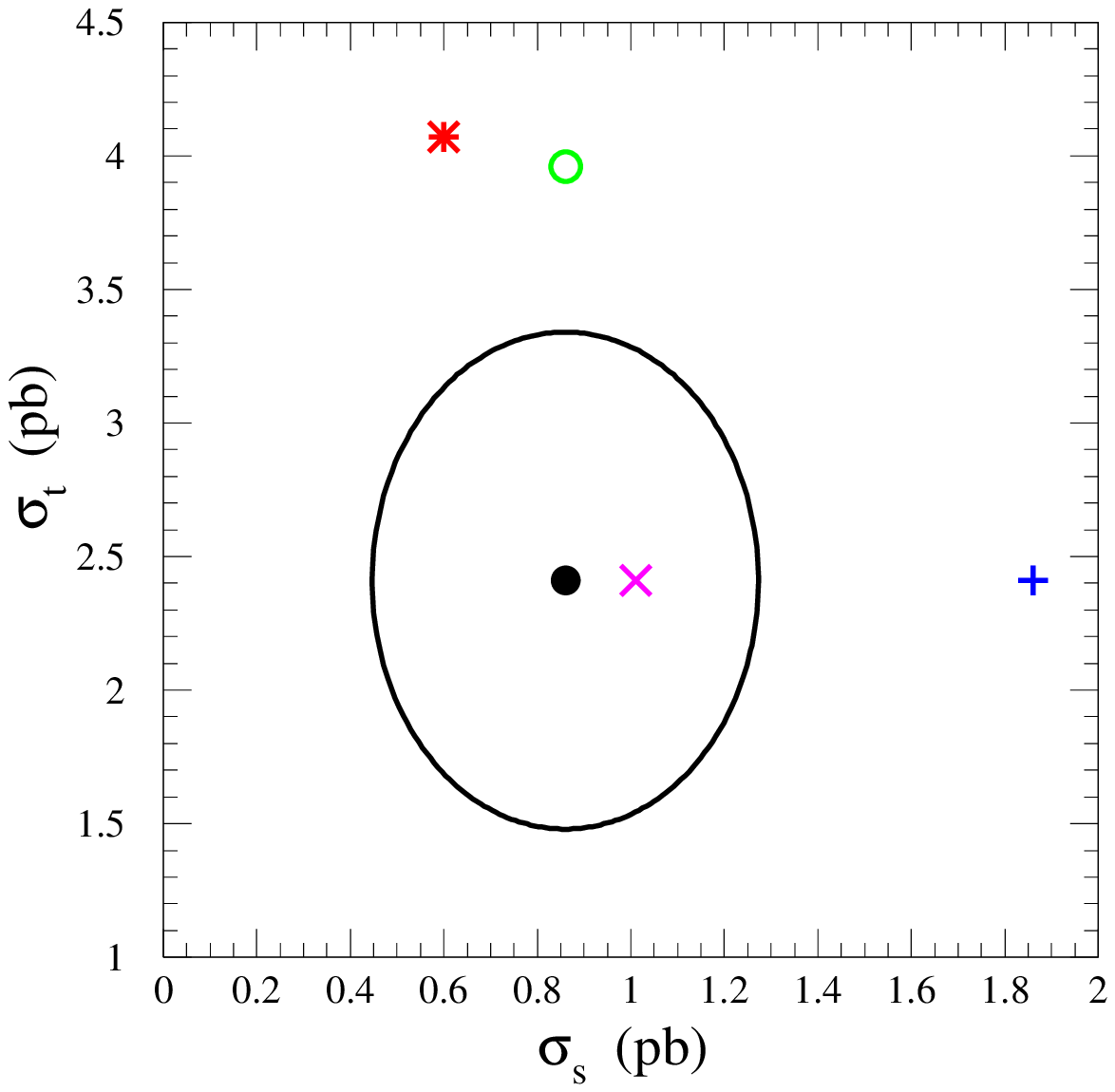}
  \includegraphics[height=.4\textheight]{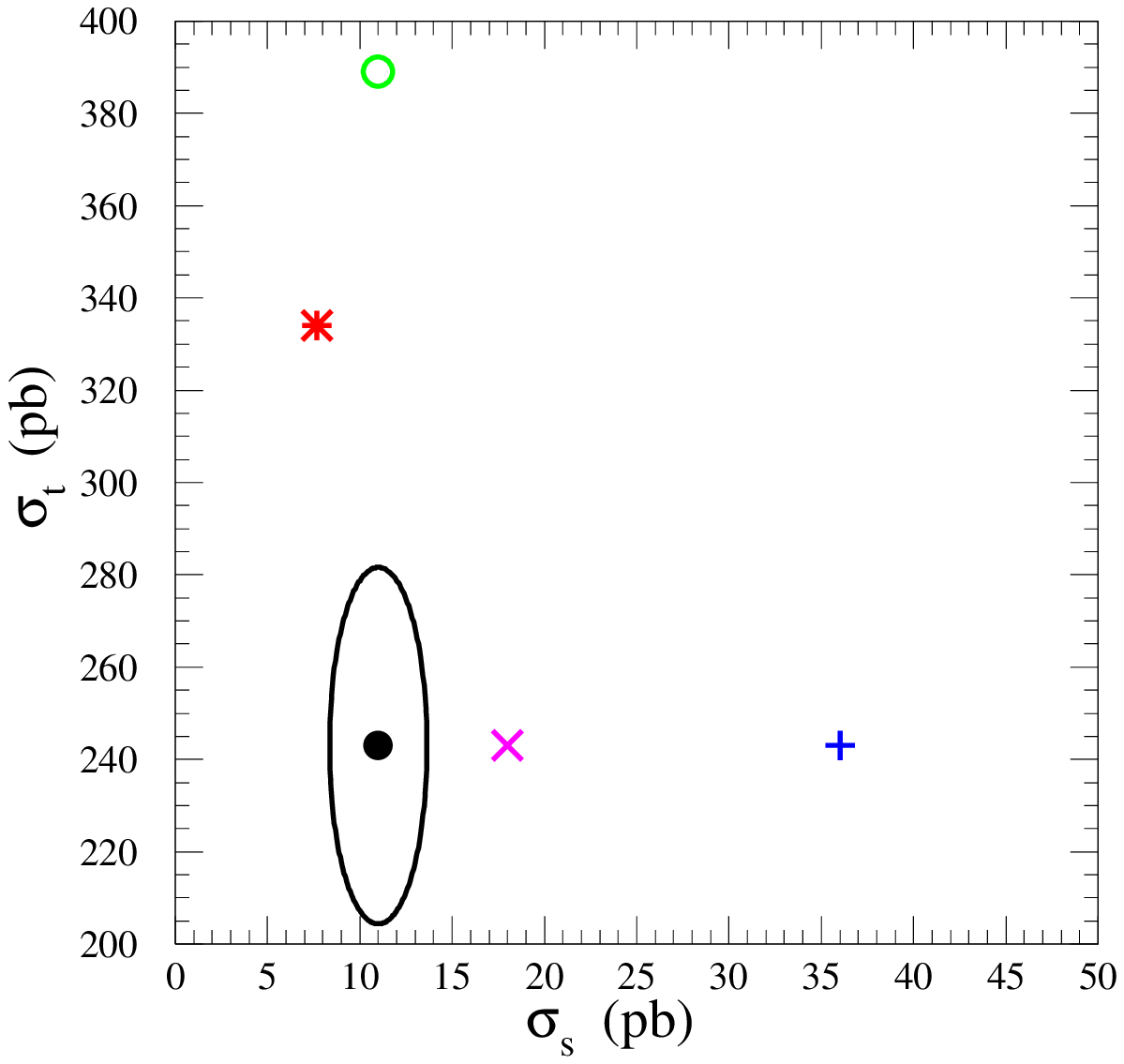}
  \caption{Single top cross sections in the $s$- and $t$-channels in the SM (including theoretical and expected statistical uncertainties) and a few models of physics beyond the SM, at the Tevatron run II and LHC (from Ref.~\protect{\cite{Tait:2000sh}}).}
\label{fig:st-theory}
\end{figure}

The $t W^-$ mode is more or less insensitive to new bosons, because the
$W$ is manifest in the final state.  From this line of thinking, we see
that all three modes are really complimentary views of the top quark,
and thus measured separately they provide more information than would be
obtained by lumping them together into a singular single top process.
This point is emphasized (at Tevatron run II and LHC)
for a few different models in Fig.~\ref{fig:st-theory},
where we also show the SM predictions, and some estimates for the theoretical 
and statistical uncertainties in the $s$- and $t$-channels.



%
%
%
\newcommand{\oalphas}{O(\alpha_{s})}

\subsubsection*{Next-to-Leading Order Corrections to Single Top Quark Production
and Decay}
\label{sec:singletoptheorynlo}

\textbf{Contributed by:~Q.-H.~Cao, R.~Schwienhorst, J.A.~Benitez, R.~Brock, C.-P.~Yuan}\\

In a few recent papers~\cite{Cao:2004ky,Cao:2004ap,Cao:2005pq},
we first developed methods for calculating the next-to-leading (NLO)
order QCD corrections to the production and decay of the top quark in
the $s$- and $t$-channel single-top events produced at hadron colliders,
and then studied the implication of NLO corrections to the phenomenology
of single-top physics at the Tevatron Run-II. In this section,
we first briefly review the method of our calculations and then summarize
the main results of our phenomenological studies.

\paragraph{Method of Calculations}

We adopted the phase space slicing method with one cut-off scale to
organize the NLO calculations~\cite{Giele:1991vf,Giele:1993dj,Keller:1998tf}.
When the invariant mass of the two colored partons in the $2\to3$
tree level production processes is smaller than some theoretical cutoff
scale $\sqrt{s_{min}}$ , collinear and/or soft singularities are
taken care of using the dimensional regularization method, and they
are canceled by similar singularities in the $2\to2$ virtual processes
after redefining the normalized parton distribution functions (PDF).
For the remaining phase space region of the $2\to3$ processes, we
numerically evaluate the final state parton distributions.
By this way, we calculate the differential distributions of final
state partons in the production processes, including both $2\to2$
and $2\to3$ kinematics. A similar procedure was also adopted to handle
the decay of top quark via $t\to bW(\to l\nu)(g)$ at the NLO in QCD.
Again, the soft singularities cancel among the virtual and real gluon
emission contributions and there is no remaining collinear singularity
after integrating out the sliced regions of phase space that correspond
to soft and/or collinear singularities in the tree level process $t\to bWg$.
In our calculation, we have ignored the bottom quark mass, for its
contribution to the matrix element is negligible in single-top processes.
In order to obtain the fully spin-correlated matrix elements, we take
the complete set of Feynman diagrams for the production and
decay of top quark in single-top processes with effective form factors
obtained from summing up both virtual and real emission contributions
(coming from the sliced phase space regions with the invariant mass
of a set of two external partons less than $\sqrt{s_{min}}$). We
have also introduced a new method in our calculation which is called
the modified narrow width approximation. In contrast to taking the
usual narrow width approximation to approximate the internal top-quark
propagator by a delta-function, so as to take the top quark width
to be exactly equal to zero, we have generated a Breit-Wigner resonance 
distribution of top quark mass according to its predicted SM total decay width
at NLO. We then use that generated mass to calculate
the production and decay matrix elements in order to respect gauge
invariance and to clearly separate the production and decay contributions
beyond Born level. By doing so, we are able to generate differential
distributions of final state particles where the reconstructed top
quark invariant mass peaks around the true value of the top quark mass,
and with a Breit-Wigner shape whose width is the top quark total decay
width. Hence, it improves the prediction of NLO calculations in some
kinematic distributions.

\paragraph{Phenomenology of s- and $t$-channel Single Top Quark Events at NLO}

Although all the results of our studies regarding the phenomenology
of $s$- and $t$-channel single-top events predicted by our NLO calculations
have been published in Refs.~\cite{Cao:2004ky,Cao:2004ap,Cao:2005pq},
it is useful to summarize a few key findings from our studies in this
section. 

In order to calculate the fully differential cross sections at
NLO and compare to experimental data, we have to impose kinematic
cuts on the final state partons. Moreover, if the number of signal
events is large, then one would like to impose a tight kinematic cut to
further suppress the backgrounds. However, in some cases, such as
the single-top search at the Tevatron in Run~II, the signal rate
is not large. It is thus not desirable to impose a tight kinematic
cut because that would not only suppress the background rate but also the signal
rate and thus not improve the signal significance compared
to imposing a loose kinematic cut. Furthermore, we must define a jet
as an infrared-safe observable. In our studies, we adopt the cone-jet
algorithm~\cite{Alitti:1990aa}, as explained in Ref.~\cite{Cao:2004ap,Cao:2005pq}.
More specifically, we adopt the $E$-scheme cone-jet approach (4-momenta
of particles in a cone are simply added to form a jet) with radius
$R=\sqrt{\Delta\eta^{2}+\Delta\phi^{2}}$ in order to define $b$,
$q$ and possibly extra $g$, $\bar{q}$, or $\bar{b}$ jets, where
$\Delta\eta$ and $\Delta\phi$ are the separation of particles in
the pseudo-rapidity $\eta$ and the azimuthal angle $\phi$, respectively.
For reference, we shall consider both $R=0.5$ and $R=1.0$. The same
$R$-separation will also be applied to the separation between the
lepton and each jet. 

Below, we discuss a few aspects of the single-top phenomenology studies
based on our calculations for the Tevatron in Run~II, a 1.96~TeV
$p\bar{p}$ collider. Here, we take $m_{t}=178$ GeV and $M_{W}=80.33$ GeV.

\paragraph{Kinematic Acceptance}

The kinematic cuts imposed on the final state objects are:\begin{eqnarray}
P_{T}^{\ell}\ge15\,{\rm GeV} & , & \left|\eta_{\ell}\right|\le\eta_{\ell}^{max},\nonumber \\
\met\ge15\,{\rm GeV} & ,\nonumber \\
E_{T}^{j}\ge15\,{\rm GeV} & , & \left|\eta_{j}\right|\le\eta_{j}^{max},\nonumber \\
\Delta R_{\ell j}\ge R_{cut}\,\, & , & \Delta R_{jj}\ge R_{cut},\label{eq:cuts}\end{eqnarray}
where the jet cuts are applied to both the $b$- and light quark jets
as well as any gluon or antiquark jet in the final state. $\eta_{l}^{max}$
(and $\eta_{j}^{max}$) denotes the maximum value in magnitude of
the charged lepton (and jet) rapidity. The minimum transverse energy
of the lepton and jets is chosen to be 15~GeV. Each event is furthermore
required to have at least one charged lepton and two jets passing
all selection criteria. The cut on the separation in $R$ between
lepton and jets as well as between different jets is given by $R_{cut}$.
In Table~\ref{tab:xsec-accept}, we show the $s$- and $t$-channel
single-top production cross sections (in femtobarns) , including the top
quark decay branching ratio $t\rightarrow bW(\rightarrow e\nu)$, as
well as acceptances at leading order (LO) and NLO for several
sets of cuts. We apply the $E_{T}$ cuts listed in Eq.~(\ref{eq:cuts})
and study three separate sets of values:

\begin{enumerate}
\item loose cuts with small $R_{cut}$: $\eta_{l}^{max}=2.5$, $\eta_{j}^{max}=3.0$,
and $R_{cut}=0.5$,
\item loose cuts with large $R_{cut}$: $\eta_{l}^{max}=2.5$,$\eta_{j}^{max}=3.0$,
and $R_{cut}=1.0$, 
\item tight cuts with small $R_{cut}$: $\eta_{l}^{max}=1.0$, $\eta_{j}^{max}=2.0$,
and $R_{cut}=0.5$. 
\end{enumerate}
As clearly illustrated in Table~\ref{tab:xsec-accept}, the
acceptance for single-top signal events is sensitive to the applied
kinematic selections. A larger value for $R_{cut}$ reduces the acceptance
significantly mainly because more events fail the lepton-jet separation
cut. With tight cuts, LO and NLO acceptances are almost the same.
By contrast, with loose cuts, LO and NLO acceptances are quite different.
The important lesson here is that with a loose cut, to keep most of
the signal events, the acceptance for NLO kinematics cannot be
accurately modeled wit a multiplicative $K$-factor
(to scale the inclusive cross section from LO to NLO).

\begin{table}[htbp]
\begin{center}
\begin{tabular}{|c|c|c|c|c|c|c|c|c|}
\hline 
&
\multicolumn{4}{c|}{$s$-channel}&
\multicolumn{4}{c|}{$t$-channel}\tabularnewline
\hline 
&
\multicolumn{2}{c|}{$\sigma[fb]$}&
\multicolumn{2}{c|}{Accept. (\%)}&
\multicolumn{2}{c|}{$\sigma[fb]$}&
\multicolumn{2}{c|}{Accept. (\%)}\tabularnewline
\hline 
&
LO&
NLO&
LO&
NLO&
LO&
NLO&
LO&
NLO\tabularnewline
\hline 
(1)&
22.7&
32.3&
73&
64&
65.6&
64.0&
66&
61\tabularnewline
\hline 
(2)&
19.0&
21.7&
61&
46&
56.8&
48.1&
57&
46\tabularnewline
\hline 
(3)&
14.7&
21.4&
47&
45&
31.1&
34.0&
31&
32\tabularnewline
\hline
\end{tabular}
\end{center}
\caption{The $s$- and $t$-channel single-top production cross sections (in
$fb$) and acceptance at the Tevatron in Run~II under various scenarios. The decay
branching ratio $t\rightarrow bW(\rightarrow e\nu)$ is included.
\label{tab:xsec-accept}}
\end{table}

\paragraph{Top Quark Reconstruction}

\begin{figure}[htbp]
\begin{center}
\includegraphics[height=.38\textheight]{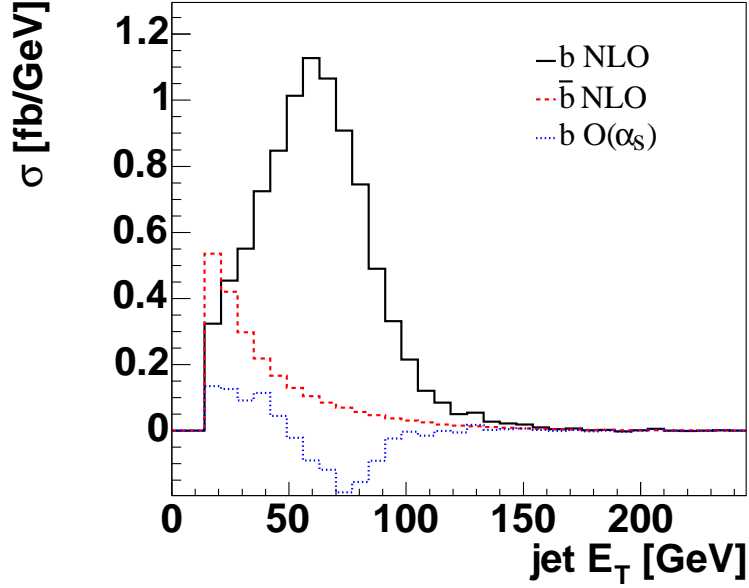}
\end{center}
\caption{Transverse momentum of the $b$- and $\bar{b}$-jets in the $t$-channel single-top
process. The dotted curve shows the $O(\alpha_{s})$ correction to the
$b$-jet $E_{T}$ distribution.}
\label{fig:ptbbbar-tchan}
\end{figure}

In order to identify single-top signal events and to test the polarization
of the top quark by studying spin correlations amongst the final state
particles, we need to reconstruct the top quark in each single top event. To
do so, we need to first identify the $b$-jet and reconstruct the
$W$ boson from the top decay. In Table~\ref{tab:efficiency-bnu}, we
show the efficiency of finding the correct $b$-jet ($\epsilon_{b}$)
in two different algorithms: the best-jet algorithm and the leading $b$-tagged
jet algorithm. The ``best-jet'' is defined to be the $b$-tagged
jet which gives an invariant mass closest to the true top mass when
it is combined with the reconstructed $W$ boson after determining
the longitudinal momentum $p_{z}^{\nu}$ of the neutrino from the $W$
decay. The leading $b$-tagged jet algorithm picks the leading $b$-tagged
jet as the correct $b$-jet to reconstruct the top quark after combining
with the reconstructed $W$ boson. As shown in Refs.~\cite{Cao:2004ap,Cao:2005pq},
we find that the best-jet algorithm shows a higher efficiency (about
$80\%$) in picking up the correct $b$-jet than the leading-jet algorithm
(about $55\%$) for $s$-channel single-top events. On the other
hand, for $t$-channel single-top events, the leading $b$-tagged
jet algorithm picks up the correct $b$-jet with a higher efficiency,
about $95\%$ for inclusive 2-jet events and $90\%$ for exclusive
3-jet events. The reason that the leading $b$-tagged jet algorithm
works well in exclusive 3-jet $t$-channel single-top events is
that there are distinct kinematic differences between $b$ and $\bar{b}$-jets.
In Fig.~\ref{fig:ptbbbar-tchan}, we show the inclusive $b$- and
$\bar{b}-$jet $E_{T}$ distributions in $t$-channel single-top events. 
To reconstruct the top quark in
signal events, we also need to reconstruct the $W$ boson,
which is done with the help of a mass constraint:
$M_{W}^{2}=(p_{l}+p_{\nu})^{2}$. Which of the two-fold solutions
in $p_{z}^{\nu}$ is chosen depends on the $b$-jet algorithm we
use. In the case of the best-jet algorithm, we find that the one with the
smaller magnitude gives the best
efficiency in $W$ boson reconstruction. In the case of leading $b$-tagged
jet algorithm however, 
we use the top quark mass constraint $M_{t}^{2}=(p_{b}+p_{l}+p_{\nu})^{2}$
to pick up the best $p_{z}^{\nu}$ value. The efficiency for picking
up the correct $p_{z}^{\nu}$ value ($\epsilon_{\nu}$), at LO and
NLO, respectively, is presented in Table~\ref{tab:efficiency-bnu}.

\begin{table}[htbp]
\begin{center}
\begin{tabular}{|c|c|c|c|c|c|c|}
\hline 
&
\multicolumn{3}{c|}{best-jet algorithm}&
\multicolumn{3}{c|}{leading $b$-tagged jet algorithm}\tabularnewline
\hline 
&
$s$-channel&
\multicolumn{2}{c|}{$t$-channel}&
$s$-channel&
\multicolumn{2}{c|}{$t$-channel}\tabularnewline
\hline 
&
incl. 2-jet&
incl. 2-jet&
excl. 3-jet&
incl. 2-jet&
incl. 2-jet&
excl. 3-jet\tabularnewline
\hline 
$\epsilon_{b}$&
$80\%$&
$80\%$&
$72\%$&
$55\%$&
$95\%$&
$90\%$\tabularnewline
\hline 
$\epsilon_{\nu}$&
\multicolumn{3}{c|}{$70\%$}&
\multicolumn{3}{c|}{$84\%$}\tabularnewline
\hline
\end{tabular}
\end{center}
\caption{Efficiencies of identifying correct $b$-jet ($\epsilon_{b}$) and
picking up correct $p_{z}^{\nu}$ ($\epsilon_{\nu}$) in both the best-jet
algorithm and the leading-jet algorithm. \label{tab:efficiency-bnu}}
\end{table}

\paragraph{Top Quark Polarization}

\begin{figure}[htbp]
\begin{center}
\includegraphics[%
  width=1.0\linewidth,
  keepaspectratio]{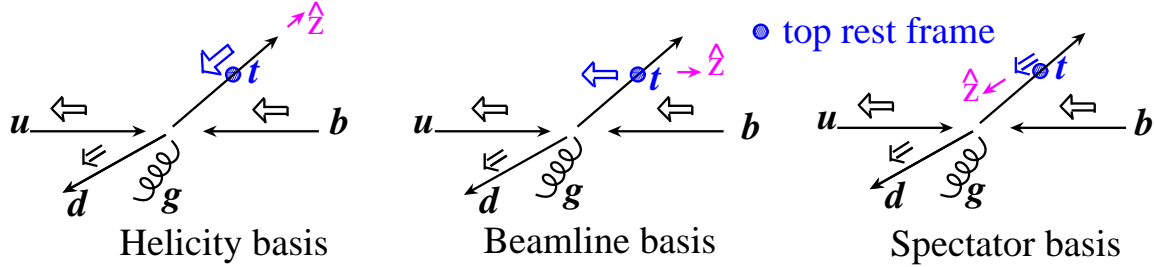}
\end{center}
\caption{Illustration of the three choices for the top quark spin basis. The
circle denotes the top quark rest frame and the blue arrows denote
the top quark spin direction.\label{fig:TopSpinBasis-tchan}}
\end{figure}

Although the top quark is produced via the left-handed charged current,
there is no reason to believe that the helicity basis will give the
best description of the top quark spin. Choosing an appropriate basis
could maximize spin correlation effects. Two definitions for the polarization
have been studied in the literature for $s$-channel processes, differing
by the reference frame used to define the polarization: one calculation
uses the helicity basis, another the so-called {}``optimal'' basis~\cite{Mahlon:1995zn,Parke:1996pr}.
Both work in the top quark rest frame, but they have different reference
axis for the top quark spin, cf. Fig.~\ref{fig:TopSpinBasis-tchan}.
In the more common helicity basis the top quark spin is measured along
the top quark direction of motion in the center-of-mass (c.m.) frame
which is chosen as the frame of the (reconstructed top quark, non-best-jet)
system after event reconstruction. In the optimal basis (beamline
basis) we can maximize the spin correlations by taking advantage of
the fact that the top quark produced through the $s$-channel single
top quark processes is almost $100\%$ polarized along the direction
of the $d$-type quark. 

When studying the top polarization in helicity basis, the c.m.
frame needs to be reconstructed in order to define the top quark momentum.
Due to additional jet radiation, the determination of the c.m. frame
at NLO is more complicated than at the Born-level. The additional
radiation will also blur the spin correlation, and the degree of reduction
depends on the chosen reference frame. Therefore, choosing
the appropriate frame will reduce this effect. In this study, two
options for reconstructing the c.m. frame are investigated:

\begin{enumerate}
\item $t\bar{b}(j)$-frame: the c.m. frame of the incoming partons. This
is the rest frame of all the final state objects (reconstructed top
quark and all other jets). In exclusive two-jet events, this frame
is the same as that at the Born-level, i.e. reconstructed from summing
over momentum of the top quark and non-best-jet. In exclusive three-jet
events, this frame is reconstructed by summing over the 4-momenta
of top quark, non-best-jet, and the third-jet from the parton level
calculation. 
\item $t\bar{b}$-frame: the c.m. frame of the top quark and non-best-jet.
In this case, even in the exclusive three-jet events, the reference
frame is constructed by summing over only the 4-momenta of the top
quark and non-best-jet. Note that this differs from the $t\bar{b}(j)$-frame
only in exclusive three-jet events.
\end{enumerate}
To better quantify the top quark polarization, it is useful to define
the degree of polarization $\mathcal{D}$ of the top quark. This is
given as the ratio\begin{equation}
\mathcal{D}=\frac{N_{-}-N_{+}}{N_{-}+N_{+}},\label{eq:DegreeOfPolarization_schan}\end{equation}
 where $N_{-}$ ($N_{+}$) is , the number of left-hand (right-hand)
polarized top quarks in the helicity basis. Similarly, in the optimal
basis, $N_{-}$ ($N_{+}$) is the number of top quarks with polarization
against (along) the direction of the anti-proton three momentum in
the top quark rest frame. Based on the degree of polarization $\mathcal{D}$,
we can easily get the spin fractions $\mathcal{F}_{\pm}$ as:\begin{eqnarray}
\mathcal{F}_{-} & = & \frac{N_{-}}{N_{-}+N_{+}}=\frac{1+\mathcal{D}}{2},\nonumber \\
\mathcal{F}_{+} & = & \frac{N_{+}}{N_{-}+N_{+}}=\frac{1-\mathcal{D}}{2}.\label{eq:FractionOfPolarization_schan}\end{eqnarray}
Note that $\mathcal{F}_{-}$($\mathcal{F}_{+}$) is the fraction of
left-handed (right-handed) polarized top quarks in the helicity basis.
Similar, in the optimal basis, $\mathcal{F}_{-}$($\mathcal{F}_{+}$)
is the fraction of top quarks with polarization against (along) the
direction of the anti-proton three momentum in the top quark rest
frame. In Table~\ref{tab:toppol-schan-2jet}, we show the prediction
on the top quark polarization in $s$-channel single-top events
at the LO and NLO for various choices of polarization basis and c.m.
frame of the hard-scattering parton system where the polarization
of the top quark is defined. One important observation is that the
measured value of the degree of polarization of the top quark strongly
depend on the algorithm for reconstructing the top quark in $s$-channel
single-top events. For example, at the parton level with known identity
of every final state particle, and before imposing any kinematic selection,
the optimal basis gives the largest degree of polarization, but after
event reconstruction it gives almost the same prediction as the helicity basis.

\begin{table}[htbp]
\begin{center}
\begin{tabular}{|cc|c|c|c|c|}
\hline 
&
&
\multicolumn{2}{c|}{$\mathcal{D}$ }&
\multicolumn{2}{c|}{\textbf{$\mathcal{F}$}}\tabularnewline
\cline{3-4} \cline{5-6} 
&
&
LO&
NLO&
LO&
NLO\tabularnewline
\hline
Helicity basis: &
Parton($t\bar{b}(j)$-frame)&
0.63&
0.54&
0.82&
0.77\tabularnewline
&
Parton($t\bar{b}$-frame)&
0.63&
0.58&
0.82&
0.79\tabularnewline
&
Recon. ($t\bar{b}(j)$-frame)&
0.46&
0.37&
0.73&
0.68\tabularnewline
&
Recon. ($t\bar{b}$-frame)&
0.46&
0.37&
0.73&
0.68\tabularnewline
\hline
Optimal basis: &
Parton &
-0.96&
-0.92&
0.98&
0.96\tabularnewline
&
Recon.&
-0.48&
-0.42&
0.74&
0.71\tabularnewline
\hline
\end{tabular}
\end{center}
\caption{Degree of polarization $\mathcal{D}$ and polarization fraction $\mathcal{F}$,
for inclusive two-jet $s$-channel single top quark events, at the parton
level (Parton) and after event reconstruction (Recon.). Here, $\mathcal{F}$
corresponds to $\mathcal{F}_{-}$ in the helicity basis for left-handed
top quarks and to $\mathcal{F}_{+}$ in the optimal basis for top
quarks with polarization along the direction of anti-proton three
momentum, respectively. The $t\bar{b}g$ frame in the helicity basis
denotes the c.m. frame of the incoming partons while $t\bar{b}$ frame
denotes the rest frame of the reconstructed top quark and $\bar{b}$
quark. \label{tab:toppol-schan-2jet}}
\end{table}

In $t$-channel single-top events, the most studied
polarization bases are the helicity basis, the beamline basis, and
the so-called {}``spectator'' basis~\cite{Mahlon:1996pn}. In the
more commonly used helicity basis, the top quark spin is measured
along the top quark direction of motion in the c.m. frame
which is chosen as the frame of the (reconstructed top quark, spectator
jet) system after event reconstruction. In the beamline basis, the
top quark spin is measured along the incoming proton direction. In
the spectator basis we can maximize spin correlations by taking advantage
of the fact that the top quark produced through the $t$-channel single
top processes is almost $100\%$ polarized along the direction of
the spectator quark. In the discussion below, we will examine the
polarization of single top quark events in these three bases. 

As same as the $s$-channel study, two options for reconstructing
the c.m. frame in the helicity basis are investigated:

\begin{enumerate}
\item $tq(j)$-frame: the c.m. frame of the incoming partons. This is the
rest frame of all the final state objects (reconstructed top quark
and all others jets). In exclusive two-jet events, this frame is the
same as the c.m. frame at the Born-level, i.e. reconstructed from
summing over momentum of the top quark and spectator jet. In exclusive
three-jet events, this frame is reconstructed by summing over the
4-momenta of top quark, spectator jet, and the third-jet from our
parton level calculation. 
\item $tq$-frame: the c.m. frame of the top quark and spectator jet. In
this case, even in exclusive three-jet events, the reference frame
is constructed by summing over only the 4-momenta of the top quark
and spectator jet. Note that this differs from the $tq(j)$-frame
only in exclusive three-jet events.
\end{enumerate}
In Table~\ref{tab:toppol-2jet-tchan}, we present our results for
inclusive two-jet events at the parton level before selection cuts
and after the loose set of cuts and event reconstruction. Our study
shows that the helicity basis (using the $tq$-frame) and the spectator
basis are equally good to study the top quark polarization. Unlike
the $s$-channel process in which the $W$-boson is not perfectly
reconstructed in the best-jet algorithm and thus the polarization
measurement is significantly degraded after event reconstruction,
using the leading $b$-tagged jet and the top mass constraint gives
excellent final state reconstruction in the $t$-channel process,
and the degree of top quark polarization is only somewhat degraded
after event reconstruction. 

\begin{table}[htbp]
\begin{center}
\begin{tabular}{|cc|c|c|c|c|}
\hline 
&
&
\multicolumn{2}{c|}{$\mathcal{D}$ }&
\multicolumn{2}{c|}{\textbf{$\mathcal{F}$}}\tabularnewline
\cline{3-4} \cline{5-6} 
&
&
LO&
NLO&
LO&
NLO\tabularnewline
\hline
Helicity basis: &
Parton($tq(j)$-frame)&
0.96&
0.74&
0.98&
0.87\tabularnewline
&
Parton($tq$-frame)&
0.96&
0.94&
0.98&
0.97\tabularnewline
&
Recon.($tq(j)$-frame)&
0.84&
0.73&
0.92&
0.86\tabularnewline
&
Recon. ($tq$-frame)&
0.84&
0.75&
0.92&
0.88\tabularnewline
\hline
Spectator basis: &
Parton&
-0.96&
-0.94&
0.98&
0.98\tabularnewline
&
Recon.&
-0.85&
-0.77&
0.93&
0.89\tabularnewline
\hline
Beamline basis:&
Parton&
-0.34&
-0.38&
0.67&
0.69\tabularnewline
&
Recon.&
-0.30&
-0.32&
0.65&
0.66\tabularnewline
\hline
\end{tabular}
\end{center}
\caption{Degree of polarization $\mathcal{D}$ and polarization fraction $\mathcal{F}$
for inclusive two-jet $t$-channel single top quark events, at the parton
level (Parton) before cuts and after selection cuts and event reconstruction
(Recon.). Here, $\mathcal{F}$ corresponds to $\mathcal{F}_{-}$ in
the helicity basis for left-handed top quarks and to $\mathcal{F}_{+}$
in the spectator and beamline bases for top quarks with polarization
along the direction of the spectator-jet and proton three momentum,
respectively. Also, the $tq(j)$-frame in the helicity basis denotes
the c.m. frame of the incoming partons, while the $tq$-frame denotes
the rest frame of the top quark and spectator jet. \label{tab:toppol-2jet-tchan}}
\end{table}

\paragraph{Single-Top Events as Background to Higgs Search}

The $s$-channel single top quark process also contributes as one
of the major backgrounds to the SM Higgs searching channel $q\bar{q}\rightarrow WH$
with $H\rightarrow b\bar{b}$. In this case it is particularly important
to understand how the $\oalphas$ corrections change kinematic distributions
around the Higgs mass region. 

\begin{figure}[htbp]
\begin{center}
\includegraphics[%
  scale=0.4]{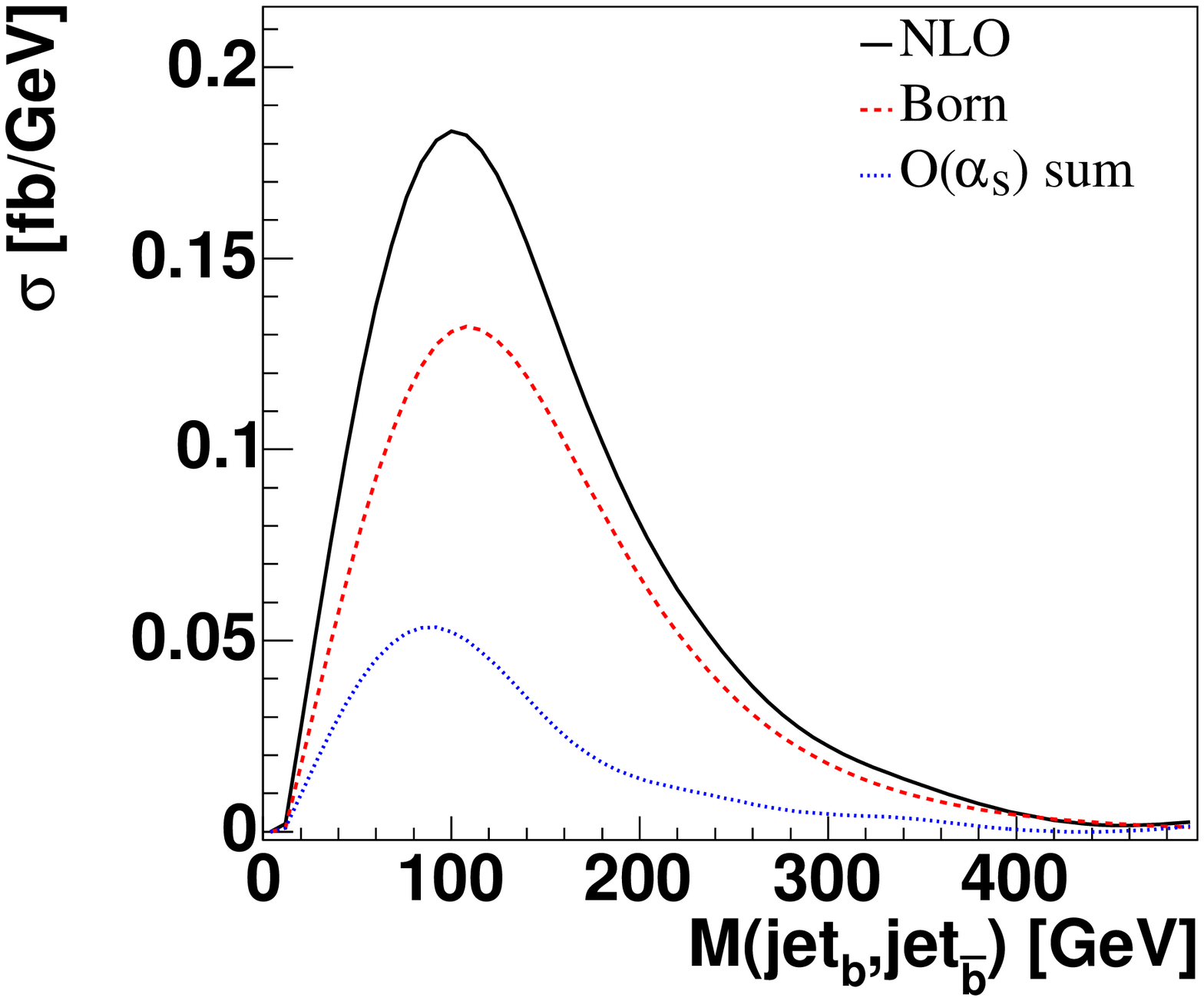}\hspace*{-4mm}\includegraphics[%
  scale=0.4]{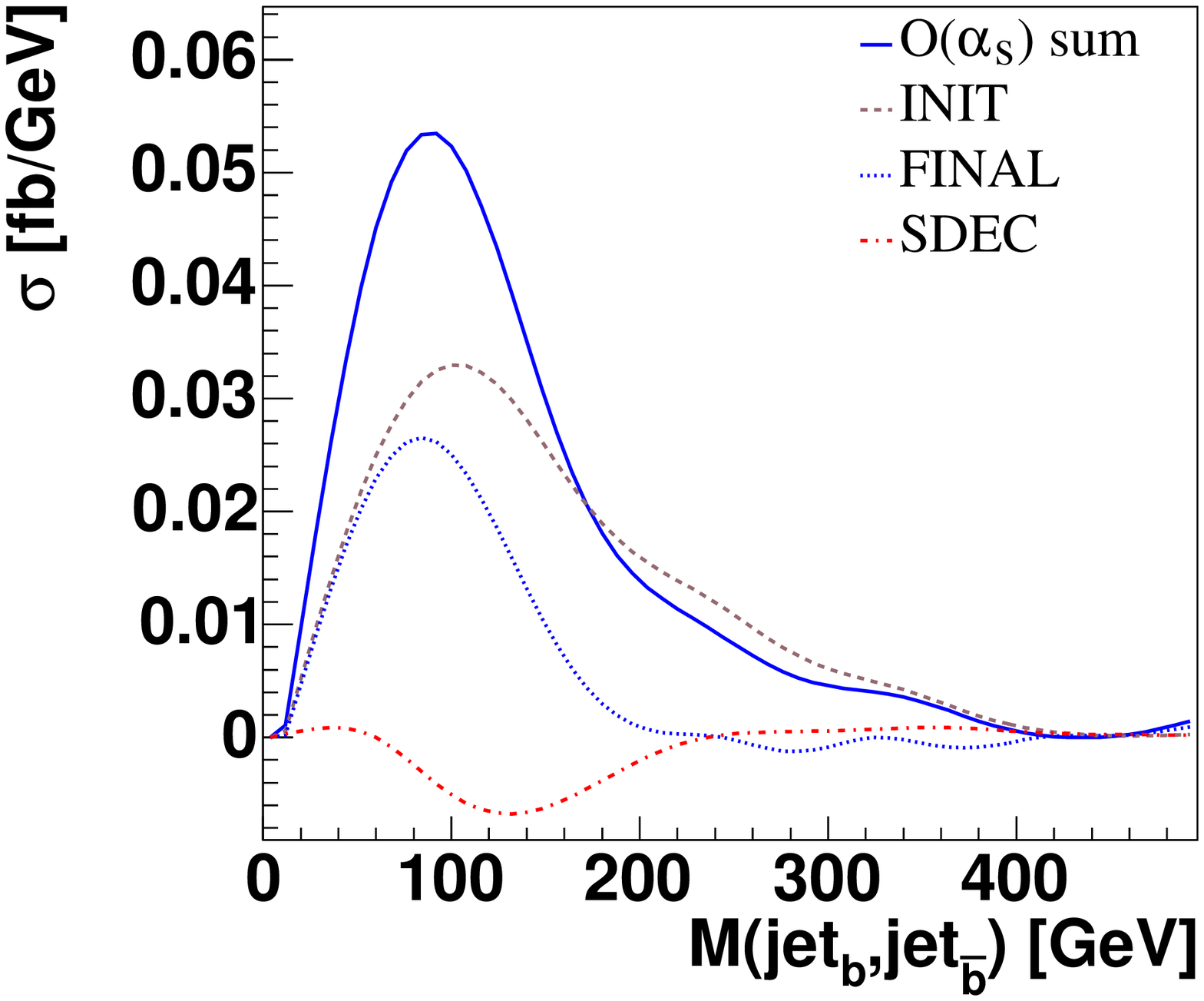}
\end{center}
\caption{Invariant mass of the ($b$-jet, $\bar{b}$-jet) system after selection
cuts, comparing Born-level to $\oalphas$ corrections. In the legend,
INIT, FINAL and SDEC denotes the contributions from initial state,
final state and top quark decay corrections, respectively.\label{fig:bJetbbarJetMass-schan}}
\end{figure}

Because of the scalar property of the Higgs boson, its decay products
$b$ and $\bar{b}$ have symmetric distributions. Fig.~\ref{fig:bJetbbarJetMass-schan}
shows the invariant mass distribution of the ($b$-jet, $\bar{b}$-jet)
system. For a Higgs signal, this invariant mass of the two
$b$-tagged jets would correspond to a plot of the reconstructed Higgs
mass. Thus, understanding this invariant mass distribution will be
important to reach the highest sensitivity for Higgs boson searches
at the Tevatron. The figure shows that at $\oalphas$, the invariant
mass distribution not only peaks at lower values than at Born level,
it also drops off faster. This change in shape is particularly relevant
in the region focused on by SM Higgs boson searches of $80\,{\rm GeV}\leq m_{b\bar{b}}\leq140\,{\rm GeV}$
which is also at the $fb$ level. In particular, the NLO contribution
from the decay of top quark, while small in its overall rate, has
a sizable effect in this region of the invariant mass and will thus
have to be considered in order to make reliable background predictions
for the Higgs boson searches. 

\begin{figure}[htbp]
\begin{center}
\includegraphics[%
  scale=0.4]{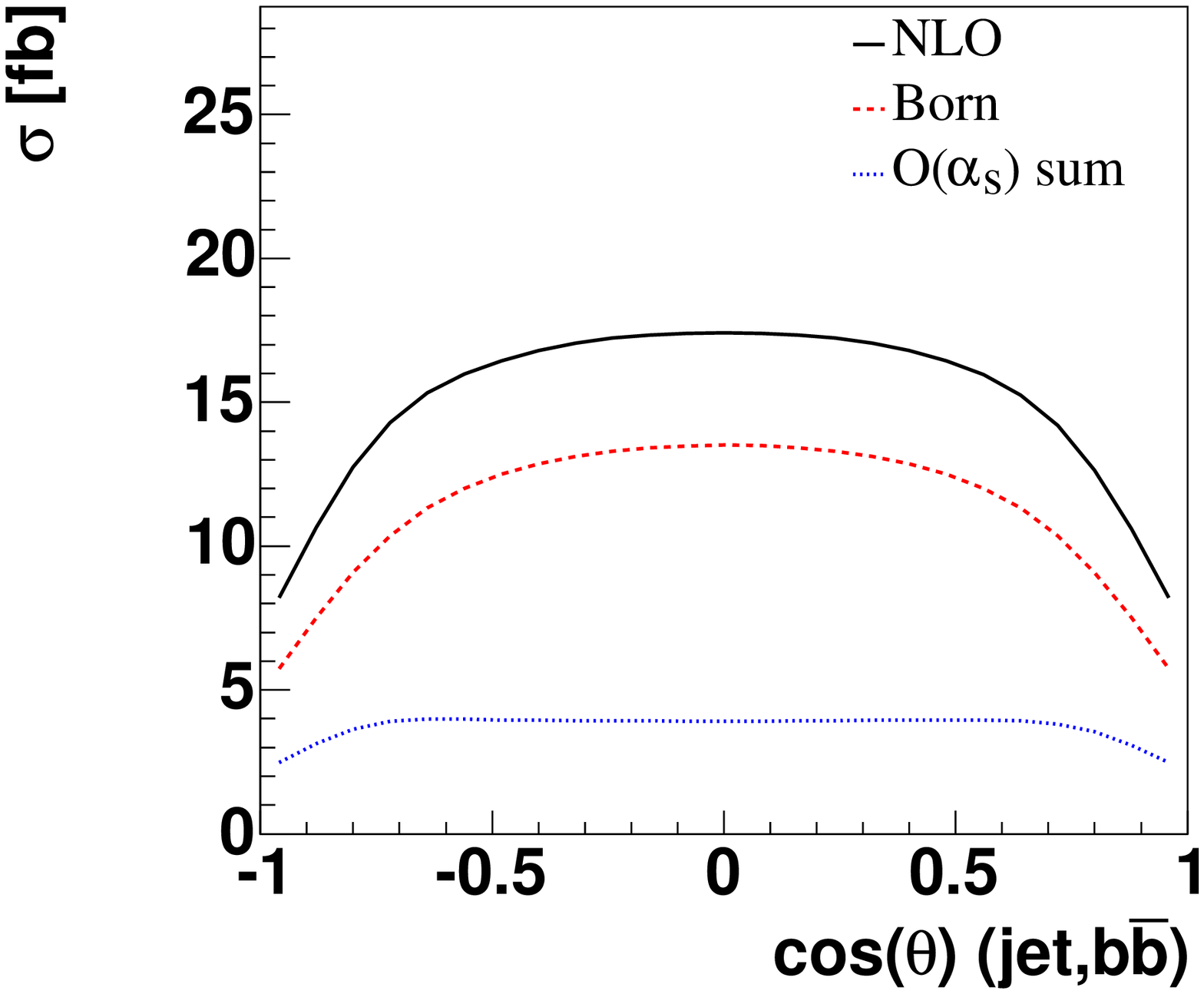}\hspace*{-4mm}\includegraphics[%
  scale=0.4]{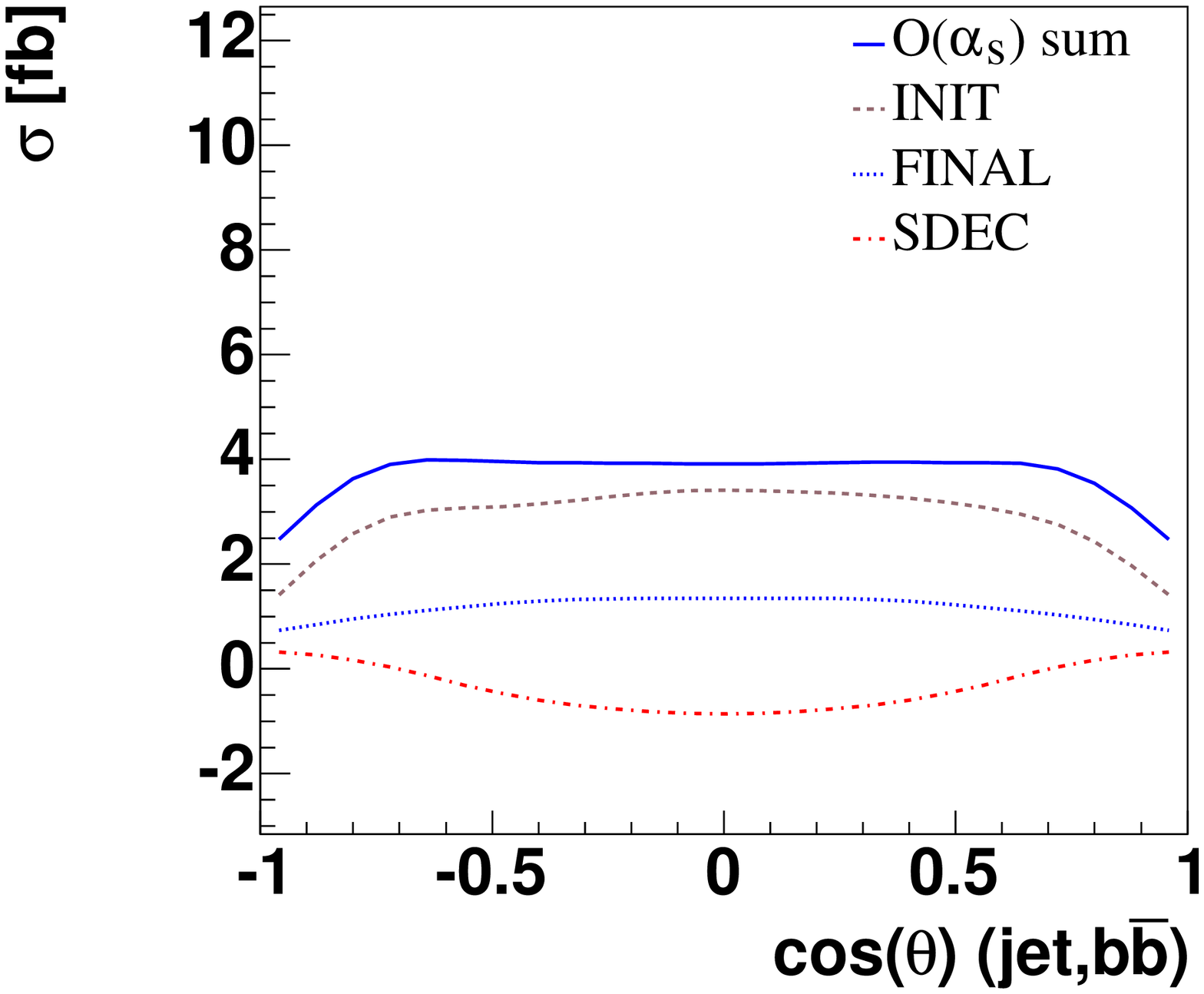}
\end{center}
\caption{Angular distance $cos\theta$ between a $b$-tagged jet and the ($b$
jet, $\bar{b}$ jet) system after selection cuts, comparing Born-level
to $\oalphas$ corrections. \label{fig:bJetbbarJetCosTheta-schan}}
\end{figure}

Other kinematic distributions are also changing in shape when going
from Born-level to $\oalphas$. Fig.~\ref{fig:bJetbbarJetCosTheta-schan}
shows the distribution of $\cos\theta$ for the two $b$-tagged jets,
where $\theta$ is the angle between the direction of a $b$-tagged
jet and the direction of the ($b$-jet, $\bar{b}$-jet) system, in
the rest frame of the ($b$-jet, $\bar{b}$-jet) system. Experiments
cannot distinguish between the $b$- and the $\bar{b}$-jets, we therefore
include both the $b$-jet and the $\bar{b}$-jet in the graph. This
distribution is generally flat at Born-level, with a drop-off at high
$\cos\theta$ due to jet clustering effects, and a drop-off at negative
$\cos\theta$ due to kinematic selection cuts. The $\oalphas$ corrections change
this distribution significantly and result in a more forward peak, 
similar to what is expected in Higgs boson production. 
In other words, a flatter $cos\theta$ distribution in $s$-channel single-top events
makes it more difficult to separate $WH$ events from the $s$-channel single top
background in an experimental analysis.

\paragraph{Connection to Higgs Boson Searches at the LHC}

One of the most important tasks at the CERN Large Hadron Collider
(LHC) is to find the Higgs boson, denoted as $H$. It has been shown
extensively in the literature that the Higgs boson production mechanism
via weak gauge boson fusion is an important channel for
Higgs boson searches. Furthermore, to test whether it is a SM Higgs
boson after the discovery, one needs to determine the coupling $H-V-V$,
where $V$ denotes either $W^{\pm}$ or $Z$, by measuring the production
rate of $q\bar{q}(VV)\to Hq^{\prime}\bar{q}^{\prime}$ via the weak boson
fusion processes. In order to suppress the large background rates,
one usual trick is to tag one of the two forward-jets resulting from emitting a vector
boson $V$ which produces the Higgs boson via $VV\to H$. Prior to the discovery
of Higgs boson, one can learn about the detection efficiency for
forward jets from studying the $s$-channel single-top process. This
is because in the $s$-channel single-top process, the forward jet
also results from emitting a $W$ boson which interacts with the 
$b$~quark from the other hadron beam to produce the heavy top quark. As
pointed out in Ref.~\cite{Yuan:1989tc}, in the effective-$W$ approximation,
a high-energy $t$-channel single top quark event is dominated by
a longitudinal $W$~boson and the $b$~quark fusion diagram. It
is the same effective longitudinal $W$~boson that dominates the
production of a heavy Higgs boson at high energy colliders via the
$W$-boson fusion process. For a heavy SM Higgs boson, the longitudinal
$W$~boson fusion process dominates the Higgs boson production rate.
Therefore, it is also important to study the kinematics of the spectator
jet in $t$-channel single top quark events in order to have a better
prediction for the kinematics of Higgs boson events via the $WW$
fusion process at the LHC.

\begin{figure}[htbp]
\begin{center}
\includegraphics[%
  width=0.55\linewidth,
  keepaspectratio]{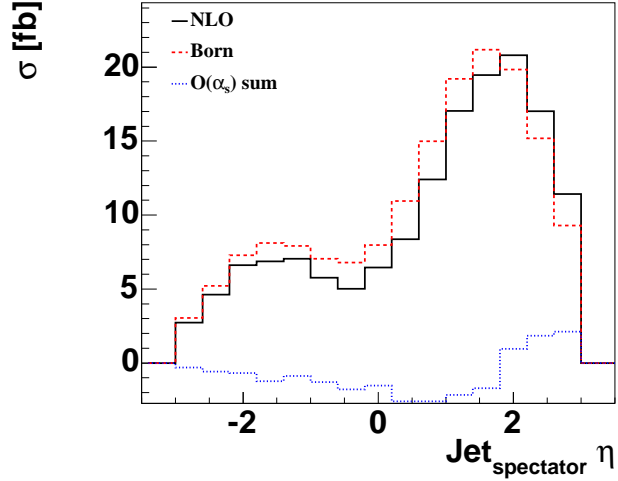}

\end{center}
\caption{Pseudo-rapidity $\eta$ of the spectator jet in $t$-channel single-top
events produced at the Tevatron in Run~II, 
after imposing kinematic selection cuts,
comparing Born-level to $\oalphas$ corrections. \label{fig:spectator-tchan}}
\end{figure}

The unique signature of the $t$-channel single top process is the
spectator jet in the forward direction, which can be utilized to suppress
the copious backgrounds, such as $Wb\bar{b}$ and $t\bar{t}$ production.
Studying the kinematics of this spectator jet is important in order to have
a better prediction of the acceptance for $t$-channel single top quark
events and of the distribution of several important kinematic variables.
Below, we discuss the impact of NLO QCD corrections on the kinematic
properties of the spectator jet. Here, we again concentrate on 
Tevatron Run~II phenomenology and show in Fig.~\ref{fig:spectator-tchan}
the pseudo-rapidity distribution of the spectator jet at
LO and NLO for comparison. 

\begin{figure}[htbp]
\subfigure[]{\includegraphics[%
  width=0.50\linewidth,
  keepaspectratio]{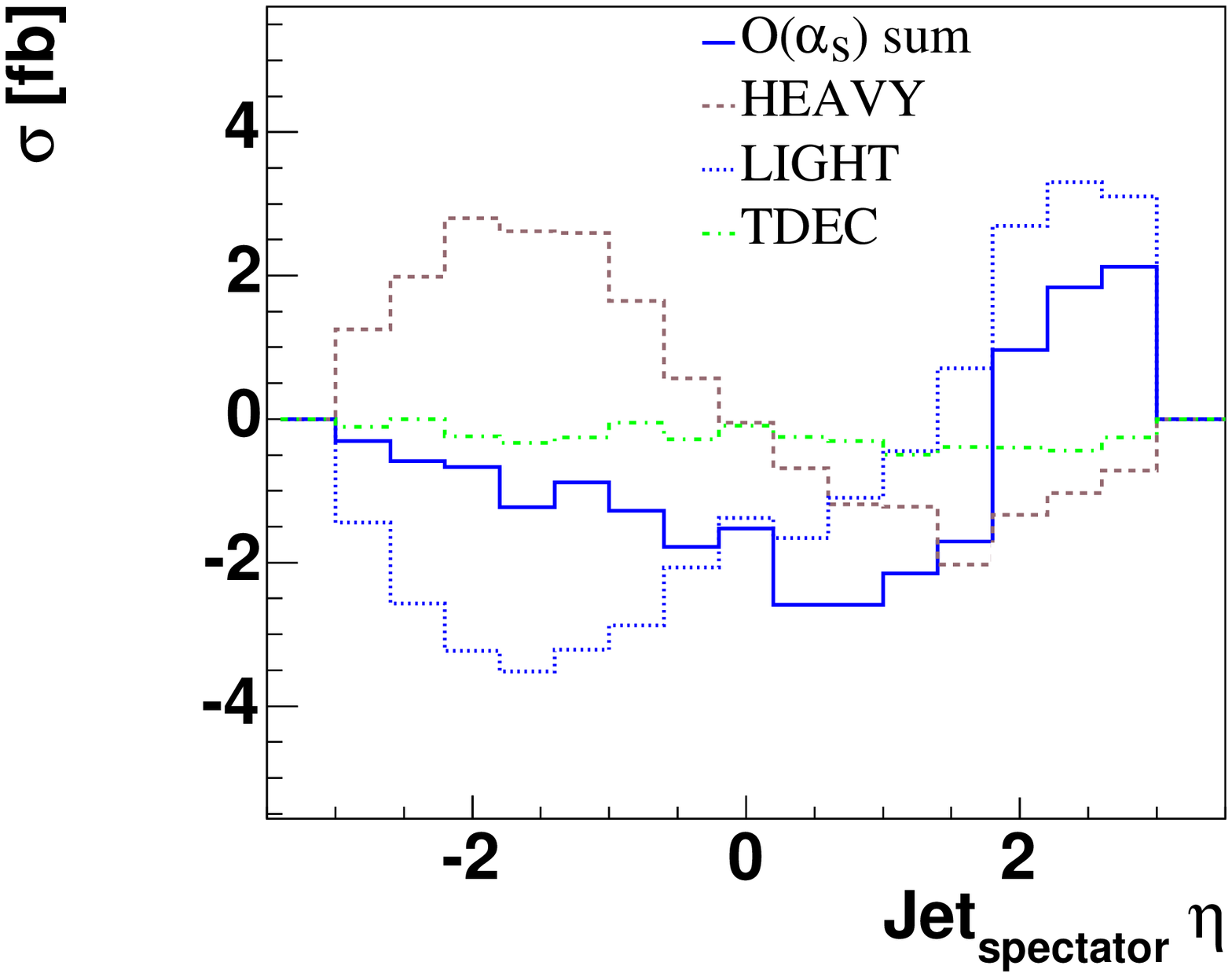}}
\subfigure[]{\includegraphics[%
  width=0.50\linewidth,
  keepaspectratio]{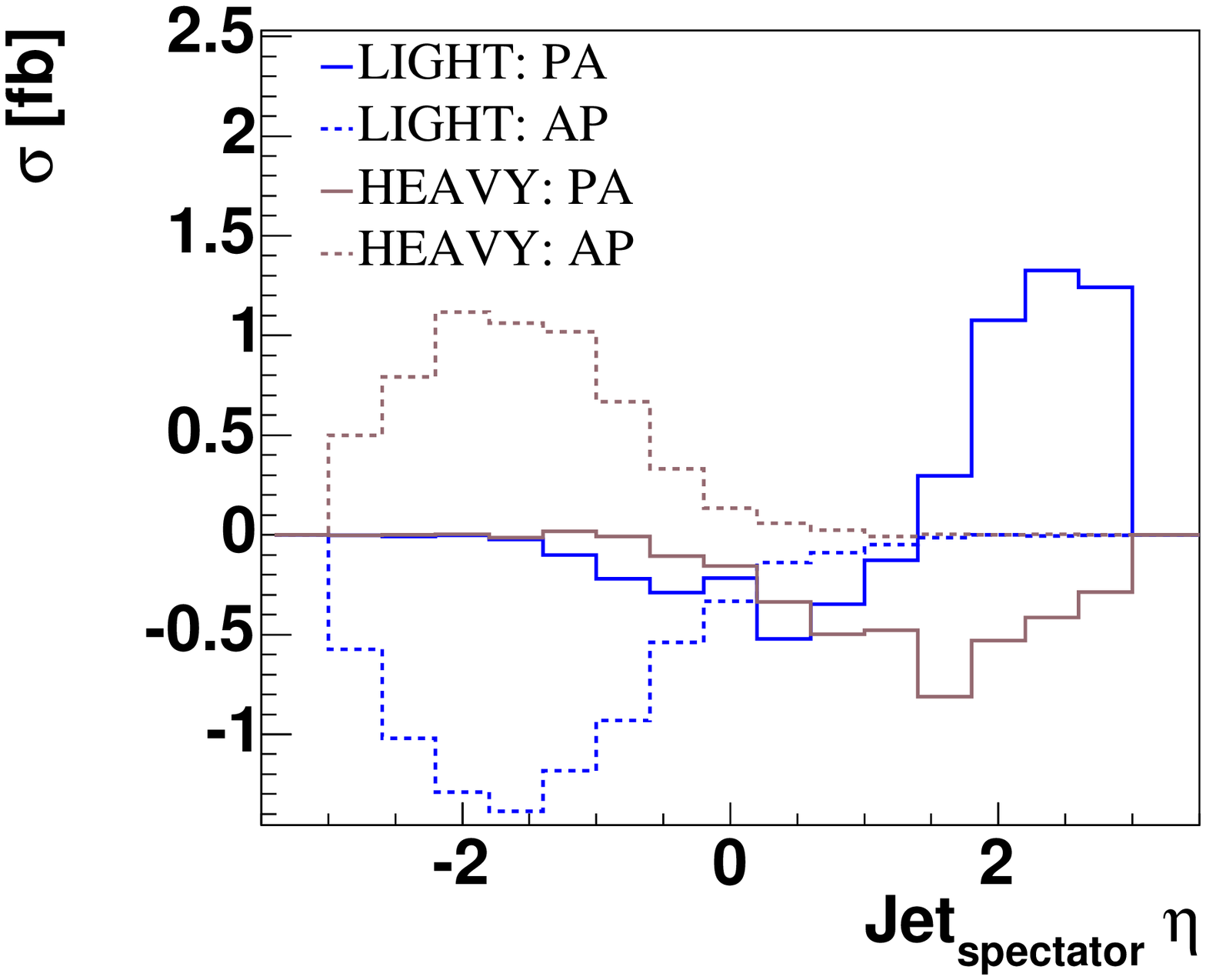}}
\caption{Each individual contribution of the $O(\alpha_{s})$ corrections
to the spectator jet pseudo-rapidity, summed (a), separately for the
case when the incoming up-type quark is from the proton and anti-proton
(b). Here, ``PA'' and ``AP'' denotes the initial state light quark
originating from proton and anti-proton, respectively. In the legend,
HEAVY, LIGHT and TDEC denotes contributions from NLO corrections to
the heavy quark line, light quark line and decay of top quark, respectively.
\label{fig:spectator_eta_nlo}}
\end{figure}

The pseudo-rapidity distribution of the spectator jet is asymmetric
at the Tevatron for being a $p\bar{p}$ collider \cite{Yuan:1989tc}.
In order to produce a heavy top quark decaying to a positively charged
lepton, the valence quark from the proton is most important, implying
that the light quark will tend to move in the proton direction. We
define the positive $z$-direction to be the proton direction in the
laboratory frame, thus the pseudo-rapidity of the spectator jet will
tend to be positive. Similarly, the spectator jet in an anti-top quark
event produced from the $t$-channel process at the Tevatron
will preferably be at
a negative pseudo-rapidity due to the large anti-up quark parton distribution
inside the antiproton. The $O(\alpha_{s})$ corrections shift the
spectator jet to even more forward pseudo-rapidities due to additional
gluon radiation. However, since the $O(\alpha_{s})$ corrections are
small compared to the Born-level contribution, the spectator jet pseudo-rapidity
distribution only shifts slightly. As Fig.~\ref{fig:spectator_eta_nlo}
shows, the LIGHT and HEAVY contributions have almost opposite behavior
(LIGHT and HEAVY denote  $O(\alpha_{s})$ contributions originating 
from the light and heavy quark line QCD corrections in the $t$-channel
single top process. 
The former shifts the spectator jet to even higher pseudo-rapidities,
while the later shifts it more to the central rapidity region. This
behavior is due to two different effects, as illustrated in Fig.~\ref{fig:spectator_eta_nlo}(b),
in which ``PA'' denotes that the light quarks come from the proton
while the bottom quarks from the anti-proton and vice versa for ``AP''.
After separating the contributions by whether the light quark is from
the proton or the antiproton, it can be seen that the HEAVY corrections
shift the proton contribution down and the antiproton contribution
up due to the slight change in acceptance caused by the additional
jet. The LIGHT corrections show the opposite tendency. For the TDEC
contribution, originating from the top quark decay,
all corrections have similar shapes and the sum of them
leaves the spectator jet pseudo-rapidity unchanged, as expected. After
summing the negative soft-plus-virtual corrections with the real emission
corrections, we obtain the result shown in Fig.~\ref{fig:spectator-tchan},
which shows that the $\oalphas$ correction shifts the spectator jet
even further to the forward direction.

Besides its forward rapidity, the spectator jet also has large transverse
momentum. Since it comes from the initial state quark after emitting
the effective $W$~boson, the transverse momentum peaks around $\sim M_{W}/2$.
By comparison, the third jet is most often much softer, we can thus
use $p_{T}$ of the jet to identify the spectator jet when considering
exclusive three-jet events.

\paragraph{Summary}

Based on a NLO calculation to consistently include corrections to
the production and decay of the top quark in the $s$- and $t$-channel
single-top processes, we perform a phenomenology study for the Tevatron
Run-II. We find that: 

\begin{enumerate}
\item When using loose kinematic cuts to maximize the acceptance of single-top
signal events, the full NLO kinematics have to be studied. Applying
a constant $K$-factor with LO kinematics does not reproduce the actual
NLO distributions. 
\item In order to reconstruct the top quark in single-top events, the best-jet
algorithm works better in the $s$-channel process, while the leading
$b$-tagged jet algorithm works best in the $t$-channel process. 
\item NLO corrections can largely change some kinematic distributions and
spin correlations. After event reconstruction with kinematic selection
cuts, we find that the degree of top quark polarization is about
the same in the optimal basis and the helicity basis ($tb$-frame) for the
$s$-channel process. For the $t$-channel process, the helicity basis
(the $tq$-frame) gives almost the same prediction as the spectator basis.
\item To accurately model the $s$-channel single-top background in
searches for Higgs boson production via $WH$ associated production, one
has to use NLO kinematics to model the decay of the top quark in
single-top events. This is because the LO top decay kinematics underestimate
the $s$-channel single-top rate as a background for Higgs searches.
\item Studying the detection efficiency of the forward light quark jet 
(the spectator jet) in
$t$-channel single-top events can help us optimize the detection efficiency
in searches for the Higgs
boson produced via weak gauge boson fusion processes at the LHC. We
find that the NLO corrections to the light quark line in $t$-channel
single-top events tend to shift the spectator jet to even more forward
rapidities. Hence, the NLO effect is important for determining the coupling
of $HVV$ by measuring the $VV\to H$ production rate.
\end{enumerate}

\paragraph*{ACKNOWLEDGMENTS}
The work of C.-P. Y. was supported in part by the U. S. National Science
Foundation under award PHY-0244919.



\def\slsh{\rlap{$\;\!\!\not$}}
\def\ep{\epsilon}
\def\ldot{\!\cdot\!}
\def\cM{{\cal M}}
\def\Li{\mathop{\mathrm{Li}}\nolimits}
\def\bentarrow{\:\raisebox{1.3ex}{\rlap{$\vert$}}\!\rightarrow}
\def\dk#1#2#3{
\begin{array}{r c l}
#1 & \rightarrow & #2 \\
 & & \bentarrow #3
\end{array}
}
\def\bothdk#1#2#3#4#5{
\begin{array}{r c l}
#1 & \rightarrow & #2#3 \\
 & & \:\raisebox{1.3ex}{\rlap{$\vert$}}\raisebox{-0.5ex}{$\vert$}%
\phantom{#2}\!\bentarrow #4 \\
 & & \bentarrow #5
\end{array}
}

\subsubsection*{Single top production and decay at next-to-leading
order}
\label{theory_nlo_ct}
\textbf{Contributed by:~J.~Campbell and F.~Tramontano} \\

In this section, we report on the recent calculations of all single
top processes at next-to-leading order and their inclusion in the 
Monte Carlo program MCFM~\cite{Campbell:2004ch,Campbell:2005bb}.
The implementation of these processes includes the leptonic decays of
the top quark (with full spin correlations) as well as the effects of gluon
radiation in the decay of the top quark. The inclusion of these effects
allows for the application of cuts on all the decay products and thus a better
comparison with experimental studies.

The lowest order processes which we consider are $s$-channel production,
\begin{equation}
\dk{u+\bar{d}} {t +\bar{b}}{\nu +e^+ + b}
\label{eq:schannel}
\end{equation}
$t$-channel production,
\begin{equation}
\dk{b+u} {t +d}{\nu +e^+ + b}
\label{eq:tchannel}
\end{equation}
and single top production in association with a $W$ boson which also decays
leptonically, 
\begin{equation}
\label{eq:wtchannel}
\bothdk{b+g}{W^-+}{t}{\nu+e^++b}{e^-+{\bar \nu}}
\end{equation}
At the Tevatron only the processes in Eqs.~(\ref{eq:schannel}) and~(\ref{eq:tchannel})
can be observed. At the LHC top quarks can be produced copiously in all channels,
with a significant amount of events from the associated channel, Eq.~(\ref{eq:wtchannel}).
Thus the study of single top events passes from the search for their observation
at the Tevatron to their study as a significant source of background events in
new physics searches at the LHC.
We note that at the Tevatron, the rates for the production of an anti-top quark
in any of these modes are identical to those for a top quark. At the LHC, the cross
sections for top and anti-top production in the $s$ and $t$ channels differ. In contrast,
the rate for $W^+{\bar t}$ is the same as that for $W^-t$ due to the equality of
the perturbatively-derived $b$ and ${\bar b}$ distribution functions.

All of these processes have previously been considered extensively at leading order, but the
first serious approximation in QCD is obtained by including $O(\alpha_S)$ radiative
corrections. Such next-to-leading order calculations can give important information about
the choice of factorization and renormalization scales. In addition, it is only at
next-to-leading order that we obtain accurate predictions of event rates which are sensitive
to the structure of jets in the final state. Such NLO calculations have so far been
available only for the case where the decays of the top quark (and the $W$ boson, in the
case of associated production) are not
included~\cite{Smith:1996ij,Bordes:1994ki,Stelzer:1997ns,Harris:2002md,
Sullivan:2004ie,Giele:1995kr,Zhu:2002uj}. 

\begin{figure}
\begin{center}
\includegraphics[angle=0,width=0.3\columnwidth]{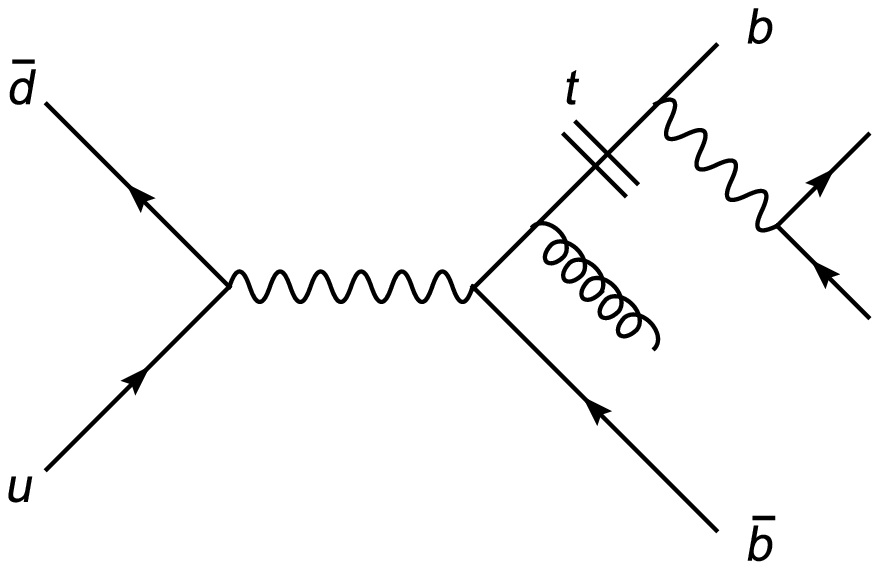}
\hspace*{0.6cm}
\includegraphics[angle=0,width=0.3\columnwidth]{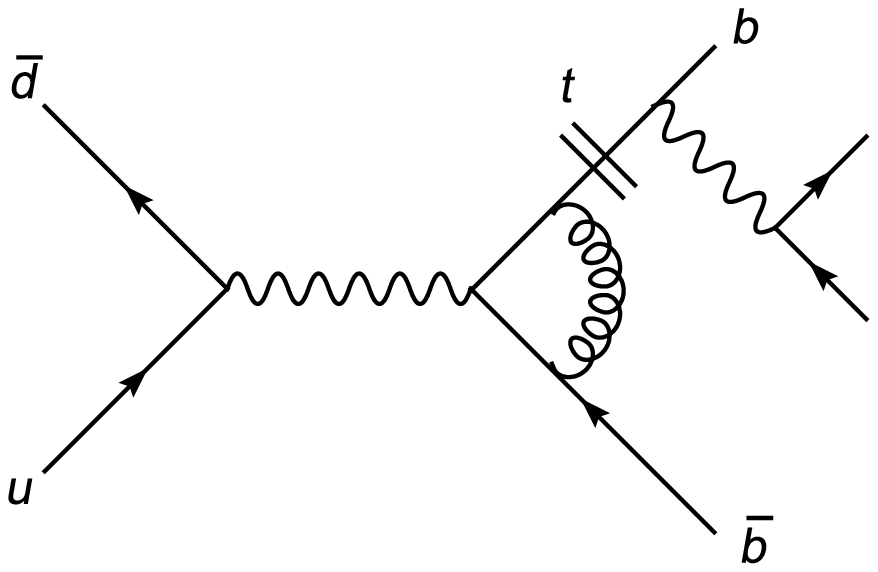} \\ \vspace*{0.8cm}
\includegraphics[angle=0,width=0.3\columnwidth]{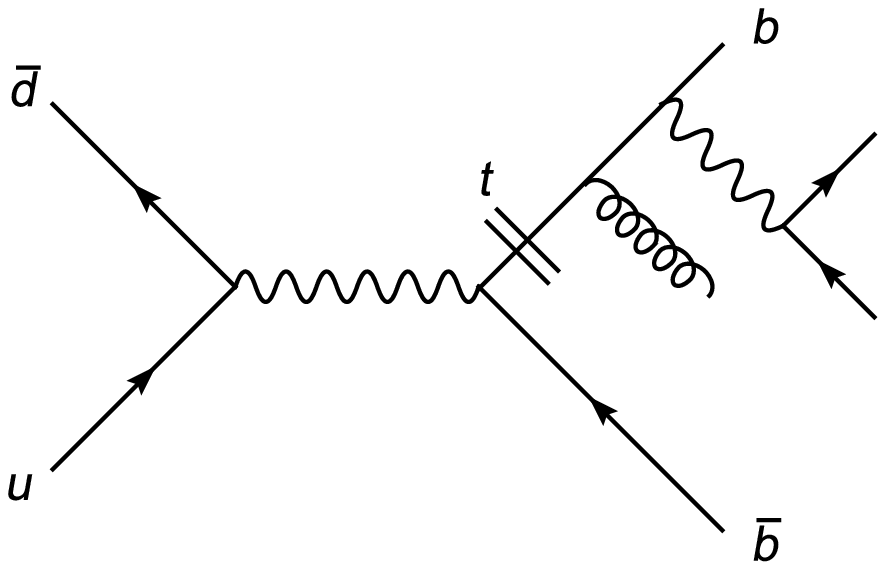}
\hspace*{0.6cm}
\includegraphics[angle=0,width=0.3\columnwidth]{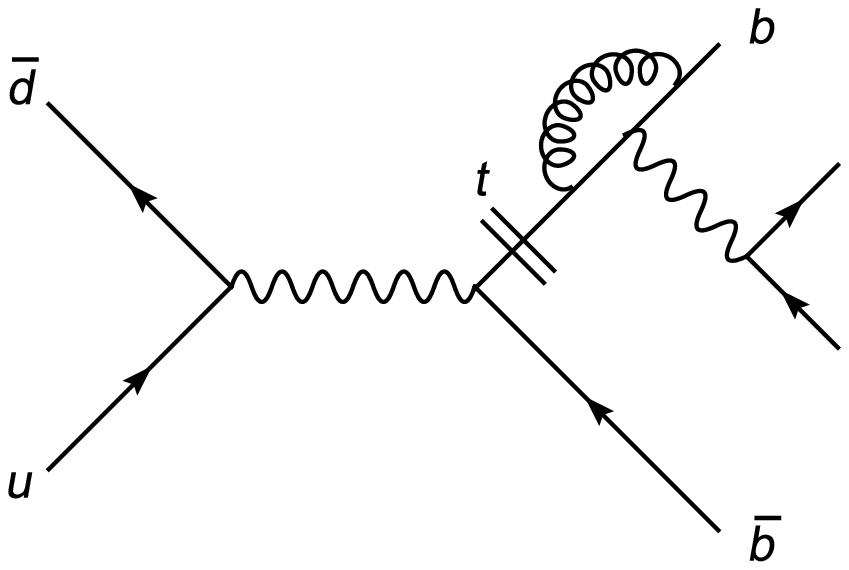}
\caption{Real and virtual radiation in the production and decay
stages of $s$-channel single top production. The double bar indicates the on-shell
top quark. \label{fig:decayrad}}
\end{center}
\end{figure}

First we describe the inclusion of radiative corrections with reference to the $s$-channel
process, although a similar procedure is followed for the other two processes. In general,
the real and virtual radiative corrections fall into two categories. The first type is
radiation in the production stage of the top quark and the second corresponds to radiation
associated with its decay. Examples of diagrams in each category are depicted in
Figure~\ref{fig:decayrad}, where the double bar indicates the separation of production and
decay stages. 

\begin{figure}
\begin{center}
\includegraphics[width=0.4\columnwidth]{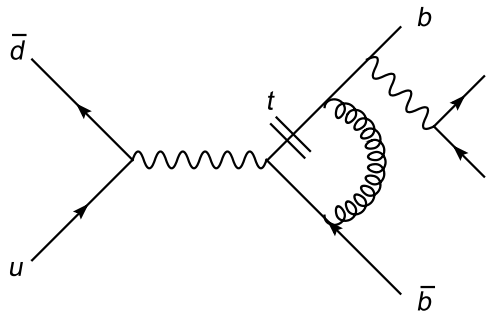}
\caption{An example of a diagram that is not included, in this case
interference between virtual radiation in the production and decay stages.
\label{fig:virtint}}
\end{center}
\end{figure}

In order to make this separation in a gauge-invariant way, the double bar represents a top quark
which is on its mass shell. Thus every diagram has exactly one top quark which is on its mass
shell and diagrams without an on-shell top quark  are suppressed by $\Gamma_t/m_t$ where
$\Gamma_t$ and $m_t$ are the width and mass of the top quark.
In this procedure, we have neglected the interference between radiation in the
production and decay stages, both in the real and virtual contributions. An example
of such an interference term in the virtual contribution is shown in Fig.~\ref{fig:virtint}.
The physical argument for neglecting these terms is based on the
characteristic time scale for the production and the decay of
the top quark~\cite{Fadin:1993kt,Fadin:1993dz,Melnikov:1993np}.
For the production, this time scale is
of order $1/m_t$ while for the decay it is $1/\Gamma_t$.
In general, this suggests that radiation in the production and decay
stages is separated by a large time and the interference effects are
expected to be of order $\alpha_s \Gamma_t/m_t$. In both total cross
sections and in distributions of selected observables, there is evidence that
this is indeed the case~\cite{Pittau:1996rp,Macesanu:2001bj}.

The implementation of the cancellation of soft and collinear singularities
between the real and the virtual contributions is performed using the dipole
subtraction method~\cite{Ellis:1980wv,Catani:1996jh,Catani:1996vz}.
For the case of single top production we have a massive quark in the final state, so
in fact we have implemented a generalization of this scheme as suggested in~\cite{Catani:2002hc}. We
have also extended these results to include a tunable parameter which controls the size of
the subtraction region, as originally proposed in~\cite{Nagy:1998bb,Nagy:2003tz}.
Further details may be found in Ref.~\cite{Campbell:2004ch}.
In order to deal with radiation in the 
decay stage of the process we have developed a specialized 
subtraction procedure, which can be applied to the decay of the top quark
in any process. We will briefly describe this procedure here.
 
We begin by constructing a counter-term for the process,
\begin{equation}
t \rightarrow W + b +g,
\end{equation}
which has the same soft and collinear singularities as the 
full matrix element. This counter-term
takes the form of a lowest order matrix element multiplied by a
function $D$ which describes the emission of soft or collinear
radiation, 
\begin{equation} \label{decayfactorization}
|{\cM}( \ldots p_t,p_W,p_b,p_g) |^2 \to
|{\cM}_0( \ldots p_t,\tilde p_W,\tilde p_b) |^2 \times 
D(p_t.p_g,p_b.p_g,m_t^2,m_W^2) \; ,
\end{equation}
In the region of soft emission, or in the region where the momenta
$p_g$ and $p_b$ are collinear, the right hand side of 
Eq.~(\ref{decayfactorization}) 
has the same singularity structure as the full matrix element. 
The lowest order matrix element $\cM_0$ in Eq.~(\ref{decayfactorization}) 
is evaluated for values of the momenta $p_W$ and $p_b$
modified to absorb the four-momentum carried away by the gluon,
and subject to
the momentum conservation constraint, $p_t \rightarrow {\tilde p_W}+\tilde p_b$. 
The modified momenta denoted by a tilde are also subject to 
the mass-shell constraints, ${\tilde p_b}^2=0$ and ${\tilde p_W}^2=p_W^2$. 
The latter condition is necessary in order that the rapidly varying 
Breit-Wigner function for the $W$ is evaluated at the same kinematic point 
in the counterterm and in the full matrix element.
We define $\tilde p_W$ by a 
Lorentz transformation, ${\tilde p_W}^\mu=\Lambda^{\mu}_{\nu} p_W^\nu$
fixed in terms of the momenta $p_W$ and $p_t$.
Because $\tilde p_W$ and $p_W$ are related by a Lorentz transformation 
the phase space for the subsequent
decay of the $W$ is unchanged.

The required transformation defining $\tilde p_W$ lies in the plane of the vectors $p_t$
and $p_W$, with the transformed momentum of the $b$ quark fixed by
$\tilde p_b=p_t-\tilde p_W$. The full details of the transformation, subtraction term
and integrated form of the dipole can be found in Ref.~\cite{Campbell:2005bb}.

\begin{figure}
\begin{center}
\includegraphics[angle=0,width=0.7\columnwidth]{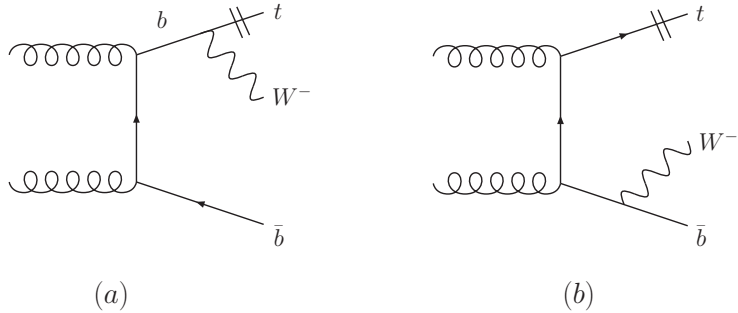}
\caption{Diagrams present in the real corrections to $W^-t$ production
which involve an additional ${\bar b}$ quark. The double bars indicate the on-shell
top quark which subsequently decays into $W^+b$. Diagram (b)
contains a resonant ${\bar t}$ propagator, while (a) does not.
\label{fig:wtbdiags}}
\end{center}
\end{figure}

In the calculation of the real radiative corrections to the associated
$Wt$ process, a further complication arises. The difficulty stems from
diagrams in which the additional radiated parton is a ${\bar b}$ quark,
such as the ones illustrated in Figure~\ref{fig:wtbdiags}.
Both of these diagrams produce a final state consisting of a $W^-$, an
on-shell top quark and a ${\bar b}$ quark. However, diagram
(b) contains a resonant ${\bar t}$ propagator and represents the
production of a $t{\bar t}$ pair with the subsequent decay of the
${\bar t}$ into the $W^-$ and ${\bar b}$ quark. Therefore the
contribution from diagrams such as this, when integrated over the total
available phase space, can be much larger than the lowest order $Wt$
cross section (an order of magnitude at the LHC).

Rather than using an invariant mass cut~\cite{Belyaev:1998dn}, or subtracting the 
problematic resonant
contribution~\cite{Tait:1999cf} we instead utilize an approach which is more suited to our
Monte Carlo implementation which includes decays. Using the $b$-quark PDF, we already
include all contributions such as diagram (a) of Fig.~\ref{fig:wtbdiags} up to a $p_T$ of the
${\bar b}$-quark equal to the factorization scale, $\mu_F$. In order to ensure
the validity of the collinear approximation used in the derivation of the $b$-PDF~\cite{Boos:2003yi},
we should choose $\mu_F \lesssim \, (m_W+m_t)/4 \, \approx 65$~GeV.
When a ${\bar b}$ quark is observed with a $p_T$ above $\mu_F$
then the doubly resonant diagrams (such as (b) of Fig.~\ref{fig:wtbdiags}) dominate. In
this region of phase space, the $t{\bar t}$ process is therefore more appropriate. Thus, in
order to disentangle these two processes, we perform our calculation of the $Wt$ process by
applying a veto on the $p_T$ of the additional $b$ quark that appears at next-to-leading
order. For the results presented here, we have chosen this veto to occur at $50$~GeV.
 In doing so, the result for the diagrams represented in Figure~\ref{fig:wtbdiags}
remains at the level of a few percent of the lowest order cross section and, for simplicity,
the doubly resonant diagrams can even be omitted.

\begin{figure}
\begin{center}
\includegraphics[angle=0,width=0.7\columnwidth]{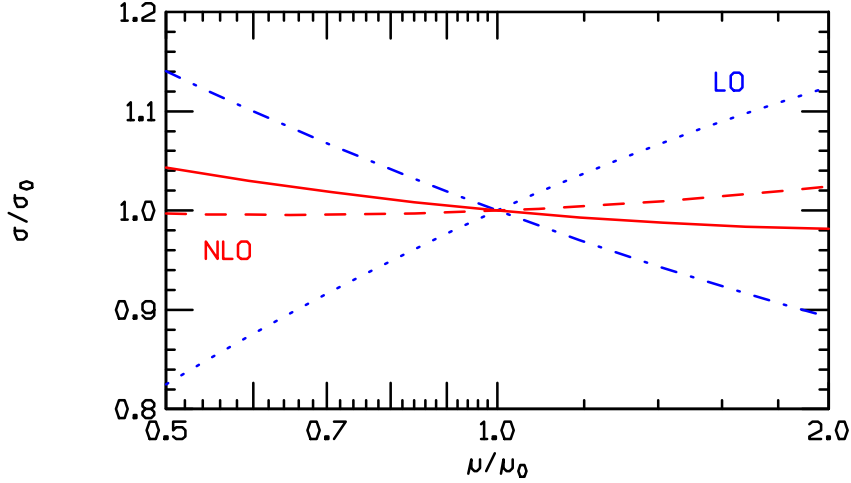}
\caption{Scale dependence of the cross section for $Wt$ production at the LHC, as
described in the text. Factorization scale dependence is shown by a dotted curve at LO
and a dashed curve at NLO. Renormalization scale dependence is shown by a dot-dashed curve
at LO and a solid curve at NLO.
\label{fig:mudep_cteq6}}
\end{center}
\end{figure}

The methods that we have described have been implemented in the  Monte Carlo program, MCFM,
allowing us to make predictions for kinematic distributions in all channels. As a simple
example of our simulation of these single top processes, we first compare the leading
order and NLO cross sections for each of the channels in Table~\ref{tab:xsecstot}. These
cross sections are calculated for a top mass of $175$~GeV and use the CTEQ6 set of
structure functions. 
\begin{table}[tb]
\caption{LO and NLO cross sections (in picobarns) for each channel of single top-quark
production at the Tevatron and LHC, for $m_t=175$ GeV. 
Cross sections are evaluated with CTEQ6L1 ($\alpha_s(M_Z)=0.130$) and CTEQ6M
($\alpha_s(M_Z)=0.118$) PDFs~\cite{Pumplin:2002vw}, using scales of $m_t$ for
the $s$- and $t$-channel processes and $50$~GeV for $Wt$.
\label{tab:xsecstot}}
\begin{center}
\begin{tabular}{lll|ll} 
& \multicolumn{2}{c}{Tevatron} & \multicolumn{2}{c}{LHC} \\ 
Process~[pb]      & $\sigma_{LO}$ & $\sigma_{NLO}$ & $\sigma_{LO}$ & $\sigma_{NLO}$  \\
\hline
$s$-channel & 0.582 & 0.872  & 7.27  & 10.4  \\
$t$-channel & 1.75  & 1.92   & 237   & 245   \\
$Wt$        & 0.104 & 0.143  & 61.3  & 68.7  
\end{tabular}
\end{center}
\end{table}
Both the $s$-channel and $Wt$ processes can receive sizeable corrections at NLO, with
the cross-sections increasing by around $40$--$50$\% at the Tevatron.
In contrast, the $t$-channel process receives only mild corrections at both colliders.
As well as the normalization of the cross section changing, its dependence
upon the factorization and renormalization scales can also be significantly
reduced at next-to-leading order. This is illustrated in Figure~\ref{fig:mudep_cteq6},
where we show the effect of varying these scales on the $Wt$ cross-section. The
renormalization and factorization scales are varied separately by a factor of two, with
the other scale kept fixed at $\mu_0=50$~GeV. The LO and NLO cross sections are each
normalized to their central values. 
One can see that there is a great reduction in the dependence of the cross
section on each of these scales. If both scales are varied together, the scale dependence
from each individually practically cancels, even at leading order. Thus one might 
incorrectly assume that the cross section is well predicted at leading order, when this
is clearly not the case.  

We now consider the search for single top processes at the Tevatron where, as mentioned
earlier, only the $s$- and $t$-channel  cross sections can possibly be observed.
However, much of the lessons learned at the Tevatron will be applicable for the
observation of the top quark in the $Wt$ channel at the LHC. We shall consider the signal
for single top production to be  the presence of a lepton, missing energy and two jets,
one of which is tagged as a $b$-jet. With this signal, the largest background comes from
the process $Wb{\bar b}$, with further substantial contributions when a charm quark is
mis-tagged as a $b$ in $u{\bar s} \to Wu{\bar c}$ and from other  mis-tagged $W+2$~jet
events. Smaller background contributions result from $t{\bar t}$ and $WZ$ production.

\begin{figure}
\begin{center}
\includegraphics[,angle=270,width=0.6\columnwidth]{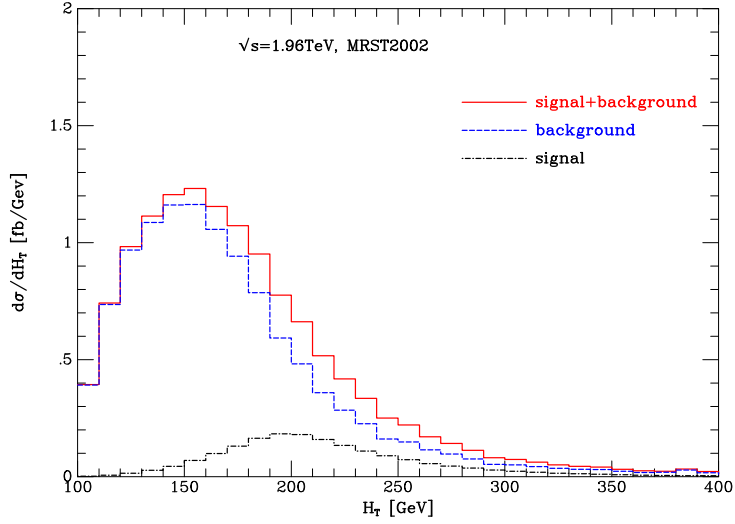}
\caption{The $H_T$ distributions of signal ($s$- and $t$-channel single top
production), background and signal plus background at the Tevatron.
\label{fig:htoverall}}
\end{center}
\end{figure}

Most of these processes can be calculated to NLO in MCFM, with cuts designed to reproduce the
ones used in the experimental searches at CDF and D0. To that end, we have used the cuts,
\begin{equation}
p_T^{e} > 20~{\rm GeV}, \qquad |\eta^{e}| < 1.1, \qquad \slsh{E_T} > 20~{\rm GeV},
\end{equation}
on the leptons and missing transverse energy, as well as,
\begin{equation}
p_T^{\rm{jet}} > 15~{\rm GeV}, \qquad |\eta^{\rm{jet}}|  < 2.8, \qquad \Delta R> 1.0,
\end{equation}
on the jets, which have been clustered using the $k_T$ algorithm. Lastly, in order to
reduce the background from events that do not contain a top quark, we apply a cut
on the reconstructed mass of the `$b+l+\nu$'-system, $140 < m_{bl\nu} < 210$~GeV.
Using these cuts, we have calculated the distribution of the variable $H_T$, the sum
of the lepton $p_T$, missing transverse energy and jet transverse momenta. This
can be useful for selecting single-top events from the large backgrounds, as indicated
in Figure~\ref{fig:htoverall} where we show the distribution of the signal and the sum of all
background processes, under some assumptions about mis-tagging and
efficiencies~\cite{Campbell:2004ch}.
Although the single top processes represent a large fraction of the events in the region
of large $H_T$, searches using this distribution as a key are heavily reliant on accurate
predictions of the shapes and normalization of the signal and backgrounds. Since almost
all of these are now known at next-to-leading order, this information can be used to
refine current analyses.

\begin{figure}
\begin{center}
\includegraphics[angle=0,width=0.7\columnwidth]{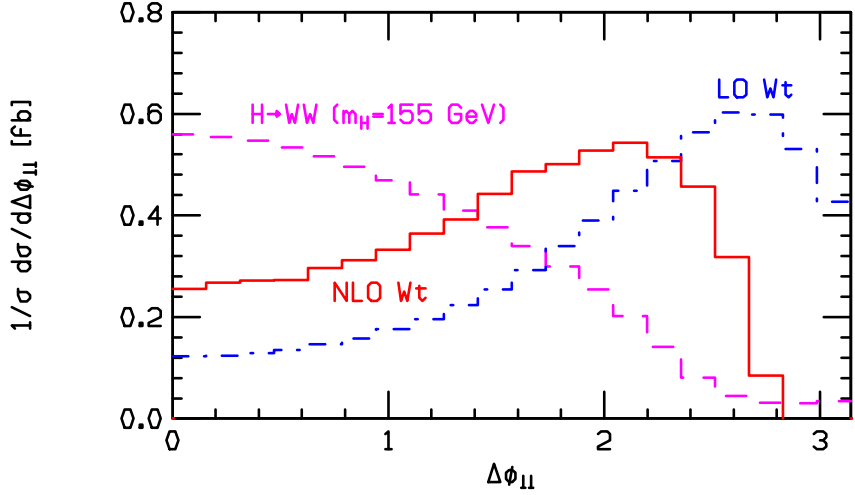}
\caption{The opening angle between the leptons in the $H \to WW$ and $Wt$ processes,
for the search for a Higgs of mass $155$~GeV. Cross sections are normalized to unity,
after suitable search cuts have been applied.
\label{fig:dphi}}
\end{center}
\end{figure}

As a final example of the utility of our calculations we consider a rather different
role that single top production can take at the LHC. In the search for
an intermediate mass Higgs boson, of mass $155 < m_H < 180$~GeV, the $Wt$ process can
be a significant background when trying to observe Higgs production via
gluon fusion~\cite{Dittmar:1996ss},
\begin{equation}
\bothdk{g+g \to H}{W^-+}{W^+}{\nu+e^+}{e^-+{\bar \nu}}
\end{equation}
The significant missing energy in the signal process means that the Higgs mass peak
cannot be fully reconstructed, so that accurate predictions for all backgrounds are
imperative. Here we do not detail all aspects of the study that we have
performed (for further details, see Ref.~\cite{Campbell:2005bb}), but merely draw
attention to the conclusions. A useful observable for discriminating between 
the signal and $Wt$ background is the opening angle in the transverse plane between
the leptons from the $W$ decays, $\Delta \phi_{ll}$.
As shown in Figure~\ref{fig:dphi}, the leptons in the
signal are predominantly produced with only a small opening angle, while
the $Wt$ background tends to produce them mostly back-to-back.
One can see that this statement is weakened at NLO since the $Wt$ peak
is shifted to smaller values and becomes more broad. Such a shape change could have
a significant impact on search strategies in this channel at the LHC.

We conclude by noting that a number of approximations have been used in order to
make the NLO calculations tractable. Notably, we have not included the mass of the
bottom quark in our computations,
ignored off-shell effects for the top quark and neglected interference effects
between radiation in the top quark production and decay stages. However, none
of these is expected to amount to much more than a few percent correction. This
should certainly not be a serious issue when considering single top searches
at the Tevatron, nor at the LHC when considering these channels as backgrounds.
However such effects may become important when studying properties of the top quark
in these channels at the LHC. In that case further study will be necessary and indeed the
development of such improved tools is already underway~\cite{Frixione:2005vw}.

{\bf Acknowledgements}
We thank K.~Ellis for collaboration on some of the calculations presented here
and acknowledge many helpful discussions with F.~Maltoni
and S.~Willenbrock.


%
%
%
%
\subsubsection*{Parton-level comparison of MadEvent Monte Carlo events to NLO calculations}
\label{sec:singletoptheorycompare}
\textbf{Contributed by:~J.~L\"uck, W.~Wagner, C.~Ciobanu} \\

A good modeling of signal and background processes with Monte Carlo generators
is essential for particle physics analyses. This is particularly true if
one aims for the observation of a new process, like single-top production. 
Qualitatively a false discovery has to be avoided, 
quantitatively the significance of a signal has to be evaluated correctly. 

In Run I, CDF used the \PY program to generate single-top 
events~\cite{Acosta:2001un,Acosta:2004un}. 
Several authors pointed out~\cite{boos2000,Sullivan:2004ie}
that the leading order contribution of single-top $t$-channel
production as modeled in \PY and {\sc herwig} does not fully 
represent the measured final states.

This is a $2\rightarrow 2$ process with a $b$ quark in the initial state:
$b+u\rightarrow d+t$ or $b+\bar{d}\rightarrow \bar{u}+t$. 
A $b$ quark parton distribution function is used.
The $b$ quark stems originally from a gluon splitting into a $b\bar{b}$
pair. 
Since flavor is conserved in the strong interaction, a $\bar{b}$ quark
has to be present in the event as well.
\PY creates the $\bar{b}$ through backward
evolution following the DGLAP scheme. Using this method, only the soft
  region of the transverse momentum of the $\bar{b}$ is modeled well.
  The high-$p_\mathrm{T}$ tail is not estimated as accurately. 
  In addition, the 
  $\eta$ spectrum comes out too far forward.
  In following we will call the $\bar{b}$ the $2^\mathrm{nd}$ $b$ quark. 

One can improve the modeling of single-top quark production by producing
Monte Carlo events with matrix element generators and apply a shower Monte
Carlo on the parton final states.
To model the single-top $t$-channel kinematics
it was proposed to generate two samples with the matrix element generator:
A $2\rightarrow 2$ sample, $b+q\rightarrow q^\prime +t$, and a 
$2\rightarrow 3$ sample with a gluon in the initial state,
$g+q\rightarrow q^\prime + t + \bar{b}$. In the second process the 
$2^\mathrm{nd}$ $b$ quark is produced directly in the hard process
described by the matrix element. This sample describes the most important
next-to-leading order contribution to $t$-channel production and is 
therefore suitable to describe the high-$p_\mathrm{T}$ tail of the 
$2^\mathrm{nd}$ $b$ quark $p_\mathrm{T}$ distribution. However, the two 
samples have to be matched together to give one unified sample of
Monte Carlo events. In their first Run II analyses CDF and D\O \ used
a matching procedure based on the $p_\mathrm{T}$ spectrum of the
$2^\mathrm{nd}$ $b$ quark~\cite{RunII:cdf_result,Abazov:2005zz}. 
CDF used the matrix element generator
MadEvent~\cite{Stelzer:1994ta,Maltoni:2002qb}, D\O \ used 
the program CompHEP~\cite{Boos:2004kh}. 
At CDF the $p_\mathrm{T}$ distributions of the $2^\mathrm{nd}$ $b$ quark
of LO and NLO sample were normalized to the ratio of the corresponding
cross sections calculated by {\sc MadEvent}, $R=2.56$. 
The intersection point of two curves was found to be 
$K_\mathrm{T}=18\,\mathrm{GeV}/c$. 
Subsequently, events of the LO ($2\rightarrow 2$) sample were accepted for the 
final sample if the $p_\mathrm{T}$ of the $2^\mathrm{nd}$ $b$ quark
was below $K_\mathrm{T}$. Events of the NLO sample were selected if
$p_\mathrm{T}(2^\mathrm{nd} b) > K_\mathrm{T}$. 

\begin{figure}[htbp]  
\begin{center}
a) \hspace*{0.48\textwidth} b) \\
\includegraphics[width=0.48\textwidth]
{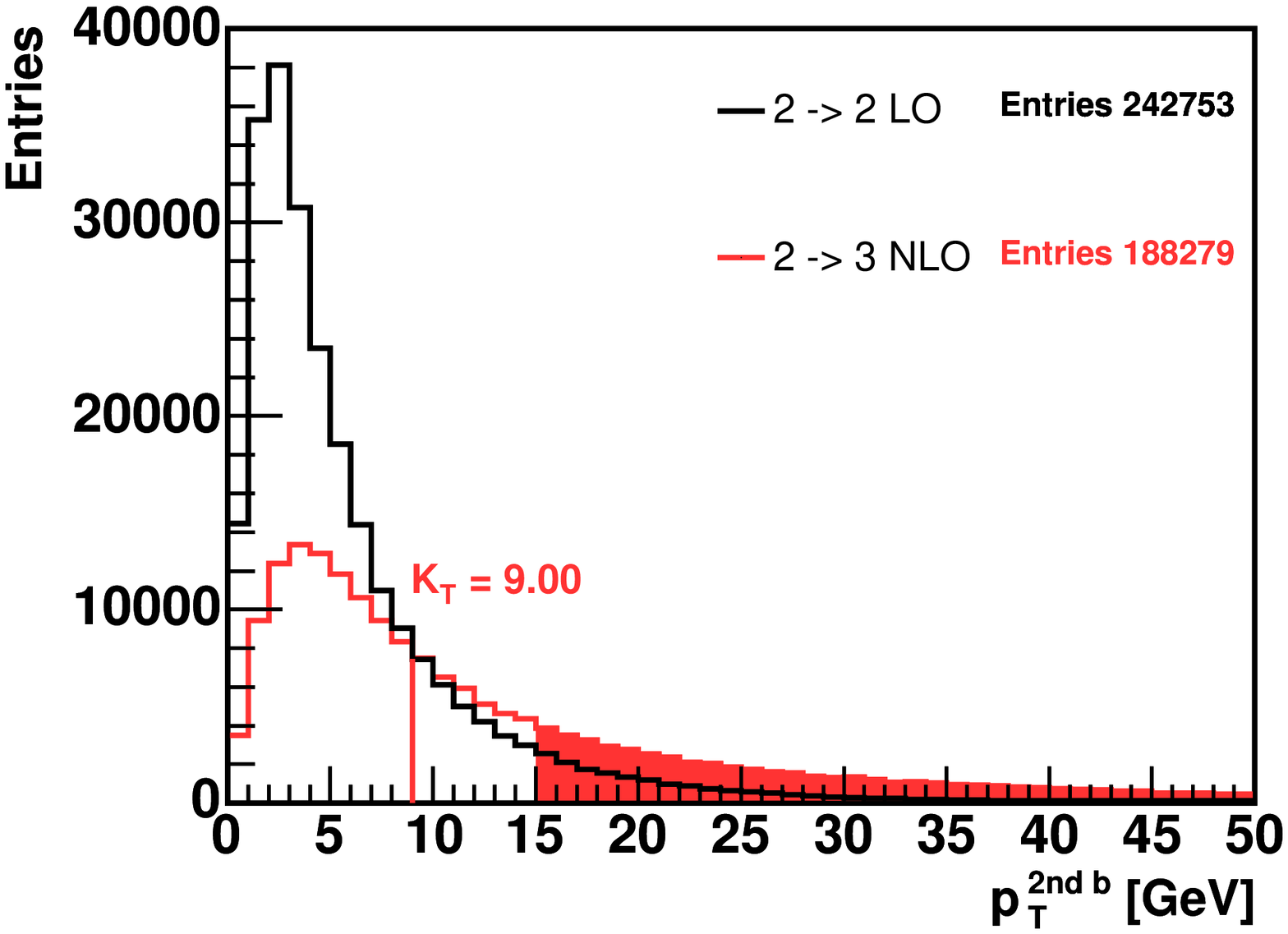}
\includegraphics[width=0.48\textwidth]
{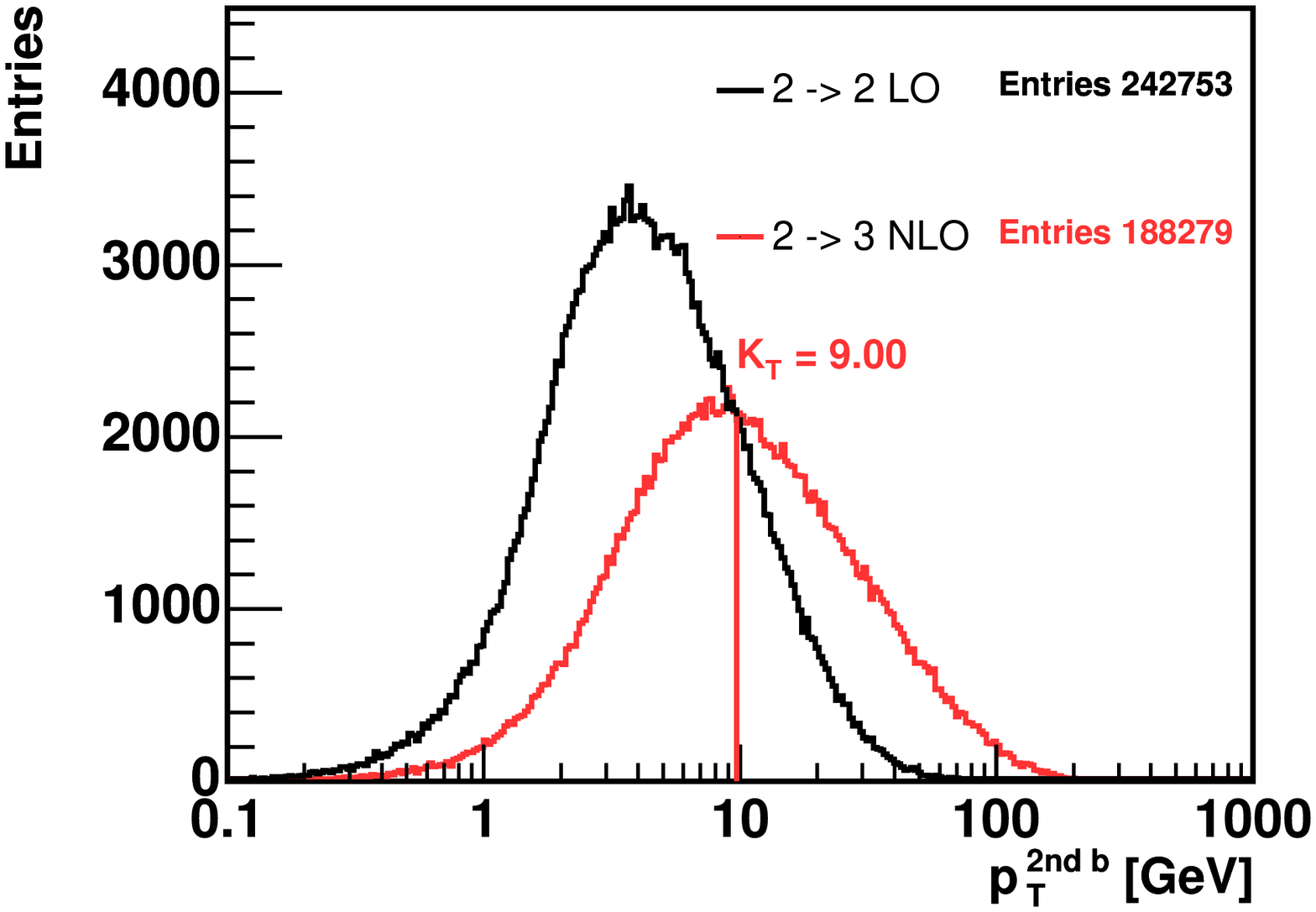} \\
\vspace*{3mm}
c) \hspace*{0.7\textwidth} \\
\includegraphics[width=0.48\textwidth]
{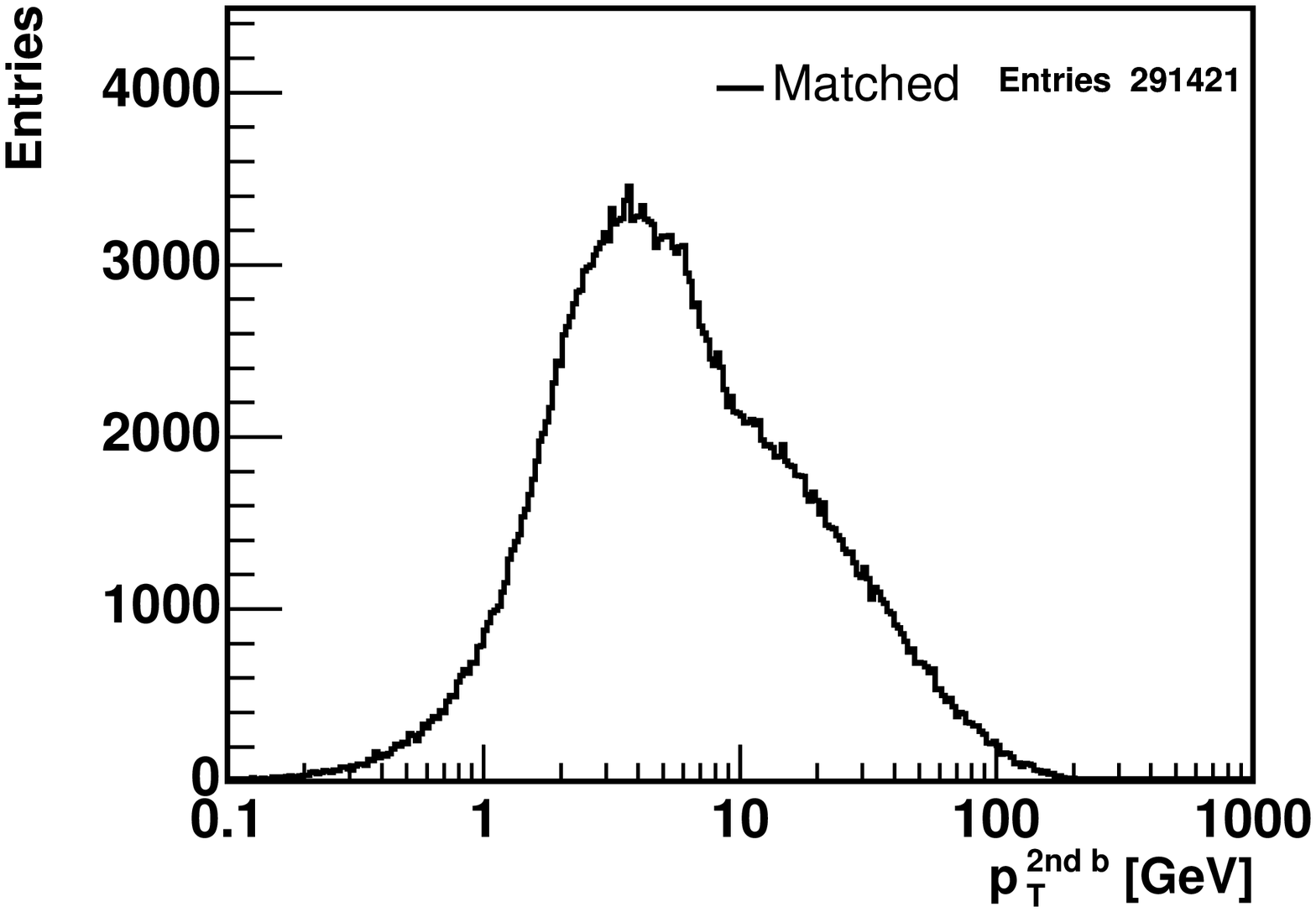}
\end{center}
\caption[tchannel-matching]{\label{fig:t-channel-matching} Matching of the 
single-top $t$-channel samples in CDF.
$p_\mathrm{T}$ distributions of the $2^\mathrm{nd}$ $b$ quark: a) on a 
linear scale, b) on a logarithmic scale, for the $2\rightarrow 2$
and the $2\rightarrow 3$ process. 
The ratio of $2\rightarrow 2$ to $2\rightarrow 3$ events is adjusted
such that the rate of $2^\mathrm{nd}$ $b$ quarks with 
$p_\mathrm{T} > 15\,\mathrm{GeV}/c$ matches the NLO prediction.
In c) the $p_\mathrm{T}$ distribution for 
the matched sample is shown.} 
\end{figure}

\begin{figure}[htbp]  
\begin{center}
\includegraphics[width=0.48\textwidth]
{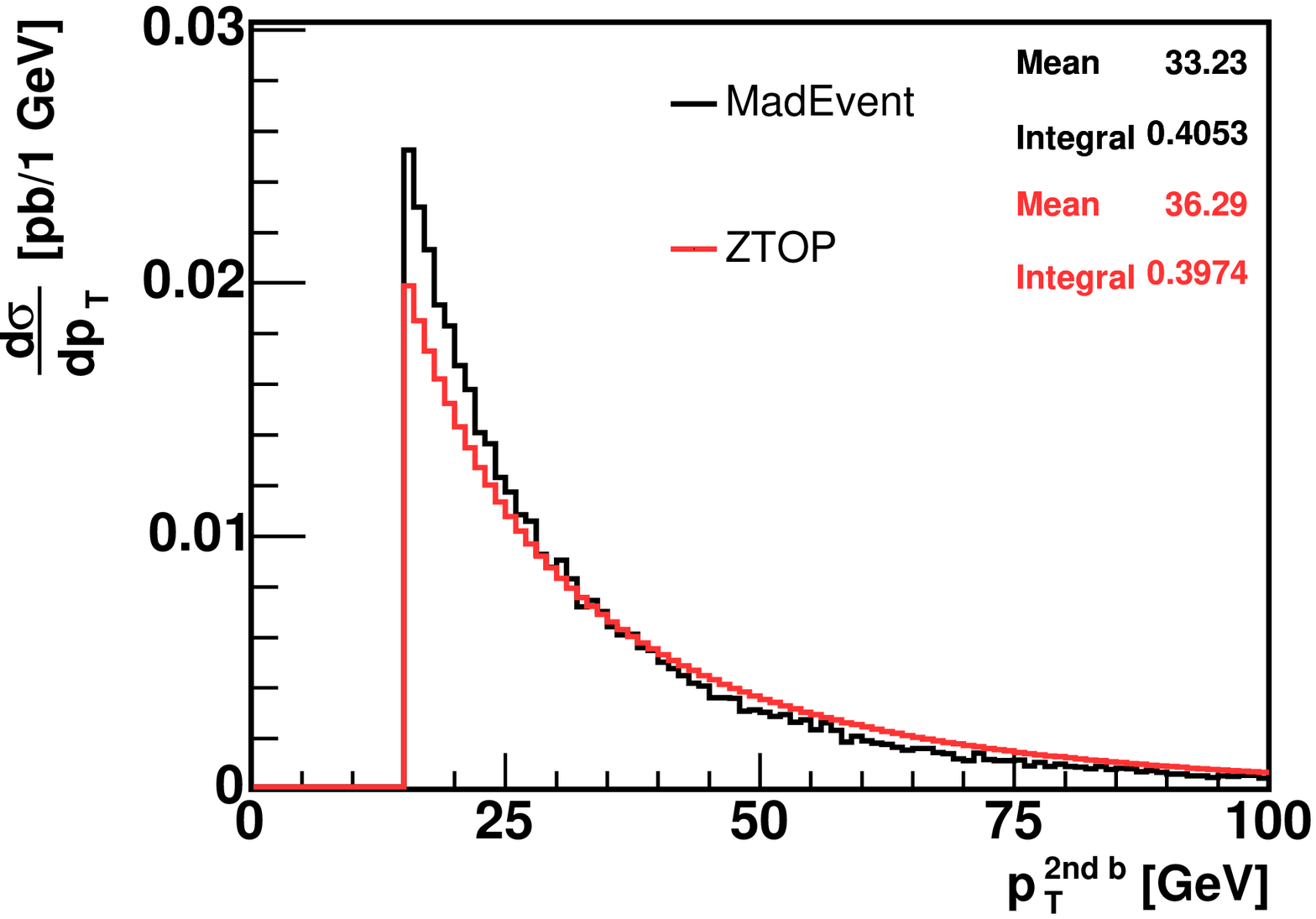}
\includegraphics[width=0.48\textwidth]
{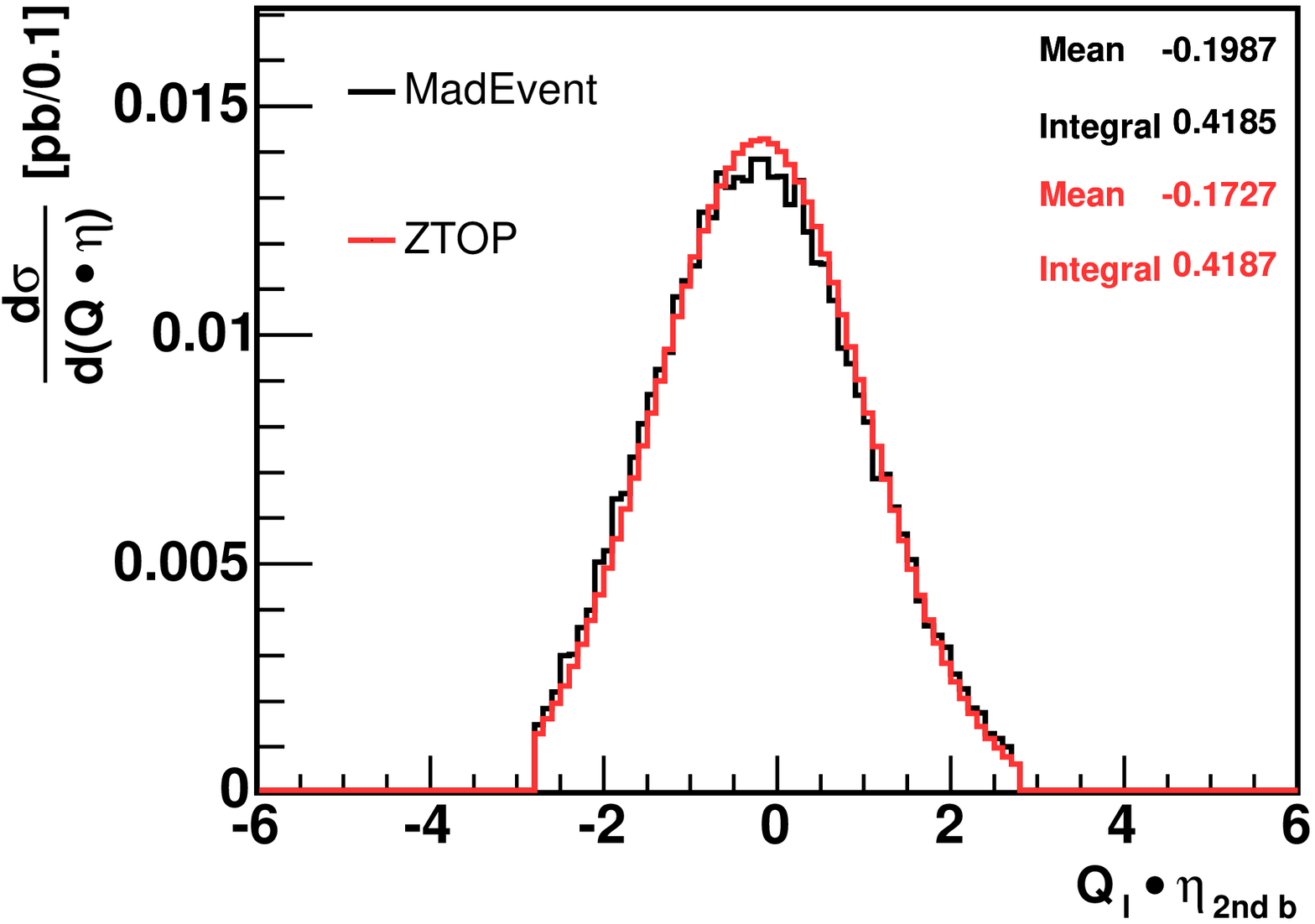}
\includegraphics[width=0.48\textwidth]
{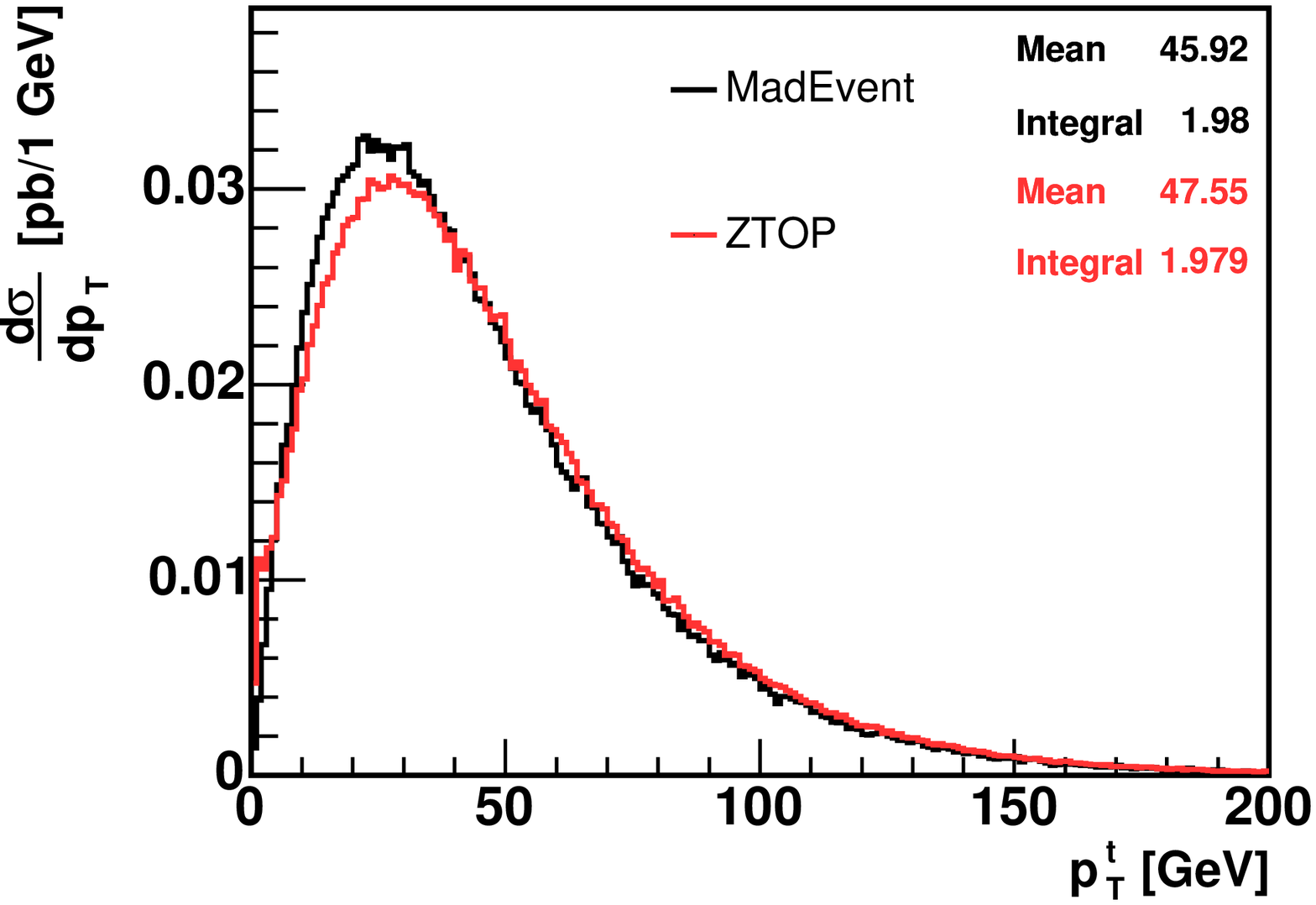}
\includegraphics[width=0.48\textwidth]
{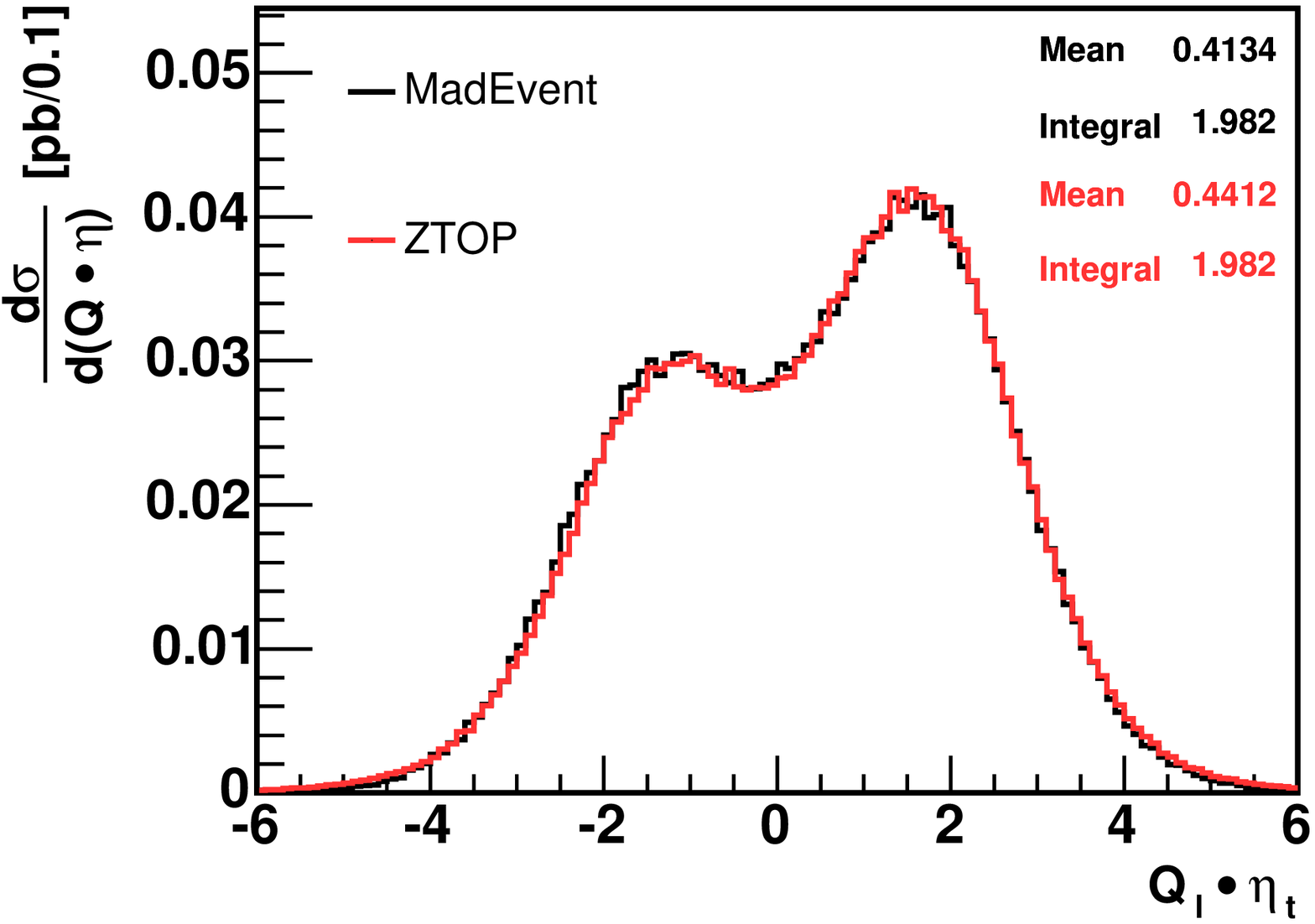}
\includegraphics[width=0.48\textwidth]
{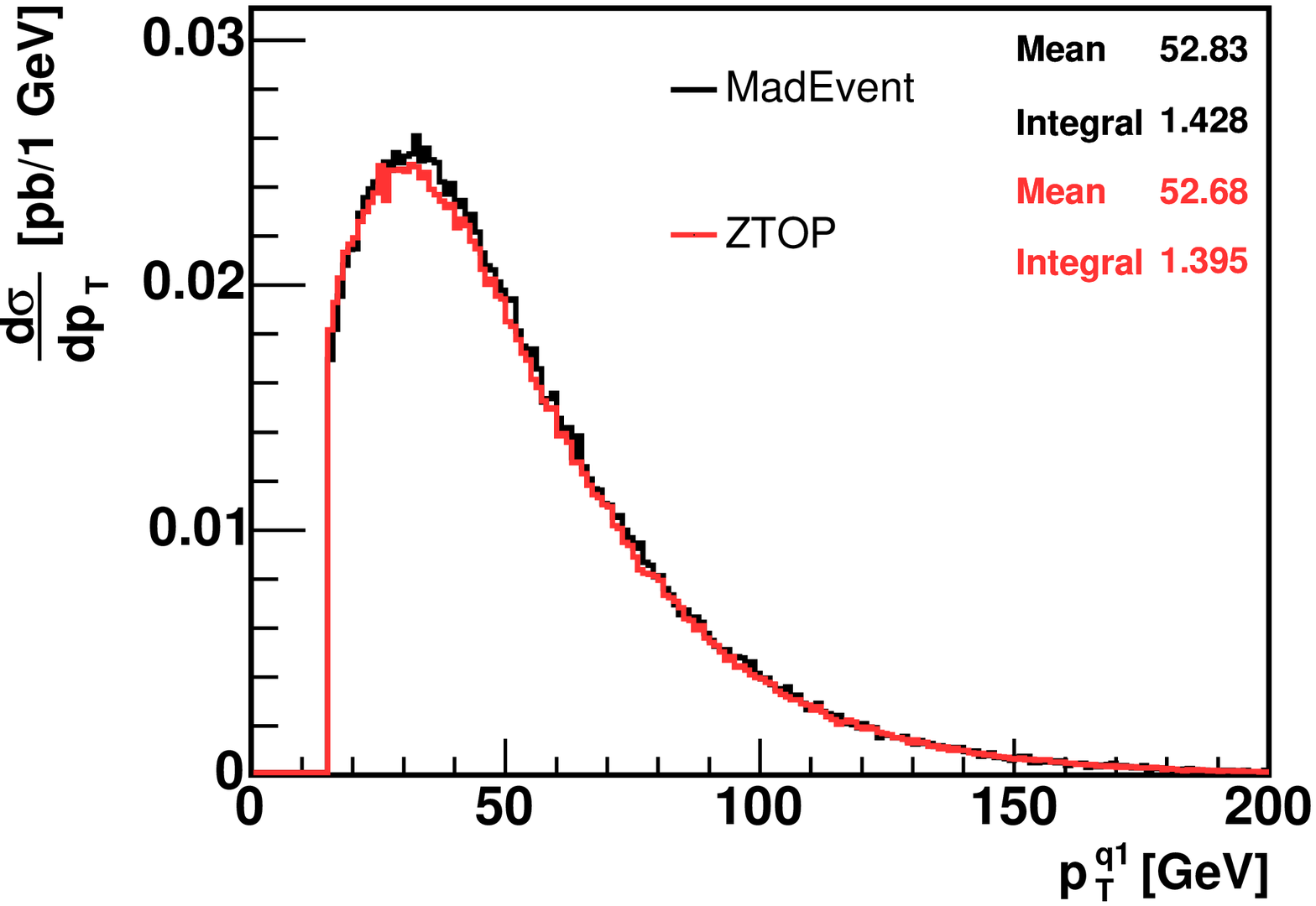}
\includegraphics[width=0.48\textwidth]
{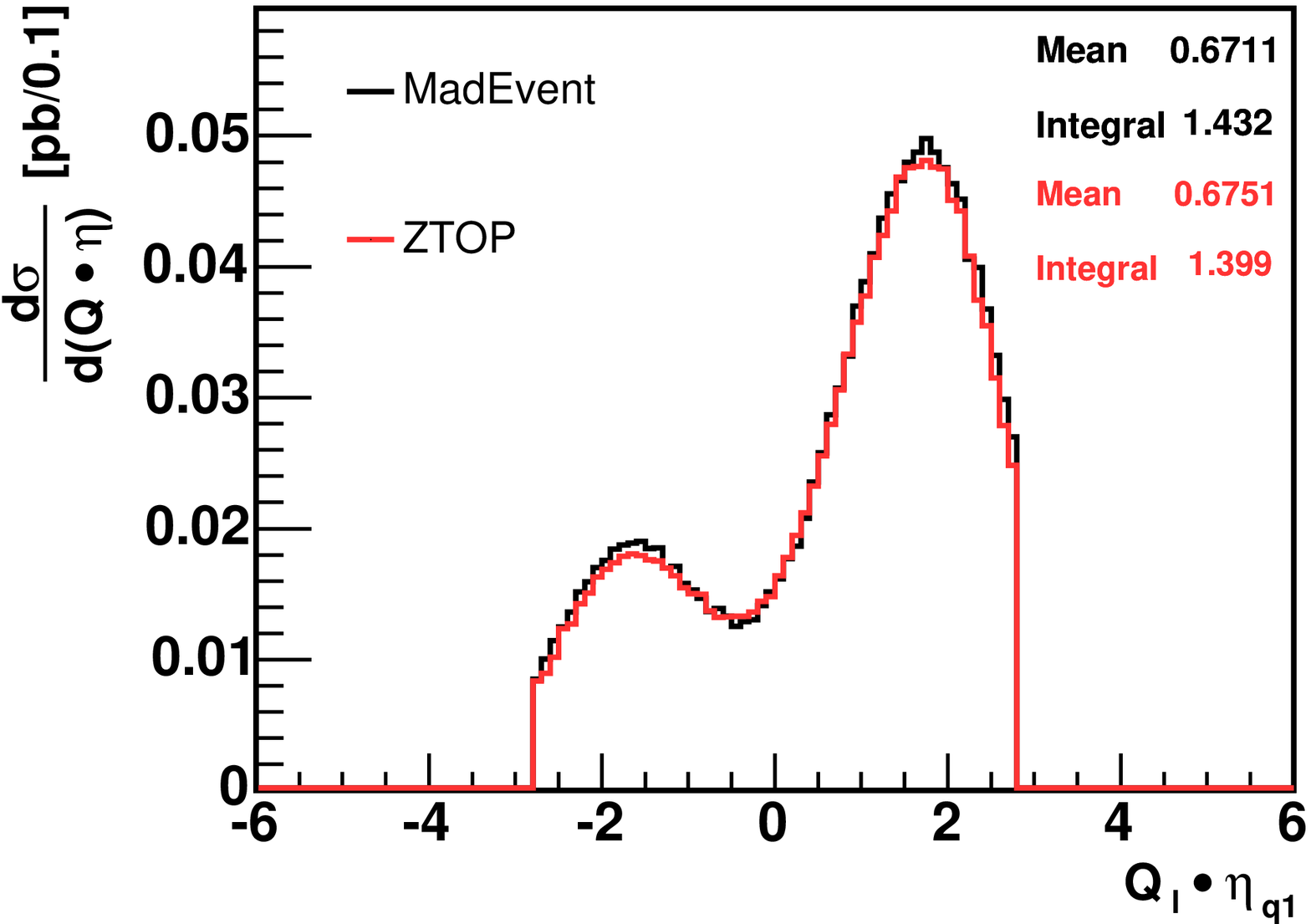}
\end{center}
\caption[tchannelEval]{\label{fig:tchannelEval}Comparison of kinematic
 distributions at parton level for the matched $t$-channel single-top
 Monte Carlo sample with NLO calculations from {\sc Ztop}. The upper
 two plots show the $p_\mathrm{T}$ and pseudo-rapidity distribution
 for $2^\mathrm{nd}$ $b$ quarks. The middle row shows the distributions
 for the top quark. The lower two plots show the $p_\mathrm{T}$ and $\eta$
 distributions for the leading light quark jet.
}
\end{figure}

\begin{figure}[htbp]  
\begin{center}
\includegraphics[width=0.48\textwidth]
{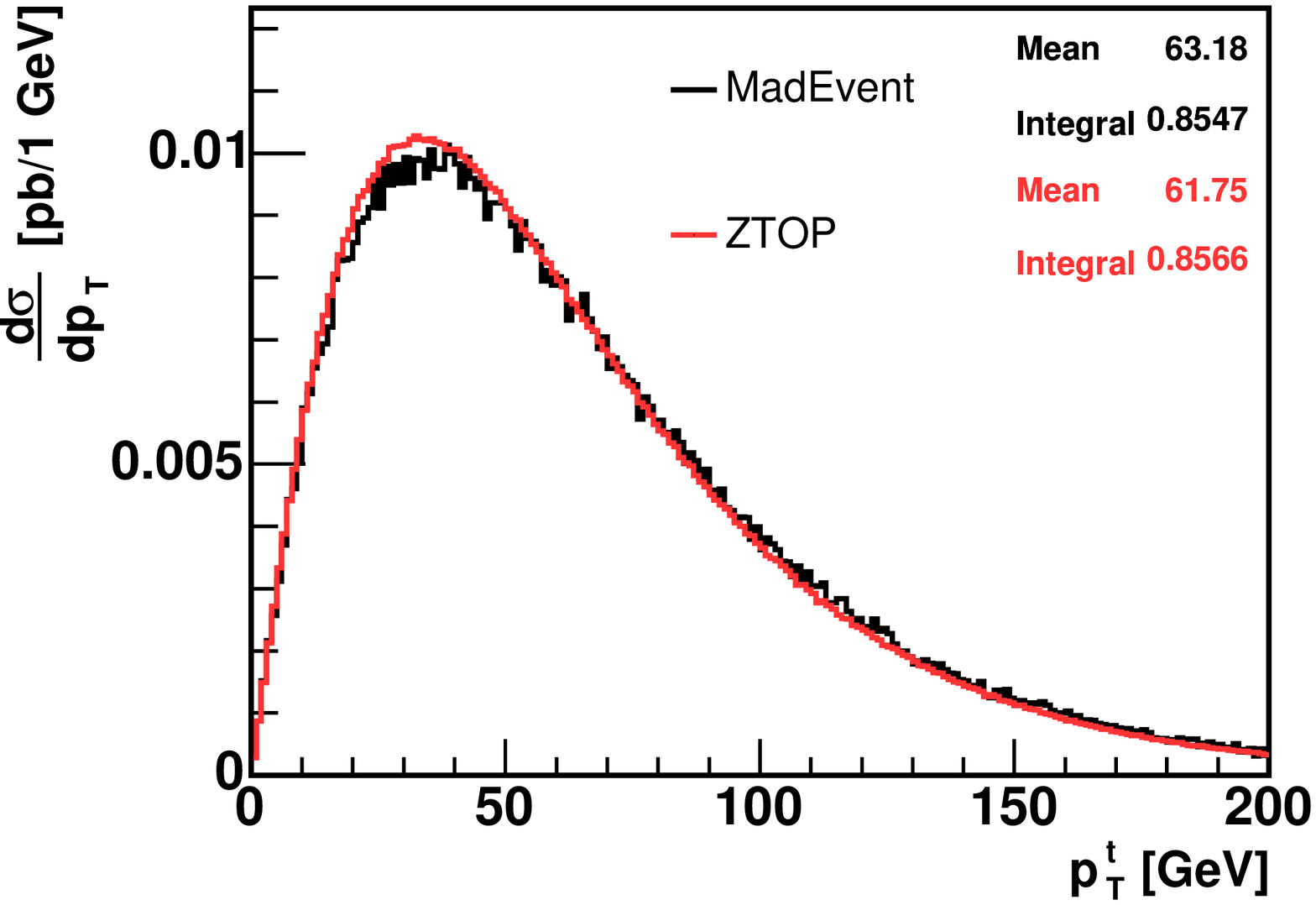}
\includegraphics[width=0.48\textwidth]
{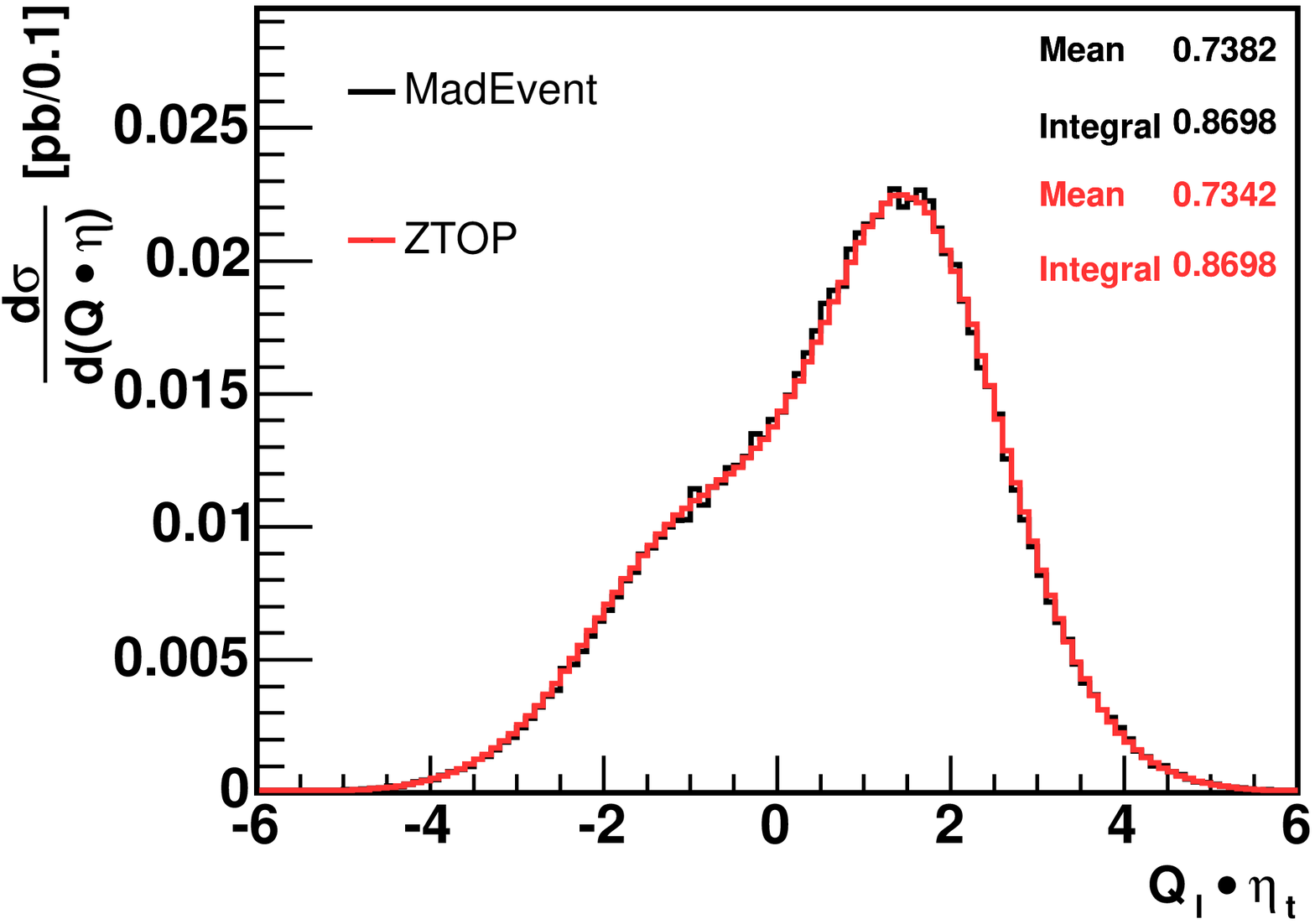}
\includegraphics[width=0.48\textwidth]
{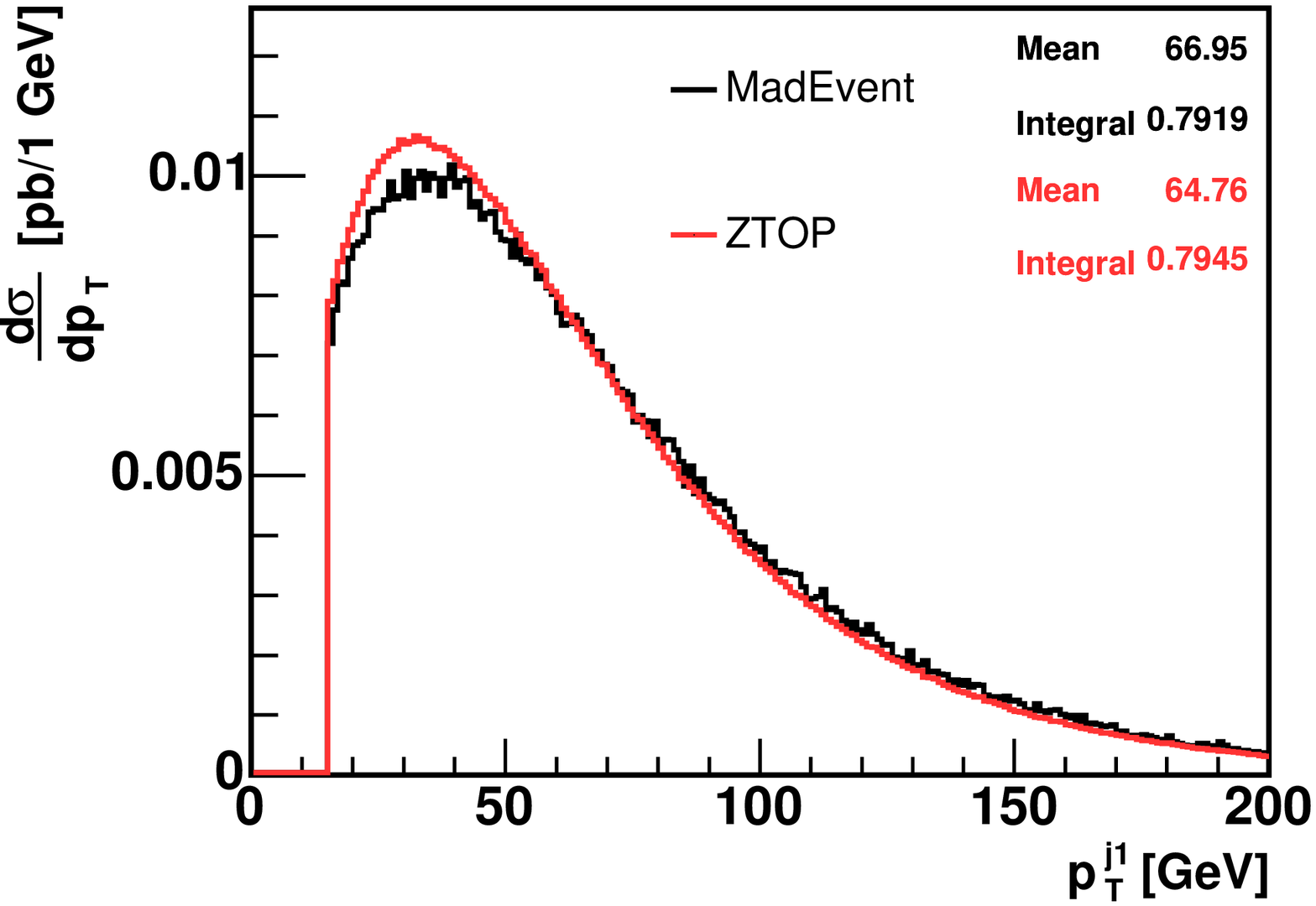}
\includegraphics[width=0.48\textwidth]
{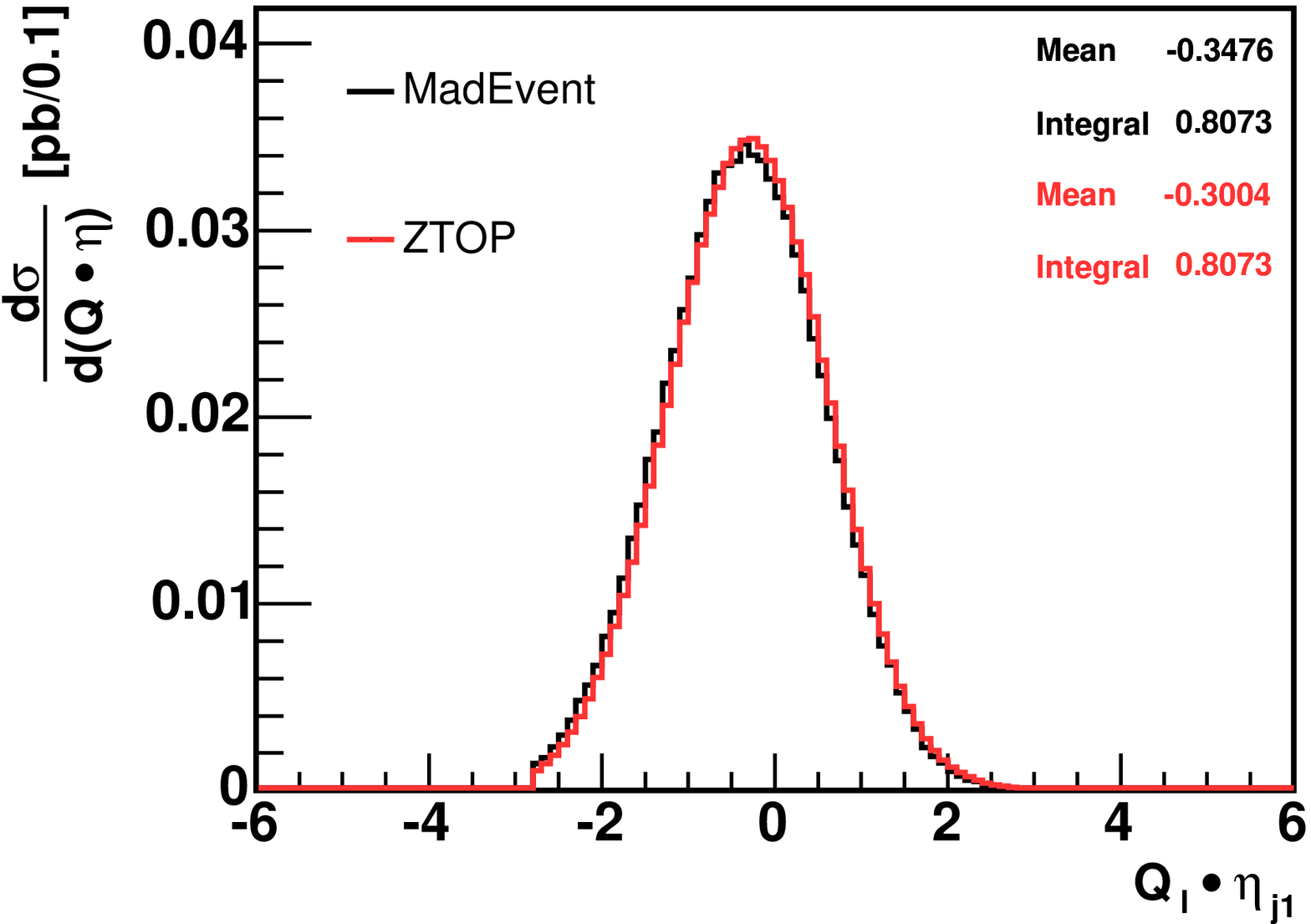}
\end{center}
\caption[schannelEval]{\label{fig:schannelEval}Comparison of kinematic
 distributions at parton level for the matched $s$-channel single-top
 Monte Carlo sample with NLO calculations from {\sc Ztop}. The upper
 two plots show the $p_\mathrm{T}$ and pseudo-rapidity distribution
 of the top quark. The lower two plots show the same distributions
 for the leading jet.
}
\end{figure}

One important question which has to be addressed is how good the matching 
procedure is and how well the final Monte Carlo sample describes 
the single-top $t$-channel kinematics. To achieve this goal we compared
the kinematic distributions of the primary partons obtained from 
the matched {\sc MadEvent} Monte Carlo sample with NLO differential cross
sections that are made available by the {\sc Ztop} 
software~\cite{Sullivan:2004ie}. We found that the shape of 
the kinematic distributions of the $2^\mathrm{nd}$ $b$ quark, 
namely the $p_\mathrm{T}$ and the pseudorapidity distributions, are modeled
quite well. However, we found a small rate difference for visible
$2^\mathrm{nd}$ $b$ quark jets with $p_\mathrm{T} > 15\,\mathrm{GeV}$
and $|\eta|<2.8$, which are the jet cuts used in the CDF single-top
analysis. Therefore, we adjusted the original matching procedure such
that the rate of visible $2^\mathrm{nd}$ $b$ quark jets in
our matched {\sc MadEvent} sample is equal to the rate predicted
by {\sc Ztop}~\cite{Lueck:2006hz}. Effectively, this results in a new
intersection point $K_\mathrm{T}=9\,\mathrm{GeV}/c$ for the matching
procedure. As a result all visible $2^\mathrm{nd}$ $b$ quarks of the 
matched sample are coming from the NLO ($2\rightarrow 3$) sample.
Figure~\ref{fig:t-channel-matching} illustrates the matching procedure.
We have evaluated the matched $t$-channel single-top Monte Carlo
sample by comparing distributions at parton level to the NLO prediction
from {\sc Ztop}. Figure~\ref{fig:tchannelEval} shows a few examples.
We also compared kinematic distributions for the $s$-channel 
production, see figure~\ref{fig:schannelEval}.
In general, we find very good agreement for the Monte Carlo modeling
of the single-top kinematics. We quantify the remaining difference
between the Monte Carlo and the NLO calculation 
by assigning weights to the Monte Carlo events. The weight is derived
from a comparison of several kinematic distributions that are combined
in a weighted average. We apply the single-top event selection to the
Monte Carlo events and sum up the weights. As a result we find an 
estimate on the deviation of the acceptance in Monte Carlo compared 
to the NLO prediction. In the $W+2$ jets bin we find a discrepancy of
$-1.8\%\pm0.9\%\,(\mathrm{MC\; stat.})$ for the $t$-channel, 
i.e. our study indicates that the Monte Carlo estimate of the
acceptance is a little higher than the NLO prediction. 
In the $s$-channel we find 
excellent agreement, no evidence for a deviation, 
$-0.3\%\pm0.7\%\,(\mathrm{MC\; stat.})$. 

The general conclusion from our study is that the {\sc MadEvent}
Monte Carlo events give an excellent representation of the 
single-top production process. Due to the matching procedure for the
$t$-channel sample the NLO effects are sufficiently taken into account.



%
%
%
\subsubsection* {A simulation method of the Electroweak Top Quark Production Events
in the NLO Approximation. Monte-Carlo Generator {\em SingleTop}}
\label{sec:singletoptheorygenerator}
\textbf{Contributed by:~E.E.~Boos, V.E.~Bunichev, L.V.~Dudko, V.I.~Savrin, A.V.~Sherstnev}\\

\paragraph{Introduction}

The CompHEP package~\cite{Boos:2004kh} has been used to prepare a 
special event generator
SingleTop to simulate the electroweak single top quark 
production with its subsequent decays at the Tevatron and LHC.
Single top is expected to be discovered at the Tevatron Run II
and will be a very interesting subject of detail studies at
the LHC (see the review \cite{Beneke:2000hk}).

\begin{figure}[hbtp]
\includegraphics[width=0.9\textwidth]{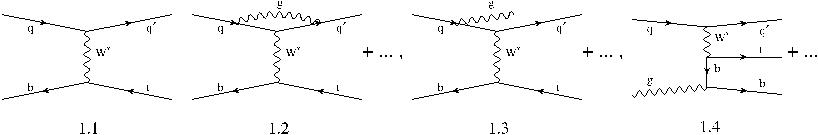}
\caption{LO and representative loop and tree NLO
diagrams of the $t$-channel single top production }
\label{SinT}
\end{figure}

The generator SingleTop includes all three single top processes and 
provides Monte-Carlo unweighted events at the NLO QCD level.
In the paper~\cite{Sullivan:2004ie} it has been argued that
the NLO distributions for $s$-channel process are the same as the LO
multiplied by a known k-factor. The LO cross sections for the $s$-channel
process are shown in the table~\ref{eq:pp->tb} and the NLO cross sections are
taken from the papers~\cite{Stelzer:1997ns,Harris:2002md} and are shown in 
the table~\ref{eq:pp->tb-nlo}
We discuss shortly here only the main process with the largest 
rate, the $t$-channel production.
The representative LO and NLO diagrams are shown in the Fig.~\ref{SinT} 
The top decay is not shown, however it is included at leading order
with all spin correlations.  

\begin{table}[htb]
\begin{center}
\caption{
The total LO cross section of $s$-channel single top quark production process
(The LHC cross section of $pp\to t\bar{b}(\bar{t}b)$ processes are equal
4.96 (3.09) pb; for the Tevatron the cross sections of $p\bar{p}\to t\bar{b}$
and  $p\bar{p}\to\bar{t}b$ processes are the same and equal 0.3 pb (the numbers in brackets)). 
}
\begin{tabular}{|cccc|}
\hline
                       & Processes, pb      &                        &                         \\
\hline
$u\bar{d}\to t\bar{b}$ & $\bar{d}u\to t\bar{b}$ & $\bar{d}c\to t\bar{b}$ & $c\bar{d}\to t\bar{b}$ \\
$u\bar{s}\to t\bar{b}$ & $\bar{s}u\to t\bar{b}$ & $\bar{s}c\to t\bar{b}$ & $c\bar{s}\to t\bar{b}$ \\
 \hline
          2.22 (0.291) &            2.22 (0.006) &          0.26(0.001)   &          0.26 (0.001)   \\
\hline
\hline
$d\bar{u}\to\bar{t}b$ & $\bar{u}d\to\bar{t}b$ & $\bar{c}d\to\bar{t}b$ & $s\bar{c}\to\bar{t}b$ \\
$s\bar{u}\to\bar{t}b$ & $\bar{u}s\to\bar{t}b$ & $d\bar{c}\to\bar{t}b$ & $\bar{c}s\to\bar{t}b$ \\
\hline
        1.285 (0.291) &         1.285 (0.006) &           0.26(0.001) &          0.26 (0.001) \\
\hline
\end{tabular}\label{eq:pp->tb}
\end{center}
\vspace{-5mm}
\end{table}

\begin{table}[htb]
\begin{center}
\caption{ The total NLO cross section~\cite{Harris:2002md} ($M_t=175$ GeV).
}
\begin{tabular}{|c|c|c|c|c|}
\hline
Collider & Process & $t$ & $\bar t$ & $t + \bar t$ \\
\hline
LHC & $t$--channel & $152.6\pm 3.1$ & $90.0\pm 1.9$ & $242.6\pm 3.6$ \\
    & $s$--channel & $6.55\pm0.14$  & $ 4.1\pm0.1 $ & $10.6\pm0.17$ \\
\hline
Tevatron &  $t$--channel & $ 0.95\pm 0.1$  & $ 0.95\pm 0.1$ & $  1.9\pm 0.1$ \\
        &  $s$--channel &  $0.44\pm0.04$  &  $0.44\pm0.04$ & $0.88\pm0.05$ \\
\hline
\end{tabular}
\label{eq:pp->tb-nlo}
\end{center}
\vspace{-5mm}
\end{table}

\paragraph{Overview of the effective NLO approach.}

We compute by means of the CompHEP the LO order process $2\rightarrow 2$
with the b-quark in the initial state and top spin correlated
$1\rightarrow 3$ subsequent decay, put it into PYTHIA~\cite{Sjostrand:2003wg} 
and switch on ISR/FSR. Then with CompHEP we compute   
the NLO tree level corrections -- $2\rightarrow 3$ processes with 
additional b- and light quarks or gluons in the final state 
including also the top decay with spin correlations.
We split the phase space region in "soft" and "hard" parts on $p_t$ 
of those additional b and light jets being from PYTHIA radiation
in the "soft" and from the CompHEP matrix element calculation
in the "hard" regions. The soft part is normalized in such a way that 
all parts being taken together give
known from  calculations the NLO cross section~\cite{Stelzer:1997ns,Harris:2002md}
which are shown in the table~\ref{eq:pp->tb-nlo}
for the LHC and Tevatron.
The splitting parameters are tuned based on the requirements that all the distributions
become smooth after the normalization. The performed
cross checks show an agreement with exact NLO calculations where the 
computed NLO distributions are 
correctly reproduced by our method. Therefore, generator ``SingleTop'' 
prepared in that way
does not have a double counting problem, produces correctly the NLO rate and distributions,
and includes all the spin correlations. 

The first release of the generator~\cite{Boos:cmsnote} did not include
the hard radiation of the light jets, while the latest version~\cite{Boos:SingleTop} 
currently used in the analysis by the Fermilab D0
and the LHC CMS collaborations includes all the mentioned properties.

\paragraph{Practical implementation of the method in generator SingleTop.}

The generator ``SingleTop'' (based on CompHEP program) realizes an 
effective NLO approach of event generation for the single top-quark 
processes by taking into account the main NLO corrections to kinematics.
The model of simulation is based on the phase space slicing method.

The method begins with the $t$-channel cross section 
in the Born approximation, taking into 
account the full set of Feynman diagrams where the 
top quark appears with additional b and 
light quarks in the final state ($2\to 3$). However, calculation of the process 
$2\to 3$ at the tree level doesn't include large logarithmic QCD corrections 
(related to the process $g\to b\bar b$)
that appears in the "soft" phase space region where the
b quark has a small $P_T$.
It is possible to calculate these corrections via standard renormalization procedure
and include them into partonic distributions of the b-quarks in the proton.
In this case the reaction $2\to 2$ (with b-quark in the initial state) would 
be the LO approach of the $t$-channel process. 
In the same way another b-quark should appear also in the final state. 
It follows from the fact that b-quark can be produced in the proton
only in $b\bar b$ pairs from the virtual gluon. One can simulate the final 
b-quark in the process $2\to2$ via ISR-mechanism. In this case b-quark could be 
produced by initial state radiation and will appear in the final state within a
branch of parton shower, from the splitting function $g\to b\bar b$. One of these 
b-quarks (from gluon splitting) is the initial hard parton and the second one 
goes to the final state. The LO cross sections for the $2\to 2$ processes 
are shown in the table~\ref{eq:pp->tj}. The LO cross sections of $2\to 3$ processes are shown in the
table~\ref{eq:pp->tbj}, the cut $P_T(b)>10$ GeV is applied. 

\begin{table}[htb]
\begin{center}
\caption{The LO cross sections of $t$-channel $2\to2$ processes.
(The total LHC cross sections of the process $pp\to tj(\bar{t}j)$ is
155.39 (89.85) pb; the Tevatron cross sections of 
 $p\bar{p}\to tj$ and $p\bar{p}\to\bar{t}j$ processes are the same and equal 
0.966~pb (the numbers in brackets))
}
\begin{tabular}{|ccc|}
\hline
                     &  Processes, pb      &            \\
\hline
$ub\to dt$           & $ub\to st$    & $\bar{d}g\to\bar{c}t$ \\
$bu\to dt$           & $bu\to st$    & $g\bar{d}\to\bar{c}t$ \\
$cb\to dt$           & $cb\to st$    & $\bar{s}g\to\bar{c}t$ \\
$bc\to dt$           & $bc\to st$    & $g\bar{s}\to\bar{c}t$ \\
$\bar{d}b\to\bar{u}t$ &               &                       \\
$b\bar{d}\to\bar{u}t$ &               &                       \\
$\bar{s}b\to\bar{u}t$ &               &                       \\
$b\bar{s}\to\bar{u}t$ &               &                       \\
\hline
      129.26 (0.869) & 15.01 (0.057) &         11.12 (0.040) \\
\hline
\hline
$\bar{u}\bar{b}\to\bar{d}\bar{t}$ & $\bar{u}\bar{b}\to\bar{s}\bar{t}$ & $d\bar{b}\to c\bar{t} $ \\
$\bar{b}\bar{u}\to\bar{d}\bar{t}$ & $\bar{b}\bar{u}\to\bar{s}\bar{t}$ & $\bar{b}d\to c\bar{t} $ \\
$\bar{c}\bar{b}\to\bar{d}\bar{t}$ & $\bar{c}\bar{b}\to\bar{s}\bar{t}$ & $s\bar{b}\to c\bar{t} $ \\
$\bar{b}\bar{c}\to\bar{d}\bar{t}$ & $\bar{b}\bar{c}\to\bar{s}\bar{t}$ & $\bar{b}s\to c\bar{t} $ \\
$d\bar{b}\to u\bar{t}           $ &                                   &                         \\
$\bar{b}d\to u\bar{t}           $ &                                   &                         \\
$s\bar{b}\to u\bar{t}           $ &                                   &                         \\
$\bar{b}s\to u\bar{t}           $ &                                   &                         \\
\hline
                    66.99 (0.869) &                     10.05 (0.057) &           12.81 (0.040) \\
\hline
\end{tabular}\label{eq:pp->tj}
\end{center}
\vspace{-5mm}
\end{table}
\begin{table}[htb]
\begin{center}
\caption{
The LO cross sections of $t$-channel $2\to3$ processes after the cut $P_T(b)>10$ GeV
(The total LHC cross sections of the process $pp\to tq\bar{b}$ ($pp\to\bar{t}qb$) is
82.3 (47.9) pb; the Tevatron cross sections of $p\bar{p}\to tq\bar{b}$ and $p\bar{p}\to\bar{t}qb$
 processes are the same and equal 0.379~pb (the numbers in brackets); the cut is explained in the text).
}
\begin{tabular}{|ccc|}
\hline\small
                             & Subprocesses       &                             \\
\hline
$ug\to dt\bar{b}$            & $ug\to st\bar{b}$ & $\bar{d}g\to\bar{c}t\bar{b}$ \\
$gu\to dt\bar{b}$            & $gu\to st\bar{b}$ & $g\bar{d}\to\bar{c}t\bar{b}$ \\
$cg\to dt\bar{b}$            & $cg\to st\bar{b}$ & $\bar{s}g\to\bar{c}t\bar{b}$ \\
$gc\to dt\bar{b}$            & $gc\to st\bar{b}$ & $g\bar{s}\to\bar{c}t\bar{b}$ \\
$\bar{d}g\to\bar{u}t\bar{b}$ &                   &                              \\
$g\bar{d}\to\bar{u}t\bar{b}$ &                   &                              \\
$\bar{s}g\to\bar{u}t\bar{b}$ &                   &                              \\
$g\bar{s}\to\bar{u}t\bar{b}$ &                   &                              \\
\hline
            68.8 (0.328) pb &   7.6 (0.03) pb &              5.9 (0.021) pb \\
\hline
\hline
$\bar{u}g\to\bar{d}\bar{t}b$ & $\bar{u}g\to\bar{s}\bar{t}b$ & $dg\to c\bar{t}b $ \\
$g\bar{u}\to\bar{d}\bar{t}b$ & $g\bar{u}\to\bar{s}\bar{t}b$ & $gd\to c\bar{t}b $ \\
$\bar{c}g\to\bar{d}\bar{t}b$ & $\bar{c}g\to\bar{s}\bar{t}b$ & $sg\to c\bar{t}b $ \\
$g\bar{c}\to\bar{d}\bar{t}b$ & $g\bar{c}\to\bar{s}\bar{t}b$ & $gs\to c\bar{t}b $ \\
$dg\to u\bar{t}b           $ &                              &                    \\
$gd\to u\bar{t}b           $ &                              &                    \\
$sg\to u\bar{t}b           $ &                              &                    \\
$gs\to u\bar{t}b           $ &                              &                    \\
\hline
             36.2 (0.328) pb &                4.9 (0.03) pb &     6.8 (0.021) pb \\
\hline
\end{tabular}
\label{eq:pp->tbj}
\end{center}
\vspace{-5mm}
\end{table}

Calculations of the process $2\to 3$ at the tree level approach doesn't include 
large logarithmic corrections (related to the process $g\to b\bar b$), but 
the exact tree level calculations correctly simulate behavior of the b-quark 
in the "hard" phase space region with the large $P_T$. 
We will demonstrate, that combination of the processes $2\to2$ and $2\to3$ allows 
us to construct MC samples at "effective" NLO level approach.
We can prepare correct events with "soft" b-quark via ISR simulation.
But in this case we lose the significant contribution of the "hard" b-quark. 
We can probably come to an appropriate result if we would use
different strategies to simulate the different kinematic regions of 
the phase space.
Unfortunately, we can't naively combine the samples with $2\to2$ and $2\to3$ 
processes because in this case we will get double counting of some phase space 
regions. To avoid the problem of the double counting we propose to use different 
methods of MC simulation in the different phase space regions and combine 
them based on some kinematic parameters.
\begin{figure}[htb]
\begin{minipage}[b]{.46\linewidth}
\begin{center}
\includegraphics[width=\textwidth]{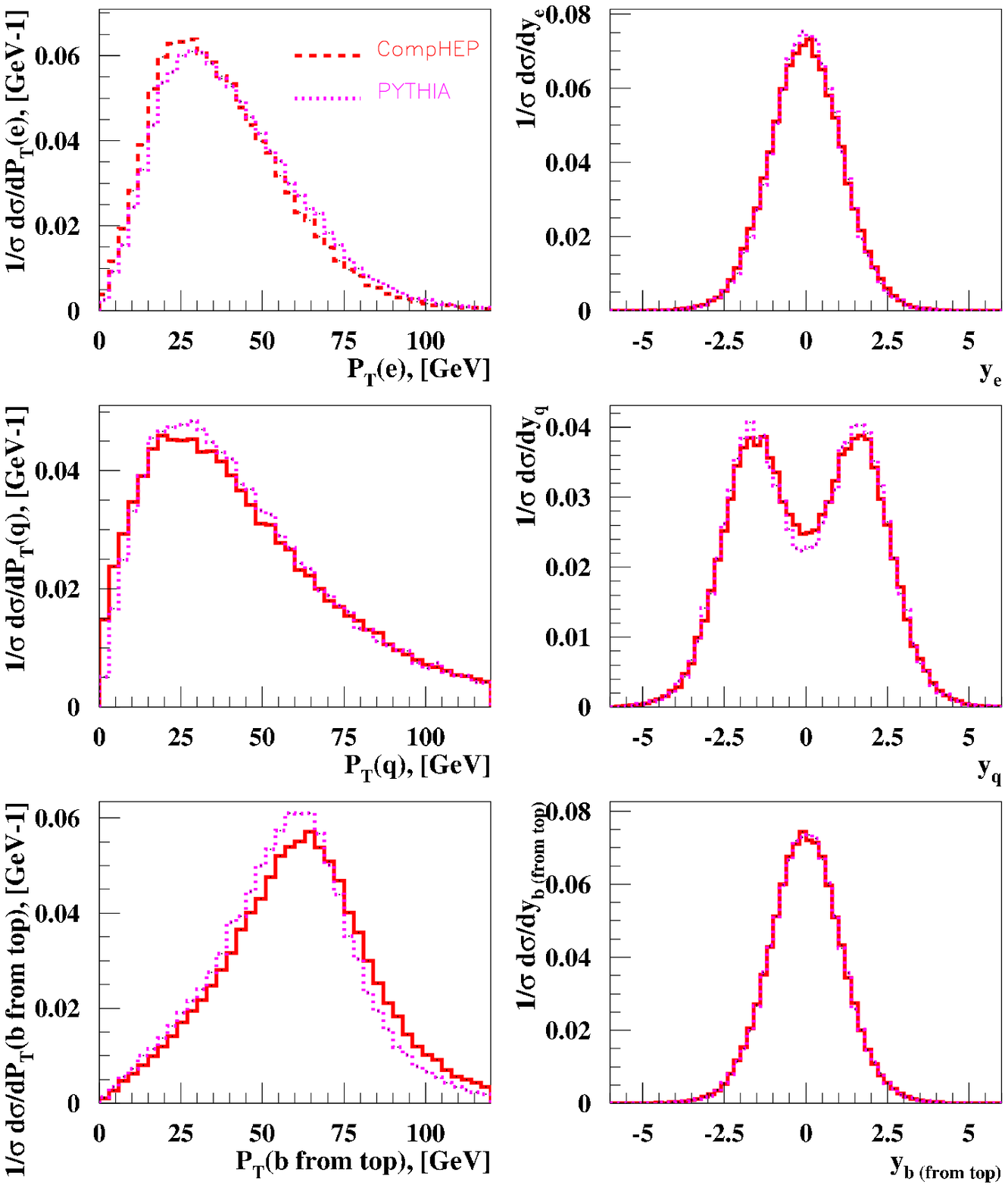}
\end{center}
\caption{
The comparison of $P_T$ and  $\eta$ distributions
for the $pp\to tq+b_{ISR}$~(PYTHIA) and $pp\to tq+b_{LO}$~(CompHEP)
simulations for the Tevatron. The distributions are normalized to unity and
no cuts applied. }
\label{fg:tqbtq_tev1}
\end{minipage}\hfill
\begin{minipage}[b]{.46\linewidth}
\includegraphics[width=\textwidth]{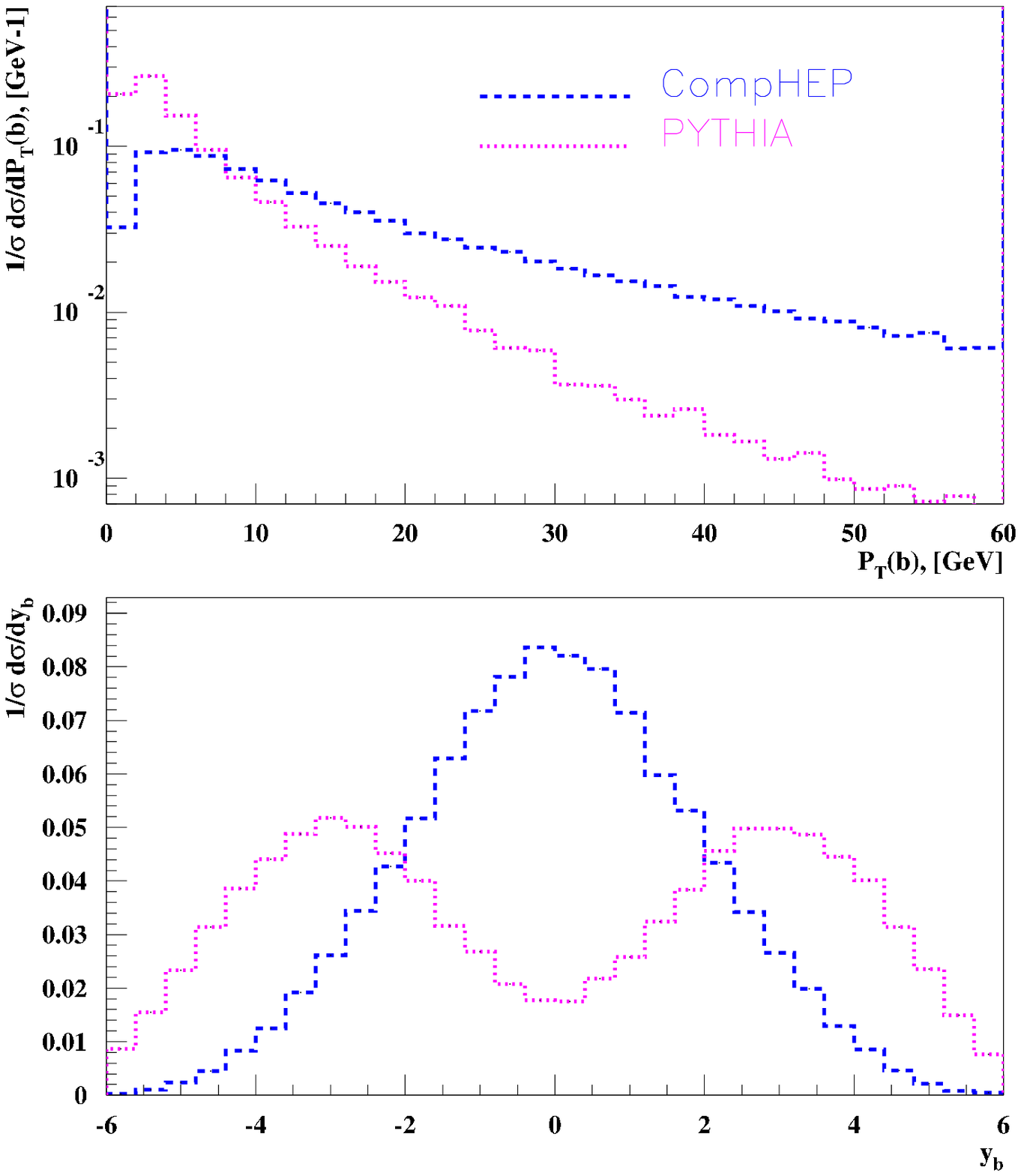}
\caption{
The comparison of $P_T$ and  $\eta$ distributions
for the $b_{ISR}$ and $b_{LO}$ in the
$pp\to tq+b_{ISR}$~(PYTHIA) and $pp\to tq+b_{LO}$~(CompHEP)
simulations for the Tevatron. The distributions are normalized to unity and
no cuts applied.  
}
\label{fg:tqbtq_tev2}
\end{minipage}
\end{figure}

\begin{figure}[htb]
\begin{minipage}[b]{.46\linewidth}
\includegraphics[width=\textwidth,height=8.5cm]{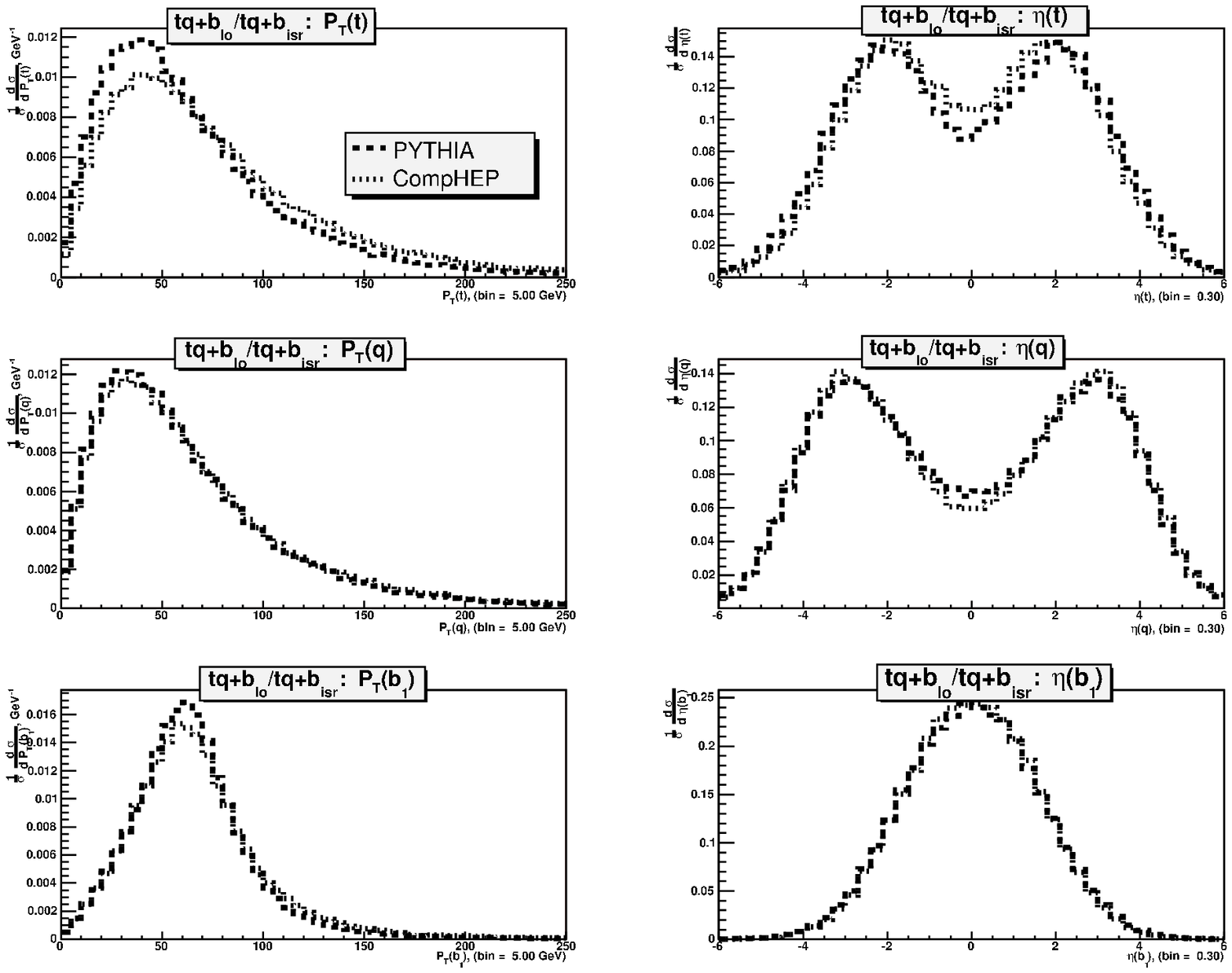}
\caption{
The comparison of $P_T$ and  $\eta$ distributions
for the $pp\to tq+b_{ISR}$~(PYTHIA) and $pp\to tq+b_{LO}$~(CompHEP)
simulations for the LHC. The distributions are normalized to unity and
no cuts applied.}
\label{fg:tqbtq_lhc1}
\end{minipage}\hfill
\begin{minipage}[b]{.46\linewidth}
\includegraphics[width=0.8\textwidth]{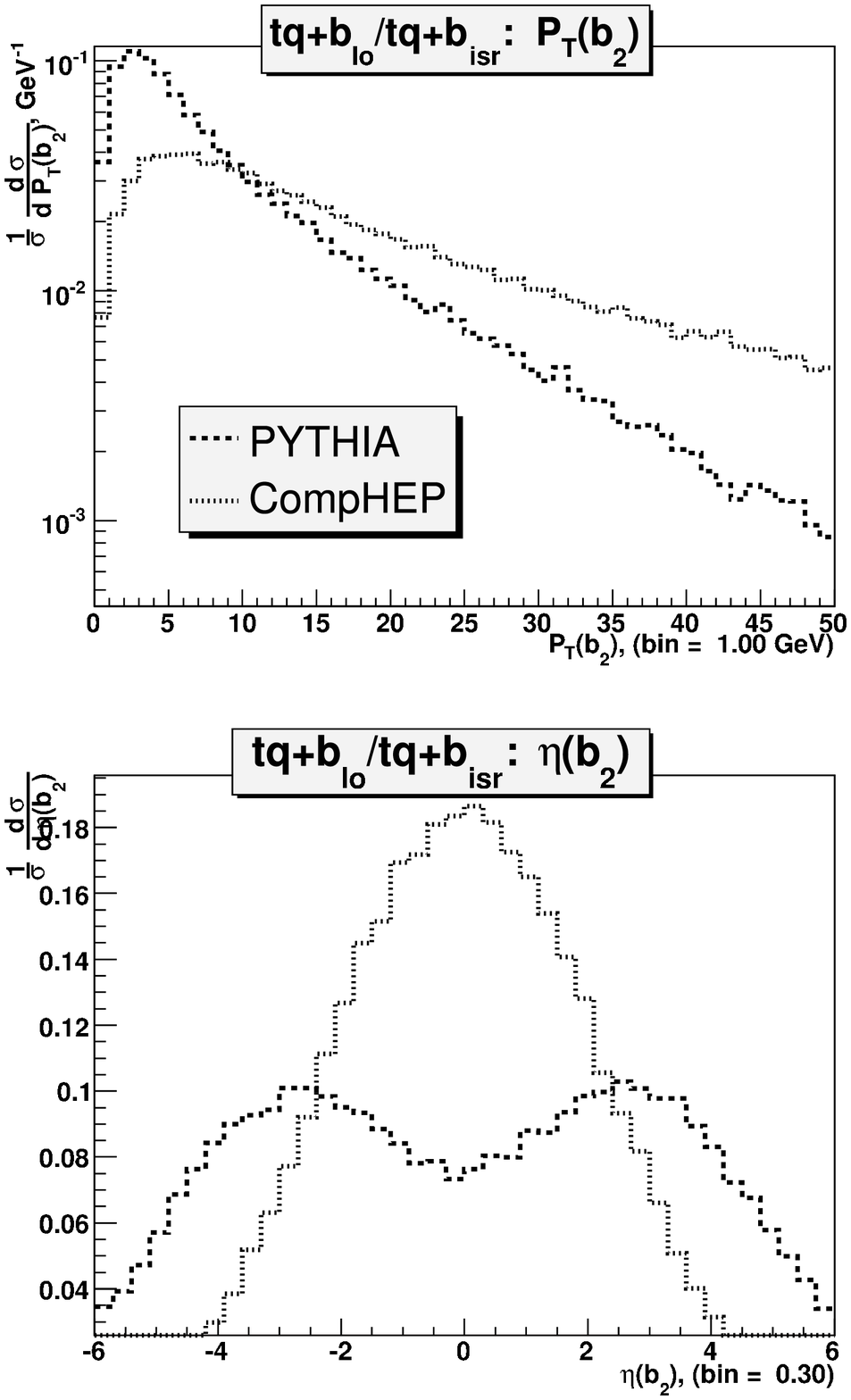}
\caption{
The comparison of $P_T$ and  $\eta$ distributions
for the $b_{ISR}$ and $b_{LO}$ in the
$pp\to tq+b_{ISR}$~(PYTHIA) and $pp\to tq+b_{LO}$~(CompHEP)
simulations for the LHC. The distributions are normalized to unity and
no cuts applied.  
}
\label{fg:tqbtq_lhc2}
\end{minipage}
\end{figure}
Figures~\ref{fg:tqbtq_tev1}-\ref{fg:tqbtq_lhc2} show the normalized distributions, that 
have been prepared for the Tevatron and LHC.
On these plots we can see that the distributions for $P_T$ and
pseudorapidity of the top and light quarks looks similar (Figs.~\ref{fg:tqbtq_tev1},~\ref{fg:tqbtq_lhc1}), 
but the distributions of the
additional b-quark (that comes from gluon-splitting) differ
significantly (Figs.~\ref{fg:tqbtq_tev2},~\ref{fg:tqbtq_lhc2}).
The distribution for pseudorapidity of additional ISR b, have a peaks at 
larger 
values than the distributions
for processes $2\to 3$ at tree level. The $P_T$ spectra for the events 
that we prepare in PYTHIA with ISR simulation are "softer" than in tree level calculations. 
The main contribution from the large logarithmic appears in the "soft" 
region of $P_T(b)$.
Therefore, it is reasonable to use transverse momentum of additional b-quark as a 
kinematic parameter for slicing the phase space to hard and soft regions.
To prepare events at NLO effective approach we apply the following 
procedure:
first, we prepare the CompHEP events $2\to 3$ (at tree level) with $P_T(b)$ 
larger than some
critical value ${P^0}_T$. Then we prepare events $2\to2$ in the "soft" region
of the phase space with $P_T(b)<{P^0}_T$. The cross section of $2\to2$
events in the "soft" region we multiply by K-factor
for taking into account loop corrections which do not change significantly the kinematic
distributions. The value for K-factor we can calculate with the requirements of normalization of
the events in the full phase space to the total NLO cross section, as demonstrated 
in the following equation:
$$
\sigma_{NLO}=K\cdot\sigma_{PYTHIA}(2\to2)|_{P_T(b)<P_T^0}+
                     \sigma_{CompHEP}(2\to3)|_{P_T(b)>P_T^0}.
$$
The K-factor here is a function of slicing parameter ${P^0}_T$, the total NLO cross 
section we know from exact NLO calculations~\cite{Stelzer:1997ns,Harris:2002md}.

In case of LHC collider we have:
$$\sigma_{CompHEP}(2\to3)|_{P_T^b>20 
   \mbox{\scriptsize\rm GeV}}\approx108.7\;\mbox{\rm pb},$$
$$\sigma_{CompHEP}(2\to3)|_{P_T^b > 10 
   \mbox{\scriptsize\rm GeV}}\approx125.7\;\mbox{\rm pb}$$
and K=0.89 for $P_T^{0}=20$ GeV, and k=0.77 for $P_T^{0}=10$ GeV. 

In case of TEVATRON collider we have:
$$\sigma_{CompHEP}(2\to3)|_{P_T^b>20 
\mbox{\scriptsize\rm GeV}}\approx0.46\;\mbox{\rm pb}$$
$$\sigma_{CompHEP}(2\to3)|_{P_T^b>10 
\mbox{\scriptsize\rm GeV}}\approx0.72\;\mbox{\rm pb}.$$
and K=1.32 for $P_T^{0}=20$ GeV, and k=1.21 for 
$P_T^{0}=10$ GeV. 

\begin{figure}[htb]
\begin{minipage}[b]{.46\linewidth}
\includegraphics[width=\textwidth]{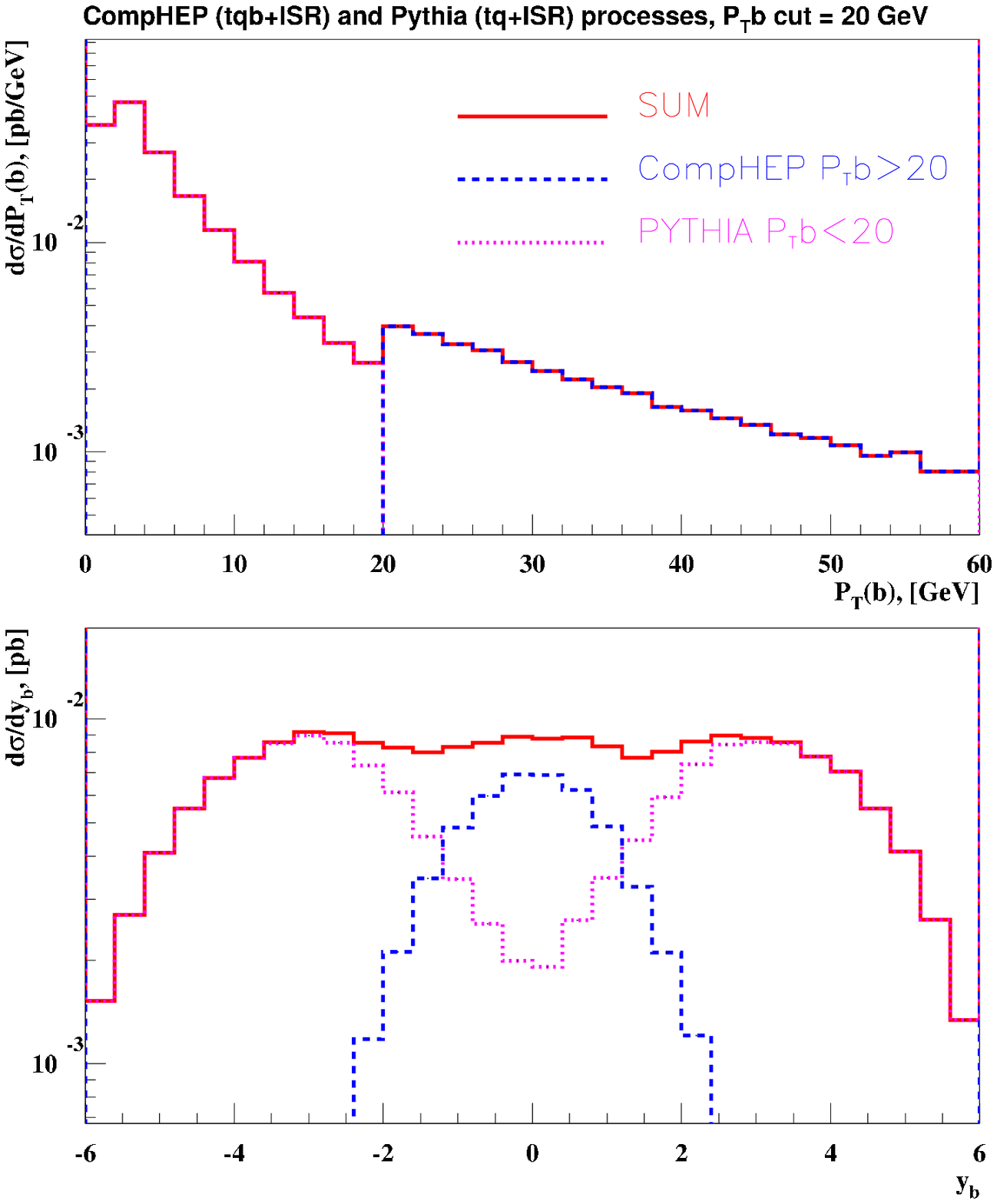}
\caption{
The combined distributions for the
"soft" $pp\to tq+b_{ISR}$~(PYTHIA) and "hard" $pp\to tq+b_{LO}$~(CompHEP) 
regions for the Tevatron collider with $P_T^0(b)=20$ GeV.}
\label{fg:ptd20_tev}
\end{minipage}\hfill
\begin{minipage}[b]{.46\linewidth}
\includegraphics[width=\textwidth]{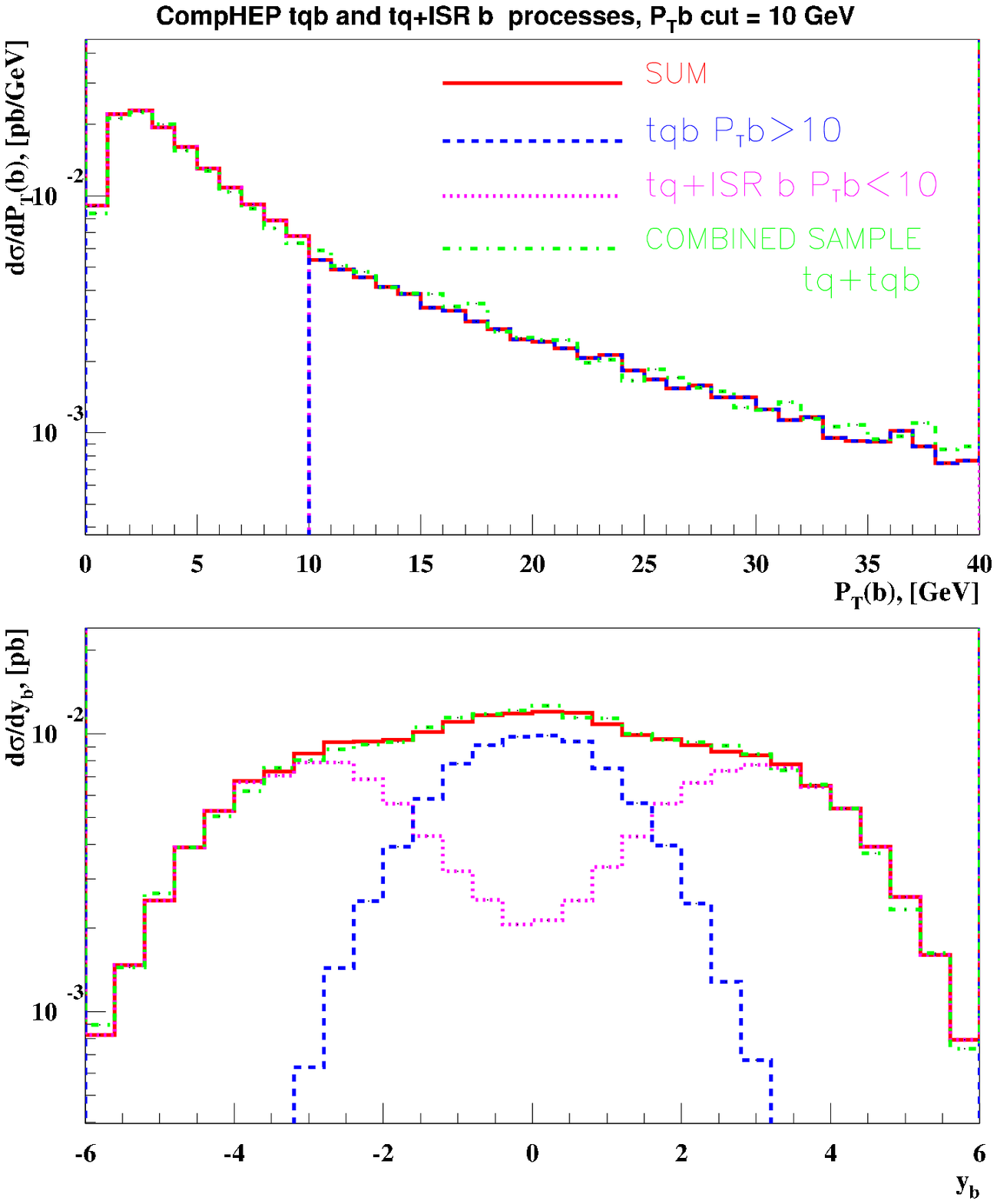}
\caption{
The combined distributions for the
"soft" $pp\to tq+b_{ISR}$~(PYTHIA) and "hard" $pp\to tq+b_{LO}$~(CompHEP) 
regions for the Tevatron collider with $P_T^0(b)=10$ GeV.}
\label{fg:ptd10_tev}
\end{minipage}
\end{figure}

\begin{figure}[htb]
\begin{minipage}[b]{.4\linewidth}
\includegraphics[width=\textwidth]{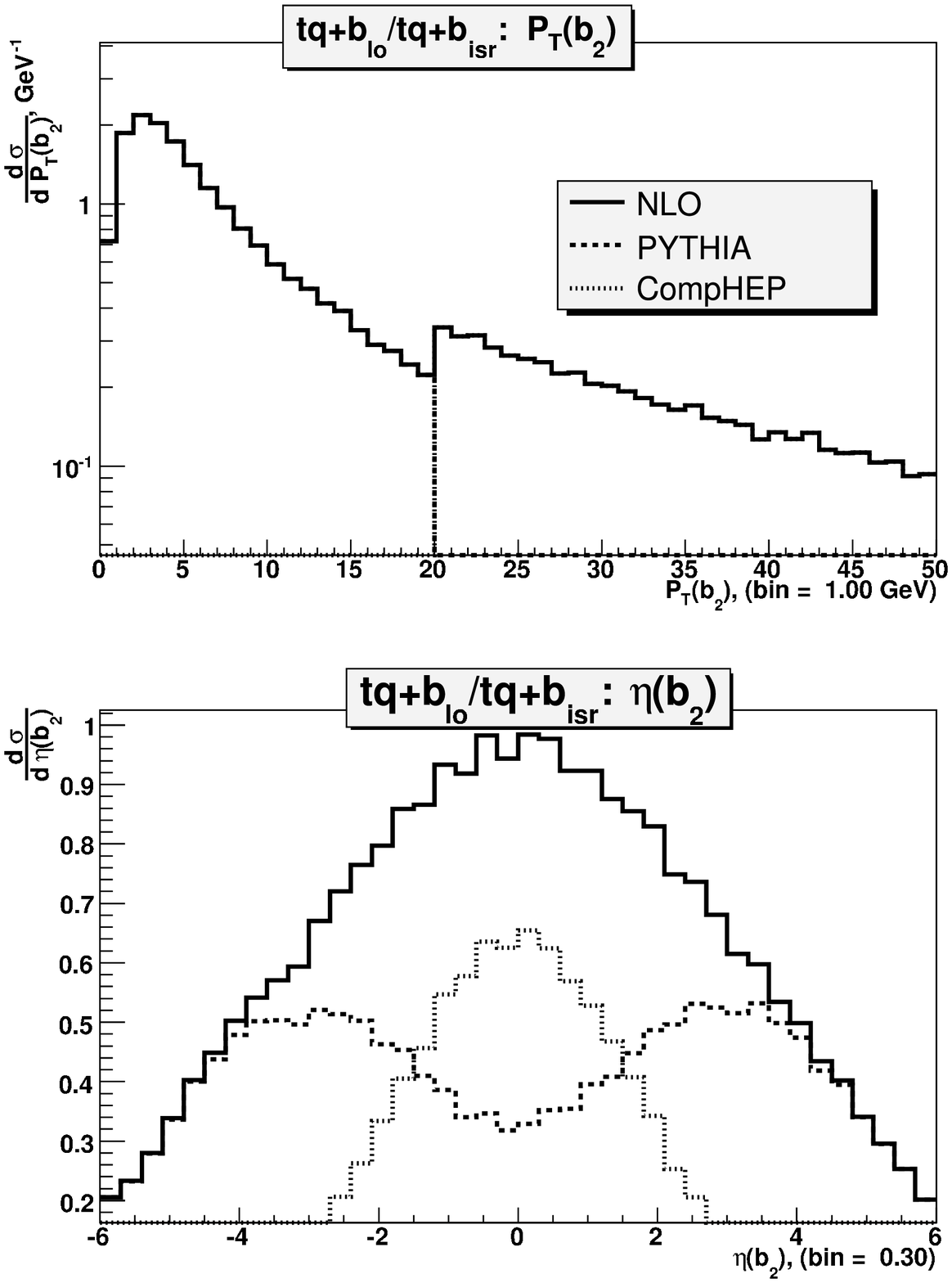}
\caption{The combined distributions for the
"soft" $pp\to tq+b_{ISR}$~(PYTHIA) and "hard" $pp\to tq+b_{LO}$~(CompHEP) 
regions for the LHC collider with $P_T^0(b)=20$~GeV.}
\label{fg:ptd20_lhc}
\end{minipage}\hfill
\begin{minipage}[b]{.57\linewidth}
\includegraphics[width=\textwidth,height=8.5cm]{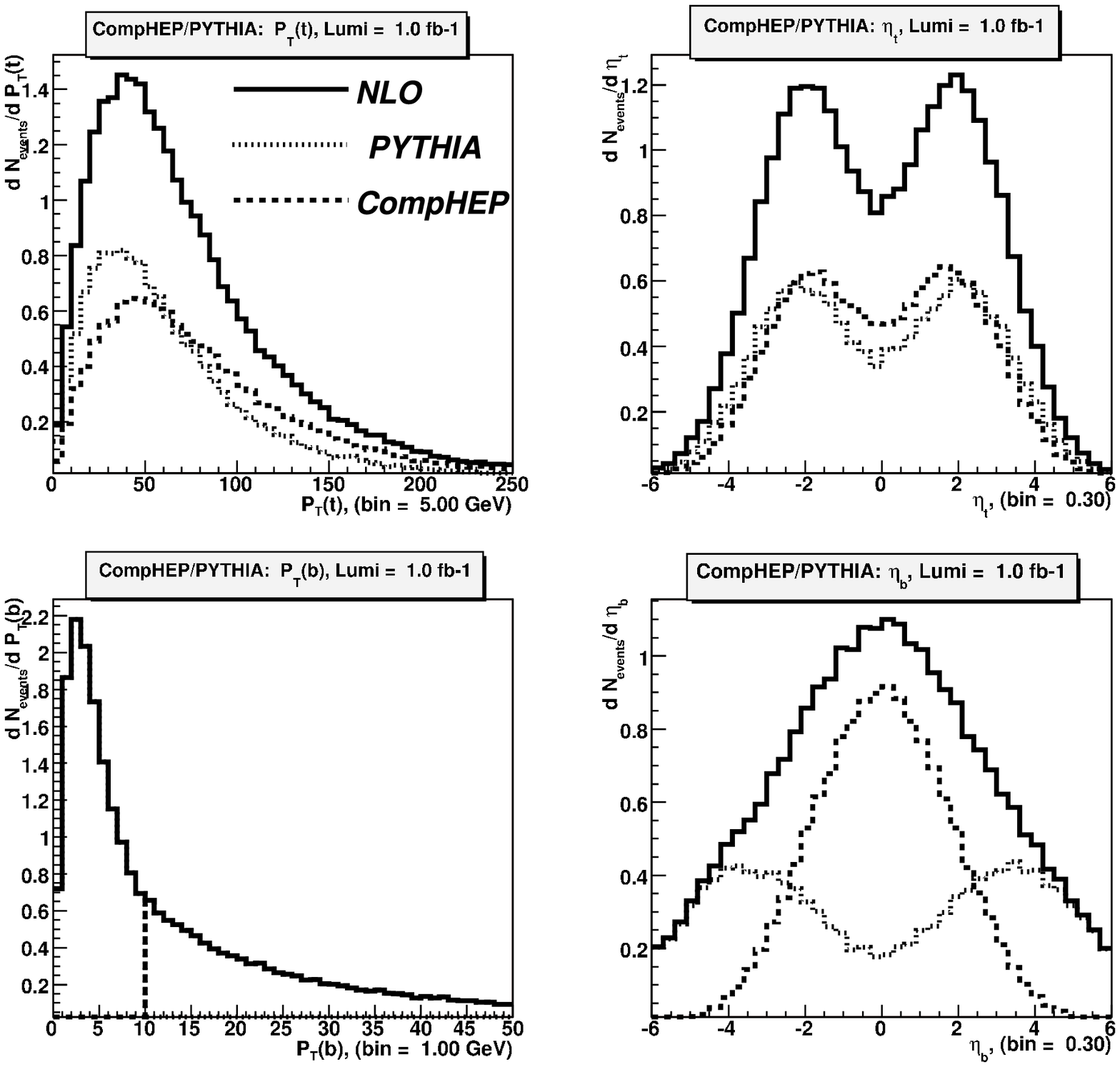}
\caption{The combined distributions for the
"soft" $pp\to tq+b_{ISR}$~(PYTHIA) and "hard" $pp\to tq+b_{LO}$~(CompHEP) 
regions for the LHC collider with $P_T^0(b)=10$~GeV.}
\label{fg:ptd10_lhc}
\end{minipage}
\end{figure}

The natural requirement for the correct slicing parameter ${P^0}_T$ is a 
smoothness of the final $P_T$ distribution in the whole kinematic region for the 
additional b-quark. On the Fig.~\ref{fg:ptd20_tev} and Fig.~\ref{fg:ptd20_lhc} shown the distributions
for the ${P^0}_T=20$ GeV and we can see the bump at the matching point. 
After series of iterations we  have found that
$P_T$ distribution becomes smooth enough with ${P^0}_T=10$ GeV. The result is
shown in the Figure~\ref{fg:ptd10_tev}. The distributions for the LHC collider
are shown in the figure~\ref{fg:ptd10_lhc} for the same value of ${P^0}_T=10$ GeV.
The algorithm described above we call "effective NLO approach".

\paragraph{Comparison of the results.}

\begin{figure}[hbt]
\includegraphics[width=\textwidth,height=0.4\textheight]{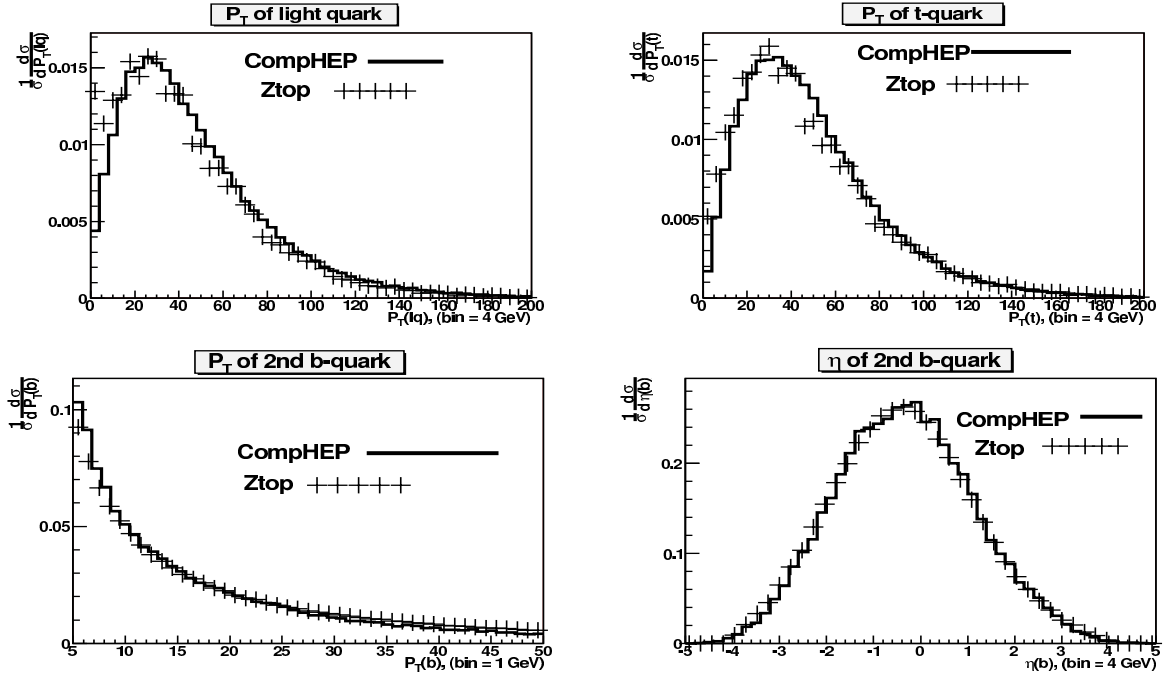}
\caption{
The $P_T$ and pseudorapidity distributions of final quarks in effective
NLO approach (``SingleTop'') and exact NLO calculations (ZTOP) for the Tevatron collider.
}
\label{fg:ztop_stop}
\end{figure}
\begin{figure}[hbt]
\includegraphics[width=\textwidth,height=0.4\textheight]{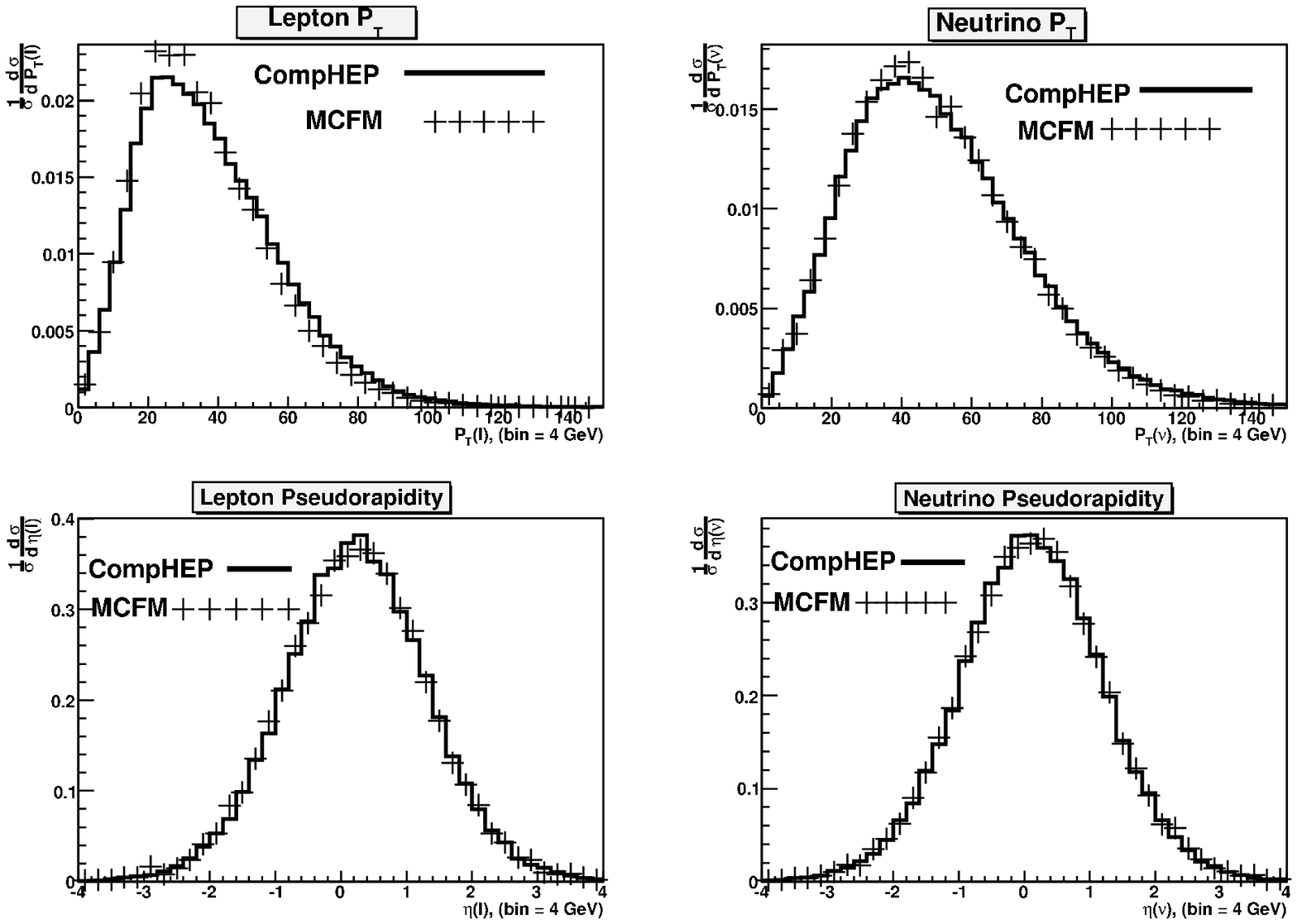}
\caption{
The $P_T$ and pseudorapidity distributions of final leptons from top-quark decay in effective
NLO approach (``SingleTop'') and exact NLO calculations (MCFM) for the Tevatron collider.
}
\label{fg:mcfm_stop}
\end{figure}

To check the correctness of our approach we compare our results with
two independent NLO calculations. The programs ZTOP~\cite{Sullivan:2004ie} and 
MCFM~\cite{Campbell:2004ch} provide the kinematic distributions at NLO level. 
The MCFM takes into account the NLO corrections in the decay of t-quark as well as 
in its production. The ZTOP includes NLO
corrections only in the production of top quark. The ZTOP and MCFM
programs provide the possibility to calculate NLO distributions, but do not simulate events
which are important in the real analysis. We should note, that due to the model 
of showering for the final partons, generator ``SingleTop'' takes into account the most part of
NLO corrections in the decay of t-quark as well as in the production. 
We compare the representative distributions 
from our effective NLO approach with exact NLO calculations. 
The results are shown in the Figures~\ref{fg:ztop_stop},~\ref{fg:mcfm_stop}. 
We can see how the events simulated in effective NLO approach
correctly reproduce the exact NLO distributions produced by ZTOP and MCFM programs.
The good agreement in distributions demonstrates the correctness of the simple approach
to model the most important part of NLO QCD corrections on the level of event simulations.

\paragraph*{ACKNOWLEDGEMENTS}
 The work is partly supported by RFBR 04-02-16476, RFBR 04-02-17448, 
Universities of Russia UR.02.02.503, and Russian Ministry of 
Education and Science NS.1685.2003.2 grants.



%
%
%
\subsubsection*{$W$+jets as a Background to Discovering Single Top Quarks}
\label{sec:singletoptheoryasym}
\textbf{Contributed by:~M.T..~Bowen, S.D.~Ellis, and M.J.~Strassler}

Standard Model production of $W$ bosons and associated jets is
currently obstructing the discovery of single-top-quark production at
the Tevatron.  This background is now known to be significantly larger
than expected a few years ago.  The systematic errors on prediction
and measurement of this background, especially in the context of $b$
tagging, have made a simple counting experiment virtually impossible,
as the uncertainties are comparable to the single-top signal.  It
seems necessary to use the kinematic distributions (``shapes'') of the
main backgrounds ($W$+jets, $t\bar{t}$, QCD) in order to separate
signal from background.
However, predicting or measuring the shape of the $W$+jets
background after $b$-tagging algorithms are applied, as required
for single-top discovery, is itself subject to significant uncertainty.
In this note, we point out a possible approach to reducing
one aspect of this problem.  

An analysis of the use of shape differences between signal and
background was performed in Reference \cite{Bowen:2004my}.  The 
use of asymmetries and
correlations involving the lepton from the top decay and the jet
associated with the $t$-channel production process were shown to
dramatically mitigate problems from the $t\bar t$ background.  The
reduction of the $W$+jets background was shown to be significant, but
still insufficient, unless systematic errors on the shape of $W$+jets
can be brought down to roughly the 20 percent level.  The challenges in doing
so were discussed in section IV of \cite{Bowen:2004my}.  The
various contributions to the sample of $W$+jets {\it with a single
$b$-tag} were compared, and it was shown that many different
subprocesses, with many different initial and final states, are of
comparable importance.  Unfortunately, each of these subprocesses has a
different shape.  Unless their relative normalizations
can be determined, it is impossible to know the shape of the total
$W$+jets single-tag background with low uncertainty.  Further, each of the
many contributions has its own independent uncertainties, stemming
from parton distribution functions (PDF's), loop corrections, and issues involving tagging and
mistagging of heavy flavor, among others.  It seems difficult to
imagine that all of these subprocesses can separately be measured in
data.  Therefore, it is important to reduce the unknowns in this
context using a combination of data, theory, and simulation.

Among the lessons of section IV of \cite{Bowen:2004my} was that roughly a third
of the events entering the $W$+jets single-tag sample do so through
the tagging of heavy flavor quarks emerging within the parton shower
of a short-distance gluon.  Consequently, a significant portion of the
normalization uncertainty in certain subsamples is due to incomplete
knowledge concerning the fragmentation of short-distance gluons to
heavy quark pairs, which leads to uncertainties in how often
parton-level processes such as $u\bar d\to W g g$ will receive a
single $b$ tag.  (This problem extends well beyond single-top-quark
production, of course; any similar process, such as
$t\bar{t}h$ or $Wh$, will
have background from gluon radiation and subsequent
splitting to heavy quark pairs.)  While Monte Carlo programs are
relied upon to carry out this splitting in most studies, they have not
been sufficiently verified up to the present time.  Any neural net
method for single-top-quark production trained on Monte Carlo
simulations will suffer a substantial uncertainty from this source,
unless the Monte Carlo can be tuned more convincingly to data.

\paragraph{Summary: Proposal to Study Gluon Splitting
in
$W$+1j Events}

To reduce the systematic error from
gluon splitting to heavy flavor requires a combination of data and Monte Carlo.  It has
already been suggested \cite{Tung:2004md} that events with a single $W$, $Z$
or photon and a single hard jet are important tools for extracting
heavy-flavor PDF's.  We wish to
emphasize further that one should view these events as tools for a
study of gluon fragmentation to heavy flavor, and for reducing correlated
uncertainties involving
PDF's, fragmentation and heavy-flavor tagging.  In particular, with
integrated luminosities at the Tevatron exceeding 1 fb$^{-1}$, $W$
events with a single hard jet represent an ample, relatively
well-understood, gluon-rich and heavy-quark-poor resource.  The study
we present below suggests that the sample of $W$ plus one high-$p_T$
jet ($W$+1j) provides
an opportunity to study in some detail, via investigation of
(sometimes multiple) secondary vertices and embedded muons, the
fragmentation of gluons into heavy flavor, and the interplay of gluon
splitting with tagging algorithms.  Our results should be
considered preliminary;  much further study is required.

The only published intersection between theory and experiment for
gluon-splitting to heavy quark pairs has been at $e^+e^-$ colliders
through the process $e^+e^-\rightarrow Z \rightarrow q\bar{q}g$, where
the gluon radiated off of one of the quarks then fragments to a
$c\bar{c}$ or $b\bar{b}$ pair.  The kinematics of SLAC and LEP
restricted the energy of this gluon to be in the 20--40 GeV range.
Further, the production of the short-distance gluons in an $e^+e^-$
collider takes place in a color environment different than that of a
hadron-hadron collider.  Thus the predictions of the
gluon-fragmentation algorithms implemented in showering generators
such as \PY \cite{Sjostrand:2000wi} and \HW \cite{Marchesini:1991ch}
remain somewhat untested for Tevatron applications.  It is therefore
important to measure gluon splitting rates directly at the Tevatron,
ideally in multiple settings.

Naively, the $W$+1j sample provides such an opportunity, since at
leading order (LO) there are no short-distance $Wb$ final states,
except through negligibly small CKM mixing angles.  Some fraction of
the final states contain charm quarks, but almost all jets with multiple
secondary vertices in this sample will come from a gluon fragmenting
to either a $b\bar{b}$ or $c\bar{c}$ final state.  The numbers below
will show that even events with a single heavy-flavor tag will be
substantially, or even dominantly, from the parton-shower of a gluon.
Disentangling the various sources for heavy-flavor tags may be
possible in this sample using the differences in impact parameter
distributions for short-distance $c$ and $b$ quarks, as well the
relative $p_T$ of muons in the decays.

Let us be more specific: we define the $W$+1j sample to be all
events with one lepton, MET, one high-$p_T$, central jet, and no other 
high-$p_T$ jets at any rapidity.    
This {\it exclusive} W+1j cross-section can 
be calculated at NLO since it is the
difference between the inclusive W+1j and W+2j cross-sections both at NLO,
which have been evaluated \cite{Campbell:2002tg}.
We recommend using higher-$p_T$ jets as they are, in general, under better 
theoretical control and are reconstructed with greater efficiency by detectors.
Further, the rate at which
gluons split to $Q\bar{Q}$ pairs increases significantly with energy, so 
the fraction of jets containing $b\bar{b}$ and $c\bar{c}$ pairs
becomes larger.  What $p_T$ cut best balances statistics and 
systematics will have to be determined by a future study.

\paragraph{Simulation of $\ell \nu  j$ Events}

The proposal above requires NLO studies
for both $W$+1j and $W$+2j to 
normalize the $W$+1j exclusive event set.  For now, we 
use the K-factor for $W$+1j inclusive production
from \cite{Campbell:2002tg} to normalize our event set.  This overestimates
the number of events the Tevatron experiments will have to work with,
but probably by less than ten percent.
Our crude simulation of the $W$+1j sample suggests
there will be enough events at the Tevatron to measure the
gluon splitting rate even with a small reduction in rate when the normalization
is calculated more accurately.  

To provide an estimate of the number of $\ell \nu j$ events the Tevatron
experiments will have to work with, we have generated an unweighted
$\ell \nu j$ event set using the LO event generator Madgraph
\cite{Maltoni:2002qb} and CTEQ5L PDF's \cite{Lai:1999wy}.  Events are
generated with the factorization and renormalization scales set to
$M_W$, and a $K$-factor of 1.1 is taken from the $W$+1j inclusive 
NLO calculation in \cite{Campbell:2002tg}.  
After accounting for the branching-ratio for
$W\to\mu\nu, e\nu$, and the generic cuts given in Table \ref{cuts1},
the numbers of events with 1 fb$^{-1}$ of integrated luminosity are
given in Table \ref{cross1}.  The events are broken into various
sub-channels, differing by the underlying source of the jet $j$.  For
simplicity, the cuts are applied to the the short-distance partons,
not to the jets.  Triggering efficiencies are not accounted for, but
are expected to be at least 80$\%$ for all channels.

Table \ref{cross1} shows the abundance of
short-distance gluons in $\ell \nu j$ events.  The numbers of
events in different channels suggests (though it does not prove) that
by looking at $\ell \nu j$ events with one and two secondary
vertices, as well as events with one and two high-$p_T$ embedded
muons, the processes $\ell \nu g\rightarrow \ell\nu b\bar{b}$
and $\ell \nu g\rightarrow \ell\nu c\bar{c}$ can be
disentangled both from each other and also from $\ell \nu c$ and
fake tags.  Indeed, given that $B$ meson decays frequently involve charm
mesons, there is the possibility of some jets with four real
secondary vertices.  

Unfortunately, the contribution from
short-distance light quark jets, and from gluons that shower
only to light quarks, can lead to reconstructed secondary vertices,
and constitutes a significant background to measuring gluon fragmentation
to $c\bar{c}$ and $b\bar{b}$.  However we expect this effect 
can be constrained in several ways, including the absence
of muons in such jets, and a different dependence on vertex
position, charge multiplicity, etc.

The other competing short-distance process, with final state $\ell \nu
c$, needs to be determined in order to allow a
measurement of gluon fragmentation, and
is interesting in its own right.  Though $\ell \nu c$ events will give
real secondary vertices and high-$p_T$ muons, they provide at most one
of each, and when both are present, the muon will intersect the
vertex. Moreover, the charge of the embedded muon will be opposite to
the charge of the isolated lepton, in contrast to events with gluon
fragmentation, where the muon from the heavy flavor decay may have
either charge.  The theoretical rate for $\ell \nu c$ production has a
large systematic error from uncertainties in the $s$ quark PDF,
because initial-state strange quarks contribute over 80$\%$ of the
rate at LO; the sample provides an opportunity to measure the $s$ PDF
and reduce such uncertainties \cite{Baur:1993zd,Tung:2004md}.  The NLO
calculation (with the heavy quark mass included) has been completed
\cite{Giele:1995kr}.  To our knowledge, the corresponding experimental
study has not yet been done.

\begin{table}[ptb]
\par
\begin{center}%
\begin{tabular}
[c]{|c|c|c|}\hline
Item & $p_{T}$ & $\left\vert \eta\right\vert $      \\\hline
 $\ell $  & $\geq15$ GeV & $\leq1.1$                          \\\hline
MET $\left(  \nu\right)  $ & $\geq15$ GeV & -   \\\hline
$j$ & $\geq40$ GeV & $\leq1.1$                     \\\hline
\end{tabular}
\end{center}
\caption{Detector cuts applied to partons in our study.}%
\label{cuts1}%
\end{table}

\begin{table}[ptb]
\par
\begin{center}%
\begin{tabular}
[c]{|c|c|}\hline
Channel & Events After Cuts   \\\hline
 $\ell \nu g$  &  24,000            \\\hline
 $\ell \nu q$  &  22,000            \\\hline
 $\ell \nu c$  &   2,200              \\\hline
\end{tabular}
\end{center}
\caption{Numbers of events with 1 fb$^{-1}$ for the subsets of
$\ell \nu  j$ with the cuts from table \ref{cuts1}.  Here
$\ell $=($e^\pm$,$\mu^\pm$), $c$ is both $c$ and $\bar{c}$, and $q$ sums
over all light quark and antiquark flavors.  There is no $\ell \nu b$
channel at LO, except through negligibly small CKM matrix elements such as
$|V_{cb}|$.}%
\label{cross1}%
\end{table}

\paragraph{Showering of $\ell \nu g$ Events Using \PY}

To get a sense for the number of $\ell \nu g\rightarrow \ell\nu
Q\bar{Q}$ events the Tevatron experiments will have to work with, we
allow the $\ell \nu g$ events to undergo parton showering, using
\PY \cite{Sjostrand:2000wi}.  We take the factorization scale to be $M_W$,
and we turn off initial-state radiation in order to focus solely on
the evolution of the short-distance gluons; note gluons in initial
state radiation are at low $p_T$ and any heavy flavor quarks in their
parton showers are rarely tagged.  The specific numbers have large
uncertainties, perhaps of order 30 percent, but they are
only intended to be illustrative.

For 1 fb$^{-1}$ of data, and the generic cuts in Table \ref{cuts1}, there
are 24,000 $\ell \nu g$ events.  After showering, these short-distance gluons have fragmented
to 620 $b\bar{b}$ pair and 1300 $c\bar{c}$ pairs.  Because the $b$ and $c$ quarks from
gluon fragmentation have smaller $p_T$ than the original
short-distance gluon, it is not obvious how many of these heavy quarks
will lead to observable secondary vertices.  Indeed a detector
simulation would be necessary to estimate this rate.  As a crude
measure, we have estimated the number of tags per jet by 
modeling the tagging of each heavy quark parton inside a jet as
{\it independent} of 
any other nearby heavy quark.  While this completely ignores complications from
$b\rightarrow c$ decays, and overlapping secondary vertices, it
provides 
some measure of the number of events the Tevatron
experiments may have to work with, and has the benefit of being straightforward
as an estimate.

Each $b$ parton from gluon fragmentation is tagged at a rate of
0.5 tanh($p_T/$36), where $p_T$ is the $b$ quark $p_T$.  Charm quarks 
are tagged at a rate of 0.15 tanh($p_T$/42), and jets originating
from light quarks and gluons without heavy quark pairs in them are mis-tagged
at a rate of 0.01 tanh($p_T$/80).

\begin{table}[ptb]
\par
\begin{center}%
\begin{tabular}
[c]{|c|c|c|}\hline
Channel                                         & 1 tag     & 2 tags                            \\\hline
$\ell \nu g\rightarrow b\bar{b}$  &   260     & 47                                  \\\hline
$\ell \nu g\rightarrow c\bar{c}$   &   150    &  3                                    \\\hline
$\ell \nu c$                                     &  280      &  -                                     \\\hline
$\ell \nu q(g)$                                &  ~300    &  -                                     \\\hline
\end{tabular}
\end{center}
\caption{Numbers of $\ell \nu j$ events with one or two tags.  The last column
is either for light quarks or gluons which do not fragment to heavy quark pairs.
We have not tried to estimate the number of double tags for the second two processes. }
\label{tags1}%
\end{table}

We have not attempted to investigate the use of the muons from $b$ and
$c$ decays, but we believe they should provide additional helpful
information with complementary systematic uncertainties.  Lepton-tagging
of heavy quark jets has already been shown to work in top physics studies
 in Run II \cite{Acosta:2005zd}.  The event rate for $W$+1j production
 is also sufficiently high to overcome the relatively small branching ratios
 of $b\rightarrow \mu X$ and $c\rightarrow \mu X$.

We also wish
to note that the excellent resolution of the Tevatron silicon trackers in
the $xy$ plane may allow measurement of the displacement in the $xy$
plane between two different secondary vertices, as well as their
distance from the primary vertex.  Thus, an event with two heavy
quarks could yield two impact parameters and either an angle or a
distance between the two displaced vertices.  Fitting to these
distributions, using a Monte Carlo to simulate the heavy flavor
decays, may well allow unknown parameters in the Monte Carlo
description of gluon fragmentation to be pinned down more precisely.

\paragraph{Final Remarks}

The proposed study of secondary tags in $W$+1j events should also usefully
supplement the ongoing $Zb$ studies at the Tevatron
\cite{Abazov:2004zd}.  Currently, the $b$ PDF is assumed to be zero at
the ``$b$ threshold'' (4.5 GeV) and is generated by letting QCD
evolution equations create it from the gluon PDF at $Q_F$ greater than 4.5 GeV.  The
uncertainties in the $b$ PDF are then almost completely tied to
uncertainties in the gluon distribution.  If one further relaxes the
assumption of the $b$ PDF being zero exactly at the $b$ threshold, the
uncertainties are even bigger. The $b$ PDF can be studied in $Z$+1j
events with secondary vertex tags
\cite{Campbell:2003dd,Abazov:2004zd}.  (The $Z$+1j study may also have
sensitivity to the $c$ PDF; though there are some experimental results
from DIS production of charm that place some constraints
\cite{Aktas:2005iw,Aktas:2004az,Breitweg:1999ad,Chekanov:2003rb}, the
uncertainties are still large.)  A background to this study is $Zg$,
where the short-distance gluon then splits to heavy flavor.
Cross-checking results between $Z$+1j and $W$+1j samples should help
reduce systematic and statistical uncertainties in our understanding
of these processes.



\newcommand{\gtlt}{\stackrel{>}{\scriptstyle{<}}}


\subsubsection*{Angular correlations in single-top and $Wjj$}
\textbf{Contributed by:~Z.~Sullivan}\\

Recent studies of single-top-quark production
\cite{Acosta:2004bs,Abazov:2005zz} have emphasized the importance of reducing
the $Wjj$ backgrounds.  These backgrounds are strongly sensitive to
achievable $b$-tagging efficiencies and jet-energy resolution.
New theoretical examinations \cite{Bowen:2005rs,Cao:2004ap} have shown that
only modest improvements in $Wjj$ rejection can be made by improving cuts in
pseudorapidity or $b$-jet assignment.  Hence, additional information appears
to be required.

It has been demonstrated that a spin correlation between the final-state
lepton and non-$b$ jet in single-top-quark production might lead to a useful
angular discriminate against the $Wjj$ backgrounds at both the Tevatron
\cite{Stelzer:1998ni} and the LHC \cite{O'Neil:2002ks}.  These studies relied
on leading order (LO) theoretical predictions.  This Workshop has motivated a
recent paper \cite{Sullivan:2005ar} (summarized here) that provides a
next-to-leading order (NLO) confirmation of the LO angular correlations for
both the single-top-quark signal and $Wjj$ backgrounds.  In addition,
sensitivity to top-quark rest-frame reconstruction is quantified, and
additional angular correlations are shown to be effective discriminants.

In order to understand angular correlations, it is essential to understand
the contribution from spin correlations versus kinematic correlations.
Spin correlations in single-top-quark production and decay are a direct result
of the electroweak nature of the processes.
The matrix elements for both $s$-channel and $t$-channel single-top-quark
production are proportional to
\begin{equation}
[p_d\cdot(p_t-m_t s_t)][p_e\cdot(p_t-m_t s_t)] \,,
\end{equation}
where $p_d$ and $p_e$ are the four-momenta of the down-type quark and charged
lepton in the event, $p_t$ and $m_t$ are the top-quark four-momentum and mass,
and $s_t$ is top-quark spin four-vector.  In the top-quark rest frame $p_t =
m_t (1,0,0,0)$, and $s_t = (0, \hat{s})$.

\begin{figure}[tbh]
\centering
\includegraphics[width=0.403333\textwidth]{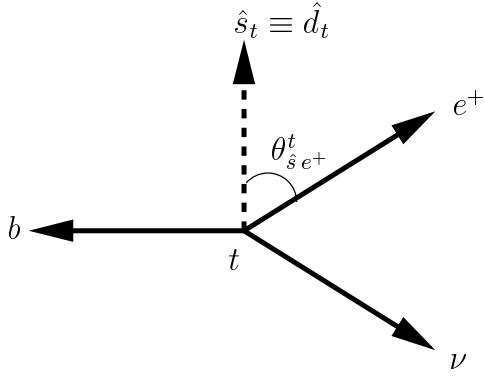}
\caption{Decay products of the top quark, and the angle
$\theta^t_{\hat{s}\,e^+}$ between the charged lepton $e^+$ and the spin
$\hat{s}_t$ of the top quark in the top-quark rest frame.  The spin is
projected in the direction of the down-type quark $d$ in the event.
\label{fig:tsblv}}
\end{figure}

In Ref.\ \cite{Mahlon:1996pn}, Mahlon and Parke showed that the direction of
the down-type quark provides a convenient axis to project the top-quark spin,
i.e., choose $\hat{s} = \hat{d}$ as in Fig.\ \ref{fig:tsblv}.  With this
choice, the matrix element reduces to $E_d E_e m_t^2 (1+\cos
\theta^t_{e^+d})$.  Since roughly 98\% of the events at the Fermilab Tevatron
are produced by pulling a $\bar d$ from the incoming antiproton, measuring
$\cos \theta^t_{e^+\bar p}$ provides the best possible measure of the spin
correlation for $s$-channel production.

\begin{figure}[tbh]
\centering
\includegraphics[width=0.325\textwidth]{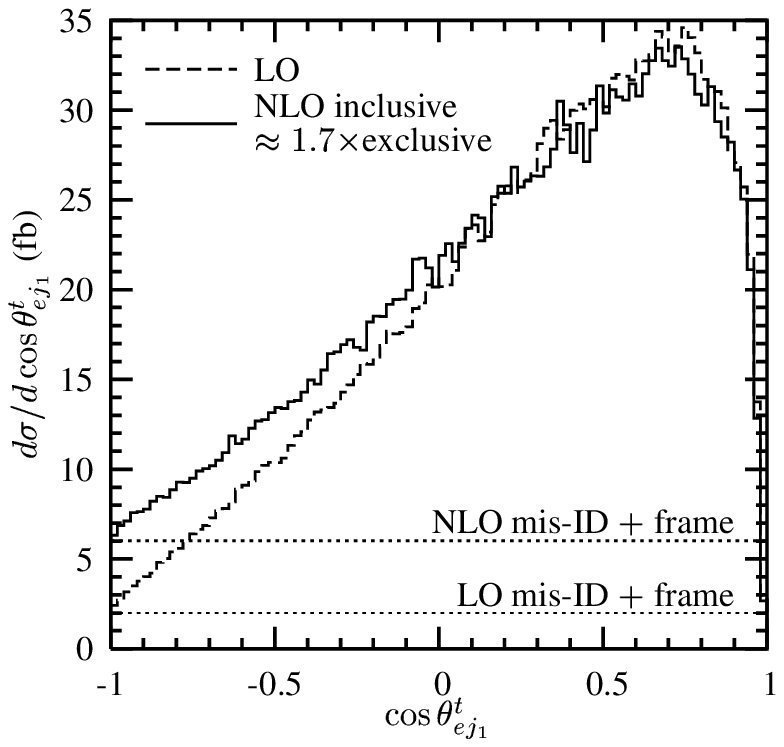}
\includegraphics[width=0.325\textwidth]{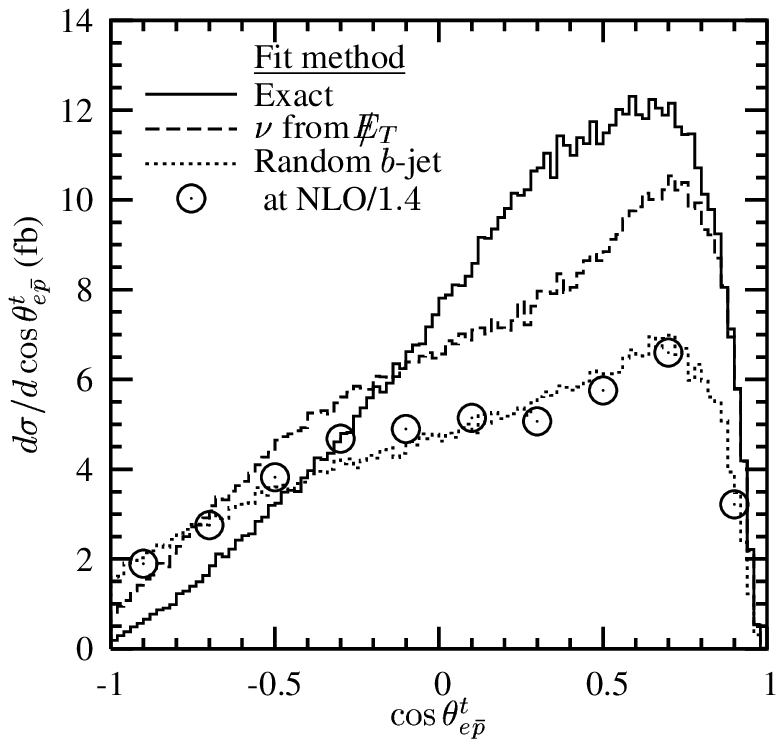}
\includegraphics[width=0.325\textwidth]{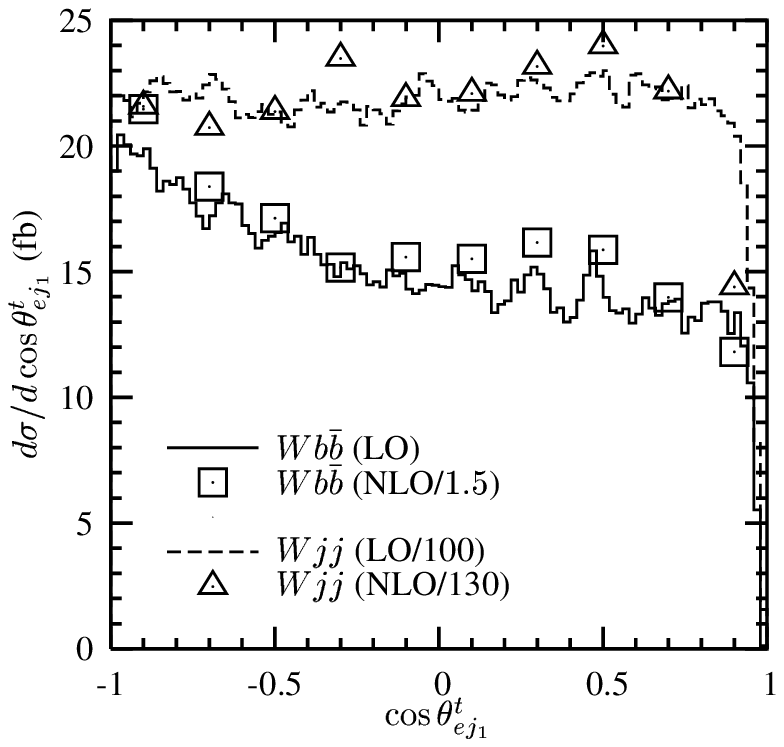}
\caption{Cosine of angle ($\cos \theta^t_{ej_1}$) between the charged lepton
and the highest-$E_T$ light-quark jet in the top-quark rest frame for (left)
$t$-channel single-top-quark production, and (right) $Wjj$ production at the
Tevatron.  (Center) cosine of angle between the lepton and antiproton for
$s$-channel single-top-quark production.
\label{fig:costex}}
\end{figure}

The only complication for $s$-channel production is reconstruction of the
top-quark rest frame.  Degeneracies in the measured neutrino momentum, and
assignment of the $b$-jet to top-quark decay, degrade top-quark
reconstruction.  These kinematic effects soften the measurable angular
correlations, as seen in the center plot of Fig.\ \ref{fig:costex}.  However,
the LO and NLO distributions agree exactly after top reconstruction up to an
NLO $K$-factor.  This has been confirmed in the fully correlated phase space,
so Monte Carlo simulations can reliably predict these angles.

Angular correlations in $t$-channel single-top-quark production are more
complicated.  The $d$ quark ends up in the highest-$E_t$ non-$b$-tagged jet
$j_1$ approximately $3/4$ of the time at the Tevatron.  The other $1/4$ of the
time a $\bar d$-quark is in the initial state, and a perfect correlation
exists with the incoming hadron (mostly the antiproton at the Tevatron).  This
adds a dilution factor, so that the the matrix element is proportional to $(1
+ \cos\theta^t_{d\, j_1} \cos\theta^t_{e^+j_1})$.  The dilution factor
$\cos\theta^t_{d\, j_1} = 1 - Q^2/(E^t_d E^t_{j_1})$ is typically around
$0.8$, because the $t$-channel exchange of the $W$ boson pushes $j_1$ forward
toward the beam line.  Hence, $\cos \theta^t_{e^+j_1}$ is a good quantity to
measure because of a combination of spin and kinematic correlations.

A complication in $t$-channel production is that additional initial-state
radiation can occasionally be misconstrued as the hard forward jet in the
event.  Since this additional radiation is uncorrelated with the final-state
lepton, it slightly flattens the distribution in Fig.\ \ref{fig:costex}.
However, it has been shown \cite{Sullivan:2004ie} that LO Monte Carlos can be
properly matched to NLO distributions.  Using matched distributions, the
softening of the correlation is seen to come solely from the misidentification
of which jet contained the down-type quark.  Spin-dependent matched
distributions reliably predict the fully correlated angular correlations.

The analytic form of the correlations for $s$- and $t$-channel production at
the LHC is the same as at the Tevatron.  However, there is one striking
difference in $t$-channel production.  Because the LHC is a $pp$ collider, $t$
production comes almost entirely from a valence $u$ quark in the initial
state, while $\bar t$ production comes mostly from valence a $d$ quark in the
initial state.  This means that the spin correlation for $t$ production is
almost 100\% with the light jet $j_1$, but for $\bar t$ production it is
almost 100\% with the beam axis.

An additional complication for $\bar t$ production is determining which proton
the $d$ quark came from.  The correlation suggests that a good choice for
reconstruction is the proton remnant closest to the charged lepton, i.e., for
$\eta_{e^+}\gtlt 0$ use $P_p=\sqrt{S}(1,0,0,\pm 1)/2$.  Despite the
fact that the best correlation is with the proton, the light jet tends to be
very forward, and hence the dilution factor for using the Mahlon-Parke basis
is close to 1.  Early studies of fully-reconstructed events using the ATLAS
detector simulation show that the single-top-quark and $Wjj$ angular
correlations are very similar to those at the Tevatron \cite{Barisonzi}.
Further, it appears that the Mahlon-Parke basis works equally well for both
$t$ and $\bar t$ production at the LHC.

\begin{figure}[tbh]
\centering
\includegraphics[width=0.3125\textwidth]{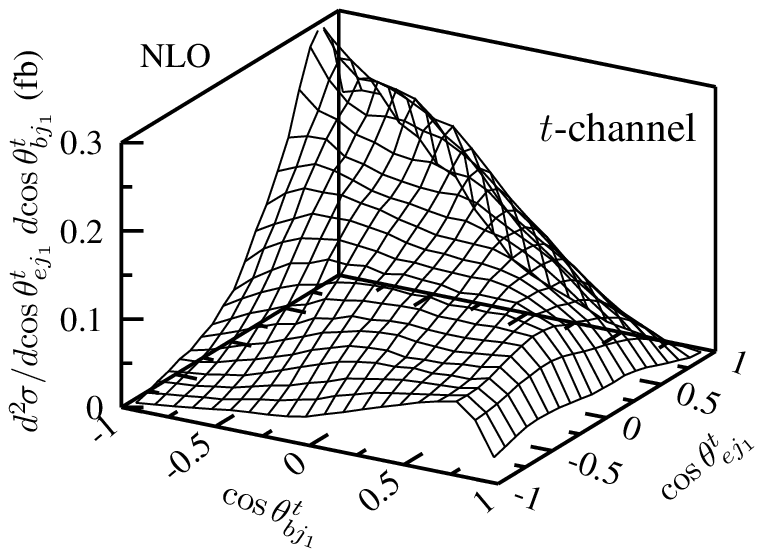}
\includegraphics[width=0.3125\textwidth]{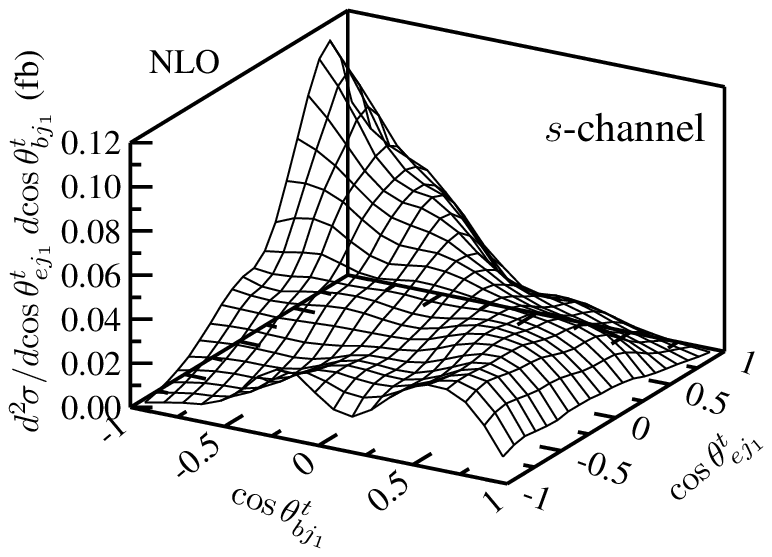}
\includegraphics[width=0.3125\textwidth]{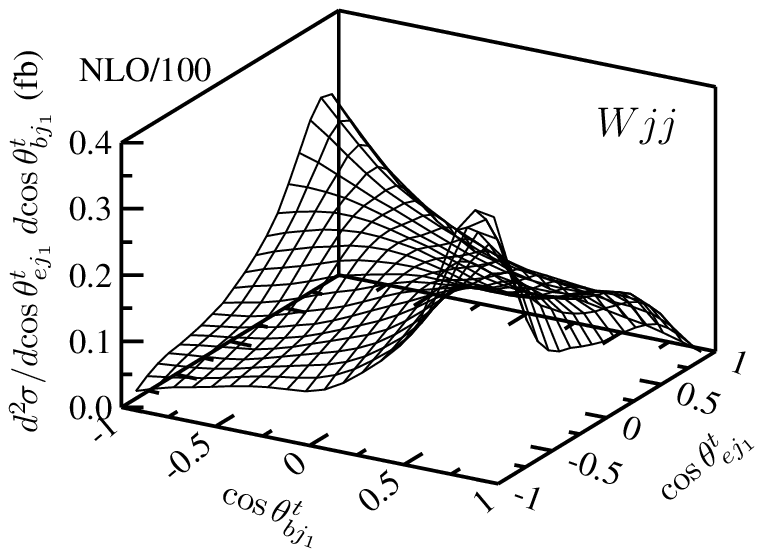}
\caption{Correlated angular distributions of the final-state particles in the
top-quark rest frame of NLO (left) $t$-channel and (center) $s$-channel
single-top-quark production, and (right) NLO $Wjj$ production.  This is a
two-dimensional projection between $\cos\theta^t_{ej_1}$, where $e$ is the
charged lepton and $j_1$ is tagged as the highest-$E_T$ light-quark jet, and
$\cos\theta^t_{bj_1}$, where $b$ is the $b$ jet from the top-quark decay.
\label{fig:tteobo}}
\end{figure}

The purpose of studying angular correlations is to find cuts to reduce the
$Wjj$ backgrounds.  As seen in Fig.\ \ref{fig:costex}, the two general classes
$Wjj$ backgrounds are found to be well-represented by a LO calculation plus an
NLO $K$-factor.  This has been confirmed in the fully correlated angular
distributions as well.  In Fig.\ \ref{fig:tteobo}, the correlation between
$\cos\theta^t_{ej_1}$ and $\cos\theta^t_{bj_1}$ demonstrates the power of
using angular information.  The flat distribution in $\cos\theta^t_{ej_1}$ for
$Wjj$ is seen to be an artifact of integrating over two broad peaks in the
correlated phase space.  A simple cut, such as $\cos\theta^t_{ej_1} >
\cos\theta^t_{bj_1}$, can remove roughly $1/2$ of the background with little
signal loss in either single-top channel.

The signal in Fig.\ \ref{fig:tteobo} peaks in one corner due to an additional
angular correlation not used in previous analyses.  In the real top-quark rest
frame the $b$ jet recoils against the $W$, and hence the charged lepton in the
event.  The strong spin correlation between the lepton and light jet $j_1$
leads to an almost degenerate phase space for single-top-quark production,
with the $b$ jet recoiling against the lepton-$j_1$ system.  The angle between
the $b$ and $j_1$ is further enlarged, because the initial production mode is
a two-body state with the top-quark recoiling against the light jet.  The $b$
picks up some of the top-quark's momentum, and the combination of kinematic
boost and spin correlation pushes the jets far apart.

The large angle between the $b$ and the charged lepton leads to the
possibility of using $\cos\theta^t_{ej_i}$ as a way to choose which jet came
from top decay.  This is useful for $s$-channel production where it is not
clear which $b$-jet to assign.  The following procedure picks the correct
assignment better than 80\% of the time, and effectively removes the
$b$-assignment uncertainty:
\begin{enumerate}
\item Construct two candidate top-quarks from the two highest-$E_T$ jets $j_i$.
\item Call the $b$-jet from top decay the one with the smallest
$\cos\theta^t_{ej_i}$ in its own candidate top-quark rest frame.
\end{enumerate}
This is effectively equivalent to making the cut $\cos\theta^t_{eb} <
\cos\theta^t_{ej_1}$ on the correlated angular distributions.  The $Wjj$
background is very close to flat in the plane of these two angles, and this
cut will reduce the background by another factor of two.  This sort of cut
emphasizes the importance of having complete and accurate angular
correlations, since it will cause the supposedly flat $Wjj$ distribution
$\cos\theta^t_{ej_1}$ in Fig.\ \ref{fig:costex} to look exactly like the
signal in that projection.  Fortunately, the fully correlated distributions
maintain the distinction.

Another useful distribution that arises from the large $\cos\theta^t_{bj_1}$
angle is the dijet mass.  The dijet mass for the signal is pushed to large
values, because the initial $E_{Tj}$ tends to be large, and the jets are
roughly back-to-back.  This in contrast with the $Wjj$ backgrounds, in which
the momentum is roughly split between the $W$ boson and the two jets, and
leads to a softer dijet mass.  Significant improvements in signal to
background can be made by adding a minimum dijet mass cut of order 100 GeV.

Use of the fully correlated angular distributions will require detailed
simulations of fully reconstructed events.  Early indications from LHC are
that the angular distributions are barely disturbed by detector effects
\cite{Barisonzi}.  This is not surprising from a quark-jet duality point of
view, but it is less clear what the ultimate sensitivity to top-quark
rest-frame reconstruction will be.  Many new physics analyses will require
complex cuts on phase space to separate signal from background.
Single-top-quark production at the Tevatron presents an important opportunity
to confirm that the NLO matched samples and full correlated angular
distributions agree with real data.


%
%
%
%
\subsubsection* {The ``Best Variables" Method and Implementation of Neural Networks in Physics Analysis }
\label{sec:singletoptheorybestvar}
\textbf{Contributed by:~E.E.~Boos, V.E.~Bunichev, L.V.~Dudko, A.A.Markina}\\

\paragraph{The Basic Idea}

In High Energy physics a discrimination between a signal and 
corresponding backgrounds is
especially important when the data statistics are limited or the 
signal to background ratio is small.
In this case it is important to optimize all steps of the analysis.
One of the main questions which arises in a physics analysis 
is which, and how many variables should be chosen
 in order to extract 
a signal from the backgrounds in an optimal way.
The general problem is rather complicated and finding a solution
depends on having a concrete process for making the choice, because
 usually it takes a lot
of time to compare results from different sets of variables.     

One observation which helps in making the best choice of the most sensitive
variables  is to study the structure of Feynman diagrams which contribute to the signal
and background processes. Based on such analysis we can distinguish three classes of
variables which are potentially most sensitive to the differences in signal and background
processes.

   The first class of variables is based on the analysis of singularities which usually
appear in physics processes.
Let us call those kinematic variables 
in which singularities occur as "singular variables".
What is important to stress here is that most of the rates for both the signal and
the backgrounds come from the integration over the phase space region 
close to these singularities. 
One can compare the lists of singular variables and the positions of the 
corresponding singularities in Feynman diagrams for the 
signal process and for the  backgrounds. 
If some of the singular variables are different or the positions of 
the singularities
are different for the same variable 
for the signal and for the backgrounds the
corresponding  distributions will differ most strongly. 
Therefore, if one uses all such 
singular variables in the analysis, then the largest part of the phase space
where the signal and backgrounds differ most will be taken
into account.
One might think that it is not a simple task 
to list all the singular variables when the phase space
is very complex, for instance, for 
reactions with many particles involved.
However, in general, all singular variables can be of
only two types, either $s$-channel: $$M_{f1,f2}^2 = (p_{f1} + p_{f2})^2,$$ 
where $p_{f1}$ and $p_{f2}$ are the four
momenta of the final particles  $f1$ and $f2$,
or $t$-channel: $$\hat{t}_{i,f} = (p_f-p_i)^2,$$ where $p_f$  
and  $p_i$ are the momenta of the final particle (or cluster)
and the initial parton. For the $\hat{t}_{i,f}$ all the needed variables
can be easily found in the
massless case: $\hat{t}_{i,f} = - \sqrt{\hat{s}} e^{Y} p_T^f e^{-|y_f|}$, 
where $\hat{s}$ 
is the total invariant mass of the
produced system, and {\it Y} is the rapidity of the total system (rapidity
of the center mass of the colliding partons), $p_T^f$ and $y_f$ 
are transverse momenta and pseudorapidity of the
final particle {\it f}.
The idea of using singular variables as the most discriminative ones is  
described in~\cite{Boos:1999dd} and the corresponding method was demonstrated in 
practice in~\cite{Dudko:aihenp},~\cite{Dudko:2000wx},~\cite{Boos:2003gv}.

   Singular variables correspond to the structure of the denominators
 of Feynman diagrams. Another type of interesting variables corresponds to
the numerators of Feynman diagrams and reflects the spin effects and 
the corresponding
 difference in angular distributions of the final particles. 
In order to discriminate between a signal and the backgrounds, 
one should choose in addition to singular variables mentioned above 
those angular variables whose distributions are different 
for the signal and backgrounds. The set of these singular and 
angular variables will be the most efficient set for a Neural Network 
(NN) analysis.

The third type of useful variables which we call "Threshold"
variables are related to the fact that various signal and background
processes may have very different thresholds. Therefore the distributions
over such kind of variables also could be very different keeping in mind
that effective parton luminosities depend strongly on 
$\hat{s}$. The variable $\hat{s}$ would be a very efficient variable of 
that kind. However, the problem is that in case of neutrinos in the final 
state one can not measure $\hat{s}$ and should use the effective
$\hat{s}$ which is reconstructed by solving t-,W-mass equations 
for the neutrino longitudinal momenta. That is why we propose to use
not only the effective variable $\hat{s}$ but different $H_T$ variables
as well.  

To apply the method it is important to use a proper Monte-Carlo model 
of signal and background events which includes all needed spin 
correlations between production and decays. 
We illustrate the method by considering single top quark production 
at hadron colliders, the Tevatron and the LHC.
The complete recipe how
to model the single top production processes with 
NLO precision is described in the
section~\ref{sec:singletoptheorygenerator}.
Comparing to a parton level analysis 
the detector smearing generically smooth out the distributions, and makes 
possible separation worse. However, 
kinematic properties of the processes basically remain the same after 
smearing, and no any new kinematic
differences between a signal and backgrounds appear after smearing which could 
help in signal and background separation.

\paragraph{Demonstration of the Method}

\begin{figure}[h]
\begin{center}
{
\unitlength=1.0 pt
\SetScale{1.0}
\SetWidth{0.7}      
\tiny    
{} \qquad\allowbreak
\begin{picture}(79,65)(0,0)
\Text(13.0,57.0)[r]{$d$}
\ArrowLine(14.0,57.0)(31.0,49.0) 
\Text(13.0,41.0)[r]{$\bar{u}$}
\ArrowLine(31.0,49.0)(14.0,41.0) 
\Text(39.0,53.0)[b]{$W^+$}
\DashArrowLine(48.0,49.0)(31.0,49.0){3.0} 
\Text(66.0,57.0)[l]{$b$}
\ArrowLine(48.0,49.0)(65.0,57.0) 
\Text(44.0,41.0)[r]{$t$}
\ArrowLine(48.0,33.0)(48.0,49.0) 
\Text(66.0,41.0)[l]{$\bar{b}$}
\ArrowLine(65.0,41.0)(48.0,33.0) 
\Text(66.0,25.0)[l]{$W^-$}
\DashArrowLine(65.0,25.0)(48.0,33.0){3.0} 
\Text(39,0)[b] {1.1}
\end{picture} \ 
}
{\def\chepscale{0.9} 
\unitlength=\chepscale pt
\SetWidth{0.7}      
\SetScale{\chepscale}
 \scriptsize   
\begin{picture}(100,60)(0,0)
\Text(13.9,39.8)[r]{$u$}
\ArrowLine(14.7,39.8)(38.7,39.8)
\Text(50.3,40.4)[b]{$g$}
\DashLine(38.7,39.8)(62.7,39.8){3.0}
\Text(87.5,49.3)[l]{$g$}
\DashLine(62.7,39.8)(86.7,49.3){3.0}
\Text(87.5,30.3)[l]{$g$}
\DashLine(62.7,39.8)(86.7,30.3){3.0}
\Text(35.6,30.3)[r]{$u$}
\ArrowLine(38.7,39.8)(38.7,20.8)
\Text(13.9,20.8)[r]{$\bar d$}
\ArrowLine(38.7,20.8)(14.7,20.8)
\DashLine(38.7,20.8)(62.7,20.8){3.0}
\Text(87.5,11.4)[l]{$W+$}
\DashArrowLine(62.7,20.8)(86.7,11.4){3.0}
\Text(50,0)[b] {2.1}
\end{picture} \
\begin{picture}(100,60)(0,0)
\Text(13.9,49.3)[r]{$u$}
\ArrowLine(14.7,49.3)(62.7,49.3)
\Text(87.5,49.3)[l]{$g$}
\DashLine(62.7,49.3)(86.7,49.3){3.0}
\Text(59.6,39.8)[r]{$u$}
\ArrowLine(62.7,49.3)(62.7,30.3)
\Text(87.5,30.3)[l]{$g$}
\DashLine(62.7,30.3)(86.7,30.3){3.0}
\Text(59.6,20.8)[r]{$u$}
\ArrowLine(62.7,30.3)(62.7,11.4)
\Text(13.9,11.4)[r]{$\bar d$}
\ArrowLine(62.7,11.4)(14.7,11.4)
\Text(87.5,11.4)[l]{$W+$}
\DashArrowLine(62.7,11.4)(86.7,11.4){3.0}
\Text(50,0)[b] {2.2}
\end{picture} \
\begin{picture}(70,60)(0,0)
\Text(13.5,49.3)[r]{$d$}
\ArrowLine(14.0,49.3)(28.0,39.8) 
\Text(13.5,30.3)[r]{$\bar d$}
\ArrowLine(28.0,39.8)(14.0,30.3) 
\Text(34.8,40.4)[b]{$g$}
\DashLine(28.0,39.8)(42.0,39.8){3.0} 
\Text(56.5,49.3)[l]{$d$}
\ArrowLine(42.0,39.8)(56.0,49.3) 
\Text(40.2,30.3)[r]{$d$}
\ArrowLine(42.0,20.8)(42.0,39.8) 
\Text(56.5,30.3)[l]{$W+$}
\DashArrowLine(42.0,20.8)(56.0,30.3){3.0} 
\Text(56.5,11.4)[l]{$\bar u$}
\ArrowLine(56.0,11.4)(42.0,20.8) 
\Text(35,0)[b] {2.3}
\end{picture} \ 

}
\end{center}
\caption{\small Typical Feynman diagrams for the $Wjj$ processes.}
\label{fg:feyn_wjjd}
\end{figure}
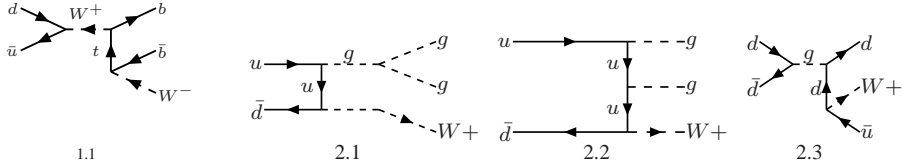

   Implementation of the above method in real analysis can be found in the papers 
describing Single Top quark search in D0 
(Run I and Run II)~\cite{Abazov:2001ns},~\cite{Abbott:2000pa},~\cite{Abazov:2005zz},~\cite{Dudko:d03612},\cite{Dudko:d03856}
and CMS (to be published in CMS Physics Technical Design Report).
In this section, we demonstrate how the above method works in case of the mostly 
simple single top quark production process, the 
$s$-channel production ($p\bar p\to t\bar b + X $), 
and one of the main background processes ($p\bar p\to Wjj + X $) at the Tevatron.
Typical Feynman diagrams for these signal and background processes are shown in the 
Fig.~\ref{fg:feyn_wjjd}. As explained in the previous section, one should 
compare the singularities for the signal and background diagrams. 
The signal diagram Fig.~\ref{fg:feyn_wjjd} (1.1) has only one singularity, 
a pole at the mass of the top quark:
$$M_t^2=(p_{b} + p_{W})^2 \ \rightarrow \ m_t^2.$$ 
(The pole for the W-boson decay is the same for the signal and for the 
background, and therefore the corresponding variable is not a sensitive
variable here.)
There are two singularities in the first background diagram 
Fig.~\ref{fg:feyn_wjjd} (2.1):
$$
M_{g1,g2}^2=(p_{g1} + p_{g2})^2 \rightarrow 0, 
$$
$$
\hat t_{u,(g1g2)}=(p_{g1} + p_{g2} - p_{u})^2 \rightarrow 0,
$$
corresponding to underlying soft and collinear singularities when additional
partons become soft or coincident in direction.

In diagram Fig.~\ref{fg:feyn_wjjd} (2.2) there are three
 singularities, but
one ($\hat t_{u,(g1g2)}$) is the same as in the first diagram:
$$\hat t_{u,g1}=(p_{g1} - p_{u})^2  \rightarrow   0,$$
$$\hat t_{u,g2}=(p_{g2} - p_{u})^2  \rightarrow   0,$$
\centerline{\mbox{$\hat t_{u,(g1g2)}=(p_{g1} + p_{g2} - p_{u})^2 \ \rightarrow \
0.$}}\\

We construct a complete set of singular variables using relations 
from the previous section and compare physics analysis using such a set of 
variables with analysis based on more simple often used set variables.
For the comparison of different sets of variables we take the neural 
network (NN) technique as one of the most popular and efficient methods 
of signal and background separation.
The efficiency criteria for different sets is the standard training parameter 
"Training Error function":
\begin{equation}
 \chi^2 = \frac{1}{N_{test}}\sum_{i=1}^{N_{test}} (d_i-o_i)^2.
\vspace{-2mm}
\label{eq:chi}
\end{equation}
In the formula $N_{test}$ is the number of test patterns, 
 $d_i$ is the desired NN output
 (1 for the signal and 0 for background), and
 $o_i$ is the NN output.
The lowest training $\chi^2$ has led to best separation of signal and background 
by the constructed NN. Compare the $\chi^2$ for different sets of input variables
we can conclude which set of variables is more efficient.

     The processes under consideration have been calculated using 
CompHEP~\cite{Boos:2004kh}
at the parton level, then decayed and 
processed with \PY~\cite{Sjostrand:2003wg}
in order to include initial-state and final-state radiation, 
and to fragment the final state partons into jets. 
Detector smearing of the jet energies has been included in our 
model by means of the SHW~\cite{Dudko:shw} program. For the NN 
training we use JETNET package~\cite{Peterson:1993nk}.

 The first set of variables consists of the complete set of singular variables 
for the $W+jets$ and $s$-channel
 signal processes:\\ 
\centerline{${\rm Set 1:} \ \ M_{j1,j2}, \ M_{top}, \ \hat{s}, \ Y_{\rm tot}, \ p_{Tj1}, \
y_{j1},\ p_{Tj2}, \ y_{j2}, \ p_{Tj12}, \ Y_{j12} $}\\
where $Y_{\rm tot}$ is the total rapidity of the center of mass of the initial
partons reconstructed from the final state particles, using  
the reconstructed neutrino momentum via equation $M_W^2 = (p_{\nu} +
p_{lepton})^2$. Next one is a simpler set:\\
\centerline{${\rm Set 2:} \ \ p_{Tj1}, 
\ p_{Tj2}, \ H_{\rm all}, \ H_{T \rm all} $}\\
Here \(H_{\rm all}=\sum E_f \), and
\( H_{T \rm all}=\sum P_{Tf} \), where the sums are over all final-state
particles and jets.

\begin{figure}[h]
\begin{minipage}[b]{.46\linewidth}
\mbox{\epsfysize=8cm\epsffile{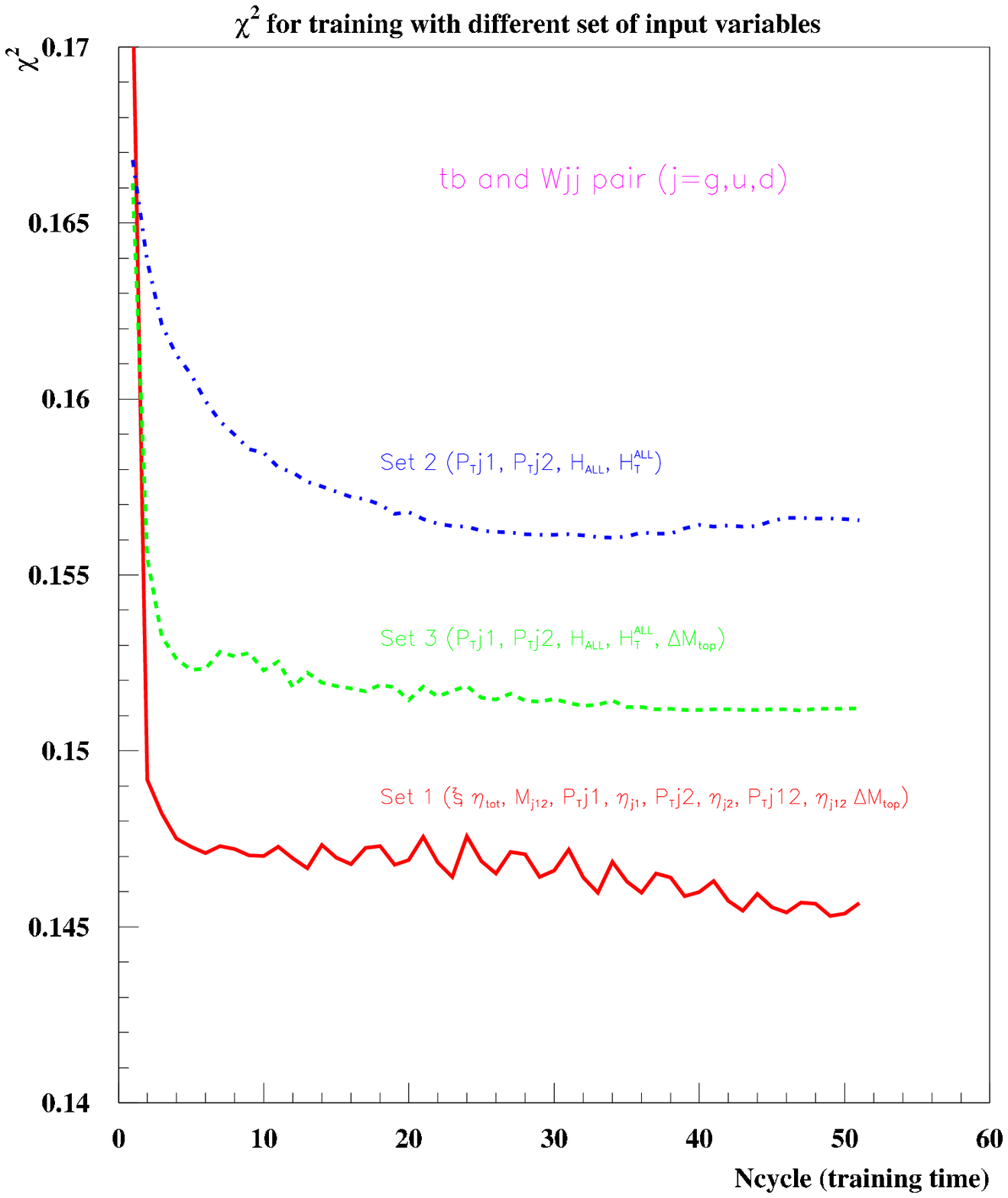}}
\caption{\small  Improvement of NN training for different sets of input variables.}
\label{fg:nnhi1}
\end{minipage}\hfill
\begin{minipage}[b]{.46\linewidth}
\mbox{\epsfysize=8cm\epsffile{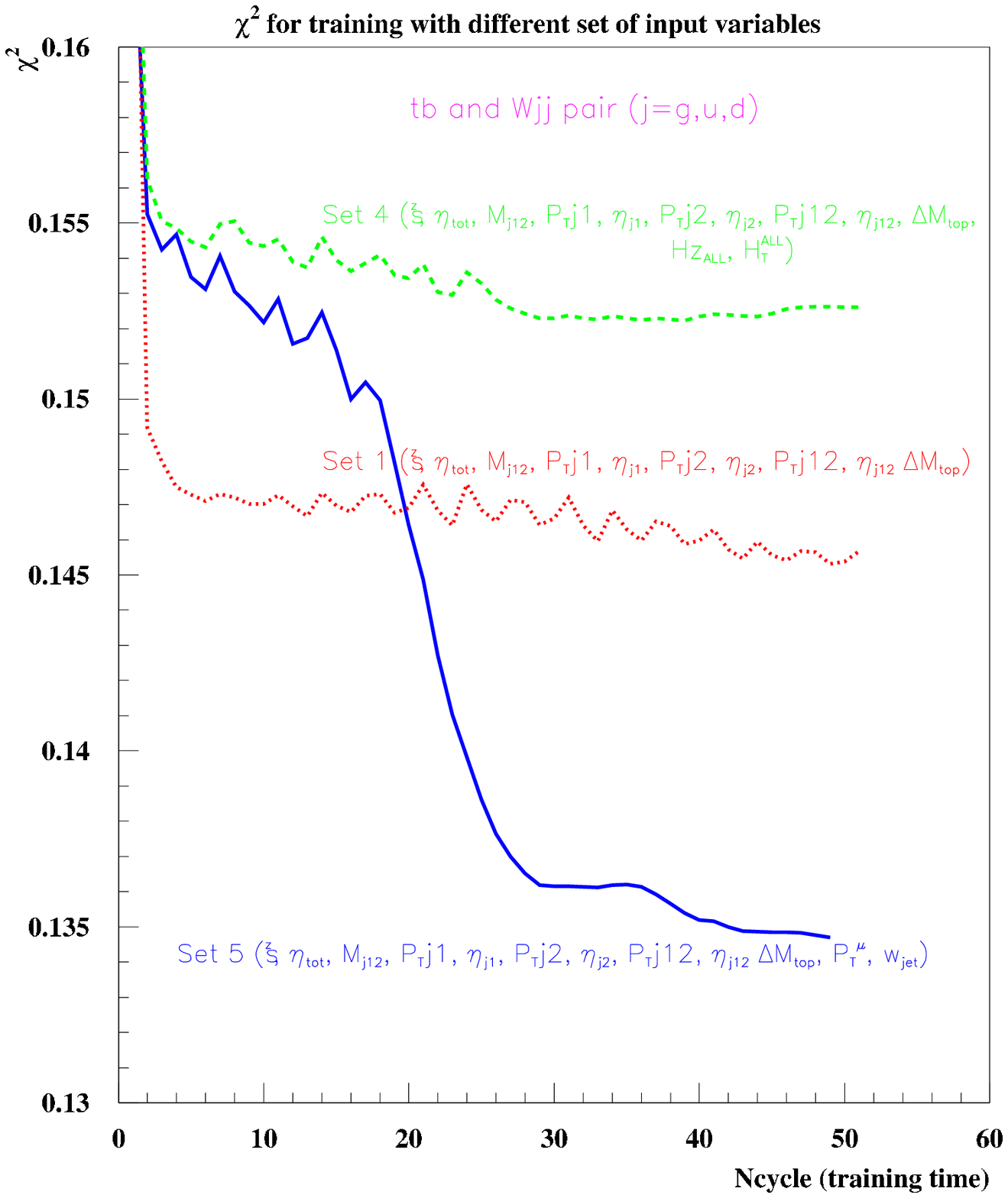}}
\caption{\small  Improvement of NN training for different sets of input variables.}
\label{fg:nnhi2}
\end{minipage}
\end{figure}

The third set includes one singular variable ($M_{top}$) in the
previous set:\\ 
\centerline{$ {\rm Set 3:} \ \ P_{Tj1}, 
\ P_{Tj2}, \ H_{\rm all}, \ H_{T \rm all}, \ M_{top} $}\\
The results for the $\chi^2$ are shown in Fig.~\ref{fg:nnhi1} (Ncycle is
the number of the Neural Net training cycles, it is proportional to the training
time).
The best network is defined as the one with lowest
\(\chi^2\), because the output from such a network is 
closer to the desired output.
From this plot, one can see that  the \(\chi^2\) for Set~1 
of singular variables is lower then for the other two  described above, and 
therefore the corresponding NN is better analysis tool.

\begin{figure}[h]
\begin{center}
\mbox{\epsfysize=10cm\epsffile{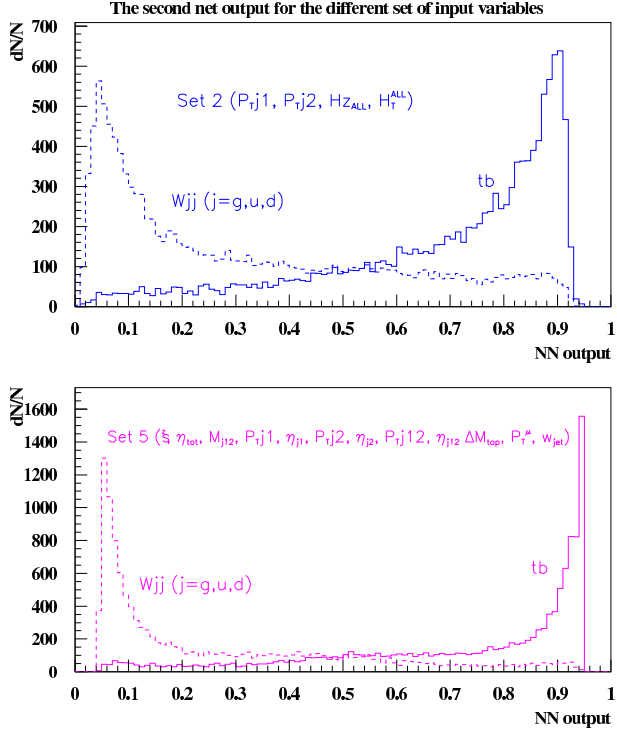}}
\end{center}
\vspace{-5mm}
\caption{\small Output of Neural Networks trained on the Set~2 and Set~5 of sensitive variables.}
\label{fg:nn25out}
\end{figure}

We tried to check the Set~1 for completeness by  
adding more kinematic
variables, to see if there would be any improvement.
We added the scalar sum of the final particles energy $H_{\rm all}$, and the 
scalar sum of their transverse energy $H_{T \rm all}$, and called this Set~4. 
We see that the \(\chi^2\) gets worser relative
to Set~1 of
 the original network.
This means that the additional kinematic variables do not add 
sufficient information to counter the increase in the number 
of degrees. 
But nevertheless, we can still search for other possible variables that contain
information that will be useful for separating
signal from background. In our case, where the signal is single-top quark 
production, we have 
a different probability for reconstructing a tagging muon in a jet from b decay
than for misidentifying of a $\mu$-tagged light jet
as a {\it b} jet.
In fact, the NN method can be regarded in some sense as a way of
 b-tagging.  
We introduce this information into  the NN through the
transverse momentum of the tagging muon, $p_T^{(tag \ \mu)}$, which is set to zero
for untagged events. In addition, we include two more useful variables, the
width ($w_{jet}$) of two jets with highest $E_T$. The final set of variables
is Set~1 together with the three additional variables:\\
\centerline{${\rm Set 5:} \ \ {\rm Set 1} \ + \ p_T^{tag \ \mu}, 
\ w_{jet1}, \ w_{jet2}$}\\
The $\chi^2$ for this final set is shown in Fig~\ref{fg:nnhi2}.
The comparison of NN outputs for the Set~2 and Set~5 is shown in the Fig.~\ref{fg:nn25out}.
It is the lowest on the plot, therefore we can choose this set of
variables for the analysis and get serious improvement in comparison with the simple Set~2. 

   It is possible to get further improvement, if we add the 
angular variables which we mentioned as
the second class of sensitive variables. The necessary information 
on this type of variables can be found in the 
papers~\cite{Mahlon:1999gz},~\cite{Boos:2002xw},~\cite{Sullivan:2005ar} (and references therein).

\paragraph{Implementation of Neural Network Technique in Physics Analysis}

Neural networks widely used in different fields of science and technology.
The main advantages of this method are the following: with this method it is possible to analyze large number of sensitive variables; it takes into account non-linear correlations in the analysis space; it is universal and can be applied in 
the same way for different tasks.
Based on the experience of single top search in D0 at the Run I and Run II analysis,
we summarize shortly, how to apply NN technique for extracting a signal from the backgrounds. 
At the D0 Run I single top search analysis this method of NN implementation provided in 2 times better physics result than the classical analysis method.

   We use the simple and most efficient in our case NN which we can teach (train) with a set of known examples 
-- feedforward NN with supervised training. In this case, the first step in the analysis is to prepare the correct model of the signal and background events. This step was described in the section~\ref{sec:singletoptheorygenerator}. At the next step,
we need to prepare the set of variables which mostly reflect the difference in signal and background properties, this step is described in the previous sections~(\ref{sec:singletoptheorybestvar}). For the NN training we have to use only the variables which were simulated properly in the model and exclude the variables which distributions are different when we compare the complete model (signal and background) and real DATA (if it is available).

   There are several background processes for the single top production. The kinematic properties for 
some of the backgrounds are significantly different. For example, QCD $W+jets$ production and $t\bar t$ production processes have different singularities, spin correlations and energy thresholds. In such a case it is more efficient to train different networks with different set of input variables for each background process. The same difference we can see for the signal processes. For the single top production we distinguish three signal processes ($t$-channel, $s$-channel and $tW$ production), each of them require a special approach and has unique properties which can help to extract it from the backgrounds. Therefore, the most effective separation of signal and background processes we can get by the set of NNs where each network is trained to recognize only one pair: one of the signal and one of the background processes. For the single top analysis we have three signal processes and
five main background processes, in this case the most effective separation we can get by the set of fifteen NNs.
It is not trivial to analyze fifteen outputs of NN and usually people can use some additional method to combine the network outputs and get the simple discriminator of the events. The reasonable method is to combine these NN outputs to additional NN (we call it Super NN) with five inputs (outputs of initial NN, each signal process consider separately) and one output. Such a network should be trained on the complete set of background processes which are mixed or weighted proportional by its contribution to the total background. As a result we will have three Super NN -- each one for every signal process.

   The further optimization of NN inputs is possible with the standard recommendations for the NN training. The first
 recommendation is to normalize input variables to the same region $[0,1]$ or $[-1,1]$. The second recommendation is
to use logarithmic scale for the variables with a long tail in distribution. 

   The next step in the NN analysis is to find the most effective architecture of the NN and set of the training parameters.
The criteria of $\chi^2$ (equation~\ref{eq:chi}) can help to find the optimal number of hidden nodes and set of training parameters. The optimal number of hidden nodes usually is within the region $[n,2n+1]$, where $n$ is the number of input variables. One hidden layer is usually the proper choice for most of the tasks in HEP. 

To avoid an overfitting problem one can use the standard solution  and split the samples for the training and testing parts, then
train the NNs on the training events and check the $\chi^2$ (equation~\ref{eq:chi}) for the testing events. Additional check for the trained networks can be performed by the comparison of NN output distributions for the simulated events and real DATA. If the distributions are not the same we can conclude that the NNs were overfitted or we do not model properly some input variables. After these checks the NNs are ready to calculate the expected number of signal and background events from the model and count the events which are passed the NN filters from the DATA flow.

  A detailed description of the Neural Network analysis of single top quark production at D0 can be found in the 
papers~\cite{Dudko:aihenp}-\cite{Dudko:d03856}.

\paragraph*{ACKNOWLEDGEMENTS}
 The work is partly supported by RFBR 04-02-16476, RFBR 04-02-17448, 
Universities of Russia UR.02.02.503, and Russian Ministry of 
Education and Science NS.1685.2003.2 grants.



\clearpage

%
%
%
%
\subsection {Tevatron Single Top Quark Searches}
\label{sec:singletoptev}

\subsubsection*{Physics goals}
\label{sec:st_tev_physicsgoals}

\textbf{Contributed by: Schwienhorst}\\

The main goal for the Tevatron experiments is to observe electroweak production of single top
quarks for the first time. The focus of current searches is observing any single top quark 
production, including both the $s$-channel and $t$-channel modes. 
Once the production of single top quarks has been observed, the emphasis shifts to measurements
of top quark properties and the $tWb$ coupling. The initial observation will serve as a 
measurement of the production cross section as well as the CKM matrix element $V_{tb}$, thus providing
a test of CKM matrix unitarity.
The initial samples of single top quark events can also be used to measure top quark properties such
as top quark spin correlations. With further increasing datasets, 
emphasis will be on separating $s$-channel from $t$-channel production in order
to probe details of the $tWb$ coupling.

The single top quark final state is also sensitive to models of new physics. 
Stringent limits on several different models can be 
set even before an actual observation of single top quark production.

\subsubsection*{Experimental signal signature}
\label{sec:st_tev_signal}
\textbf{Contributed by: Garcia-Bellido}\\

The two main production modes at the Tevatron are the $t$- and $s$-channel
processes, shown in Fig.\ref{fig:singletop} (a) and (b), respectively. 
The final state signature is thus characterized by a high energy isolated lepton
and missing transverse energy from the decay of the W from the top quark into
$\ell\nu$, and two or three jets. One of the jets originates from a $b$ quark
from the top quark decay and is usually central (low pseudorapidities) and energetic. 
In the $s$-channel, the other energetic jet 
is also from a $b$ quark, and shares similar kinematics with the $b$ from the
top. Thus $b$ quark identification, or $b$ tagging, in the $s$-channel is equally
likely between the $b$ from the top quark decay and the $b$ from the original
interaction.   
In the $t$-channel there usually is, apart from the $b$ jet from top quark
decay, a moderately energetic light flavor jet and a high pseudorapidity  
low energy $b$ quark jet from gluon splitting. 
This very forward or backward $b$ jet is a unique feature of this signal, but it
is rarely reconstructed and even more difficult to tag.   

At the Tevatron the final state is CP invariant, thus equal numbers of top and
anti-top quarks are produced.

\subsubsection*{Backgrounds}
\label{sec:st_tev_backgrounds}
\textbf{Contributed by: Garcia-Bellido}\\

The main processes that can mimic the final state topology arising from single
top quark production are: (i) $W$+jets events, where the $W$~boson decays
semileptonically and two or more associated jets are produced; 
(ii) $t\bar{t}$ events, where one or both top quarks decay leptonically;
and (iii) QCD or multijet events.

The W+jets background is by far the most problematic to get rid of at the
Tevatron. It consists of a leptonically decaying $W$~boson and at least two
associated quarks or gluons. $W$+jets events contain less energy in the event than
the single top quark signals since they do not contain a heavy object like
the top quark. But the cross section is very large in comparison to single top
quarks, and the flavor composition of the associated jets is sufficiently
complex, to make this background hard to model and even harder to get rid of as
one applies $b$ tagging techniques, since they tend to shift distributions to be
more signal-like and wash away any low energy features.  
This background has been estimated using simulated events, by {\sc alpgen} for
example, and is usually scaled to data to get the overall normalization right. 

Top pair production has a cross section around twice as big as single top quark
production. But the average energy in the event is larger, due to the presence
of two top quarks, and events tend to be more spherical and have more jet
multiplicity than single top quark events. 
The two top quarks produce two $W$~bosons and two $b$ jets, the latter
with very similar kinematics to the signal and therefore likely to be $b$ tagged
as well.  
The same final state signature as in single top quark processes is obtained if
only one of the $W$~bosons decay leptonically and the other hadronically, or if
both do, but only one lepton is reconstructed. 
This background can be properly simulated using {\sc alpgen} or {\sc pythia}.

The QCD background typically enters as misreconstructed events, where a jet
is wrongly identified as an electron, or a muon from a heavy flavor jet appears
isolated in the detector. Multijet events may also contain heavy flavor jets or
just light jets that are misidentified by the $b$ tagging algorithms. The
transverse energy of QCD events is much less than signal events, and the mass of
the system of the $b$-tagged jet, the lepton and the neutrino does not peak at
$m_t$, but the cross section is overwhelmingly large. 
This background is usually obtained directly from data,
and after some initial basic criteria can be reduced in size to the same level
as the signal. 

\subsubsection*{Description of the D\O\ search for single top quarks}
\label{sec:st_tev_d0analysis}
\textbf{Contributed by: Jain}\\

This section describes the search for single top quarks in the $s$-channel and 
$t$-channel modes, using the \dzero detector~\cite{D0detector} at the Tevatron. 
The data was recorded with a lepton+jets trigger, where the lepton is either an 
electron or a muon. The integrated luminosity was 226~pb$^{-1}$ for the 
electron channel and 229~pb$^{-1}$ for the muon channel. We perform a 
cut-based analysis using kinematic variables that discriminate between signal 
and background, and a multi-variate analysis using neural networks. We observe 
no significant deviation in data~\cite{Abazov:2005zz} from the  Standard Model 
prediction, and hence, set upper limits at 95$\%$ CL, on the single top production 
cross section, in the $s$-channel and $t$-channel modes, of 10.6 pb and 11.3 pb, 
respectively, in the cut-based analysis, and 6.4 pb and 5.0 pb, respectively, in the 
neural network analysis. 

\paragraph{Initial Event Selection and Yields}
\label{selection}

We apply a loose initial selection in order to maximize the acceptance for the 
single-top quark signal while rejecting the W+jets and misreconstructed events. 
In the electron channel, we require exactly one isolated electron
with the transverse momentum, $p_{T} > 15$~GeV, and the detector 
pseudorapidity, $|\eta_{\rm det}|< 1.1$. In the muon
channel, events are selected by requiring exactly one isolated muon
with $p_{T} > 15$~GeV and $|\eta_{\rm det}|<2.0$. For both channels,
events are also required to have $\MET>15$~GeV, and between two to four jets, 
with the jet $p_{T}>15$~GeV and $|\eta_{\rm det}|<3.4$. The leading jet is 
required to be more central ($|\eta_{\rm det}|<2.5$), and have $p_{T}>25$~GeV. 
Jets are defined using a cone algorithm with radius 
${\cal{R}} = 0.5$. In addition, misreconstructed events which are difficult to 
model, are rejected by requiring that the direction
of $\MET$ is not aligned or anti-aligned in azimuth ($\phi$) with the lepton or 
the jets. This selection has a negligible effect on the efficiency of signal events. 

The fraction of signal-like events is further enhanced through the
selection of $b$-quark jets that are identified by a secondary vertex tagging 
algorithm, that reconstructs displaced vertices from long-lived particles. 
In the $t$-channel search, we additionally require that one of the jets is not $b$ 
tagged, to account for the light flavor jet from the original interaction. 

For both $s$-channel and $t$-channel searches, we separate the data
into independent analysis channels based on the final-state 
lepton flavor (electron or muon) and the $b$-tag multiplicity 
(=1~tag or $\geq$2~tags) to take advantage of the different final state topologies. 
In each channel, we find that the expected yield for the single top quark signal is 
small compared to the overwhelming backgrounds.  We, therefore, use additional 
kinematic variables that allow us to discriminate between signal and background. 
The number of events for each signal, background, and data after the initial event 
selection are shown in Table~\ref{yields} for the  combined electron, muon, 
single-tagged, and double-tagged analysis sets. 
%
%
\begin{table}[htbp]
\begin{center}
\begin{tabular}{lr@{$\,\pm\!\!\!\!$}lr@{$\,\pm\!\!\!\!$}l} 
\hline \hline
Source           & \multicolumn{2}{c}{$s$-channel search} &
\multicolumn{2}{c}{$t$-channel search} \\ \hline
$tb$             &   5.5 & 1.2  &   4.7 & 1.0  \\ 
$tqb$            &   8.6 & 1.9  &   8.5 & 1.9  \\ 
$W$+jets         & 169.1 & 19.2 & 163.9 & 17.8 \\
$\ttbar$         &  78.3 & 17.6 &  75.9 & 17.0 \\
Multijet         &  31.4 & 3.3  &  31.3 & 3.2  \\ \hline
Total background & ~~~287.4 & 31.4 & ~~~275.8 & 31.5 \\
Observed events  & \multicolumn{2}{c}{283}  &  \multicolumn{2}{c}{271} \\
\hline \hline
\end{tabular}
\caption{Estimates for signal and background yields, and the number
of observed events in data after initial event selection for the combined lectron,
muon, single-tagged, and double-tagged analysis sets. The $W$+jets yields
include the diboson backgrounds.  The total background for the $s$-channel
($t$-channel) search includes the $tqb$ ($tb$) yield. The quoted yield
uncertainties include systematic uncertainties taking into account correlations
between the different analysis channels and samples.}
\label{yields}
\end{center}
\end{table}
\paragraph{Discriminating Variables}
\label{discrim_vars} 

\begin{figure}[!h]  
\vspace{0.2in}
\centerline{
\includegraphics[width=.4\textwidth]{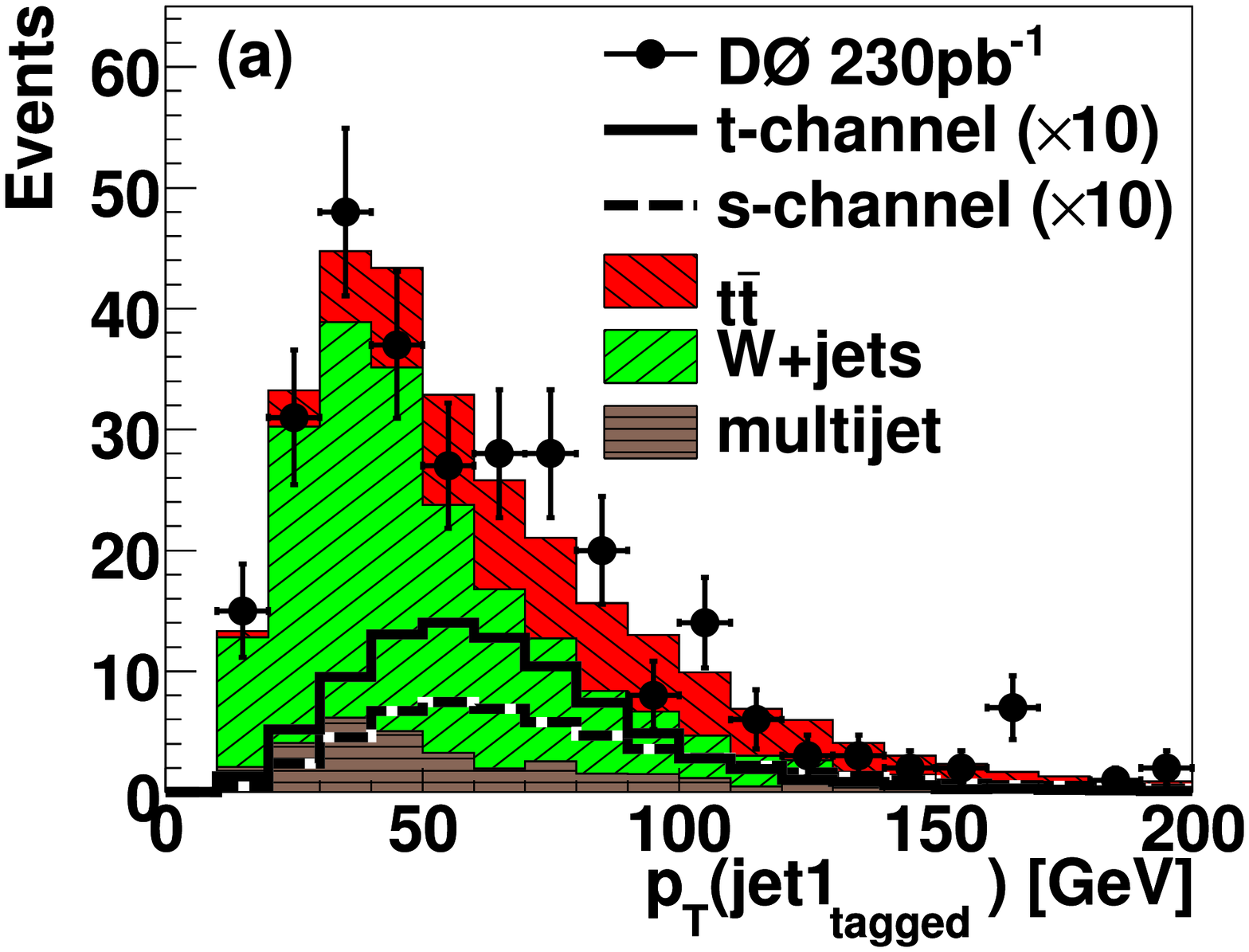}
\includegraphics[width=.4\textwidth]{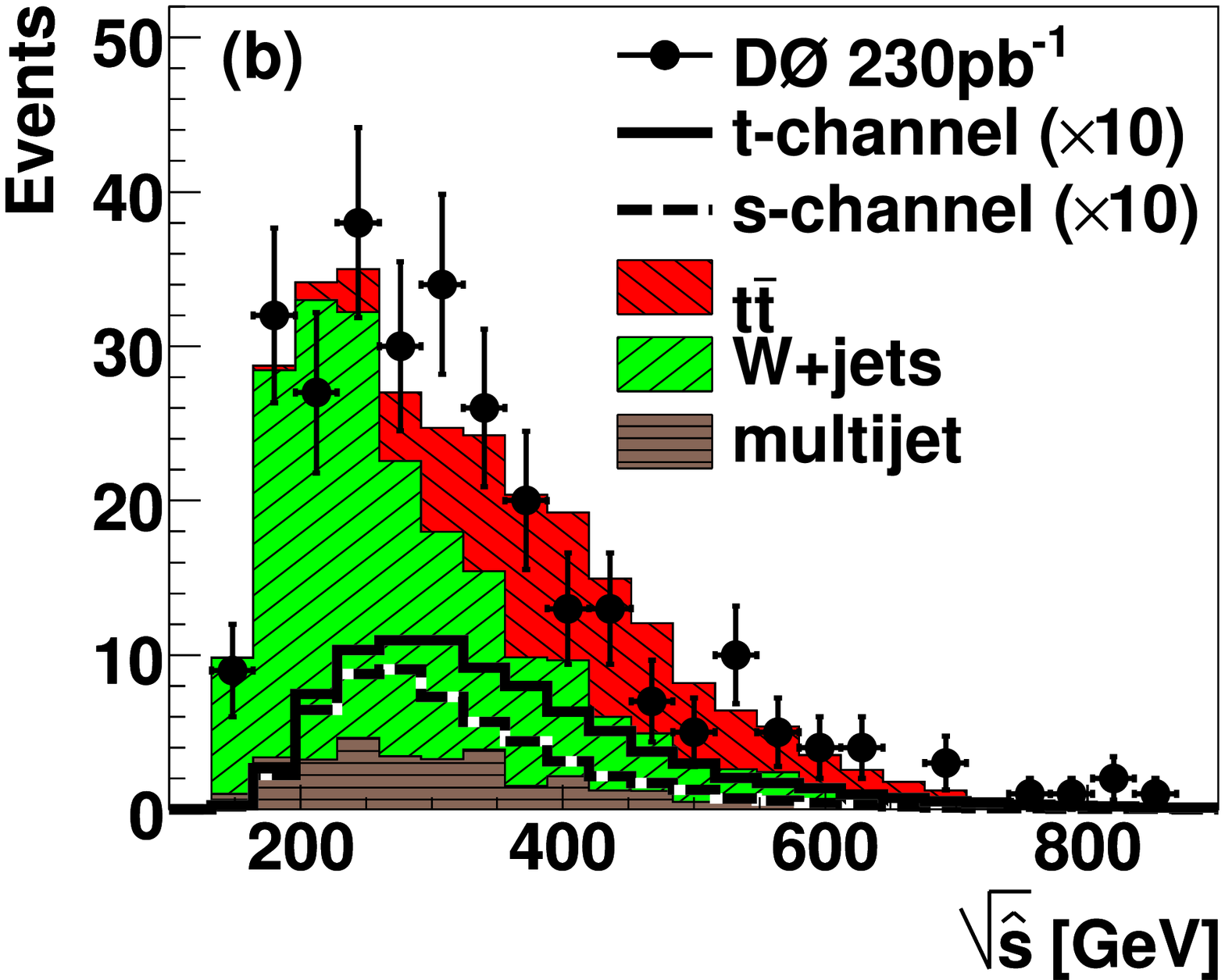}
}
\vspace{0.1in}
\centerline{
\includegraphics[width=.4\textwidth]{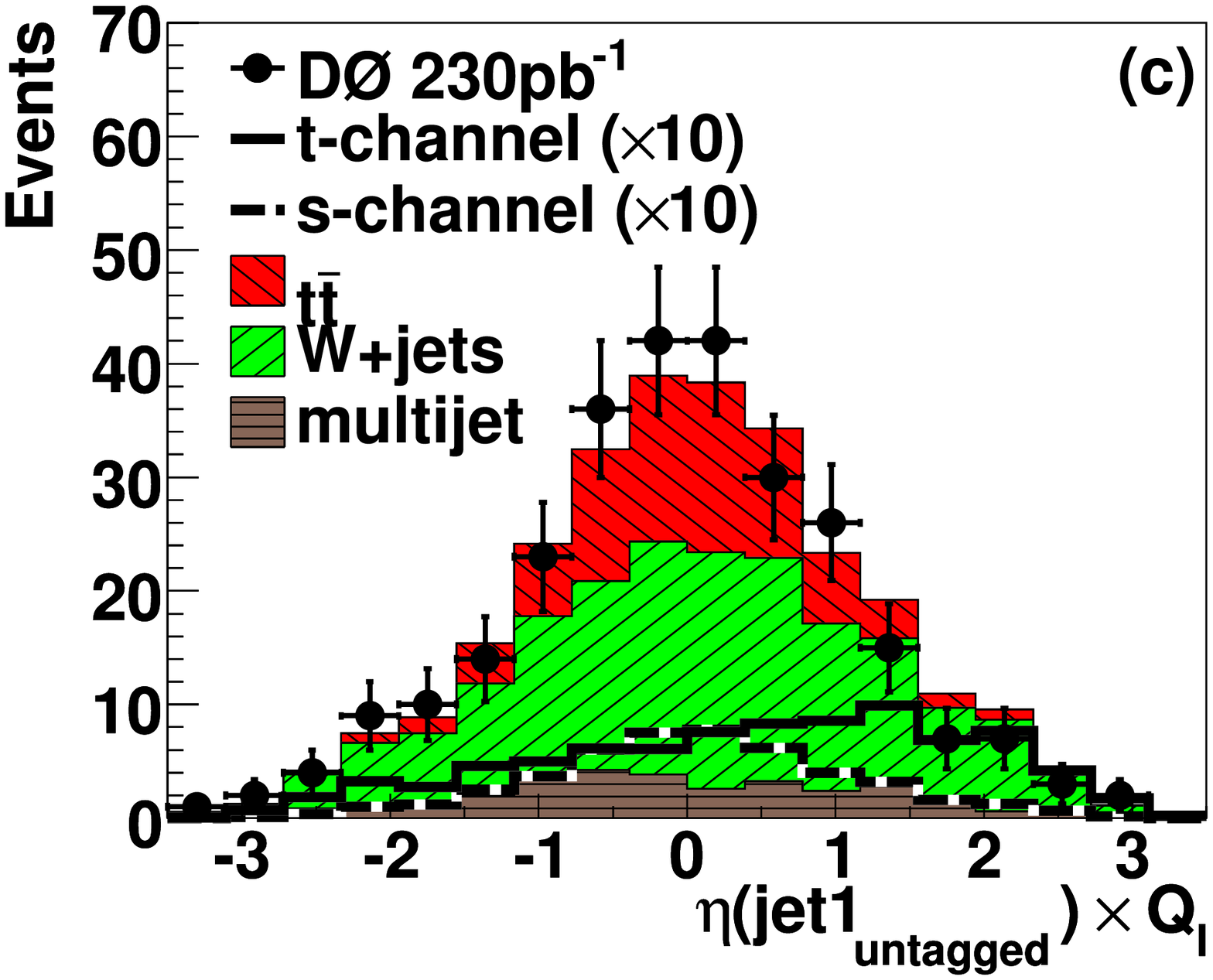}
\includegraphics[width=.4\textwidth]{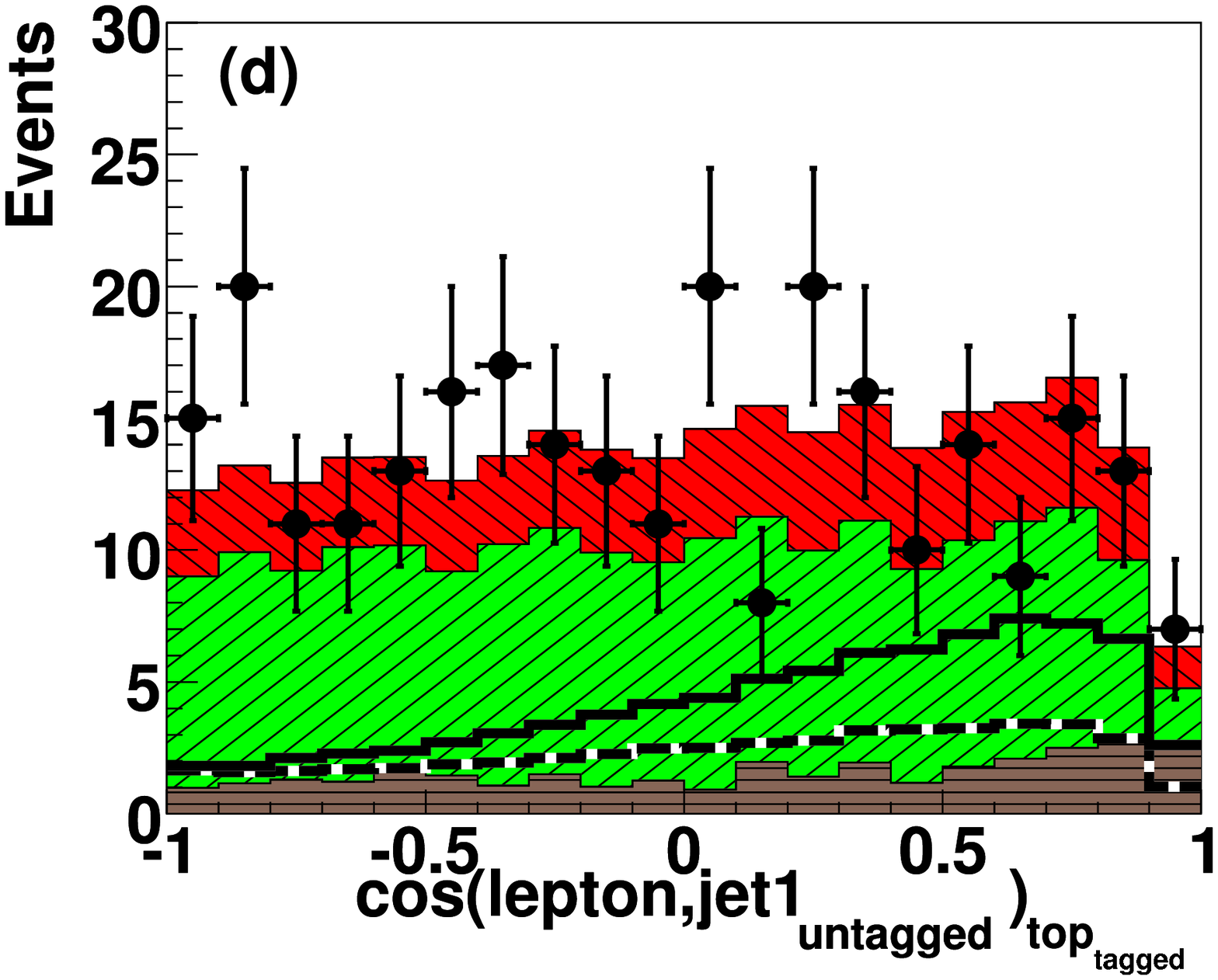}
}
\caption[input_vars]{Comparison of signal, background, and 
data for the combined electron, muon, single-tagged, and double-tagged analysis 
sets for representative discriminating variables. Shown are
(a) the transverse momentum of the leading tagged jet, 
(b) the invariant mass of all final state objects,
(c) the pseudorapidity of the leading untagged jet multiplied by the
lepton charge,
(d)  the top quark spin correlation in the optimal basis for the $t$-channel. 
Signals are multiplied by ten.
}
\label{inputvars}
\end{figure}

The variables that discriminate between the signal top quark signal and 
backgrounds were chosen based on an 
analysis of Feynman diagrams of these processes~\cite{boos-dudko}, 
and on a study of single top quark production at NLO
~\cite{Cao:2004ap,Cao:2005pq}. The variables fall into three categories: 
individual object kinematics, global event kinematics, and variables based on 
angular correlations. The list of variables is shown in 
Table~\ref{tab:variable-sets}. Figure~\ref{inputvars} shows distributions of a 
few representative variables comparing the single top quark signal to the sum of 
backgrounds, and the data. 
%
%
\begin{table*}[!h!tbp]
\begin{center}
\begin{footnotesize}
\begin{tabular}{lp{0.60\textwidth}cccc} 
\hline \hline
\multicolumn{6}{r}{Signal-Background Pairs} \\
& & \multicolumn{2}{c}{$tb$} &  \multicolumn{2}{c}{$tqb$} \\
\multicolumn{1}{c}{Variable}&\multicolumn{1}{c}{Description} & $W\bbbar$  & ${\ttbar}$ &  $W\bbbar$ & ${\ttbar}$ \\
\hline
\multicolumn{6}{c}{\bf{Individual object kinematics}} \\
$p_T({\rm jet1}_{\rm tagged})$     & 
Transverse momentum of the leading tagged jet     & $\surd$ & $\surd$ & $\surd$ & --- \\            
$p_{T}({\rm jet1}_{\rm untagged})$ & 
Transverse momentum of the leading untagged jet   & --- & --- & $\surd$ & $\surd$ \\            
$p_{T}({\rm jet2}_{\rm untagged})$ & 
Transverse momentum of the second untagged jet    & --- & --- & --- & $\surd$ \\            
$p_{T}({\rm jet1}_{\rm non-best})$ & 
Transverse momentum of the leading non-best jet   & $\surd$ & $\surd$ & --- & --- \\            
$p_{T}({\rm jet2}_{\rm non-best})$ & 
Transverse momentum of the second non-best jet    & $\surd$ & $\surd$ & --- & --- \\            
\multicolumn{6}{c}{\bf{Global event kinematics}} \\
$\sqrt{\hat{s}}$ &
Invariant mass of all final state objects 
                                       & $\surd$ & --- & $\surd$ & $\surd$ \\    
$p_T({\rm jet1},{\rm jet2})$        & 
Transverse momentum of the two leading jets& $\surd$ & --- & $\surd$ & --- \\            
$M_T({\rm jet1},{\rm jet2})$        & 
Transverse mass of the two leading jets     & $\surd$ & --- & --- & --- \\
$M({\rm alljets})$           & 
Invariant mass of all jets           & $\surd$ & $\surd$ & $\surd$ & $\surd$ \\ 
$H_T({\rm alljets})$         & 
Sum of the transverse energies of all jets      & --- & --- & $\surd$ & --- \\
$p_T({\rm alljets}-{\rm jet1}_{\rm tagged})$ & 
Transverse momentum of all jets excluding the leading tagged jet    & --- & $\surd$ & --- & $\surd$ \\
$M({\rm alljets}-{\rm jet1}_{\rm tagged})$ & 
Invariant mass of all jets excluding the leading tagged jet   & --- & --- & --- & $\surd$ \\  
$H({\rm alljets}-{\rm jet1}_{\rm tagged})$ & 
Sum of the energies of all jets excluding the leading tagged jet & --- & $\surd$ & --- & $\surd$ \\ 
$H_T({\rm alljets}-{\rm jet1}_{\rm tagged})$ & 
Sum of the transverse energies of all jets excluding the leading tagged jet         & --- & --- & --- & $\surd$ \\ 
$M(W,{\rm jet1}_{\rm tagged})$ & 
Invariant mass of the reconstructed top quark using the leading tagged jet & $\surd$ & $\surd$ & $\surd$ & $\surd$ \\
$M({\rm alljets - jet_{best}})$ & 
Invariant mass of all jets excluding the best jet          & --- & $\surd$ & --- & --- \\            
$H({\rm alljets}-{\rm jet}_{\rm best})$ & 
Sum of the energies of all jets excluding the best jet    & --- & $\surd$ & --- & --- \\      
$H_T({\rm alljets}-{\rm jet}_{\rm best})$ & 
Sum of the transverse energies of all jets excluding the best jet     & --- & $\surd$ & --- & --- \\
$M(W,{\rm jet}_{\rm best})$ & 
Invariant mass of the reconstructed top quark using the best jet  & $\surd$ & --- & --- & --- \\            
\multicolumn{6}{c}{\bf{Angular variables}} \\
$\eta({\rm jet1}_{\rm untagged}) \times Q_{\ell}$ & 
Pseudorapidity of the leading untagged jet  $\times$ lepton charge     & --- & --- & $\surd$ & $\surd$ \\
$\Delta \cal{R}({\rm jet1},{\rm jet2})$ &
Angular separation between the leading two jets     & $\surd$ & --- & $\surd$ & --- \\ 
$\cos({\rm \ell},{\rm jet1}_{\rm untagged})_{\rm top_{tagged}}$      &
Top quark spin correlation in the optimal basis for the $t$-channel~\cite{Mahlon:1995zn,Parke:1996pr,Mahlon:1996pn}, reconstructing the top quark with the leading tagged jet & --- & --- & $\surd$ & --- \\            
$\cos({\rm \ell},Q_{\ell}$$\times$$z)_{\rm top_{best}}$ &
Top quark spin correlation in the optimal basis for the $s$-channel~\cite{Mahlon:1995zn,Parke:1996pr,Mahlon:1996pn}, reconstructing the top quark with the best jet  & $\surd$ & --- & --- & --- \\ 
$\cos({\rm alljets},{\rm jet1}_{\rm tagged})_{\rm alljets}$          &            
Cosine of the angle between the leading tagged jet and the alljets system in the alljets rest frame     & --- & --- & $\surd$ & $\surd$ \\ 
$\cos({\rm alljets},{\rm jet}_{\rm non-best})_{\rm alljets}$  &
Cosine of the angle between the leading non-best jet and the alljets system in the alljets rest frame  & --- & $\surd$ & --- & --- \\ 
\hline \hline
\end{tabular}
\end{footnotesize}
\caption[tab:variable-sets]{List of discriminating variables.
A tick mark in the final four columns indicates in which signal-background pair
of the Neural Net analysis the variable is used.  A best-jet is defined as the jet in 
each event
for which the invariant mass of the system of reconstructed $W$~boson
and jet is closest to 175~GeV. Jets that have not been identified by the 
$b$~tagging algorithm are called ``untagged'' jets.
}
\label{tab:variable-sets}
\end{center}
\end{table*}
%
%
%
\paragraph{Cut-Based Analysis}
\label{cut_based_analysis} 

Here, we start from the list of discriminating variables, choose the best subsets, 
and find the optimal cuts~\cite{Abazov:2005zz} on each variable therein, by 
maximizing the signal to background ratio, and improving the expected cross 
section limits. (We define expected limits as the limit obtained if the observed 
counts were equal to the background prediction). The cuts scanned for the 
optimization are determined by the value of the respective variables in the signal 
Monte Carlo events, following the approach described in~\cite{rgs}. 

The event yields for each signal, background, and data, after the optimized cuts 
are shown in Table~\ref{tab:cut_based_yields} , for the  combined electron, 
muon, single-tagged, and double-tagged analysis sets. 
\begin{table}[!h!tbp]
\begin{center}
\begin{tabular}{lccc|} 
\hline \hline
Source           &$s$-channel search & $t$-channel search \\ \hline
$tb$             &   4.5 &  3.2   \\ 
$tqb$            &   5.5 &   7.0   \\ 
$W$+jets         & 102.9 & 72.6  \\
$\ttbar$         &  27.6 &  55.9 \\
Multijet         &  17.2 &  17.0   \\ \hline
Total background & 153.1  &148.7  \\
Observed events  & 152 & 148     \\
\hline \hline
\end{tabular}
\caption{Signal and background yields, and the numbers
of observed events in data, after selections in the cut-based analysis, for the 
combined electron, muon, single-tagged, and double-tagged analysis sets. 
The $W$+jets yields include the diboson backgrounds.  The total background 
for the $s$-channel ($t$-channel) search includes the $tqb$ ($tb$) yield.}
\label{tab:cut_based_yields}
\end{center}
\end{table}

\paragraph{Neural Network Analysis}
\label{nn-analysis} 

Here, we combine the discriminating variables and perform a multi-variate 
analysis. We use the {\sc mlpfit}~\cite{mlpfit} neural network package. We 
choose to create networks for each search ($s$-channel and $t$-channel mode) 
by training on the single top quark signal against the two dominant backgrounds: 
$W$+jets and {\ttbar}.  For $W$+jets, we train using
a $Wb\bar{b}$ Monte Carlo sample as this process best
represents all $W$+jets processes.  For {\ttbar},
we train on ${\ttbar}{\rar}\ell$+jets which is 
dominant over the dilepton background. 

\begin{figure}[!h]  
\vspace{0.2in}
\centerline{
\includegraphics[width=0.4\textwidth]
{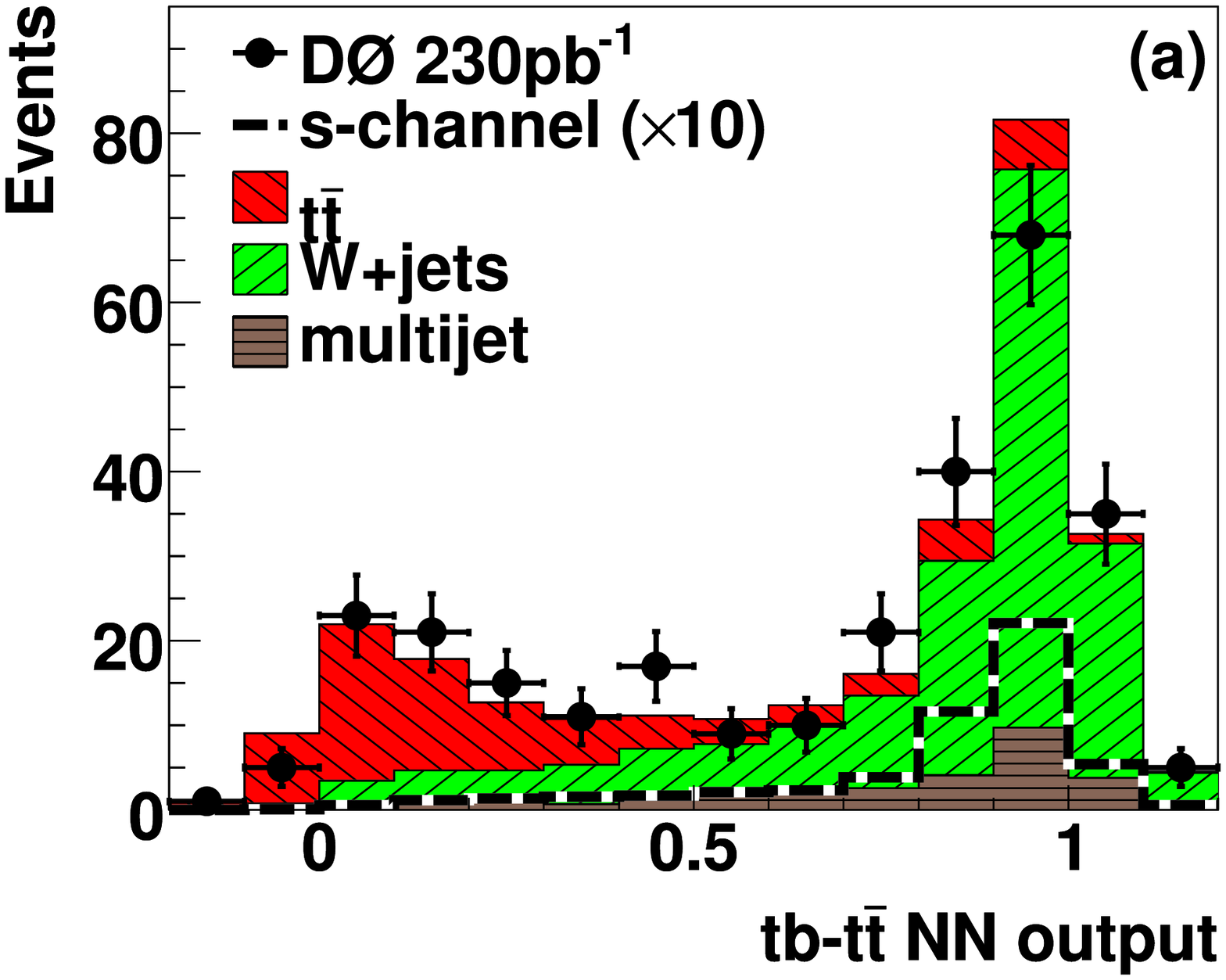}
\includegraphics[width=0.4\textwidth]
{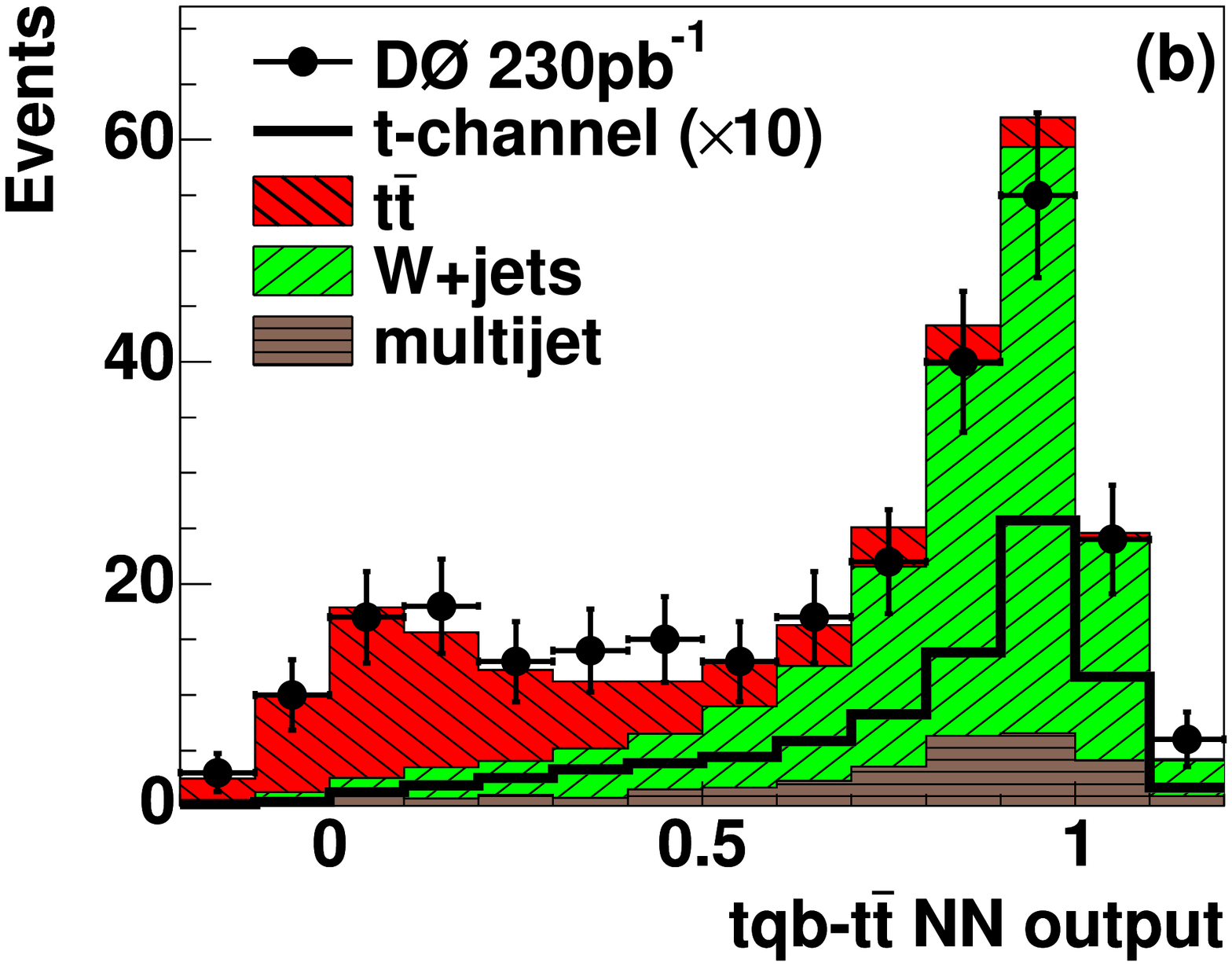}
}
\vspace{0.1in}
\centerline{
\includegraphics[width=0.4\textwidth]
{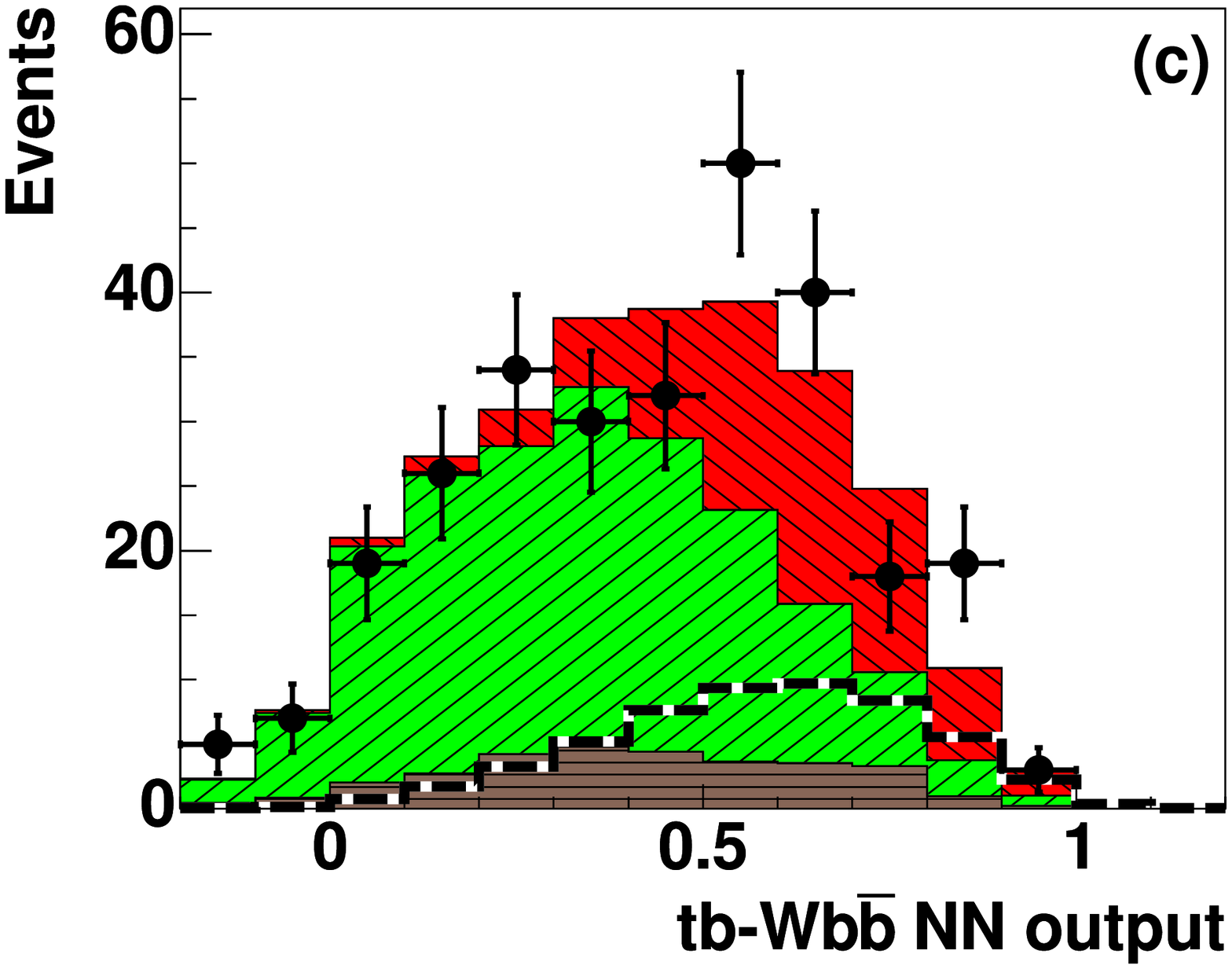}
\includegraphics[width=0.4\textwidth]
{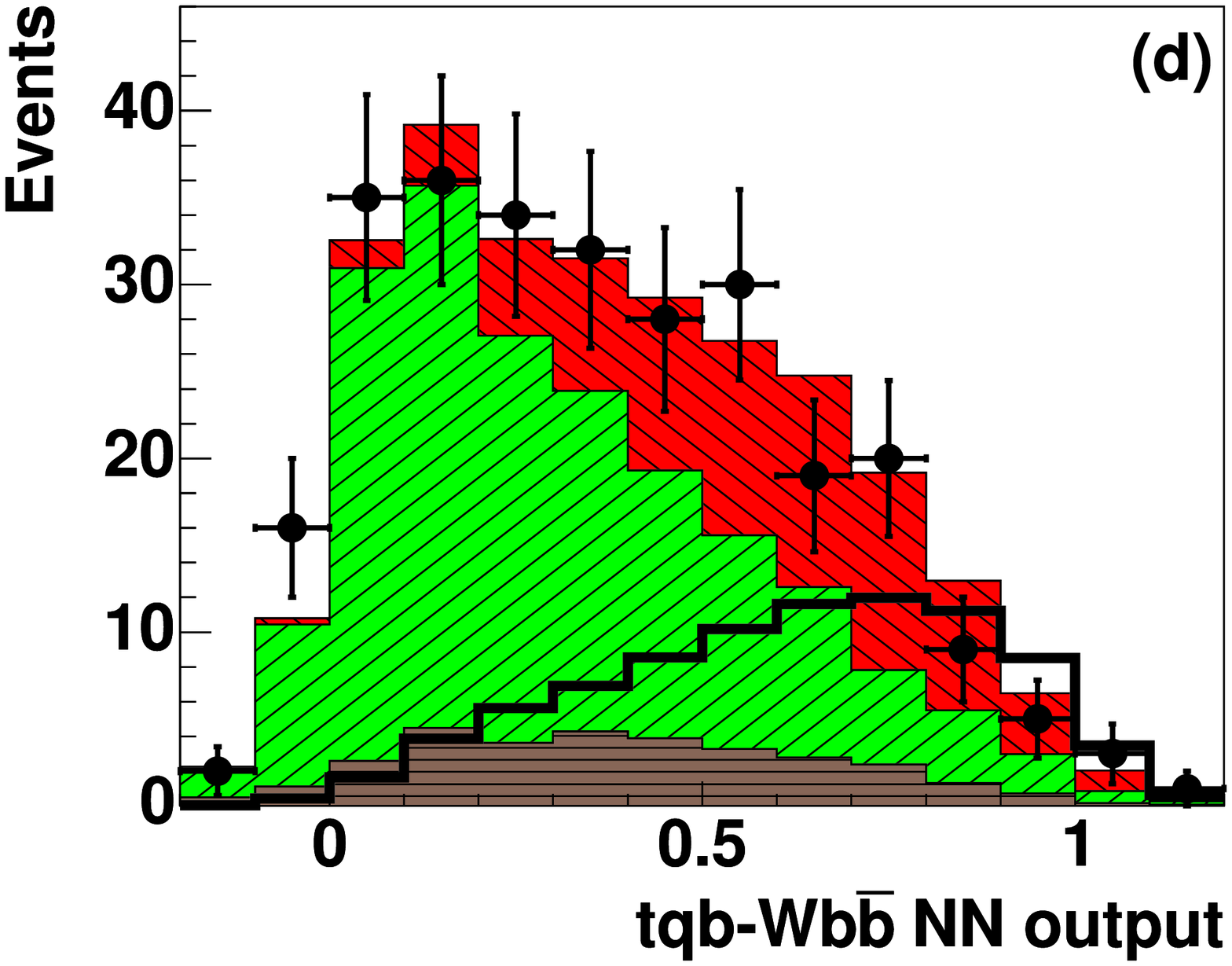}\\
}
\caption[nn-yield-compare]{Comparison of signal, background, and 
data for the neural network outputs, for the combined electron, muon, 
single-tagged, and double-tagged anlaysis sets.  Shown are the outputs for 
(a) the $tb$-$\ttbar$ filter, 
(b) the $tqb$-$\ttbar$ filter,
(c) the $tb$-$W\bbbar$ filter, and
(d) the $tqb$-$W\bbbar$ filter. Signals are multiplied by ten.}
\label{nn-yield-compare}
\end{figure}

Figure~\ref{nn-yield-compare}
shows the outputs of the neural networks for
the data and the expected backgrounds, as well as the signals for the combined 
electron, muon, single-tagged, and double-tagged analysis sets. We see that the 
$\ttbar$ networks separate signal and $\ttbar$ backgrounds
efficiently. The $Wb\bar{b}$ networks are less efficient for
the $W$+jets backgrounds because the event kinematics are similar
between signal and background.  

%
%

\paragraph{Systematic uncertainties}
\label{systematics} 

Systematic uncertainties are evaluated for the Monte Carlo signal
and background samples, separately for the electron and muon channels and for 
each 
$b$-tag multiplicity. 
The most important sources of systematic uncertainty are listed in
Table~\ref{tab:systematics}. 

%
%
\begin{table}[h!tbp]
\begin{center}
\begin{tabular}{lc}
\hline \hline
Source of               &  Uncertainty \\
systematic uncertainty  &  range (\%)\\
\hline
Signal and background acceptance \\
~~$b$-tag modeling      &  5 -- 20  \\ 
~~jet energy calibration&  1 -- 15  \\ 
~~trigger modeling      &  2 -- 7~  \\
~~jet fragmentation     &  5 -- 7~  \\
~~jet identification    &  1 -- 13  \\
~~lepton identification &   4  \\ 
Background normalization \\
~~theory cross sections &  2 -- 18  \\
~~$W$+jets flavor composition& 5 -- 16   \\
Luminosity              &   6.5 \\ 
\hline\hline
\end{tabular}
\caption{Range of systematic uncertainty values for the various Monte
Carlo signal and background samples in the different
analysis channels.}
\label{tab:systematics}
\end{center}
\end{table}

\paragraph{Cross section limits}
\label{sec:limits} 

We see from Table~\ref{tab:cut_based_yields} and 
Figure~\ref{nn-yield-compare} that the observed lepton+jets data agrees with 
the predicted Standard Model backgrounds within statistical uncertainty. We, 
therefore, set upper limits on the 
single top quark production 
cross section separately, in the $s$-channel and
$t$-channel searches. The limits are derived using Bayesian 
statistics~\cite{IainTM2000}. The likelihood function is proportional to the 
Poisson probability to obtain the number of observed counts. In the cut-based 
analysis, we use the total number of 
counts, and in the neural network analysis, we use the two-dimensional 
distributions of the $t\bar{t}$ versus $Wb\bar{b}$ network outputs, and 
construct a binned likelihood. 

The prior probability for the signal cross section is assumed to be flat. The prior 
for
the signal acceptance and background yields is a multivariate
Gaussian, with a vector of means given by the estimates of 
the yields, and covariance matrix computed from the associated 
uncertainties to take into account all correlations. The effect on the shape of 
neural network outputs from uncertainties like  $b$-tag modeling, jet energy 
calibration,
jet identification, and trigger modeling, is also considered in the binned likelihood. 

\begin{figure}[!h]
\centerline{
\includegraphics[width=0.4\textwidth]
{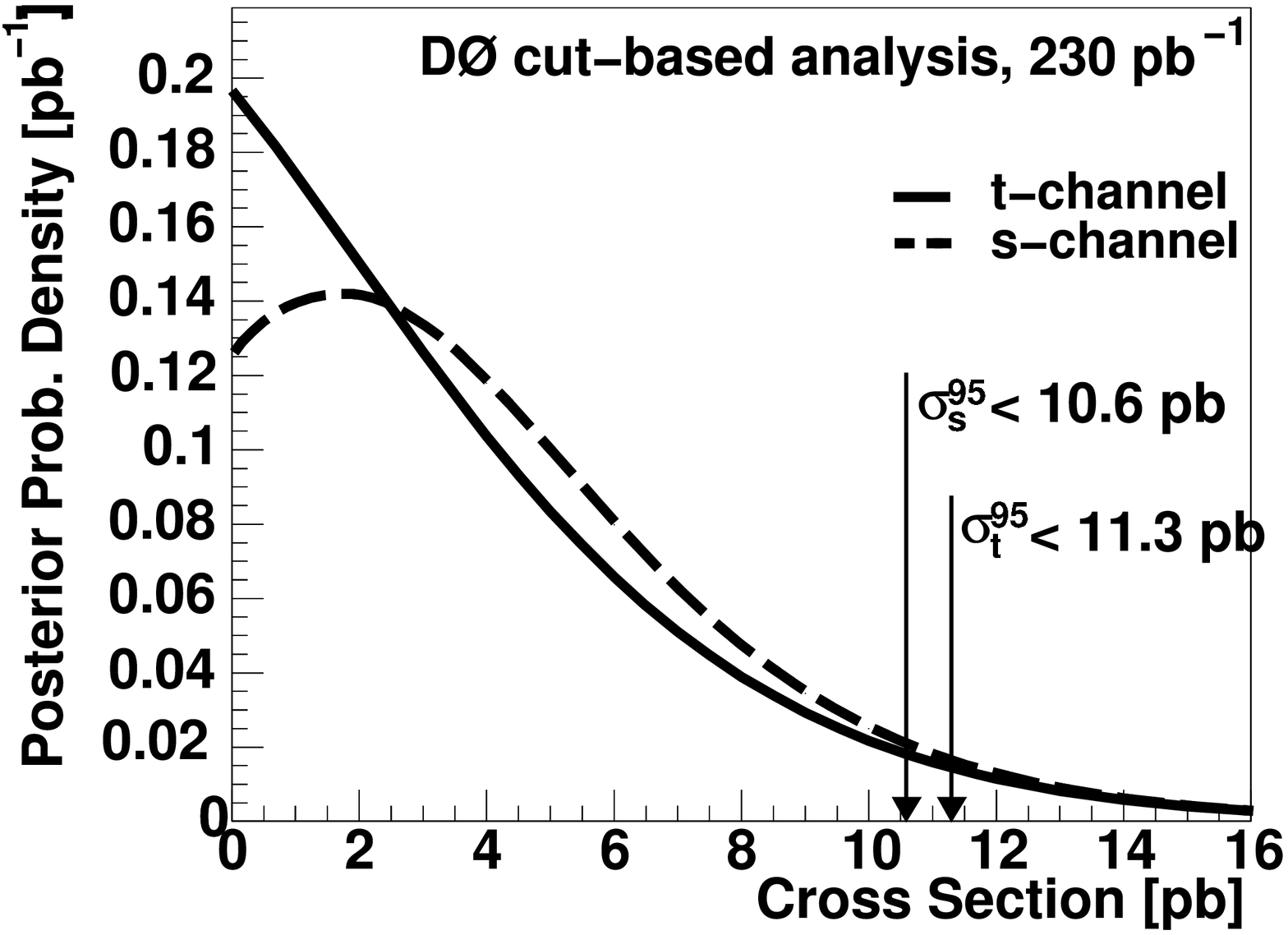}
\includegraphics[width=0.40\textwidth]
{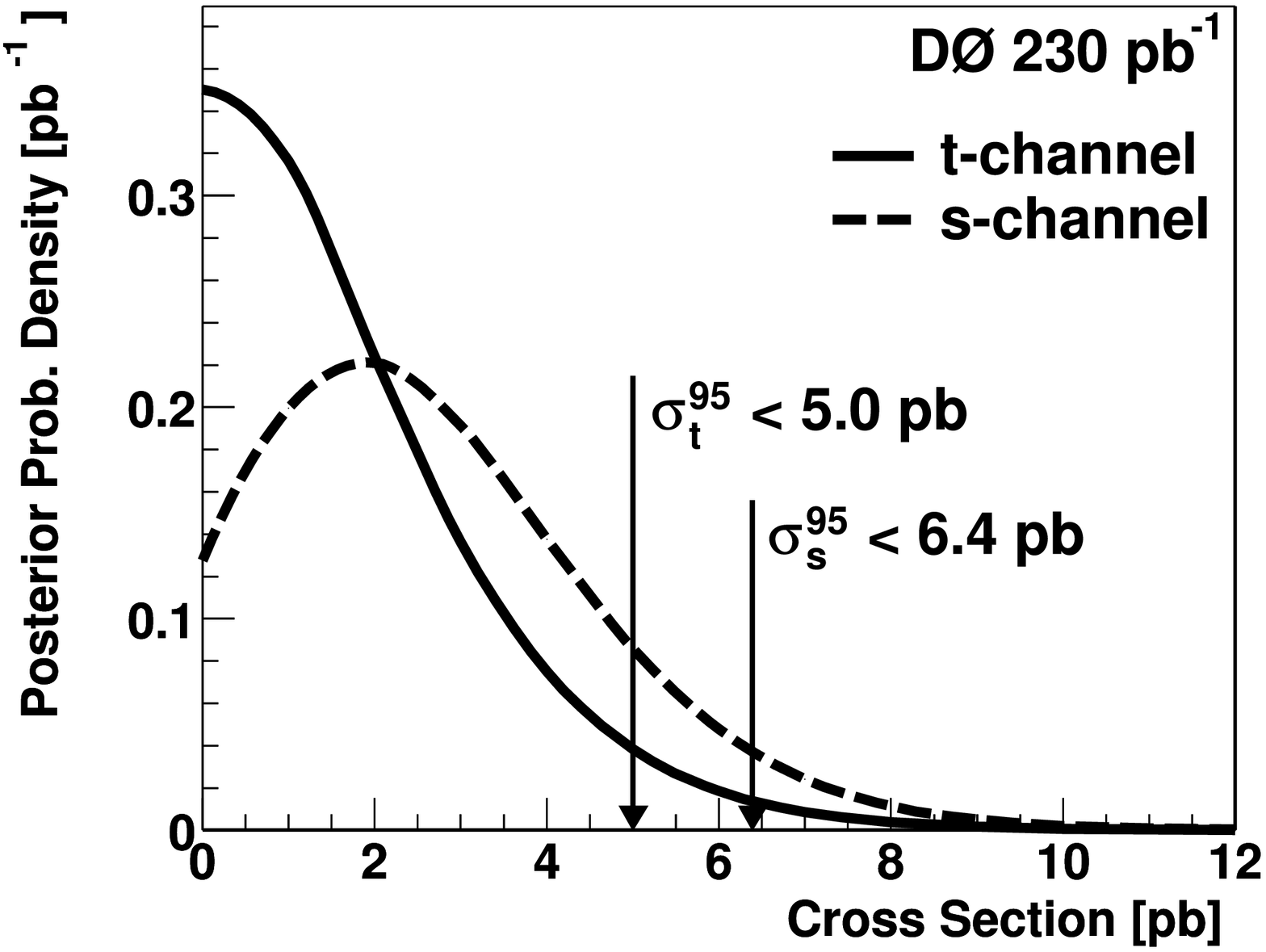}
}
\caption{The observed posterior probability density 
as a function of the single top quark cross section in the $s$-channel and the 
$t$-channel modes, using the combined electron, muon, single-tagged, and 
double-tagged analysis sets, for the cut-based (left) and neural network (right) 
analyses.}
\label{cb-posterior-1d}
\end{figure}

The expected and observed upper limits at 95\% confidence level, after the initial 
event selection, and from the cut-based and neural network analyses, are shown in 
Table~\ref{limits} for the combined electron, muon, single-tagged and 
double-tagged channels. We see that the limits improve upon applying cuts on the 
discriminating variables, but that tighter limits are obtained when the variables 
are combined using neural networks. The observed posterior probability densities 
as a function of the $s$-channel and $t$-channel cross section are shown in 
Fig.~\ref{cb-posterior-1d} for the cut-based and the neural network analyses. 

\begin{figure}[!h]
\centerline{
\includegraphics[width=0.7\textwidth]
{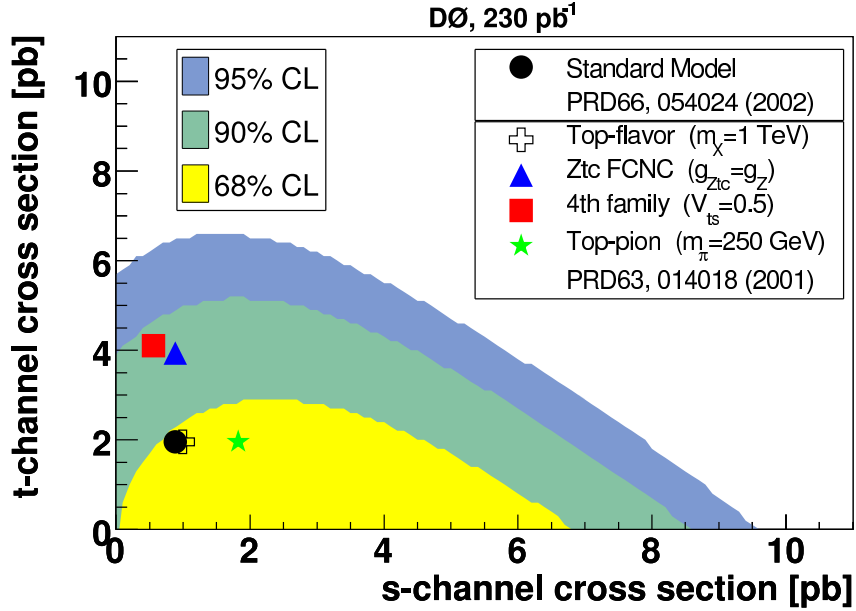}
}
\caption[nn-posterior-2d-nonsm]{ Exclusion contours at 68\%, 90\%, and 95\%
confidence level on the observed posterior density distribution as a function of
both the $s$-channel and $t$-channel cross sections in the neural networks 
analysis. Several representative non-standard model
contributions from Ref.~\cite{Tait:2000sh} are also shown.}
\label{nn-posterior-2d-nonsm}
\end{figure}

We also plot contours of the observed posterior 
density at different level of confidence, in the two-dimensional plane of the 
$t$-channel versus the $s$-channel single top production cross sections, for the 
neural network analysis, as shown in figure~\ref{nn-posterior-2d-nonsm}. 
In order to illustrate the sensitivity of this analysis to probe models of
physics beyond the standard model, the expected SM cross section as well as
several representative non-SM contributions are also shown.~\cite{Tait:2000sh}
%
%
\begin{table}[!h!tbp]
\begin{center}
\begin{tabular}{lcccc}
\hline\hline
& \multicolumn{2}{c}{Expected Limits}
 & \multicolumn{2}{c}{Observed Limits} \\
 &$s$-channel & $t$-channel & $s$-channel & $t$-channel \\
\hline
Initial selection & 14.5 & 16.5 & 13.0 & 13.6\\
Cut-based & 9.8 & 12.4 & 10.6 & 11.3 \\
Neural networks & 4.5 & 5.8 & 6.4 & 5.0 \\
\hline\hline   
\end{tabular}
\vspace{-0.1 in}
\end{center}
\caption[counting-limits-cuts]{Expected and observed upper limits (in picobarns) 
at 
95\% confidence level, on the production cross sections of single top quarks in 
$s$-channel ($tb$) and $t$-channel ($tqb$) searches, for the combined electron, 
muon, single-tagged and double-tagged channels.}
\label{limits}
\end{table}
%
%
%
%
\paragraph{Conclusions}
\label{sec:conclusions} 
 
To summarize, we find no evidence  for single top quarks in 
$\approx$230~pb$^{-1}$ of lepton+jets data collected by the \dzero detector 
at $\sqrt{s}=1.96$~TeV. The upper limits on the single top production cross 
section in the  $s$-channel and $t$-channel modes, at 95$\%$ CL, are 10.6 pb 
and 11.3 pb, respectively, using event counts in a cut-based
analysis, and 6.4 pb and 5.0 pb, respectively, using binned likelihoods in a  
neural network analysis.


\subsubsection*{Description of the First CDF Run II Analysis}
\label{sec:st_tev_cdfanalysis}
\textbf{Contributed by: Ciobanu, Stelzer, Wagner} \\

This section describes the first search for single top quark
production in Run II of the Tevatron performed by CDF.
Two analyses were carried out using an early data sample of 162 pb$^{-1}$ 
of proton-antiproton collisions. The first analysis (``$A1$'') was a combined 
search for $s-$ and $t-$channel single top, while the second 
analysis (``$A2$'') was a separate search for the $t-$channel and the 
$s-$ channel individually.
No significant evidence for a single top signal was found and an upper limit of 
17.8 pb on the combined single top production cross section, at 95\% 
confidence level was set. Upper limits of 10.1 pb and 13.6 pb were set on 
the production cross sections of $t-$channel, and $s-$channel single top, 
respectively \cite{Acosta:2004bs}.

The event selection for $A1$ exploits the kinematic features of the signal
final state, which contains a top quark, a bottom quark, and possibly additional
light quark jets. To reduce multijet backgrounds, the $W$ originating from the top
quark is required to have decayed leptonically. We demand therefore a high-energy
electron or muon ($E_{T}(e)>$ 20 GeV, or $P_{T}(\mu)>20$ GeV/$c$) and large missing
energy from the undetected neutrino $\not\kern-.35em {E_{T}}>$ 20 GeV. We reject dilepton events 
from $t\bar{t}$ and $Z$ decays by requiring the dilepton mass to satisfy:
76 GeV/$c^{2}<M_{\ell \ell}<106$ GeV/$c^{2}$. 
Exactly two jets 
with $E_{T}>15$ GeV and $|\eta|<2.8$ are required to be present in the event. A large fraction of 
the backgrounds is removed by demanding
at least one of these two jets to be tagged as a $b$-quark jet 
by using displaced vertex information from the silicon vertex detector (SVX).
The backgrounds surviving these selections can be classified as ``non-top'' 
and $t\bar{t}$. The non-top backgrounds are: $Wb\bar{b}$, $Wc\bar{c}$, $Wc$, 
mistags (light quarks misidentified as heavy flavor jets), non-$W$ (events where a jet is 
erroneously identified as a lepton), and diboson $WW$, $WZ$, and $ZZ$.

Finally, we require the invariant mass of the reconstructed top
quark to be within the range: 140 GeV/$c^{2}<M_{\ell \nu b}<210$ GeV/$c^{2}$.
We will refer to the above set of selection cuts as the ``$A1$ selection''.

\begin{figure}[h!]
\begin{center}
\mbox{ \includegraphics[width=0.5\textwidth]{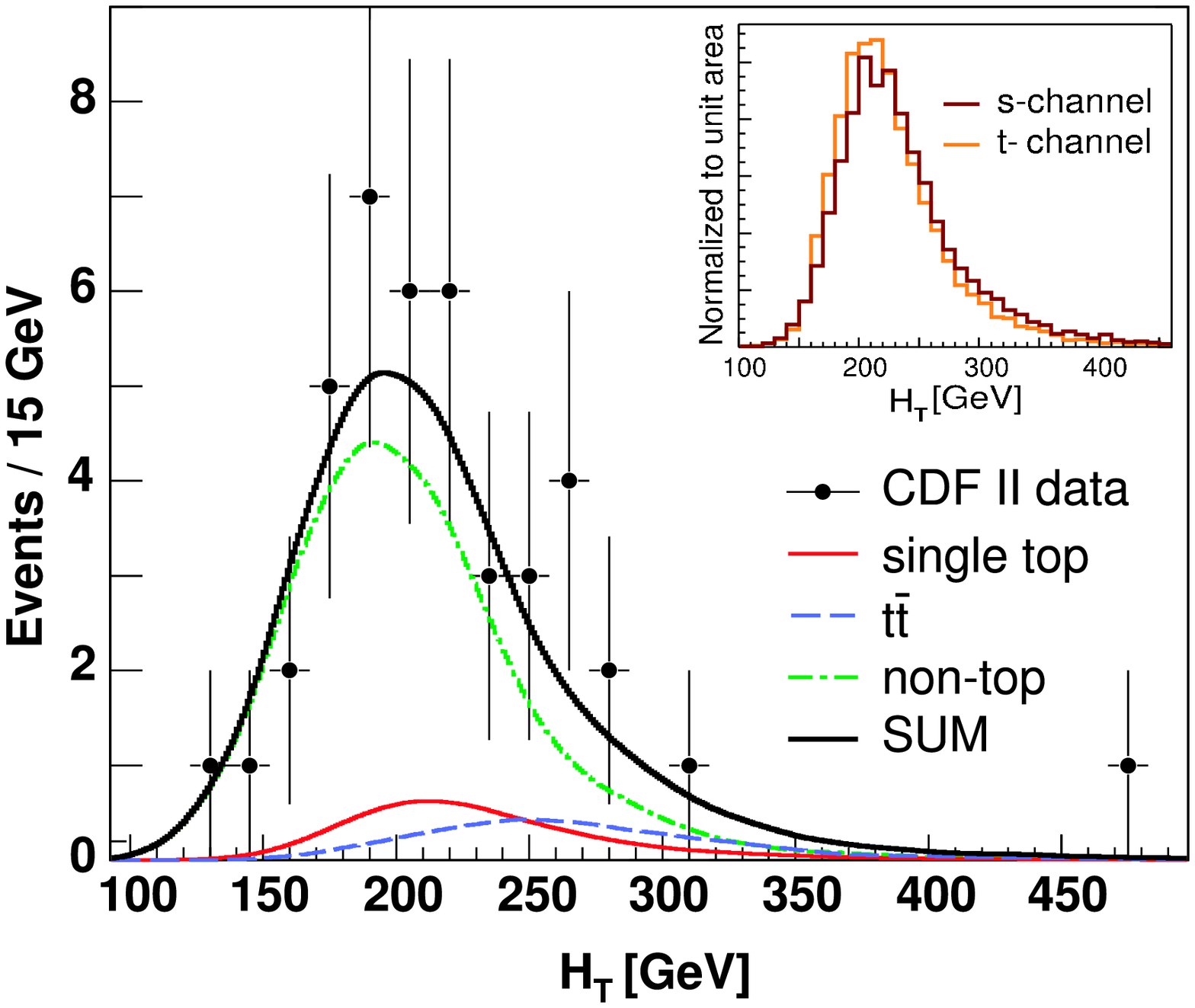}\includegraphics[width=0.5\textwidth]{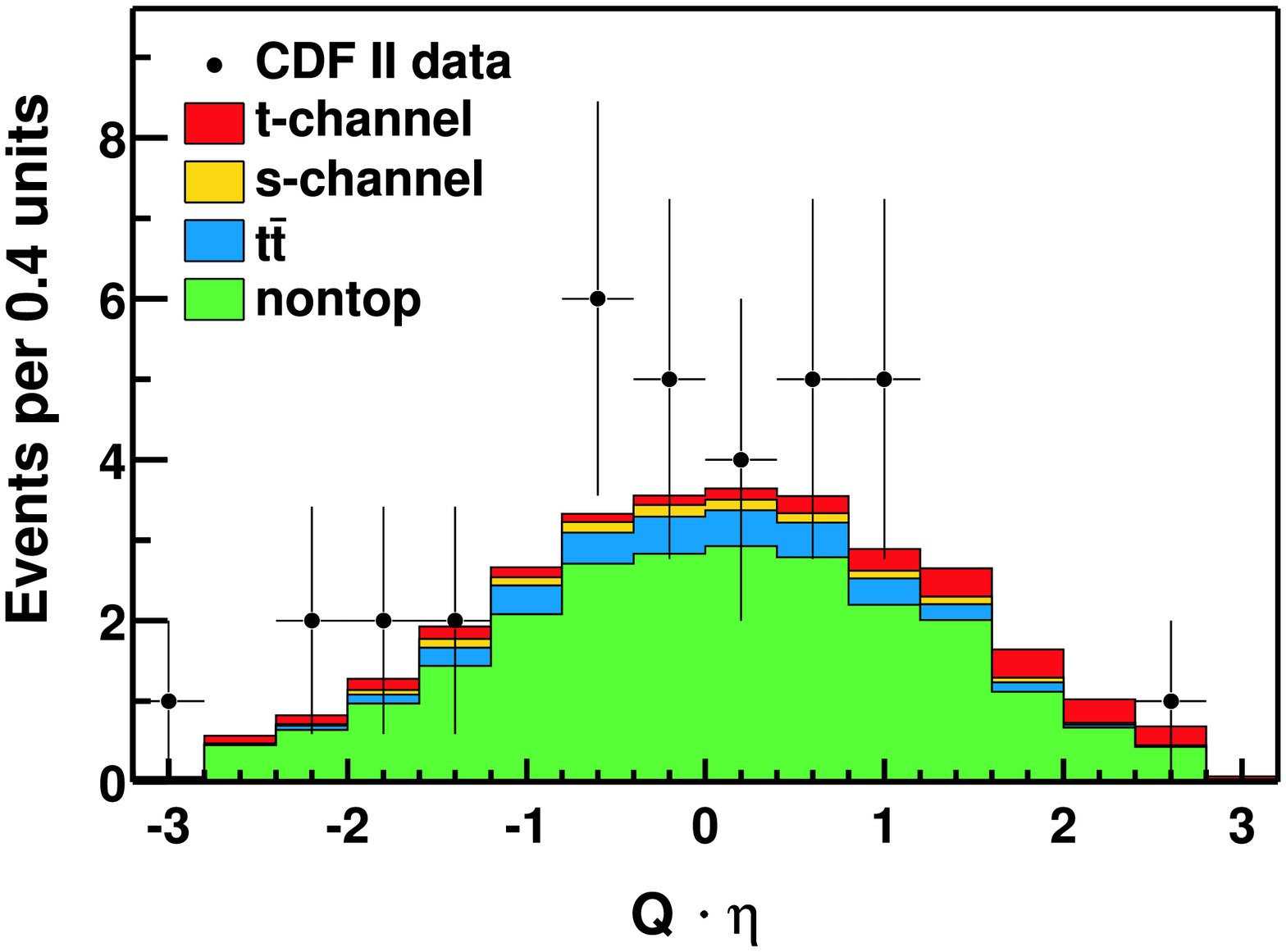} }
\end{center}
\caption{The $H_{T}$ and the $Q \cdot \eta$ distributions for CDF II data (points) 
compared with the Monte Carlo predictions. In both cases the distributions are normalized
to the expected number of events from Table \ref{table1}.}
\label{figura1}
\end{figure}

The second analysis $A2$ starts from the $A1$ selection and forms two distinct 
subsets of events. The first subset is formed by retaining events with 
exactly one $b$-tagged jet, and also demanding that at least one of 
the two jets have $E_{T}>30$ GeV. These requirements optimize the 
$t$-channel signal content of the sample with respect to the
backgrounds. The second subset is formed by selecting the double $b$-tagged 
events, i.e. the events where both jets are SVX $b$-tagged. This selection 
was found to be optimal for identifying $s$-channel signal events. 
The expected signal and background yields in 162 pb$^{-1}$ of data are 
summarized in Table \ref{table1}.
\begin{table}[htbc]
\begin{center}
\begin{tabular}{lclclc}
\hline\hline
 & Combined search $A1$ & & \multicolumn{3}{c} {Separate search $A2$}\\
Process                 &  & & Single-tag & & Double-tag       \\
\hline
$t$-channel             & $2.8\pm0.5$ &  & $2.7\pm0.4$ & & $0.02\pm0.01$   \\
$s$-channel             & $1.5\pm0.2$ &  & $1.1\pm0.2$ &  & $0.32\pm0.05$  \\
$t\bar{t}$              & $3.8\pm0.9$ &  & $3.2\pm0.7$ &  & $0.60\pm0.14$   \\
non-top                 & $30.0\pm5.8$ &  & $23.3\pm4.6$ &  & $2.59\pm0.71$ \\
\hline
Total                   & $38.1\pm5.9$   & & $30.3\pm4.7$ &  & $3.53\pm0.72 $ \\
\hline
Observed                & 42     & & 33 &  & 6\\
\hline\hline
\end{tabular}
\end{center}
\caption{\label{table1} Expected signal and background contributions 
and total number of events observed in 162 pb$^{-1}$ after all 
selection cuts, described in the text, have been applied.}
\end{table}
\paragraph{Methodology}
For the combined search $A1$ the kinematic distribution of the total transverse 
energy in the event $H_{T}$ is employed which looks similar for both signal channels while
it looks different for the background processes. 
The CDF data and the Monte Carlo $H_{T}$
distributions (using the contributions from Table \ref{table1}) are shown in 
Fig. \ref{figura1}. 
We employ a maximum likelihood method to estimate the signal content in the data. 
The likelihood function is expressed as:
\begin{equation} 
{\cal L}(\sigma_{{\mathrm{s+t}}}; \delta_1, \dots, \delta_9) =  
\prod_{i=1}^{N_{\mathrm{bin}}}\frac{e^{-\mu_{i}}\cdot \mu_{i}^{d_{i}}}{d_{i}!} 
\cdot  \prod_{j=1}^{9}G(\delta_{j}),
\label{like} 
\end{equation} 
where $i$ indexes the $H_{T}$ bins, and $j$ indexes the nuisance parameters $\delta_{j}$
(two background rates and seven sources of systematic uncertainty) accounted for 
by using Gaussian functions $G(\delta_{j})$. The 
\begin{figure}[h!]
\begin{center}
\includegraphics[width=0.8\textwidth]{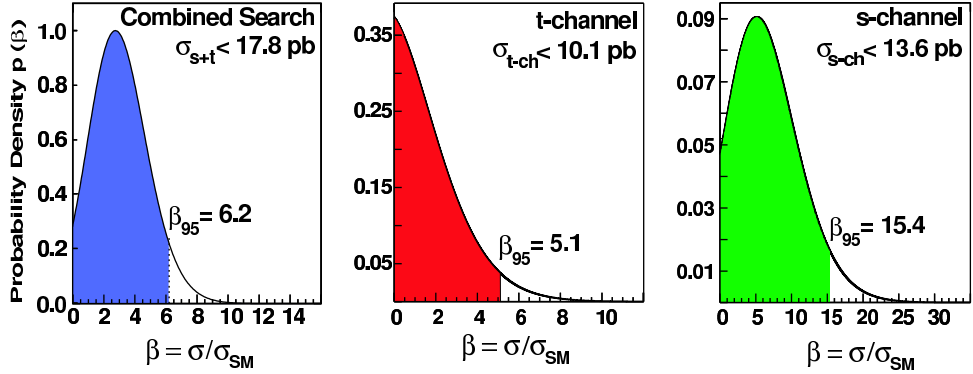}
\end{center}
\caption{\label{figura2}The posterior probability density obtained by integrating the
likelihood of Eq.~(\ref{like}) with respect to the nuisance parameters $\delta_{j}$.
In all three cases a flat prior is assumed to restrict signal cross-sections $\sigma$ to 
physical (positive) values.}
\end{figure}
number of data events in bin $i$ are denoted $d_{i}$, while $\mu_{i}$ is the expected number 
of events in bin $i$ and incorporates the full correlations between 
systematic effects modifying both the $H_{T}$ shape and the signal rate. 
The $\delta_{j}$ parameters are integrated out numerically,
and the resulting function (marginalized likelihood) is used to set the 95 \% confidence
level on the single top cross section.

For the individual search $A2$, the $t$-channel analysis is performed in the single $b$-tag 
sample.
In this subsample, we employ the kinematic distribution $Q\cdot \eta$, i.e. the product of
the lepton charge and the pseudorapidity of the non-$b$-tagged jet. The $t$-channel signal events 
are expected to exhibit an asymmetry toward the positive $Q\cdot \eta$ region. No such asymmetry is
observed in the data (right plot of Fig.~\ref{figura1}). 
The likelihood function used in the separate search closely resemble Eq.~(\ref{like}). 
To obtain sensitivity to the $s$-channel process a Poisson term for the number of
double b-tagged events is added to the likelihood.
The posterior probability density function for the combined search and the separate searches
are shown in Fig.~\ref{figura2}.

In summary, we find no significant evidence for electroweak single top quark production
in 162 pb$^{-1}$ of integrated luminosity recorded with CDF in Run II. We set upper
limits of 10.1 pb at the 95\% C.L. for the $t$-channel cross section, 13.6 pb
for the $s$-channel and 17.8 pb for the combined search.


\subsubsection*{CDF single-top analysis with neural networks based on $\mathbf{695\;pb^{-1}}$}
\label{sec:cdf_st_NN_analysis}
\textbf{Contributed by: Wagner}\\

CDF has updated its single-top search using a dataset corresponding to
$695\,\mathrm{pb^{-1}}$. Two analyses are performed based on neural networks
or likelihood functions, respectively. We described here briefly the neural 
network analysis, that has the better {\it a priori} sensitivity.

The event selection is very close to the one described in the previous
section. The updated analysis uses in 
addition electrons measured in the forward calorimeter. The cut on
the reconstructed invariant mass $M_{\ell\nu b}$ is omitted, since this
variables is fed into the neural network. The numbers of expected and
observed events are listed in Table~\ref{tab:07expect}. 
\begin{table}[!th]
\begin{center}
\begin{tabular}{lc}
\hline
Process                 & $N$ events \\
\hline
$t$-channel              & $16.7\pm1.7$ \\
$s$-channel              & $11.5\pm0.9$ \\
$t\bar{t}$               & $40.3\pm3.5$ \\
diboson, $Z$             & $17.2\pm0.8$ \\
$W+b\bar{b}$             & $170.7\pm 49.2$ \\                
$W+c\bar{c}$             & $64.5\pm 17.3$ \\
$Wc$                     & $69.4\pm 15.3$ \\
$W+q\bar{q}$, mistags    & $164.3\pm 29.6$ \\
non-$W$                  & $119.5\pm 40.4$ \\
\hline
Total                    & $674.1\pm 96.1$ \\
\hline
Observed                 & 689 \\
\hline
\end{tabular}
\end{center}
\caption{\label{tab:07expect} Expected number of signal and background events 
and total number of events observed in 695 pb$^{-1}$ in the
$W+2$ jets dataset.}
\end{table}

\begin{figure}[!th]  
\begin{center}
\includegraphics[width=0.65\textwidth]
{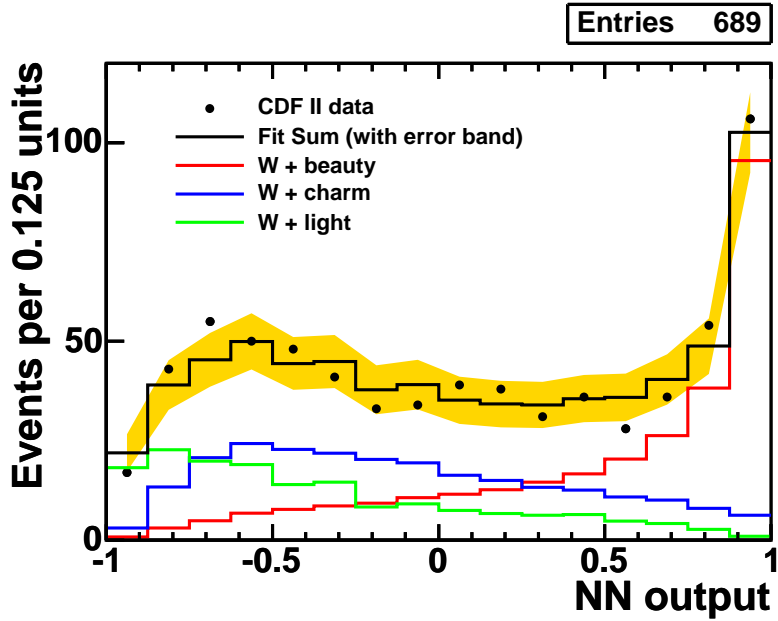}
\end{center}
\caption[nnbtag]{\label{fig:nnbtag}Output distribution of the neural
net $b$ tagger for 689 candidate events in the $W+2$ jets bin.
Overlayed are the fitted components of beauty-like, charm-like
and mistag templates.} 
\end{figure}

Using a neural network 14 kinematic or event shape variables are combined to a 
powerful discriminant. 
One of the variables is the output of a neural net $b$ tagger.
In Fig.~\ref{fig:nnbtag} the distribution of this $b$ tag
variable is shown for the 689 data events in the $W+2$ jets bin.
The neural net $b$ tagger gives an additional handle to reduce the
large background components where no real $b$ quarks are contained,
mistags and charm-backgrounds. Both of them amount to about 50\%
in the $W+2$ jets date sample even after the requirement that one jet
is identified by the secondary vertex tagger of CDF.

\begin{figure}[!th]  
\begin{center}
a) \hspace*{0.48\textwidth} b) \\
\includegraphics[width=0.48\textwidth]
{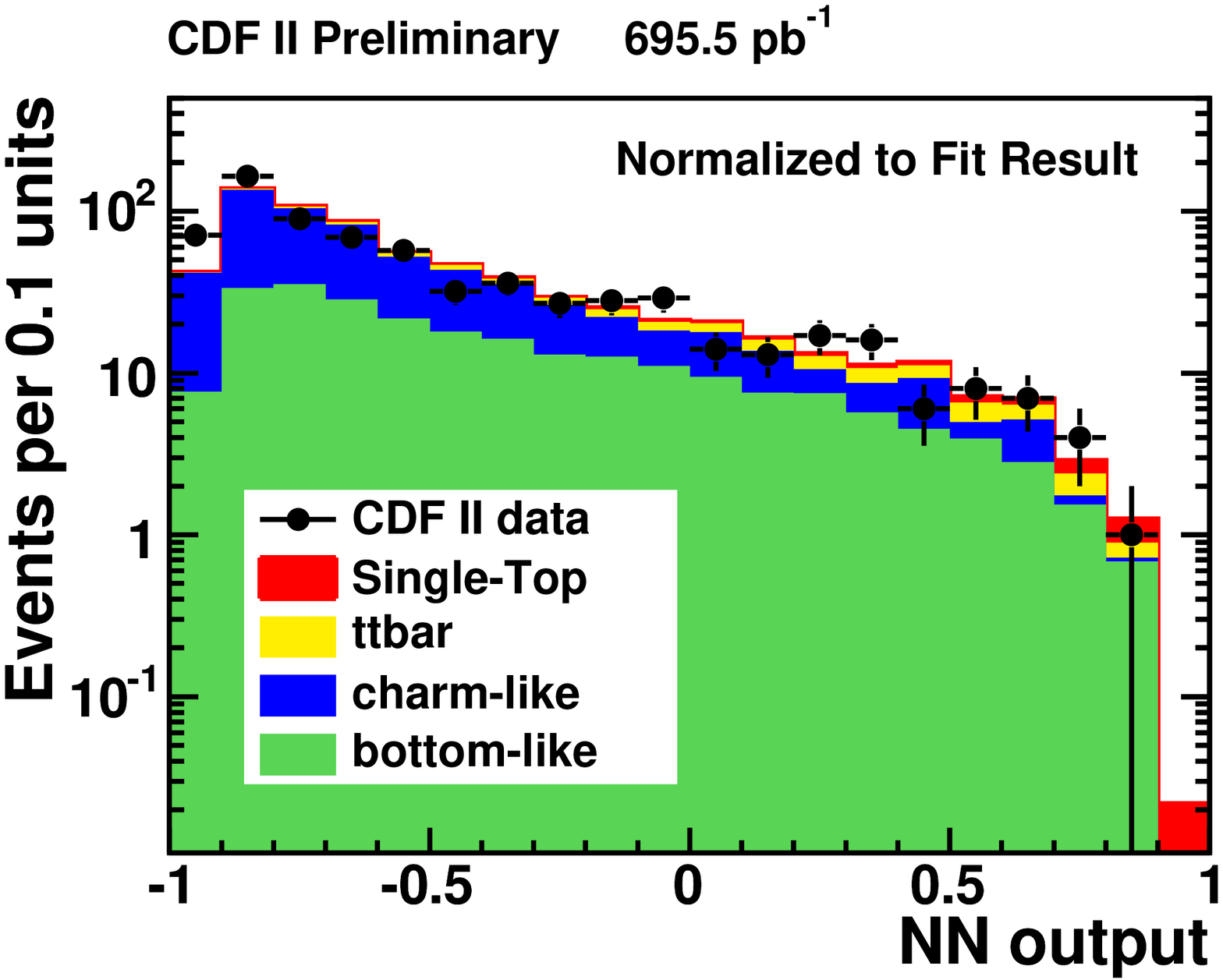}
\includegraphics[width=0.48\textwidth]
{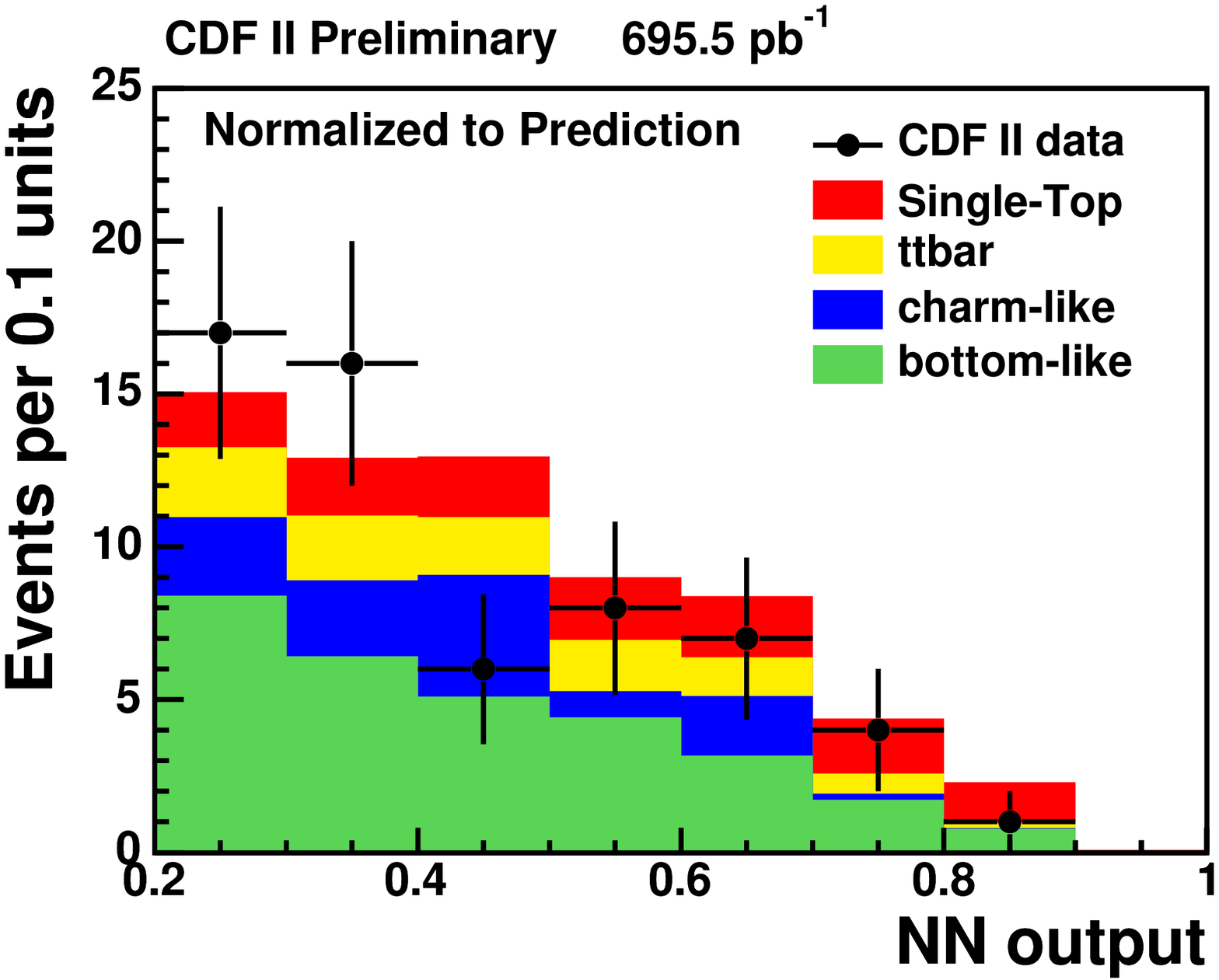} \\
\vspace*{3mm}
c) \hspace*{0.48\textwidth} d) \\
\includegraphics[width=0.48\textwidth]
{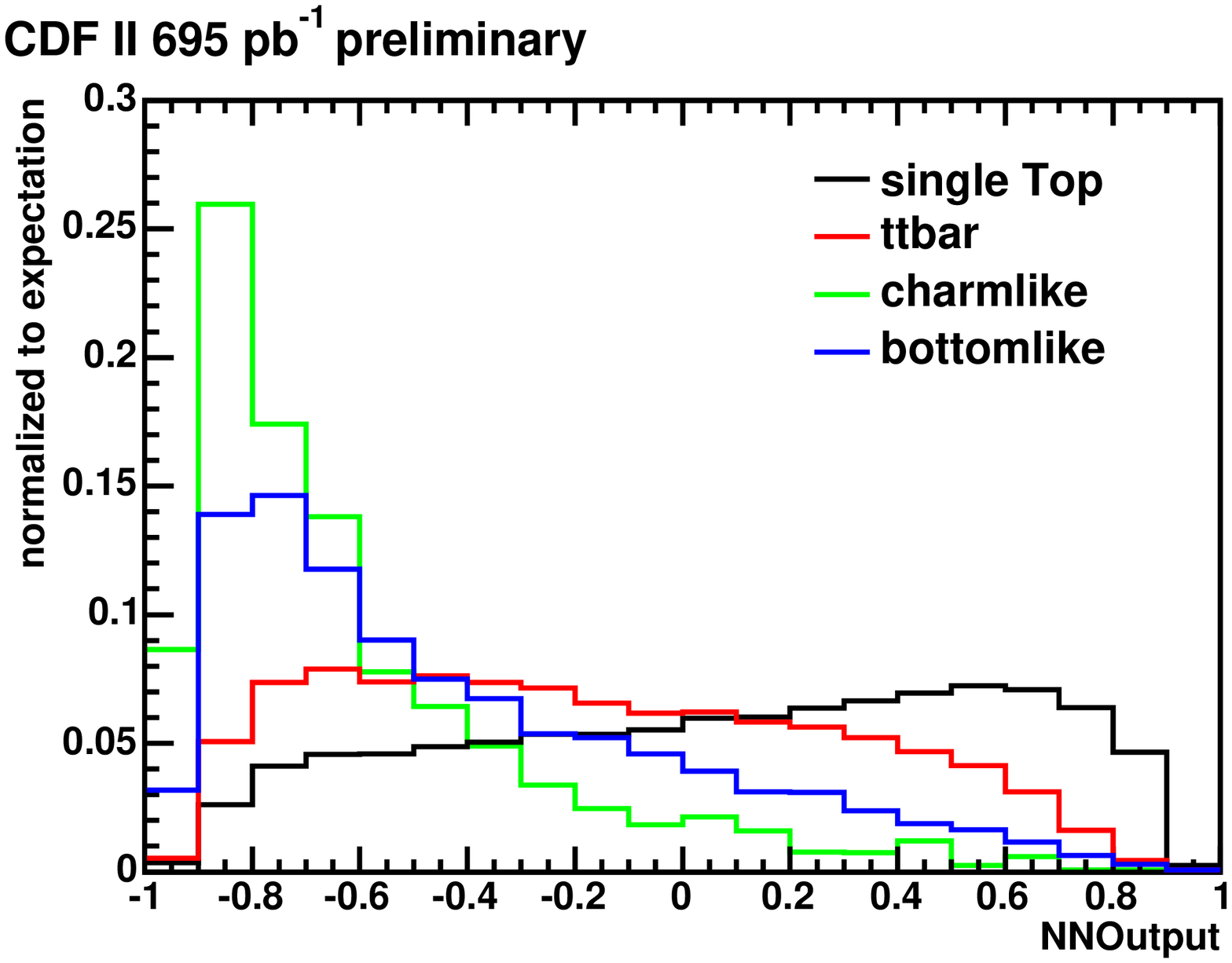}
\includegraphics[width=0.48\textwidth]
{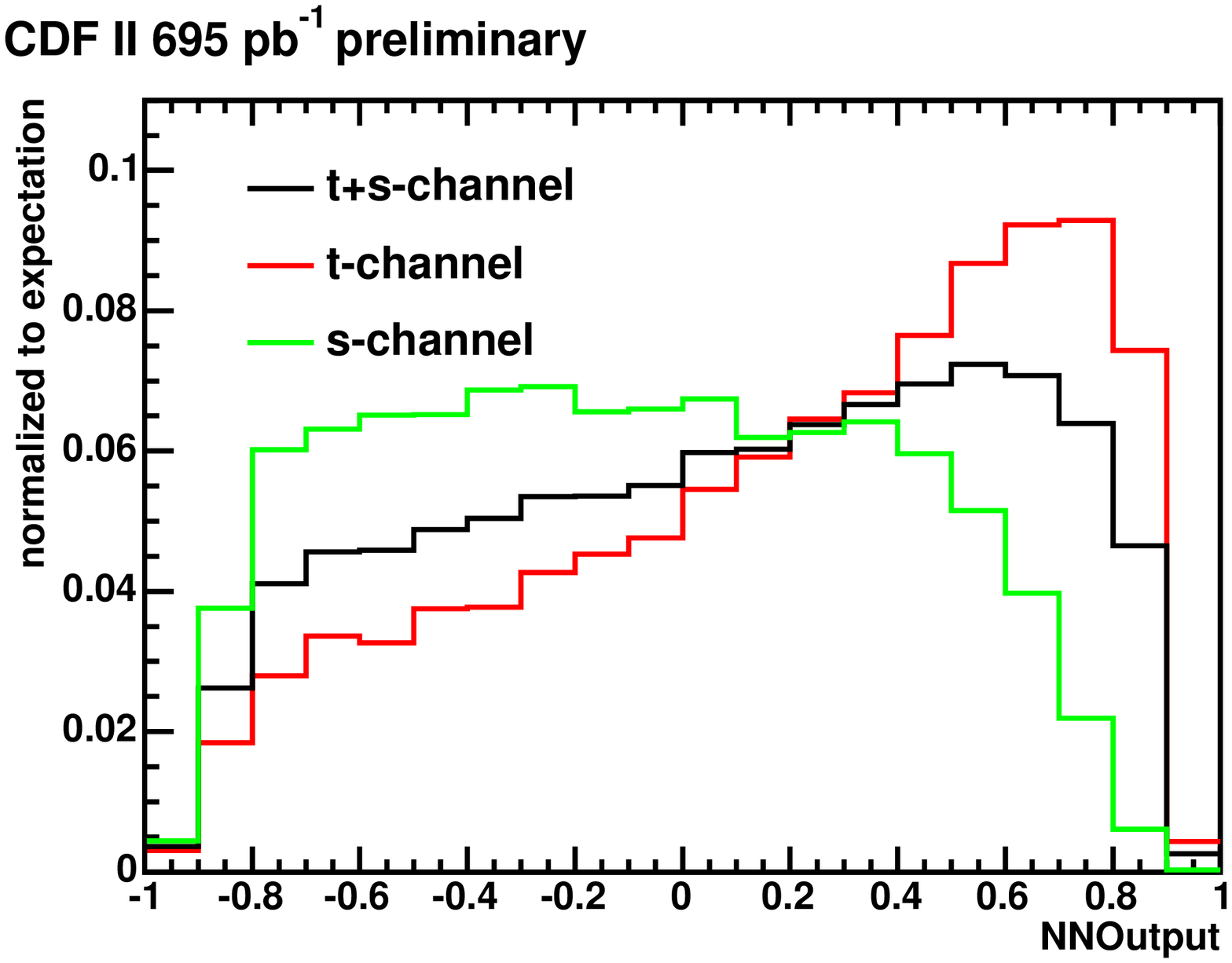} \\
\end{center}
\caption[nnstdata]{\label{fig:nnstdata}Single-Top search with neural networks
  at CDF. The analysis is based on $695\,\mathrm{pb^{-1}}$.
  a) data compated to the fit result, b) data compared to the 
  standard model expectation in the signal region with neural network outputs
  larger than 0.4. c) and d): For comparison the Monte Carlo template 
  distributions normalized to unit area are shown.}
\end{figure}

Figure~\ref{fig:nnstdata} shows the 
observed data compared to the fit result (a) and the expectation 
in the signal region (b) for the single-top neural network. 
For comparison the Monte Carlo
template distributions normalized to unit area are also shown (c, d).
The data are fitted with a binned likelihood function.
The $t$- and the $s$-channel are treated as one single-top signal
assuming the ratio of the two processes to be the one predicted by
the standard model. The most probable value of the likelihood function
is $0.8^{+1.3}_{-0.8}\,(\mathrm{stat.})\,^{+0.2}_{-0.3}\,(\mathrm{syst.})\;\mathrm{pb}$. At present, this result yields no significant evidence for single-top
production. The corresponding upper limit on the cross section is
$3.4\,\mathrm{pb}$ at the 95\% confidence level. The expected standard
model value is $2.9\pm0.4\,\mathrm{pb}$.

\begin{figure}[!th]  
\begin{center}
\includegraphics[width=0.7\textwidth]
{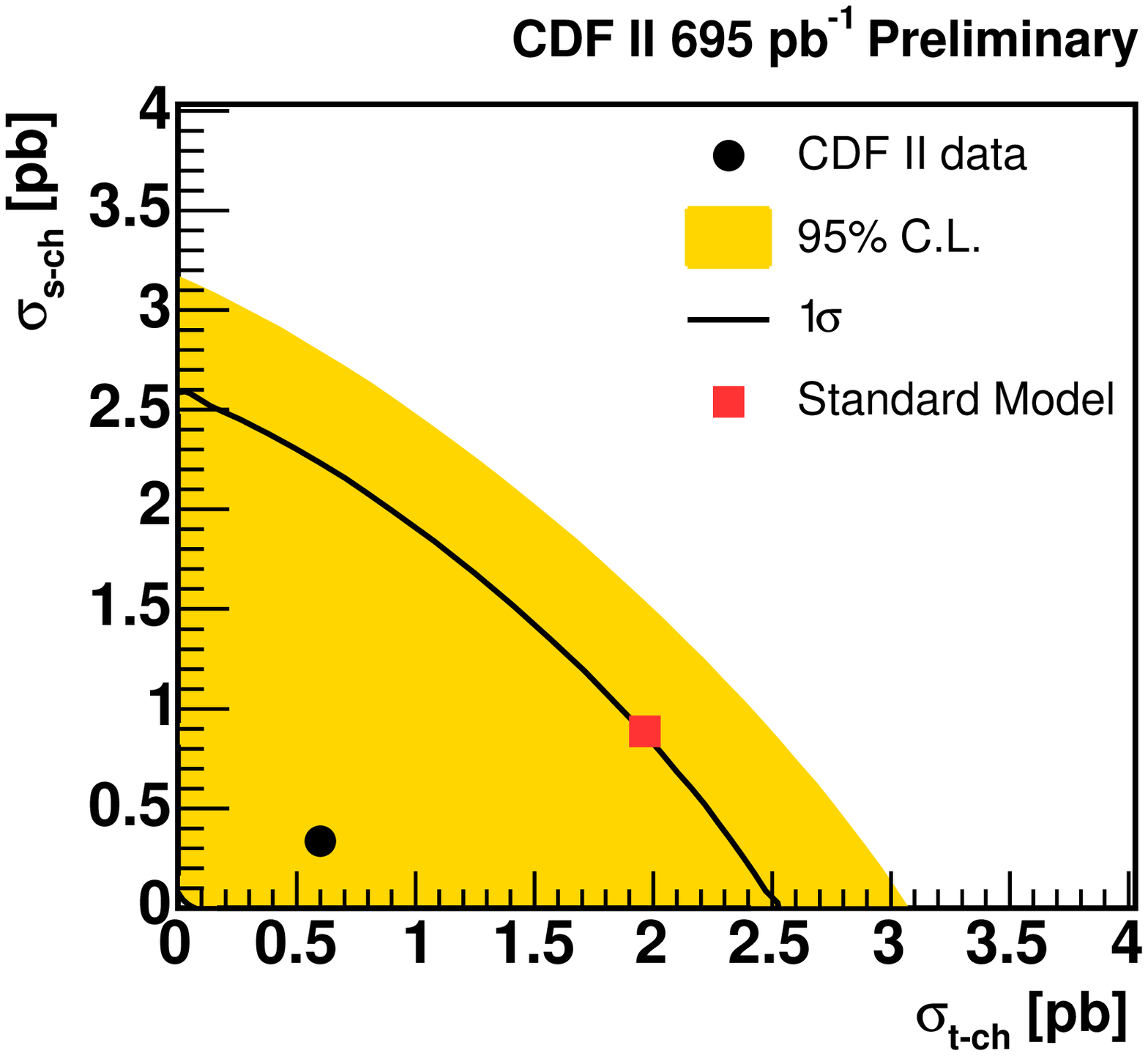}
\end{center}
\caption[2dnn]{\label{fig:2dnn}Result of a simultaneous fit for
  the $t$- and $s$-channel production cross section to two-dimensional
  neural network discriminants.}
\end{figure}

To separate $t$- and $s$-channel production two additional networks are
trained and a simulanteous fit to both discriminants is performed.
The fit result is illustrated in Fig.~\ref{fig:2dnn} and
summarized in Table~\ref{tab:2dnnResults}.
\begin{table}[!th]
\begin{center}
\begin{tabular}{lcc}
\hline              
                    & $t$-channel & $s$-channel \\ \hline
Observed most probable value & 
\begin{minipage}{0.2\textwidth}
\vspace*{2mm}

$0.6^{+1.9}_{-0.6}\,(\mathrm{stat.})$ \\
$\;\ \ \ ^{+0.1}_{-0.1}\,(\mathrm{syst.})\;\mathrm{pb}$

\vspace*{2mm}
\end{minipage}
&
\begin{minipage}{0.2\textwidth}
\vspace*{2mm}

$0.3^{+2.2}_{-0.3}\,(\mathrm{stat.})$ \\
$\;\ \ \ ^{+0.5}_{-0.3}\,(\mathrm{syst.})\;\mathrm{pb}$

\vspace*{2mm}
\end{minipage} \\
Observed 95\% C.L. upper limit & 3.1 pb & 3.2 pb \\
Expected 95\% C.L. upper limit & 4.2 pb & 3.7 pb \\
\hline
\end{tabular}
\end{center}
\caption{\label{tab:2dnnResults}Fit results for the separate search for
  $t$- and $s$-channel single-top production. The expected limits are 
  calculated from pseudo-experiments which included single-top quark events
  at the standard model rate.} 
\end{table}
Again, there is no evidence for single-top production yet.
However, the upper limits are already quite close to the predicted 
standard model values. 

\subsubsection*{Prospects for discovery}
\label{sec:st_tev_prospects}
\textbf{Contributed by: Jain, Wagner}\\

\begin{figure}[!h]
\begin{center}
\includegraphics[width=0.7\textwidth]{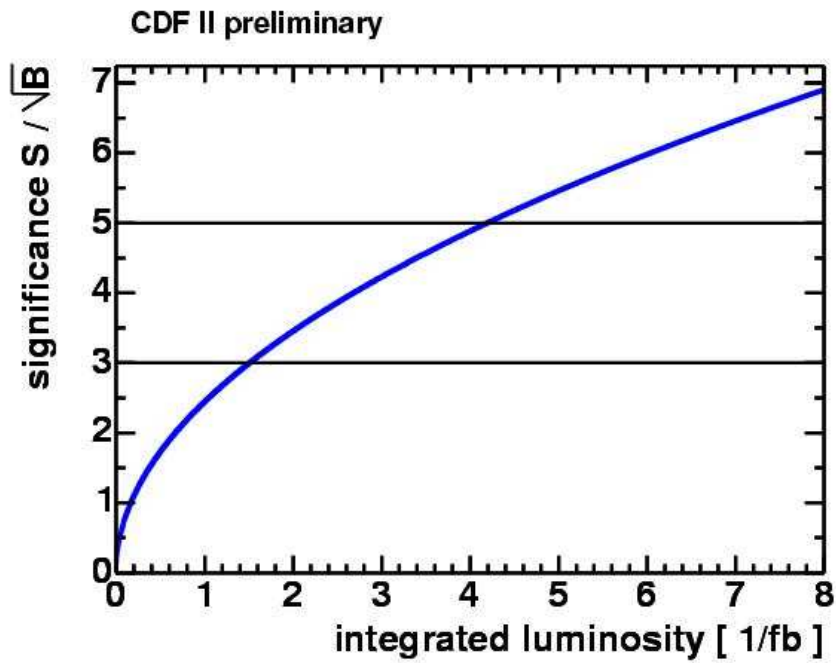}
\end{center}
\caption{The significance for Standard Model single top production in the 
combined $s+t$ mode, projected at different integrated luminosities, using 
CDF's initial 162 pb$^{-1}$ lepton+jets dataset. 
To discriminate signal and background a neural network is used.
With about 1.5 fb$^{-1}$ of 
data we expect to have a 2$\sigma$ signal needed to claim evidence for
single top production.}
\label{cdf_sensitivity}
\end{figure}

Both \dzero and CDF are currently working on increasing the acceptance and 
purity of the analysis as well as on several analysis methods which improve the 
search for single top quark production using different multivariate techniques. 
The sensitivity of the analysis for the combined $s+t$ mode, projected using 
CDF's  162 pb$^{-1}$ dataset and employing neural networks, is shown in 
Fig.~\ref{cdf_sensitivity}. Here, the significance is defined as $S/\sqrt{B}$, 
which can be interpreted as the statistical significance of the excess in the 
observed data above Standard Model predictions. 
A neural network was used to distinguish signal from background events.
The cut on the network output was adjusted to optimize the value of
$S/\sqrt{B}$ of the remaining events. 
No systematic uncertainties are included in this study.
Based on statistical uncertainties only, CDF expects to see an excess
corresponding to a $3\,\sigma$ Gaussian fluctuation with a dataset
of $1.5\,\mathrm{fb^{-1}}$.
%
%

\begin{figure}[!h]
\begin{center}
\includegraphics[width=0.7\textwidth]{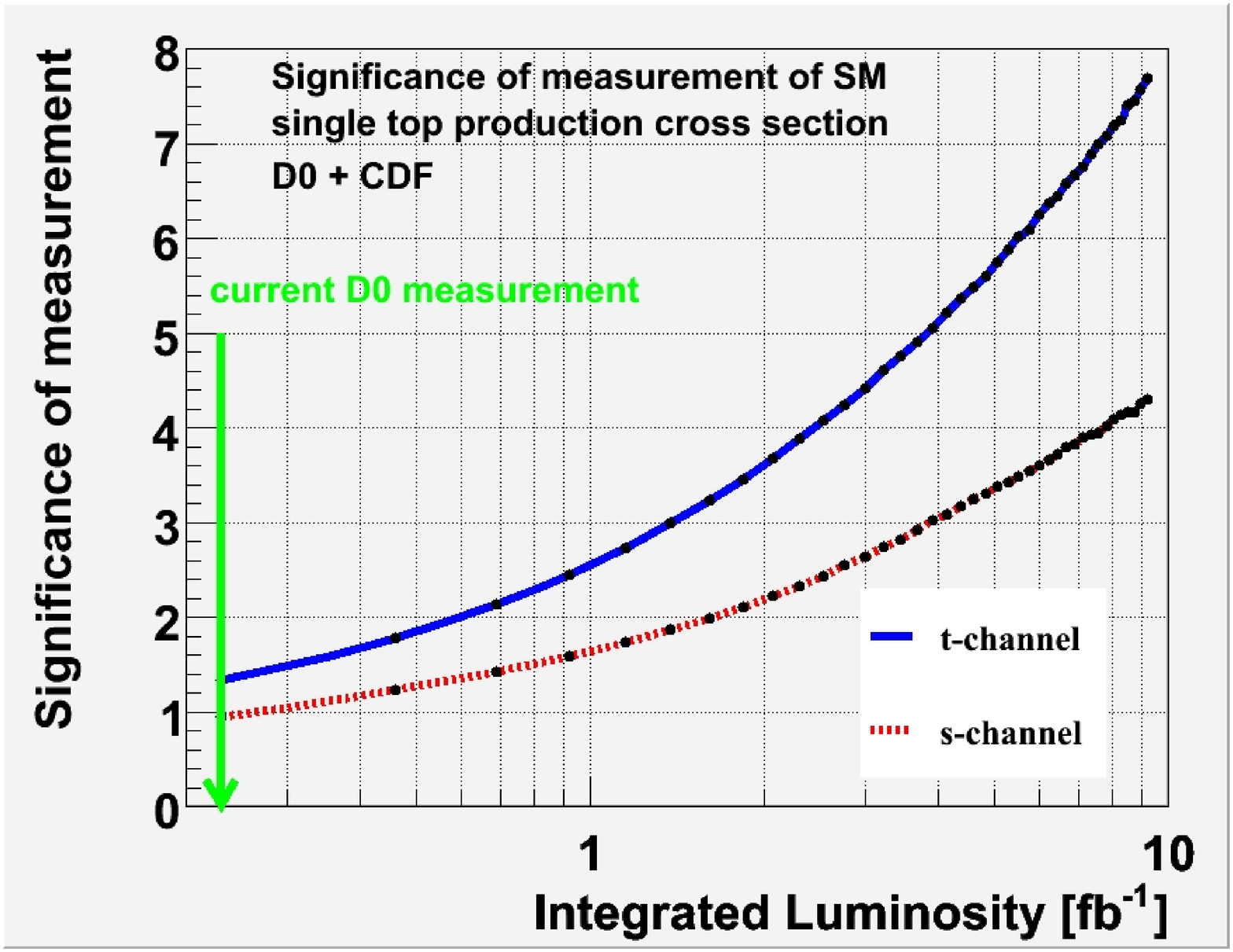}
\end{center}
\caption{The significance of a measurement of Standard Model single top quark 
production cross section, in the $s$-channel and $t$-channel modes,  projected at 
different integrated luminosities, using \dzero's initial 230 pb$^{-1}$ 
lepton+jets dataset.}
\label{d0_sensitivity}
\end{figure}

The sensitivity of \dzero's search for single top quarks at different integrated 
luminosities is shown in Fig.~\ref{d0_sensitivity}, for the $s$-channel and 
$t$-channel searches separately, by projecting twice the current \dzero datset of 
$\approx$230 pb$^{-1}$ in order to simulate the effect of combining the data 
from the two experiments (\dzero and CDF). Here, the significance is defined as 
the ratio of the peak of the Bayesian posterior probability density to the width of 
the distribution. This can be interpreted as the significance of a {\it measurement} 
of single top production cross section, where a measurement of the cross section 
can be defined by the peak of the probability distribution and its uncertainty by 
the corresponding width.  All systematic effects are ignored as mentioned before. 
It can be seen that it is possible to observe the production of single top quarks in 
the $t$-channel mode with a 2.5$\sigma$ significance at 1 fb$^{-1}$, but that it 
is possible to observe  them in the combined $s+t$ mode even earlier. 
%
%
%



\clearpage

%
%
%
%

%
%
\newcommand{\FB}{$\rm fb^{-1}$}
\newcommand{\HT}{$\rm H_T~$}
\newcommand{\MT}{$\rm m_t~$}
\newcommand{\ETA}{$\rm \eta~$}
\newcommand{\GEVCc}{\mbox{$\mathrm{\; GeV}/c^2$}}
\newcommand{\GEVC}{\mbox{$\mathrm{\; GeV}/c$}}
  \newcommand{\GEV}{$\rm GeV/c~$}
\newcommand{\MGEV}{$\rm GeV/c^2~$}
\newcommand{\TTB}{$\rm t\bar{t}~$}
\newcommand{\TTBLJET}{$\rm t\bar{t}\to l\nu b~jjb~$}
\newcommand{\TTBDIL}{$\rm t\bar{t}\to l^+l^-\nu\bar{\nu}b\bar{b}~$}
\newcommand{\TTBDILEP}{$\rm t\bar{t}\to l^+l^-\nu\bar{\nu}b\bar{b}~$}
\newcommand{\TTBDITAU}{$\rm t\bar{t}\to \tau^+\tau^- \nu\bar{\nu}b\bar{b}~$}
\newcommand{\TTBTAUJET}{$\rm t\bar{t}\to \tau\nu b~jjb~$}
\newcommand{\WQQ}{$\rm WQ\bar{Q}~$}
\newcommand{\WJETS}{$\rm W+jets~$}
\newcommand{\WZ}{$\rm WZ\to l\nu b\bar{b}$}
\newcommand{\WT}{$\rm W+t~$} 
\newcommand{\SCH}{$\rm W^*~$}
\newcommand{\TCH}{$\rm Wg~$}
\newcommand{\TANBETA}{$\rm \tan\beta~$}
\newcommand{\HIGGS}{$\rm H^{\pm}~$}
\newcommand{\SIXFIG}[8]{\begin{figure}[pht] 
                        \begin{center} 
                        \begin{sloppypar} 
                        \parbox{12cm}{ 
                              \mbox{ 
                                    \epsfig{file=topew/figures/#1.eps,width=6.cm} 
                                    \epsfig{file=topew/figures/#2.eps,width=6.cm} 
                                     } 
                               }  
                        \parbox{12cm}{ 
                              \mbox{ 
                                    \epsfig{file=topew/figures/#3.eps,width=6.cm} 
                                    \epsfig{file=topew/figures/#4.eps,width=6.cm} 
                                    } 
                                   } 
                        \parbox{12cm}{ 
                              \mbox{ 
                                    \epsfig{file=topew/figures/#5.eps,width=6.cm} 
                                    \epsfig{file=topew/figures/#6.eps,width=6.cm} 
                                    } 
                              \caption{{\footnotesize { #7} } 
                                \label{#8}} 
                           }                            
                      \end{sloppypar} 
                     \end{center} 
                    \end{figure} 
                    \vskip0.1cm} 
\newcommand{\TRIFIG}[5]{\begin{figure}[pht] 
                        \begin{center} 
                        \begin{sloppypar} 
                        \parbox{15cm}{ 
                              \mbox{ 
                                    \epsfig{file=topew/figures/#1.eps,width=5.cm} 
                                    \epsfig{file=topew/figures/#2.eps,width=5.cm} 
                                    \epsfig{file=topew/figures/#3.eps,width=5.cm} 
                                     } 
                              \caption{{\footnotesize { #4} } 
                                \label{#5}} 
                                   }   
                      \end{sloppypar} 
                     \end{center} 
                    \end{figure} 
                    \vskip0.1cm} 
\newcommand{\SPEQUAFIG}[6]{\begin{figure}[pht] 
                        \begin{center} 
                        \begin{sloppypar} 
                        \parbox{13cm}{ 
                              \mbox{ 
                                    \epsfig{file=topew/figures/#1.eps,width=6.5cm} 
                                    \epsfig{file=topew/figures/#2.eps,width=6.5cm} 
                                     } 
                                   }  
                        \parbox{13cm}{ 
                              \mbox{ 
                                    \epsfig{file=topew/figures/#3.eps,width=6.5cm} 
                                    \epsfig{file=topew/figures/#4.eps,width=6.5cm} 
                                    } 
                           } 
                           \caption{{\footnotesize { #5} } 
                         \label{#6}} 
                         \end{sloppypar} 
                     \end{center} 
                    \end{figure} 
                    \vskip0.1cm}
                    
\newcommand{\QUAFIG}[8]{\begin{figure}[pht] 
                        \begin{center} 
                        \begin{sloppypar} 
                        \parbox{13cm}{ 
                              \mbox{ 
                                    \epsfig{file=topew/figures/#1.eps,width=6.5cm} 
                                    \epsfig{file=topew/figures/#2.eps,width=6.5cm} 
                                     } 
                              \caption{{\footnotesize { #3} } 
                                \label{#4}} 
                                   }  
                        \parbox{13cm}{ 
                              \mbox{ 
                                    \epsfig{file=topew/figures/#5.eps,width=6.5cm} 
                                    \epsfig{file=topew/figures/#6.eps,width=6.5cm} 
                                    } 
                              \caption{{\footnotesize { #7} } 
                                \label{#8}} 
                           } 
                      \end{sloppypar} 
                     \end{center} 
                    \end{figure} 
                    \vskip0.1cm}

\newcommand{\DBLFIG}[6]{\begin{figure}[phtb] 
                        \begin{center} 
                        \begin{sloppypar} 
                        \parbox{6.5cm}{ 
                              \mbox{ 
                                      \epsfig{file=topew/figures/#1.eps,width=6.0cm} 
                                    } 
                                \caption{{\footnotesize { #2} } 
                                \label{#3}} 
                                        } 
                              \hspace{0.5cm} 
                              \parbox{6.5cm}{ 
                              \mbox{ 
                              \epsfig{file=topew/figures/#4.eps,width=6.0cm} 
                                   } 
                              \caption{{\footnotesize { #5} } 
                                \label{#6}} 
                           } 
                      \end{sloppypar} 
                     \end{center} 
                    \end{figure} 
                    \vskip0.1cm} 

\newcommand{\FIGVIII}[3]{\begin{figure}[pht]
                        \begin{center}
                        \begin{sloppypar}
                        \parbox{6.5cm}{  
                        \mbox{  
                        \epsfig{file=topew/figures/#1.eps,width=7.0cm}  
                                } 
                                }
                        \hspace{0.5cm} 
                        \parbox{5.0cm}{ 
                                \caption{{\footnotesize { #2} } 
                                \label{#3} 
                                } 
                                } 
                        \end{sloppypar} 
                        \end{center} 
                    \end{figure} 
                    \vskip0.1cm  
                    } 

\subsection {LHC Single Top Quark Searches}
\label{sec:singletoplhc}

    

\subsubsection*{Introduction}

At the LHC, the production of single top quarks accounts for a third 
of the top pairs production. With more than two millions single top 
events per experiment produced every year during a low luminosity run, 
a precise determination of all contributions to the total single top 
cross section seems achievable. These measurements will constitute the 
first direct measurement of $\rm V_{tb}$ at the few percent level of 
precision, and also constitute a powerful probe for new physics, via 
the search for evidence of anomalous couplings to the top quark, or 
the measurements of additional bosonic contributions to the single top 
production.

The single top production mechanisms proceeds through three different 
sub-processes resulting in disctinct final states, topologies and backgrounds. 
This Section establishes both ATLAS and CMS potentials for the cross section 
measurements of those three contributions. The event selections are 
presented extensively for both experiments and the performance are assessed 
in terms of statistical precision and systematic uncertainties. Both 
approaches address the experimental issues as the lepton identification, 
the jet reconstruction and the b-tagging performance as well as the strategies 
needed to evaluate Standard Model backgrounds from the data when possible.

\subsubsection*{Single top studies at ATLAS}
\textbf{Contributed by: Chevallier, Lleres, Lucotte, }\\
\paragraph{Phenomenomenology of single top and SM backgrounds}

\subparagraph{Single top production}
\hfill\break
In the Standard Model framework, the single-top production is due to three different 
mechanisms: the W-boson gluon fusion mode, noted \TCH, which includes the 
t-channel 
contribution; the associated production of a top quark and a W-boson, noted \WT; 
and the s-channel coming from the exchange of a charged boson $\rm W^*$. We note however 
that these definitions are valid only at Leading Order (LO) level of corrections. The 
total NLO cross section for all three mechanisms amounts to about 300~pb at the LHC. 
Among those channels, the dominant contribution comes from the \TCH processes, which account 
for about 240~pb. The \WT contribution amounts for about 60~pb while the s-channel \SCH mode 
is expected with a cross section of about 10~pb~\cite{Sullivan:2004ie,Campbell:2004ch,Campbell:2005bb}. 
We note that in pp collision, the cross section for single-top processes are not 
charge symmetricaly produced: the s-channel $\rm t\bar{b}$ final state cross section is 
thus expected to be produced with a factor of 1.60$\rm \pm 0.01$ higher than the 
$\rm \bar{t}b$ final state. This ratio amounts to 1.67$\rm \pm 0.01$ in the 
t-channel. This feature is of special interest since it generates a charge asymmetry 
in the leptonic final state that can be exploited in the analysis to reduce
 the contamination from the top quark pair production, which constitutes the main 
background to the single-top events selection. 
\begin{table}[phtb]
 \begin{center}
  \begin{tabular}{lc}
    \hline
 processes         &  $\rm\sigma\times BR$ (fb)  \\[1mm]
    \hline\hline
$\rm W^*\to t\bar{b}\to l^+\nu b\bar{b}$ & 1,300    \\[1mm]
$\rm W^*\to \bar{t}b\to l^-\bar{\nu} b\bar{b}$ & 800    \\[1mm]
$\rm W\to t\bar{b}q \to l^+\nu b\bar{b}q$ & 32,040    \\[1mm]
$\rm W\to \bar{t}{b}q\to l^-\bar{\nu}b\bar{b}q$ & 18,900   \\[1mm]
$\rm Wt\to qq'l\nu b, l\nu b qq'b$ & 9,320    \\[1mm]
 \hline
 \end{tabular}
  \caption[]{Cross sections convoluted by BR for single-top production. Those numbers are used in the LHC analyes. For further references, see previous Sections}
  \label{TAB-XSECTION}
 \end{center}
\end{table}

\noindent 
 In the Standard Model, the top quark decays almost decays exclusively into a W~boson 
and a b~quark. In the following, we use only the leptonic decay of the W's. The s-channel 
contribution from letponic tau decays has been taken into account and is considered among 
signal events. For the associated production, we consider the two cases where the 
leptons originates either directly from the W produced in parallel to the top quark, or 
from the W-boson appearing in the top quark decay channel. Table~\ref{TAB-XSECTION} reports 
the cross sections corresponding to all three mechanisms depending on the charge of the final 
W-boson.\\
 Significant sources of uncertainties affect the theoretical predictions of the 
production cross sections: the \SCH channel is known with a precision of $\rm 7.5\%$ 
at NLO, while the \TCH channel has an uncertainty of $\rm 3.5\%$. An uncertainty of 
$\rm 8\%$ is quoted for the \WT channel. More details can be found in the previous 
sections.

 At the time of the present analysis, only LO single-top generators were available 
for Monte Carlo studies. We use the TopRex~\cite{SGTOP-LHC-PHENO-TOPREX} generator for the event 
production and selection efficiency determination, and normalize a posteriori the event 
yields to the NLO cross sections. It is obvious that this approach does not account for 
the possible biases in final state jet (or lepton) momentum distributions. The use of 
a NLO generator as MC@NLO~\cite{SGTOP-LHC-PHENO-MCNLO} appears necessary to validate the 
selection as it becomes available. 

\subparagraph{Top pair production}
\hfill\break
At the LHC, the top pair production constitutes a dominant background to the 
single-top analyses. The total production cross section is 
$\rm \sigma(t\bar{t})=835^{+52}_{-39}~pb$~\cite{Beneke:2000hk},  about 3 times 
larger than the corresponding total single-top cross section, and more than 80~times 
that of the \SCH channel.\\
 The main channel affecting the analysis is the "lepton+jets" channel, 
with a final state composed of two b~jets, a high~\PT lepton and missing 
energy; the di-lepton channel ($\rm t\bar{t} \to l\nu b l\nu \bar{b}$) where 
a high~\PT lepton is lost in acceptance also constitute a major background. Finally, 
top pairs with one or both W decaying into a $\rm tau$~lepton where the $\rm \tau$ decays 
into an electron or a muon, may also survive the selection 
($\rm t\bar{t} \to\tau \nu b jjb$ or ($\rm t\bar{t} \to\tau \nu b \tau\nu b$)
The cross sections used in the following analyses are reported in Table~\ref{TAB-XSECTION-TOP}. 
Production cross sections are calculated up to NLO~\cite{SGTOP-LHC-PHENO-MCNLO}.\\ 
\begin{table}[phtb]
 \begin{center}
  \begin{tabular}{lc}
    \hline
 processes         &  $\rm\sigma\times BR$ (fb)  \\[1mm]
    \hline\hline
 \TTBLJET ($\rm l=e,\mu$) & 242,420    \\[1mm]
 \TTBDIL  ($\rm l=e,\mu$) & 38,096     \\[1mm]
 \TTBDITAU                & 9,520      \\[1mm]
 \TTBTAUJET               & 121,210    \\[1mm]
 \hline
 \end{tabular}
  \caption[]{Cross sections convoluted with the Branching Ratio for top pair production used in our analysis}
  \label{TAB-XSECTION-TOP}
 \end{center}
\end{table}

\noindent
Even at NLO, the theoretical uncertainty is dominated by the choice of the 
renormalization scale: a scale variation of $\rm \mu/2 ~to~2\times \mu$ 
results in an uncertainty of about 100~pb, representing an uncertainty of 
about 12\%. As these events constitute our main background, it will therefore 
be necessary to use cross section directly from measurements on data to assess 
properly the contamination of our final sample.

Regarding the Monte Carlo studies carried on \TTB events, we use the (LO) 
TopRex generator and apply a scale factor on the production 
cross sections. Thus, the same remarks as for single-top mechanisms apply 
here. Further studies including the comparison of TopRex and the NLO generator 
generator MC@NLO~\cite{SGTOP-LHC-PHENO-MCNLO} have already 
started~\cite{SGTOP-LHC-PHENO-HUBAUT}. 

\subparagraph{W+jet production}
\hfill\break
\hfill\break
\WQQ events where Q stands for b or c quarks involve the presence 
of long-lifetime particle jets that are also present in our signal sample.
The corresponding cross section has been computed at LO and is about 
the same order of magnitude that for the signal. However, NLO 
calculations~\cite{SGTOP-LHC-PHENO-MCFM-WQQ} are available. They have been performed by 
imposing some realistic constraints to the partons present in the final 
state. Numbers together with the requirements applied on the final partons 
are reported in Table~\ref{TAB-XSECTION-WQQ} for the various final states.
\begin{table}[phtb]
 \begin{center}
  \begin{tabular}{lccc}
    \hline
processes &  \multicolumn{3}{c}{Cross sections} \\[1mm]
          &  $\rm\sigma_{NLO}$ (fb) & $\rm\sigma_{LO}$ (fb) & Specific requirements  \\[1mm]
    \hline\hline
$\rm W^+jj\to e^+\nu jj$  &    669,000$\pm 10$  &  773     & $\rm p_T^l\ge 15,p_T^j\ge 20$\\[1mm]
$\rm W^-jj\to e^-\nu jj$  &    491,000$\pm 10$  &  558     & $\rm p_T^l\ge 15,p_T^j\ge 20$\\[1mm]
$\rm Zjj\to e^-e^+ jj$    &    105,000$\pm 5$   &  116     & $\rm p_T^l\ge 15,p_T^j\ge 20$\\[1mm]
  \hline
  $\rm W^+b\bar{b}\to e^+\nu b\bar{b}$       
                          &    3,060$\pm 60$    &  1300    &  $\rm p_T^l\ge 15,p_T^j\ge 20$\\[1mm]
  $\rm W^-b\bar{b}\to e^+\nu b\bar{b}$       
                          &    2,110$\pm 50$     &  900    &  $\rm p_T^l\ge 15,p_T^j\ge 20$\\[1mm]
  $\rm Zb\bar{b}\to e^+e^- b\bar{b}$       
                          &    2,280$\pm 30$     &  1800   &  $\rm p_T^l\ge 15,p_T^j\ge 20$\\[1mm]
   \hline
  \end{tabular}
  \caption[]{Cross sections for W+jets  and Z+jets events~\cite{SGTOP-LHC-PHENO-MCFM-WQQ}}
  \label{TAB-XSECTION-WQQ}
 \end{center}
\end{table}

\noindent
As no event generator including NLO calculations is presently available, we use 
the (LO) TopRex generator for the event production and normalize the corresponding 
cross sections to the NLO values. This method imposes us to reproduce the criteria  
applied in the phenomenological approach~\cite{SGTOP-LHC-PHENO-MCFM-WQQ}, in order to normalize 
properly our selection efficiencies. 

 W+light jets events constitute a major source of background because 
of a cross section several orders of magnitude above the signal. In our case, 
this processes can mimic the signal if two light jets are wrongly tagged  as 
a b-jet. Some calculations provide the NLO cross section~\cite{SGTOP-LHC-PHENO-MCFM-WQQ} 
for specific final states including W+j, W+jj and W+jjj events, with a leptonic 
decay for the W: in these calculations, requirements that reproduces typical 
LHC acceptance and energy thresholds are imposed on leptons and jets composing 
the final states. To estimate the NLO cross sections for our selection, we use 
the same method as for the \WQQ events, reproducing (when possible) the effects 
of the applied cuts at the parton level.
All available generators are presently LO generators and the numbers used 
for this analysis are quoted in \ref{TAB-XSECTION-WQQ}. Background production makes 
use of the HERWIG~\cite{SGTOP-LHC-PHENO-HERWIG} generator for $\rm W+jets$. 
It appears necessary to use of more appropriate generators (ALPGEN, AcerMC, MC@NLO) 
will be needed for future checks.

\subparagraph{Di-boson production}
\hfill\break
\hfill\break
Similarly, diboson events with light constitute backgrounds to our signal because 
of the presence a high-\PT lepton as well as b-jets in the final states. 
The $\rm WZ\to l\nu b b\bar{b}$ production cross sections have been computed 
at the NLO level for specific final states including a high-\PT lepton (electron 
or muon) and is found to be $\rm \sigma\times BR = 426~fb$.
The $\rm ZZ\to l^-l^+~b\bar{b}$ has a cross section of $\rm 340~fb$. The WW 
production where a light jet is mistagged as a b-jet has also to be considered. 
The corresponding cross section is 18,500 fb. Samples have been generated 
using the PYTHIA generator.

\paragraph{Discriminant variables in single-top event analyses}
\hfill\break
\label{VARIABLES}

 The three single-top processes result in quite distinct final states 
and topologies, leading to the definition of specific analyses in each case. The 
discrimination beetween them makes use of difference in jet multiplicity, 
number of b-tagged jets required, as well as angular distributions between 
lepton and/or jets present in the final states. Besides, important 
difference subsist in the level of backgrounds that are faced in the 
various analyses, leading to the development of tools dedicated to the 
rejection of specific backgrounds.

 We present in this section the basic set of relevant variables that 
are used to differentiate single-top events from main SM backgrounds. 
The selection of single-top events is based upon the presence of an 
isolated high-\PT lepton and a high missing transverse energy to reject 
non-W events. Events are required to contain at least two high-\PT jets, 
among which exactly one or two have to be identified as coming from the
 hadronization of a b quark. This set of requirements allows to reduce 
significantely QCD, and more generally, the jet production contamination. 
Global and topological variables may also be used to discriminate further 
top pair and W+jets events from our signal. We use in our case the total 
transverse energy of the events as well as the reconstructed top mass.

\subparagraph{Lepton selection}
\hfill\break\hfill\break
 In the ATLAS detector, the electron acceptance is defined in the pseudo-rapidity range 
$\rm |\eta|\le 2.5$. Beyond that range, the absence of tracking information makes the 
lepton identification more complex. The electron transverse energy is determined 
with a precision of :
$$\rm \sigma(E)/E = 12\%/\sqrt{E/GeV}\oplus 24.5\%/E_T/GeV \oplus 0.7\%$$
Fig.~\ref{PDF_LEPTON_PT} displays a comparison of the lepton \PT distribution for single-top events 
and all various backgrounds. 
Leptons present in the QCD $\rm pp\to b\bar{b}$ samples originate mainly from the semi-leptonic 
decays of b hadrons and are thus much softer than those coming from a W boson decay. Leptons 
originating from $\rm \tau$ decays in \TTBDITAU and \TTBTAUJET events also have much lower \PT 
spectra. All those backgrounds are therefore very sensitive to the lepton \PT threshold used in 
the analysis. On the upper range of the distribution, W-boson produced leptons tend to be harder 
in top events than in W+jets events.
\TRIFIG{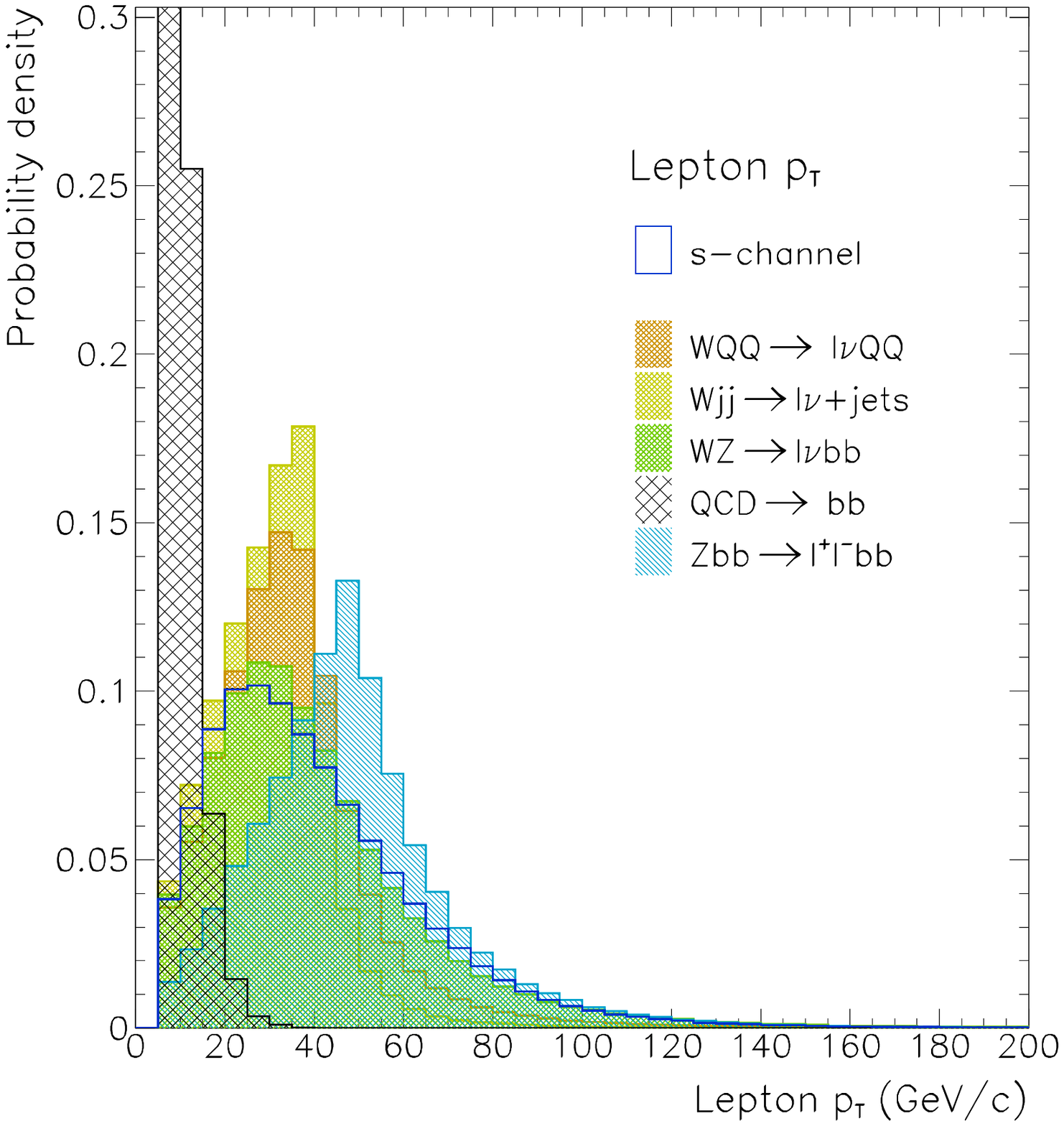}{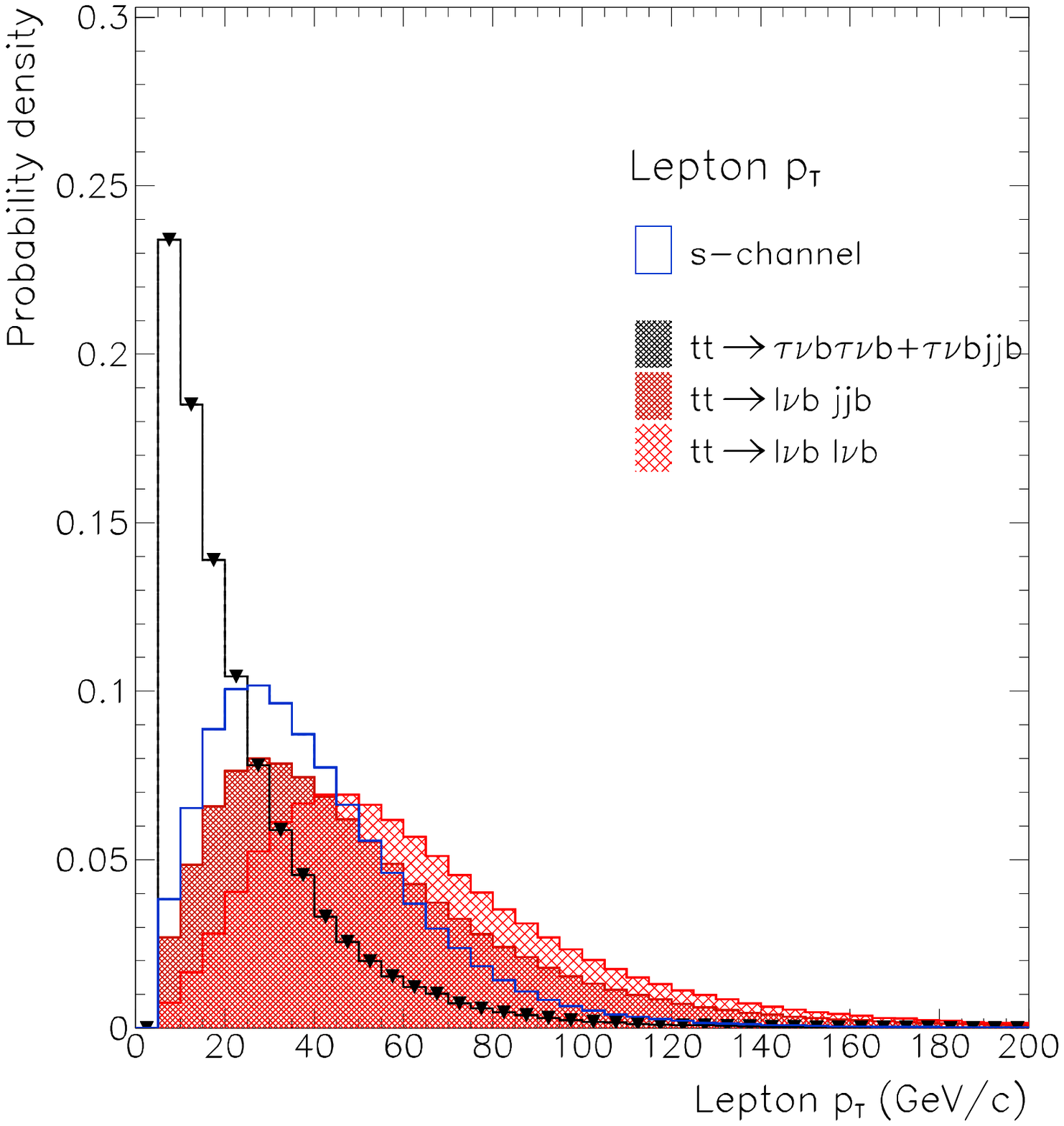}{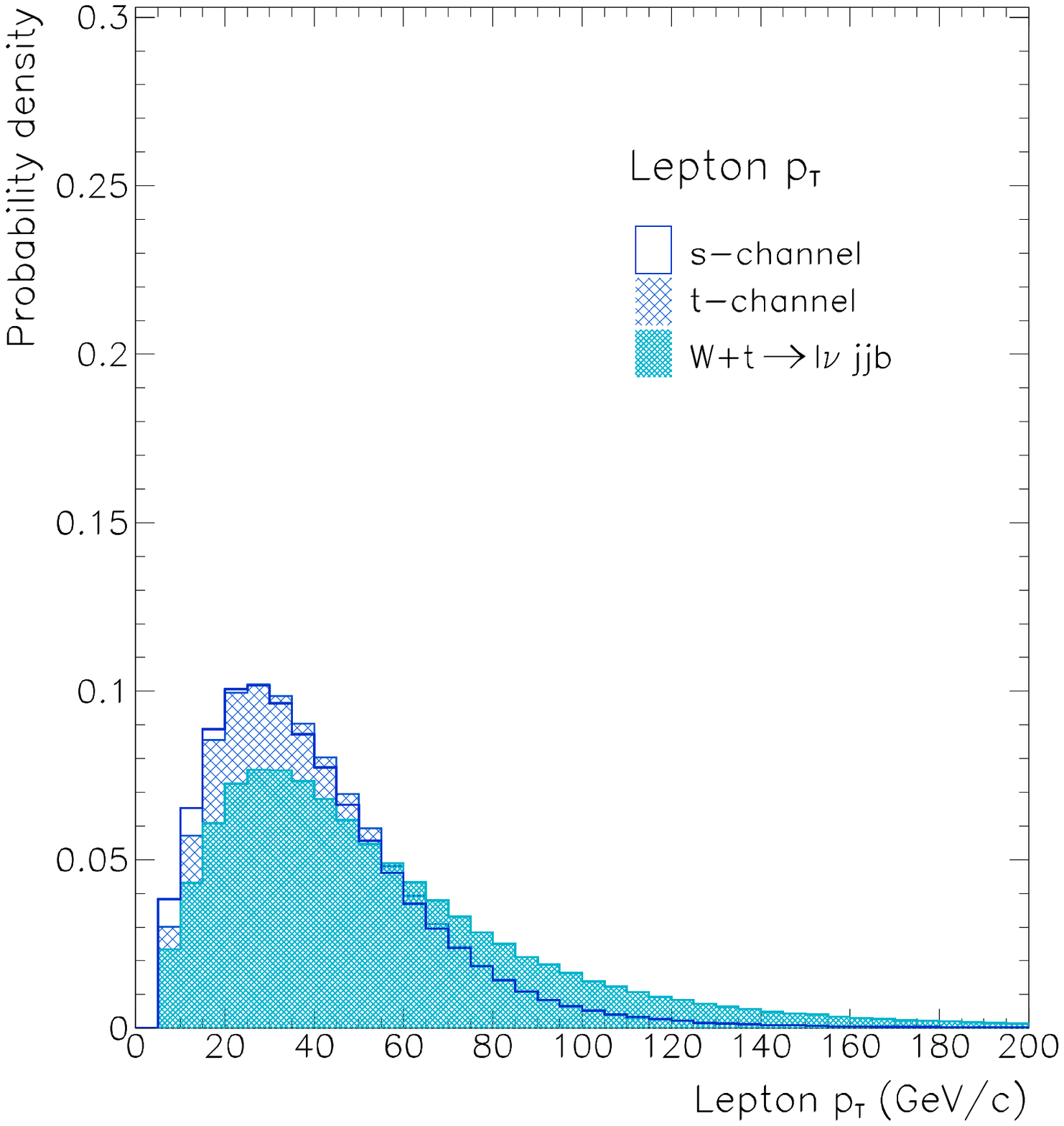}{Lepton transverse momentum 
probability density for signal and backgrounds.}{PDF_LEPTON_PT}

\noindent
 The average \PT is about 40~\GEV for the s-channel, 50~\GEV for 
\TTB events and has to be compared with the mean value of 30~\GEV for WZ and \WQQ productions. 
A threshold of 25~\GEV is set to select a high-\PT lepton. This value corresponds to the lepton 
trigger threshold that is used to detect such events, and allows to reduce significantly non-W 
as well as tau decays' s top pair events contamination.

The lepton is required to be isolated. The lepton isolation is defined as the distance 
to the closest jet by $\rm \Delta R=\sqrt{\Delta\Phi^2+\Delta \eta^2}$. Note that jets 
are defined by the use of a cone algorithm whose 
performance are described in Ref.~\cite{ATL-PHYS-98-131}. The isolation of a high-\PT lepton with 
respect to the closest jet depends upon the topology of events. In a high jet multiplicity 
environment like \TTB and single top events, the $\rm \Delta R(lepton,jet)$ value tend to be lower 
than in a simple W+jets event. A cut at $\rm \Delta R(lepton,jet) \ge 0.4$ is set for the selection.

To remove events with two leptons like $\rm Z\to l^+l^-$ and dileptonic top pairs, a veto is 
performed in any pairs of leptons with opposite signs and above 10~\GEV. Note that this lepton 
veto may introduce some systematic effects due to the mis-identification of the lepton sign 
as well as a lower lepton identification efficiency at a lower \PT threshold. These effects 
have to be addressed in a full event reconstruction stage.
Note that the sign of the selected lepton provides the nature of the single-top event: 
a positron or positive muon will sign a $\rm t\bar{b}$ final state, while an electron or 
muon will sign a $\rm \bar{t}b$ decay.

\subparagraph{Missing energy \MET}
\hfill\break\hfill\break
 The missing energy physically originates from the presence of a neutrino in the W-boson 
decays. Missing transverse energy is shown in Fig.~\ref{PDF_MISSING_ET} for 
signal and backgrounds. 
\TRIFIG{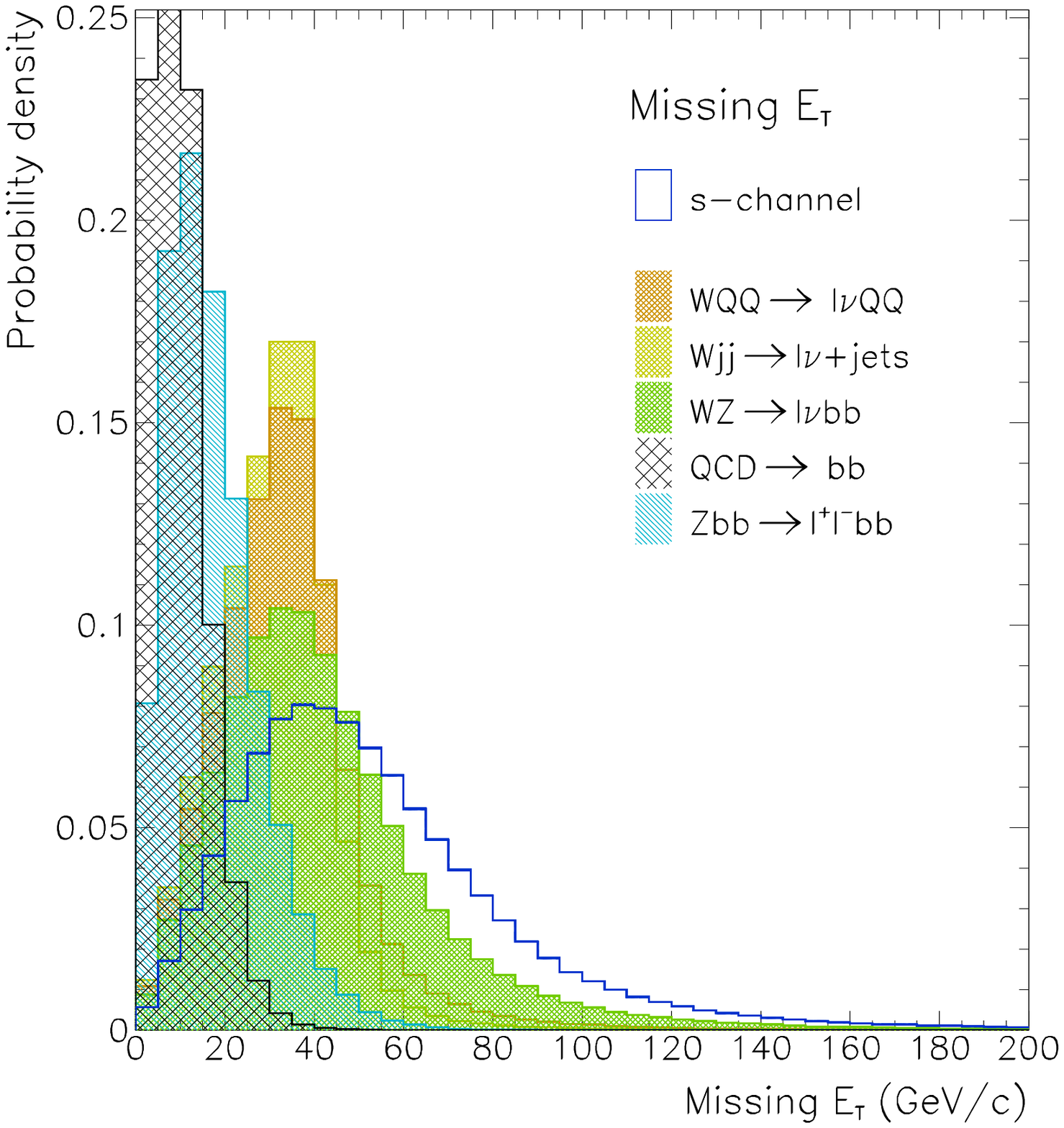}{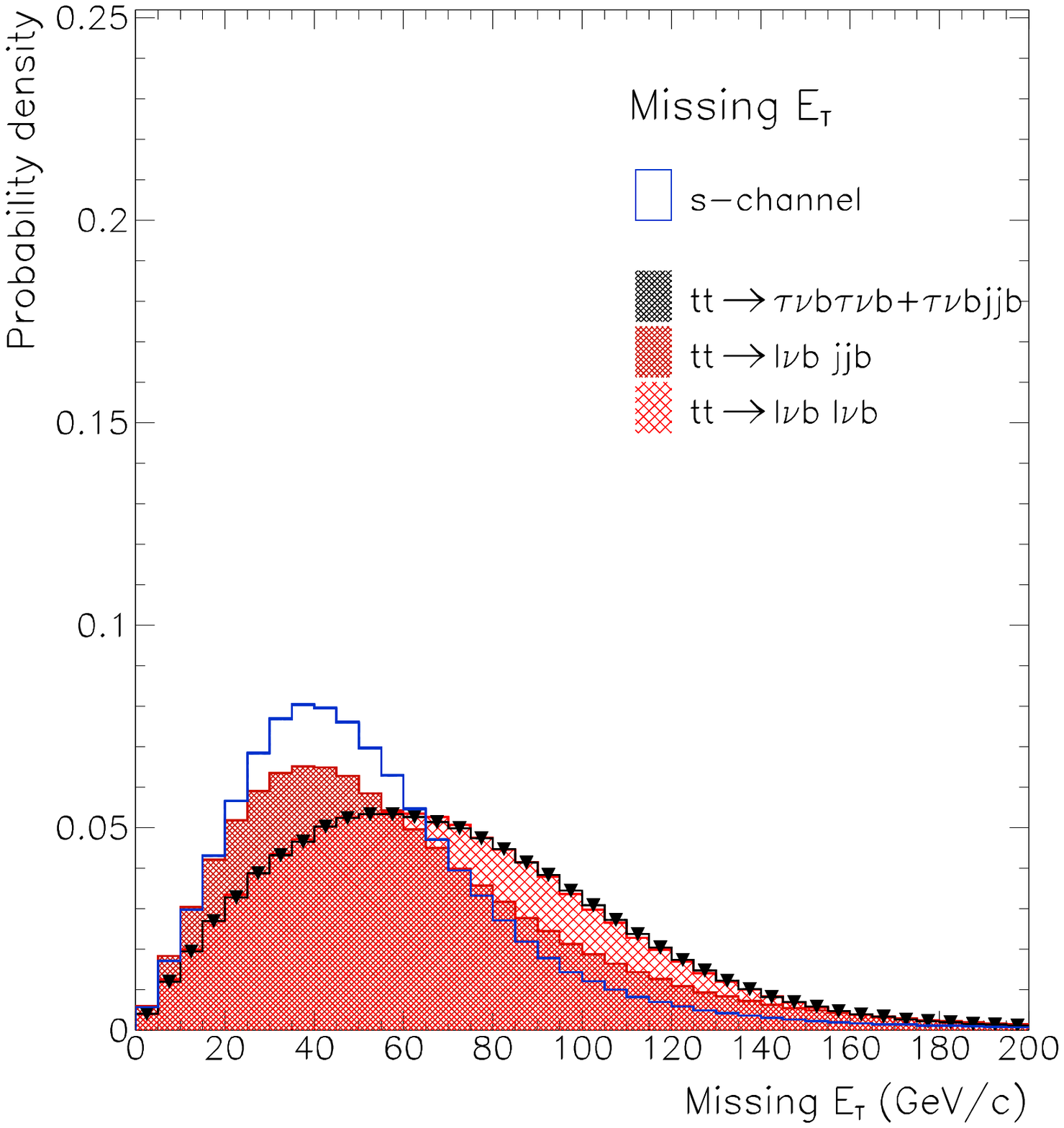}{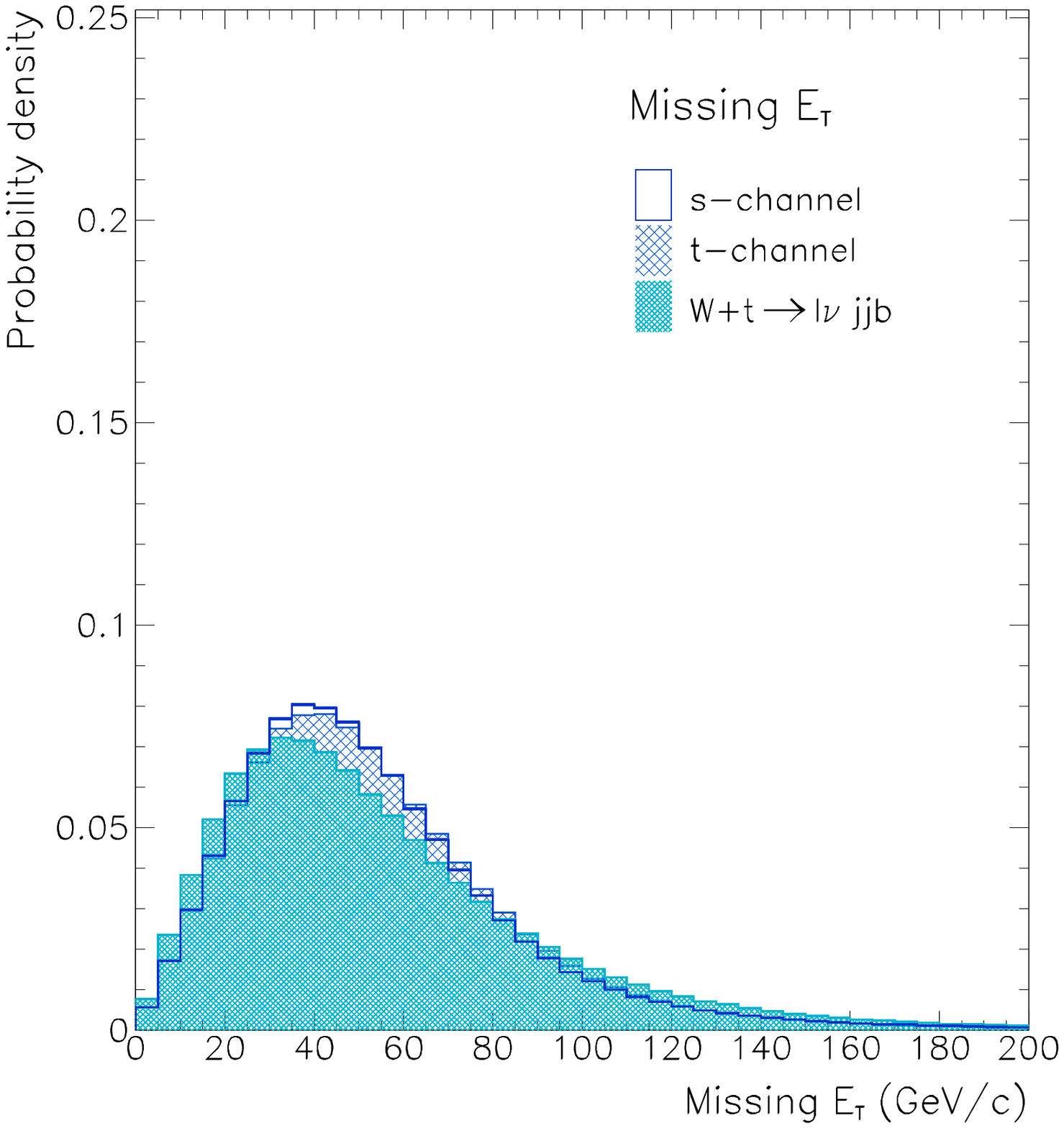}{Transverse missing 
energy for signal and backgrounds.}{PDF_MISSING_ET}

\noindent
Significant differences can be seen in the distributions 
which carry a significant discriminating power: average values are around 30~\GEV 
for W+jets production and about 55~\GEV for single-top productions; those values 
are raised above or higher than 60~\GEV for "lepton+jet" and "dilepton" top pair 
events. 

A threshold at 25~\GEV is thus applied so as to select a leptonic W~decays.  
The use of the full spectrum may however help the discrimination against backgrounds 
with softer \MET like WZ, WQQ, and W+jets events, as well as against events with 
harder \MET spectrum like top pairs. A likelihood approach could thus benefit the 
selection.  

 This variable is extremely sensitive to the performance of the hadronic and 
electromagnetic energy measurement of the detector. Angular and energy resolution, 
the identification capabilities of noisy calorimeter cells, the modelling of the 
underlying events and the pile-up effects thus appear as key factors in the 
missing energy measurement. Again, full reconstruction studies are required to 
assess the magnitude of those effects.
\TRIFIG{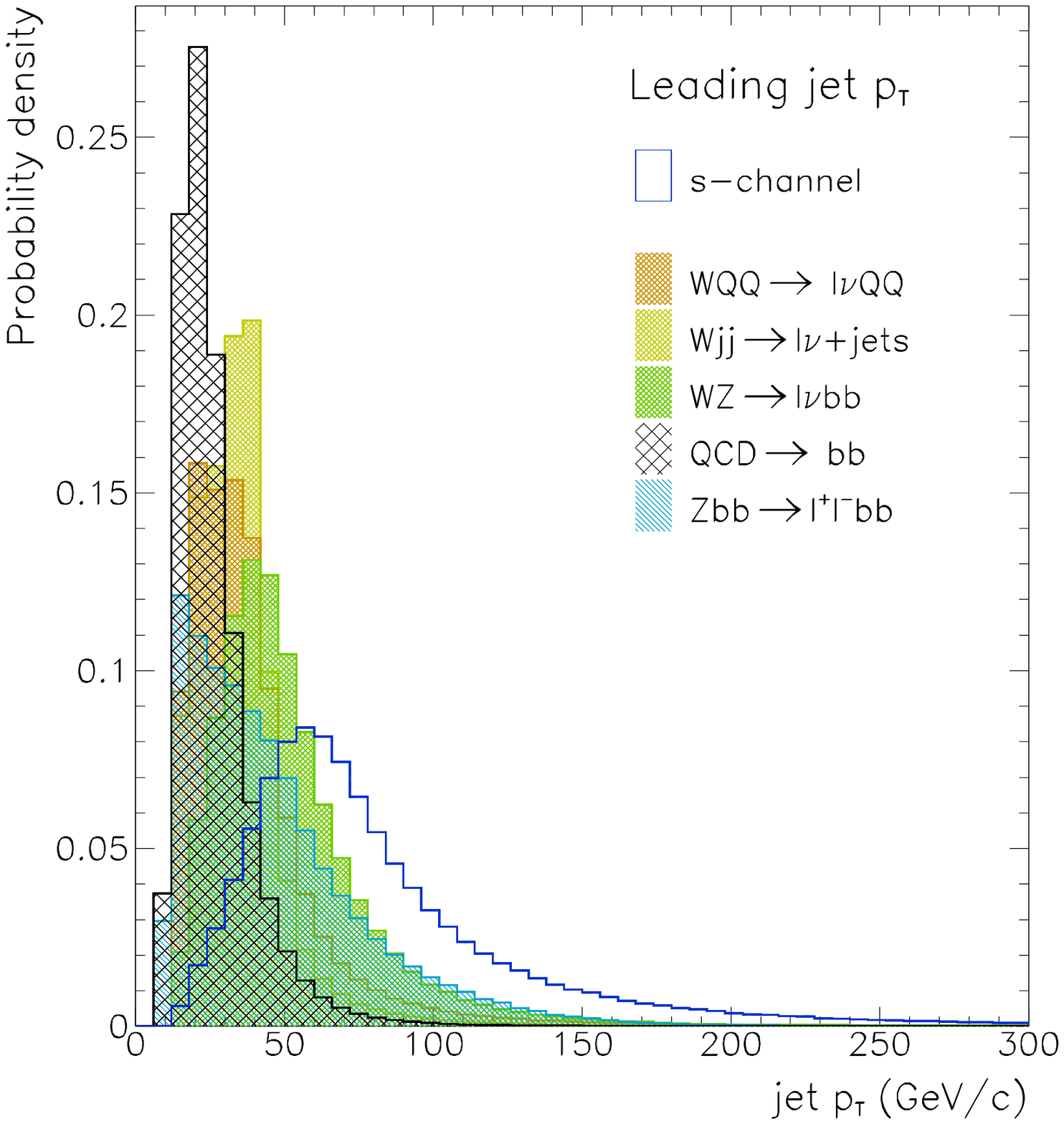}{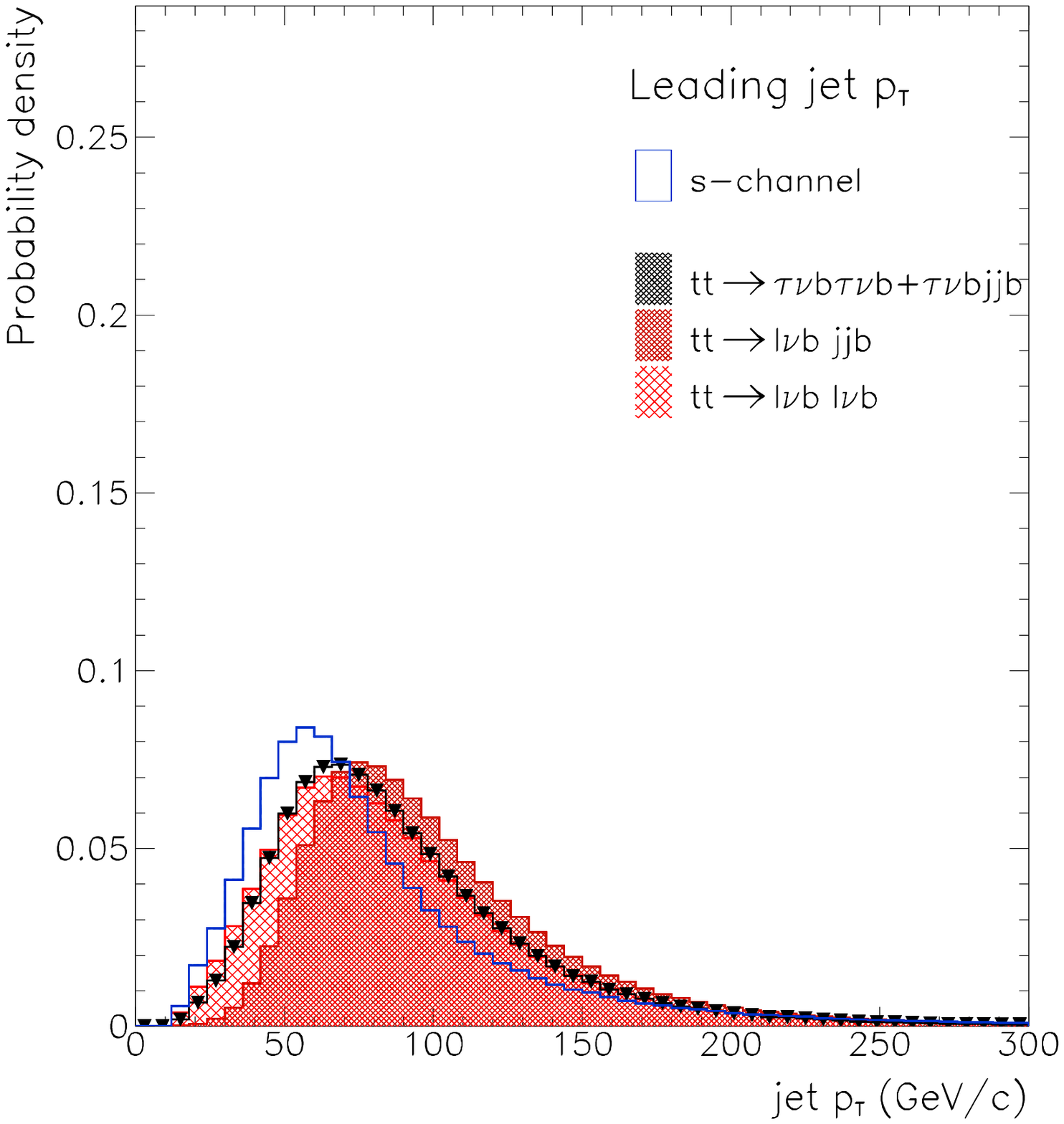}{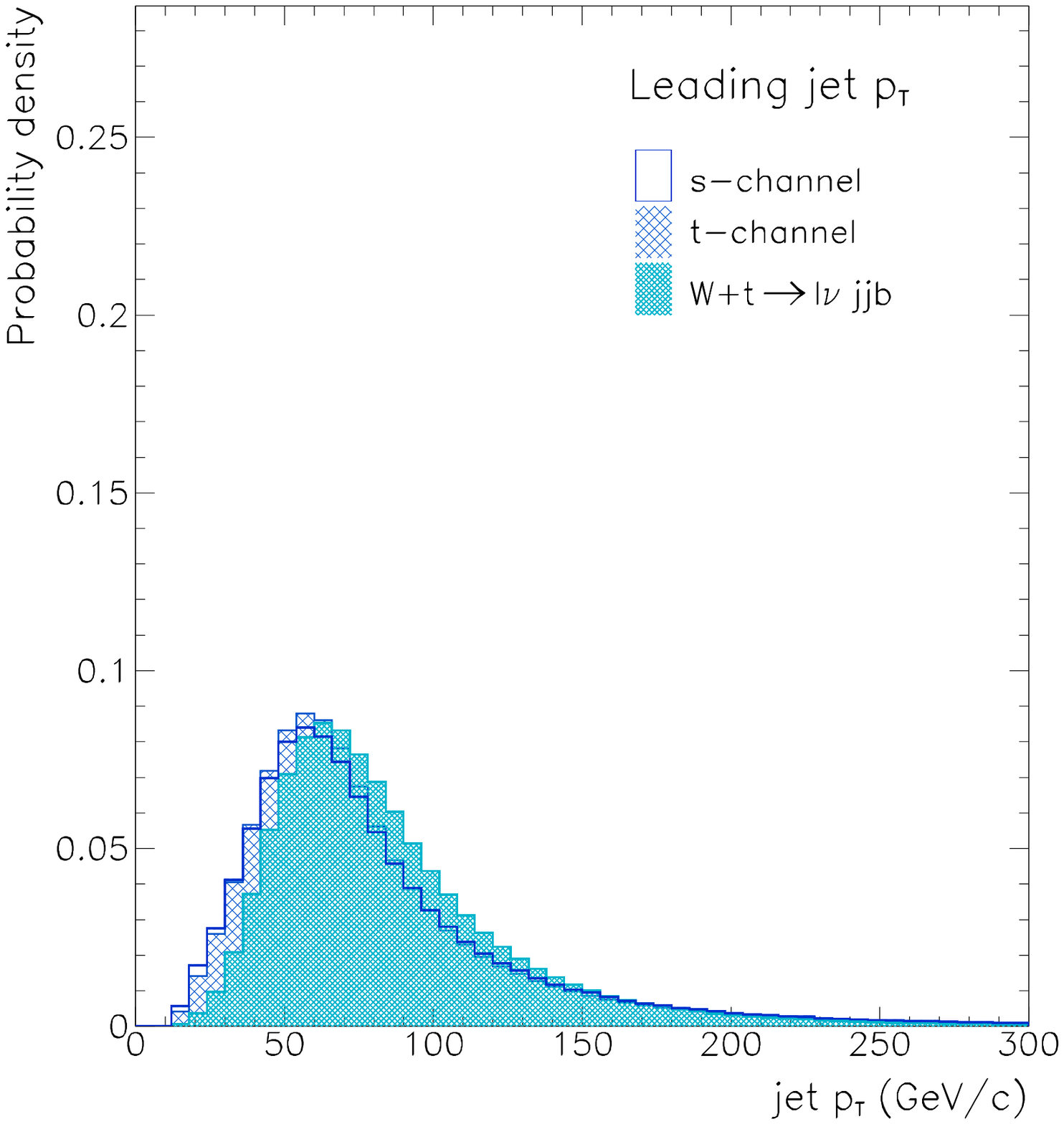}{Transverse momentum for 
the leading jet for signal and backgrounds.}{PDF_JET_PT1}

\noindent

\subparagraph{Light jets}
\hfill\break\hfill\break
 A jet is identified as a group of clusters falling within a fixed cone algorithm 
defined with a radius of $\rm \Delta R=\sqrt{\Delta\Phi^2+\Delta\eta^2}=0.4$. 
In ATLFAST~\cite{ATL-PHYS-98-131}, jets are defined in the pseudo-rapidity range 
$\rm |\eta | \le 5.0$ 
with a \PT above 15~\GEV. The jet energy resolution is given by: 
$$\rm \sigma(E)/E = 50\%/\sqrt{E/GeV}\oplus 3\%~for~|\eta_{jet}| \le 3$$  
$$\rm \sigma(E)/E = 100\%/\sqrt{E/GeV}\oplus 7\%~for~|\eta_{jet}| \ge 3.$$
Distributions for the two highest~\PT jets also are shown in Fig.~\ref{PDF_JET_PT1}.
Those Figures show that \TTB events have harder 
\PT~spectra than the other processes, with average values of about 100~\GEV and 70~\GEV 
respectively for the leading and 2nd highest jet energy. Those values are respectively 80 and 
50~\GEV for all three single-top processes. For \WQQ and \WJETS events, the average energies   
are found at much lower values, around 35-40~\GEV for the leading jet and 20-30~\GEV for the 
second highest jet energy. 
\TRIFIG{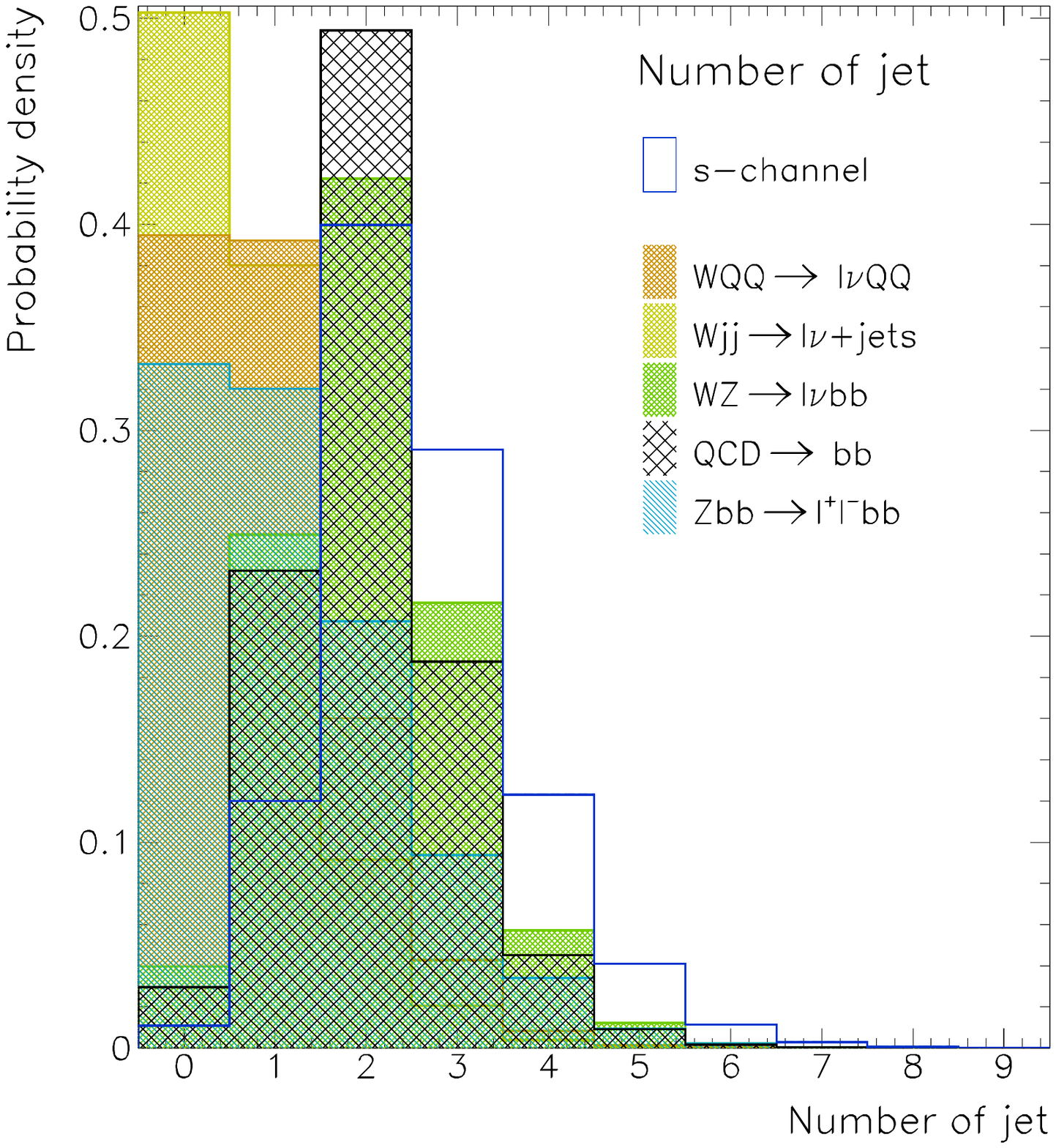}{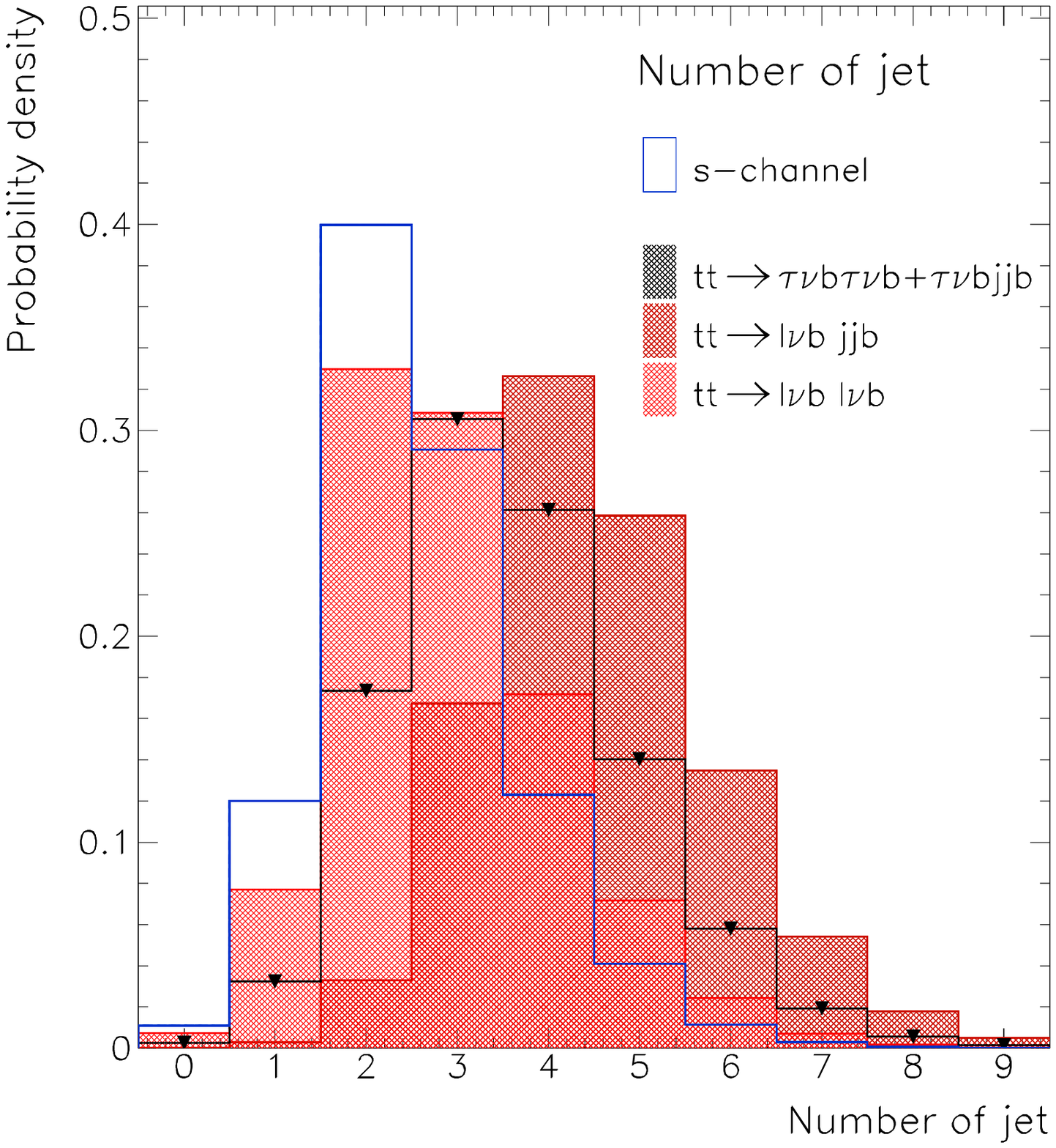}{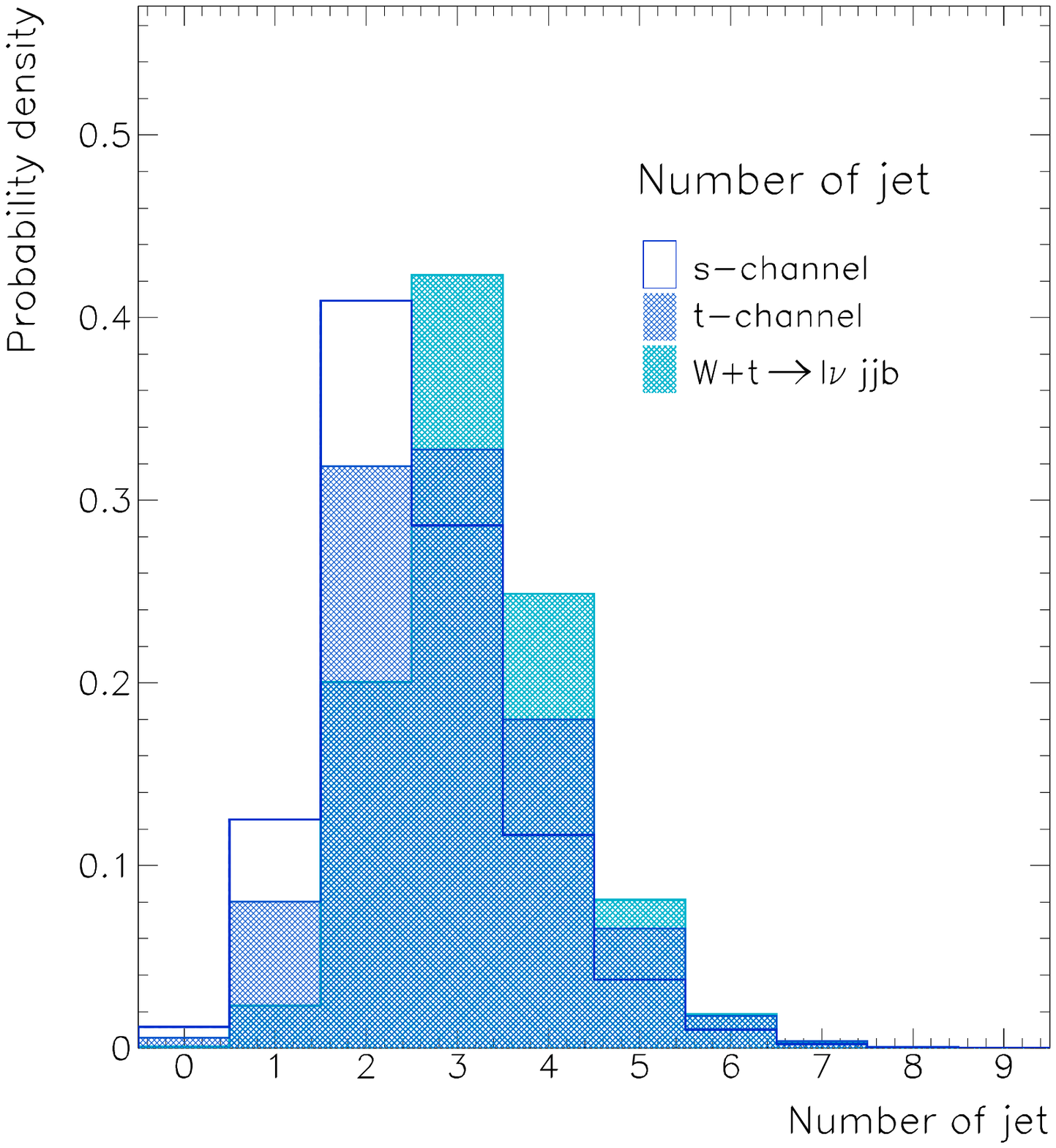}{Number of jets for signal and 
backgrounds.}{PDF_JET_NJET}

\noindent 
The preselection requires at least two jets above a threshold of 25~\GEV in order to reduce 
the QCD, W+jets as well as WZ/WW contamination. Again, the use of the full spectrum may 
revealed useful as an input to a likelihood function.
Jet multiplicity plays a crucial role in the discriminating the single-top s-channel  
from all backgrounds as shown in Fig.~\ref{PDF_JET_NJET}, where all jets above 15~\GEV 
are represented. About 40\% single-top s-channel events have exactly two jets and 70\% have 
two or three jets. Jet multiplicity is smaller for both \WJETS and \WQQ events with only 
about 30\% events reconstructed with more than one jet. On the contrary, more than four jets are expected in the "lepton+jets" and "tau+jets" 
\TTB events in about 70\% cases. 
Significant differences can also be seen among the three single-top production 
mechanisms. In the associated \WT sample, about 45\% events have exactly three jets, 
as expected from the hadronic decay of the W-boson associated to the Top quark.
In this sample, about 40\% events have more than three jets. In the W-gluon fusion  
events, the top decay gives a (b-)jet and a leptonic W as well as a b- and a non-b 
hadrons that can form eventually two extra jets. 

It thus appears that the analysis must be performed in bins of jet multiplicity. 
At the preselection stage, selected events are required to have exactly two or three 
jets above 15~\GEV with, among them, at least two above 25~\GEV. This requirement 
is crucial to reduce the \TTB contamination level. 
 
Two issues must be addressed at this stage. The use of NLO generators for both signal 
and backgrounds may affect significantly those results: it seems mandatory to 
use them as they become available so as to quantify the effects on selection efficiencies. 
The second issue concerns the gluon Initial State Radiation (ISR) and Final State 
Radiation~(FSR) modelling and its impact on the selected jet multiplicity: ISR affect 
crucially the number of jets that can be selected in the events while FSR have an impact 
on the jet energy due to the gluon emission in or outside the jet initiated by the 
parton. The selection efficiency thus depends closely upon the ISR and FSR modelling. 
These effects are adressed in Section~\ref{SYST} devoted to the estimates of the systematic 
uncertainties affecting the analysis. 
These two remarks emphasize the role of the jet definition: the choice of a cone algorithm 
with a larger radius ($\rm \Delta R= 0.7$ for eg.) or the use of a $\rm k_T$-algorithm to form 
the jet will affect the result of such analysis.

\subparagraph{b-tagged jets}
\hfill\break\hfill\break
 A jet can be identified as a b-jet only in the pseudo-rapidity range 
$\rm |\eta| \le 2.5$ corresponding to the tracking acceptance. 
In ATLFAST~\cite{ATL-PHYS-98-131}, the parametrization makes use of a combined tagging efficiency 
of 60\% for b initiated jets above 35~\GEV in \PT.  The corresponding mistag rate is 1\% 
for u,d,s quark jets (factor 100 rejection) and 10\% of tagging efficiency for 
c-quark jets. 
\TRIFIG{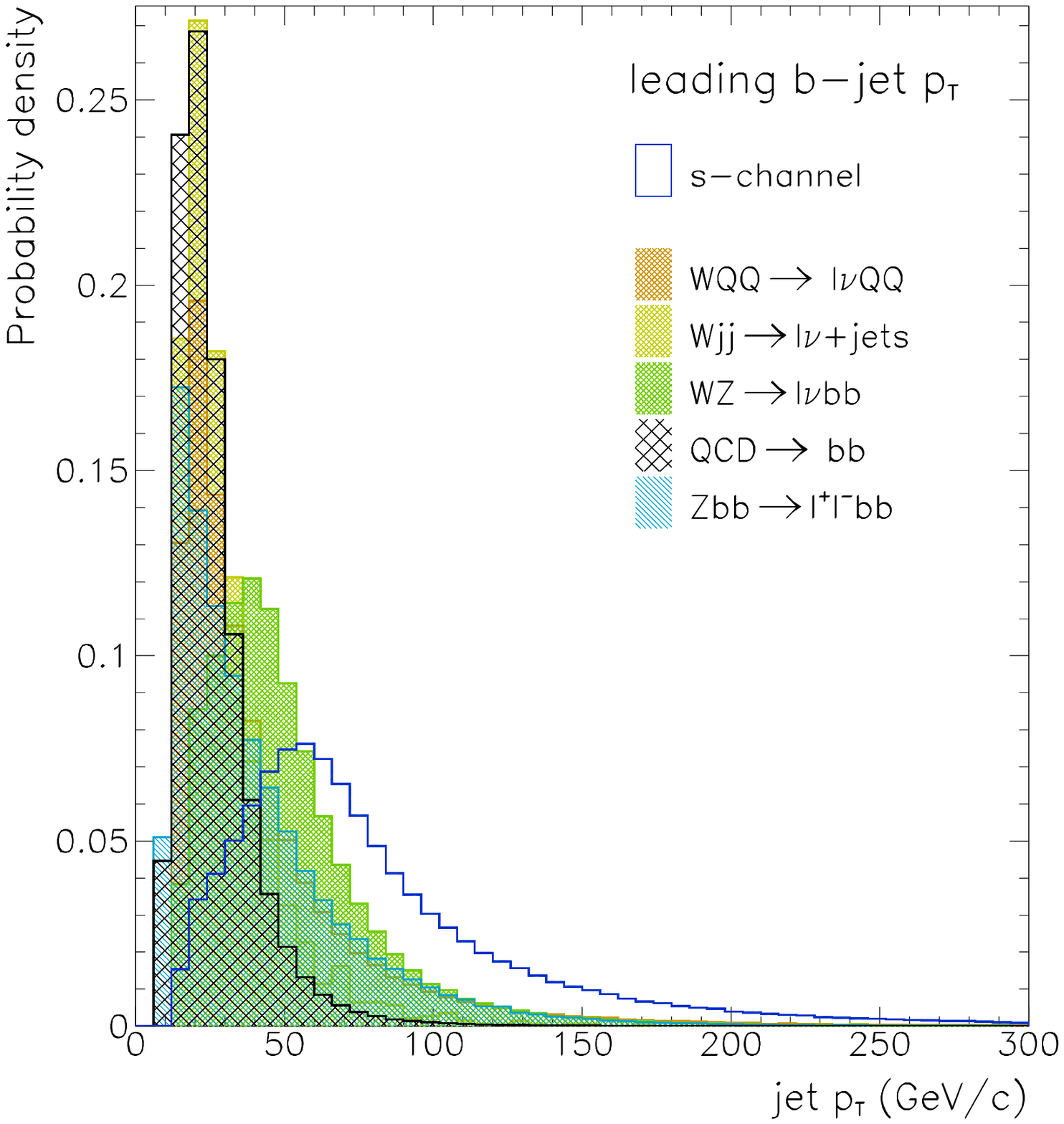}{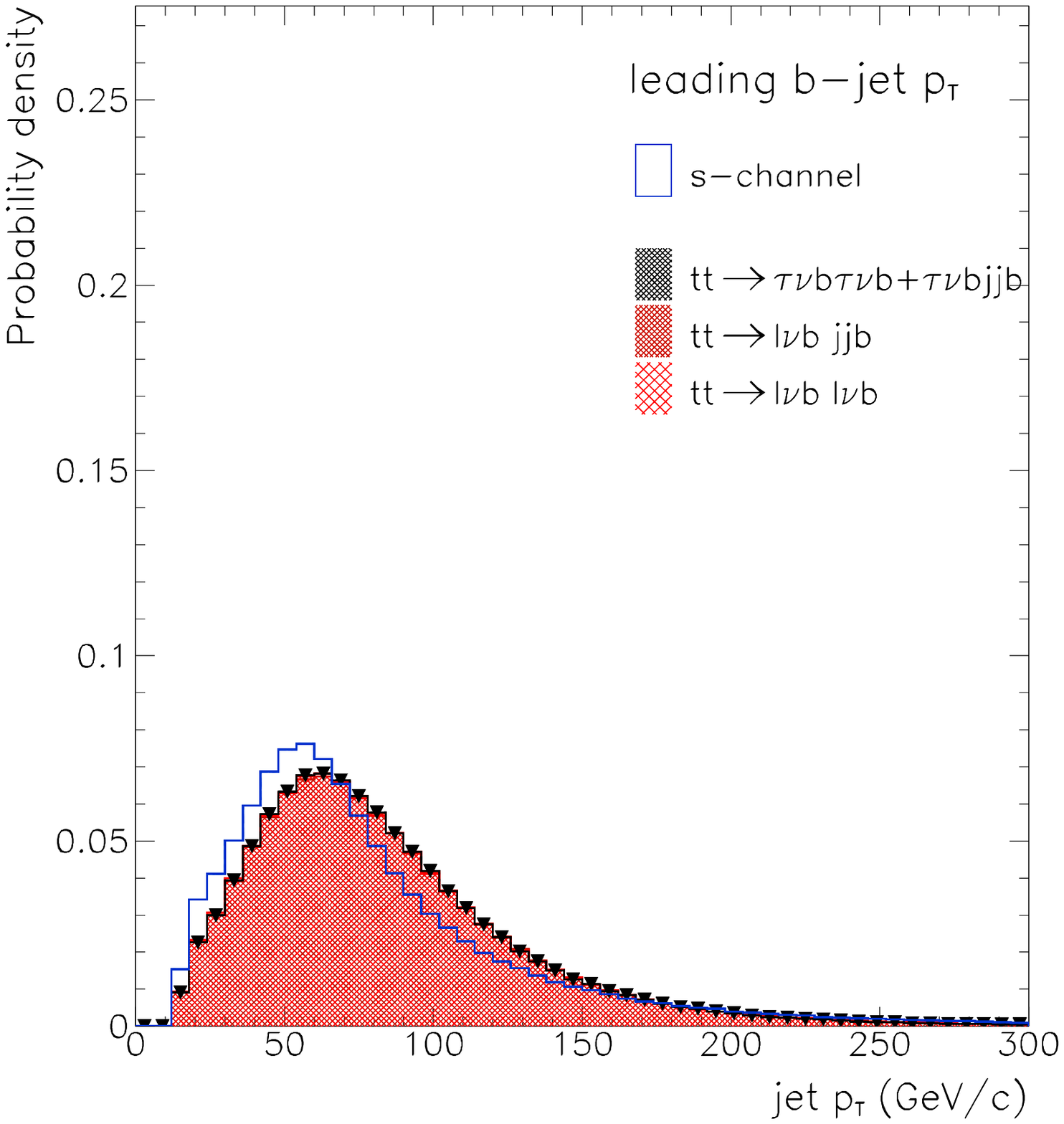}{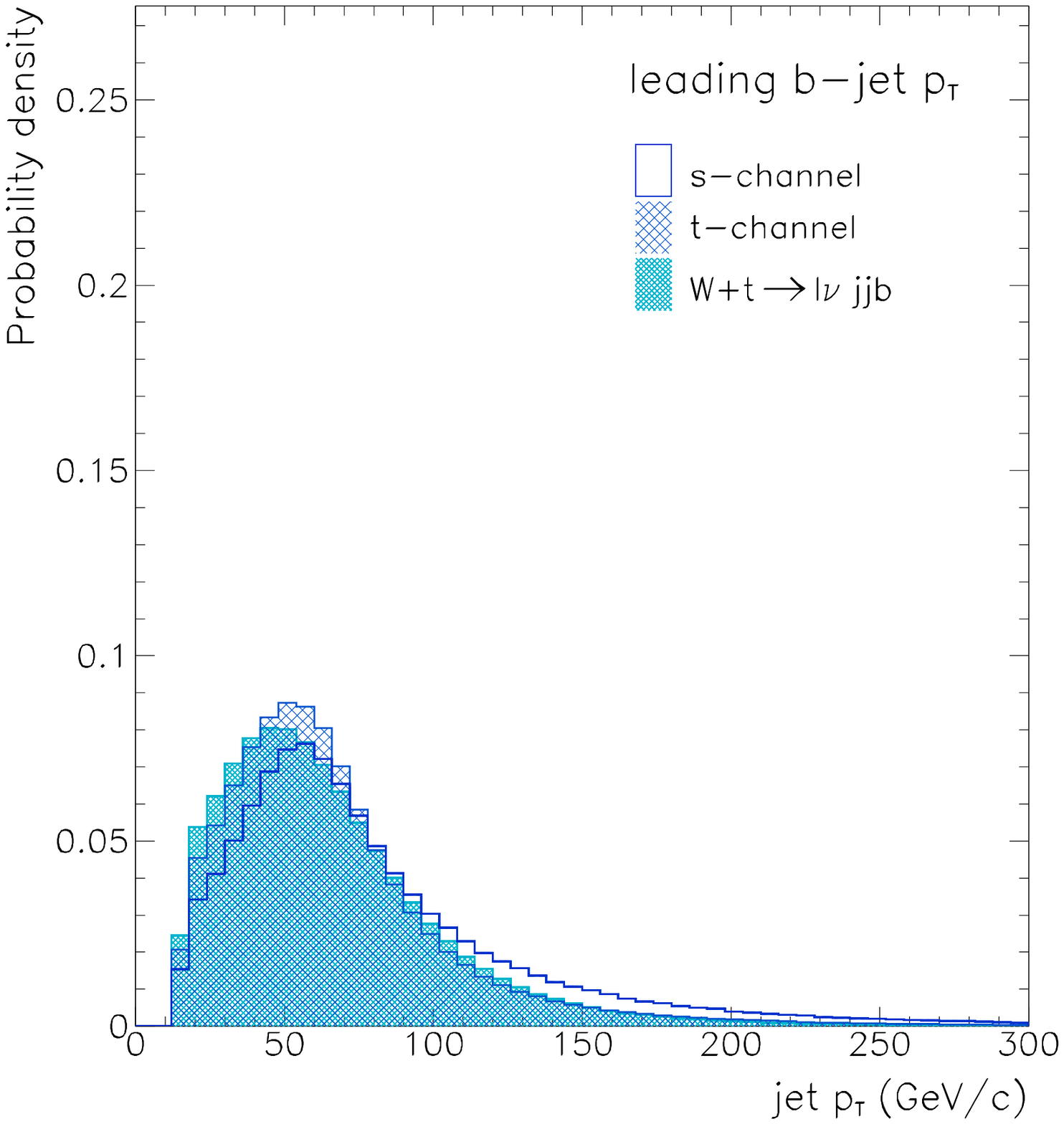}{Leading b-jet transverse 
momentum for signal and backgrounds.}{PDF_BJET_PT1}

\noindent
Figs.~\ref{PDF_BJET_PT1} and \ref{PDF_BJET_PT2} display the \PT distributions for the two 
leading b-tagged jets in signal and background events. The average is about 80~\GEV (resp. 
 40~\GEV) for the leading jet (resp. 2nd leading b-jet).
Jets present in QCD $\rm pp\to b\bar{b}$ events have a significantly softer spectrum than 
all others sources of backgrounds with an average value well below 30~\GEV (resp 29~\GEV). 
However, the cross section being several orders of magnitude, it is important to set the 
threshold above as high as possible to prevent from a high contamination. It has been 
checked that out of $\rm 5\times 10^6$ events, 17 events have 1 b-tagged jet (while none 
pass the 2-btag requirement) for a 30~\GEV threshold. This number falls to 11 at 40~\GEV 
and 7 at 50~\GEV. This gives confidence that a 35~\GEV threshold is relevant for our selection.  
\WQQ events also contain softer b-tagged jets (the same holds for c-jets in W+jets sample) 
than single-top events, as well as b-jets originated from Z decays, with an average \PT of 
60~\GEV (resp. below 30~\GEV). A high threshold in the highest b-jet \PT can therefore 
help reject significantly the QCD and W+jets background. 
A looser cut may be applied to the 2nd b-jet to further eliminate remaining \WQQ events.   
\TRIFIG{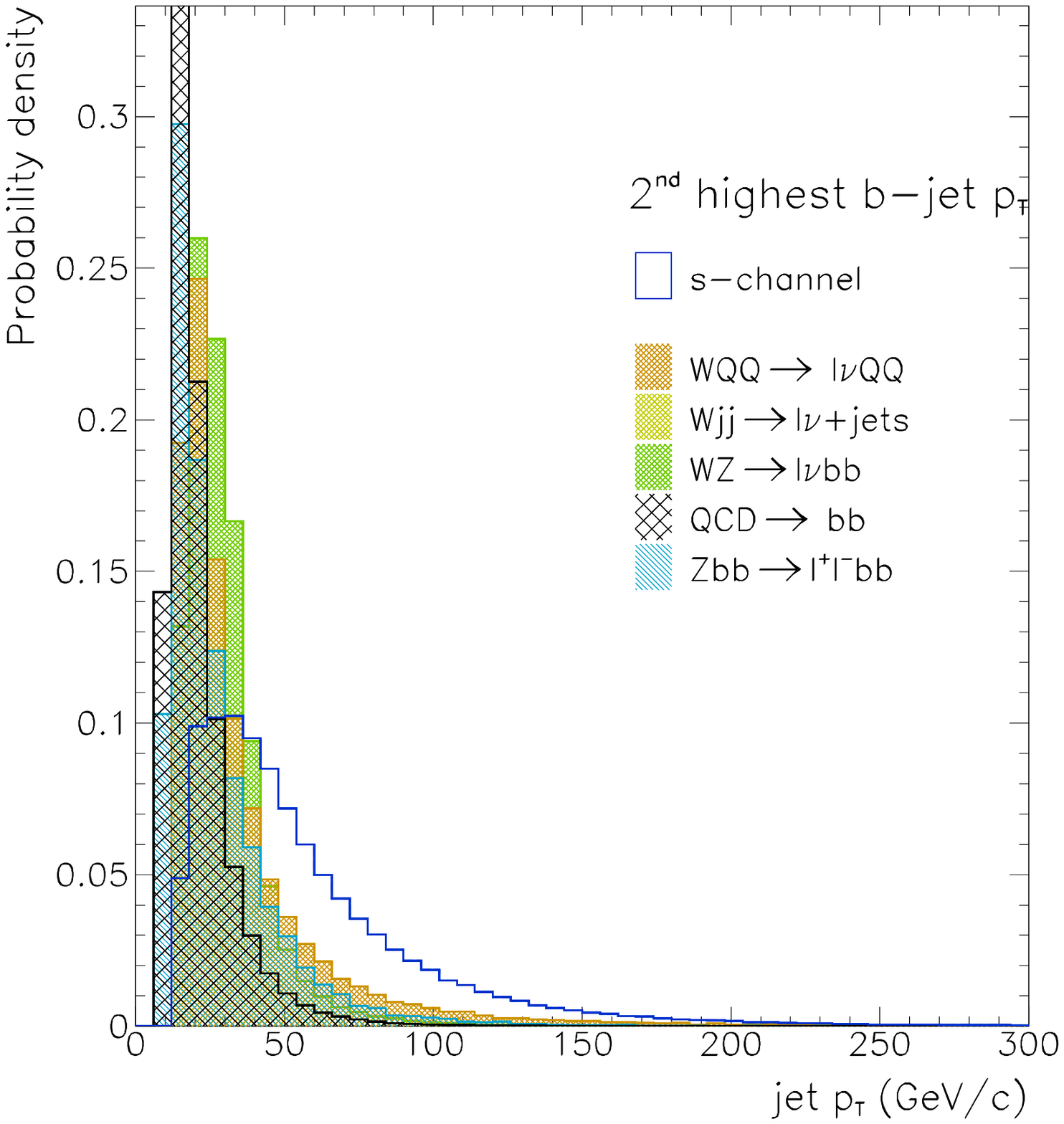}{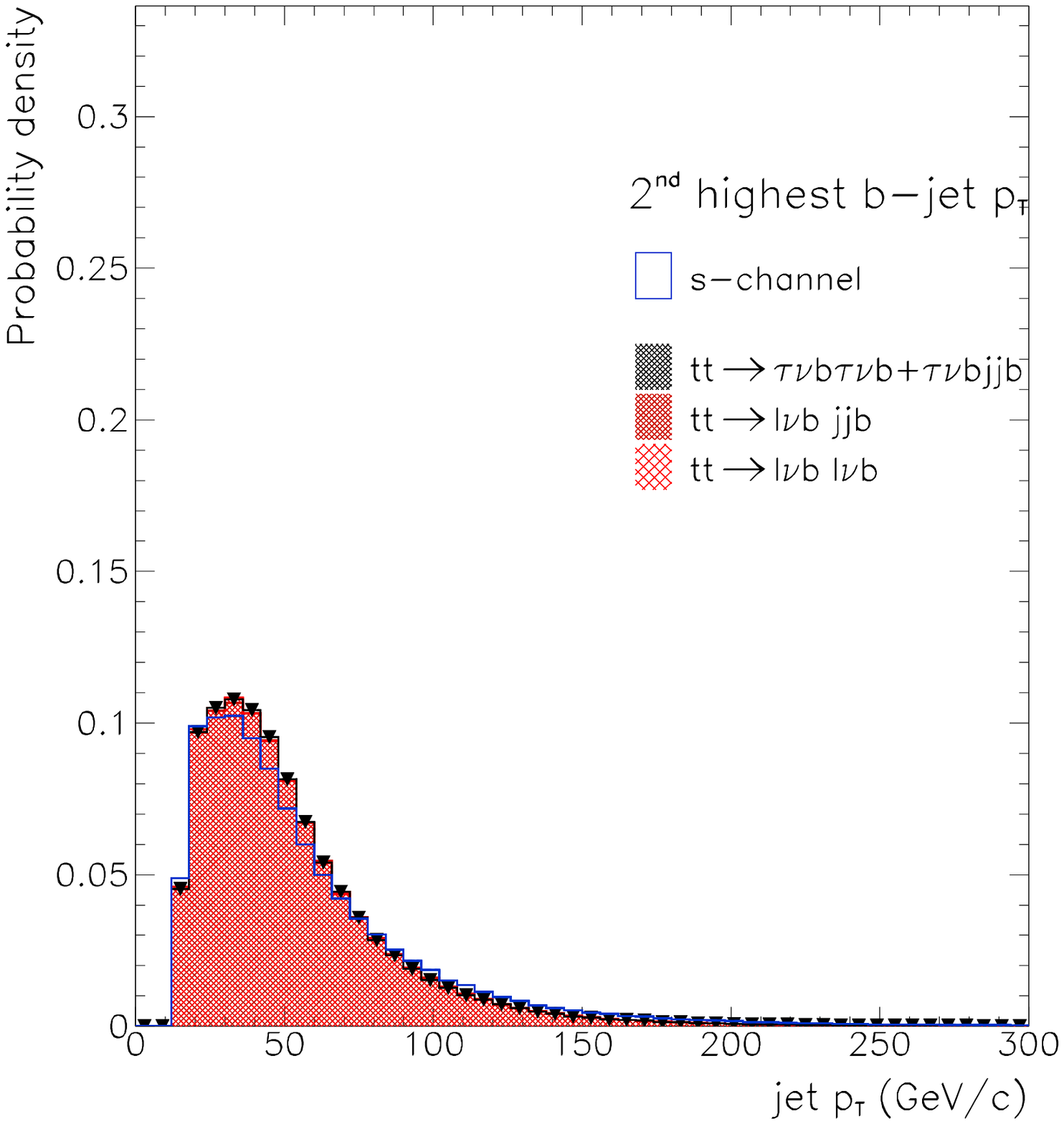}{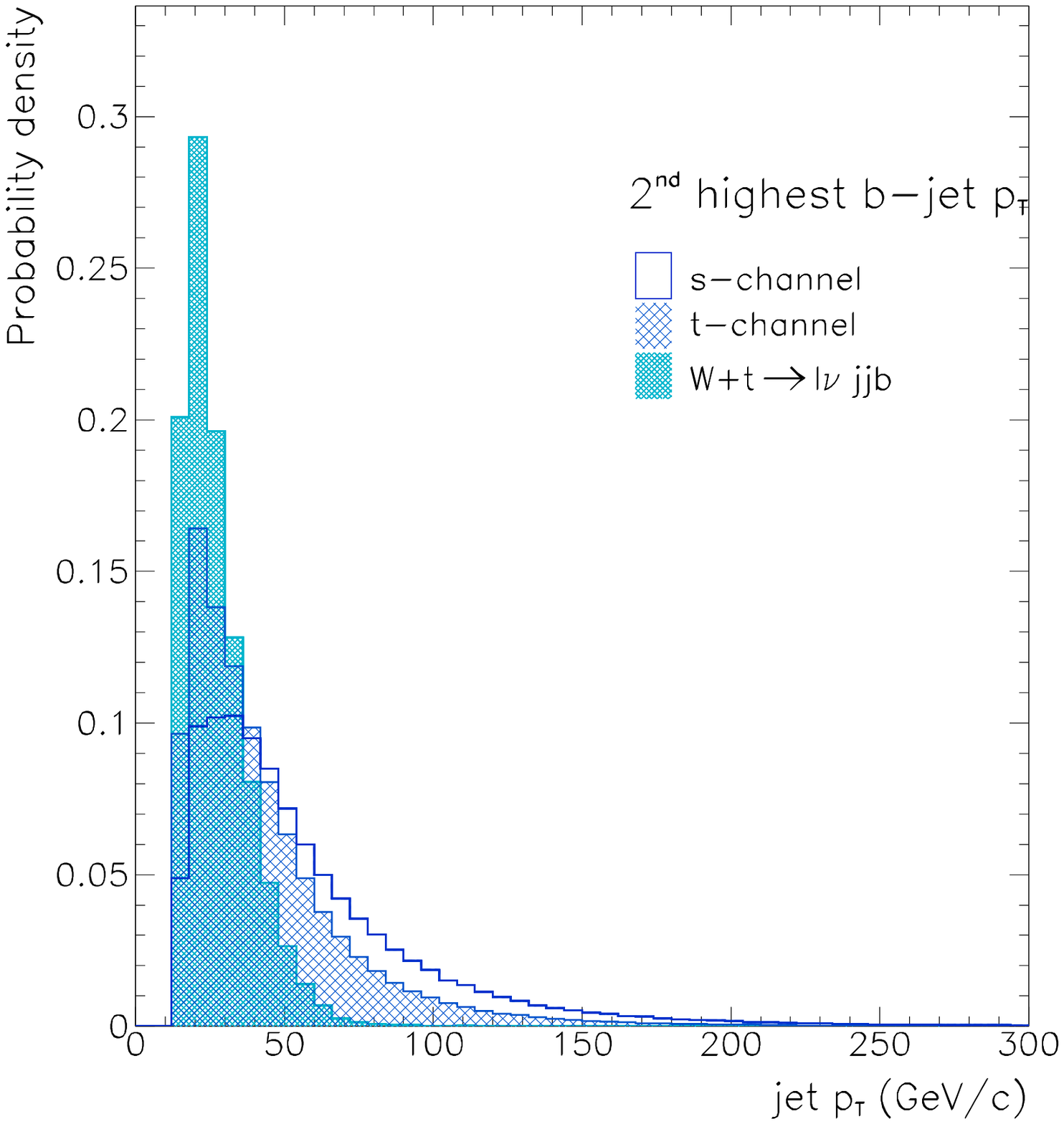}{2nd highest b-jet transverse 
momentum for signal and backgrounds.}{PDF_BJET_PT2}

\noindent
The expected number of b-tagged jet in the acceptance is shown in Fig.~\ref{PDF_BJET_NJET} 
for signal and all backgrounds. About 90\% \WQQ events have only one b-tagged jet, the other 
being either out of acceptance or below the \PT threshold. No QCD events out of 5,000,000 
pass the requirement on two b-tagged jets. The situation is dramatically 
different in \TTB and \SCH events which both contain more than 13\% events with two 
b-tagged jets. Requiring more than one b-tagged jet is therefore mandatory to improve the rejection of QCD 
and W+jets backgrounds. 
Regarding the two other single-top mechanisms, the number of b-tagged jets is 
not as high as for the s-channel events. If one indeed expects two b-hadrons 
in the \TCH channel, the second b-jet is missed in a significant fraction of time 
because the b parton is produced along the beam pipe, mostly outside the tracking 
acceptance and with a low \PT, as can be seen in Fig.~\ref{PDF_BJET_PT2}. The 
probability to see a second b-tagged jet in this sample is less than 15\%. For 
\WT events, no second b is expected, which results in more than 97\% events with 
only one b-tag, the remaining 2nd b-tagged jet coming mostly from charm decay. 

 The b-jet multiplicity strongly depends upon the b-tagging capabilities of the detector. 
A high efficiency and a low mistag rate will affect the discrimination against 
non-top background, making an impact in the analysis sensitivity. Section~\ref{SYST}
will adress the effects of deviations from the nominal expected performance on the 
systematic uncertainties affecting the selection efficiencies.
\TRIFIG{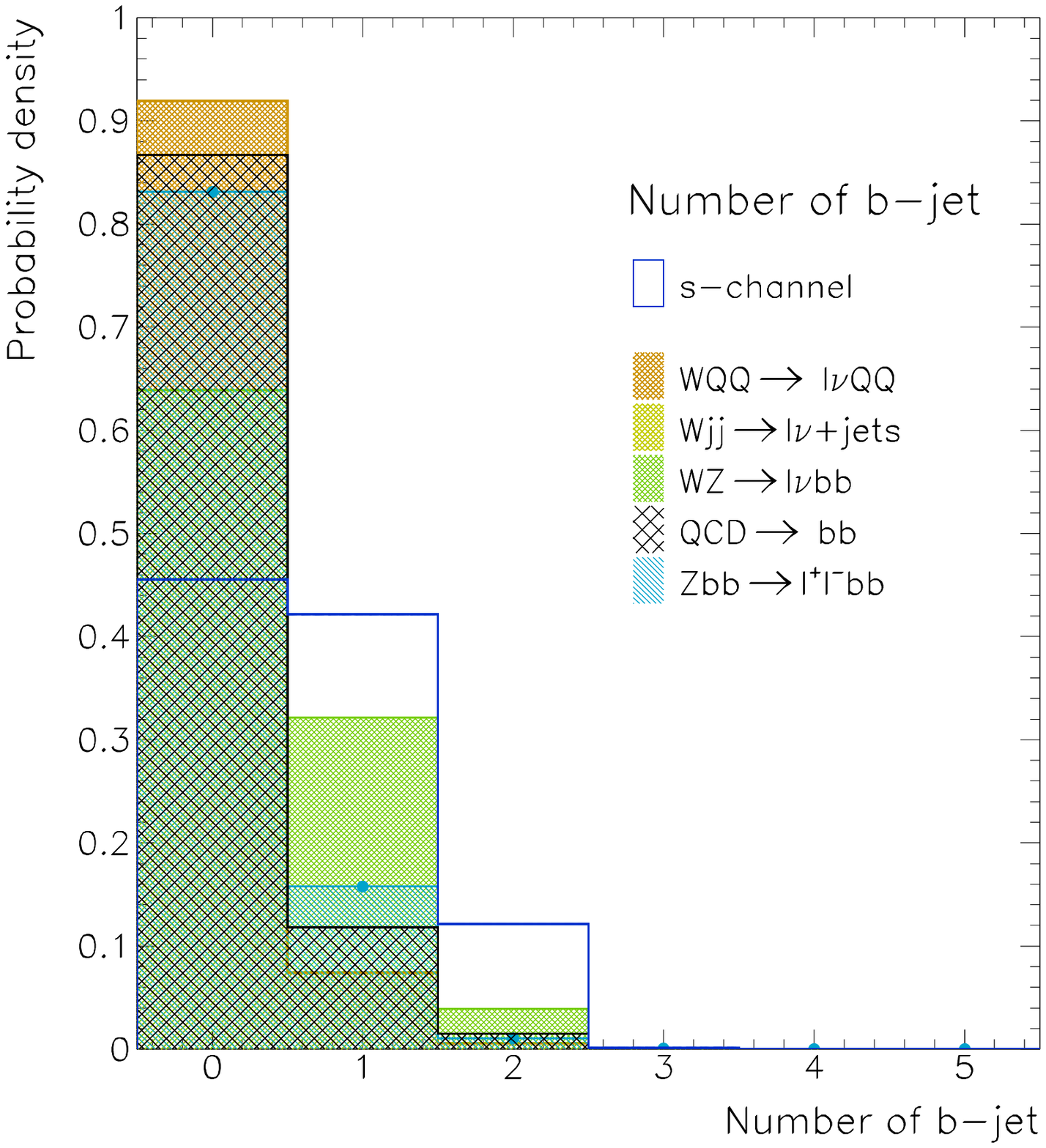}{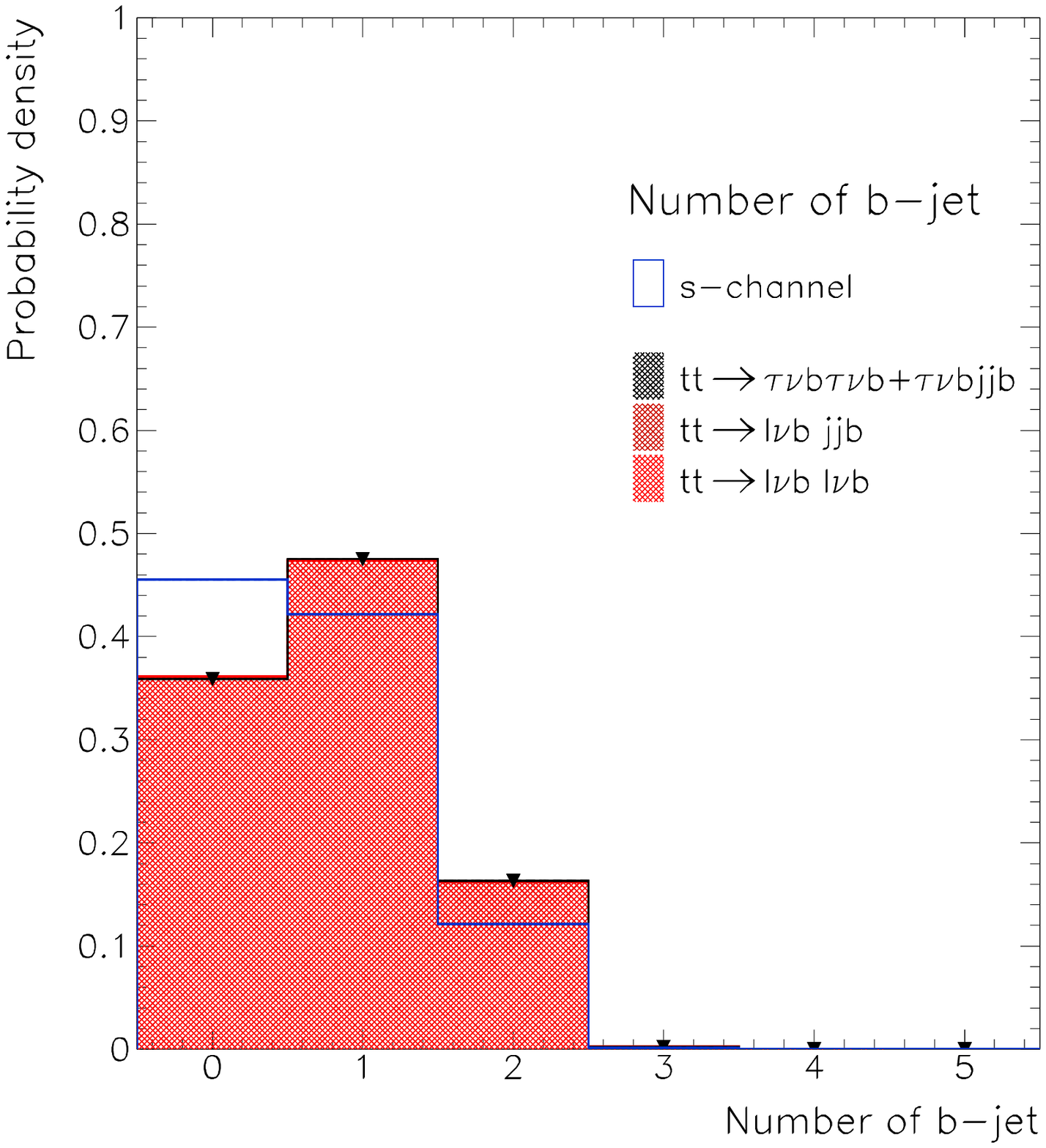}{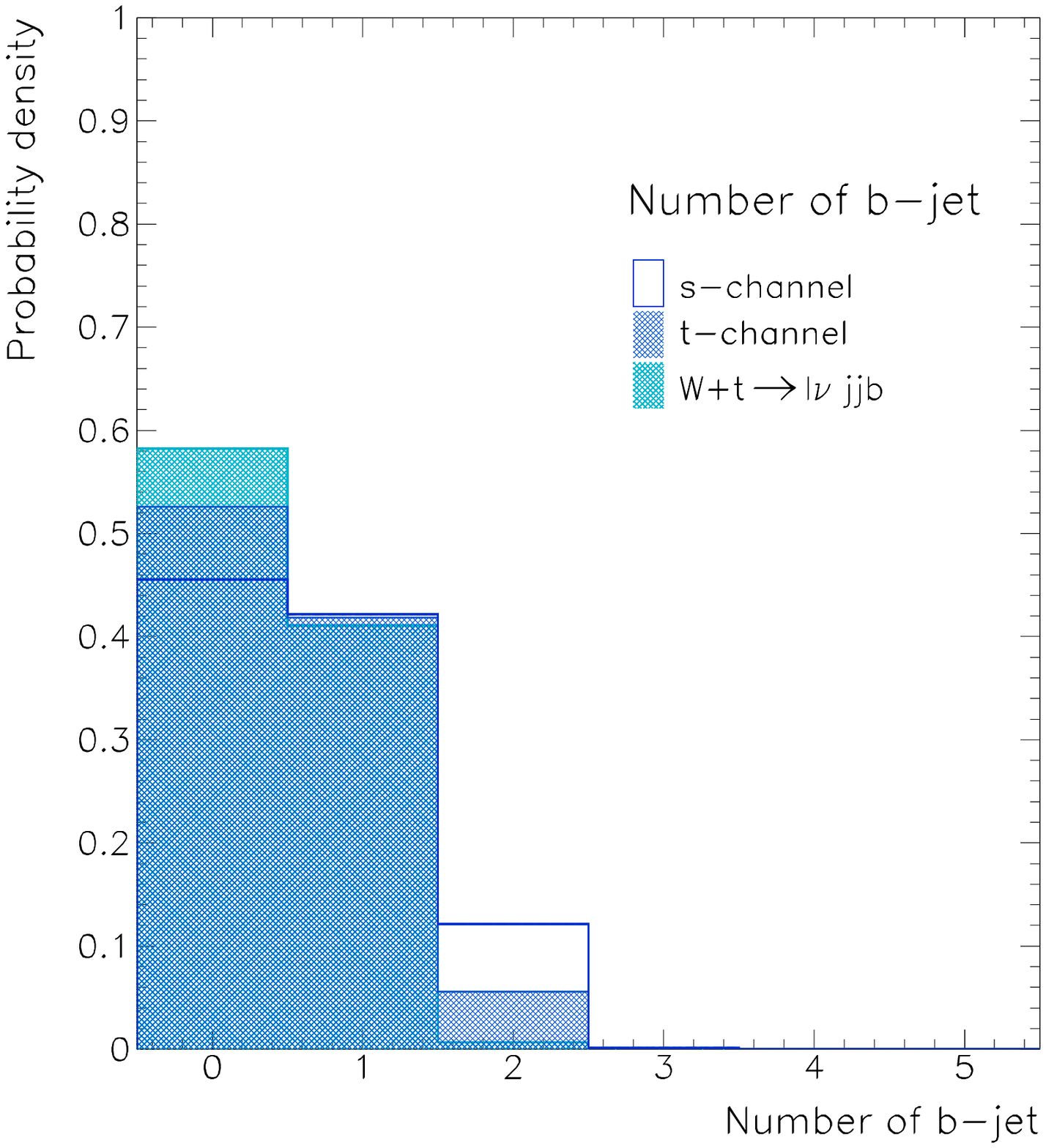}{Number of b-tagged jets for 
signal and backgrounds.}{PDF_BJET_NJET}

\noindent

\subparagraph{Total transverse energy $\rm H_T$}
\hfill\break\hfill\break
The total transverse energy of the event is shown to have a significant 
discriminant power against both top pair and \WJETS production. While the 
\TTB events tend to contain harder jets, the latter are characterized by the 
presence of softer jets in the final state compared to those for the signal. 
One usually uses the scalar sum of transverse energy computed over jets as well 
as leptons and missing energy. Obviously this variable is correlated to the number 
of jets and therefore careful treatment must be applied. In our selection \HT 
is defined as:
$$\rm H_T = \Sigma_{jet} ~E_T^{jet} + E_T^l + mE_T.$$ 
Probability density for this quantity is represented in Fig.~\ref{PDF_HT} for 
signal and the various backgrounds. The \HT distribution peaks at around 180~\GEV 
for \WQQ events while the average value for the \SCH channel is about 230~\GEV. 
For \TTB events in the various channels the distributions peak around 300~\GEV. 
A window in \HT seems therefore  to bring a significant rejection power against 
both \WQQ and \TTB backgrounds.
\TRIFIG{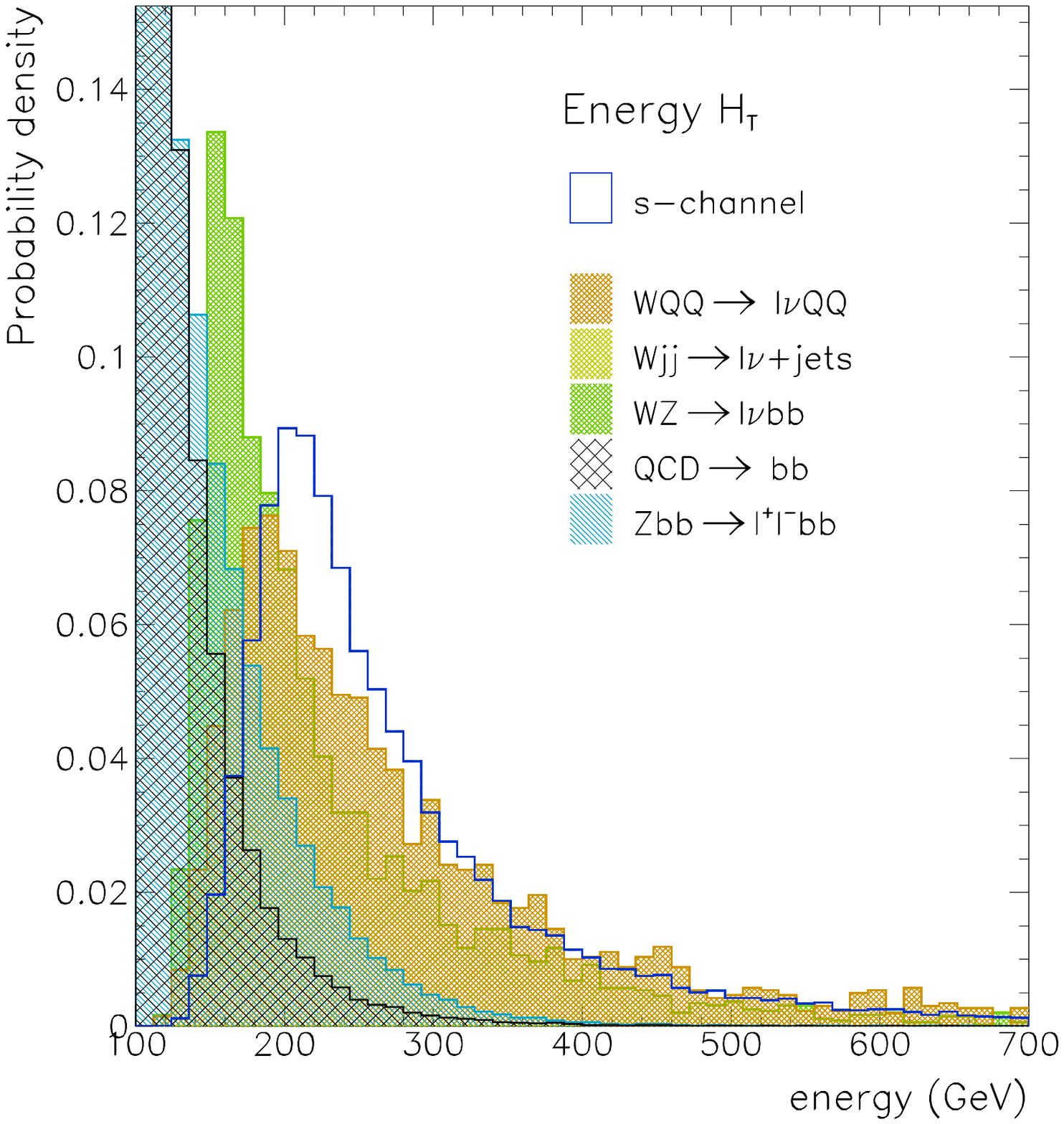}{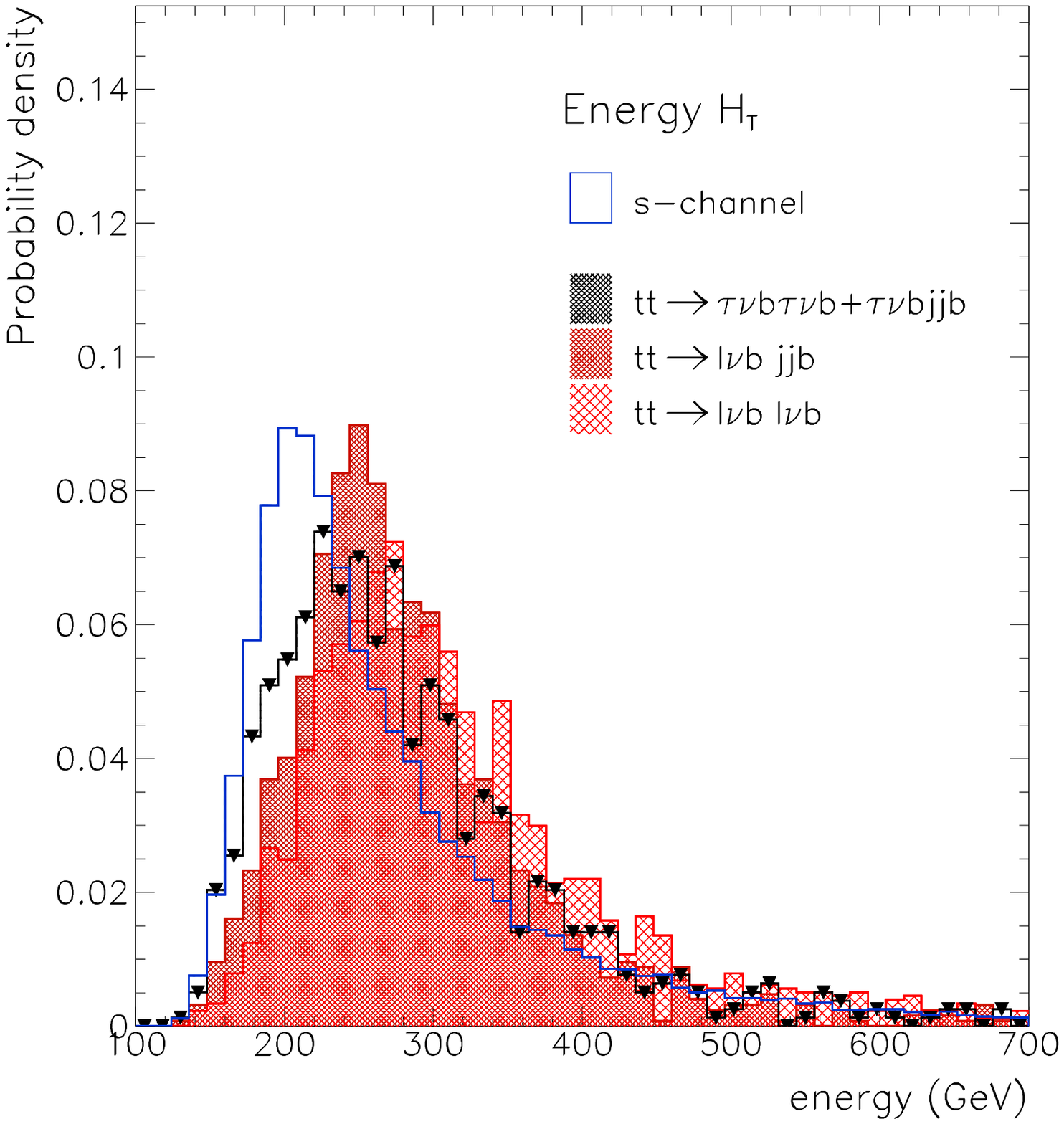}{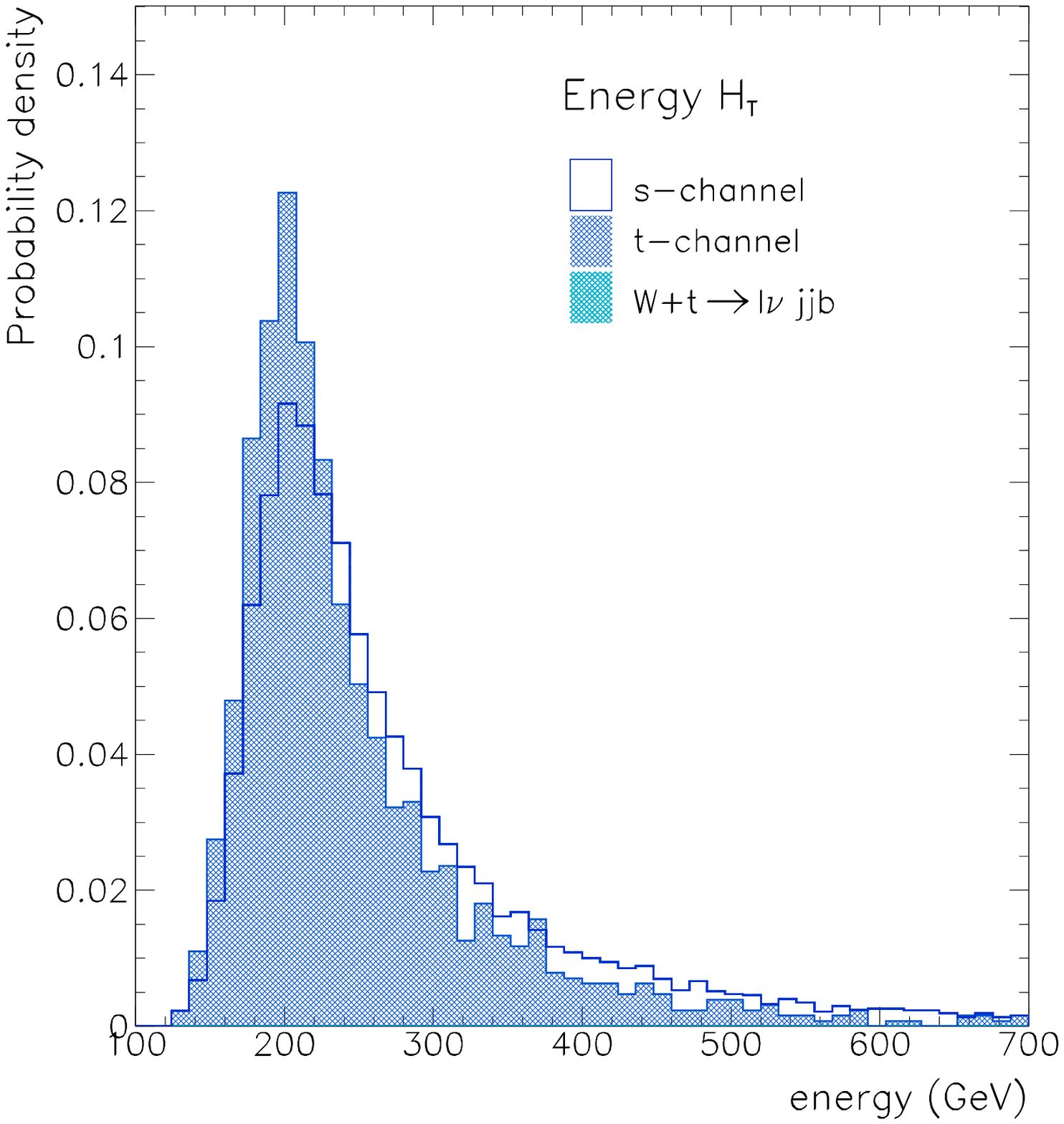}{Distribution of the energy \HT 
for signal and backgrounds.}{PDF_HT}

\noindent

\subparagraph{Reconstructed Top mass $\rm m_{t}$}
\hfill\break\hfill\break
With the LHC statistics, one can consider reconstructing a top mass 
from its decay products in order to reduce further the non-top background 
contamination of the selected sample. In our case where the W-boson decays 
leptonically, one faces an ambiguity arising from the determination of the 
neutrino longitudinal momentum: while the neutrino transverse energy can be inferred 
from the transverse missing \ET, the longitudinal momentum is unknown. It is however 
possible to obtain the $\rm p_z^{\nu}$ by using the W-mass as a constraint. 
The longitudinal momentum can thus be written as:
$$\rm  p_z(\pm \nu) = {-b \pm \sqrt{b^2-4ac}\over 2a} $$ 
where :
$$\rm a = E^{2}(l)-p_{z}^{2}(l)~,~ b= -2\left( {m_W^2\over 2} + \vec{p_T(l)}.\vec{p_T({\nu})}\right).p_z(l)$$ 
and 
$$\rm c= E^2(l). p_T^2(\nu) - \left( {m_W^2\over 2} + \vec{p_T(l)}.\vec{p_T(\nu)}\right)^2$$
Usually the twofold ambiguity is lifted by chosing the solution that gives 
the lowest $\rm p_z^{\nu}$. In our case though we choose not to apply this criterium 
but apply a criterium at a later stage of the selection. One has to notice that this 
method may have no solution: this corresponds to events where the transverse reconstructed W mass 
is larger than the W boson mass due to resolution effects. In this case 
we keep the real part of the solution, following the D\O~prescription.
\TRIFIG{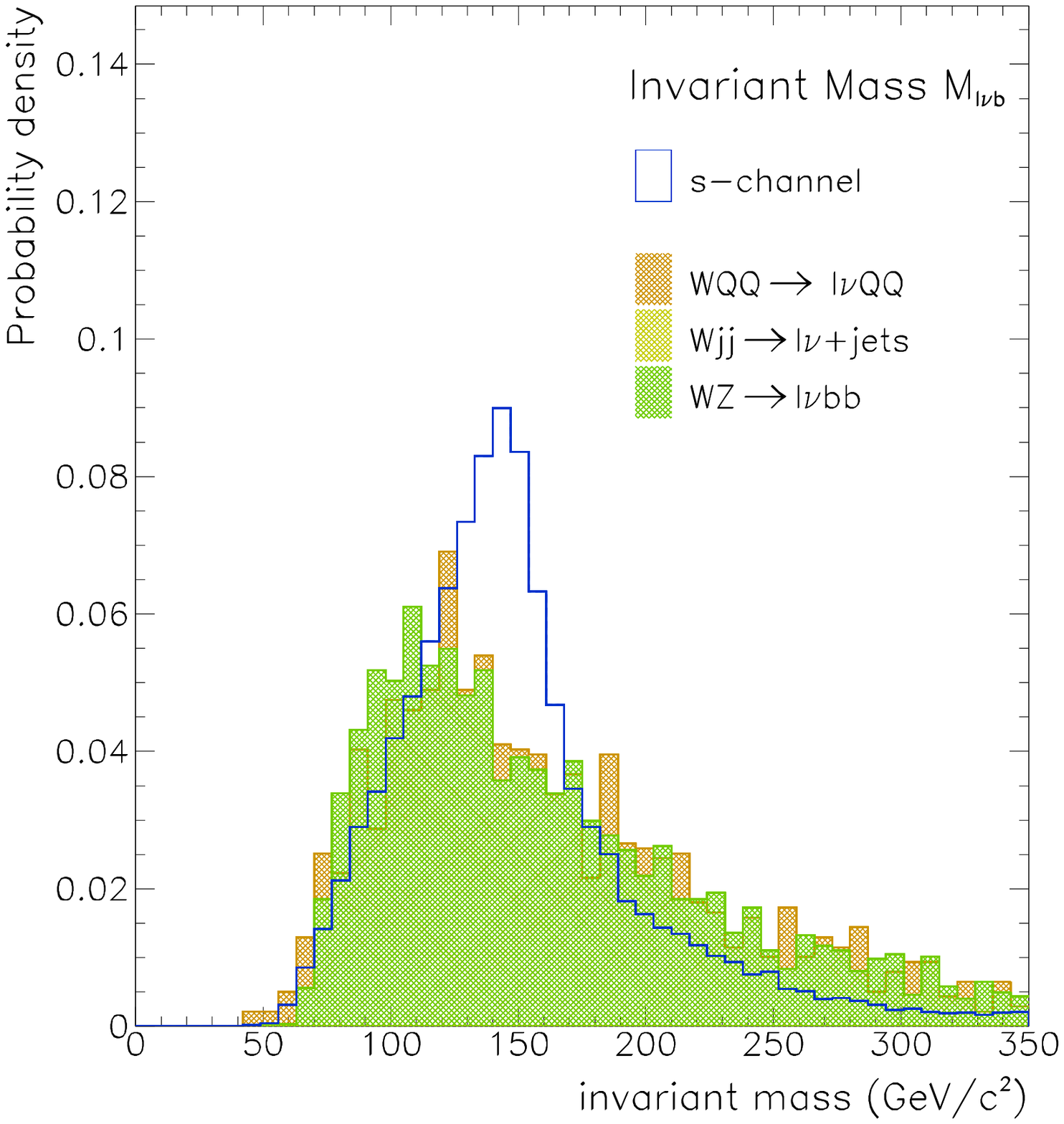}{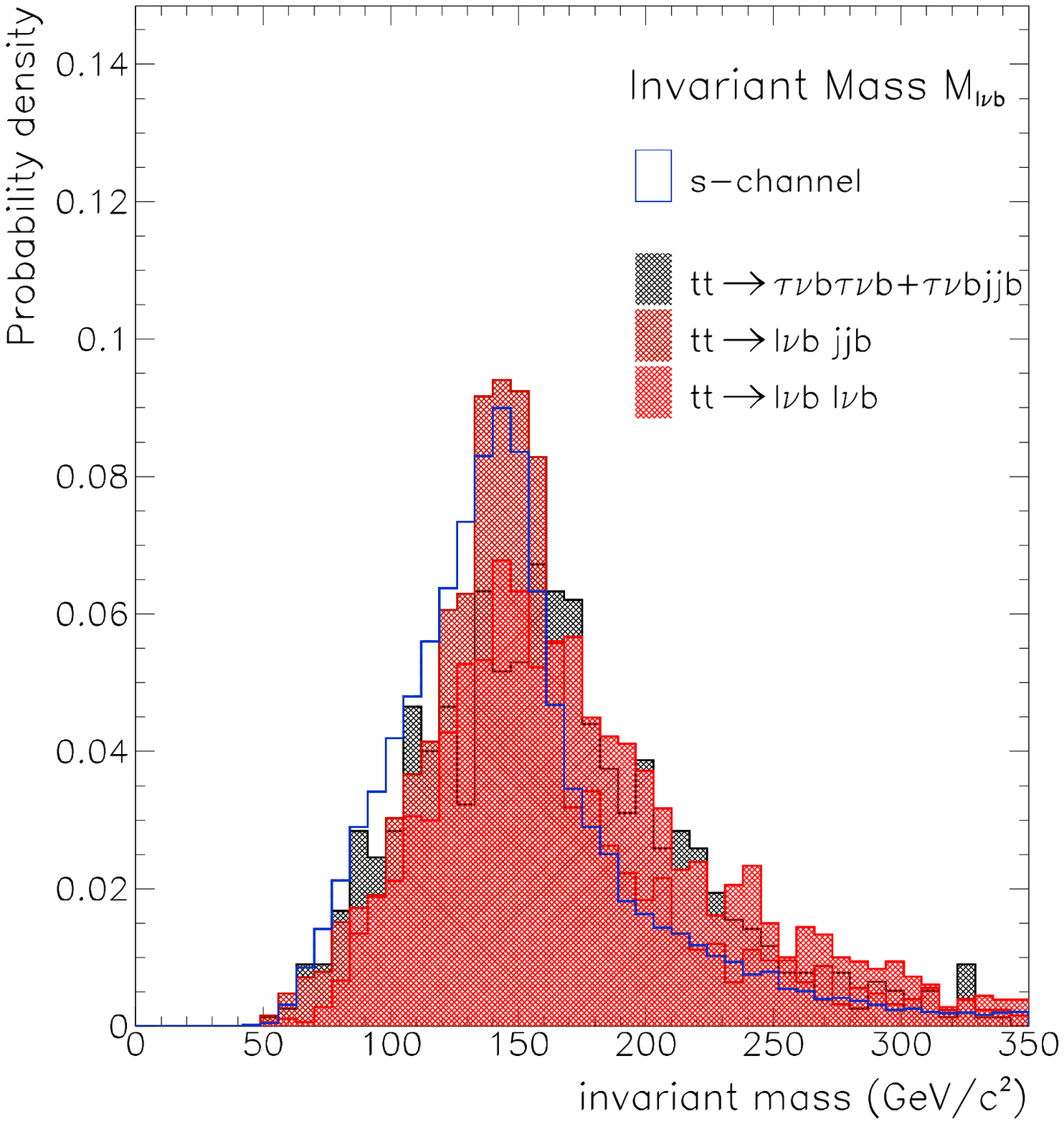}{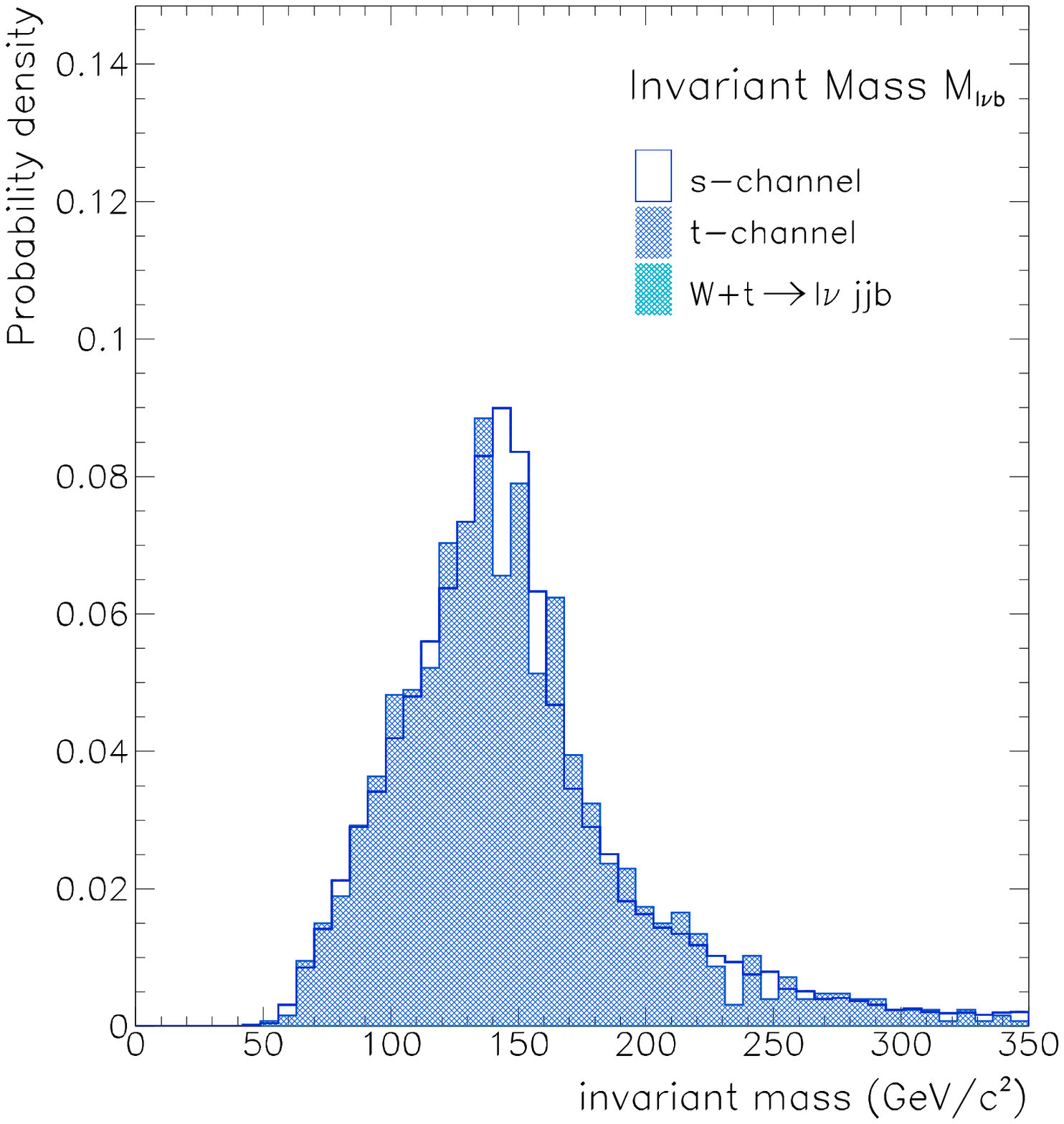}{Distribution of the 
reconstructed top mass for signal and backgrounds.}{PDF_MT}

\noindent
Once the solutions to $\rm p_Z^{\nu}$ are found there are four possible combinations 
to reconstruct the top quark momentum and mass: two depending on the neutrino solution 
and two due to the presence of the b-tagged jets. We choose to keep the solution leading 
to the highest \PT top~\cite{SGTOP-LHC-NOTE-TOPMASS}. Figs.~\ref{PDF_MT} show the probability 
densities associated to the reconstructed top mass for signal and the various backgrounds,

\paragraph{s-channel cross section measurement}
\hfill\break
\hfill\break
The measurement of the s-channel may appear as the most delicate of the 
three main single-top processes, because of its relatively low 
cross section compared to the two others. It is however one of the most 
interesting because the production of tb final state events is directly 
sensitive to contributions from extra W bosons or charged Higgs bosons 
as predicted in two Higgs doublet model (2HDM) ot type II~\cite{Gunion:1989we}.
The present analysis is extensively detailed in Ref.~\cite{SGTOP-LHC-LUCOTTE}.

\subparagraph{Preselection}
\hfill\break\hfill\break
In ATLAS, the s-channel analysis is based upon the following criteria: 
selected event must have at least one high-\PT lepton in the central region 
with a \PT above 25~\GEV and a total transverse missing energy above 25~\GEV. 
The event must pass a secondary lepton veto cut, applied to any lepton above 
10~\GEV with a sign opposite to that of the selected high~\PT lepton. The sign 
of the high~\PT lepton is used to determine the "flavour" of the final 
top and b-quark pair: a positive (negative) charge lepton signs a $\rm t\bar{b}$ 
($\rm \bar{t}b$) final state.
\begin{table}[phtb]
 \begin{center}
  \begin{tabular}{lrr}
    \hline
    processes      &  $\rm t\bar{b}$ final state  & $\rm \bar{t}b$ final state \\[1mm]
    \hline\hline
    s-channel      &           $1,200\pm 7$       &  $840\pm 4$                \\[1mm]
    t-channel      &           $1,860\pm 35$     & $1,120\pm 20$             \\[1mm]
    W+t channel    &           $\le$8             &   $\le$5                    \\[1mm]
        \hline
$\rm t\bar{t}$ background &                       &                       \\[1mm]
$\rm tt\to e\nu b~jjb$    & $2,220\pm 75$         & $2,220\pm 75$               \\[1mm]    
$\rm tt\to e\nu b~e\nu b$
                   &           $2,790\pm 40$      &   $2,790\pm 40$            \\[1mm]
$\rm tt\to \tau\nu b,\tau\nu b$
                   &             $360\pm 28$      &  $360\pm 28$                \\[1mm]
$\rm tt\to \tau\nu b,jjb$
                   &              $60\pm 10$      &   60$\pm 10$                \\[1mm]
    \hline
Z/W+jets background  &                              &                             \\[1mm]
WQQ                &           $2,250\pm 50$      & $1,410\pm 30$               \\[1mm]  
$\rm Wjj\to e\nu,jj$&          $1,710\pm 170$     &  $1,260\pm 120$              \\[1mm]
$\rm WZ\to e\nu b\bar{b}$&              $90\pm 10$&    $60\pm 5$                \\[1mm]
$\rm Zb\bar{b}\to e^+e^-b\bar{b}$ & $7\pm 3$      &   $7\pm 3$ \\[1mm]
\hline
  \end{tabular}
  \caption[]{Number of pre-selected events in the "2b0j" sample expected for an 
  integrated luminosity of 30~\FB. Uncertainties come from Monte Carlo statistics only}
  \label{TAB-PRESELECTION-NEVT}
 \end{center}
\end{table}

\noindent
The event must have exactly two or three jets above 15~\GEV. Among those, 
two must be above 25~\GEV. Finally, the events are then required to have, among those 
two or three selected jets, two b~tagged jets with a \PT above a threshold of 30~\GEV. 
Selected events are thus classified as "2b0j" (2 b-tagged jets and no extra 
light jet above 15~\GEV) or as "2b1j" events (2 b-tagged jets plus one extra light 
jet and no 4th jet above 15~\GEV). 
Note that the requirement of two b-tagged jets is crucial to reduce the 
contamination of W+jets events that have a cross section several orders of magnitude that 
of the signal. To a lesser extent, this is also true for \TTB events since among the 2-jet 
and 3-jet events, only a few of them have 2 b-tagged jets. 
Table~\ref{TAB-PRESELECTION-NEVT} reports the number of expected events with 30~\FB. 
About 1,200 (840) \SCH events are pre-selected in the $\rm \bar{t}b$ ($\rm t\bar{b}$) 
final states. 
The dominant background comes from the top pair production in the dilepton and 
"lepton+jets" channels, 
followed by the WQQ contamination. The remaining W+jets contamination is due to the 
high cross section for such events, and is expected, at this stage, to be slightly 
above the signal expectation.  The resulting S/B ratio is about 11\% (9\%) in the 
$\rm t\bar{b}$ ($\rm t\bar{b}$) final state. It is obvious that the combination of 
both final states is required to improve the sensitivity. 

For 2-jet samples, the signal efficiency is slightly above 3.0\%. No QCD events 
are selected out of $\rm 5\times 10^6$. Top pair events are selected with an efficiency 
below 0.1\% in the "lepton+jets" channel while tau+jets events are almost negligible. 
On the contrary "dilepton" (including "ditau") top pair events are selected with a 
higher efficiency ranging from 0.25 to 0.5\%. Overall, this results in an almost equal 
contamination originating from "lepton+jets" and "dilepton" channels, due to the difference 
in branching ratios. 
As expected, the \WQQ contamination is greatly reduced by a 2-b tag requirement with  
a 0.2\% selection efficiency. At the same time, only 1.2\% WW and WZ diboson 
events are selected. \WJETS events are removed because of the presence of non-b 
softer jets, with a final yield depending upon the mistag rate, for which we take 
to be equals to 1\% in the present analysis. 
Regarding the three-jet samples, the signal efficiency is about 1.9\%.  While the double 
tag requirement keeps the \WJETS contamination relatively low, the signal is swamped 
in the \TTB background with a much lower S/B below 1\%.
\DBLFIG{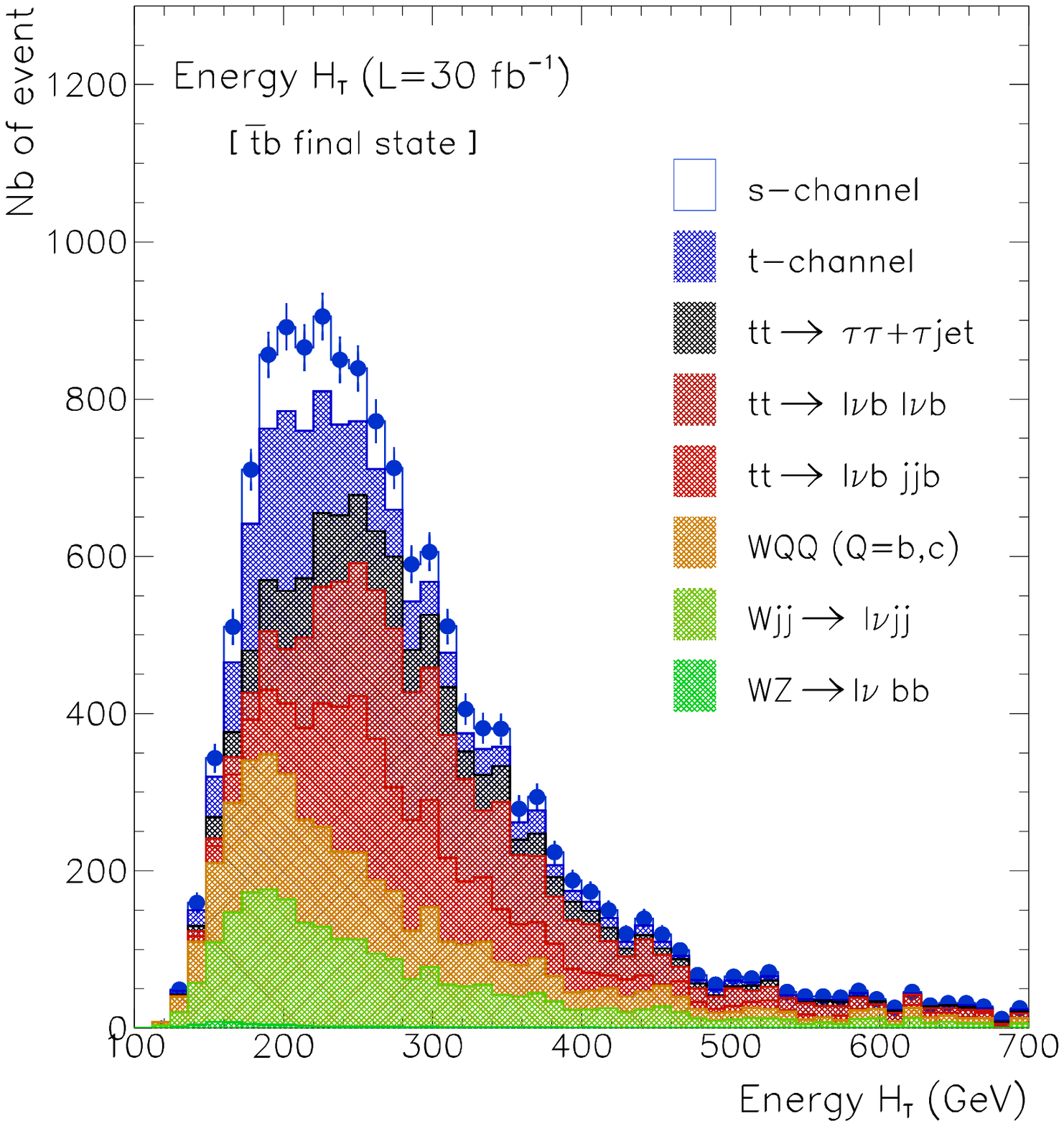}{Event yield for the $\rm H_T$ distribution for 30 $fb^{-1}$.}{FIG-PRESEL-HT}{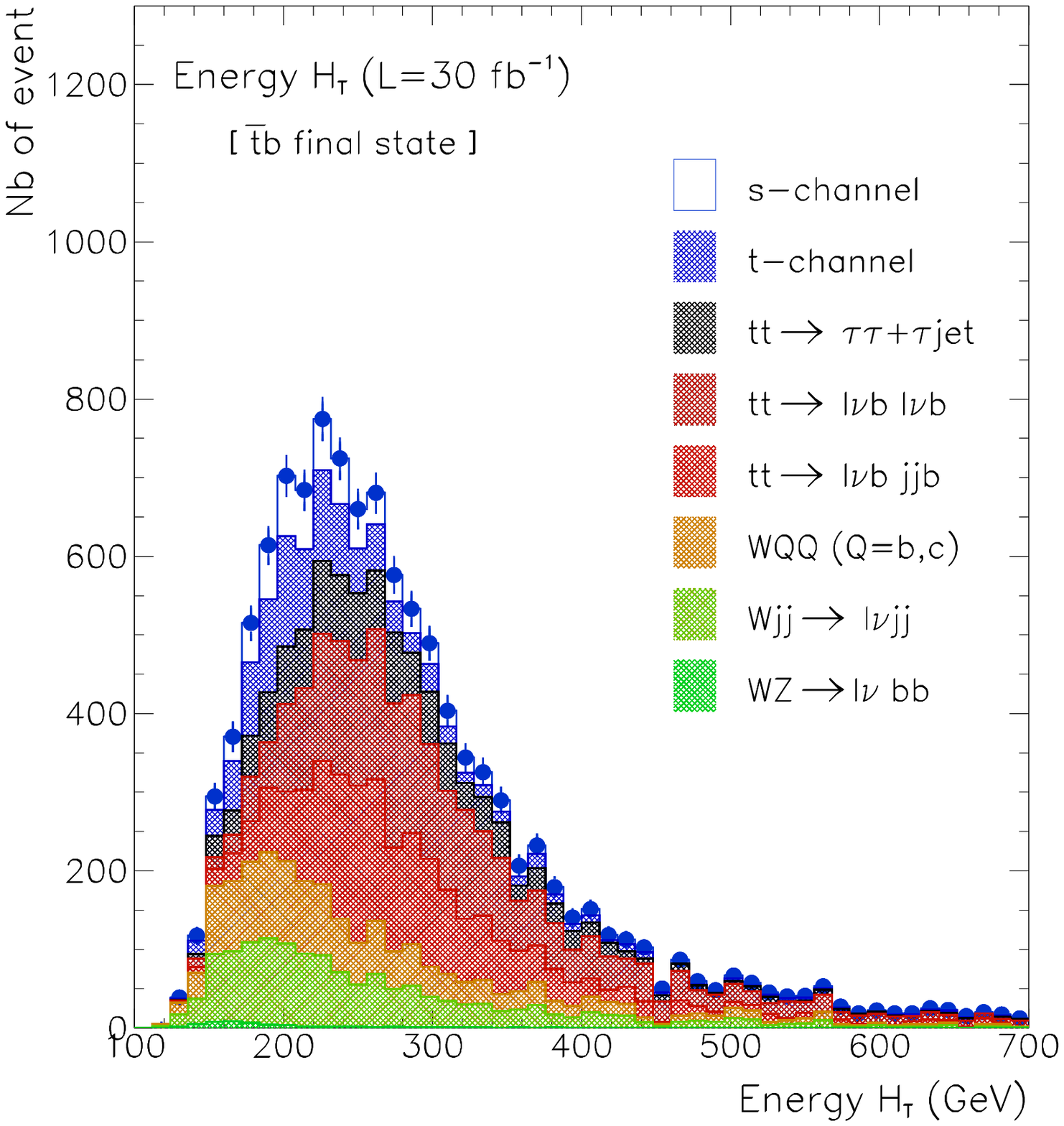}{Event yield for the $\rm M_{l\nu b}$ distribution for 30 $fb^{-1}$.}{FIG-PRESEL-MTOP}

\noindent
%
 Results have been interpreted as a function of the integrated luminosity.  
In 2-jet events, a $\rm 5~\sigma$ discovery requires about 5~\FB. 
The use of the 3-jet samples does not bring any significant improvement since at 
least 60~\FB are needed to reach the same yields. 
 The statistical sensitivity to the cross section measurements has also been evaluated 
from the ratio $\rm \sqrt{S+B}/S$ which provides the sensitivity of the signal to 
signal and background statistical fluctuations. 
A statistical sensitivity of 7\% can be achieved by combining both $\rm t\bar{b}$ and 
$\rm \bar{t}b$ final state analyses with an integrated luminosity of 30~\FB.
To reduce further the systematic uncertainties associated to the background 
estimates, we can choose to apply further requirements on topological variables. 
Figure~\ref{FIG-PRESEL-HT} displays the distribution of the total transverse energy 
of the events while Fig.~\ref{FIG-PRESEL-MTOP} shows the recontructed (leptonic) top 
mass after the preselection.

\subparagraph{HT window optimization}
\hfill\break
\hfill\break
After the pre-selection stage, the remaining sample is characterized by a low 
ratio signal over background of about 10\% with dominant backgrounds originating  
from the top pair and \WQQ production. In order to purify the sample, we apply further 
requirements based on the total transverse energy $\rm H_T$ measured in the event and 
on the top mass reconstructed from the b~jet and the 
leptonic decays of the W~boson. 
\DBLFIG{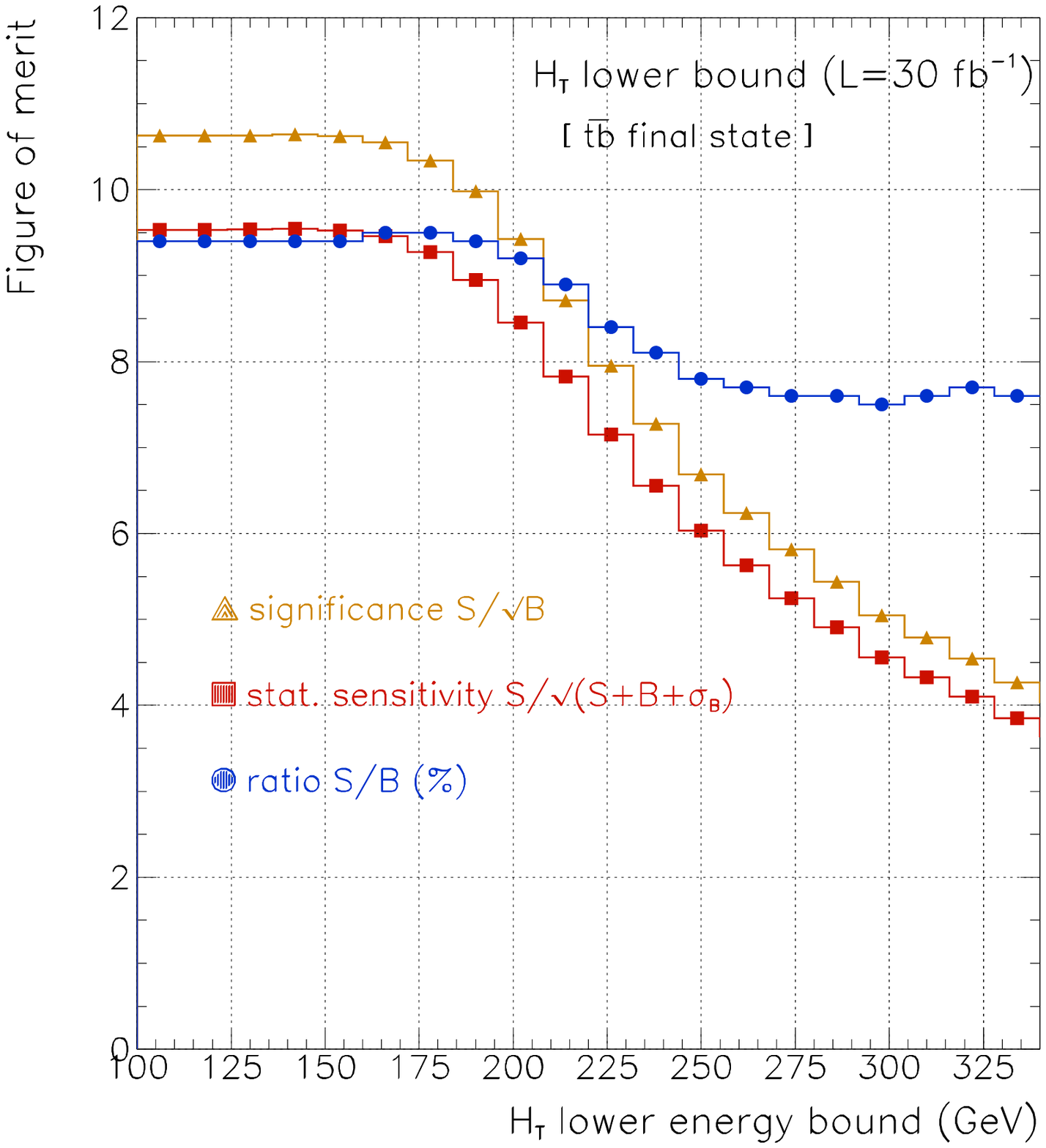}{Optimization for the $\rm H_T$ 
lower boundary: are shown the ratio S/B, statistical significance 
and the sensitivity as function of the threshold}{HT_SENSITIVITY-LOW}{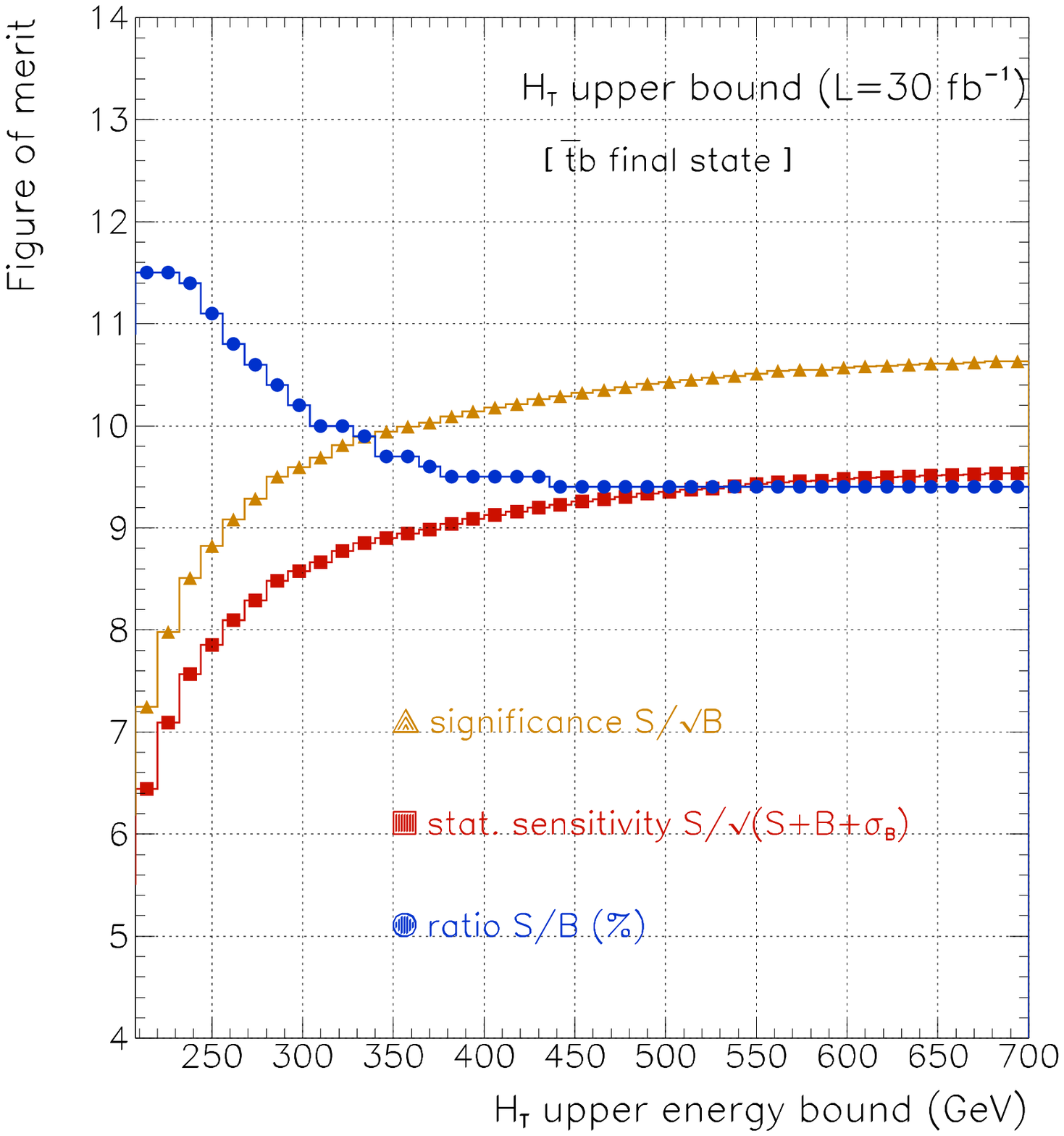}
{Optimization for the $\rm H_T$ upper boundary: are shown the ratio S/B, statistical 
significance and the sensitivity as function of the threshold}{HT_SENSITIVITY-HIGH}

\noindent
In order to optimize the upper and lower bounds applied on $\rm H_T$, one can 
use three estimators: the ratio S/B, which reflects the sample purity as function 
the threshold values; the statistical significance $\rm S/\sqrt{B}$; and the 
sensitivity defined as $\rm S/\sqrt{S+B+\sigma_B}$, which includes the systematic 
uncertainty in background estimate, set at $\rm \sigma_B = 12\%\times B$. 
Figs.~\ref{HT_SENSITIVITY-LOW} and Fig.~\ref{HT_SENSITIVITY-HIGH} show the sensitivity 
as function of the $\rm H_T$ energy cut for both the lower and upper bounds. 

The optimal choice for the window results from a compromise between a minimal loss 
in statistical sensitivity and a maximal improvement in the purity: the lower threshold 
is set at 170~\GEV while the upper bound is set at 300~\GEV. The signal efficiency 
is decreased by about 40\% for the signal. The corresponding loss is about 50\% for 
\TTB events and above 70\% for \WQQ and \WJETS events, resulting in a slight 
S/B ratio increase.  
\DBLFIG{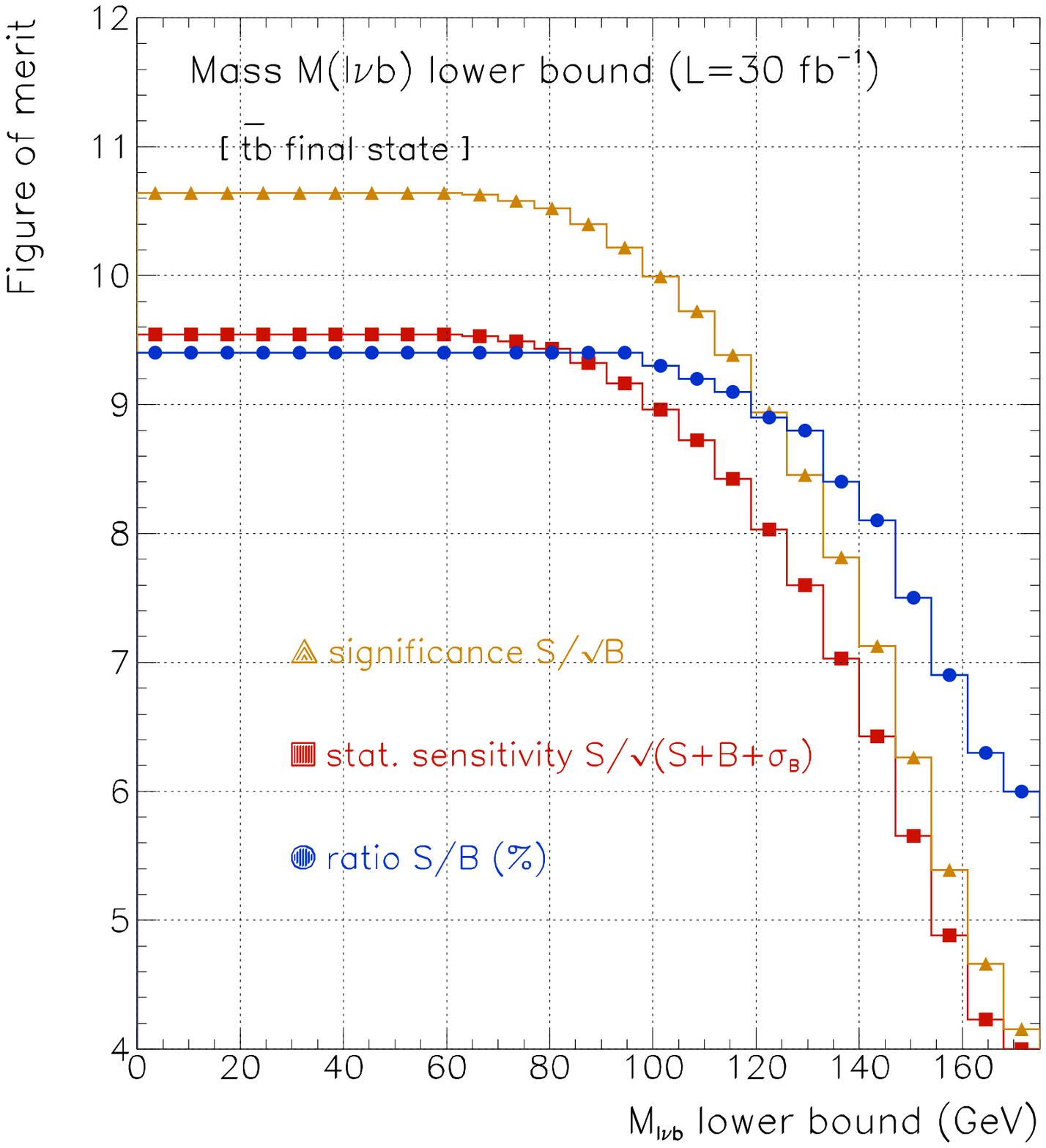}{Optimization for $M(l\nu b)$ window lower boundary, using the 
ratio S/B, the significance and the sensitivity as function of the threshold }{MTOP_SENSITIVITY-LOW}
{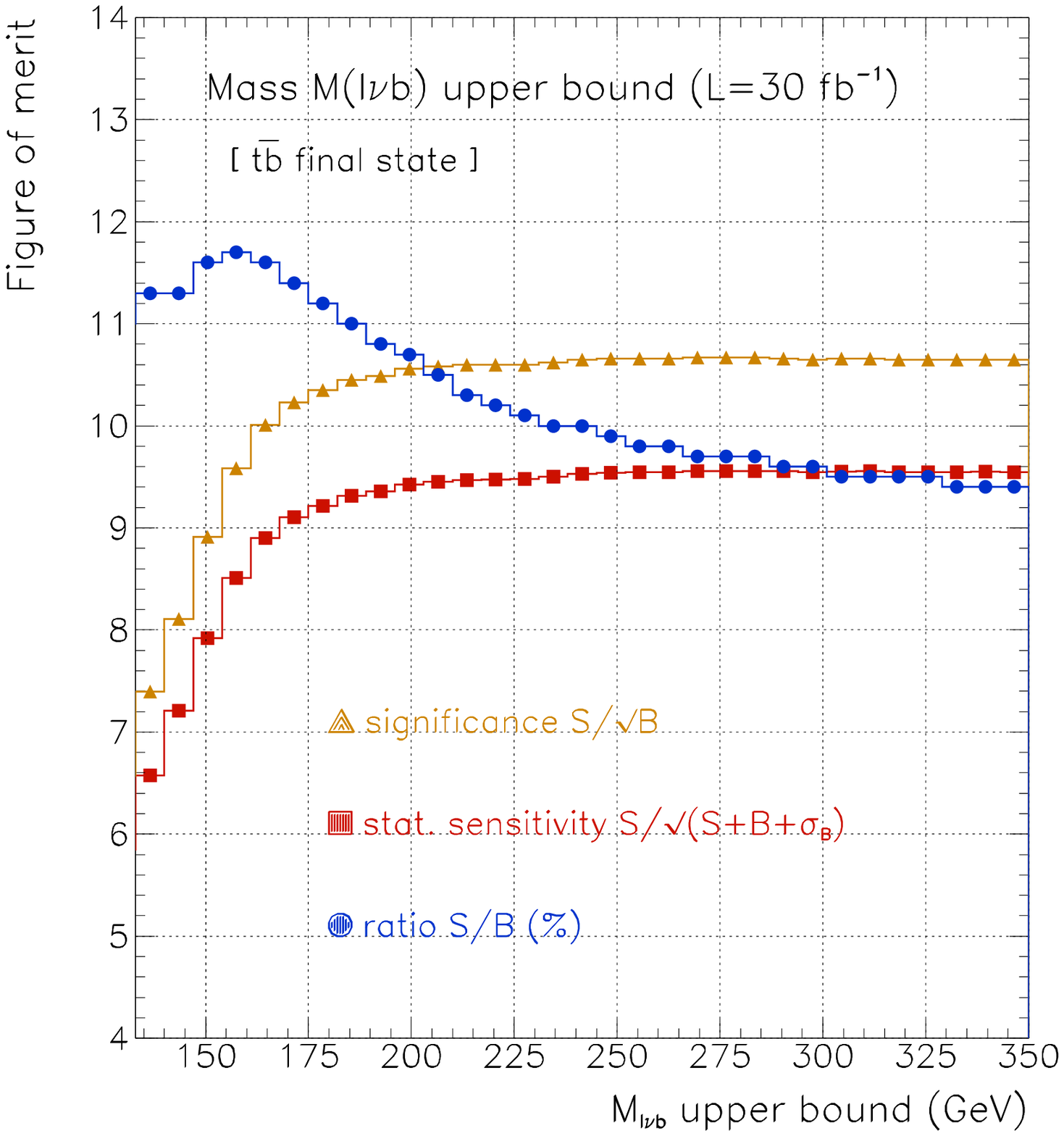}{Optimization for $M(l\nu b)$ window upper boundary, using the 
ratio S/B, the significance and the sensitivity as function of the threshold}{MTOP_SENSITIVITY-HIGH}

\noindent

\subparagraph{Top mass window optimization}
\hfill\break
\hfill\break
The optimization of the lowest and upper bounds has been performed in the same 
way as for the $\rm H_T$ quantity. Fig.~\ref{MTOP_SENSITIVITY-LOW} and 
Fig.~\ref{MTOP_SENSITIVITY-HIGH} show the sensitivity 
as function of the $\rm M(l\nu b)$ cut respectively for the lower and the upper bounds. 
The choice of a reconstructed mass in the $\rm [ 120,200 ]~GeV/c^2$ range increases the 
ration S/B by 40\% to about 14\% at the loss of half the acceptance in signal 
efficiency.   
We also estimate the top purity in our sample by using the MC truth information and 
comparing the true top momentum $\rm p_T^{true}$ and phi $\rm \Phi_{top}^{true}$ 
with the corresponding reconstructed values $\rm p_T^{rec}$ and 
$\rm \Phi_{top}^{rec}$. For a match defined by the two requirements 
$\rm |p_{T}^{rec}-p_{T}^{true}|\le 20~GeV/c$ and $\rm |\Phi^{rec}-\Phi^{true}|\le 0.4$ 
an overall purity above 60\% is measured using the highest top \PT criterium. 
Further studies on this topics are still on going in order to optimize the performance. 
\begin{table}[phtb]
 \begin{center}
  \begin{tabular}{lrr}
    \hline
    processes      &  $\rm t\bar{b}$ final state  & $\rm \bar{t}b$ final state \\[1mm]
    \hline\hline
    s-channel      &           $385\pm 2$         &  $275\pm 1$                \\[1mm]
    t-channel      &           $666\pm 30$        &  $410\pm 20$                \\[1mm]
    W+t channel    &           --                 &   --                      \\[1mm]
        \hline
$\rm t\bar{t}$ background &                       &                          \\[1mm]
$\rm tt\to e\nu b~jjb$    &   $750\pm 35$         & $750\pm 35$              \\[1mm]    
$\rm tt\to e\nu b~e\nu b$
                   &           $395\pm 20$        &  $395\pm 20$             \\[1mm]
$\rm tt\to \tau\nu b,\tau\nu b$
                   &           $105\pm 7$         &  $105\pm 7$              \\[1mm]
$\rm tt\to \tau\nu b,jjb$
                   &              $20\pm 2$       &   20$\pm 2$             \\[1mm]
    \hline
W+jets background  &                              &                          \\[1mm]
WQQ                &           $460\pm 20$        &  $290\pm 15$             \\[1mm]  
$\rm WZ\to e\nu,bb$&           $18 \pm 1$         &  $12\pm 1$                \\[1mm]
$\rm Wjj\to e\nu,jj$&          $350\pm 20$        &  $260\pm 15$               \\[1mm]
\hline
  \end{tabular}
  \caption[]{Number of selected events in the "2b0j" sample expected for an 
  integrated luminosity of 30~\FB for both final states. Uncertainties come from Monte Carlo statistics only}
  \label{TAB-SELECTION}
 \end{center}
\end{table}
\noindent

\subparagraph{Topological selection: statistical precision}
\hfill\break
\hfill\break
Table~\ref{TAB-SELECTION} reports the number of selected events  
after the $\rm H_T$ and the top mass criteria have been applied. The 
signal efficiency is reduced by 2/3 after both criteria have been applied. 
At the same time, non-top backgrounds are reduced by 80\%. In the top pair 
background, the contamination from "dilepton" events is decreased by 90\% while 
the "lepton+jet" is decreased by 70\%. Note that no significant \WT events survive 
the topological selection.  

 The total number of events expected for an integrated luminosity of 30~$\rm fb^{-1}$. 
For the $\rm t\bar{b}$ final state, about 385 signal events survive with 2,760 background 
events, resulting in an improved S/B ratio of $\rm S/B = 13.9\%$. For the $\rm \bar{t}b$ 
final state, 275 signal events are remaining for a total background of 2,242, resulting 
in a S/B ratio of 12.3\%. In both cases, the main background is due to the "lepton+jets" 
top pair production (about 30\% of the total), followed by the \TCH single-top (27\%). 
Heavy flavour \WQQ events now constitute less than 20\% of the reminaing background, 
which is about the same order than \WJETS events. Other top pair backgrounds (including tau 
decays) and WZ production appear at a negligible level.

 The statistical sensitivity to the cross section measurement has been re-evaluated 
after the topological selection. It is obvious that the application of any further 
selection criterium resulting in a decrease of the number of expected signal events 
may result in a poor statistical sensitivity. 
\DBLFIG{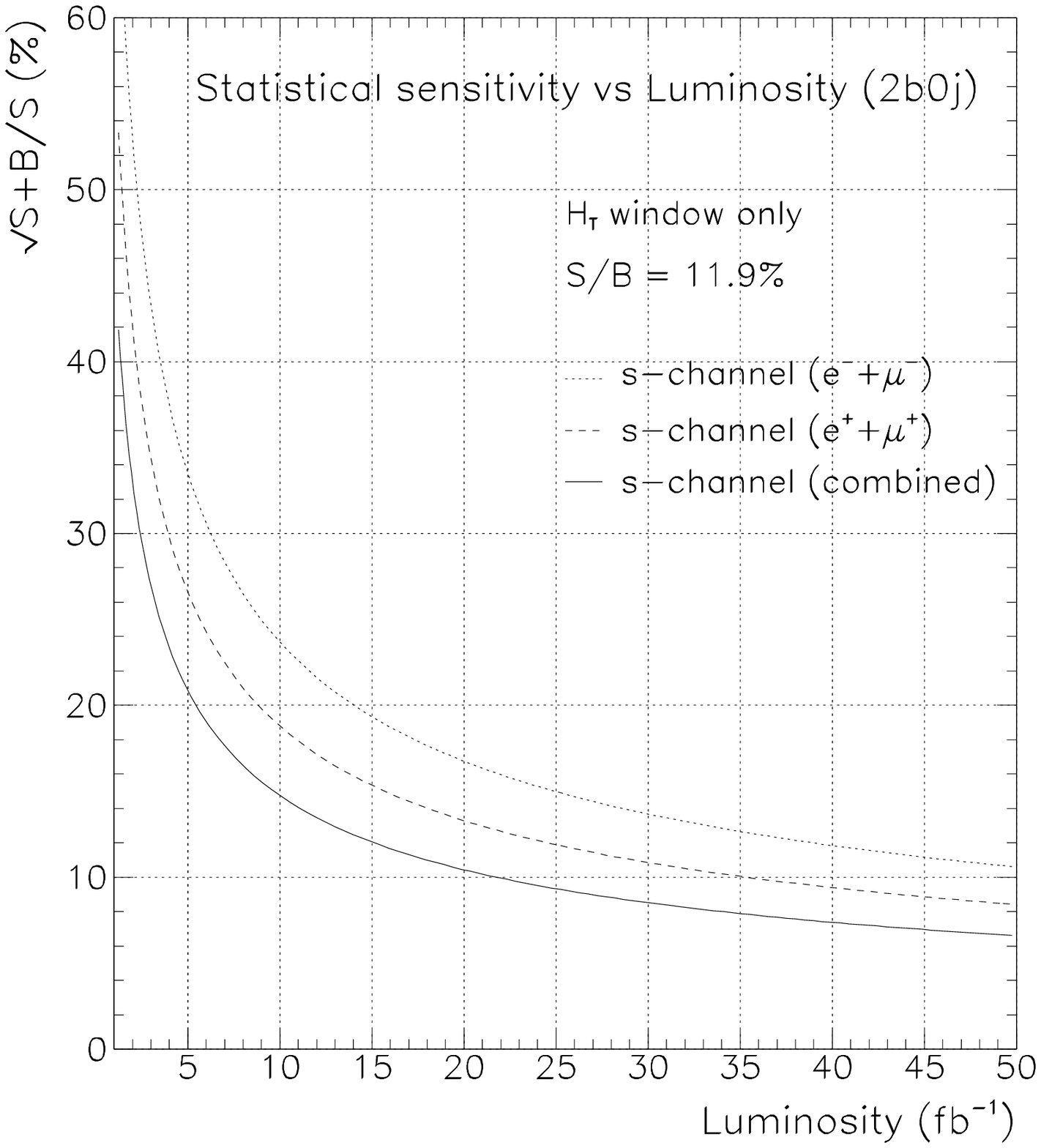}{Statistical sensitivity as a function of the integrated 
luminosity as only the $H_T$ requirement is applied.}{HT_ONLY_HT_STAT}{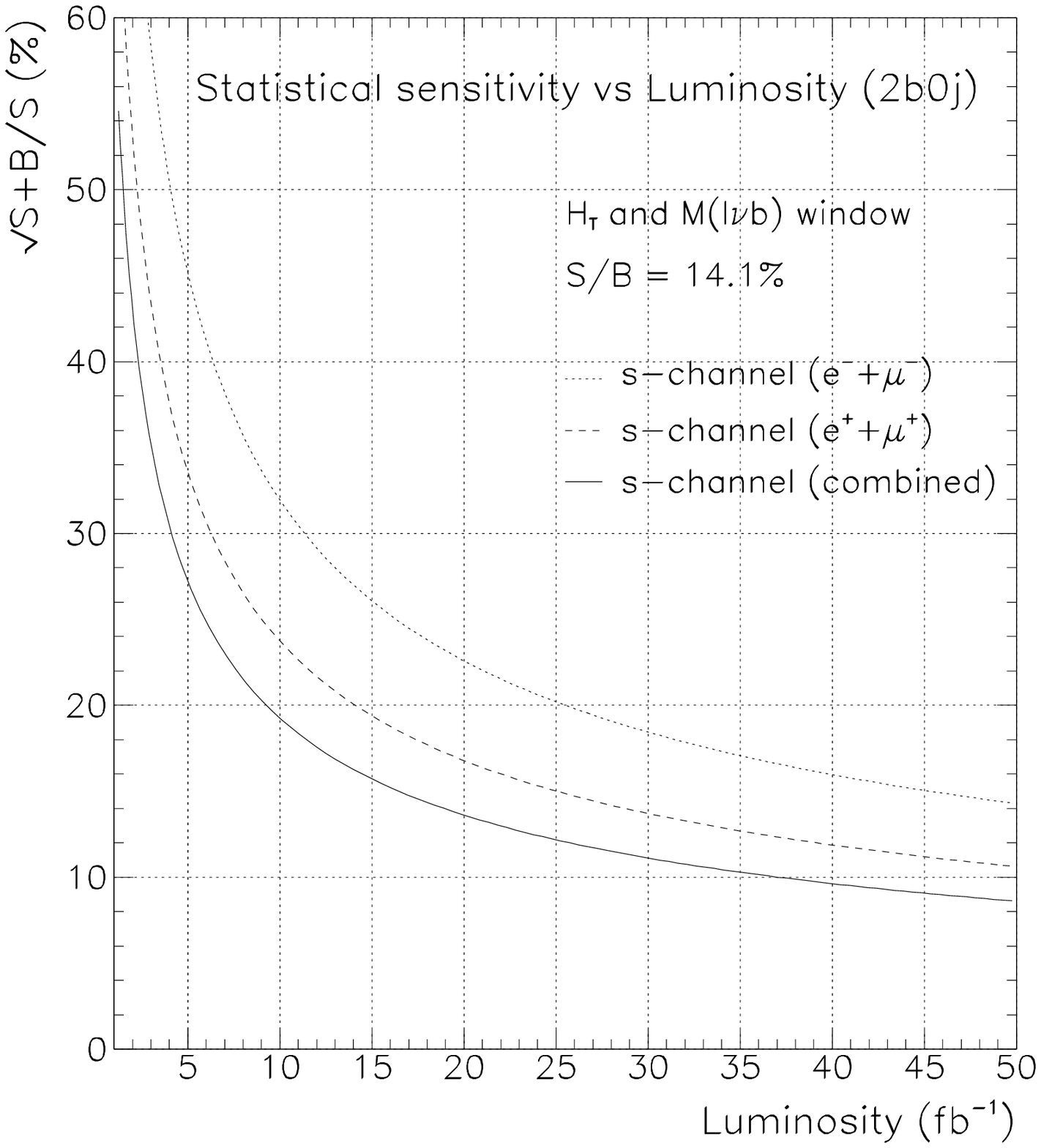}
{Statistical sensitivity as a function of the integrated luminosity after the 
topolofical selection.}{HT_FULL_STAT}

\noindent

\subparagraph{Systematic uncertainties}
\hfill\break
\hfill\break
\label{SYST}
Common experimental systematic uncertainties originate from three main 
sources: the jet energy scale, the b-tagging efficiency and mistag rate, and 
the modelling of ISR and FSR effects. These sources affect the signal 
as well as the background the background selection efficiencies.

\subparagraph{Jet energy scale}
\hfill\break
\hfill\break
 Uncertainty in the jet energy scale affects all jet \PT distributions, hence 
resulting in a bias in the jet selection efficiency. This also has a significant 
impact on the jet veto performance that is used in our analysis as well as in 
the determination of the missing energy, $\rm H_T$ and the reconstructed 
top mass that are used in the topological selection. 
In order to quantify such effect, the energy of each jet has been shifted up and 
down in the Monte Carlo by a value corresponding to the jet \PT uncertainty, 
and half of the difference in the selection efficiency was taken as a systematic 
uncertainty. 
A variation of 3.5\% is measured in the signal efficiency 
$\rm \epsilon_{W*}$, resulting in a relative error of $\rm \pm 1.8\%$ due to 
the uncertainty of the jet energy scale. For the background processes, this effect 
is shown to have a poor impact on to the top pair production. On the other hand, 
the rejection of W+jets events, which contain softer jets, depends significantely 
from the knowledge of the jet energy scale. A total background variation of 6.8\% 
is measured, thus resulting in a systematic uncertainty of 3.4\%.

\subparagraph{b-tagging efficiency}
\hfill\break
\hfill\break
Another source of systematics comes from the imperfect knowledge of the b-tagging
efficiency and mistag rates. As can be inferred from the selection described
in Section~3 b-tagging performance is crucial for
background rejection. A variation of b-tagging efficiency thus directly results
in a variation of the relative contribution of each sample. \\
For signal events a $\rm 2.6\%$ change in the selection efficiency is seen for a
1\% variation of the b-tagging efficiency.
This change is similar  for most backgrounds, with a variation of 2.7\% for the
summed backgrounds. This results in a relatively stable S/B ratio over the
full range of variation of $\rm \epsilon_b$.

This observation results in a reduced dependence of the cross-section
measurement to the exact determination of the b-tagging efficiency. In
our case, a 5\% variation in the b-tagging efficiency will result in
a 13.5\% change in the number of selected signal and background events.
It is obvious that, the S/B ratio being stable, this number does not
reflect the uncertainty in the cross-section. We however conservatively
quote half of this number as our systematics associated to the cross-section
measurement, ie: 7.0\% (including the MC statistics).

The uncertainty on the mistag rate impacts mainly the rejection of W+jets
events~: in our case a 5\% mistag rate results in a 10\% variation of the
W+jets events. This translates to an uncertainty of 3.5\% in the total
background estimate. The total uncertainty quoted is thus 8.0\%.
This number however makes of the b-tagging and mistag rate knowledge
one of our main source of errors, which is expected from a double-tag
based analysis. \\

\subparagraph{ISR/FSR modelling}
\hfill\break
\hfill\break
Another source of uncertainty is the modelling of the event and the 
effects of initial and final gluon radiations. ISR dramatically directly affects 
the jet multiplicity of the event, while uncertainty in the FSR modelling 
affects the determination of the jet energy scale, which may result in a 
change of the selection efficiency. 
For b-jets the effects are particularly significant in the \WQQ selection, 
as seen in Fig.~\ref{SYS-ISR-BJET1}. We quantified this effect by switching 
ON and OFF the ISR and the FSR separately, and by taking 10\% of the observed 
shift in selection efficiency as a systematic. This value constitutes a (very) 
conservative approach and corresponds to the expected precision of the strong 
coupling constant $\rm \alpha_s$ determination at the LHC~\cite{SGTOP-LHC-ATL-PHYS-ALPHAS}.\\
For the signal events selection, a relative variation of 4.9\% is seen for 
the ISR alone while an effect of 6.0\% is observed for the FSR. 
We thus quote an error of 7.9\% as the sum of both effects.\\
Backgrounds are differently affected by the ISR/FSR modelling. Top pair 
backgrounds are increased as the ISR are switched OFF because of the increased 
population of 2-jet events. On the other hand, as the FSR os switched OFF, most 
of those processes are reduced compared to signal variations. FSR particularly 
affects the \WQQ events selection, since switching Off the FSR tend to increase 
the jet energy and thus the jet selection efficiency. A factor 20\% is found 
to affect the \WQQ selection. The total effect on the sum of all backgrounds is 
estimated as the quadratic sum of both ISR and FSR effects. An uncertainty of 
7.3\% for the total background. This number is clearly an overestimate of 
this effect. 
\DBLFIG{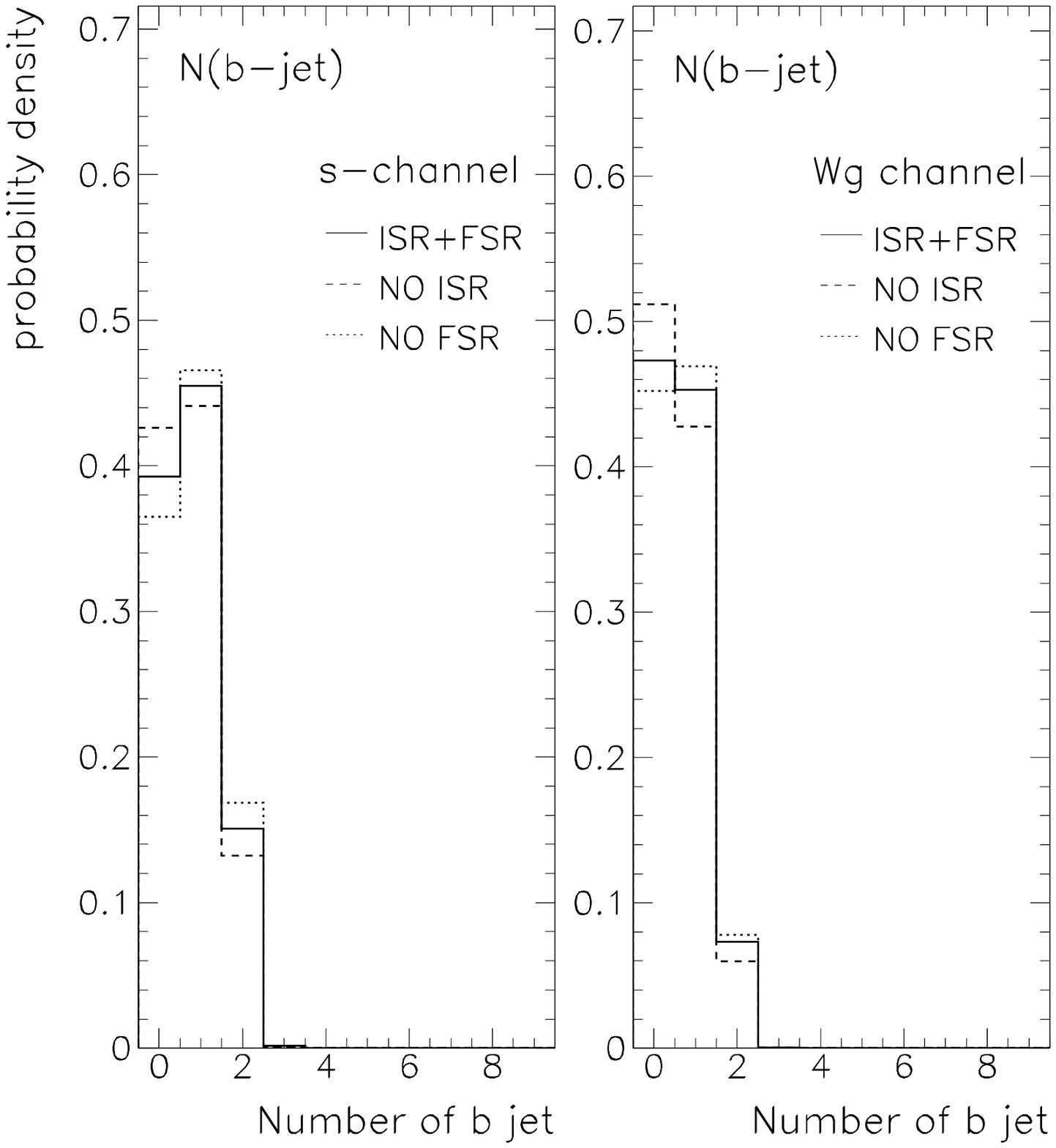}{Impact of ISR and FSR on b-jet multiplicity for s- and Wg single-top channels}{SYS-ISR-BJET1}{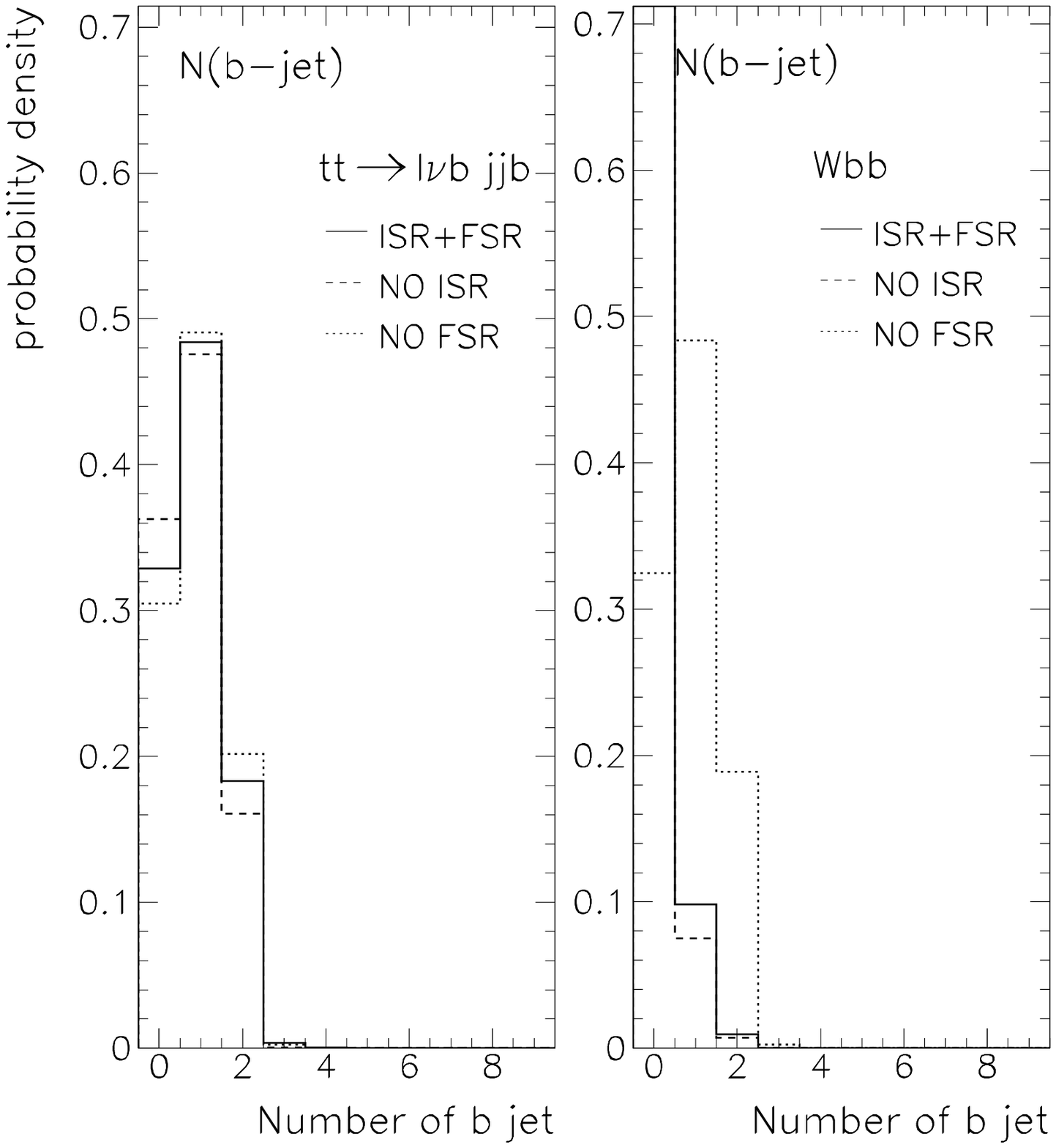}{Impact of ISR and FSR on b-jet multiplicity for top pair and WQQ productions}{SYS-ISR-BJET2}

\noindent

\subparagraph{Background estimates}
\hfill\break
\hfill\break
All of the background estimates rely upon Monte Carlo that are used
to compute the selection efficiencies. Those generators employ LO matrix
element for the hard parton scattering followed by parton showering to simulate
radiation and fragmentation. We use in our cases PYTHIA~v6.2, TopRex~v4.1
and HERWIG~v6.4 for the event generation, and normalized the event yields to
the NLO cross-section. However, even at NLO the theoretical sources of uncertainty
are significant. As a consequence, direct measurements from data itself will be required.

The sources of theoretical uncertainties come from the choice of the renomalization
and factorization scales, the choice of the parton distribution functions and the
uncertainty in the input parameters such as the top mass and the b-fragmentation
function.
The single-top cross-section is expected to decrease with the top
mass value: a 4~GeV uncertainty results in an 9\% uncertainty in the s-channel
cross-section and 5\% in the t-channel. The knowledge of the PDF (b- and
gluon-PDF for t-channel ) contributes significantely to the errors.
Regarding the top pair production, the cross-section including the NLO+NLL
corrections is quoted with an uncertainty of 12\%.
This number results of a contribution from the scale uncertainty
(about $\rm \pm 6\%$) and from the PDF where the level is at 10\%
(MRST vs CTEQ5M) for $\rm m_t=175~GeV$. The difference between the
two sets is about 3\% but is highly sensitive to the input value used
for $\rm \alpha_s(M_Z)$.

Regarding the Wbb ($\rm e\nu bb+X$) production, recent computations with
MCFM lead to an uncertainty of 20\% in the NLO cross-sections, this result 
being obtained with the use of a LHC-like selection applied
on the final lepton and jets. Regarding W+jet backgrounds, a conservative
approach has been chosen and an uncertainty of 20\% is quoted as well.
Summing all background contributions (in the fraction of selected events)
result in a total theoretical error of 11\%.\\
Note that the input top mass also has an impact in the selection efficiencies
determination, the jet \PT distributions depending upon the mass of the decaying
particle. For a higher top mass value, jet \PT distributions are shifted
towards higher values, leading to a better pre-selection efficiency for
all top events production: an effect of about 2\% is seen in the selection efficiency
of \SCH and \WT channels as one goes from 175 to 180~\MGEV. This is considered
as negligible in regards to the other sources of error.

\subparagraph{Summary: s-channel cross section measurement in ATLAS}
\hfill\break
\hfill\break
The precision on the cross section has been assessed for an integrated 
luminosity of 30~\FB at different stages of the analysis. 
After the simple preselection stage, results show a good statistical 
sensitivity but higher level of systematic uncertainties:
$$\rm {\Delta\sigma\over \sigma}=7\%_{stat}\pm 13.8\%_{exp}\pm 11\%_{bckgd~theo}\pm 5\%_{lumi} $$ 
Using both the $\rm H_T$ and reconstructed top mass results in a significantly 
reduced level of systematics at the price of a loss in statistical sensitivity:
$$\rm {\Delta\sigma\over \sigma}=12\%_{stat}\pm 12\%_{exp}\pm 11\%_{bckgd~theo}\pm 5\%_{lumi} $$ 
In all cases, systematic errors are expected to dominate the cross section 
determination. Experimental effects are dominated by the ISR/FSR 
modelling effects because of the importance of the jet multiplicity requirement 
in the selection. The other significant effect comes from the knowledge of the b-tag 
and mistag rates, since the double-tag also constitutes a central point in the 
selection. It is obvious that the error associated to the ISR/FSR modeling 
is an overestimation, and that this uncertainty will be constrained by 
comparison of Drell-Yann data and the event generators. B-tagging should 
also benefit from the use of a huge b-enriched control sample.  Finally, 
theoretical uncertainties are the same order of magnitude of the statistical 
errors. They should be reducible if we are able to estimate the background 
contamination directly from the data. Besides, the uncertainty in the 
parton structure functions should also be reduced by constraints from 
the W leptonic asymmetry measurements.

\paragraph{t-channel cross section measurement}
\hfill\break
\hfill\break

  The measurement of the t-channel cross section benefits from a 
significantely higher statistics compared to the s-channel analysis. 
The final topology is also significantely different of that of the  
s-channel, and leads to a specific selection.
The present analysis can be found in Ref.~\cite{SGTOP-LHC-LUCOTTE}.

\subparagraph{Event selection} 
\hfill\break
\hfill\break
We select t-channel events in the channel where the W~boson 
decays leptonically. This leads to requirements on the presence 
of a high~\PT lepton and a high missing transverse energy. To remove events 
with two leptons like $\rm Z\to l^+l^-$ and dileptonic top pairs, a veto is 
performed in any pairs of leptons with opposite signs and above 10~\GEV, just 
as in the s-channel analysis.
\TRIFIG{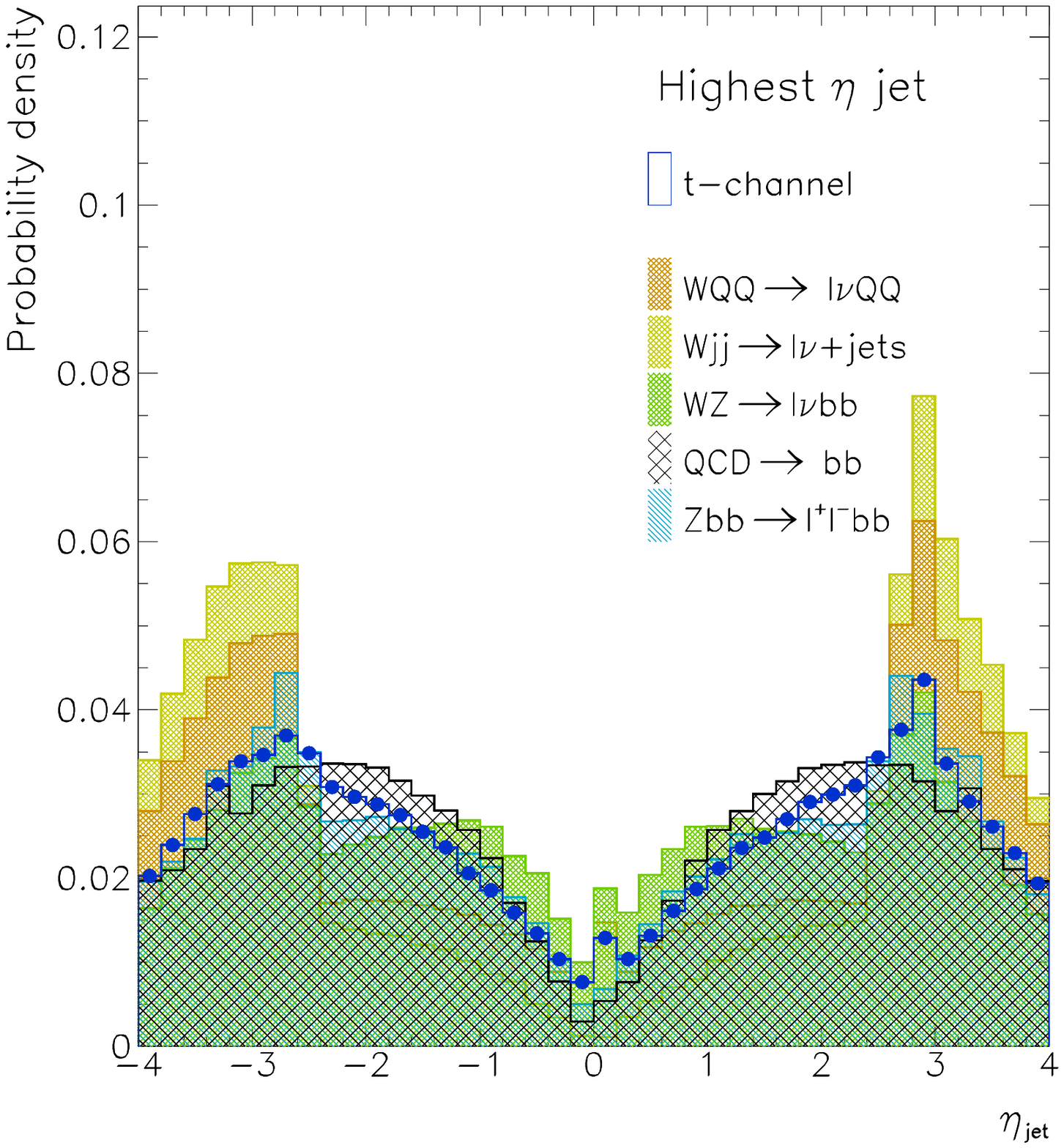}{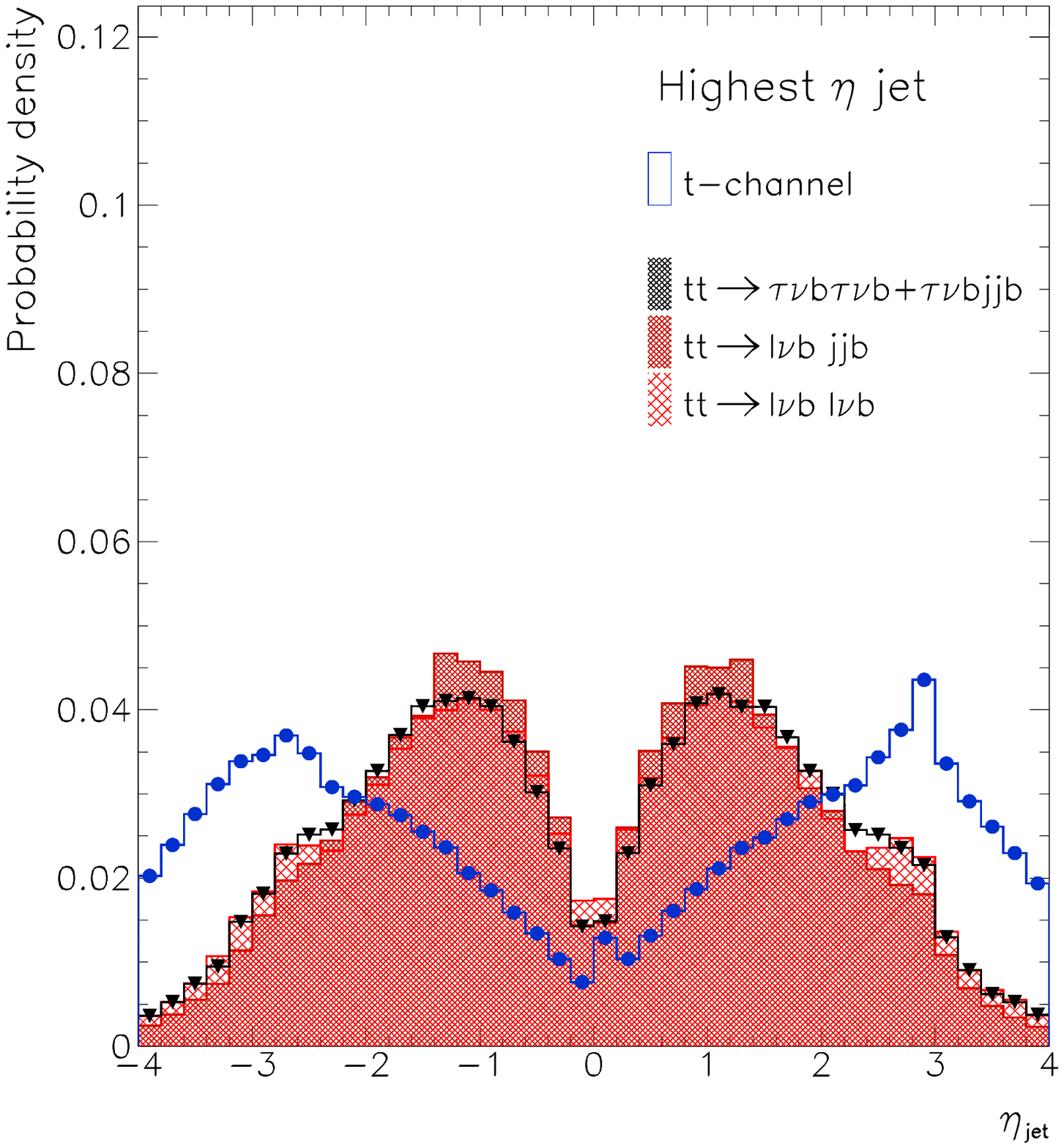}{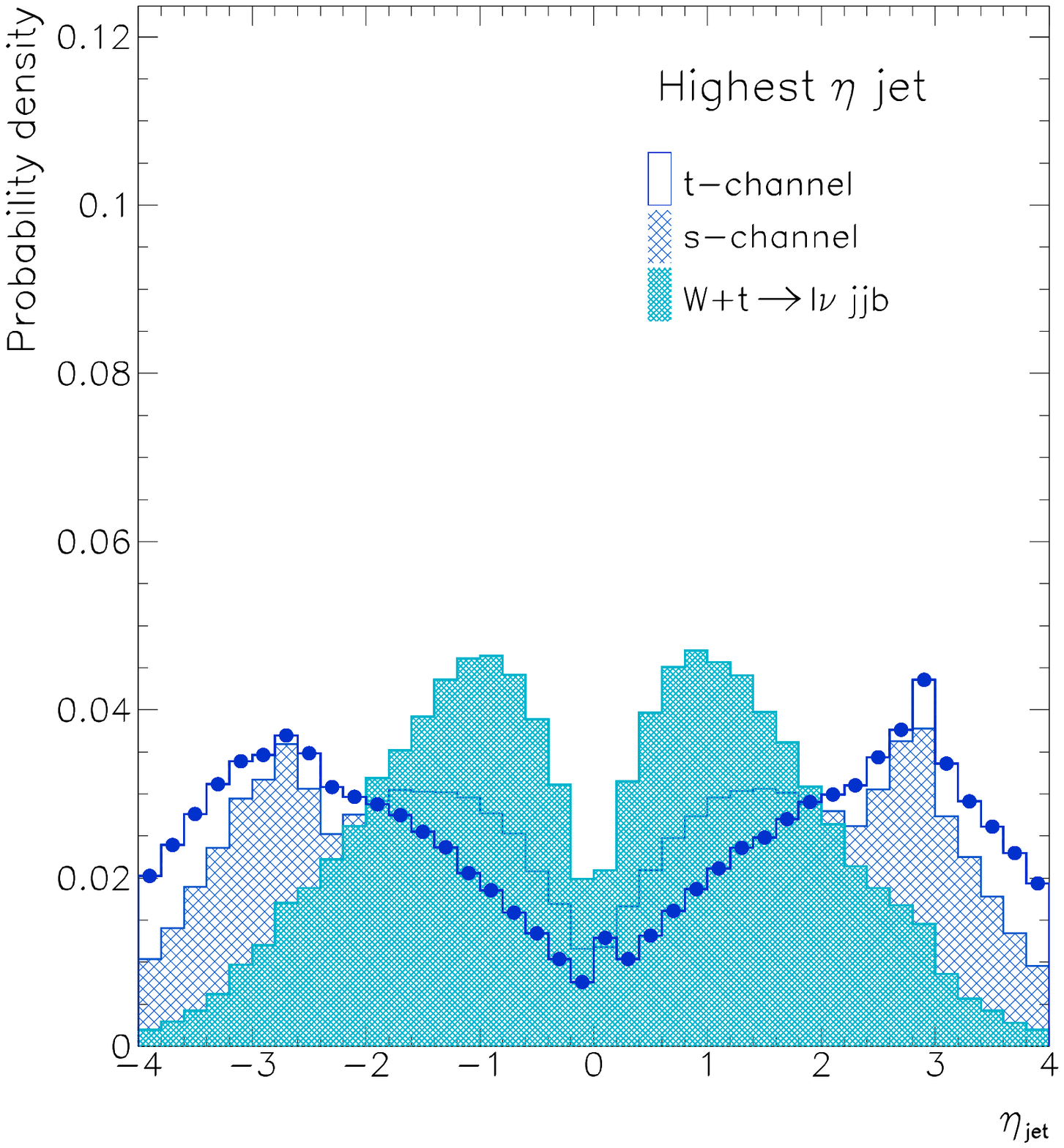}{Pseudo-rapidity of the highest jet in two jet events for signal and backgrounds.}{FIG-FWDJET}

\noindent 
The situation is different from the latter analysis in the domain 
of reconstructed jets. More than 60\% of t-channel events have two or three jets. 
Among those jets, one points towards the forward region, beyond the pseudo-rapidity 
range $\rm |\eta_{jet}|\ge 2.5$. This is a distinct feature which is used 
to discriminate from the other top quark production sources, as shown in 
Fig.~\ref{FIG-FWDJET}. This forward jet must also pass a high~\PT threshold 
in order to reduce the contamination from W+jets, WQQ and QCD, WZ and QCD 
events. Figure~\ref{FIG-FWDJET-PT} displays the momentum of the selected forward 
jet in 2~jet final state events.

Among the two or three jets, at least one jet must be b~tagged in the central 
pseudo-rapidity region. The other b~jet present in the final state is usually 
emitted towards the very forward region, outside the tracker acceptance and thus 
out of reach of the b-tagging algorithm in most cases. 
\TRIFIG{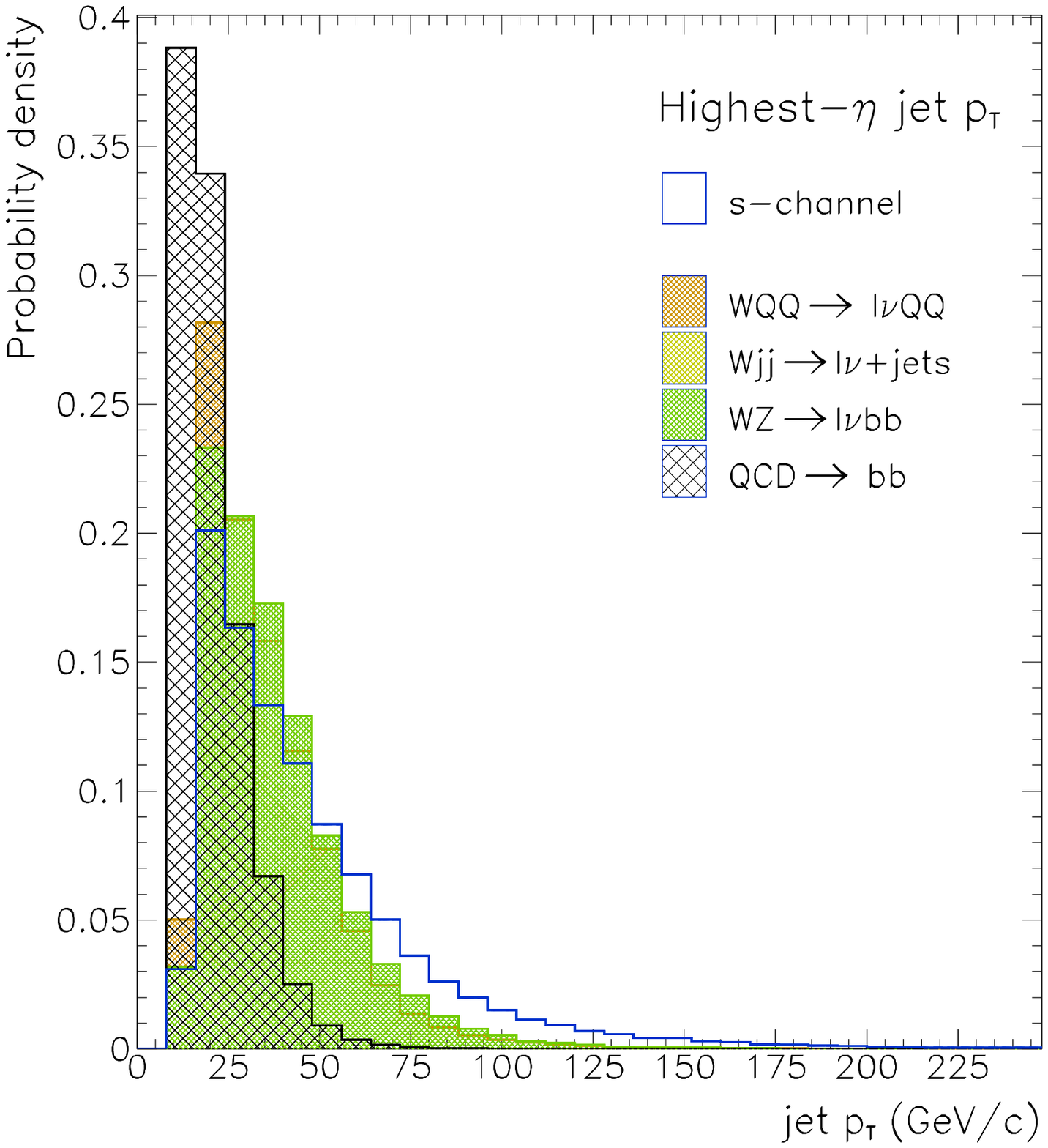}{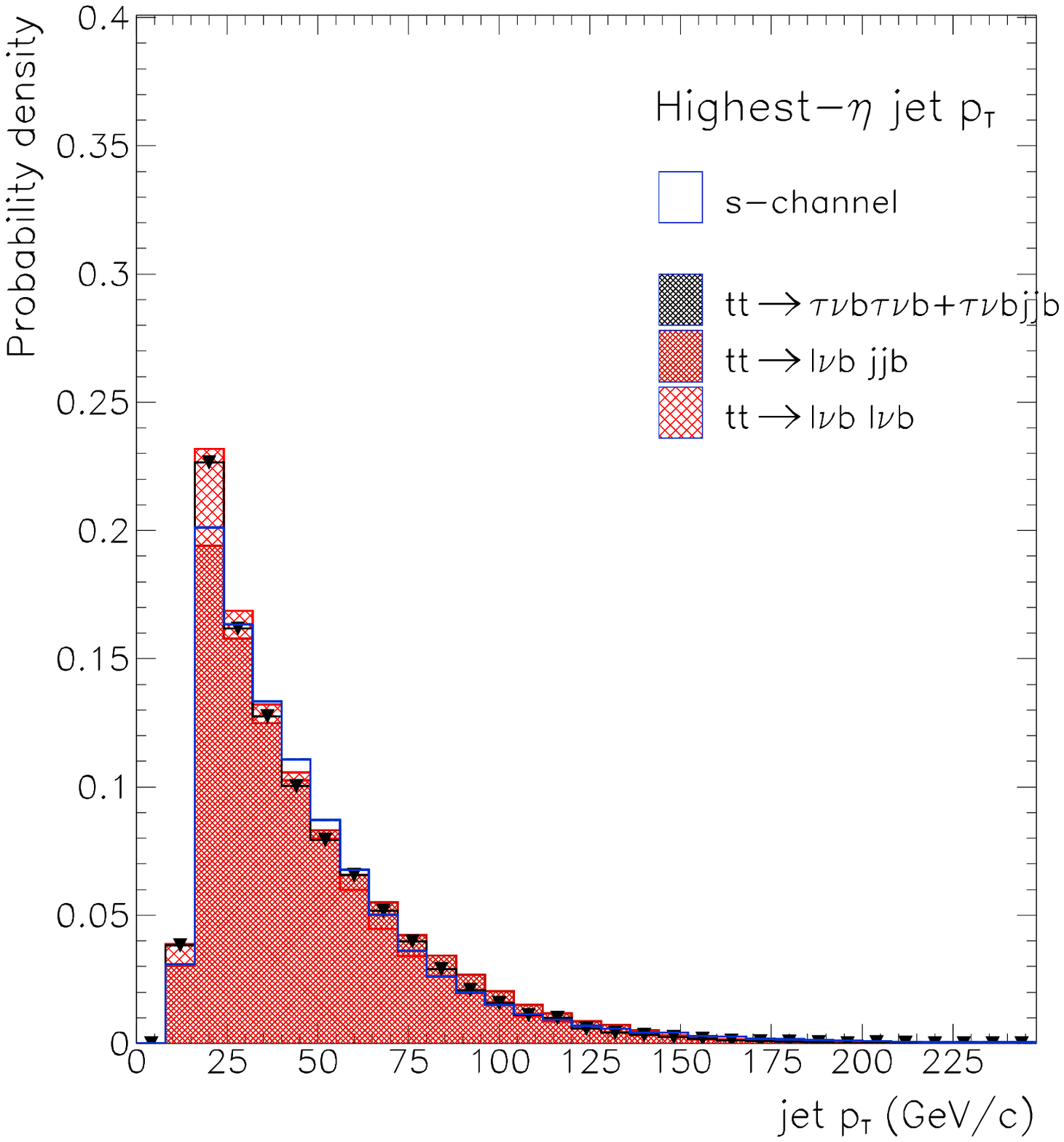}{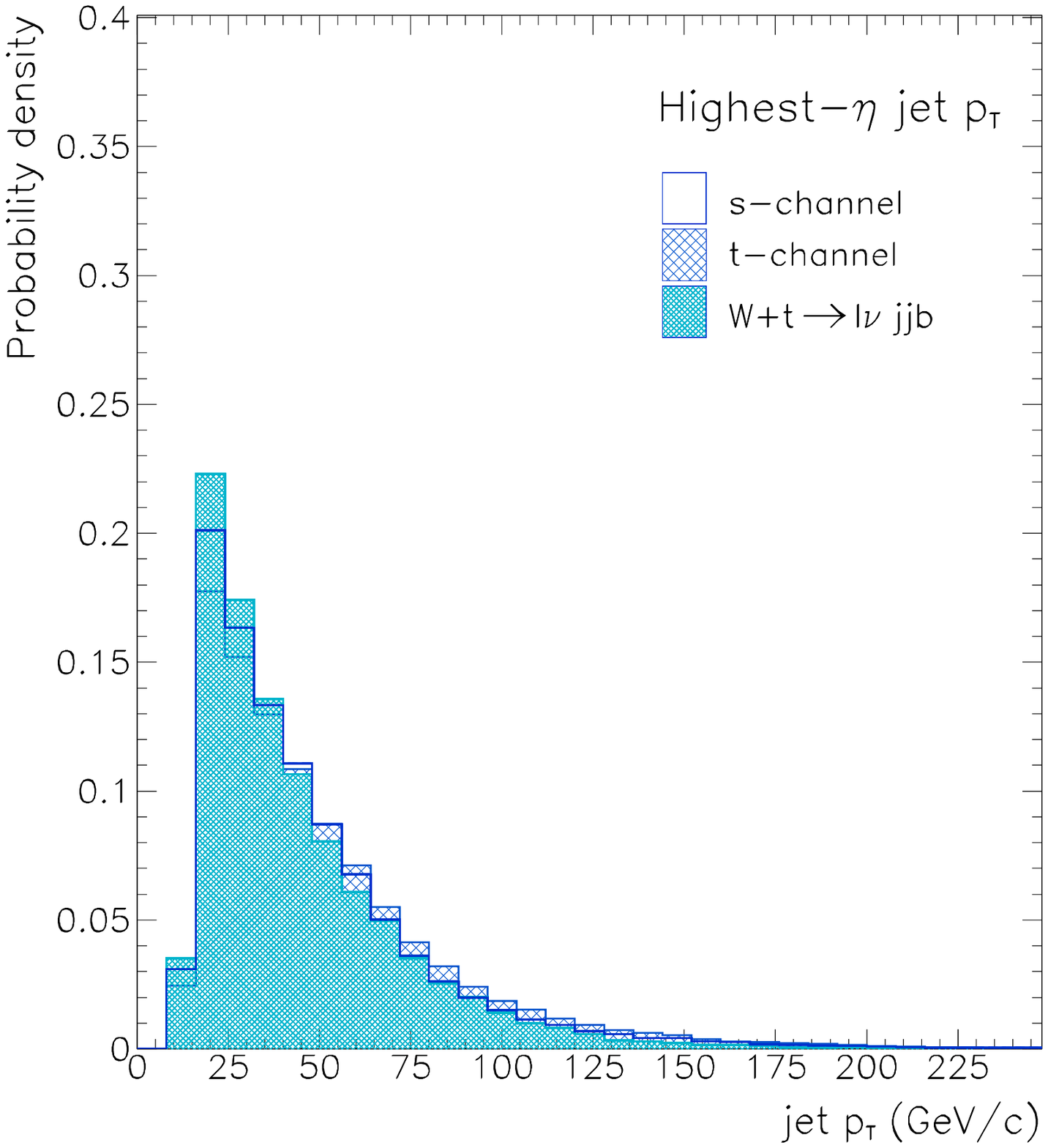}{Transverse momentum for the forward jet in two jet events and for signal and backgrounds.}{FIG-FWDJET-PT}

\noindent 

\subparagraph{Efficencies and background rejection}
\hfill\break
\hfill\break
 A preliminary analysis has been developped in ATLAS. The selection 
requires the presence of an isolated high~\PT lepton above 25~\GEV, 
missing transverse energy above 25\GEV and makes use of a secondary lepton 
veto. At least one jet must be b~tagged with a transverse momentum above 
50~\GEV. The event must contain a forward jet above the pseudo-rapidity 
$\rm |\eta_{jet}|\ge 2.5$ with a transverse momentum above 50~\GEV.  
Selected events are then splitted into two 2~jet and 3~jet final states.
Like in the s-channel analysis, the selection is splitted into the two final 
states $t\bar{b}$ and $\bar{t} b$ in order to reduce the contamination from 
the charge symmetric top pair production. Table~\ref{TAB-TCH-EVENT} reports the 
event yields expected in the two final states for an integrated luminosity 
of 10~\FB.

 In two-jet final state, signal events are selected with an efficiency of 
about 1\%, leading to a total of 3,000 events in 10~\FB. The dominant background 
comes from the WQQ production despite the central high~\PT b~jet requirement.  
The efficiency for those events is well below the per mill level. Remaining 
backgrounds consists in top pair events in both the ``dilepton'' and ``lepton+jets'' 
channels, although the low multiplicity cut removes most of them. Finally, 
the contamination from the other (s- and Wt-) single-top channels represents 
less than 5\% of the selected events. At the end, the ratio S/B is above 3 for an 
integrated luminosity of 10~\FB. Figure~\ref{FIG-HT-TCH-1B1J} displays the 
event yields for the $\rm H_T$ distribution and an integrated luminosity 
of 10~\FB.
\begin{table}[phtb]
 \begin{center}
  \begin{tabular}{lccc}
    \hline
    processes      &  2~jet final state ($\rm t\bar{b}q$) & \multicolumn{2}{c}{3~jet final state ($\rm t\bar{b}q$)} \\[1mm]
    \hline\hline
    t-channel      &           $3,130\pm 40$      &  $3,410\pm 40$  & $54\pm 2$   \\[1mm]
    s-channel      &           $80\pm 1$          &  $ 40\pm 1$     & negl.   \\[1mm]
    W+t channel    &           $50\pm 2$          &  $120\pm 4$     & negl.  \\[1mm]
        \hline
$\rm t\bar{t}$ background &                       &                          \\[1mm]
$\rm tt\to e\nu b~jjb$    &      $205\pm 10$      &  $1,890\pm 35$  & $17\pm 1$   \\[1mm]  
$\rm tt\to e\nu b~e\nu b$ &      $215\pm 10$      &  $560\pm 15$    & $11\pm 1$   \\[1mm]
$\rm tt\to \tau\nu b,\tau\nu b$& $ 15\pm 1.5$     &  $30\pm 2$      & negl.   \\[1mm]
$\rm tt\to \tau\nu b,jjb$ &      $ 10\pm 2.5$     &  $60\pm 6$      & negl.   \\[1mm]
    \hline
Z/W+jets background &                             &           &                \\[1mm]
WQQ                 &          $230\pm 15$        &  $60\pm  5$    & $7\pm 2$  \\[1mm]  
$\rm Wjj\to e\nu,jj$&          $120\pm  8$        &  $30\pm  3$    & negl.     \\[1mm]
\hline
  \end{tabular}
  \caption[]{Number of selected events in the "1b1j", "2b1j" and "1b2j" samples expected for an 
  integrated luminosity of 10~\FB for $\rm t\bar{b}q$ final state. Uncertainties come from Monte Carlo 
statistics only}
  \label{TAB-TCH-EVENT}
 \end{center}
\end{table}

\noindent
In three jet final states, the situation is less favorable because of a higher 
contamination from high multiplicity events like top pair production. Two 
situations are considered depending on the number of b~tagged jet contained 
in the event.\\ 
For events with exactly one b~tagged jet and two light jets, the signal 
efficiency is slightly above 1\%. The dominant background comes from the 
top pair production in both the ``lepton+jets'' and the ``dilepton'' channels. 
The contamination from those events amounts to about 40\% of the selected 
sample. As expected, the single-top Wt channel now also constitutes a 
significant background, representing 2\% of the total. The third jet 
requirement removes most of the W+jet and WQQ backgrounds. The ratio 
S/B is about 1.2. Figure~\ref{FIG-HT-TCH-1B1J} displays the corresponding 
event yields for the $\rm H_T$ distribution and an integrated luminosity 
of 10~\FB.
\DBLFIG{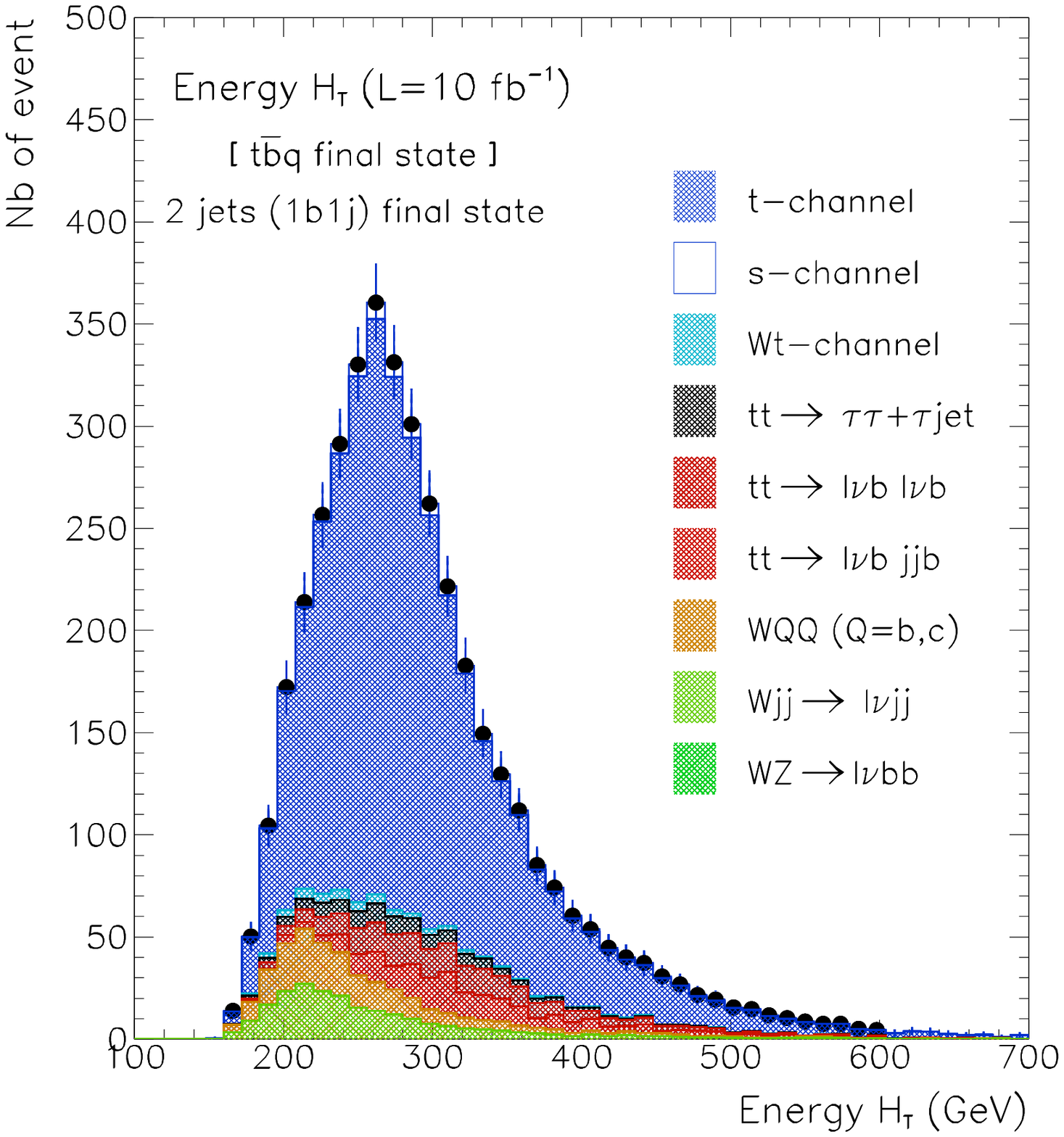}{Event yields for the $H_T$ distribution for a $t\bar{b}q$ 2-jet final state events for 10~\FB.}{FIG-HT-TCH-1B1J}{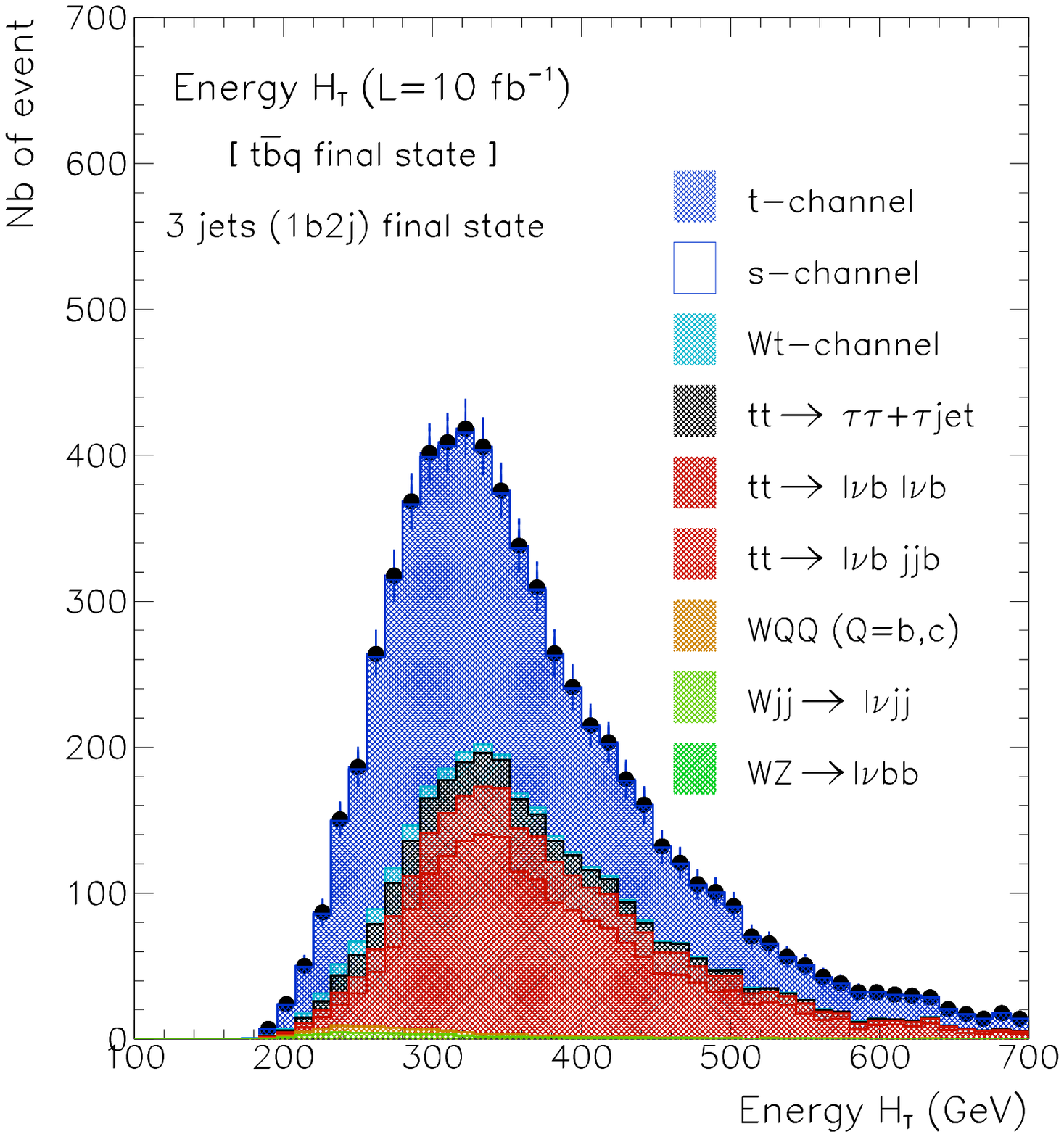}{Event yields for the $H_T$ distribution for a $t\bar{b}q$ 3-jet final state events for 10~\FB.}{FIG-HT-TCH-1B2J}

\noindent
For events composed with 2 b~tagged jets and one light (forward) jet, 
the signal efficiency is decreased to 0.17\%. This is due to the 
fact that the second b~jet present in such events is expected to point 
towards the very forward region, thus being out of the tracker acceptance. 
About 50~events are expected in 10~\FB. In this case, dominant backgrounds 
are the top pair events. The ratio S/B is about 1.5, making this channel 
the least significant in terms of statistical precision.
\FIGVIII{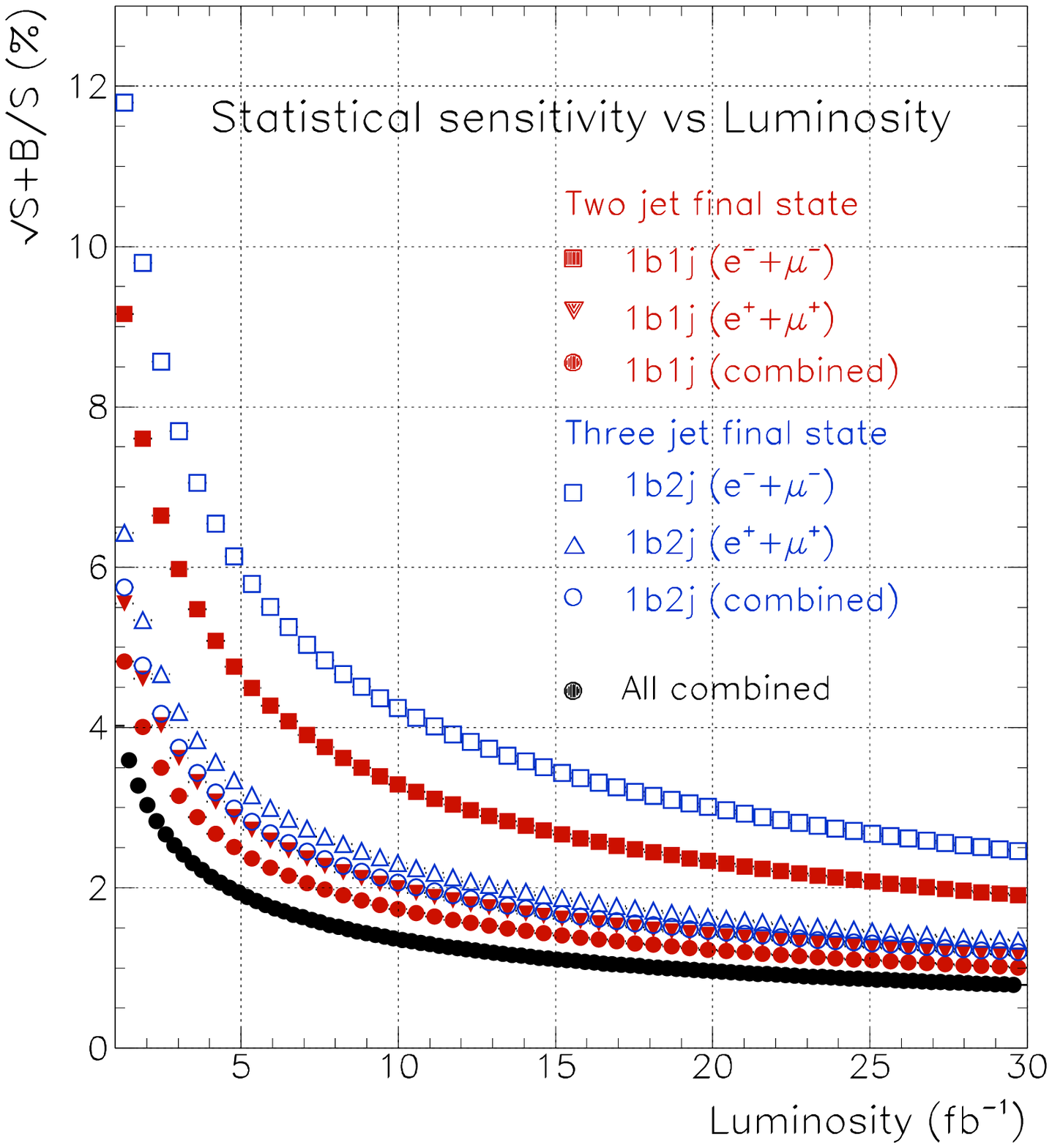}{Statistical precision as a function of the luminosity 
for the t-channel analysis in two- and three- jet events and for $t\bar{b}q$ and 
$\bar{t}b q$ final states. The result of the combination of all channels is also 
shown.}{FIG-TCH-STAT}

\noindent
In this preliminary analysis, no use is made of the $\rm H_T$ nor the reconstructed 
leptonique top mass. We may consider using those variables to purify the selected 
sample. This may be the case if a better control of the systematic uncertainty associated 
to the background estimates is required.

\subparagraph{Summary: t-channel cross section measurement in ATLAS}
\hfill\break
\hfill\break
With a cross-section corresponding to about a third of that for the top pair
production, the t-channel processes will be the first single-top production
accessible with the early data at the LHC. Contrary to the situation at the
TeVatron, the main background comes from the top pair production, well above
the W+jets and WQQ events. The statistical precision is about 4\% for an
integrated luminosity of 1~\FB and well below 1\% with 30~\FB.

This measurement will however be limited by the systematic errors.
The dominant uncertainties comes from the jet energy scale and the ISR/FSR modeling,
which affect directly the selection efficiencies for both signal and backgrounds. The
b-tagging systematic error is expected to be reduced compared to
the s-channel analysis where two b-tags were required. The uncertainty
associated to the background estimate is again a major source of error and,
as in the s-channel case, will require the use of data itself.
With a simple selection, the precision on the cross-section is expected to be:
$$\rm {\Delta\sigma\over \sigma}=1.0\%_{stat}\pm 11.0\%_{exp}\pm 6\%_{bckgd~theo}\pm 5\%_{lumi}~for~L=30~fb^{-1}$$
which shows how sensitive the selection is to the experimental
and background estimates effects. Same remarks as for the s-channel
measurement apply.

\paragraph{Wt associate production cross section measurement}
\hfill\break
\hfill\break

\noindent
The \WT-channel is the second largest source of single top production. Due to the presence of a
second W in the final state, Wt events are topologically similar to \TTB background events and
are therefore difficult to separate.

\subparagraph{Event selection}
\hfill\break
\hfill\break
As for the s and t-channels, we select \WT events by requiring a single high \PT lepton
and a high missing transverse energy. Such a selection criterion implies that one W boson 
decays leptonically and that the second W boson must decay into two jets.
Therefore, the selected events have exactly three jets with one of them tagged as a b-jet.
This allows to reject part of \TTB background. In addition, by requiring a 2-jet invariant
mass within a window around the W mass, it is possible to eliminate most events that
do not contain a second W, \rm {i.e.} all backgrounds other than \TTB. Indeed, as shown in 
Fig.~\ref{FIG-WTCH-MJ}, a sharp peak in the 2-jet invariant mass distribution
is observed for the \WT and \TTB events.

\TRIFIG{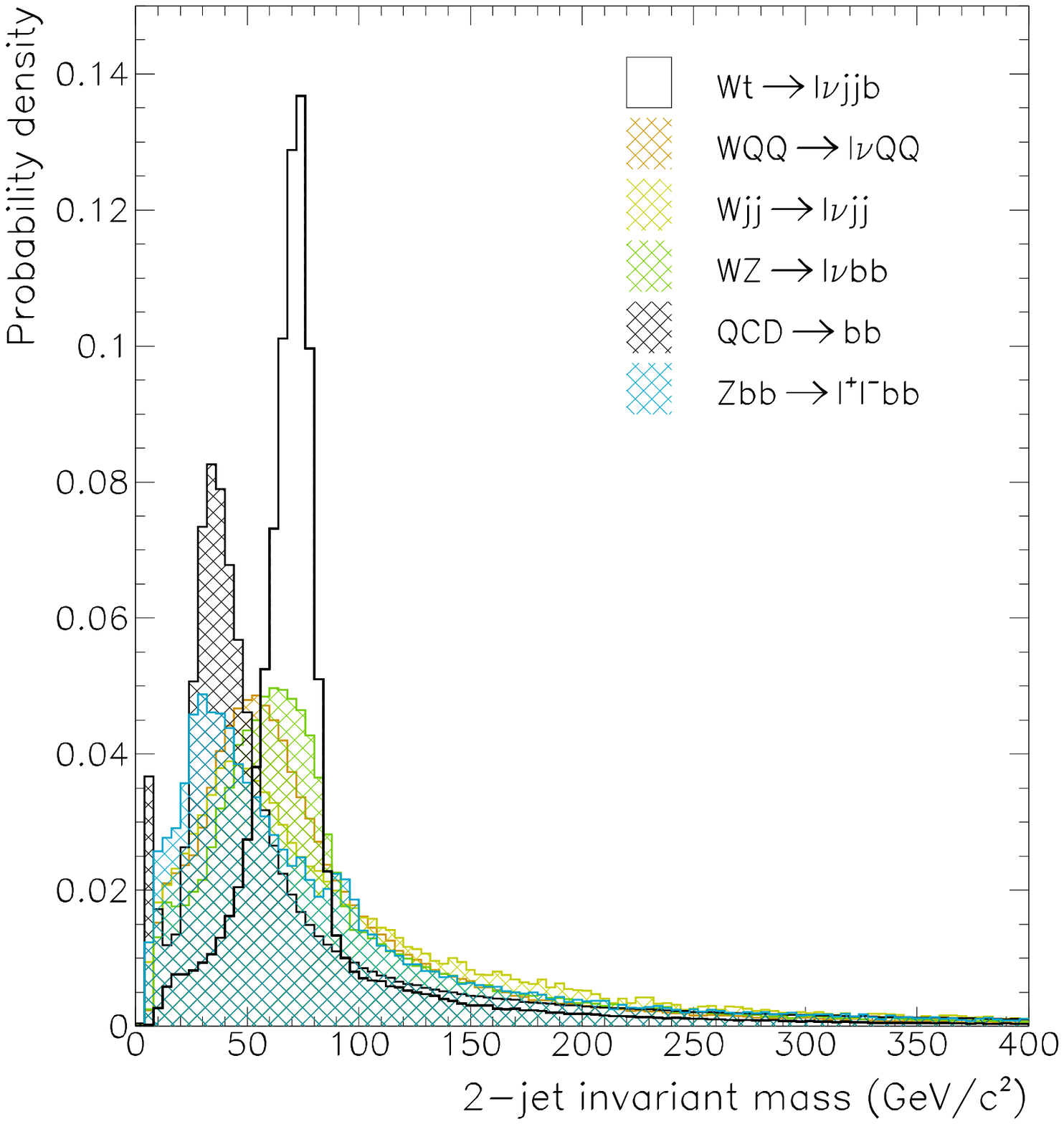}{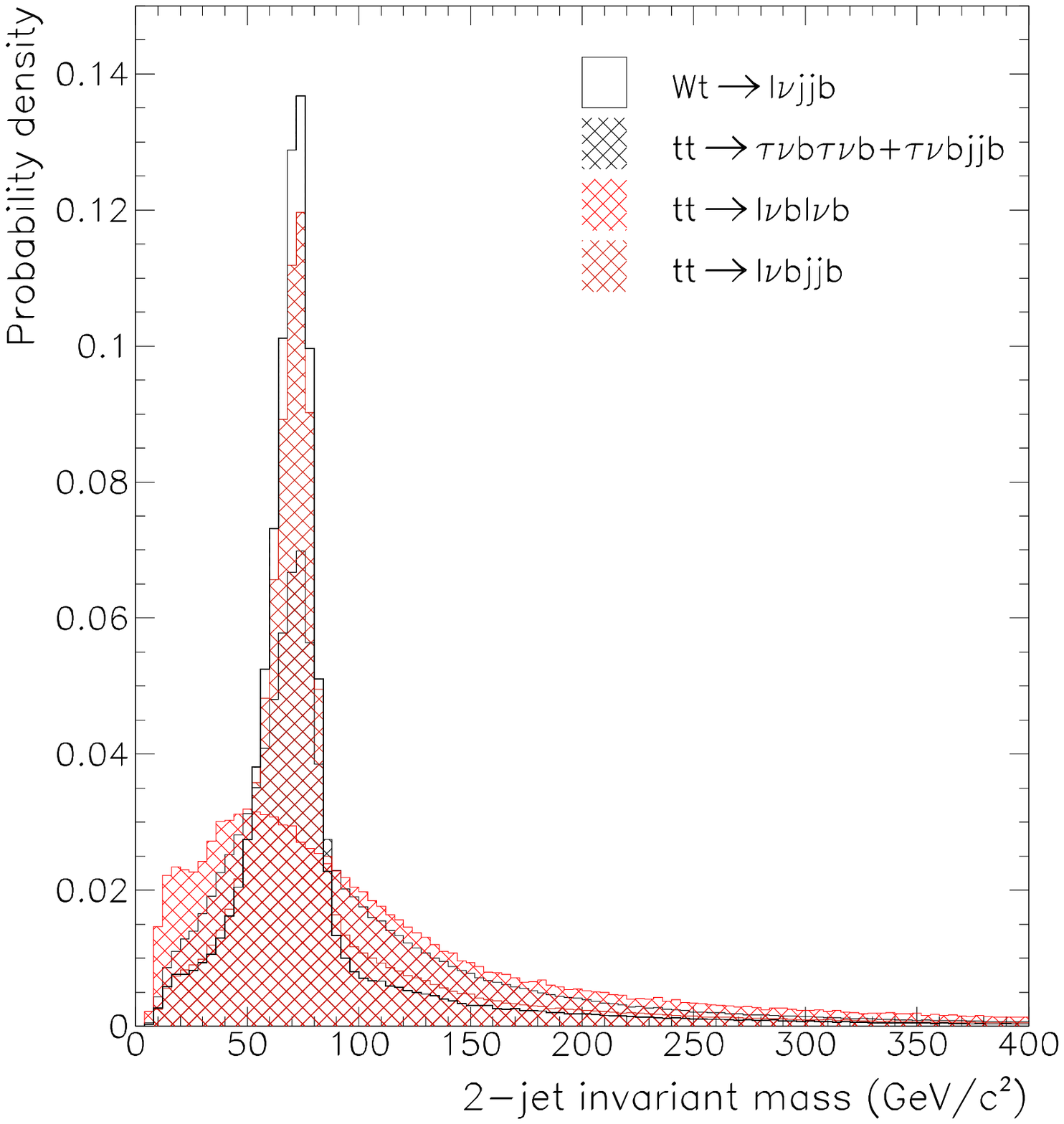}{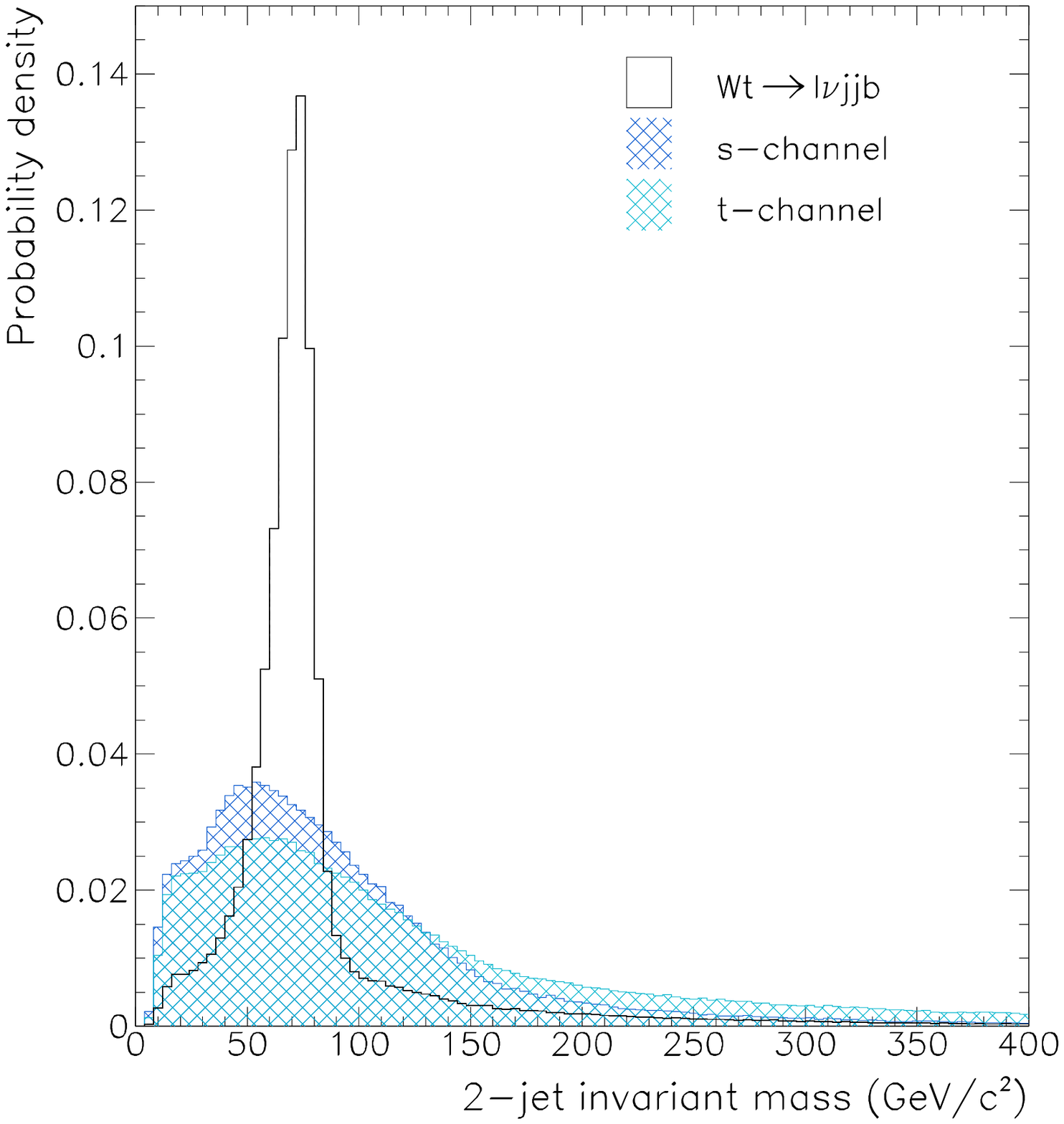}{2-jet invariant mass for signal and backgrounds.}{FIG-WTCH-MJ}
\noindent

\subparagraph{Efficencies and background rejection}
\hfill\break
\hfill\break
The selection for the preliminary analysis presented here requires the presence of an isolated
high \PT lepton above 25~\GEV and a missing transverse energy above 25~\GEV. In addition, 
a secondary lepton veto cut is applied to any lepton above 10 GeV/c with a sign oposite to
that of the selected high \PT lepton. 
The event must contain, among three jets of \PT above 25~\GEV, one b-tagged jet with a transverse 
momentum greater than 50~\GEV. An additionnal constraint on the 2-jet invariant mass (55-85 ~GeV/c$^2$)
is required.

\noindent
The efficiencies are reported in Table~\ref{TAB-WTCH-EFF}. The numbers of events expected for an integrated 
luminosity of 30~\FB and the expected individual signal-to-background ratios are also
tabulated in Table~\ref{TAB-WTCH-EFF}. The calculated values include only the 
electron/positron contribution of the leptonic components.

\noindent
The \WT events are selected with an efficiency of about 4.6\%. Top pair events are
selected with a global efficiency of around 1.7\% (3.3\% for the ``lepton+jets'' channel, which is the
main \TTB background). As expected, 
the other sources of background are greatly reduced by the selection criteria ; we obtain efficiencies
less than 0.05\% for W/Z+jets channels and 0.2\%-0.3\% for the two other single top production
channels.

\noindent
The predicted global signal-to-background ratio for the \WT-channel is 0.1 and the main background
contribution comes from the top pair production in the "lepton+jets" channel.

\begin{table}[phtb]
\begin{center}
\begin{tabular}{lccc}
\hline
Process                                &   Efficiency       &  Nb of events     &  Individual S/B ratio   \\[1mm]
\hline
\hline
W+t-channel                            &  $4.58\pm 0.02$    &  $12,852\pm 46$   &                         \\[1mm]
s-channel                              &  $0.20\pm 0.01$    &  $62\pm 1$        &  $206$                  \\[1mm]
t-channel                              &  $0.34\pm 0.01$    &  $2,572\pm 42$    &  $5$                    \\[1mm]
\hline
$\rm t\bar{t}$ background              &                    &                   &                         \\[1mm]
$\rm t\bar{t} \to e\nu b~jjb$          &  $3.33\pm 0.01$    &  $121,834\pm 331$ &  $0.1$                  \\[1mm]  
$\rm t\bar{t} \to e\nu b~e\nu b$       &  $0.27\pm 0.01$    &  $794\pm 18$      &  $16$                   \\[1mm]
$\rm t\bar{t} \to \tau\nu b~\tau\nu b$ &  $0.07\pm 0.01$    &  $206\pm 9$       &  $62$                   \\[1mm]
$\rm t\bar{t} \to \tau\nu b~jjb$       &  $0.22\pm 0.01$    &  $7,985\pm 121$   &  $1.6$                  \\[1mm]
\hline
W/Z+jets background                    &                    &                   &                         \\[1mm]
$\rm Wb\bar{b} \to e\nu~b\bar{b}$      &  $0.006\pm 0.001$  &  negl.            &  -                      \\[1mm]  
$\rm Wjj\to e\nu~jj$                   &  negl.             &  negl.            &  -                      \\[1mm]
$\rm WZ\to e\nu~b\bar{b}$              &  $0.044\pm 0.003$  &  negl.            &  -                      \\[1mm]
$\rm Zb\bar{b}\to e^+e^-~b\bar{b}$     &  $0.014\pm 0.002$  &  negl.            &  -                      \\[1mm]
\hline
\end{tabular}
\caption[]{Efficiency, number of events expected for an integrated luminosity of 30~\FB and 
individual signal-to-background ratio for single top processes and background channels. 
Uncertainties come from Monte Carlo statistics only.}
\label{TAB-WTCH-EFF}
\end{center}
\end{table}

\noindent
The signal-to-background ratio can be slightly improved (by a factor of 10\%) by applying further cuts on 
the total transverse momentum and on the centrality variable defined as:
$$\rm centrality = {\Sigma_{jet} ~p_T^{jet} \over \Sigma_{jet} ~p^{jet} }$$ 
As we can clearly see in Fig.~\ref{FIG-WTCH-C}, centrality values are much larger for the \WT events than for most
of background events.

\TRIFIG{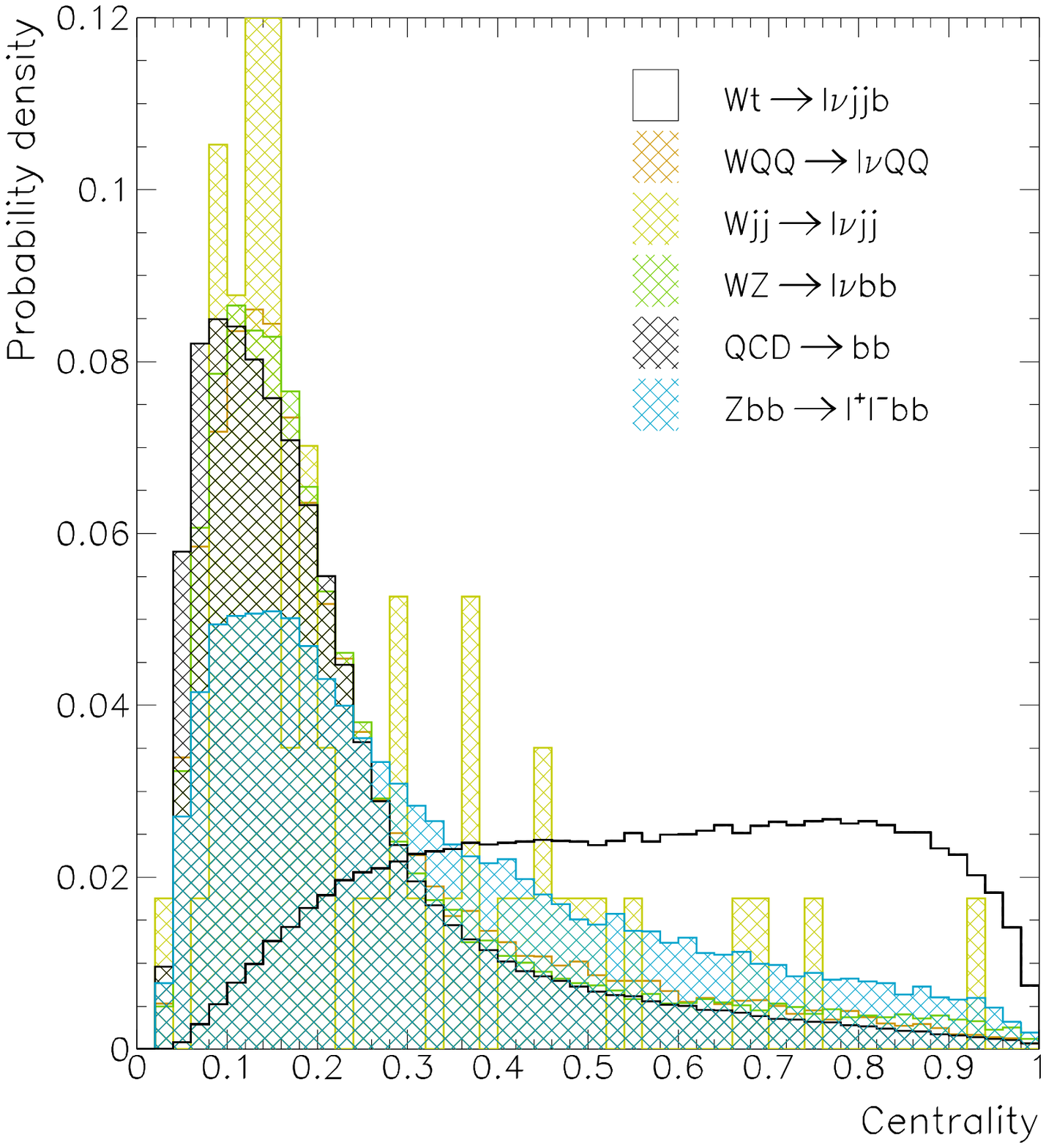}{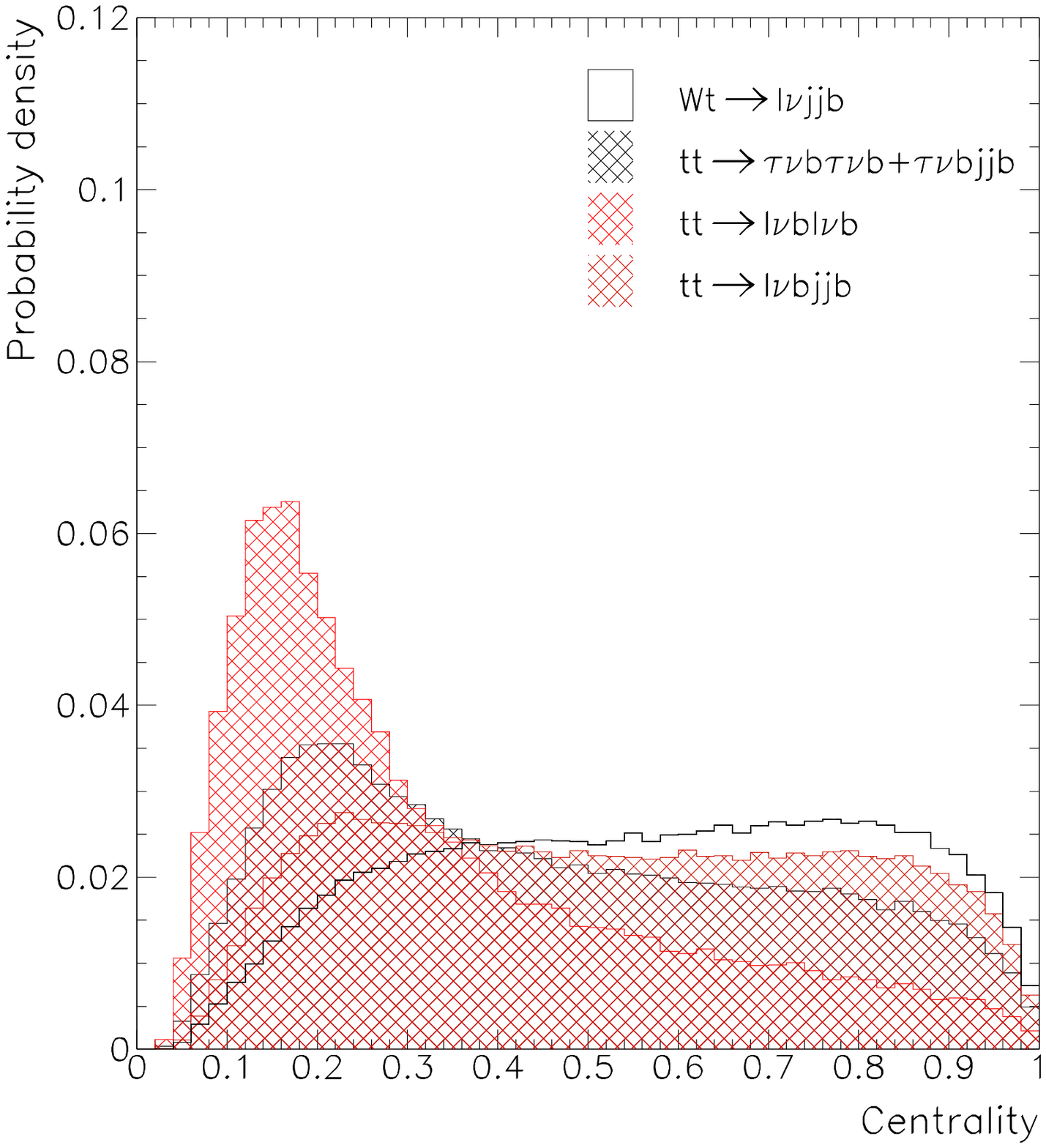}{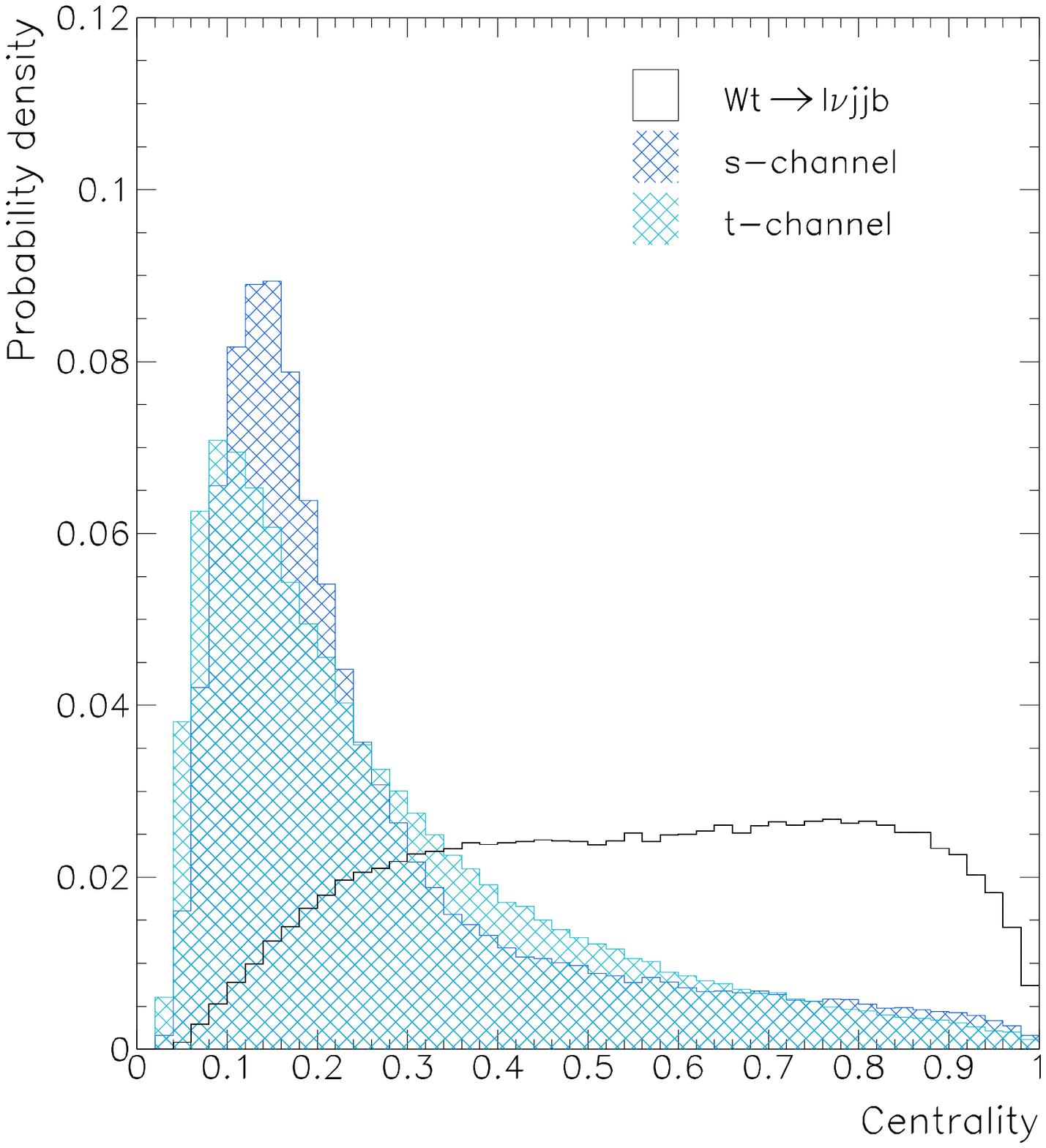}{Event centrality for signal and backgrounds.}{FIG-WTCH-C}

\noindent
Figs.~\ref{FIG-WTCH-PTS} and \ref{FIG-WTCH-CS} display the event yields 
for the total transverse momentum and centrality for an integrated luminosity of 30~\FB, 
respectively.
\DBLFIG{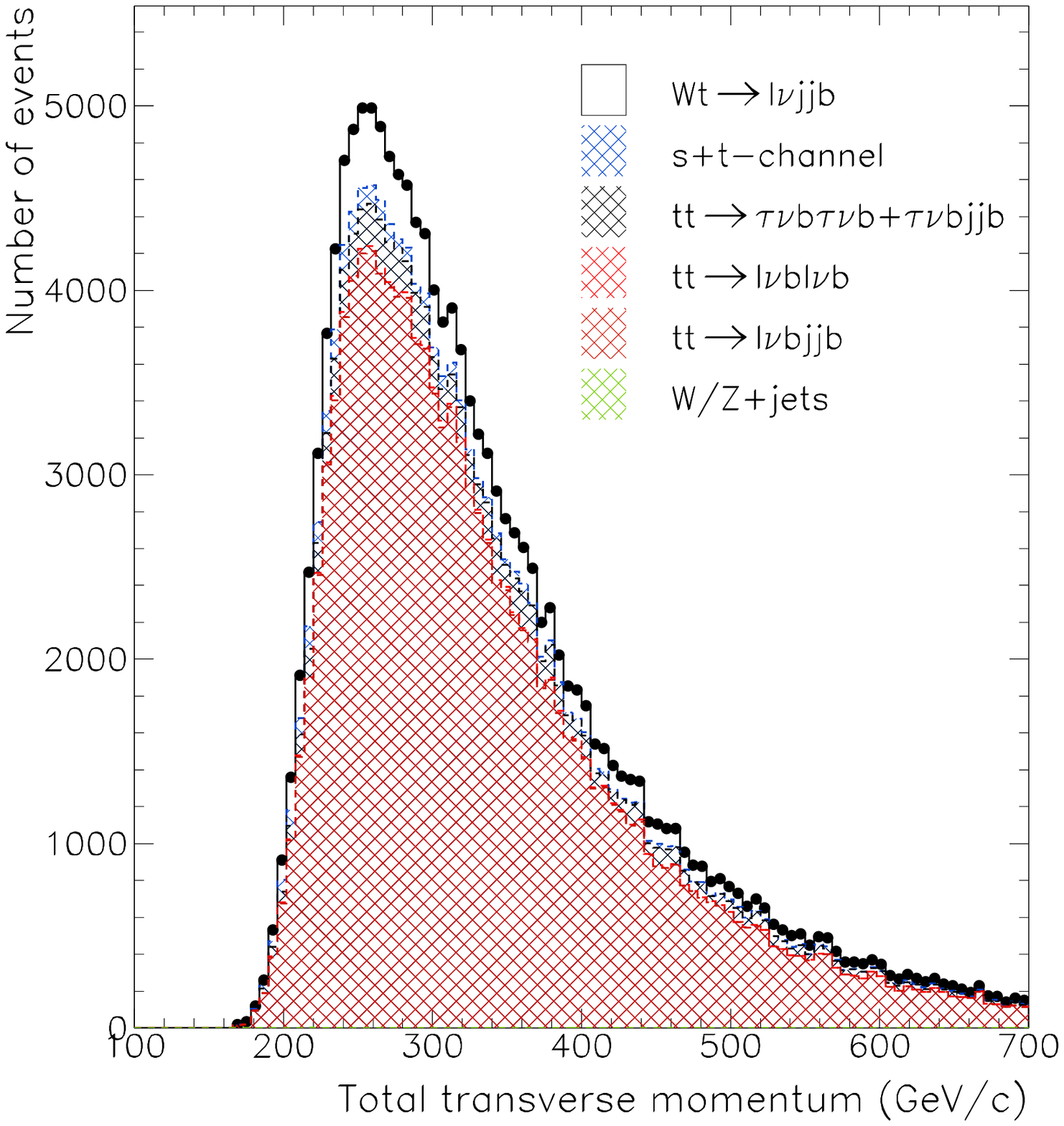}{Event yields for the total transverse momentum distribution for 30~\FB.}{FIG-WTCH-PTS}{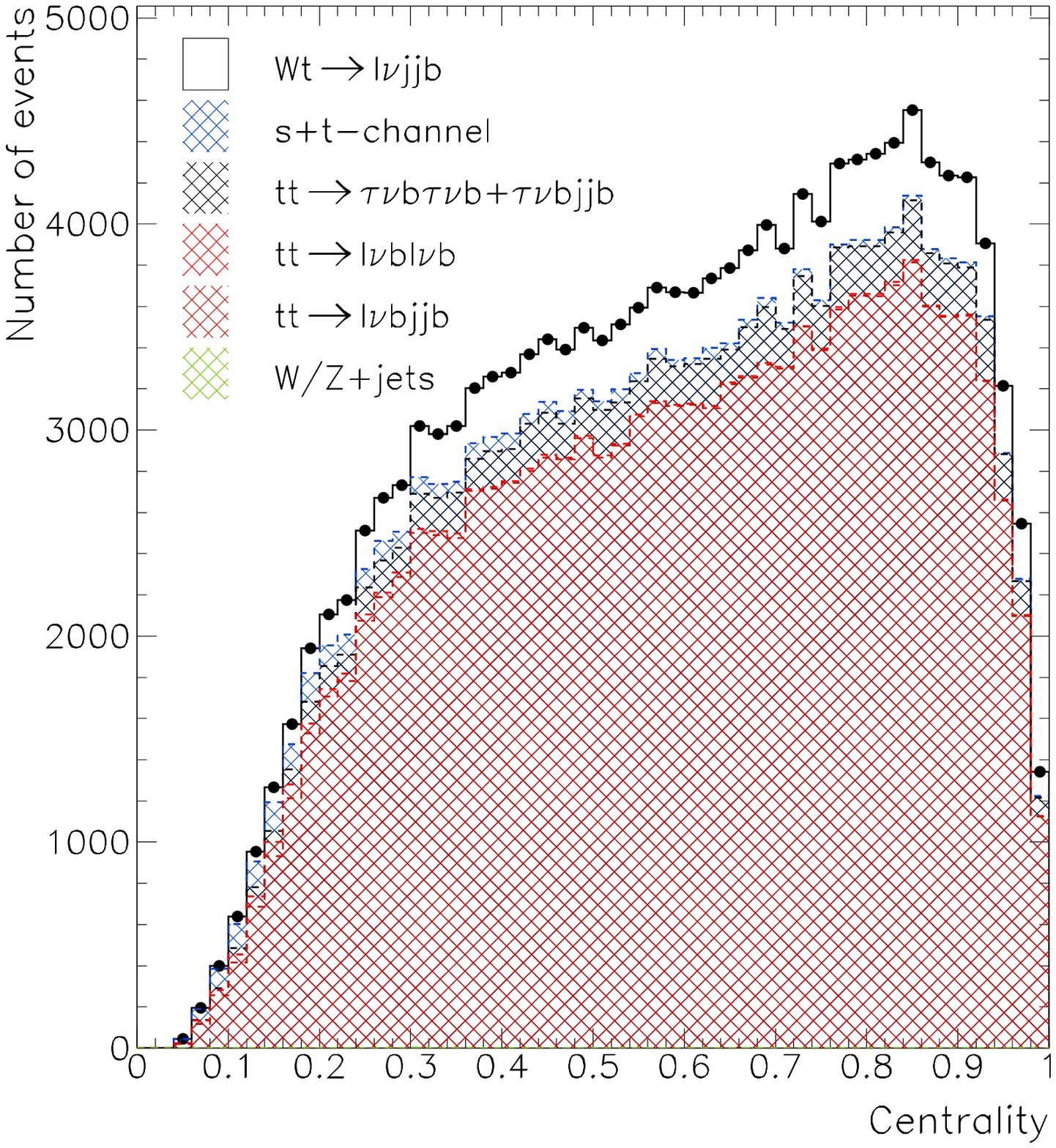}{Event yields for the centrality distribution for 30~\FB.}{FIG-WTCH-CS}

\noindent 
The performance in terms of statistical sensitivity has been determined for the 
three jet final state events and is shown as a function of the integrated luminosity 
in Fig.~\ref{FIG-WT-STAT}. A 10\% sensisitivity can be achieved with 1~\FB by combining 
both electron and muon channels.
\FIGVIII{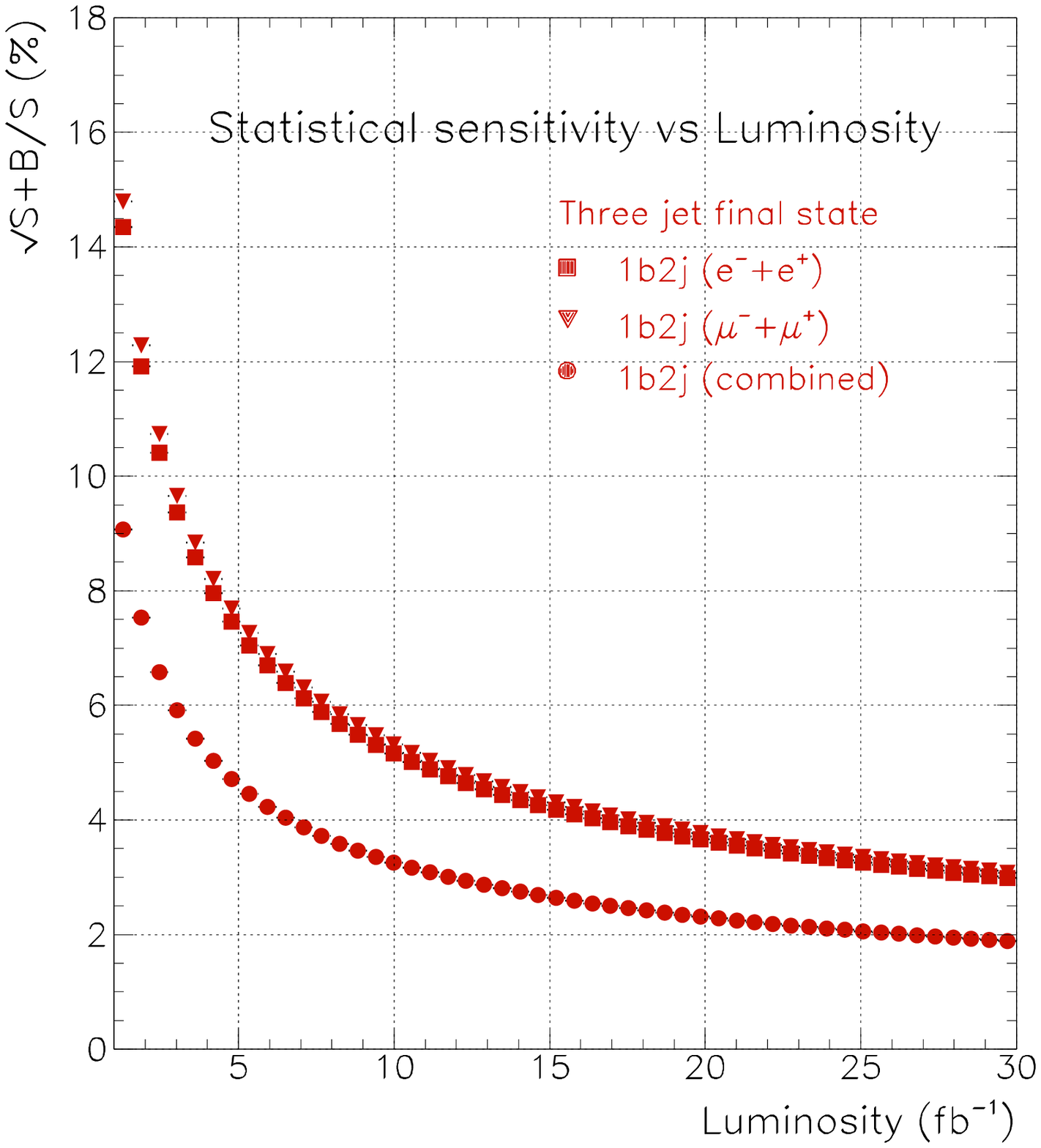}{Statistical precision as a function of the luminosity 
for the Wt-channel analysis in three- jet events.}{FIG-WT-STAT}

\subparagraph{Summary: Wt channel cross-section measurements}
\hfill\break
\hfill\break
The \WT channel analysis benefits from the relative high cross-section
of about 70~pb. However, due to high similarities with top pair events,
the selection is hampered by a high level of background contamination. This
characteristics makes the Wt cross-section very difficult to measure with
the early data at the LHC.
The chosen strategy is based on the splitting of the event selection into two
jet and three jet final states. In both cases, the main background comes from
the top pair production with a S/B ratio well below 10\%, making the prior
precise determination of the top pair production cross-section mandatory.
Combining both electron and muon channels as well as all two and three jet final
states leads to a statistical precision slightly below 6\% for an integrated
luminosity of 1~\FB. This translates into a precision of about 2-3\% at the
end of the low luminosity run.




%
%
%
%
\subsubsection*{Single top studies at CMS}
\label{sec:singletopcms}
\textbf{Contributed by: Giammanco, Slabospitsky}\\

This Section summarizes the CMS analyses published in the Physics 
TDR Vol.II and in Ref.~\cite{CMS-singletop-T} and~\cite{CMS-singletop-W}.
All results presented here assume 10 $fb^{-1}$ of
integrated luminosity, including the detector uncertainties 
that will be available at that time (as estimated in Ref.~\cite{PTDRv1}).

\paragraph{Signal and background event simulation}
\label{ewtop:generation}
\hfill\break
\hfill\break
Two generators, SingleTop~\cite{singletop} (based on the CompHEP
package~\cite{Boos:2004kh}) and TopRex~\cite{SGTOP-LHC-PHENO-TOPREX} were
used to generate events for all three single-top production
processes. The background processes, namely, $W\bbbar$,
$W\bbbar+j$, and $W+2j$ were generated with CompHEP, TopReX,
MadGraph~\cite{Maltoni:2002qb}, and Alpgen\cite{Mangano:2002ea}
programs as indicated in the Table~\ref{tab:st-1}.  The hard process
events containing all needed information were passed to
PYTHIA~6.227~\cite{Sjostrand:2001yu} for showering, hadronization and
decays of unstable particles.  The $\ttbar$ and $W + \mathrm{jets}$ background
events were generated with the same PYTHIA version.  All simulations
were done with $M_t = 175 \GEVCc$ and $M_b = 4.7-4.8 \GEVCc$, proper
considerations of the spin correlations, and the finite $W$-boson and
t-quark widths.
The list of the signal and background process cross sections as well
as generators used are given in the Table~\ref{tab:st-1}.  Both the
full simulation chain (OSCAR and ORCA) and a
fast simulation (FAMOS) were used.
\begin{table}[h]
\begin{center}
\begin{tabular}{|l|l|l|}
\hline
Process                & $\sigma \times$BR, pb & generator \\ \hline
t-ch.     ($W\to \mu\nu$)            & 18~(NLO)   & SingleTop (NLO)  \\ \hline
t-ch.     ($W\to \ell\nu$)           & 81.7~(NLO) & TopReX (NLO)  \\ \hline
s-ch.     ($W\to \ell\nu$)           & 3.3~(NLO)  & TopReX (NLO)    \\ \hline
$tW$ (2 $W\to \ell\nu$)                & 6.7~(NLO)  & TopReX (NLO)     \\ \hline
$tW$ (1 $W\to \ell\nu $)               & 33.3~(NLO) & TopReX (NLO)     \\ \hline 
\ttbar (inclusive)                      & 833~(NLO) & PYTHIA (LO)    \\ \hline
$W \bbbar$  ($W\to \ell\nu$)           & 100~(LO)   & TopReX (NLO)    \\ \hline
$W \bbbar + \mathrm{jets}$ ($W\to \mu$)& 32.4~(LO)  & MadGraph (NLO)   \\ \hline
$W+2j$  ($W\to \mu\nu$)                & 987~(LO)   & CompHEP (NLO)    \\ \hline
$W + 2j$ ($W\to \ell\nu$)              & 2500~(LO)  & ALPGEN (LO)   \\ \hline
$Z/\gamma^{*}(\to \mu^+\mu^-) \bbbar$  & 116~(LO)   &  CompHEP (NLO)    \\ \hline
\end{tabular}
\end{center}
\caption{
Cross section values (including branching ratio and kinematic cuts)
 and generators for the signal and background processes (here
$\ell = e,\mu,\tau$). Different generator-level cuts are applied.}
\label{tab:st-1}
\end{table}

\paragraph{Reconstruction algorithms and triggers}
\label{ewtop:algorithms}
\hfill\break
\hfill\break

A detailed description of the reconstruction algotithms and triggers
used in the single top studies can be found in 
Ref.~\cite{PTDRv1}. A short description is included below.
Muons are reconstructed by using the standard algorithm combining
tracker and muon chamber information; isolation criteria are based on tracker and calorimeter information.
 The electrons are
reconstructed by combining tracker and ECAL
information.  The jets are
reconstructed from the hadronic calorimeter signals by the Iterative Cone algorithm with the cone size of
0.5; for the calibration both the
Monte Carlo (in the t-channel analysis) and the $\gamma + jets$ (in
the $tW$- and $s$-channel) methods are used.  For b-tagging a probability algorithm
based on the impact parameter of the tracks is used.

The {\bf transverse missing energy} is reconstructed as follows:
\begin{eqnarray}
\vec{\MET} = - \left(\sum{\vec P}_T^\mu +  \sum {\vec E}_{T}^{tower} + \sum
( {\vec E}_{T,jet}^{calib}) - \sum ({\vec E}_{T,jet}^{raw}) \right)
\label{eq:ewtop_etmis}
\end{eqnarray}
where $E_{T}^{tower}$ is the sum of transverse energy of towers,
$E_{T,jet}^{calib}$ ($E_{T,jet}^{raw}$) is the transverse energy of
calibrated (uncalibrated) jets.
For the final states with one isolated lepton 
the neutrino (\MET) longitudinal component,
$P_{z, \, \nu}$ , is extracted from the quadratic equation:
\begin{eqnarray}
M^2_{W} = 2 \left( E_{\mu} \sqrt{ P^2_{z, \, \nu} + (\MET)^2}
 - \vec{P}_{T, \, \mu }\cdot  \vec{\MET}
 - P_{z, \, \mu }  P_{z, \, \nu} \right)
\label{eq:ewtop_03}
\end{eqnarray}
This equation has two solutions:
\begin{eqnarray}
P^{(1,2)}_{z, \, \nu} =
\frac{A P_{z, \, \mu} \pm 
\sqrt{\Delta} }{P^2_{T, \, \mu}},
\:\:\: {\rm where} \;\;
A = \frac{M_{W}^{2}}{2} + \vec{P}_{T, \, \mu }\cdot  \vec{\MET}, 
\:\:  \Delta = E_{\mu}^{2}(A^{2} - (\MET)^2 P^{2}_{T , \mu})
\label{eq:ewtop_04}
\end{eqnarray}
Among the two solutions of Eq.~(\ref{eq:ewtop_03}) the minimal value
of $|P_{z, \, \nu}|$ is used for $W$-boson momentum reconstruction.

About 30\% of the events have negative $\Delta$ values due to the
finite detector resolution and to the presence of extra missing
energy.  In this case for t-channel analysis the parameter $M_{W}$ in
Eq.~(\ref{eq:ewtop_04}) is increased until $\Delta$ becomes zero.
Using this value of $M_{W}$, $P_{z, \nu}$ is calculated from
Eq.~(\ref{eq:ewtop_04}).  For the $tW$ and s-channels analyses, only
the real part of $P_{z, \, \nu}$ is used for further analysis.

The {\bf transverse mass of the $W$-boson } is defined as
\begin{eqnarray}
 M_{T}^{W} = \sqrt{ 2(P_{T , \mu}\MET -  \vec{P}_{T, \, \mu } \cdot\vec{\MET})}.
\label{eq:ewtop_06} 
\end{eqnarray}
The {\bf sum of the transverse momentum vectors } of all reconstructed
objects 
\begin{eqnarray}
 \vec{\Sigma}_T \equiv \vec{P}_{T, \, \ell } +  \vec{\MET}  +
    \sum \vec{E}_{T, jet},
  \label{eq:ewtop_sigma} 
\end{eqnarray}
is found to be very effective for signal/background separation.

The {\bf ``jet charge''} ($Q_j$) is defined as the sum of the charges
of the tracks inside the jet cone, weighted over the projections of
the track momenta along the jet axis.

The {\bf lepton isolation} criterion used is to sum the \pt\ of all
the tracks in a cone of $\Delta R<0.2$ around the lepton track, and to
reject the event if this sum is greater than 5\% of the lepton \pt.

The present study is based on leptonic decay channels ($e \nu_e$ or
$\mu \nu_{\mu}$) of the $W$-boson. The signal is triggered by the
trigger on leptons.  The HLT \pt\ thresholds from the CMS
DAQ-TDR are assumed: 19 \GEVC (29 \GEVC) for the single muon
(electron); with $|\eta_{\mu}|\le2.1$ and $|\eta_{e}|\le2.4$.

\begin{figure}[!Hhtb]
\centering
\includegraphics[width=0.49\textwidth]{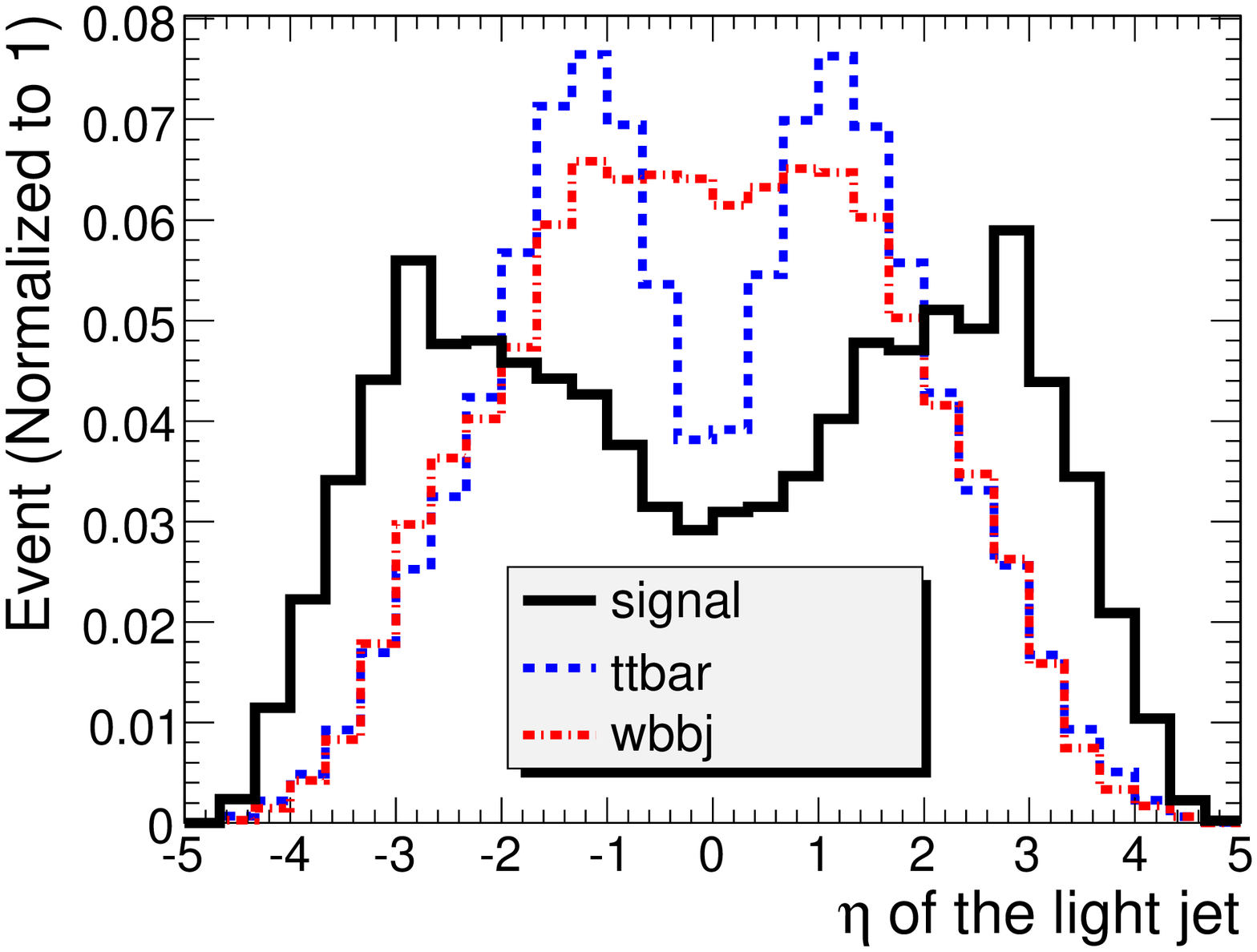}
\includegraphics[width=0.49\textwidth]{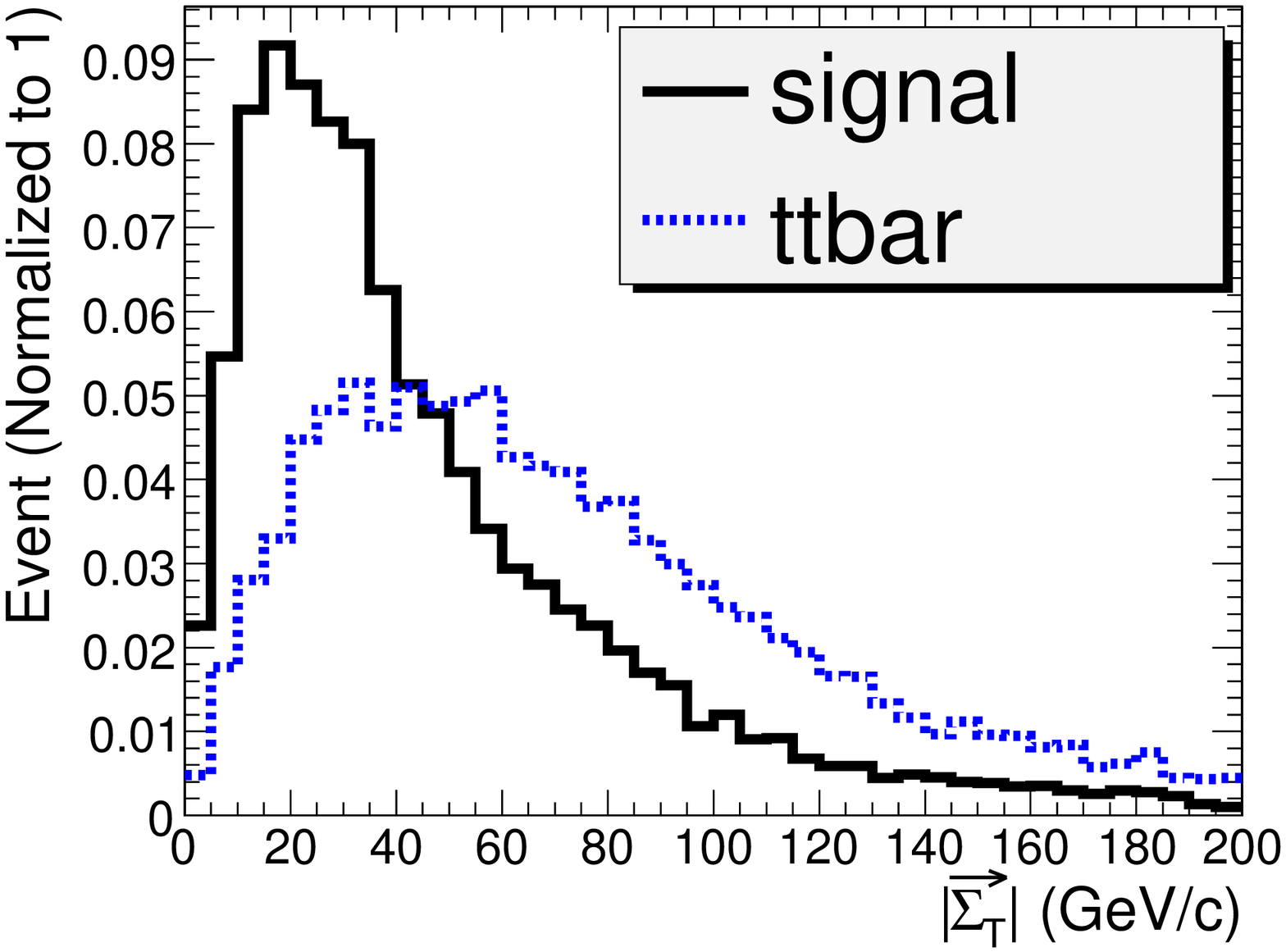}
\caption{The distributions of 
pseudorapidity ($\eta$) of the light jet (left), and of $|\vec{\Sigma}_T|$ (right).
\label{fig:st_t-TCHJets}}
\end{figure}

\paragraph{t-channel cross section measurement}
\label{sm_top:Single_Top_tchannel}
\hfill\break
\hfill\break
The analysis presented in Ref.~\cite{CMS-singletop-T} makes use of muonic decays of the top.
The final state in t-channel includes one isolated muon, missing
energy (neutrino), one or two jets from $b$-quarks, and
one ``forward'' hadronic jet. A specific feature of single top events
is production of a light jet in the forward/backward direction (see
Figs.~\ref{fig:st_t-TCHJets}) providing an additional possibility for
background suppression.
The additional $b$-quark is produced with
small transverse momentum, so this analysis requires only two jets, one of them $b$-tagged.

The selection requires:
\begin{itemize}
\item only one isolated muon with \pt $> 19 \GEVC$ and $|\eta_{\mu}| < 2.1$ (HLT selection);
\item $\MET > 40 \GeV$;
\item at least two hadronic uncalibrated jets, with \pt$ > 20 \GEVC$;
\item at least one of the selected jets should pass the $b$-tag;
\item the second (light) jet should be in the forward region ($|\eta(L)| > 2.5$);
\item after calibration these two jets must have \pt$^{\rm calib} \geq 35 \GeV$ and no
other hadronic jets with \pt$^{\rm calib} \geq 35 \GEVC$ is allowed (jet veto).
\end{itemize}

The GARCON program~\cite{GARCONMAIN} is used for further optimization
of the cuts. The signal-over-background ratio times significance is
chosen as an optimization criterion, obtaining:
\begin{itemize}
\item
b-jet: \pt$ > 35.0 \GEVC$, $|\eta| < $ 2.5 and $b$-tag discriminator 
$ > 2.4$;
\item
light forward: \pt$ > 40.0 \GEVC$ and $|\eta| > $ 2.5;
\item $|\vec{\Sigma}_T|$ cut window: (0.0, 43.5)~GeV;
 $50 < M_{T}^{W} < 120 \GEVCc$
\item reconstructed top mass window:
 $110 \GEVCc < M_{\rm rec}(b W) < 210 \GEVCc$.
\end{itemize}

\begin{table}[h]
\begin{center}
\begin{tabular}{|l|c|c|c|c|c|}
\hline
  & signal&$\ttbar$  &  $W \bbbar j$ & $Wj$ & $Wjj$ \\ \hline
 N(events) at 10 $fb^{-1}$  &$1.8 \times 10^5$ &$8.33 \times 10^6$&
 $3.24\times 10^5$ &$9.7 \times 10^7$ & $9.9 \times 10^5$ \\ \hline \hline
  isolated muon 
          & 0.73 &0.14   &0.52  & 0.16    &0.81    \\ \hline
 $p_{TB} \times p_{Tj} \times \MET$
  & 0.036 &$6.4\times10^{-3}$ &$3.4\times10^{-3}$ & $9\times10^{-6}$ &$3\times10^{-3}$
   \\ \hline
 veto on $3^{rd}$ jet
  & 0.021 &$5.8\times10^{-4}$ &$1.6\times10^{-3}$& $4\times10^{-6}$ &$1.1\times10^{-3}$
  \\ \hline
 $0.0< \Sigma_T < 43.5 \GeV$
  & 0.018 &$4.1\times10^{-4}$ &$1.2\times10^{-3}$& $4\times10^{-6}$ &$6.8\times10^{-4}$
  \\ \hline
 $50 < M_{T}^{W^\ast} < 120$
  & 0.015 &$2.2\times10^{-4}$ &$9.6\times10^{-4}$& $1\times10^{-6}$
 &$5.4\times10^{-4}$\\ \hline
 $110 < M_{\rm rec}(bW)^\ast < 210$
  & 0.013 &$1.4\times10^{-4}$ &$5.8\times10^{-4}$& $0$          &$4.1\times10^{-4}$\\ \hline
\hline
 Number of events   & 2389  &1188    &195    &0        &402     \\ \hline
 \multicolumn{6}{l}{$^\ast$ in \GEVCc}
\end{tabular}
\end{center}
\caption{Number of events (t-channel) and cumulative efficiencies
         for each cut used in the analysis of t-channel single top production.
         The symbol ``$p_{TB} \times p_{Tj} \times \MET$'' means:
         $p_{TB} > 35 \GEVC$, $p_{Tj} > 40 \GEVC$, $|\eta_j| > 2.5$,
         $\MET > 40 \GeV$.}
\label{tab:st_t-EW_efficiency}         
\end{table}

\begin{figure}[!Hhtb]
\includegraphics[width=0.49\textwidth]{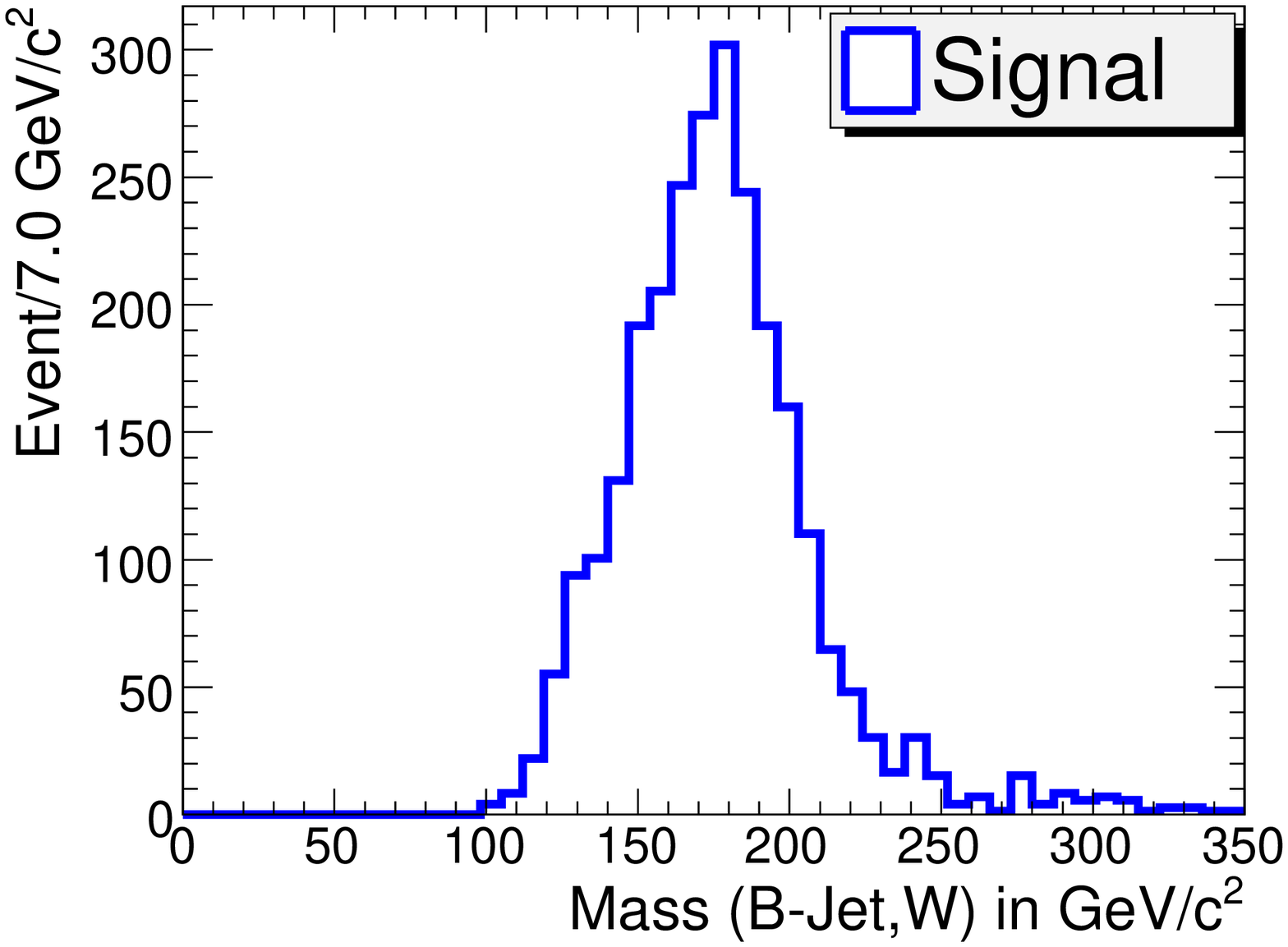}
\includegraphics[width=0.49\textwidth]{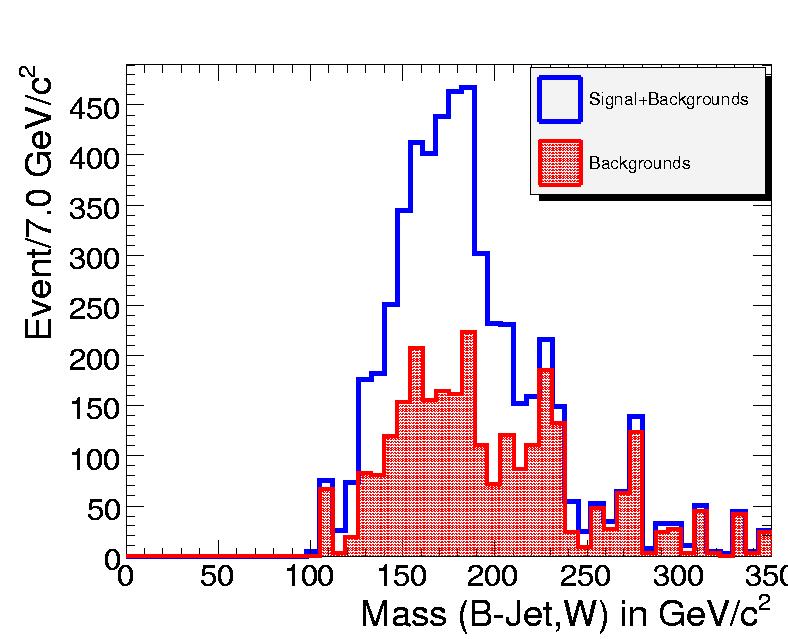}
\caption{The distribution on the reconstructed top mass, for signal
 only (left) and  with background included (right).}
\label{fig:st_t-TCHMass}
\end{figure}

The efficiencies of these cuts and the resulting number of events are
given in the Table~\ref{tab:st_t-EW_efficiency}.
The resulting signal-to-background ratio and the significance are:
${N_S}/{N_B} = 1.34$ and $S_{stat} = {N_S}/{\sqrt{N_S + N_B}} = 37.0$.
The final distribution of the reconstructed top mass is shown in
Fig.~\ref{fig:st_t-TCHMass}.

\begin{table}[h]
   \begin{center}
\vspace{2mm}
\begin{tabular}{|l|c|c|c|c|c|c|}
\hline
 sample           & selected & $\Delta N_{th}$ &
 JES & $\Delta N_{\rm b-tag}$ & $\Delta N_{\rm syst}$&
 $\Delta N_{\rm stat}$ \\\hline
 t-channel    & 2389 & 96  & 71  & 96  & 153 &  49 \\ \hline \hline
 \ttbar         & 1188 & 59  & 73  & 48  & 105 &  34  \\ \hline
 $W \bbbar j$ &  195 & 33  &  6  &  8  &  35 &  14  \\ \hline
 $W j j    $    &  402 & 20  &  0  & 16  &  26 &  20  \\ \hline
\end{tabular}
\end{center}
\caption{Number of selected events (t-channel) at 10 $fb^{-1}$ with
uncertainties due  to different sources.
$\Delta N_{\rm syst}$ represents 
the theoretical, JES and $b$-tagging  uncertainties.
$\Delta N_{\rm stat}$ is expected statistical uncertainty. }
\label{tab:st_t-TCH_sigbkg}
\end{table}

The systematic uncertainties (see Section~\ref{ewtop:syst}) evaluated
for 10 $fb^{-1}$ are given in Table~\ref{tab:st_t-TCH_sigbkg}.  In
summary, the statistical error is 2.7\%, the total systematic error
excluding the 5\% luminosity uncertainty is 8\%, resulting in a total
error of 10\%.


\paragraph{$W$t associated production cross section measurement}
\label{sm_top:Single_Top_tWchannel}
\hfill\break
\hfill\break
The $pp \to tW$ process contains two $W$-bosons and a $b$-quark in the
final state. 
The final states considered in Ref.~\cite{CMS-singletop-W} are $\ell^+\ell^-\MET b$ and $\ell^\pm\MET b$jj for the
di-leptonic and semi-leptonic modes, respectively.  The dominant
background arises from $\ttbar$ production.  Other backgrounds are
t- and s-channel single top production, 
$W \bbbar$, $W + \mathrm{jets}$, $WW + \mathrm{jets}$, and to a lesser extent 
QCD multi-jet background.

\subparagraph{Jet quality requirements and extra jet reduction}
\label{sec:extra_jets}
\hfill\break
\hfill\break
The most significant difference between $tW$ events and $\ttbar$ events
is the number of jets in the final state.  
However, most of the time there are also additional jets due to the
underlying event, pile-up or calorimeter noise.  These ``extra jets''
were identified and excluded from the counting by consideration of
five jet quality variables (see~\cite{CMS-singletop-W}).  It was
found that the most discriminating variables are $\ET^{max}$ (the
maximum tower \ET in a cone of 0.5) and $N_{track}$ (the number of
associated tracks).  A Fisher discriminant~\cite{FISHER} ($F$) is
constructed from the jet quality variables to separate real jets from
extra jets.  Each jet is classified 
value $F$ into one of three categories: good ($F < -0.5$), loose ($|F|
< 0.5$) and bad ($F \ge 0.5$) jets.  This method yields $84.3\%$
efficiency on true jets and rejects $86.9\%$ of extra jets.
Only ``good'' jets and ``loose'' jets are used in preselection and
event reconstruction.  The jet multiplicity after the extra jet reduction
in semi-leptonic channels 
reveals that the number of good jets peaks at the 2 and 3
jet bins for signal events, and at the 3 and 4 jet bins for $\ttbar$
backgrounds.

\subparagraph{Event selection and reconstruction}
\hfill\break
\hfill\break
The kinematic cuts used for this study are presented in
Table~\ref{tab:st_Wt-dil_kine} and Table~\ref{tab:st_Wt-semil_kine}.
For the semi-leptonic channel,
two non-$b$-like jets with $m_{jj} < 115 \GEVCc$ are used for
reconstruction of the $W$-boson (that decays hadronically).  
In events with a 4th jet that survives jet veto cuts,
it is required that the invariant mass of the 4th jet
with any of the selected non-$b$-like jets must be outside a window of $M_W
\pm 20 \GEVCc$.  For the leptonic decays of the $W$-boson it is required
that $M_T^W < 120 \GEVCc$.

\begin{table}[htb]
    \begin{center}
      \begin{tabular}{|l|l|} \hline
        {\bf Leptons} & {\bf Jets} \\   \hline
 $|\eta(e)| < 2.4$, $|\eta(\mu)| < 2.1$ 
    & leading jet: $|\eta|<2.4$, \pt $> 60 \GEVC$, disc $>$ 0 \\
 \pt$(e, \mu) > 20 \GEVC$ 
    & at most one extra jet \\
        no other lepton with \pt$ > 5 \GEVC$ &
        No other jets with \pt$ > 20 \GEVC$ \\
        \hline
 {\bf Missing \ET}: $\MET > 20 \GeV$ & \\ 
        \hline
      \end{tabular} \\
    \end{center}
    \caption{Kinematic cuts used in the di-leptonic channel. 
The final electron and muon should have the opposite charges.}
 \label{tab:st_Wt-dil_kine}
  \end{table}

\begin{table}[htb]
    \begin{center}
      \begin{tabular}{|l|} \hline
        {\bf Leptons} \\        \hline
        \begin{tabular}{l} 
          \pt$(e) > 30 \GEVC$, \pt$(\mu) > 20 \GEVC$, 
          $|\eta(e)| < 2.4$, $|\eta(\mu)| < 2.1$ \\
          no other lepton \pt$ > 10 \GEVC$ \\
        \end{tabular} \\        \hline
        {\bf Jets} (after removing all bad quality jets) \\ \hline
        \begin{tabular}{l}
         $b$-like jet: good quality, disc$>$2, $|\eta|<2.5$, \pt$ > 35 \GEVC$ \\
         non-$b$-like jet: good quality, $|\eta|<3.0$, disc$<$0 if $|\eta|<2.5$, \pt$>35 \GEVC$\\
          Jet counting: one $b$-like jet and 2 non-$b$-like jets \\
  Jet veto: no other ``good'' or ``loose'' jets with \pt$ > 20 \GEVC$ and $|\eta| < 3$\\
        \end{tabular} \\
        \hline
        {\bf Missing \ET}: $\MET > 40 \GeV$ \\ 
        \hline
      \end{tabular} \\
    \end{center}
    \caption{Kinematic cuts used in the semi-leptonic channel.
The presence of a good fourth jet would veto the whole event. } 
 \label{tab:st_Wt-semil_kine}
\end{table}

To find the correct pairing of $b$-jet and reconstructed $W$-boson
(coming from top decay) the following variables were used: the \pt\
of ($b$, $W $) systems; the separation of the $b$-jet with each of the
$W $ in $(\eta,\phi)$ space; the ``charges'' of jets
(see Section~\ref{ewtop:algorithms}) and $W$-bosons (see
Ref.~\cite{CMS-singletop-W} for details).
A Fisher discriminant based on these variables
is used for discriminating leptonic top events from
hadronic top events.  A cut of of 0.56 is optimal in separating these
2 types of events, and 72\% of the events are correctly paired.

To further enhance the signal to background ratio the following
``global'' cuts are applied:
\begin{itemize}
\item \pt of the reconstructed $t W $ system: $|\vec{\Sigma}(t+W )| < 60 \GEVC$.
\item Scalar sum of transverse energies $H_T$: $H_T < 850 \GeV$.
\item Reconstructed top quark mass: $110 \GEVCc < m(t) < 230 \GEVCc$.
\item \pt of the reconstructed top quark: $20 \GEVC <$ \pt$(t) < 200 \GEVC$.
\end{itemize}

\subparagraph{Efficiencies and expected yields}
\hfill\break
\hfill\break
The efficiencies estimated with Monte Carlo samples are 
converted to the effective cross sections by multiplying the
production cross sections of each process.  The effective cross
sections, as well as the expected yields with 10 $fb^{-1}$ of data for
all signal and background samples, are shown in
Table~\ref{tab:st_Wt-effi_dil} and~\ref{tab:st_Wt-effi_semil}.  The
signal to background ratio is found to be 0.37 for di-leptonic channel and 0.18 for semi-leptonic channel.

\begin{table}[htb]
    \begin{center}
    \begin{tabular}{|c| r| r|r| r|r| r| r|} \hline
        & $t W$ dil. &  $\ttbar$ dil. & $\ttbar$ oth. 
      & WW dil. & WW oth. &  t ch. lept. \\ \hline
Production & 6.667 &    92.222 &        737.778 & 11.111 & 88.889 &     81.667\\ \hline 
HLT & 4.865 &   74.090 &        346.151 &       7.674 & 27.259 &        41.409\\ \hline 
2 $\ell$ & 1.944 &      25.150 &        21.012 &        2.574 & 0.226 & 2.309\\
Lepton \pt & 0.675 &  7.919 & 0.703 & 0.543 & 0.012 & 0.098\\ \hline 
$\le 1$ extra jet & 0.459 &     6.574 & 0.664 & 0.416 & 0.010 & 0.067\\ 
Jet \pt, $\eta$ & 0.307 &     5.234 & 0.556 & 0.339 & 0.004 & 0.033\\ 
$\ge 1\ b$-jet & 0.184 &        3.864 & 0.379 & 0.017 & 0.000 & 0.018\\ 
$\MET > 20$ & 0.170 &   3.640 & 0.349 & 0.017 & 0.000 & 0.016\\ 
$\le 2$ jet & 0.150 &   2.734 & 0.221 & 0.015 & 0.000 & 0.012\\ \hline 
Final select. & 0.057 & 0.145 & 0.000 & 0.006 & 0.000 & 0.000\\ \hline 
Expected events & 567 & 1450 &  $\le 55$ &      61 &    $\le 10$ &      $\le 20$\\ \hline
    \end{tabular}
    \end{center}
    \caption{Summary of cross section times branching ratio times efficiencies
    at each stage of the analysis for the di-leptonic channel.
    All values are in picobarns 
    The last row is the expected number of events for 10 $fb^{-1}$. 
    Multi-jet background has been estimated separately. 
    When only a limit on the number of events is stated, this is due to MC statistics.}
\label{tab:st_Wt-effi_dil}
\end{table}

\begin{table}[htb]
\begin{footnotesize}
    \begin{center}
      \begin{tabular}{|l|c|c|c|c|c|c|c|c|c|} \hline
        & tW  &  $\ttbar$ & t ch. & s ch.
        & Wbb &  W2j & W3j & W4j  & Multi-jet \\
        \hline
        Total cross section  & 60 & 833 & 245 & 10 
                & 300 & 7500 & 2166 & 522 & $9.73\times 10^9$ \\
        \hline
        HLT     & 18.9 & 263.9 & 39.5 & 1.52 
                & 34.0 & 1006 & 300 & 73 & $1.86\times 10^5$ \\
        Presel. \& isolation & 9.05 & 179.4 & 12.0 & 0.54 
                & 2.15 & 52 & 35 & 12 & 1325 \\
        \hline
        \begin{tabular}{l} jet \& lepton \pt, \\ jet veto \end{tabular} & 1.28 & 18.5 & 1.31 & 0.046 
                & 0.061 & 0.60 & 4.9 & 1.0 & 4.23 \\
        $b$-tagging & 0.669 & 6.13 & 0.476 & 0.013 
                & 0.016 & 0.10 & 0.99 & 0.26 & 0.85 \\
        kinematic cuts 
                & 0.223 & 0.999 & 0.047 & 0.002 
                & 0.003 & 0.017 & 0.101 & 0.008 & 0.105 \\
        Signal box cuts & 0.170 & 0.771 & 0.035 & 0.001 
                & 0.001 & 0.013 & 0.054 & 0.008 & 0.051 \\      \hline
        Events in 10 $fb^{-1}$  & 1699 & 7709 & 351 & 14 
                & 10 & 130 & 539 & 80 & 508 \\  \hline
      \end{tabular} 
    \end{center}
    \end{footnotesize}
    \caption{Summary of cross section times branching ratio times efficiencies
    at each stage of the analysis for the semi-leptonic channel.
    All values are in picobarns. 
    The last row is the expected number of events for 10 $fb^{-1}$.}
    \label{tab:st_Wt-effi_semil}
\end{table}

\subparagraph{The ratio method}
\hfill\break
\hfill\break
The {\it ratio method} is developed to reduce systematic uncertainties related 
to the dominant $\ttbar$ background.
We define a $\ttbar$-rich control region and use ratio of efficiencies
to estimate the yield of $\ttbar$ in the signal region.
The kinematics of $tW$ and $\ttbar$ are similar so $tW$ is present in the 
control region, therefore the ratio of efficiencies for $tW$ is also used.
The signal and background yield is determined by the following equation:
\begin{eqnarray}
S &=& { R_{t \bar t} (N_s-N_s^o) - (N_c-N_c^o)} \over {R_{t \bar t} -
 R_{tW}},\\
B &=& {(N_c - N_c^o) - R_{tW}(N_s - N_s^o)} \over {R_{t \bar t} - R_{tW}}
 + N_s^o.
\label{eq:st_Wt-ratio_method}
\end{eqnarray}
Here $R_x$ is the ratio of efficiencies 
$R_x = \varepsilon_x$(control region)$/\varepsilon_x$(signal region) for $x = t \bar t, 
tW$;
$N_s$ ($N_c$) is total number of events in the signal (control) region;
$N_s^o$ ($N_c^o$) is the estimated number of non-$\ttbar$ background 
events in the signal (control) region.

For the ratio method to work it is important to find a 
control region with similar kinematics except with one more jet.  
It is expected that systematic uncertainties from PDF, JES and b tagging
cancel to a large extend, while the luminosity uncertainty drops out
for the $\ttbar$ background.
The lepton selection and jet quality requirements in the control region 
is identical to the signal region.  The differences are outlined below.

{\bf Di-leptonic}. 
A second jet is required with \pt $= 20 - 80$~GeV, $|\eta| < 2.4$
  and $b$-tagged (disc $>$ 0). No other jets with \pt$ > 20$ GeV are allowed. 
The background region is found to be filled by 97.9\% di-leptonic 
$\ttbar$, 
0.4\% other $\ttbar$ decays, 1.6\% di-leptonic $tW$, 
and $0.1\%$ for leptonic t channel single 
top while WW+jets yield is negligible.

{\bf Semi-leptonic}. It
        requires 2 jets with \pt $> 30$ GeV, 2 more jets with \pt $> 20$ GeV,
        and no bad jets with \pt $> 20$ GeV.
It is required that one of the 2 high-\pt jets is b-tagged
        (disc $>$ 2), and that both low-\pt jets be not tagged 
        (disc $<$ 0). The $b-W $ pairing is done in the same way, with a 72\%
        correct pairing. 
It is found that the $\ttbar$ purity in the control region is 93.9\%.
The non-$\ttbar$ events are mainly
composed of W+jets (2.8\%), $tW$ (2.0\%) and t-channel single top (1.2\%).
The ratio of efficiencies are found to be $R_{tW}$ = 0.319 and
$R_{t \bar t}$ = 3.31.

The $\ttbar$ cross section does not show up in the ratio
method. 
The effect is 0.8\% for t-channel single top and 3.1\% for $W$+jets.
It is found to be negligible for other backgrounds.
The systematic uncertainties for both channels are shown in
table~\ref{tab:st_Wt-systematics}.

Particular care was dedicated to the estimation of the effect of pileup.
A difference of 30\% between normal pileup and no pileup is used as an
estimate of the systematic uncertainty. \\
\noindent $\diamond$ {\it Dileptonic mode} 
The analysis is found to be rather sensitive to the pileup, as the
relative shift of the ``measured'' cross section is $+20.4\%$ for no
pileup, and $-16.2\%$ for double pileup, while is the difference between
the check sample and the reference sample $4.6\%$ (which has purely
statistical origin).  The value of 6.1\% is used as the systematic
uncertainty. \\ 
\noindent $\diamond$ {\it Semi-leptonic mode}
The extracted cross section varies by $+$35\% for no pileup and
$-$63\% for double pile-up so a systematic uncertainty of 10.3\% is
obtained. 
This is clearly an overestimation of the effect.

  \begin{table}[htb]
    \begin{center}
      \begin{tabular}{|c|c|c|c|} \hline
{\bf Source} & {\bf Uncertainty} & {\bf $\Delta\sigma/\sigma$ (di-lept.)}
        & {\bf $\Delta\sigma/\sigma$ (semi-lept.)} \\   \hline
        {\bf Statistical uncertainty} & --- & 8.8\% & 7.5\% \\
        \hline
        Integrated luminosity & 5\% & 5.4\%  & 7.8\%  \\
        $\ttbar$ cross-section & 9\% & {\it negligible} & {\it negligible} \\
        t-channel  cross-section & 5\% & {\it negligible} & 0.8\% \\
        W+jets cross-section & 10\% & {\it not applicable} & 3.1\% \\
        WW+jets cross-section & 10\% &  1\% & {\it not applicable}  \\
        \hline
        Jet energy scale & 5\%-2.5\% 
           & 19.7 \% & 9.4\% \\
        \hline
        b tagging efficiency &  4\% - 5\%  & 8.7 \% & 3.6\%  \\
        \hline
        PDF & 1$\sigma$ & +4\%/-6.0\% & 1.6\%  \\
        Pileup & 30\% & 6.1 \% & 10.3\% \\
        MC statistics & --- & 9.9\% & 15.2\% \\
        \hline
        {\bf Total uncertainty} & &
     $\pm$23.9\%(syst.)                & $\pm$16.8\%(syst.)                 \\
                                & &
                       $\pm$~9.9\%(MC) &                    $\pm$15.2\%(MC) \\
        \hline
  \end{tabular}
    \end{center}
    \caption{Summary of uncertainties of cross section measurement.}
    \label{tab:st_Wt-systematics}
  \end{table}

The results from the ratio method were used in the significance
calculation.  In addition, the uncertainty on the background
expectation, evaluated for di-leptonic ($\Delta_B/B = \pm 9.6\%$) and
semi-leptonic ($\Delta_B/B = +3.6\%/-4.4\%$), was taken into
account. The resulting significance is 4.2 for the di-leptonic channel
and 5.1 for the semi-leptonic channel.  Combining the two channels
gives a total significance of $6.4$.

\paragraph{s-channel cross section measurement}
\label{sm_top:Single_Top_schannel}
\hfill\break
\hfill\break
The present analysis of the s-channel single top production is based
on leptonic channels, i.e. the top is identified and reconstructed by
its semileptonic decays into $\ell \nu b$ final states, with
$\ell=e,\mu$.  For this study, a fast simulation of the CMS detector
with FAMOS was used, see~\cite{CMS-singletop-T, CMS-singletop-W} for
details.

The signal events are triggered by the single lepton triggers.  Since
this production mode suffers from low statistics, one could envisage
the introduction of a combined trigger $e\times jet$, with threshold
19 \GEVC for the electron (in order to make the electronic sample more
coherent with the muonic sample) and 45 \GEVC for the jet. This value
has been chosen to be the same as the threshold for the $\tau$-jet in
the already existing $e\times \tau-jet$ trigger.

\subparagraph{Preselection}
\label{singletop_schannel:preselection}
\hfill\break
\hfill\break
The preselection criteria are as follows:
\begin{itemize}
\item The event has to fire at least one of the previously described triggers
(including the proposed $e\times j$).
\item The event must contain one isolated lepton ($\mu$ or $e$) with \pt$\ge19\GEVC$
and $|\eta|\le2.1 (\le2.4$) for muons (electrons)
and no other lepton above $10\GEVC$.
\item Exactly two uncalibrated jets must have \pt$\ge 30\GEVC$ and 
$|\eta|\le2.5$ and no other jet has to be present with \pt$\ge20\GEVC$.
\item Both jets should have a positive b-tagging discriminator
value.
\item The event should have $\MET > 30\GeV$.
\item The transverse mass of the $W$-boson $M_T^W$ should be less than $100 \GEVCc$.
\end{itemize}
Details on the effect of the preselection cuts
are given in
Table~\ref{tab:st_s-SCH_presel_eff}.  As before, the multi-jet QCD
contribution is neglected.

\begin{table}[h]
\begin{footnotesize}
   \begin{center}
\begin{tabular}{|l|c|c|c|c|c|}
\hline
 Cut     & s-ch.  & t-ch.  & $\ttbar$ & $Wb\bar{b}$ & $Wt$ (1 $W\to l\nu$) \\ \hline
``HLT''  &$37.5\pm 0.2\%$ & $42.5\pm 0.1\%$ & $30.1\pm 0.1\%$ & $29.4\pm 0.1\%$
   & $46.5\pm 0.1\%$ \\
Isolation  & $33.7\pm 0.2\%$ & $39.0\pm 0.1\%$ & $21.7\pm 0.1\%$ & $28.2\pm 0.1 \%$
 & $42.3 \pm 0.1\%$   \\
$\MET$ cut  & $27.3\pm 0.2\%$ & $31.9\pm 0.1\%$ & $17.4\pm 0.1\%$ & $22.6\pm 0.1 \%$
 & $34.4\pm 0.1\%$  \\
$M_T^W$ cut & $23.2\pm 0.2\%$ & $26.3\pm 0.1\%$ & $13.6\pm 0.1\%$ & $18.4\pm 0.1 \%$
  & $29.2 \pm 0.1\%$\\
$N_j\ge 2j$  & $11.9\pm 0.1\%$ & $11.5\pm 0.1\%$ & $11.9\pm 0.1\%$
  & $0.88\pm 0.03 \%$   &  $18.5 \pm 0.1\%$ \\
$N_j = 2j$  & $8.9\pm 0.1\%$ & $8.2\pm 0.1\%$ & $1.84\pm 0.04\%$ & $0.76\pm 0.03 \%$
  &  $7.09\pm 0.05\%$ \\
$b$-tag  & $3.07\pm 0.07\%$ & $0.72\pm 0.02\%$ & $0.28\pm 0.02\%$ &$0.14\pm 0.01 \%$
 &  $0.34 \pm 0.01\%$ \\\hline
$N_{\rm ev}$ &$1010 \pm 10$ & $5880 \pm 70$ & $23300 \pm 200$& $1400\pm 35$
  & $1150 \pm 40$ \\ \hline
\end{tabular}
\end{center}
\end{footnotesize}
\caption{Efficiencies of the preselection cuts, with respect to the
initial number of events.
For all process (except of $t \bar t$) the final $W$ decays into
charged lepton ($\ell=e,\mu,\tau$) and neutrino.  ``HLT'' includes the
$1\mu$, $1e$ and $e\times j$ triggers.  $N_{ev}$ is the number of
events surviving these cuts (the uncertainties are only those due to
the limited Monte Carlo statistics).}
\label{tab:st_s-SCH_presel_eff}
\end{table}

\subparagraph{Genetic Algorithm analysis}
\label{singletop_schannel:genetic}
\hfill\break
\hfill\break
The following observables have been chosen in order to further
discriminate between signal and background after preselection:
(i) the jet $b$-tagging discriminants;
(ii) the calibrated jet transverse momenta; 
(iii) the mass of the reconstructed top;
(iv)  $|\Sigma (t, \, \bar b)|$; 
(v)  the scalar sum of the transverse momenta of all the
reconstructed objects.
The reconstructed top quark is formed by the reconstructed $W$ and one
of the two $b$-jets, chosen according to the value of the ``jet
charge'' ($Q_j$, see Section~\ref{ewtop:algorithms}).
 Since in top decays the $W$ and the
original $b$ quark have opposite sign of the charge, the jet with
$Q_j$ ``most opposite'' to the $W$ is used for top reconstruction,
leading to a probability of 67\% to identify the correct pairing.

The cuts on these variables are optimized by means of the GARCON
program~\cite{GARCONMAIN}.
The surviving events after these cuts are shown in cascade in
Table~\ref{tab:st_s-SCH_final_eff}.  With this selection, after an
integrated luminosity of 10 $fb^{-1}$ one gets $N_S/N_B\approx 0.13$.

\begin{table}[h]
\begin{footnotesize}
   \begin{center}
\begin{tabular}{|l|c|c|c|c|}
\hline
 Cut     & s-channel & t-channel & $\ttbar$ & $Wb\bar{b}$  \\ \hline
b-tag($j_1$)$>0.4$, b-tag($j_2$)$>0.1$  & $85\%$ & $75\%$ & $78\%$ & $85\%$ \\  
\pt$ (j_1)>50 \GEVC$, \pt$ (j_2)>50 \GEVC$ & $68\%$ & $53\%$ & $70\%$ & $37\%$ \\
$120 < M(l\nu b) < 220 \GEVCc$ & $52\%$ & $34\%$ & $46\%$ & $26\%$ \\           
$25 <$ \pt $(l\nu b) < 160 \GEVC$  & $48\%$ & $32\%$ & $43\%$ & $26\%$ \\               
$\Sigma_T < 20 \GEVC$  & $35\%$ & $15\%$ & $10.6\%$ & $12.5\%$ \\
$H_T < 340 \GEVC$ & $27\%$ & $10.7\%$ & $5.4\%$ & $11.1\%$ \\   \hline
number of surviving events &  $273\pm 4$ & $630\pm 14$ &
    $1260\pm 60$ & , $155\pm 12$ \\ \hline
\end{tabular}
\end{center}
\end{footnotesize}
\caption{Final cuts and their efficiencies, with respect to the preselected samples,
for the signal and the main backgrounds. For s- and t-channel and $W b \bar b$
samples the final $W$-boson decays into lepton ($e,\mu,\tau$) and neutrino.
  $t\bar t$ samples includes all $W$-boson decay modes.}
\label{tab:st_s-SCH_final_eff}
\end{table}

\subparagraph{Systematic uncertainties}
\label{SCH_syst}
\hfill\break
\hfill\break
In addition to contributions described before, the
following sources of systematic uncertainty are considered:

\begin{itemize}
 \item {\bf Top mass.}
The variation of $m_t$ within $\pm 2 \GEVCc$ around top mass $m_t=175 \GEVCc$
leads to the relative systematic error on the selection efficiency
$\sigma_\mathrm{syst}^{m_\mathrm{t}}=$0.5\% for the s-channel single
top. \\
\item {\bf Parton Distribution Functions.}
To extract the dependence on the PDF uncertainty, two different PDF
sets  were used: CTEQ61and
CTEQ6M~\cite{JHEP_0207_012}.  The result is
$\sigma_\mathrm{syst}^\mathrm{PDF}=$0.7\%. \\
\item {\bf Initial/Final State Radiation modeling.}
The model parameters were varied in the ranges
$\Lambda_\mathrm{QCD}$=0.25$\pm$0.1 \GeV and $Q^2_\mathrm{max}$ from
0.25 to 4 $\hat s$.
 The extreme values of the efficiencies are taken
as systematic error: $\sigma_\mathrm{syst}^\mathrm{rad}= 0.5\%$.
\end{itemize}

The estimation of these errors of theoretical origin has at present been done only for the signal selection. But we expect them to be significant also for the background, in particular ISR/FSR modeling should be very important for the $t\bar t$ rejection.

\begin{table}[h]
   \begin{center}
\begin{tabular}{|l|c|c|c|c|c|c|c|}
\hline
 sample           & selected & $\Delta\sigma$ &   JES    & b-tag    & $M_{top}$ & PDF
   & ISR/FSR \\\hline
$S$: s-channel  &    273   & ---            & $\pm 3$  & $\pm 11$ & $\pm 1.5$
   & $\pm 2$ & $\pm 1.5$ \\\hline
$B$: t-channel  &    630   & $\pm 25$       & $\pm 8$  & $\pm 25$ & --- & --- & --- \\
$B$: $t\bar t$    &   1260   & $\pm 63$       & $\pm 75$ & $\pm 50$ & --- & --- & --- \\
$B$: $Wb\bar b$   &    155   & $\pm 8 $       & $\pm 7$  & $\pm 6$  & --- & --- & ---\\ \hline
\end{tabular}
\end{center}
\caption{Number of selected events after 10 $fb^{-1}$ and systematic uncertainties.}
\label{tab:st_s-SCH_finaltable}
\end{table}

\subparagraph{Background normalization (ratio method)}
\label{sec:ratio_method}
\hfill\break
\hfill\break
The $t\bar t$ events in Table~\ref{tab:st_s-SCH_finaltable} are, in $41\%$
of the cases, $t\bar t \to l^+\nu b l^-\bar\nu \bar b$ events with a
lepton missed, and in the remain cases $t\bar t \to l^+\nu b q\bar q'
\bar b$ events with two jets missed ($t\bar t \to q\bar q' b q\bar q'
\bar b$ events give a negligible contribution).  These two categories
of events are very differently affected by the Jet Energy Scale
variation.
In general, any variation going in the direction of more jets gives a
better rejection of the $t\bar t \to l^+\nu b q\bar q' \bar b$
component with respect to the signal, while the $t\bar t \to l^+\nu b
l^-\bar\nu \bar b$ events, having two quarks, are affected almost in
the same way as the signal.

\noindent $\bullet$ {\it $t\bar t \to \ell^{\pm} + X$ enriched control sample} \\
In this case three jets are required instead of two
and only the muon channel is used. 
The selection efficiency for $t\bar t \to  \ell^{\pm} $ events is found to be $1.08\%$.
The ratio $R_{c1}$ between the efficiencies in the main sample and in this control
sample is $R_{c1}=0.0149$, whose variations under JES and b-tagging efficiency
systematic shifts are $\Delta R_{c1}=\pm 0.0015 (JES) \pm 0.0003 (b-tag)$.

\noindent $\bullet$ {\it $t\bar t \to \ell^{+} \ell^{-} + X$ enriched control sample} \\
This sample is obtained requiring 
two leptons with different flavours with the opposite sign.
The selection efficiency for $t\bar t \to 2l$ events is found to be
$0.822\%$.  The ratio $R_{c2}$ between the efficiencies in the main
sample and in this control sample is $R_{c2}=0.0681$, whose variations
under JES and b-tagging efficiency systematic shifts are $\Delta
R_{c2}=\pm 0.0010 (JES) \pm 0.0004 (b-tag)$.

\subparagraph{Results}
\label{sec:SCH_results}
\hfill\break
\hfill\break
The number of selected signal ($N_S$) and background ($N_B$)
events and their estimated uncertainties are listed in
Table~\ref{tab:st_s-SCH_finaltable}.
The cross section is extracted as
\begin{eqnarray}
\label{eq:st_s-cross-section-ratiomethod}
\sigma = \frac{N_{tot}-b^0-R_{c1}(N_{c1}-b^0_{c1})-R_{c2}(N_{c2}-b^0_{c2})}{\epsilon L},
\end{eqnarray}
where $b^0$ is the sum of the non-top backgrounds in the main sample,
$N_{c1}$ and $N_{c2}$ are the total events selected in the two control
regions, and $b^0_{c1}$ and $b^0_{c2}$ are their contamination by
non-top backgrounds, single top and other $t\bar t$ decays.
The statistical error is evaluated to be 18\%.
The total systematic uncertainty is 31\%,
where the largest contribution arises form the effect of
the JES uncertainty on the $\ttbar$ single-lepton background.  The use of ``Energy Flow''
techniques, including the charged tracks information, is expected to
significantly reduce this uncertainty.  The total error, including
also the $5\%$ luminosity uncertainty and the statistical error, is 36\%.

\paragraph{The contribution from  multi-jet backgrounds}
\label{ewtop:qcd}
\hfill\break
\hfill\break
A special treatment has been devoted to QCD events with jets, due to the
huge cross section.  The currently available samples have very small
statistics and typically no events remain after the application of
pre-selection cuts. Therefore, in order to estimate the impact
of the QCD-background the cuts are applied separately, assuming they
are uncorrelated.

For t-channel study these cuts are: (a)  one isolated muon 
(\pt $> 19 \GEVC$); (b) $\MET >40$ \GeV and only two jets; one $B$-jet and
one light forward jet. It was found a satisfactory suppression of the
 multi-jet events as compared to other background process
($ N_{\rm QCD} / N_{\rm bckg} = 6924 / (8.9\times 10^4) = 0.078$
(see~\cite{CMS-singletop-T}) and the QCD-background
was not considered in the analysis of the
t- and s-channel single top production.

More detailed investigation of this problem was done for
$tW$-channel~\cite{CMS-singletop-W}.  The selection cuts are arranged
into cut groups whose efficiencies are estimated with the Monte Carlo
samples.  The product of efficiencies is an indicator of the total
efficiency.

Three cut groups are used in the di-leptonic channel: lepton, $\MET$,
jet.  The same procedure is applied on signal sample to find the ratio
of total efficiency to the product of efficiencies.  The ratio is used
to correct the product of efficiencies found in multi-jet sample and
the result is 5.6 events.  Four cut groups are used in the
semi-leptonic channel: jets, leptons, kinematics and finally signal
region and b tagging.  The b tagging requirement is taken out from
jets group to have reasonable statistics for the efficiency
measurement.  By comparing the product of efficiencies with total
efficiency of applying cut groups in series, the cut groups are found
to be anti-correlated which would result in an over-estimate of the
yield.  The result of 508 events is kept to be
conservative~\cite{CMS-singletop-W}.

\paragraph{Systematic uncertainties}
\label{ewtop:syst}
\hfill\break
\hfill\break
The following sources of systematic uncertainty are common for all
three channels: (i) the {\bf theoretical errors }to the total rates of
the signal is $\Delta_{\rm th} \approx 4\%$, rising to $10\%$ for
$tW$.  The uncertainties in the background events are assumed to be:
5\% for $\ttbar$~\cite{Beneke:2000hk}, 17\% for $W\bbbar j$, 7\% for
$W+\mathrm{jets}$, $5\%$ for $W j j$~\cite{SGTOP-LHC-PHENO-MCFM-WQQ}, and $5\%$
for $W\bbbar$. (ii) the {\bf jet energy scale (JES) uncertainty:}
using a calibration method based on \ttbar events, the
JES uncertainty after 10~fb${}^{-1}$ integrated luminosity is expected
to be $\pm 5\%$ ($\pm 2.5\%$) for jets with \pt$\approx 20 \GEVC$
(\pt $> 50 \GEVC$).  In the region between 20 and $50 \GEVC$ a linear
dependence is assumed. (iii) {\bf b-tagging identification
uncertainty:} of $\pm 4\%$ on the overall selection efficiencies is
expected on the $b$-tagging efficiencies.
(iv) the {\bf luminosity uncertainty}, expected to be
$5\%$.

\paragraph{Conclusions}
\hfill\break
\hfill\break
The selection strategies developed in CMS for all the three single top
production modes, and their effectiveness, are shown taking into
account the expected statistics after 10 fb$^{-1}$.  All analyses will
be systematics dominated.  For the s-channel and $tW$-associated
cases, control samples have been proposed in order to constrain the
dominant $\ttbar$ background.

The resulting signal-to-background ratio and the significance for the
t-channel are: ${N_S}/{N_B} = 1.34$ and $S_{stat} = {N_S}/{\sqrt{N_S
+ N_B}} = 37.0$, with a statistical error of 2.7\%, and a systematic
error excluding the 5\% luminosity uncertainty of 8\%, resulting in a
total error of 10\%.

For $tW$-channel we expect to reach the significance of 4.2 (5.1) for
the di-lepton (semi-leptonic) channel, increasing to $6.4$ after
combining the two channels.  The total uncertainty is
$\pm$23.9\%(syst.) $\pm$9.9\%(MC) for di-lepton and
$\pm$16.8\%(syst.) $\pm$15.2\%(MC) for semi-leptonic channels.
The total systematic uncertainty for the s-channel is 31\%.

The analyses presented are still ongoing, and major updates are foreseen soon.
The experience gained during the effort for the Physics TDR Vol.II 
tells us that a good control of jets is crucial in single top physics, 
due to the need for a jet counting at relatively low energy, 
where the CMS calorimetry alone is probably not adequate for 
precision measurements.
``Energy Flow'' algorithms, not yet available in CMS, are expected to 
sizably improve the precision, by complementing the calorimetry 
with the informations from the very precise CMS Tracker; 
muon chambers and electromagnetic calorimeter may also give 
a significant improvement, through muon and electron/photon 
identification and correction inside jets.

Along this direction of improvement a first step is already being pursued, with the use of tracks and vertexes: an observable
\begin{eqnarray}
\alpha = \frac{\Sigma_{i} p_T^i}{E_T(jet)},
\end{eqnarray}
is defined for each jet, where the sum runs over all the tracks 
inside the jet cone, fulfilling the following requirements:
\begin{itemize}
\item have at least 5 hits in the Tracker;
\item \pt$>2$ GeV;
\item compatibility of the track with the primary vertex: $|z_{track}-z_{vtx}|<0.4$ cm.
\end{itemize}
A lower cut on this observable (e.g. $\alpha > 0.2$) gives a 
good rejection of noise even at very low $E_T(jet)$, and thanks to 
the last requirement (tracks compatible with the primary vertex) 
the dependence on pile-up is greatly reduced.
Very preliminary results show that with the help of this 
new ``jet cleaning'' criterion, $t\bar t$ rejection is 
greatly improved in all single top analyses.

\clearpage

%
%
%
%
\subsection{From the Tevatron to the LHC}
\label{sec:singletoptevlhc}
{\rm R. Schwienhorst}\\
{\it\small Department of Physics $\&$ Astronomy, Michigan State University,
East Lansing, MI 48824, USA}\\
{\rm A. Lucotte}\\
{\it\small Laboratoire de Physique Subatomique \& Cosmologie (LPSC), 53, avenue des Martyrs
38026 GRENOBLE CEDEX}\\

In the transition from the Tevatron to the LHC, several aspects of single top quark physics
change. At the Tevatron, the main goal is to observe the electroweak mode of top quark
production for the first time. That will be followed by initial measurements. Hence, the
emphasis is on extracting the signal from the backgrounds, using optimized methods.
By contrast, by the time the LHC analyses are starting,
single top quark production should already have been discovered, and the focus shifts to
precision measurements, and to using single top events as tools to
probe the EW sector and to look for new physics.

Table~\ref{tab:singletopNLOXS} shows how the production cross sections change from the 
Tevatron to the LHC for the different single top quark
production modes. 
\begin{table}[htbc]
\begin{center}
\begin{tabular}{lccc}
\hline\hline
accelerator  &  $s$-channel ($pb$) & $t$-channel ($pb$) & $Wt$ ($pb$) \\
\hline
Tevatron ($t$) & 0.44 & 0.99 & 0.1 \\
LHC ($t$)      & 6.6  & 156  & 34 \\
LHC ($\overline{t}$) & 4.1 & 91 & 34 \\
\hline\hline
\end{tabular}
\end{center}
\caption{\label{tab:singletopNLOXS} Cross sections (in $pb$) at NLO for single top quark production
at the Tevatron and the LHC~\cite{Sullivan:2004ie,Campbell:2005bb}.}
\end{table}
The $s$-channel cross section increases roughly by a factor of ten from the Tevatron to the
LHC. Since the backgrounds increase by a similar amount, it will be challenging at both
colliders to observe $s$-channel production separately. It should nevertheless be possible
to measure the $s$-channel cross section separately and thus compare the $s$-channel to the $t$-channel.
Such a comparison is very sensitive to physics beyond the SM, as Fig.~\ref{nn-posterior-2d-nonsm} 
shows.

Compared to the $s$-channel, the increase in production cross section is much more
dramatic for the $t$-channel. Here, the larger center-of-mass energy means that we are accessing
a part of phase space where the gluon and $b$-quark parton distribution functions are much larger,
resulting in an increase of the production cross section by two orders of magnitude. 
Even at the Tevatron, the large cross section makes this channel the main target for the initial
observation of single top. At the LHC, the cross section is so large that it should be possible
to collect large samples of single top quark events which can be used to study the top quark
electroweak coupling in detail. 

Similar to the $t$-channel, the production cross section for associated production also increases 
by more than two orders of magnitude. While the cross section at the Tevatron is too small for
this process to be observed, it is sufficiently large at the LHC to not only observe this mode
of single top quark production but also to study the $tWb$ coupling in detail. 

At the Tevatron, comparing the $s$-channel and $t$-channel production cross sections will
be a test of the SM prediction and a good probe for Physics beyond the SM. 
At the LHC, it will be possible to compare all three production modes with each other, thus
providing an even more sensitive probe, in particular to modifications of the $tWb$ 
coupling~\cite{Tait:2000sh}.

\subsubsection*{Summary of commonalities between TeV and LHC} 
The most important commonality between the Tevatron and the LHC is of course the physics 
process and the final state signature, in particular 
for $s$-channel and $t$-channel single top quark production. 
Many of the lessons learned from theoretical studies of single top quark production at
one collider translate to the other collider as well. This is in particular true for the 
comparisons of single top quark production at LO and 
NLO~\cite{Harris:2002md,Sullivan:2004ie,Cao:2004ky,Cao:2004ap,Cao:2005pq,Campbell:2004ch,Campbell:2005bb} 
and dedicated studies of 
correlations in the single top final state~\cite{Bowen:2005rs,Sullivan:2005ar}.
Similarly, the improvements in producing simulated single top events for a detector simulation
benefit both the Tevatron and LHC analyses.

Experimentally, this results is similar basic event selection cuts, though the Tevatron cuts 
are kept somewhat looser in order to maximize the signal acceptance. 
At the LHC, single top events are produced more 
copiously, thus allowing for somewhat tighter cuts to extract the signal.

The backgrounds to this final state signature are also similar, although they come
in different proportions. At the Tevatron, the most important background is from $W$+jets
production, with a smaller contribution from $\ttbar$ production. At the LHC, the situation
is reversed, and the $\ttbar$ background dominates over the $W$+jets background.
Nevertheless, since both backgrounds need to be modeled well at both colliders.

Due to the complexity of the final state, the focus on detector performance and
understanding is also similar between Tevatron and LHC. Selecting signal events with
high efficiency requires excellent reconstruction efficiency
for electrons, muons, jets, missing transverse energy, and $b$-quark tagging. For the $t$-channel
signal, it is especially important to reconstruct jets in the forward region with high
efficiency.
Separating the signal from the large backgrounds requires understanding and good energy resolution
for electrons, muons, jets, and missing transverse energy. The main difference between the signal
and the large background from $W$+jets production is the presence of a top quark in the final state,
and reconstructing the top quark mass accurately aids greatly in rejecting the $W$+jets background.

\subsubsection*{Summary of differences between TeV and LHC}
The main difference between Tevatron and LHC single top searches is the expected number of signal
events. Both the signal cross sections and the expected total integrated luminosity
are smaller at the Tevatron than at the LHC.
Thus he single top searches at the Tevatron are statistics limited, and even the complete projected
Run~II dataset will only yield a small set of tens of single top quark events. By contrast, the
LHC should be able to yield many hundreds of single top quark events. This has several consequences. 
\begin{itemize}
  \item Tevatron analyses are employing multi-variate analysis techniques to extract the 
  single top quark signal. These techniques significantly improve the sensitivity to SM single top quark
  production, which is important for the initial discovery. They are not as useful for later
  measurements of top quark and $tWb$ coupling measurements because they bias kinematic distributions
  and limit the sensitivity to possible new physics. 
  \item The LHC samples will have much higher event statistics, especially in the $t$-channel,
  making it easier to extract the 
  single top signal in a cut-based analysis. 

  \item In order to extract the signal with high significance, it will be very important to 
  model the backgrounds accurately at the Tevatron. There will be a sizable fraction of background
  events in the signal region, and understanding the size and shape of the backgrounds
  limits the sensitivity of the search. 
  \item At the LHC, it should be easier to extract a relatively clean sample of single top
  events. It will also be easier to find orthogonal samples and sidebands which can
  be used to estimate the background accurately.

  \item At the Tevatron, the statistical uncertainty will be large compared to the systematic
  uncertainty. Thus, cross section measurements are aimed at maximizing the signal acceptance
  and place less importance on minimizing systematic uncertainties.
  \item The statistical uncertainty will be small at the LHC, making it important to understand
  systematic effects. In particular the uncertainty on the different
  background contributions will be a limiting factor, together with the
  jet energy scale uncertainty and initial- and final-state radiation. 
\end{itemize}

\subsubsection*{Conclusions}
Selecting single top quark events with high efficiency, especially $t$-channel events 
with their unique final state signature, requires jet identification
in the forward detector region.
In order to take advantage of the angular correlations in single top
quark events, requiring leptons in the pseudorapidity region $\eta>1$
is also important (see Sec.~\ref{sec:singletoptheoryasym}). 
These are both areas where the Tevatron experience can be applied directly to the LHC.
Moreover, at the LHC, reconstructing jets in the forward region is not only important for 
$t$-channel single top but also for searches for Higgs boson production through vector
boson fusion. This is one example where both Tevatron and LHC single top analysis
experience translates directly to other searches.

Since the backgrounds to single top quark production are similar at the Tevatron
and the LHC, experiences about background modeling at the Tevatron will be relevant
at the LHC.

Most likely, SM single top quark production will have already been
discovered at the Tevatron before the LHC analyses begin. Information
from the Tevatron about the measured experimental 
cross section and basic kinematic properties can thus be used to
optimize the LHC searches, especially if there is a hint of new
physics from the Tevatron.

Similarly, the accurate top quark mass measurements from the Tevatron also help in improving the 
signal model for the LHC. The top quark mass will be measured accurately in top quark pair
events. This information can be used in the single top searches, both in the modeling
of the single top signal, and in reducing systematic uncertainties in the measurement of 
the CKM matrix element $V_{tb}$. Reducing the top quark mass
uncertainty by 1~GeV will reduce the uncertainty on the
$t$-channel cross section at the Tevatron (LHC) by 1.6\% (0.75\%)~\cite{Sullivan:2004ie}.  
Other measurements which can be done at the Tevatron that will improve the systematic uncertainty
at the LHC are of parton distribution functions, in particular for heavy quarks.

The Tevatron single top analyses employ advanced analysis methods. While such methods are likely
not going to be required to extract the single top quark signal at the LHC, they will be
used extensively in other LHC searches, for example for the SM Higgs boson or
searches for new physics beyond the SM. Several of these searches for new physics
involve top quarks, for example searches for a charged Higgs boson that arises
in supersymmetric models.
Here the Tevatron experience in modeling of backgrounds and correlations in complex
final states will be very relevant. And the Tevatron and LHC experiences together in
selecting and reconstructing SM top quark events will be useful in searches for any
new physics involving the top quark.


\clearpage


\clearpage

%

\providecommand{\boldsymbol}[1]{\mbox{\boldmath $#1$}}

\newcommand{\lyxdot}{.}

\renewcommand{\vec}[1]{{\bf #1}}\def\D0{D\O~}


%
%
\section{Precise predictions for $W$ boson observables}
\label{sec:ewktheory}

\subsection{Introduction}
\label{subsec:th_ewkintro}
{\bf Contributed by:~D.~Wackeroth}

\begin{figure}\label{fig:th_ewk_afbzp}
\begin{center}
\includegraphics[height=.2\textheight]{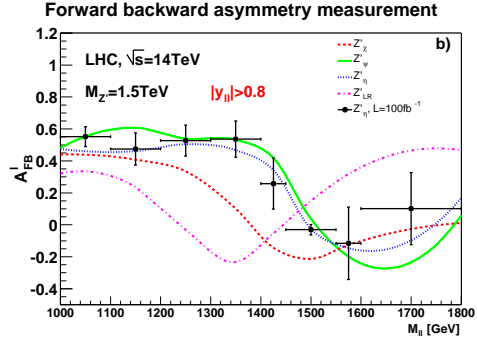}
\end{center}
\caption{The forward-backward asymmetry, $A_{FB}(M(ll))$, 
  of single $Z'$ production in $pp\to Z'\to l^+l^-$ at the LHC for a
  number of models with heavy, non-standard gauge bosons. Taken from
  Ref.~\cite{Dittmar:2003ir}.}
\end{figure}

Electroweak gauge boson production processes are one of the best, most
precise probes of the Standard Model (SM). The electroweak physics
program involving single $W$ and $Z$ boson production at hadron
colliders has many facets:
\begin{itemize}
\item 
  The comparison of direct measurements of the $W$ boson mass ($M_W$)
  and width ($\Gamma_W$) in $W$ pair production at LEP2 and single $W$
  production at the Tevatron, with indirect measurements from a global
  fit to electroweak precision data measured at LEP1/SLD, represents a
  powerful test of the SM.  Any disagreement could be interpreted as a
  signal of physics beyond the SM.  At present, direct and indirect
  measurements of $M_W$ and $\Gamma_W$ agree within their respective
  errors~\cite{lepewwg2}: $M_W$(LEP2/Tevatron)$=80.392\pm 0.029$
  GeV~\footnote{The most recent measurement by CDF finds $M_W=80.413
    \pm 0.048$ GeV (see
    http://fcdfwww.fnal.gov/physics/ewk/2007/wmass/).} versus
  $M_W$(LEP1/SLD)$=80.363\pm 0.032$ GeV and
  $\Gamma_W$(LEP2/Tevatron)$=2.147\pm 0.060$ GeV versus
  $\Gamma_W$(LEP1/SLD)$=2.091\pm 0.003$ GeV. Continued improvements in
  theory and experiment will further scrutinize the SM.
\item The precise measurements of $M_W$ and the top quark mass ($m_t$)
  provide an indirect measurement of the SM Higgs boson mass, $M_H$,
  and a window to physics beyond the SM, as discussed in
  Section~\ref{sec:topew_mass_intro} and illustrated in
  Fig.~\ref{fig:mtmw}. Future more precise measurements of $M_W$
  together with $m_t$ will considerably improve the present indirect
  bound on $M_H$: At the LHC, for instance, with anticipated
  experimental precisions of $\delta M_W=15$~MeV and $\delta
  m_t=1$~GeV, $M_H$ can be predicted with an uncertainty of about
  $\delta M_H/M_H=18\%$~\cite{Baur:2002gp}.
\item The measurement of the mass and width of the $Z$ boson and the
  total $W$ and $Z$ production cross sections can be used for detector
  calibration and as luminosity monitors~\cite{Dittmar:1997md},
  respectively.
\item The $W$ charge asymmetry and $Z$ rapidity distributions severely
  constrain quark Parton Distribution Functions (PDFs).
\item New, heavy gauge bosons may leave their footprints in
  forward-backward asymmetries, $A_{FB}$, and the distribution of the
  invariant mass of the lepton pair, $M(ll)$, produced in $Z$ boson
  production at high $M(ll)$. In
  Figure~\ref{fig:th_ewk_afbzp}~\cite{Dittmar:2003ir} the effects of a $Z'$ on
  $A_{FB}(M(ll))$ at the LHC are shown, assuming a number of different
  models of extended gauge boson sectors, and compared with simulated
  data assuming a specific model. As can be seen, measurements of
  $A_{FB}$ at the LHC will be able to distinguish between different
  new physics scenarios provided, of course, the SM prediction is well
  under control.
\end{itemize}
In order to fully exploit the potential of the Tevatron and LHC for
electroweak (EW) precision physics, the predictions have to be of the
highest standards as well. The omission of EW radiative corrections in
the comparison of predictions with data could result in fake signals
of non-standard physics. For instance, in Ref.~\cite{Baur:2004ig} it
has been shown that the effects of weak non-resonant corrections on
the tail of the transverse mass distribution of the lepton pair,
$M_T(l\nu)$, produced in $p\bar p \to W\to l\nu$ at the Tevatron, from
which $\Gamma_W$ can be extracted, are of the same order of magnitude
as effects due to non-SM values of the $W$ width. Another example is
$WZ$ production at the LHC, which is a sensitive probe of the
non-abelian structure of the SM EW sector.  As demonstrated in
Ref.~\cite{Accomando:2005xp}, for instance, effects of non-standard
weak gauge boson self-couplings can be similar in size and shape to
the effects of EW corrections, and, thus, not including the latter
could be mistaken as signals of new physics.  Consequently, in recent
years a lot of theoretical effort has gone into improving the
predictions for $W$ and $Z$ production processes in order to match (or
better exceed) the anticipated experimental accuracy. This not only
requires the calculation of higher-order corrections but also their
implementation in Monte Carlo (MC) integration programs for realistic
studies of their effects on observables. A list of publicly available
MC programs that include higher-order QED/EW corrections is given in
Table~\ref{tab:th_ewk_b} and a more detailed description of available
calculations and different approaches can be found in
Section~\ref{subsec:th_ewkcodes}.

The importance of fully understanding and controlling EW radiative
corrections to precision $W$ and $Z$ boson observables at hadron
colliders is illustrated in Table~\ref{tab:th_ewk_a} on the example of
a precise $W$ mass and width measurement. It demonstrates how
theoretical progress is driven by improvements in the experimental
precision.
\begin{table}
\begin{center}
\begin{tabular}{|c|c|c|} \hline
Theory includes: & Effects on observable: & Experimental precision: \\ \hline
final-state QED &  shift in $M_W$:  &  Tevatron RUN I: \\ 
(approximation)~\cite{Berends:1984xv} &  
-65$\pm$ 20 MeV for $W\to e\nu$ & $\delta M_W^{exp.} = 59$ MeV  \\ 
&  -168$\pm$ 20 MeV for $W\to \mu \nu$ &  $\delta \Gamma_W^{exp.} = 87$ MeV  \\ \hline
full EW ${\cal O}(\alpha)$ corrections & shift in $M_W$: &  Tevatron RUN II:\\
to resonant $W$ production & $\approx 10$ MeV & $\delta M_W^{exp.} = 27$ MeV \\ 
(pole approximation)~\cite{Wackeroth:1996hz,Baur:1998kt}  & \\ \hline
full EW ${\cal O}(\alpha)$ 
 corrections & affects distributions at high $Q^2$ and & Tevatron RUN II: \\
& direct $\Gamma_W$ measurement &  $\delta \Gamma_W^{exp.} = 25-30$ MeV\\ 
& shift in $\Gamma_W$: $\approx$ 7 MeV~\cite{Baur:2004ig} &  \\ \hline
multiple final-state  & shift in $M_W$: & LHC: \\ 
photon radiation & $2 (10)$ MeV in the $e(\mu)$ case~\cite{CarloniCalame:2003ux} & $\delta M_W^{exp.}$=15 MeV\\ \hline
\end{tabular} 
\caption{Impact of EW radiative corrections on $W$ boson observables,
in particular $M_W$ and $\Gamma_W$ extracted from the 
$M_T(l\nu)$ distribution, confronted with present and anticipated experimental
accuracies~\cite{Baur:2002gp,Abe:1995np,Abachi:1996ey,tevewwg,Abazov:2003sv}.}
\label{tab:th_ewk_a}
\end{center}
\end{table}
For predictions to be under good theoretical control it requires a
good understanding of the residual theoretical uncertainties.
Therefore, the EW theory working group of this workshop addressed the
following questions: What is the residual theoretical uncertainty of
the best, presently available predictions for $W$ boson production at
hadron colliders ?  Do we need more theoretical improvements to be
able to fully exploit the EW physics potential of the Tevatron and the
LHC ?  Our goal is to provide an estimate of the remaining theoretical
uncertainties for a number of $W$ boson observables relevant for:
\begin{itemize}
\item
$W$ mass and width measurements,
\item
luminosity monitoring, 
\item
new physics searches at high invariant masses, and
\item
extraction of quark PDFs.
\end{itemize} 
As a first step, in the spirit of the LEPI/II CERN yellow books, we
perform a tuned numerical comparison of the following publicly
available codes that provide precise predictions for $W$ observables
including electroweak ${\cal O}(\alpha)$ corrections: {\sc HORACE},
{\sc SANC}, and {\sc WGRAD2}.  First results of a tuned comparison of
$W$ and $Z$ production cross sections and kinematic distributions can
be found in Ref.~\cite{Buttar:2006zd}. As an indicator of the
intrinsic theoretical uncertainty of predictions obtained with these
codes due to missing higher-order corrections, we study the impact of
different choices for the EW input parameter scheme and of leading
higher-order (irreducible) QCD and EW corrections connected to the
$\rho$ parameter.  We also discuss the effects of multiple 
photon radiation using {\sc HORACE}. A detailed comparison of
available calculations for $Z$ boson production is work in progress.

In the following, we first review the status of predictions for $W$
and $Z$ boson observables at hadron colliders and summarize the
dominant effects of electroweak corrections. We then present the
results of a tuned numerical comparison of the MC programs {\sc
  HORACE}, {\sc SANC}, and {\sc WGRAD2}, and discuss the effects of
multiple photon radiation.  After a discussion of the impact of
small-$x$ effects, non-perturbative dynamics, the Sudakov form factor
$S_{NP}$, PDF uncertainties, and heavy quark effects on the transverse
momentum distribution of the vector boson ($q_{T}$), we conclude with
an estimate of the theoretical uncertainties and a recommendation of
required theoretical improvements.

\subsection{Theoretical status}
\label{subsec:th_ewkstatus}
{\bf Contributed by:~D.~Wackeroth}

Fully differential cross sections for single $W$ and $Z$ boson
production at hadron colliders are known at next-to-next-to-leading
order (NNLO)
QCD~\cite{Anastasiou:2003ds,Anastasiou:2003yy,Melnikov:2006di,Melnikov:2006kv}
(and references therein).  Predictions for the $W$ transverse momentum
distribution, $q_T(W)$, an important ingredient in the current $W$
mass measurement at the Tevatron, include an all-order resummation of
leading logarithms arising from soft gluon
radiation~\cite{Balazs:1997xd,Ellis:1997ii}.  The complete EW ${\cal
  O}(\alpha)$ corrections to $pp,p \bar p \to W \to l \nu$ and $pp,p
\bar p \to Z,\gamma \to l^+ l^-$ have been calculated in
Ref.~\cite{Dittmaier:2001ay,Baur:2004ig,Arbuzov:2005dd,CarloniCalame:2006zq}
and~\cite{Baur:2001ze}, respectively. Predictions including multiple
final-state photon radiation have been presented in
Ref.~\cite{CarloniCalame:2003ux,Placzek:2003zg,CarloniCalame:2005vc}.
Most of these higher-order calculations have been implemented in MC
programs and a list of some of the publicly available codes providing
precise prediction for $W$ and $Z$ boson observable at hadron
colliders can be found in Table~\ref{tab:th_ewk_b}.
\begin{table}
\begin{center}
\begin{tabular}{|ll|} \hline
{\sc HORACE}: & Multiple final-state photon radiation in $W$ and $Z$ production as solution of  \\
& QED DGLAP evolution for lepton structure functions~\cite{CarloniCalame:2003ux,CarloniCalame:2005vc}, \\
& matched with exact EW ${\cal O}(\alpha)$ corrections to $W$ production~\cite{CarloniCalame:2006zq}.\\ 
&{\tt \small http://www.pv.infn.it/$\sim$hepcomplex/horace.html}
\\ \hline
{\sc PHOTOS}: &  QED corrections in {}``any'' particle decay, multiple-photon radiation, \\
& NLO precision for $Z$ decays, full exact phase-space treatment. \\
&{\tt \small http://cern.ch/wasm/goodies.html}
\\ \hline
{\sc RESBOS}: & QCD corrections to $W$ and $Z$ production, 
soft gluon resummation, and \\
& final-state QED ${\cal O}(\alpha)$ corrections~\cite{Balazs:1997xd,Cao:2004yy}.\\
& {\tt \small http://www.pa.msu.edu/$\sim$balazs/ResBos}\\ \hline
{\sc SANC}: & EW ${\cal O}(\alpha)$ corrections to $W$ and $Z$ production: 
automatically generates \\ 
& Fortran code for one-loop corrections at parton level~\cite{Arbuzov:2005dd,Andonov:2004hi}.\\
& {\tt \small http://sanc.jinr.ru} and {\tt \small http://pcphsanc.cern.ch}\\ \hline
{\sc WGRAD2}: & QED ${\cal O}(\alpha)$ and 
weak one-loop corrections to $W$ production~\cite{Baur:2004ig}. \\ 
& {\small \tt http://ubpheno.physics.buffalo.edu/$\sim$dow/wgrad.html} \\
\hline
{\sc WINHAC}:& Multiple final-state photon radiation in
$W$ production via YFS exponentiation \\
& of soft photons~\cite{Placzek:2003zg}. {\tt \small http://placzek.home.cern.ch/placzek/winhac}\\
\hline
{\sc ZGRAD2}: & QED ${\cal O}(\alpha)$ and weak one-loop corrections 
to $Z$ production \\
& with proper treatment of higher-order terms around the $Z$ resonance~\cite{Baur:2001ze}.  \\
& {\small \tt http://ubhex.physics.buffalo.edu/$\sim$baur/zgrad2.tar.gz}\\
\hline
\end{tabular}
\caption{Publicly available MC programs that provide precise predictions
including QED and/or electroweak corrections for $W$ and/or $Z$ boson production at hadron colliders. A more
detailed description is provided below.}
\label{tab:th_ewk_b}
\end{center}
\end{table}
$W$ and $Z$ boson observables are strongly affected by EW corrections.
Their main characteristics can be summarized as follows:
\begin{itemize}
\item Photon radiation off the final-state charged lepton can
  considerably distort kinematic distributions and usually makes up
  the bulk of the effects of EW corrections.  For instance, $W$ and
  $Z$ boson masses extracted respectively from the transverse mass and
  invariant mass distributions of the final-state lepton pair are
  shifted by ${\cal O}(100)$ MeV due to final-state photon radiation.
  This is due to the occurrence of mass singular logarithms of the
  form $\alpha \log(Q^2/m_l^2)$ that arise when the photon is emitted
  collinear to the charged lepton.  In sufficiently inclusive
  observables these mass singularities completely cancel (KLN
  theorem). But in realistic experimental environments, depending on
  the experimental setup, large logarithms can survive. The more
  inclusive treatment of the photon emitted in $W^+\to e^+ \nu_e$
  decays results in a significant reduction of the final-state QED
  effects when lepton identification cuts are applied whereas in the
  muon case large logarithms survive.  Because of their numerical
  importance at one-loop, the higher-order effects of multiple
  final-state photon radiation have to be under good theoretical
  control as
  well~\cite{CarloniCalame:2003ux,Placzek:2003zg,CarloniCalame:2005vc}.
\item The impact of initial-state photon radiation is negligible after
  proper removal of the initial-state mass singularities by universal
  collinear counterterms to the quark PDFs.  This mass factorization
  introduces a dependence on the QED factorization scheme: in complete
  analogy to QCD both the QED DIS and $\overline{\rm MS}$ scheme have
  been introduced in the literature~\cite{DeRujula:1979jj,Baur:1998kt}. Recently,
  quark PDFs became available that also incorporate QED radiative
  corrections~\cite{Martin:2004dh}, which is important for a
  consistent treatment of initial-state photon radiation at hadron
  colliders.
\item At high energies, i.e.~in tails of kinematic distributions, for
  instance $M(ll) \gg M_Z$ and $M_T(l \nu)\gg M_W$, Sudakov-like
  contributions of the form $\alpha \log^2(Q^2/M_V^2)$ ($M_V=M_{W,Z}$
  and $Q$ is a typical energy of the scattering process) can
  significantly enhance the EW one-loop corrections.  These
  corrections originate from remnants of UV singularities after
  renormalization and soft and collinear initial-state and final-state
  radiation of virtual and real weak gauge bosons.  In contrast to QED
  and QCD the Bloch-Nordsiek theorem is
  violated~\cite{Ciafaloni:2000df}, i.e.~even in fully inclusive
  observables these large logarithms are present due to an incomplete
  cancellation between contributions from real and virtual weak gauge
  boson radiation. Moreover, the $W$ and $Z$ boson masses serve as
  physical cut-off parameters and real $W$ and $Z$ boson radiation
  processes are usually not included, since they result in different
  initial and/or final states.  The EW logarithmic corrections of the
  form $\alpha^L \log^N(\frac{Q^2}{M_V^2}), 1 \le N \le 2L$ ($L=1,2
  \ldots$ for 1-loop,2-loop,$\ldots$) to 4-fermion processes are known
  up to 2-loop $N^3LL$ order and are available in form of compact
  analytic formulae (see, e.g.,
  Refs.~\cite{Melles:2001ye,Jantzen:2005az,Denner:2006jr} and
  references therein).
\end{itemize}
First studies of effects of combined EW and QCD
corrections~\cite{Cao:2004yy}, higher-order EW Sudakov-like logarithms
(see, e.g., Ref.~\cite{Baur:2006sn}) and multiple final-state photon
radiation~\cite{CarloniCalame:2003ux,Placzek:2003zg,CarloniCalame:2005vc}
suggest that for the anticipated precision at the LHC these effects
need to be included in the data analysis.  Moreover, the model for
non-perturbative QCD contributions~\cite{Konychev:2005iy}, small $x$
effects~\cite{Berge:2004nt} and the impact of heavy-quark
masses~\cite{Berge:2005rv} need to be well understood for a detailed
description of the $q_T(W)$ distribution (see
Section~\ref{subsec:qtresum} for more details).  Several groups are
presently working on the combination of EW and QCD radiative
corrections in one MC program, the interface of higher-order EW
calculations, i.e.~multiple photon radiation from final-state leptons
and EW Sudakov logarithms, with fixed ${\cal O}(\alpha)$ calculations,
and the calculation of mixed QED/QCD two-loop corrections of ${\cal
  O}(\alpha \alpha_s)$, which are not yet available.  The ultimate
goal is to provide one unified MC program that includes all relevant
QED, EW and QCD radiative corrections to $W$ and $Z$ boson production
that matches the anticipated experimental capabilities of the Tevatron
and LHC for EW precision physics.

\subsection{Description of higher-order calculations and MC programs}
\label{subsec:th_ewkcodes}

\subsubsection*{{\sc HORACE}}
\label{subsec:th_ewkhorace}
{\bf Contributed by:~C.~M.~Carloni Calame, G.~Montagna, O.~Nicrosini, and A.~Vicini}

{\sc HORACE}~\cite{Horace-prd:2004,Horace-jhep:2005,CarloniCalame:2006zq} is a Monte
Carlo generator for precision simulations of charged-current and
neutral-current Drell-Yan processes
$pp\hskip-7pt\hbox{$^{^{(\!-\!)}}$} \to W \to l \nu_l$ and
$pp\hskip-7pt\hbox{$^{^{(\!-\!)}}$} \to \gamma,Z \to l^+ l^-$
($l=e,\mu$) at hadron colliders.

In its original version~\cite{Horace-prd:2004,Horace-jhep:2005} {\sc HORACE}
is based on a pure QED parton shower approach to account for
final-state-like QED corrections, both at $\cal{O}(\alpha)$ and at
higher orders, in leading logarithmic approximation.  For the
calculation of multiple photon corrections, the QED parton shower
algorithm developed in Refs.~\cite{Babayaga-npb:2000,Calame-plb:2001}
is used.

The predictions of {\sc HORACE} for multi-photon effects have been compared
with those of the independent generator {\sc WINHAC} in
Ref.~\cite{CarloniCalame:2004qw}, finding good agreement.  As shown in
Refs.~\cite{Horace-prd:2004,Horace-jhep:2005}, higher-order QED
contributions are necessary for a number of precision studies at
hadron colliders, particularly in view of high-precision measurements
of the $W$ boson mass at the Tevatron Run II and at the LHC.

Recently {\sc HORACE} has been improved and, in its present version,
includes: (i) the exact $\cal{O}(\alpha)$ electroweak corrections to
the charged-current process $pp\hskip-7pt\hbox{$^{^{(\!-\!)}}$} \to W
\to l \nu_l$, and (ii) higher-order QED contributions in the parton
shower approach (initial- and final-state corrections).  In order to
avoid double counting of leading logarithmic contributions, already
included in the parton shower, a matching procedure between fixed
order and resummed calculation has been developed.  The theoretical
and computational details about the matching are too lengthly to be
described here and can be found in Ref.~\cite{CarloniCalame:2006zq}.

Because it is well known that quark mass singularities, originating
from initial-state photon radiation, can be factorized out of the
partonic cross section and reabsorbed into a redefinition of the PDFs,
in analogy to gluon emission in QCD, a subtraction to all orders of
initial-state collinear singularities arising from photon radiation
has been developed and implemented in {\sc HORACE}.  After subtraction of
quark mass singularities, the QED initial-state radiation turns out to
be small with respect to the effects of final-state radiation.

At the time of writing, exact $\cal{O}(\alpha)$ electroweak
corrections to $Z$ production are not accounted for in {\sc HORACE}, but
their inclusion in association with parton shower effects is foreseen
in a future release.

\section*{Acknowledgments}
The work of C.M.~Carloni Calame is partially supported by a Royal
Society Short Visit grant.  

\subsubsection*{{\sc PHOTOS}}
\label{subsec:th_ewkphotos}
{\bf Contributed by:~P.~Golonka and Z.~Was}

PHOTOS~\cite{Barberio:1990ms,Barberio:1994qi,Golonka:2005pn} is a
universal Monte Carlo event generator simulating QED final-state radiative
corrections in decays of particles and resonances. Having a form of
an independent module, it cooperates with other event generators in
the simulation chains of many experimental collaborations, including
the ones for the LHC (for details, see references in available PHOTOS
literature). Over 15 years of its history the core of the photon-emission
algorithm has not changed significantly; however, areas of its applicability,
numerical stability, and precision have been improved in the span
of last few years. Recent needs of experimental collaborations to
use PHOTOS for high-precision estimates in certain channels motivated
us to review the performance of the PHOTOS algorithm in certain areas
of interest. Let us review here, in chronological order, the most
important papers that cover the versions of PHOTOS code and related
improvements in physical content,

The best documented 2.0 version of PHOTOS \cite{Barberio:1994qi},
allowed generating configurations with up to two photons in every
elementary decay process%
\footnote{PHOTOS scans the whole tree of the event record and its action is
applied for every branching which can be interpreted as an individual
decay (of a final but also intermediate step in the decay cascade).%
}. It was supposed to be used as a {}``crude'' tool, certainly not
for high-precision studies. In particular, the effects of interference
were treated with rough approximation or were not included at all.

In 2003, the version 2.07 of PHOTOS was released as a part of the
TAUOLA-PHOTOS-F package \cite{Golonka:2003xt}. In terms of precision,
it contained a process-dependent correction weight for $W$ decays,
see Ref.~\cite{Nanava:2003cg}. 

In 2004 and 2005, the universal, process-independent (approximated)
interference weight, better control of numerical stability (allowing
to use PHOTOS for decays of particles at the LHC energy scales), and
multiple-photon, {}``exponentiated'' emission were introduced. At
the same time, systematic comparison tests of PHOTOS as a high-precision
tool in certain decay channels began. Initially, such tests were conducted
for $Z$, $W$ and $\tau$ decays \cite{Golonka:2005pn}. These achievements,
including the method for tests, based on MC-TESTER \cite{Golonka:2002rz},
are documented in \cite{PhDGolonka}.

In 2006, we firstly focused our studies on the performance of PHOTOS
at NLO precision and leptonic $Z$ decays. PHOTOS has been extended
to include the NLO effects. As a result, predictions of PHOTOS simulations
match perfectly those produced by generators based on the full matrix-element
calculation (differences are not recognizable in samples of 100 mln
generated events) \cite{Golonka:2006tw}. Similar upgrade of PHOTOS
to complete NLO for $W$ decays might also have been straightforward;
nevertheless, it would probably not be needed.

The NLO effects of scalar QED \cite{Nanava:2006vv} were also installed
for B-meson decays into pairs of scalars. This may be of interest
not only for the Belle and BaBar communities, but for LHCb as well.
This proves the flexibility of PHOTOS design as well: even though
the scalar QED is not the ultimate theory of photon emission from
pions, the separation of the matrix-element and phase-space points
to a possible implementation of shape factors (to be obtained from
experimental data). Note also that PHOTOS generation covers the complete
phase-space for multi-photon configurations.

From the technical side, the mainstream version of PHOTOS is maintained
as a single, compact block of FORTRAN77/95 code, which communicates
with other generators by means of HEPEVT event record. However, a
version in C++ \cite{MScGolonka} exists since 1999, yet its popularity
is limited due to ongoing discussions of the standards for C++ event
record. Recent developments are straightforward to include in the
C++ version, if interest is expressed.


\newpage
\subsubsection*{{\sc SANC}}
\label{subsec:sanc}
{\bf Contributed by:~A.~Arbuzov, D.~Bardin, S.~Bondarenko, P.~Christova, L.~Kalinovskaya, and R.~Sadykov}

In the evaluation of the electroweak (EW) radiative corrections (RC)
to the Drell-Yan-like processes we exploit the automatized system
{\sc SANC}~\cite{Andonov:2004hi}~\footnote{{\sc SANC} is available at {\tt http://sanc.jinr.ru}
and {\tt http://pcphsanc.cern.ch}}. The system provides complete
one-loop results for the EW corrections at the partonic level both for
the neutral and charged-current processes. The {\sc SANC} system
automatically generates FORTRAN codes for corrected differential
distributions.  We subdivide the EW RC into the virtual ones, the ones
due to soft photon emission, and the ones due to hard photon
emission. An auxiliary parameter $\bar\omega$ separates the soft and
hard photonic contributions.  For the real photon emission integration
over the phase space can be performed either (semi-)analytically or by
means of a Monte Carlo integrator.

To get the cross section at the hadronic level we convolute the
partonic cross section with quark density functions. To avoid double
counting of the quark mass singularities we subtract them (using a QED
DIS-like subtraction scheme) from the density functions. Linearization
of the subtraction procedure is done as described in
Ref.~\cite{Arbuzov:2005dd}.

In order to have the possibility to impose cuts, we use the Monte
Carlo integration routine based on the Vegas
algorithm~\cite{Lepage:1977sw}.  In this case we perform a 4(6)-fold
numerical integration to get the hard photon contribution to the
partonic (hadronic) cross section.
One-loop virtual EW corrections are calculated using the $R_\xi$ gauge
and the on-mass-shell renormalization scheme. They are used as form factors
standing before different structures of the matrix element. The latter is
automatically generated with help of the helicity amplitude method. 
To get the total EW correction we sum up the contributions of the soft and hard photon 
emission and the ones of the virtual loops. The  cancellation of the 
dependence on the auxiliary parameter $\bar\omega$ 
in the sum is achieved numerically. 

For the case of charged-current Drell-Yan process an extended
description of our approach can be found in
Ref.~\cite{Arbuzov:2005dd}. Some results of a tuned comparison with
other programs were presented in Ref.~\cite{Buttar:2006zd}.

\section*{Acknowledgments}
The work of the SANC team is partially supported by INTAS grant
$N^{o}$ 03-51-4007 and by RFBR grant 04-02-17192 (AA).

\subsubsection*{{\sc WGRAD2/ZGRAD2}}
\label{subsec:th_ewkwzgrad}
{\bf Contributed by:~U.~Baur and D.~Wackeroth}

{\sc WGRAD2}~\cite{Baur:1998kt,Baur:2004ig} and {\sc
ZGRAD2}~\cite{Baur:2001ze} are parton-level Monte Carlo
programs that include the complete ${\cal O}(\alpha)$ electroweak
radiative corrections to $p\,p\hskip-7pt\hbox{$^{^{(\!-\!)}}$} \to
W^\pm\to\ell^\pm\nu X$ ({\sc WGRAD2}) and
$p\,p\hskip-7pt\hbox{$^{^{(\!-\!)}}$} \to\gamma,\, Z\to\ell^+\ell^- X$
($\ell=e,\,\mu$) ({\sc ZGRAD2}). For the numerical evaluation, the
Monte Carlo phase space slicing method for next-to-leading-order (NLO)
calculations described in Ref.~\cite{Baer:1990ra,Harris:2001sx} is
used. Final-state charged lepton mass effects are included in the
following approximation. The lepton mass regularizes the collinear
singularity associated with final state photon radiation. The
associated mass singular logarithms of the form $\ln(\hat
s/m_\ell^2)$, where $\hat s$ is the squared parton center of mass
energy and $m_\ell$ is the charged lepton mass, are included in the
calculation, but the very small terms of ${\cal O}(m_\ell^2/\hat s)$
are neglected.

As a result of the absorption of the universal initial-state mass
singularities by redefined ({\it renormalized})
PDFs~\cite{Baur:1998kt,DeRujula:1979jj}, the cross sections become
dependent on the QED factorization scale $\mu_{\rm QED}$. In order to
treat the ${\cal O}(\alpha)$ initial-state photonic corrections to $W$
and $Z$ production in hadronic collisions in a consistent way, the
MRST2004QED set of parton distribution functions~\cite{Martin:2004dh}
should be used, which currently is the only set of PDFs which includes
QED corrections.  Absorbing the collinear singularity into the PDFs
introduces a QED factorization scheme dependence. The squared matrix
elements for different QED factorization schemes differ by the finite
${\cal O}(\alpha)$ terms which are absorbed into the PDFs in addition
to the singular terms. {\sc WGRAD2} and {\sc ZGRAD2} can be used both
in the QED ${\rm \overline{MS}}$ and DIS schemes, which are defined
analogously to the usual ${\rm \overline{MS}}$~\cite{Bardeen:1978yd}
and DIS~\cite{Owens:1992hd} schemes used in QCD calculations.

{\sc WGRAD2} and {\sc ZGRAD2} can be used both with an $s$-dependent
width, or a constant width, as well as different input parameter
schemes. Radiative corrections beyond ${\cal O}(\alpha)$ are partially
implemented in both programs.

\section*{Acknowledgments}

This research was supported by the National Science Foundation under
grant No.~PHY- 0244875 and No.~PHY-0456681.

\subsubsection*{{\sc WINHAC}}
\label{subsec:th_ewkwinhac}
{\bf Contributed by:~S.~Jadach and W.~P{\l}aczek}

{\sc WINHAC}~\cite{Placzek:2003zg} is a Monte Carlo event generator
for Drell--Yan processes in proton--proton, proton--antiproton and
nucleus--nucleus collisions.  It features multiphoton radiation in
$W$-boson decays within the Yennie--Frautschi--Suura (YFS) exclusive
exponentiation scheme and the ${\cal O}(\alpha)$ electroweak radiative
corrections for $W$ decays. The latter have been provided to us by the
{\sc SANC} group.  Implementation of the total ${\cal O}(\alpha)$
electroweak radiative corrections to the full charged-current
Drell--Yan process is under way in the collaboration with the {\sc
  SANC} group.

The current version of {\sc WINHAC} includes a direct interface to
{\sc PYTHIA} for the QCD and/or QED initial-state radiation (ISR)
parton shower, proton-remnants treatment and hadronization. One of the
consequences of these effects is non-zero transverse momentum of the
$W$-bosons.  In addition to unpolarized $W$-boson production, the
program provides options for generation of transversely and
longitudinally polarized $W$-boson in the Born approximation. In the
recent version we have also added an option for generation of the
Born-level neutral-current (through $Z/\gamma$) Drell--Yan process.
For the PDFs, {\sc WINHAC} is interfaced with the PDFLIB package as
well as with its recent successor LHAPDF.  In the latter case {\sc
  WINHAC} gives the possibility to compute auxiliary weights
corresponding to PDF errors provided with some PDF parametrizations;
all these weights are calculated in a single MC run.  In the case of
nucleus--nucleus collisions, an option for switching on/off nuclear
shadowing effects for PDFs is provided. Nuclear beams are defined
through the input parameters by setting atomic numbers $A$, charge
numbers $Z$ and energies of two colliding nuclei. This collider option
was applied to studies presented in Ref.~\cite{Krasny:2005cb}.  We
also provide a special parton-level version of the program, called
WINDEC, for generation of multiphoton radiation in $W$ decays that can
be interfaced with an arbitrary MC generator of the $W$-production
process.

For QED radiative corrections {\sc WINHAC} has been compared with the
Monte Carlo generator {\sc HORACE}, both for the parton level
processes and for proton--proton collisions at the LHC. Good agreement
of the two programs for several observables has been found
\cite{CarloniCalame:2004qw}.  The comparisons with PHOTOS also show
good agreement of the two generators for the QED final-state radiation
(FSR) \cite{Golonka:2005pn}.

A similar event generator for the $Z$-boson production, called {\sc ZINHAC},
is under development now. We also work on constrained MC algorithms
for the QCD ISR parton shower that could be applied to Drell--Yan
processes, see, e.g., Ref.~\cite{Jadach:2005rd}.

\subsubsection*{{\sc Calculation presented in Ref.~\cite{Dittmaier:2001ay}}}
\label{subsec:th_ewkmdmk}
{\bf Contributed by:~S.~Dittmaier and M.~Kr{\"a}mer}

Ref.~\cite{Dittmaier:2001ay} contains a detailed description of the
calculation of the ${\cal O}(\alpha)$ corrections to W~production at
hadron colliders and a discussion of results for the Tevatron and the
LHC. In particular, the full ${\cal O}(\alpha)$ calculation is
compared with a pole approximation for the W~resonance.  The case of
Z-boson production is not considered.  For the analysis performed in
Ref.~\cite{Buttar:2006zd}, the calculation of
Ref.~\cite{Dittmaier:2001ay} has been extended (i) to include
final-state radiation beyond ${\cal O}(\alpha)$ via structure
functions and (ii) by implementing the ${\cal O}(\alpha)$-corrected
PDF set MRST2004QED.  The photon-induced processes $\gamma q\to q' l
\nu_l$ and $\gamma \bar{q'} \to \bar{q} l \nu_l$ have been calculated
as described in Ref.~\cite{Diener:2005me}.  The evaluation of the
$q\bar{q'}$ channel has been technically improved by employing a
generalization of the dipole subtraction approach
\cite{Dittmaier:1999mb} to non-collinear-safe observables, as
partially described in Ref.~\cite{Bredenstein:2005zk}.

\newpage
\subsection{Results of a tuned comparison of {\sc HORACE}, {\sc SANC} and {\sc WGRAD2}}
\label{subsec:th_ewkcomp}
{\bf Contributed by:~A.~Arbuzov, D.~Bardin, U.~Baur, S.~Bondarenko,
C.~M.~ Carloni Calame, P.~Christova, L.~Kalinovskaya, G.~Montagna, 
O.~Nicrosini, R.~Sadykov, A.~Vicini, and D.~Wackeroth}

\begin{center}
{\it Setup for the tuned comparison} 
\end{center}
\noindent
For the numerical evaluation of the cross sections at the Tevatron 
($\sqrt{s}=1.96$ TeV) and the LHC ($\sqrt{s}=14$ TeV) we chose the
following set of Standard Model input parameters:
\begin{eqnarray}\label{eq:pars}
G_{\mu} = 1.16637\times 10^{-5} \; {\rm GeV}^{-2}, 
& \qquad & \alpha= 1/137.03599911, \quad \alpha_s\equiv\alpha_s(M_Z^2)=0.1176 
\nonumber \\ 
M_Z = 91.1876 \; {\rm GeV}, & \quad & \Gamma_Z =  2.4924  \; {\rm GeV}
\nonumber  \\
M_W = 80.37399 \; {\rm GeV}, & \quad & \Gamma_W = 2.0836 \; {\rm GeV}
\nonumber  \\
M_H = 115 \; {\rm GeV}, & \quad & 
\nonumber  \\
m_e  = 0.51099892 \; {\rm keV}, &\quad &m_{\mu}=0.105658369 \; {\rm GeV},  
\quad m_{\tau}=1.77699 \; {\rm GeV}
\nonumber  \\
m_u=0.06983 \; {\rm GeV}, & \quad & m_c=1.2 \; {\rm GeV}, 
\quad m_t=174 \; {\rm GeV}
\nonumber  \\
m_d=0.06984 \; {\rm GeV}, & \quad & m_s=0.15 \; {\rm GeV}, \quad m_b=4.6 \; {\rm GeV} 
\nonumber \\
|V_{ud}| = 0.975, & \quad & |V_{us}| = 0.222 
\nonumber \\
|V_{cd}| = 0.222, & \quad & |V_{cs}| = 0.975 
\nonumber \\
|V_{cb}|=|V_{ts}|=|V_{ub}|& =& |V_{td}|= |V_{tb}|=0  
\end{eqnarray}
The $W$ and Higgs boson masses, $M_W$ and $M_H$, are related via loop
corrections. To determine $M_W$ we use a parametrization which, for
$100~{\rm GeV}<M_H<1$~TeV, deviates by at most 0.2~MeV from the
theoretical value including the full two-loop
contributions~\cite{Awramik:2003rn} (using Eqs.~(6,7,9)).  Additional
parametrizations can also be found
in~\cite{Degrassi:1997iy,Ferroglia:2002rg}.

We work in the constant width scheme and fix the weak mixing angle by
$c_w=M_W/M_Z$, $s_w^2=1-c_w^2$.  The $Z$ and $W$-boson decay widths
given above are calculated including QCD and electroweak corrections,
and are used in both the LO and NLO evaluations of the cross sections.
The fermion masses only enter through loop contributions to the vector
boson self energies and as regulators of the collinear singularities
which arise in the calculation of the QED contribution. The light
quark masses are chosen in such a way, that the value for the hadronic
five-flavor contribution to the photon vacuum polarization, $\Delta
\alpha_{had}^{(5)}(M_Z^2)=0.027572$~\cite{Jegerlehner:2001wq}, is
recovered, which is derived from low-energy $e^+ e^-$ data with the
help of dispersion relations.  The finestructure constant,
$\alpha(0)$, is used throughout in both the LO and NLO calculations of
the $W$ production cross sections.

In the course of the calculation of $W$ observables the
Kobayashi-Maskawa-mixing has been neglected, but the final result for
each parton level process has been multiplied with the square of the
corresponding physical matrix element $V_{ij}$.  From a numerical
point of view, this procedure does not significantly differ from a
consideration of the Kobayashi-Maskawa-matrix in the renormalisation
procedure as it has been pointed out in~\cite{Denner:1990yz}.

To compute the hadronic cross section we use the MRST2004QED set of
parton distribution functions~\cite{Martin:2004dh}, and take the
renormalization scale, $\mu_r$, and the QED and QCD factorization
scales, $\mu_{\rm QED}$ and $\mu_{\rm QCD}$, to be $\mu_r^2=\mu_{\rm
  QED}^2=\mu_{\rm QCD}^2=M_W^2$.
In the MRST2004QED structure functions, the factorization of the
photonic initial state quark mass singularities is done in the QED DIS
scheme which we therefore use in all calculations reported here. It is
defined analogously to the usual DIS~\cite{Owens:1992hd} schemes used
in QCD calculations, i.e.  by requiring the same expression for the
leading and next-to-leading order structure function $F_2$ in deep
inelastic scattering, which is given by the sum of the quark
distributions. Since $F_2$ data are an important ingredient in
extracting PDFs, the effect of the ${\cal O}(\alpha)$ QED corrections
on the PDFs should be reduced in the QED DIS scheme.

The detector acceptance is simulated by imposing the following
transverse momentum ($p_T$) and pseudo-rapidity ($\eta$) cuts:
\begin{equation}
p_T(\ell)>20~{\rm GeV,}\qquad\qquad |\eta(\ell)|<2.5, \qquad\qquad
\ell=e,\,\mu ,
\label{eq:lepcut}
\end{equation}
\begin{equation}
p\llap/_T>20~{\rm GeV,}
\label{eq:ptmisscut}
\end{equation}
where $p\llap/_T$ is the missing transverse momentum originating from
the neutrino. These cuts approximately model the acceptance of the CDF
II and D\O detectors at the Tevatron, and the ATLAS and CMS detectors
at the LHC. Uncertainties in the energy measurements of the charged
leptons in the detector are simulated in the calculation by Gaussian
smearing of the particle four-momentum vector with standard deviation
$\sigma$ which depends on the particle type and the detector. The
numerical results presented here were calculated using $\sigma$ values
based on the D\O (upgrade) and ATLAS specifications. 

The granularity of the detectors and the size of the electromagnetic
showers in the calorimeter make it difficult to discriminate between
electrons and photons with a small opening angle. In such cases we
recombine the four-momentum vectors of the electron and photon to an
effective electron four-momentum vector.  To simplify the comparison
we use the same recombination procedure at the Tevatron and the LHC.
We require that the electron and photon momentum four-vectors are
combined into an effective electron momentum four-vector if their
separation in the pseudorapidity -- azimuthal angle plane,
\begin{equation}
\Delta R(e,\gamma)=\sqrt{(\Delta\eta(e,\gamma))^2+(\Delta\phi(e,
\gamma))^2}, 
\end{equation}
is $\Delta R(e,\gamma)<0.1$. 
For
$0.1<\Delta R(e,\gamma)<0.4$ events are rejected if
$E_{\gamma}>0.1 \; E_e$. Here $E_\gamma$ ($E_e$) is the energy of the
photon (electron) in the laboratory frame. 

Muons are identified by hits in the muon chambers and the requirement
that the associated track is consistent with a minimum ionizing
particle. This limits the photon energy for small muon -- photon
opening angles. For muons at the Tevatron and the LHC, we require that
the energy of the photon is $E_{\gamma}<2$~GeV for $\Delta
R(\mu,\gamma)<0.1$, and $E_{\gamma}<0.1 E_{\mu}$~GeV for $0.1<\Delta
R(\mu,\gamma)<0.4$.  We summarize the lepton identification
requirements in Table~\ref{tab:th_ewk_c}.
\begin{table}
\begin{center}
\begin{tabular}{|c|c|} \hline
\multicolumn{2}{|c|}{\bf Tevatron and LHC} \\ \hline
\multicolumn{1}{|c|}{electrons} & \multicolumn{1}{|c|}{muons} \\
\hline
combine $e$ and $\gamma$ momentum four vectors, & reject events with 
$E_\gamma>2$~GeV \\ 
if $\Delta R(e,\gamma)<0.1$ & for $\Delta R(\mu,\gamma)<0.1$ \\
\hline
reject events with 
$E_\gamma>0.1~E_e$ & reject events with 
$E_\gamma>0.1~E_\mu$ \\  
for $0.1<\Delta R(e,\gamma)<0.4$ & for $0.1<\Delta R(\mu,\gamma)<0.4$ \\ \hline  
\end{tabular}
\caption{Summary of lepton identification requirements. } 
\label{tab:th_ewk_c}
\end{center}
\end{table}
For each observable we will provide ``bare'' results, i.e.~without
smearing and recombination (only lepton separation cuts are applied)
and ``calo'' results, i.e.~including smearing and recombination.  We
will show results for kinematic distributions and total cross sections,
at LO and NLO, and the corresponding relative
corrections, $\Delta(\%)=d\sigma_{NLO}/d\sigma_{LO}-1$, at both the
Tevatron and the LHC.  If not stated otherwise, we consider the
following charged current processes: $pp(p\bar p) \to W^+ \to l^+
\nu_l$ with $l=e,\mu$.

\begin{figure}
\begin{center}
  \includegraphics[width=7.1cm,
  keepaspectratio=true]{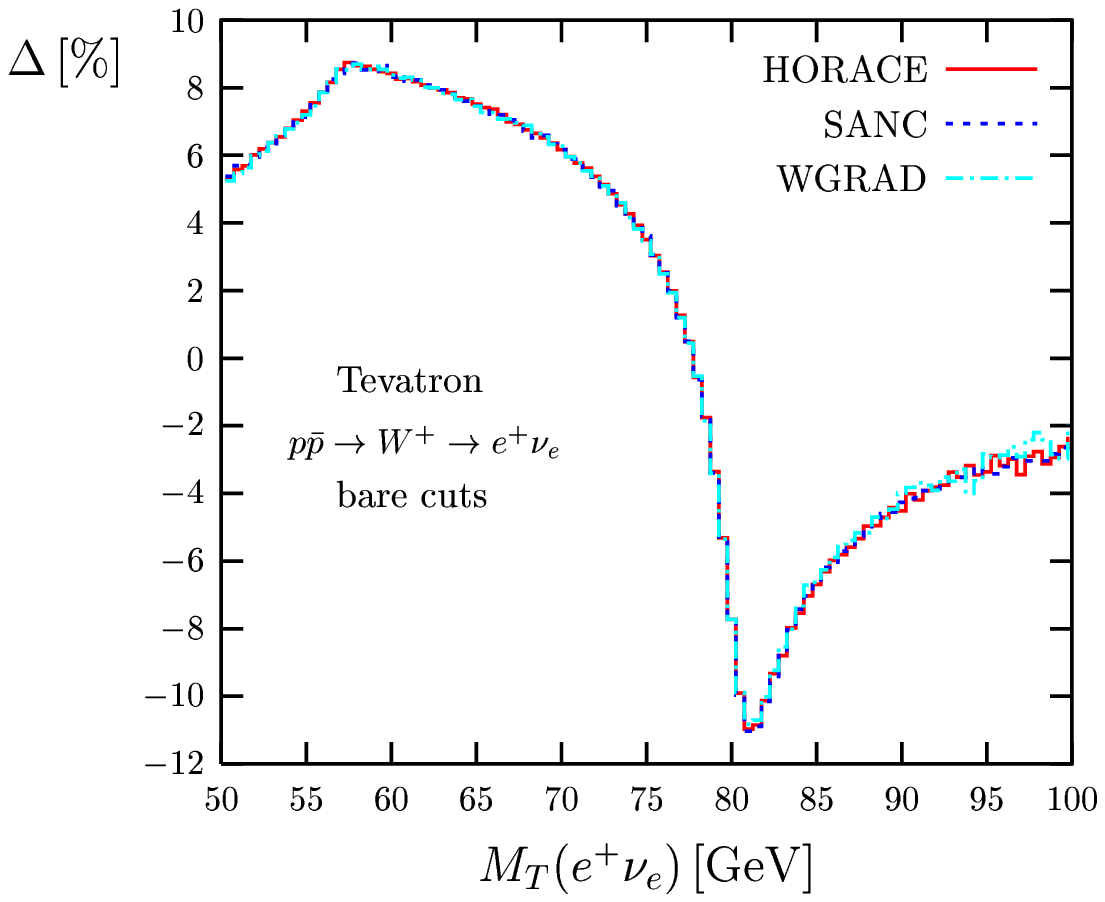}
\hspace*{-.0cm}
  \includegraphics[width=7.1cm,
  keepaspectratio=true]{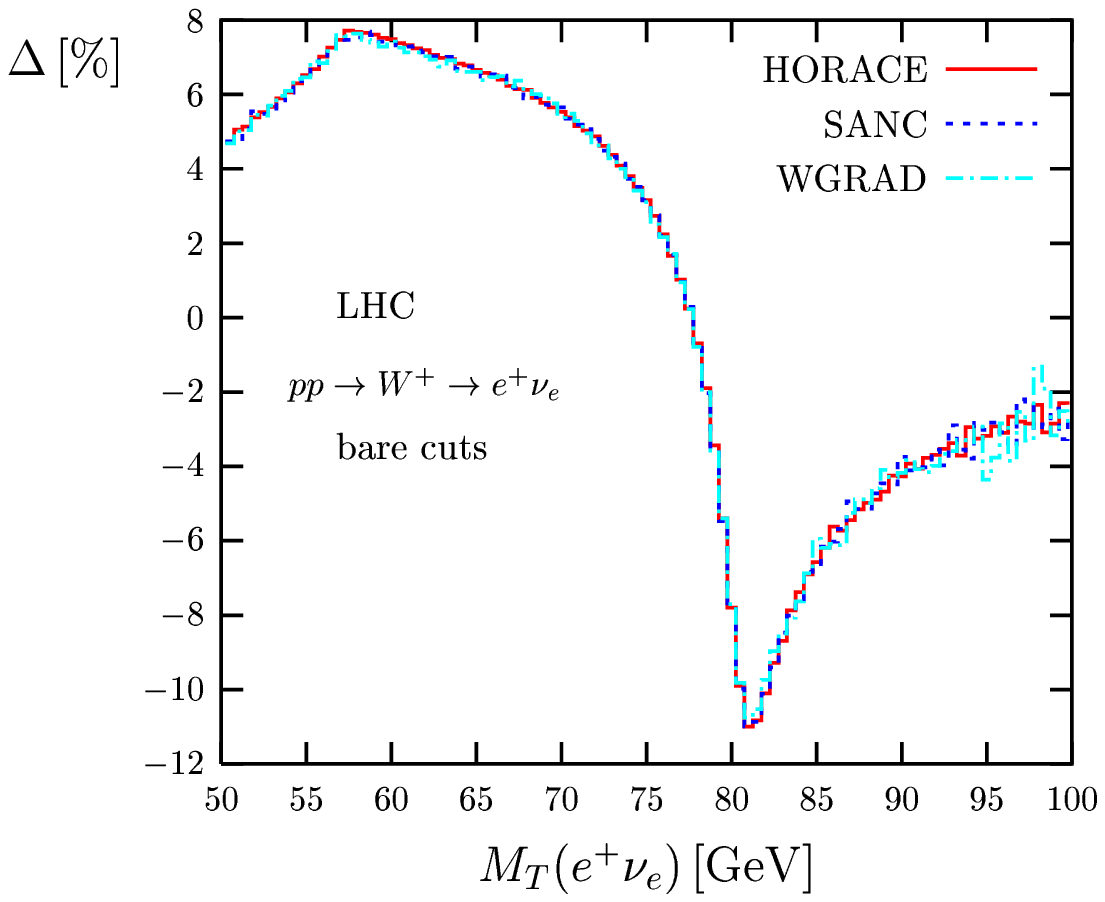}
\hspace*{-.0cm}
  \includegraphics[width=7.1cm,
  keepaspectratio=true]{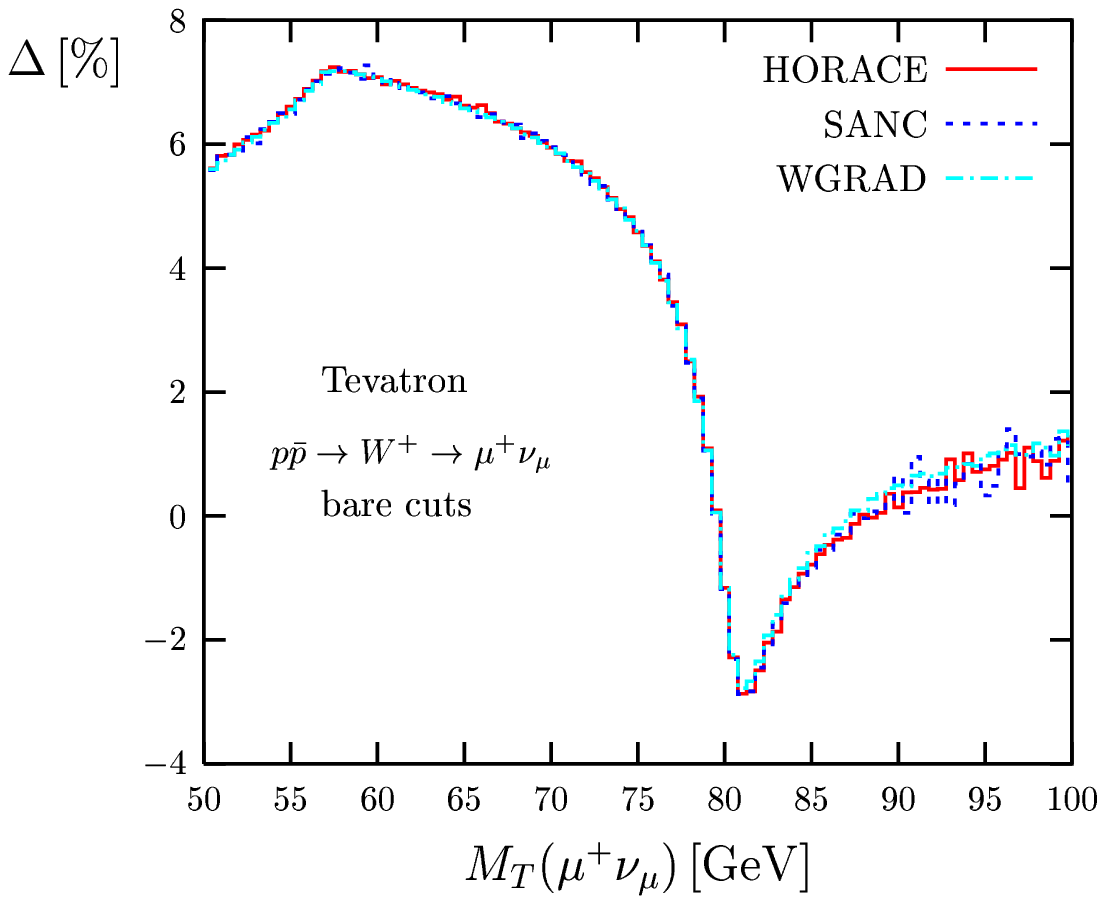}
\hspace*{-.0cm}
  \includegraphics[width=7.1cm,
  keepaspectratio=true]{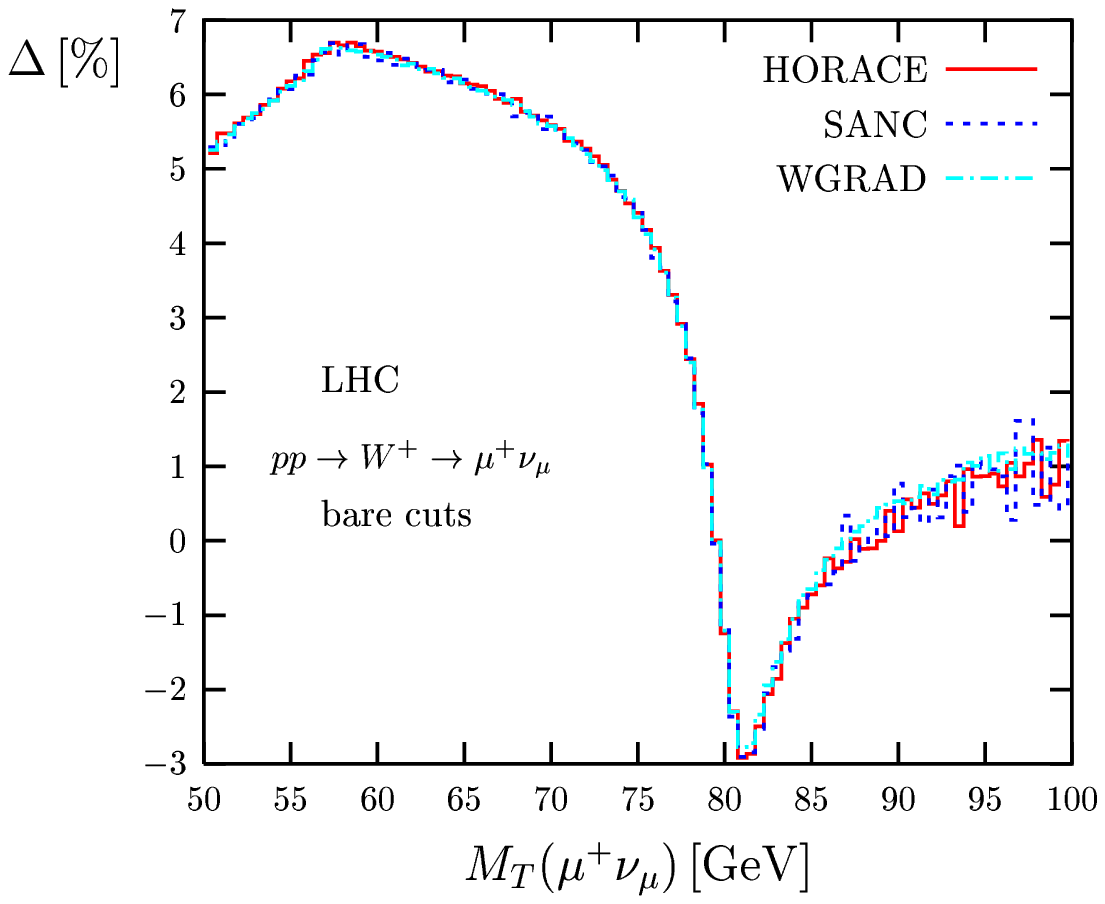}
\end{center}
\caption{The relative correction $\Delta$ due to electroweak ${\cal O}(\alpha)$ corrections to the $M_T(l\nu)$ distribution
for single $W^+$ production with bare cuts at the Tevatron and the LHC.}\label{fig:th_ewk_mt11}
\end{figure}

\begin{figure}
\begin{center}
  \includegraphics[width=7.1cm,
  keepaspectratio=true]{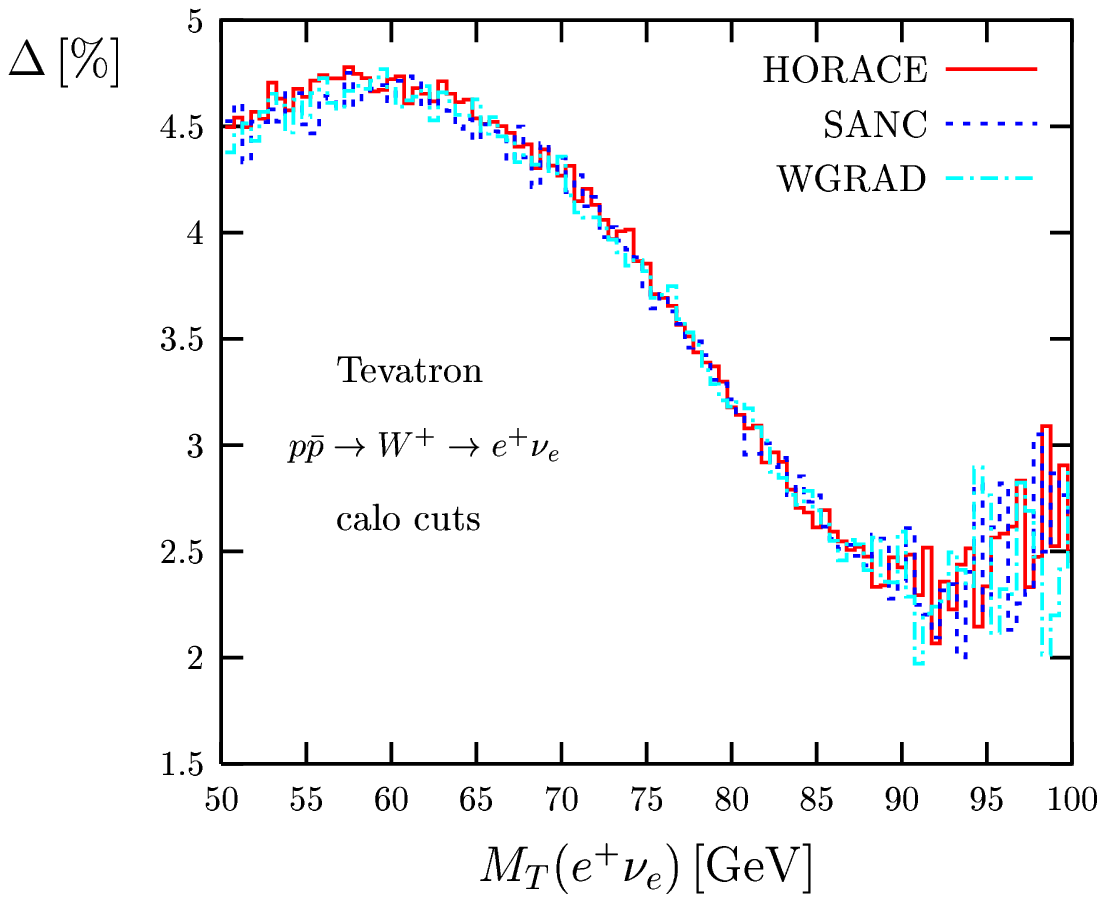}
\hspace*{-.0cm}
  \includegraphics[width=7.1cm,
  keepaspectratio=true]{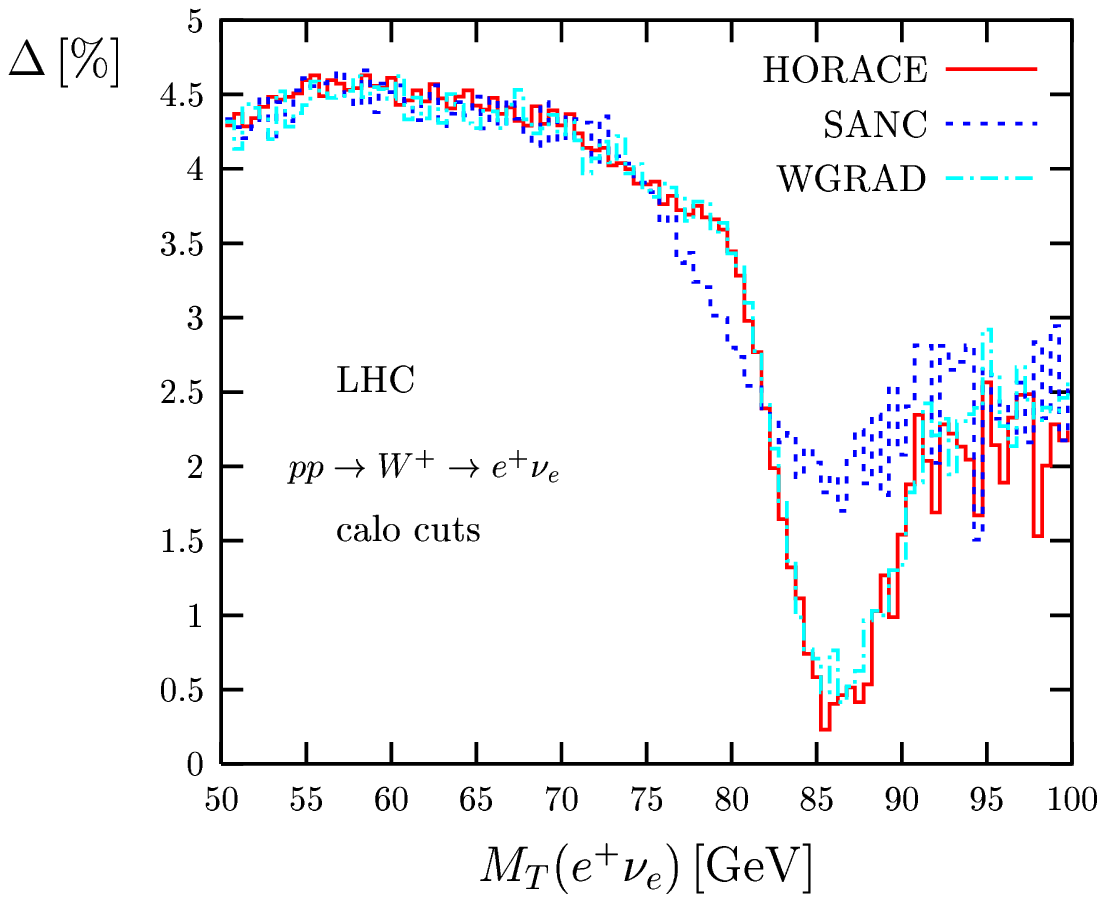}
\hspace*{-.0cm}
  \includegraphics[width=7.1cm,
  keepaspectratio=true]{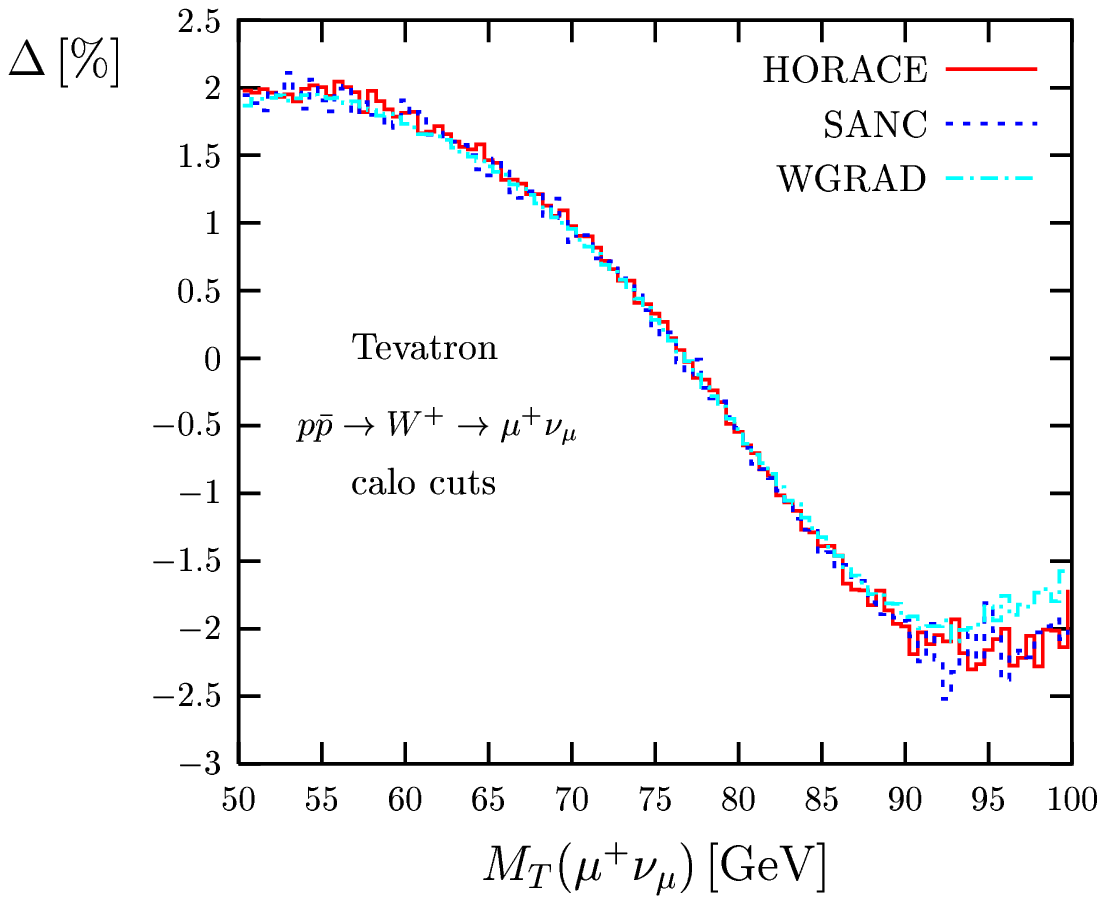}
\hspace*{-.0cm}
  \includegraphics[width=7.1cm,
  keepaspectratio=true]{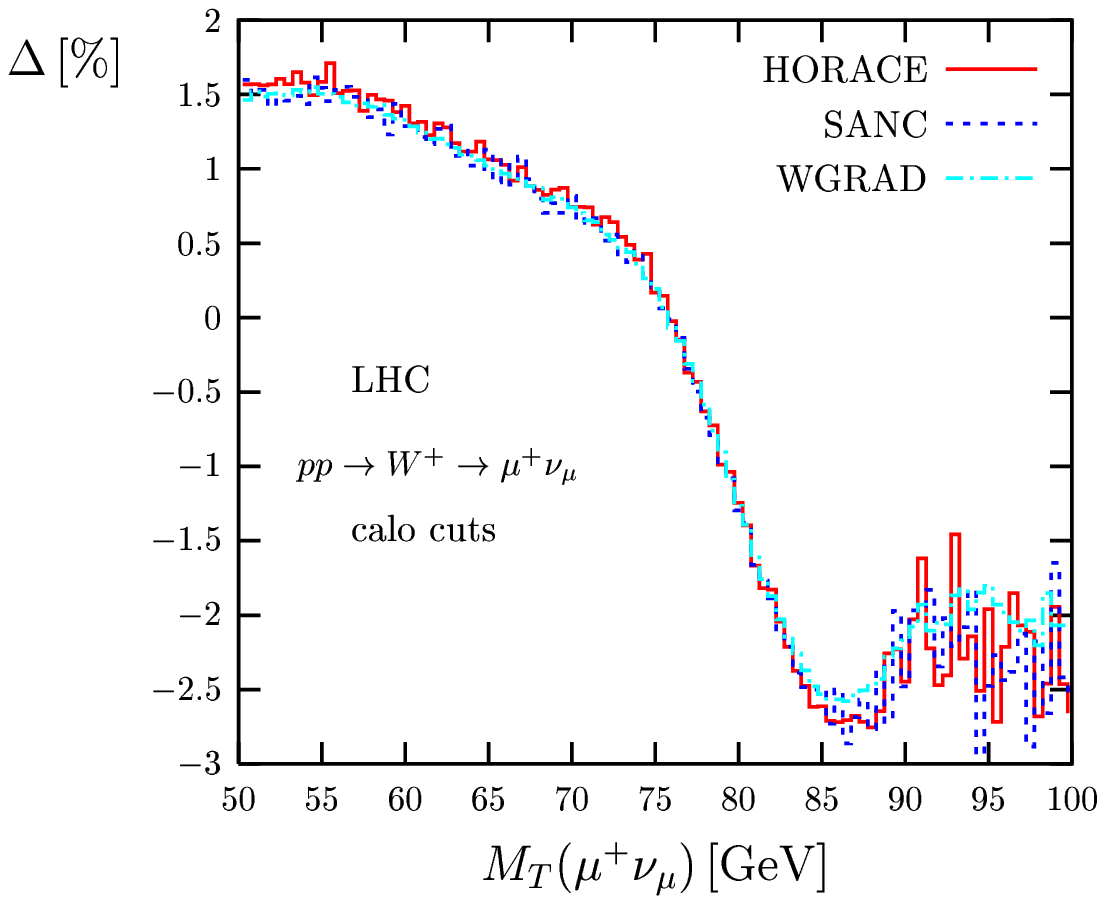}
\end{center}
\caption{The relative correction $\Delta$ due to electroweak ${\cal O}(\alpha)$ corrections to the $M_T(l\nu)$ distribution
for single $W^+$ production with calo cuts at the Tevatron and the LHC.}\label{fig:th_ewk_mt12}
\end{figure}

\begin{figure}
\begin{center}
  \includegraphics[width=7.1cm,
  keepaspectratio=true]{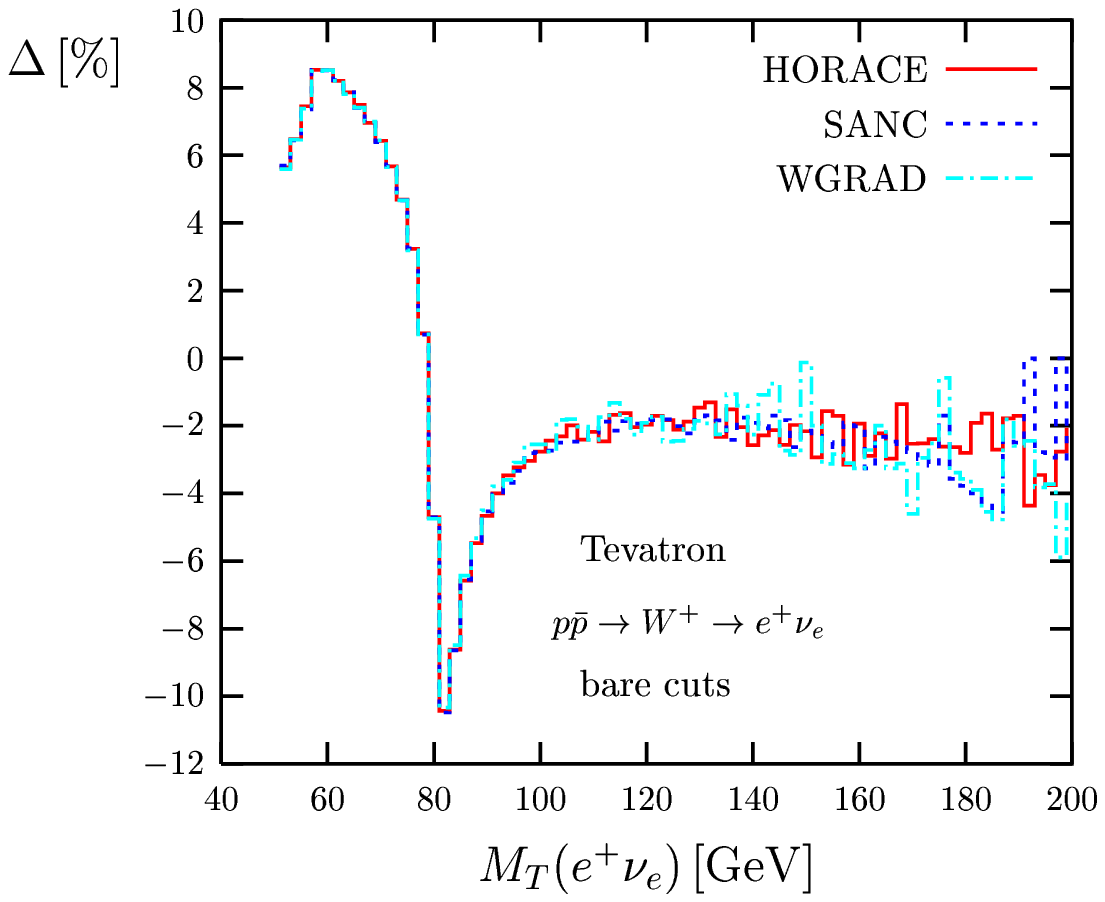}
\hspace*{-.0cm}
  \includegraphics[width=7.1cm,
  keepaspectratio=true]{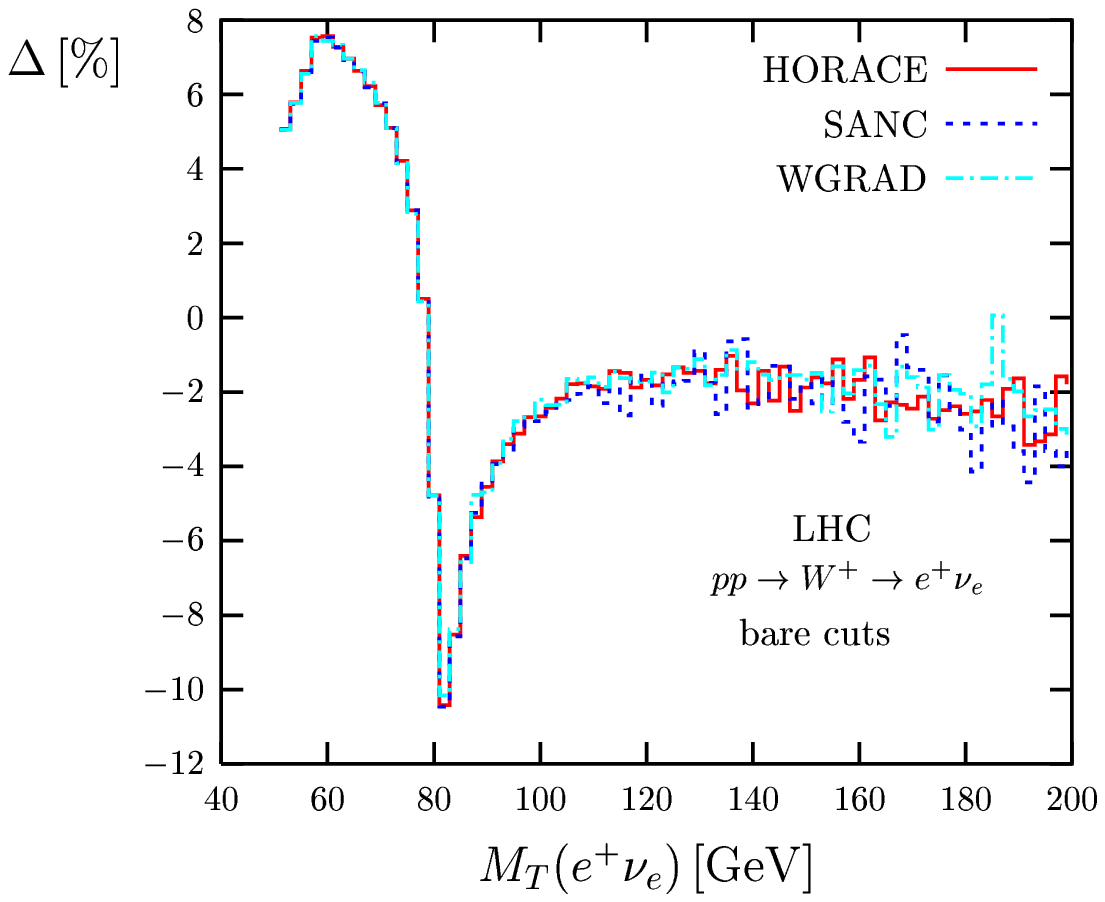}
\hspace*{-.0cm}
  \includegraphics[width=7.1cm,
  keepaspectratio=true]{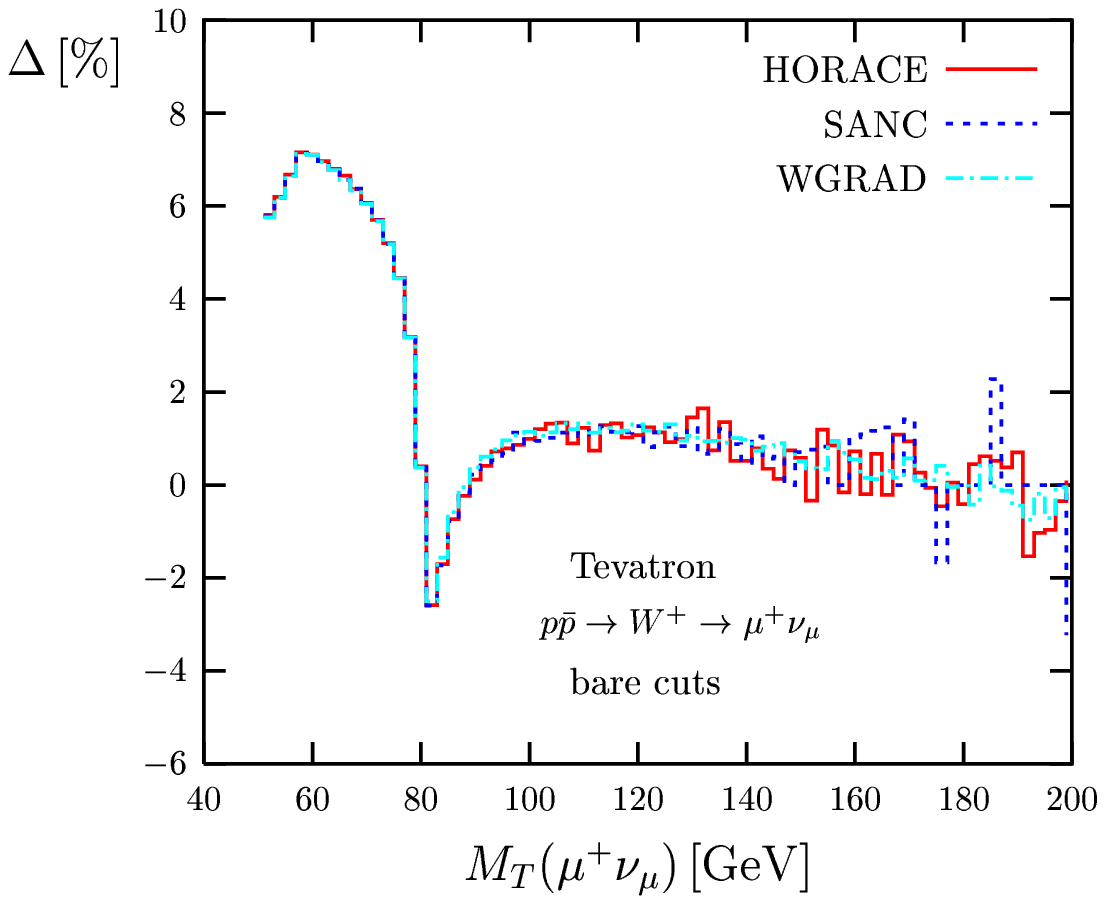}
\hspace*{-.0cm}
  \includegraphics[width=7.1cm,
  keepaspectratio=true]{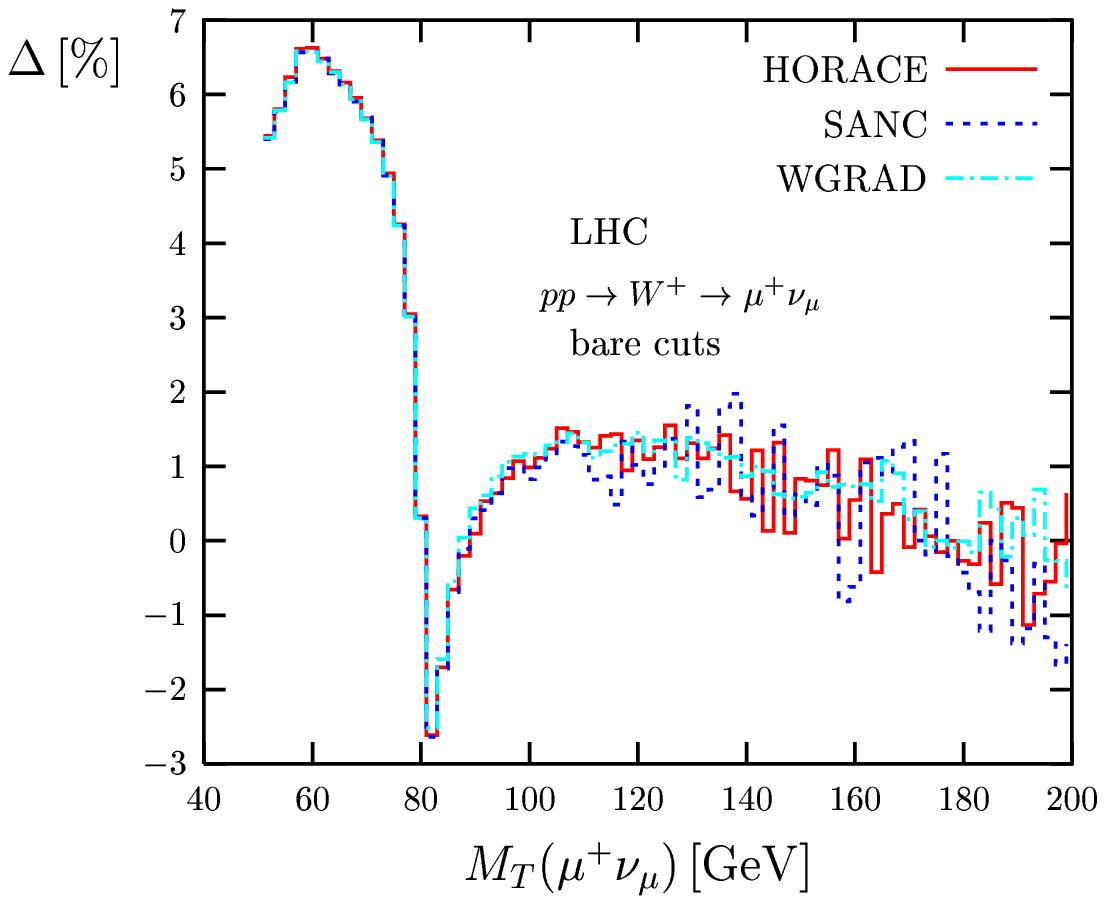}
\end{center}
\caption{The relative correction $\Delta$ due to electroweak ${\cal O}(\alpha)$ corrections to the $M_T(l\nu)$ distribution
for single $W^+$ production with bare cuts at the Tevatron and the LHC.}\label{fig:th_ewk_mt21}
\end{figure}

\begin{figure}
\begin{center}
  \includegraphics[width=7.1cm,
  keepaspectratio=true]{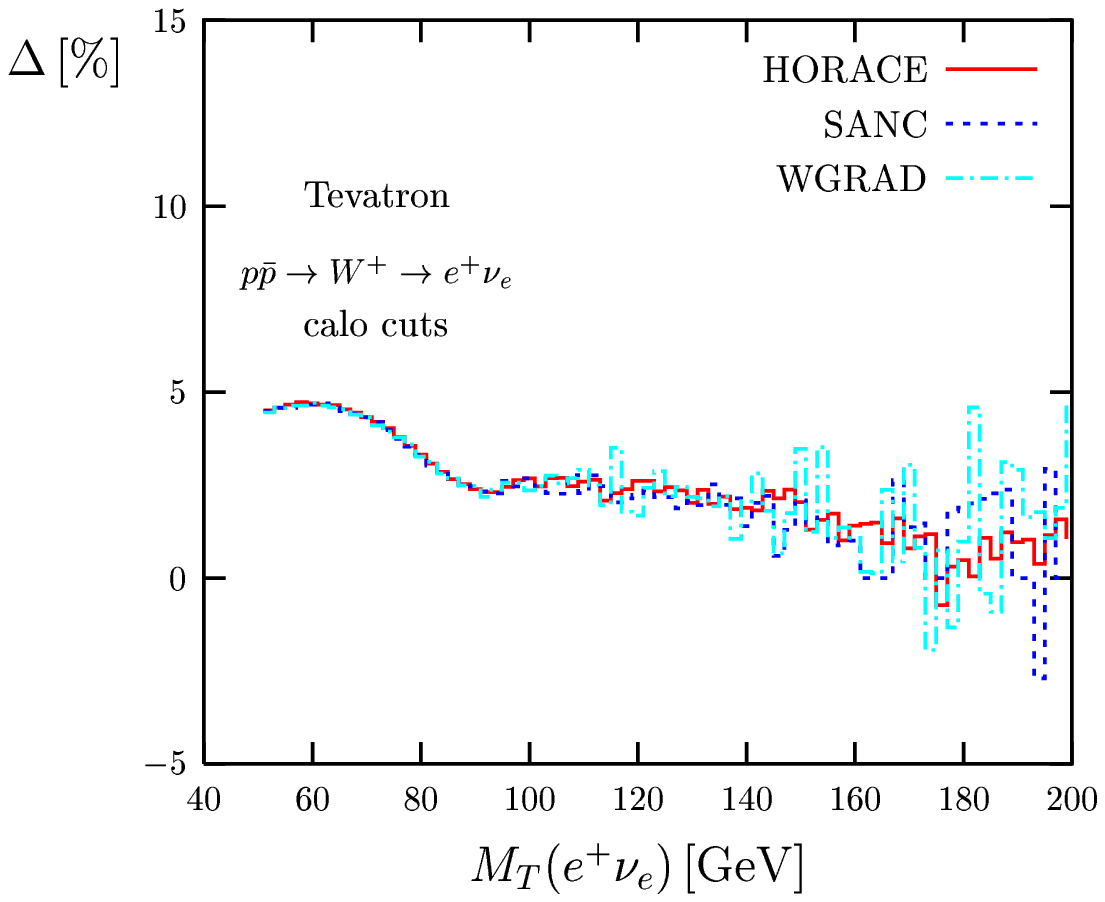}
\hspace*{-.0cm}
  \includegraphics[width=7.1cm,
  keepaspectratio=true]{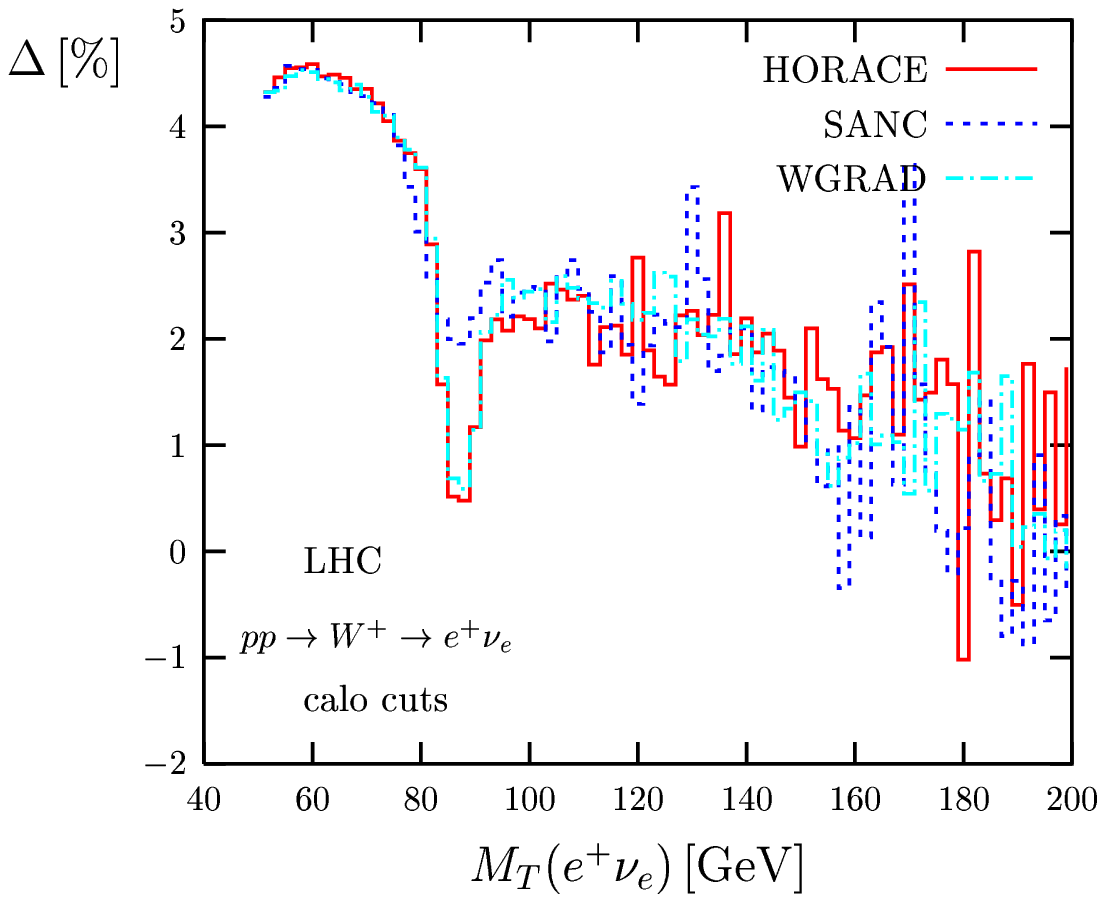}
\hspace*{-.0cm}
  \includegraphics[width=7.1cm,
  keepaspectratio=true]{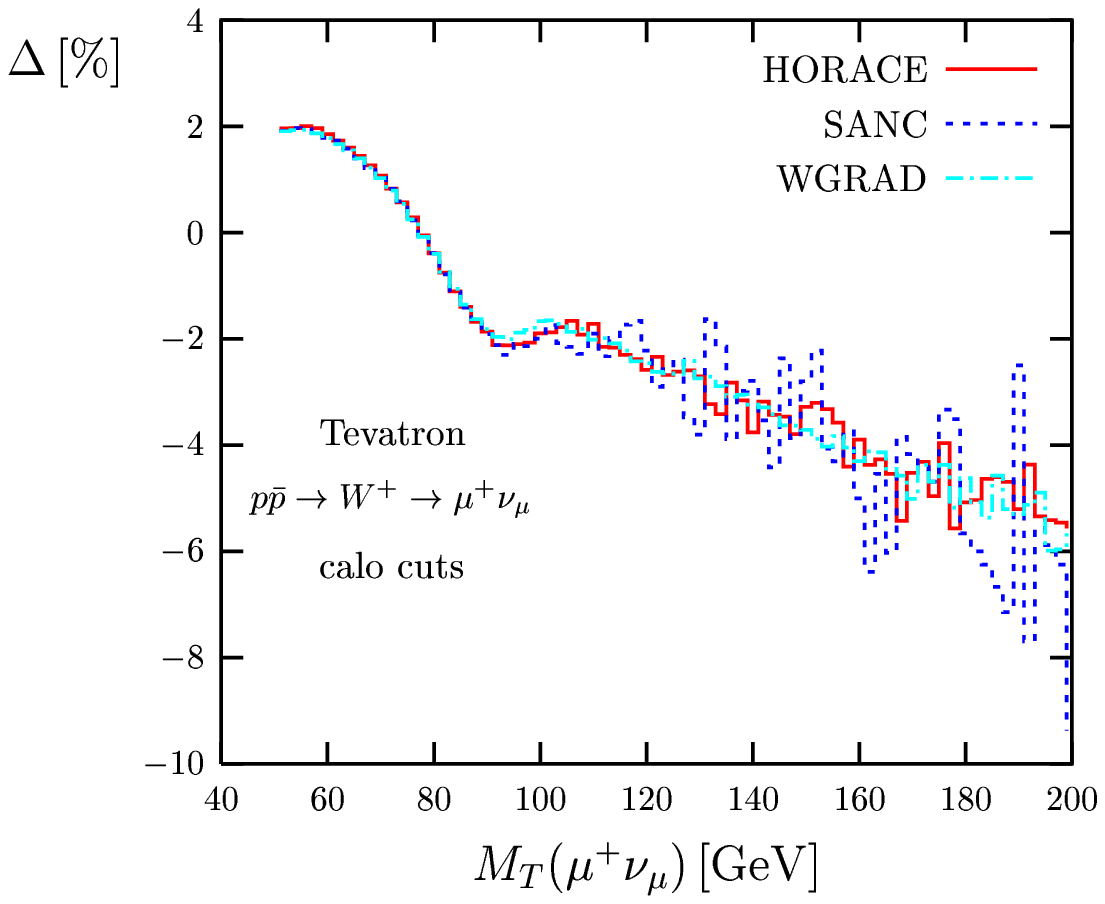}
\hspace*{-.0cm}
  \includegraphics[width=7.1cm,
  keepaspectratio=true]{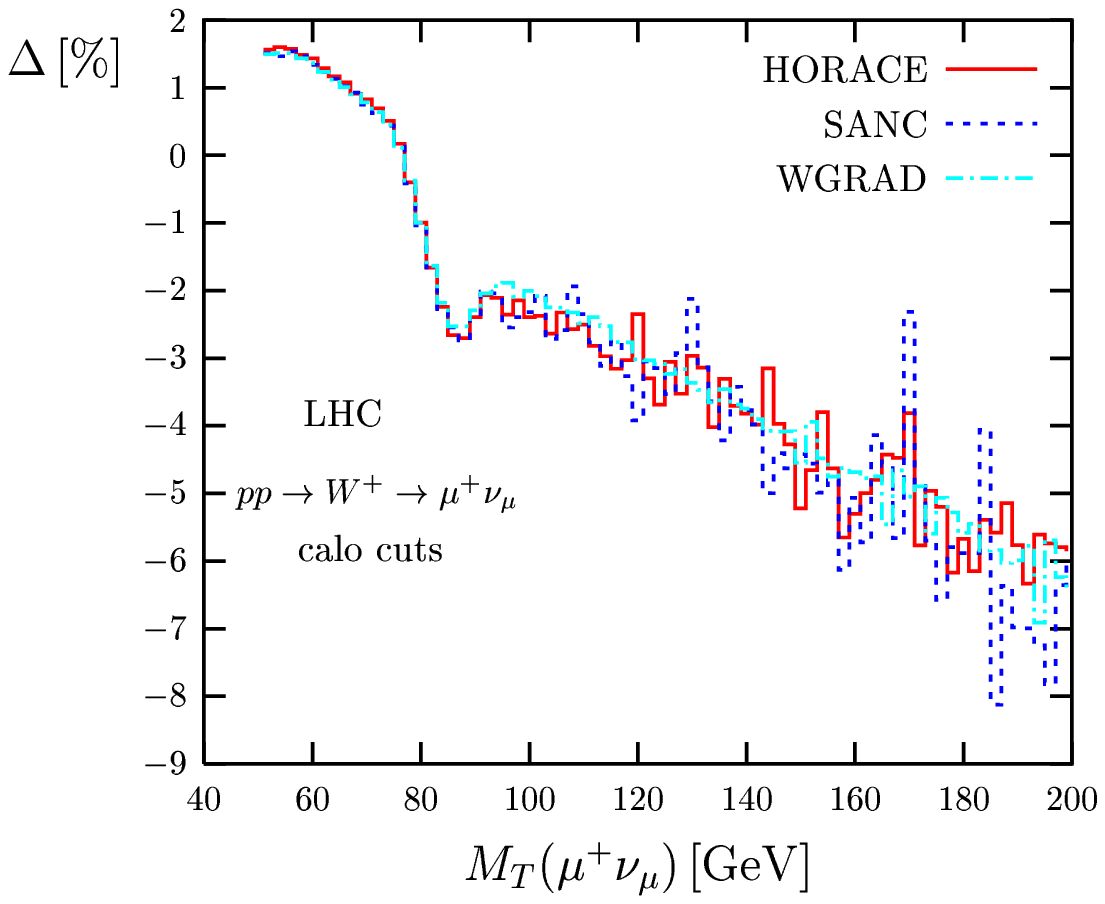}
\end{center}
\caption{The relative correction $\Delta$ due to electroweak ${\cal O}(\alpha)$ corrections to the $M_T(l\nu)$ distribution
for single $W^+$ production with calo cuts at the Tevatron and the LHC.}\label{fig:th_ewk_mt22}
\end{figure}

\begin{figure}
\begin{center}
  \includegraphics[width=7.1cm,
  keepaspectratio=true]{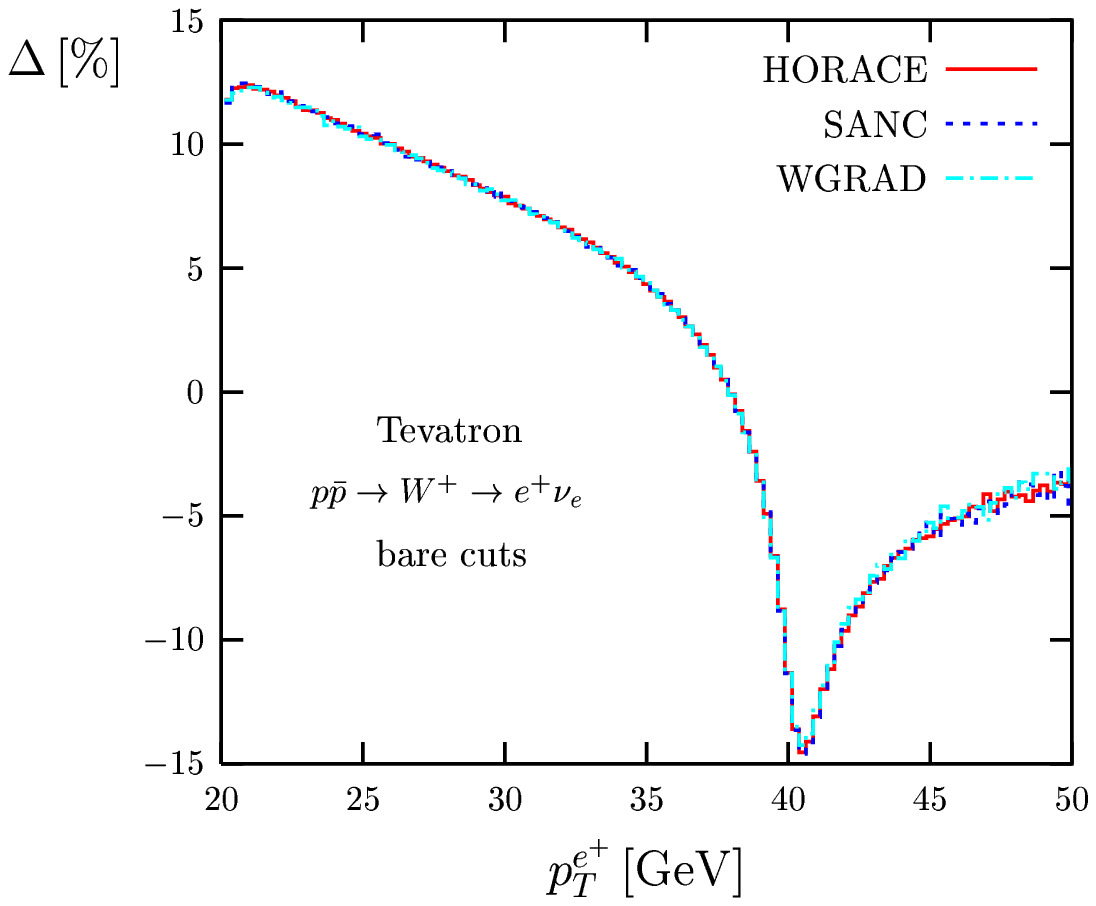}
\hspace*{-.0cm}
  \includegraphics[width=7.1cm,
  keepaspectratio=true]{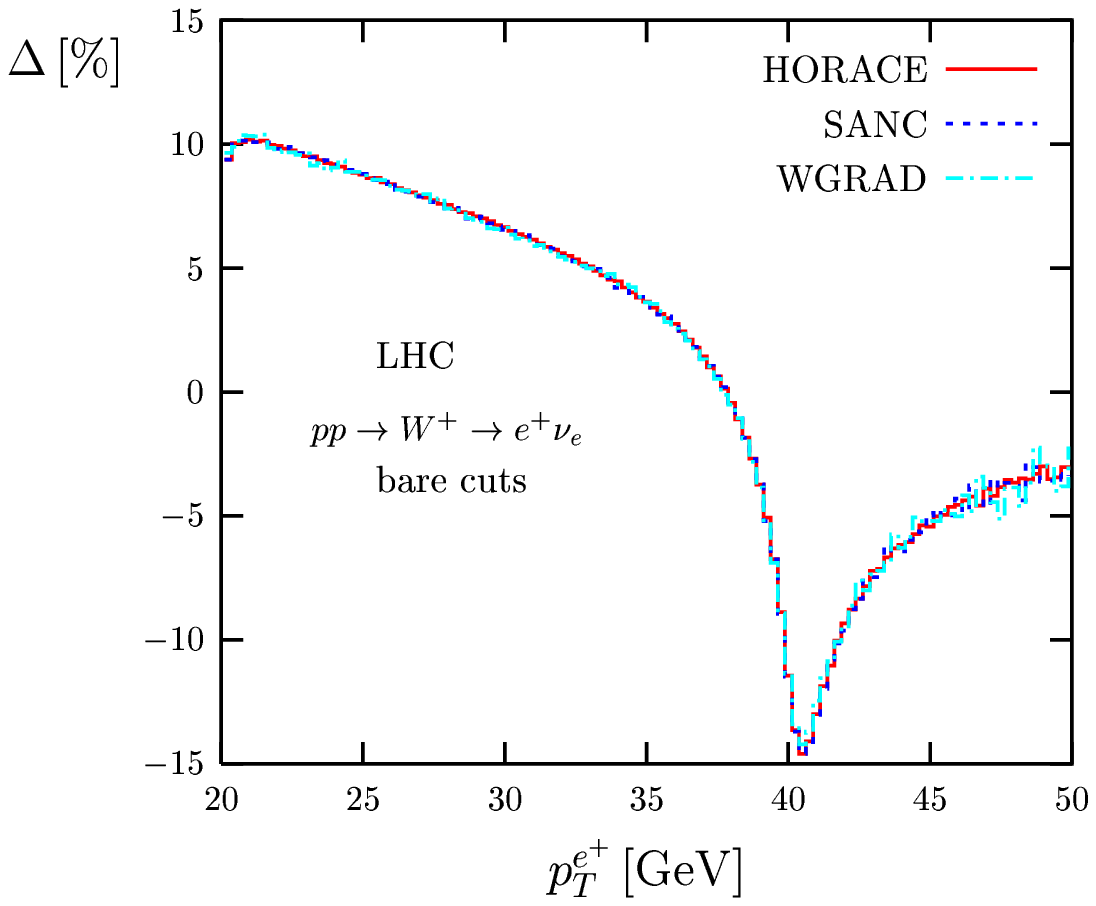}
\hspace*{-.0cm}
  \includegraphics[width=7.1cm,
  keepaspectratio=true]{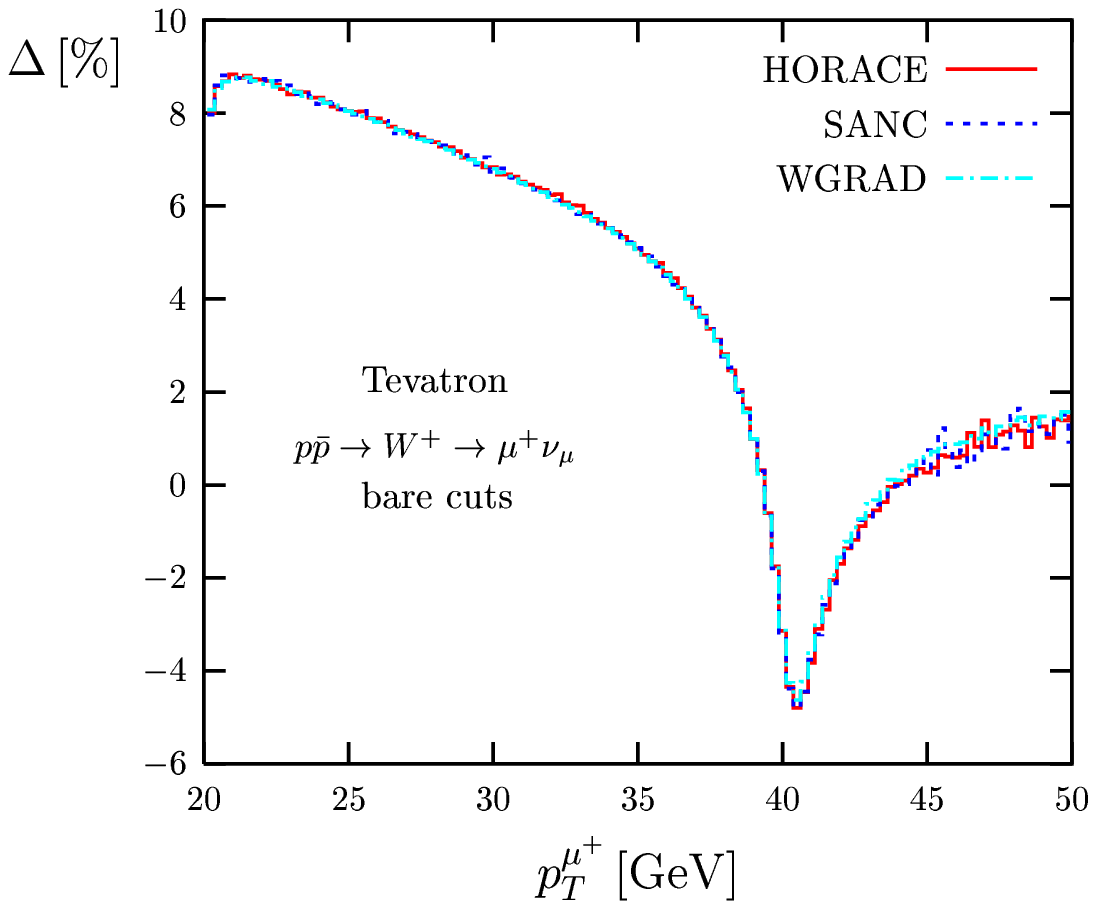}
\hspace*{-.0cm}
  \includegraphics[width=7.1cm,
  keepaspectratio=true]{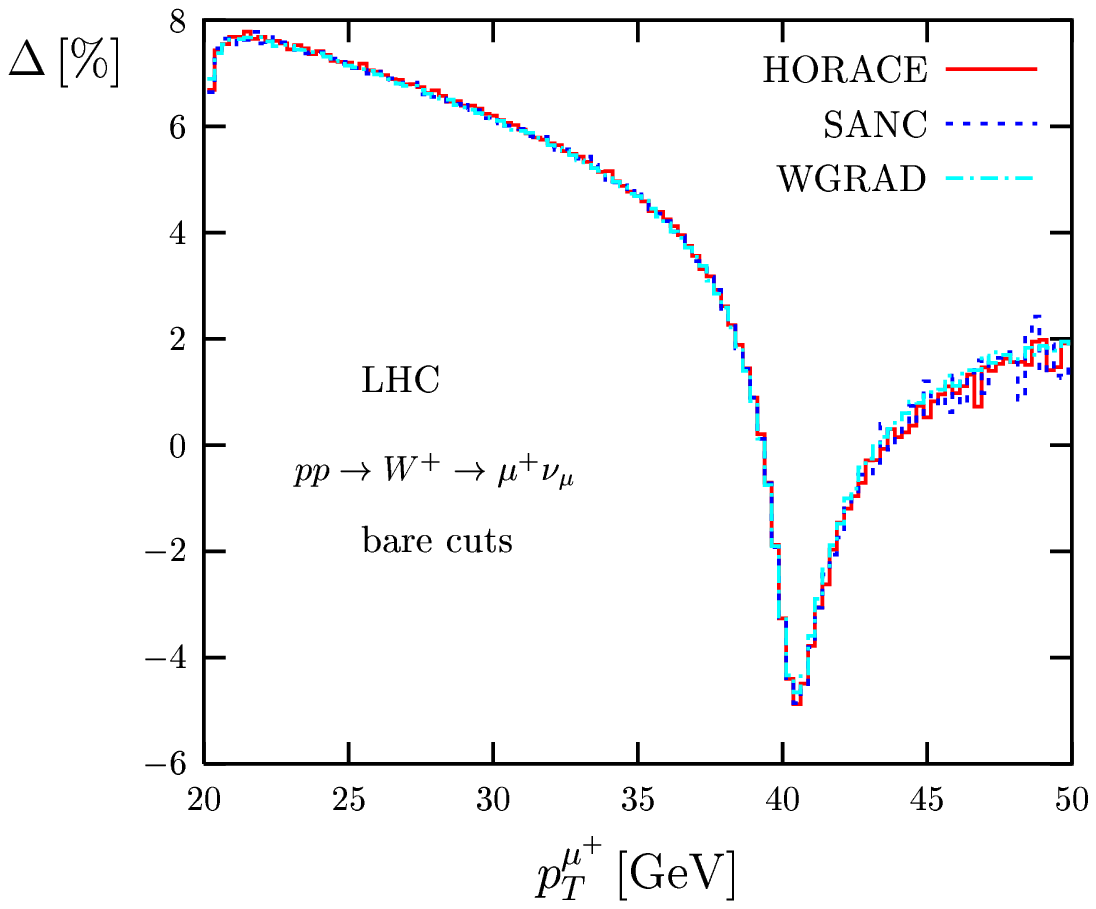}
\end{center}
\caption{The relative correction $\Delta$ due to electroweak ${\cal O}(\alpha)$ corrections to the $p_T(l)$ distribution
for single $W^+$ production with bare cuts at the Tevatron and the LHC.}\label{fig:th_ewk_pt1}
\end{figure}

\begin{figure}
\begin{center}
  \includegraphics[width=7.1cm,
  keepaspectratio=true]{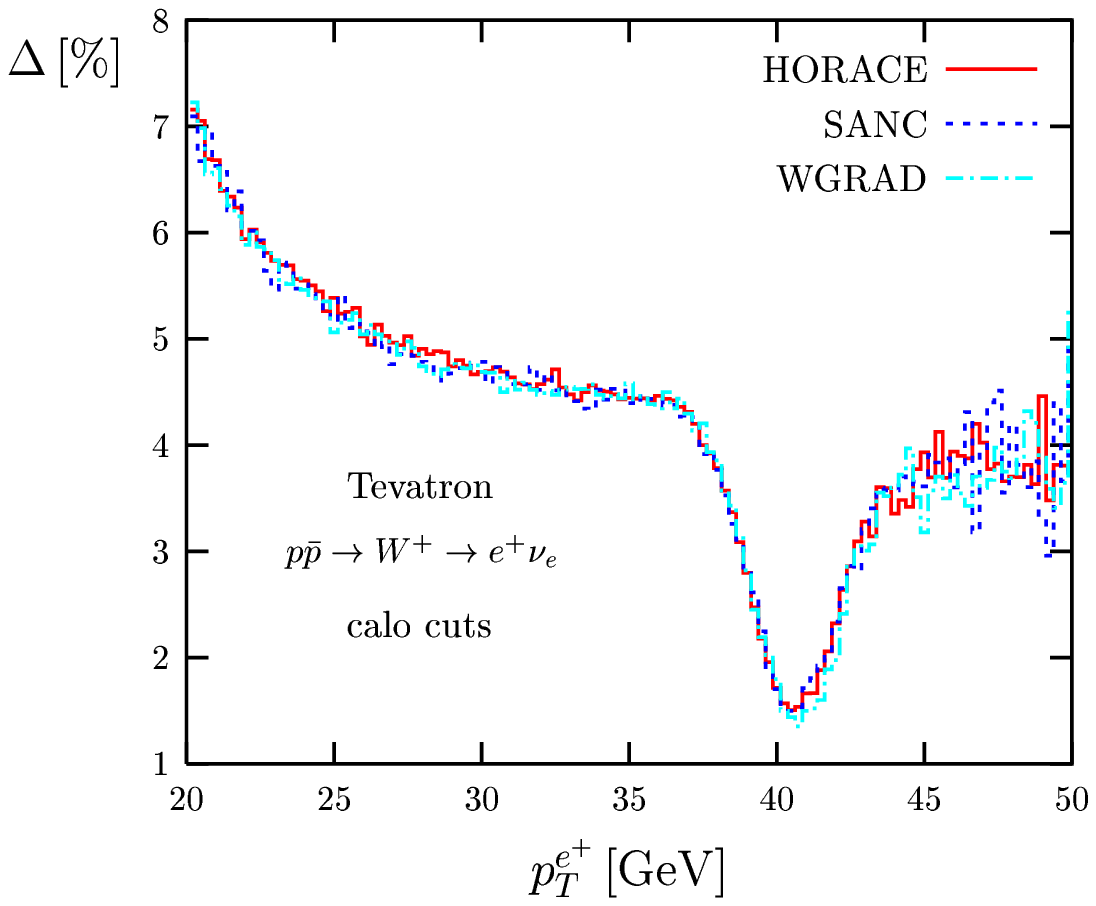}
\hspace*{-.0cm}
  \includegraphics[width=7.1cm,
  keepaspectratio=true]{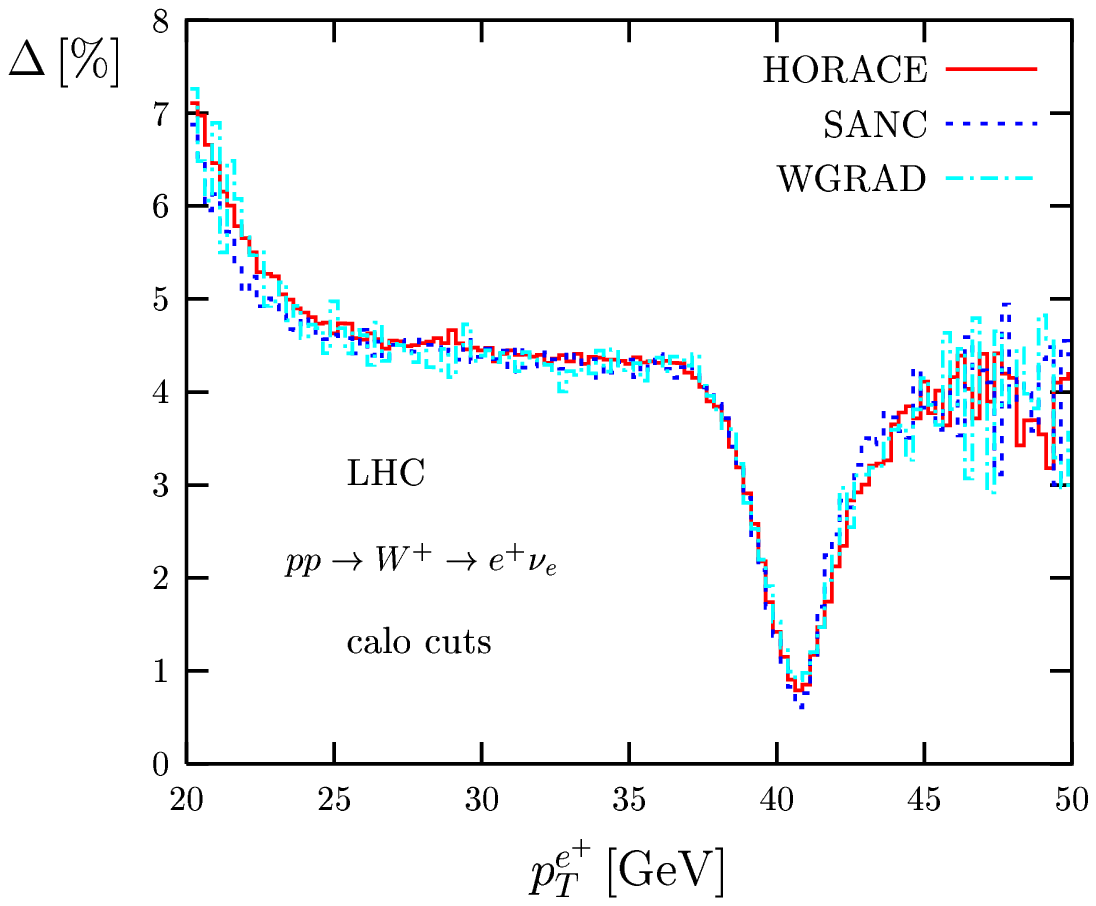}
\hspace*{-.0cm}
  \includegraphics[width=7.1cm,
  keepaspectratio=true]{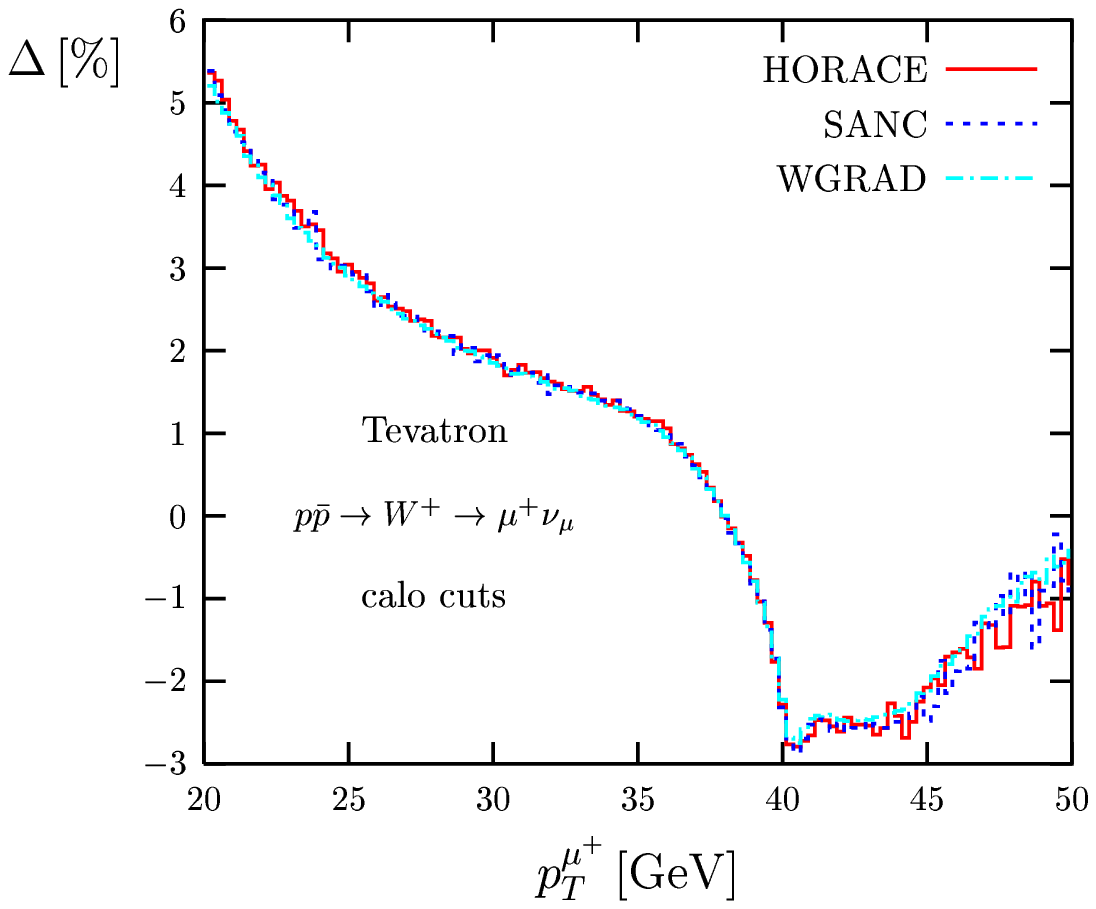}
\hspace*{-.0cm}
  \includegraphics[width=7.1cm,
  keepaspectratio=true]{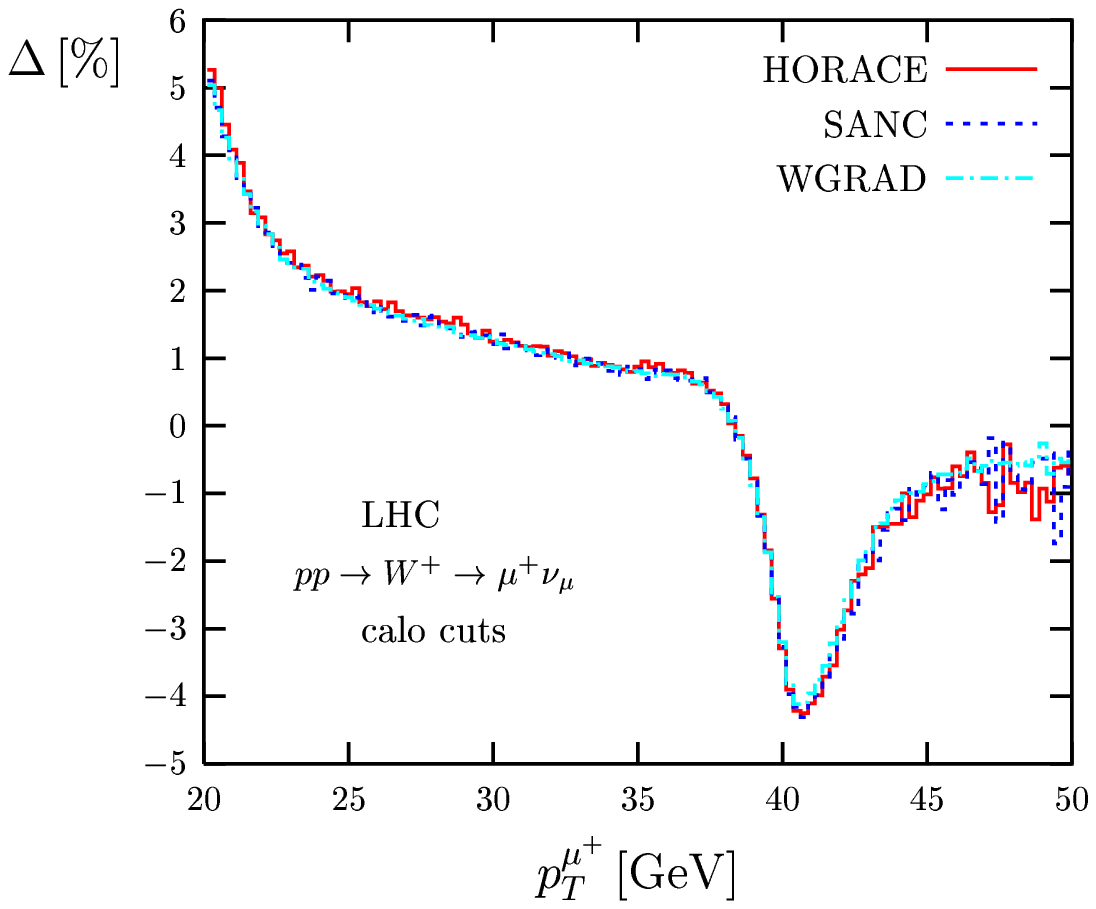}
\end{center}
\caption{The relative correction $\Delta$ due to electroweak ${\cal O}(\alpha)$ corrections to the $p_T(l)$ distribution
for single $W^+$ production with calo cuts at the Tevatron and the LHC.}\label{fig:th_ewk_pt2}
\end{figure}

\begin{figure}
\begin{center}
  \includegraphics[width=7.1cm,
  keepaspectratio=true]{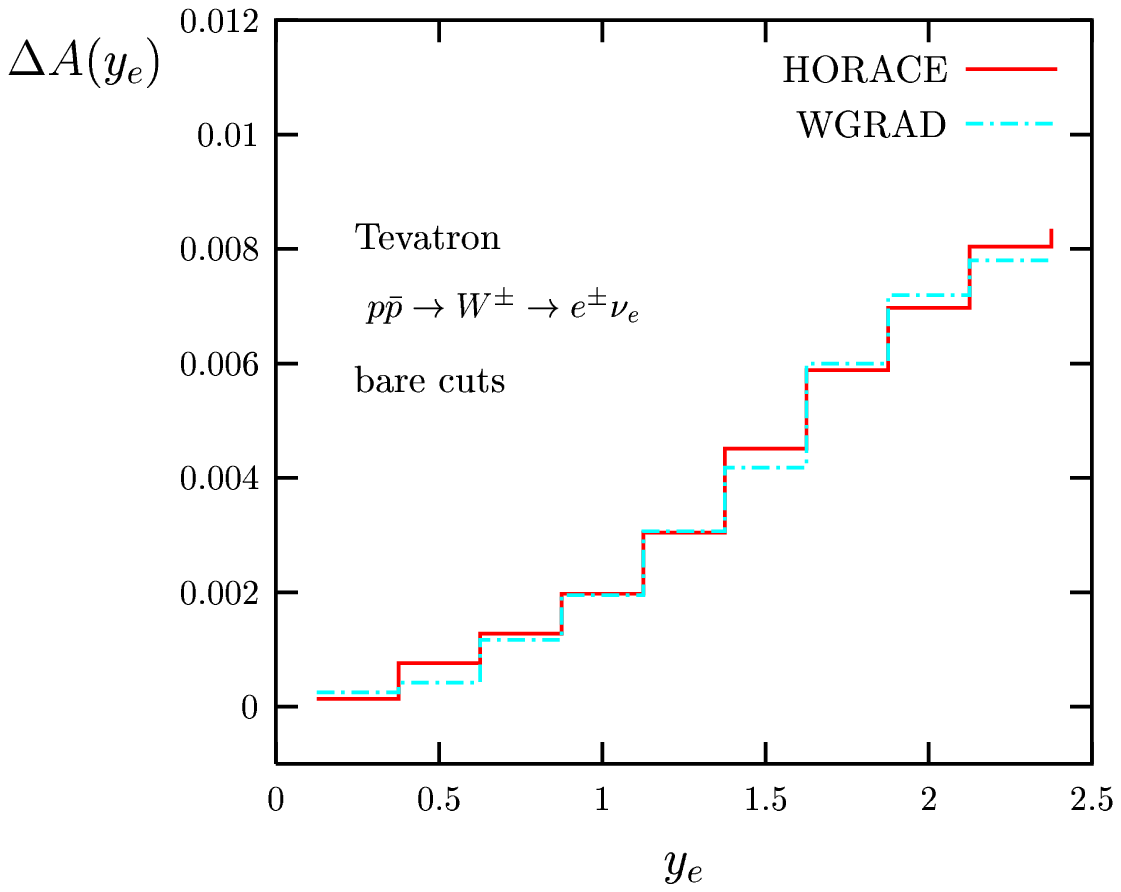}
\hspace*{-.0cm}
  \includegraphics[width=7.1cm,
  keepaspectratio=true]{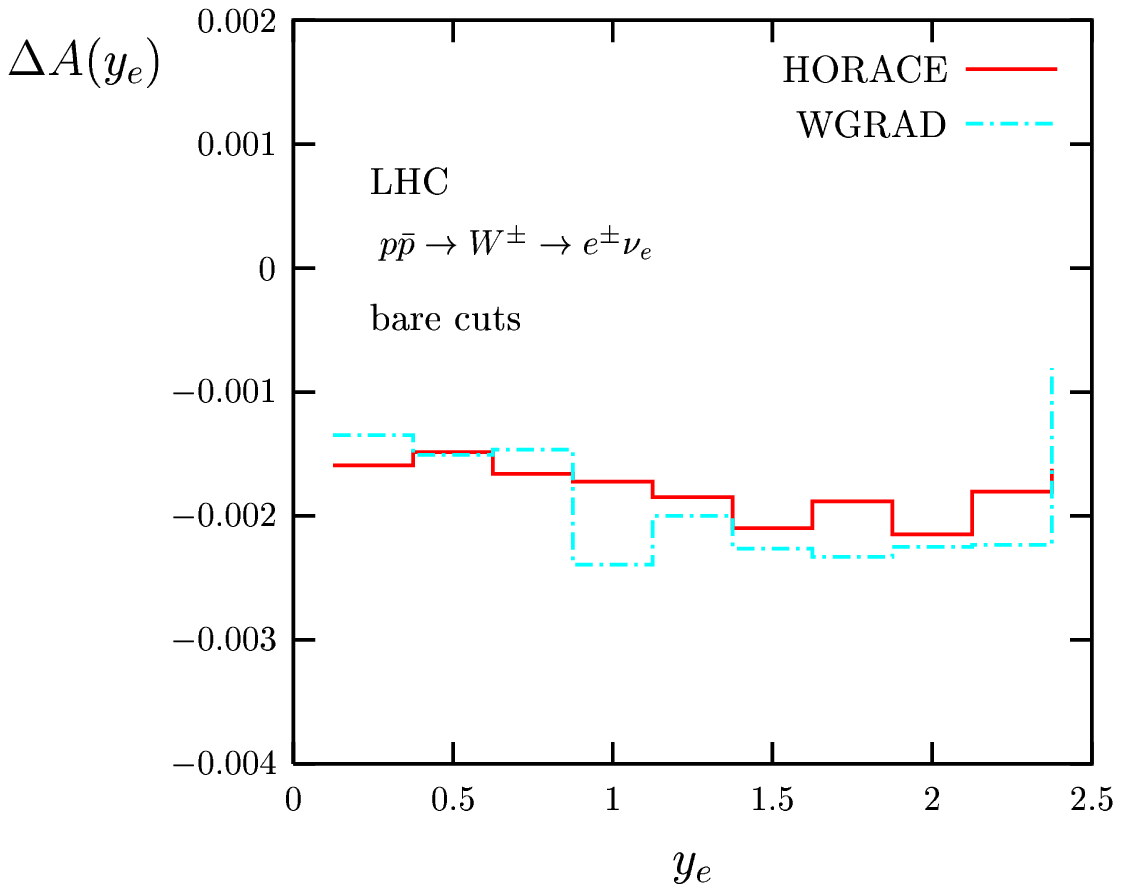}
\hspace*{-.0cm}
  \includegraphics[width=7.1cm,
  keepaspectratio=true]{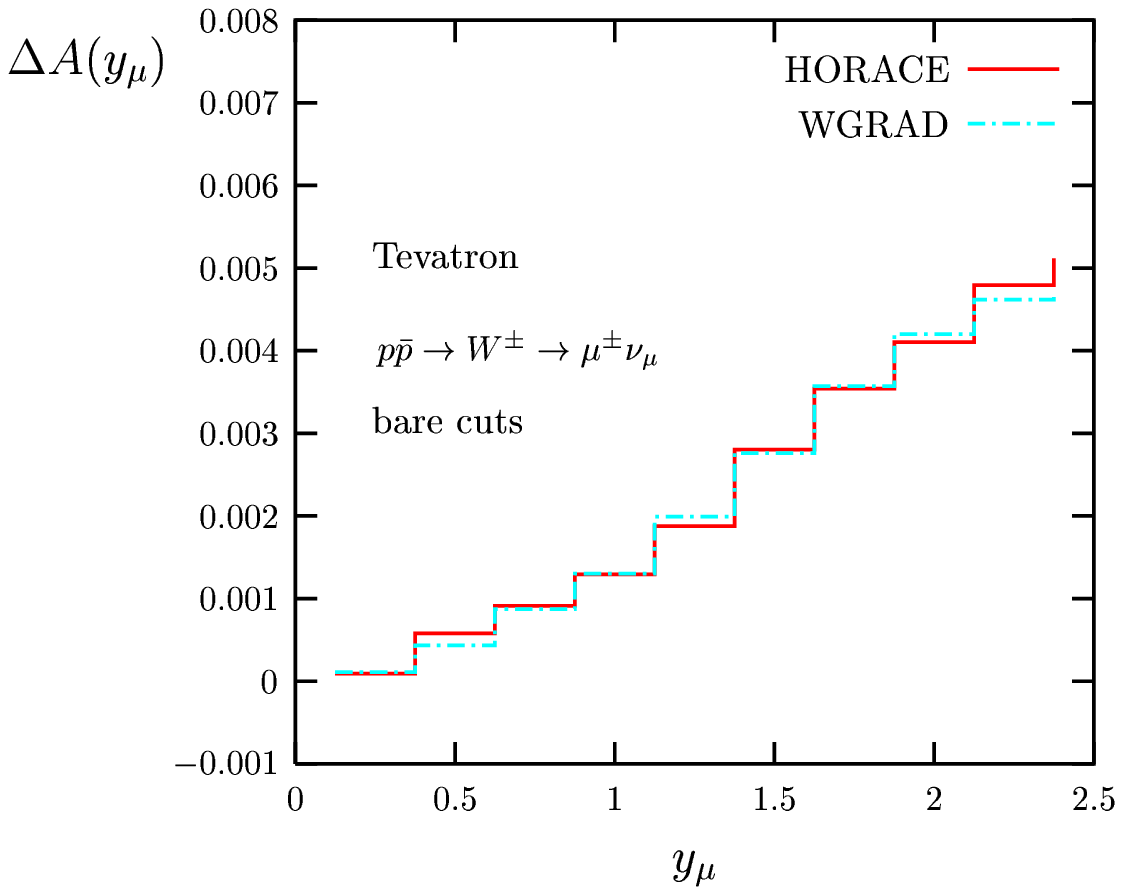}
\hspace*{-.0cm}
  \includegraphics[width=7.1cm,
  keepaspectratio=true]{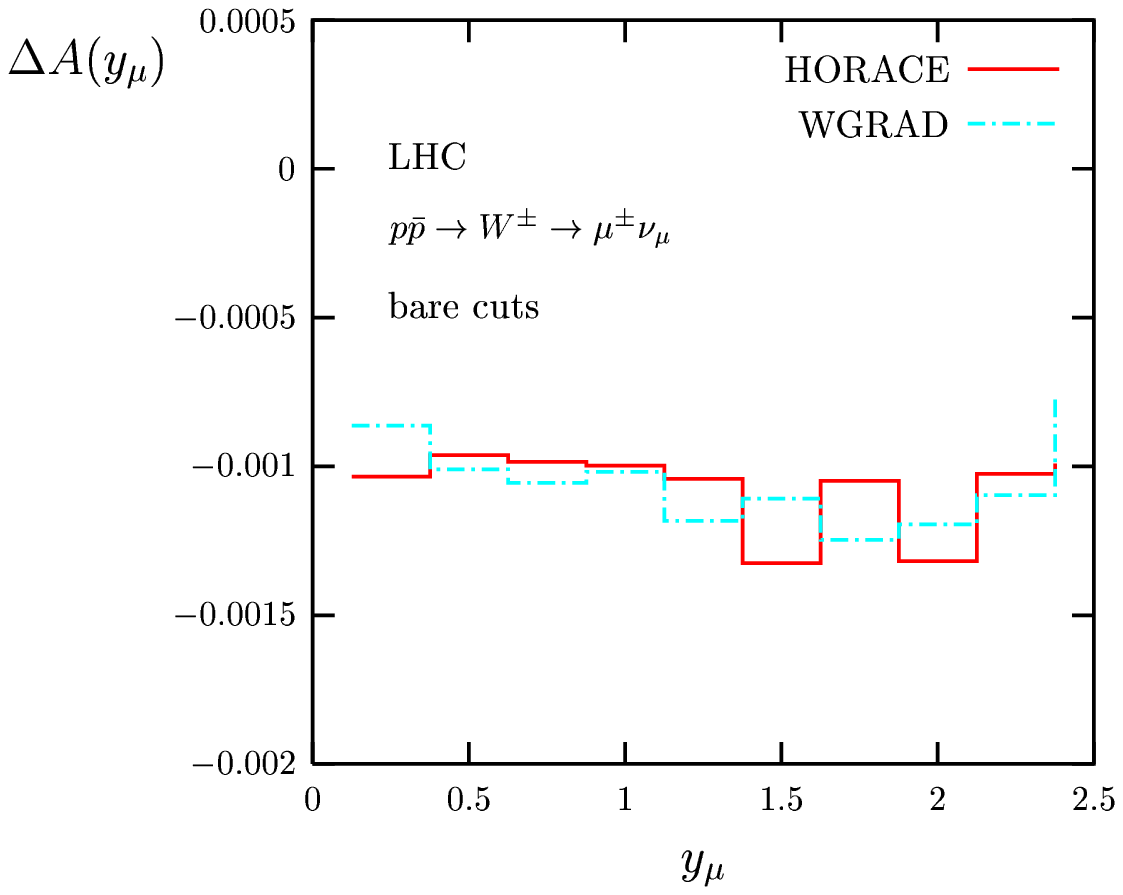}
\end{center}
\caption{The difference between the NLO and LO predictions for $A(y_l)$ due to electroweak ${\cal O}(\alpha)$ corrections
for single $W^\pm$ production with bare cuts at the Tevatron and the LHC.}\label{fig:th_ewk_yl1}
\end{figure}

\begin{figure}
\begin{center}
  \includegraphics[width=7.1cm,
  keepaspectratio=true]{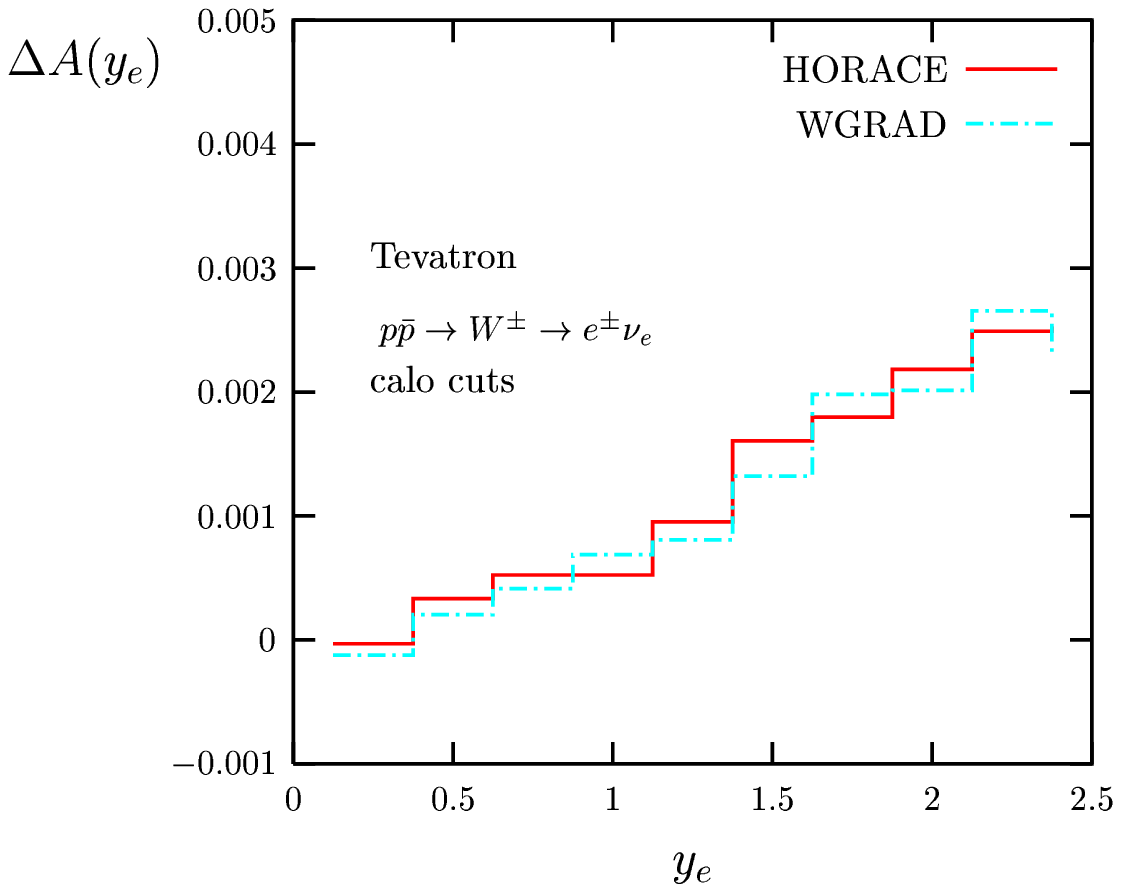}
\hspace*{-.0cm}
  \includegraphics[width=7.1cm,
  keepaspectratio=true]{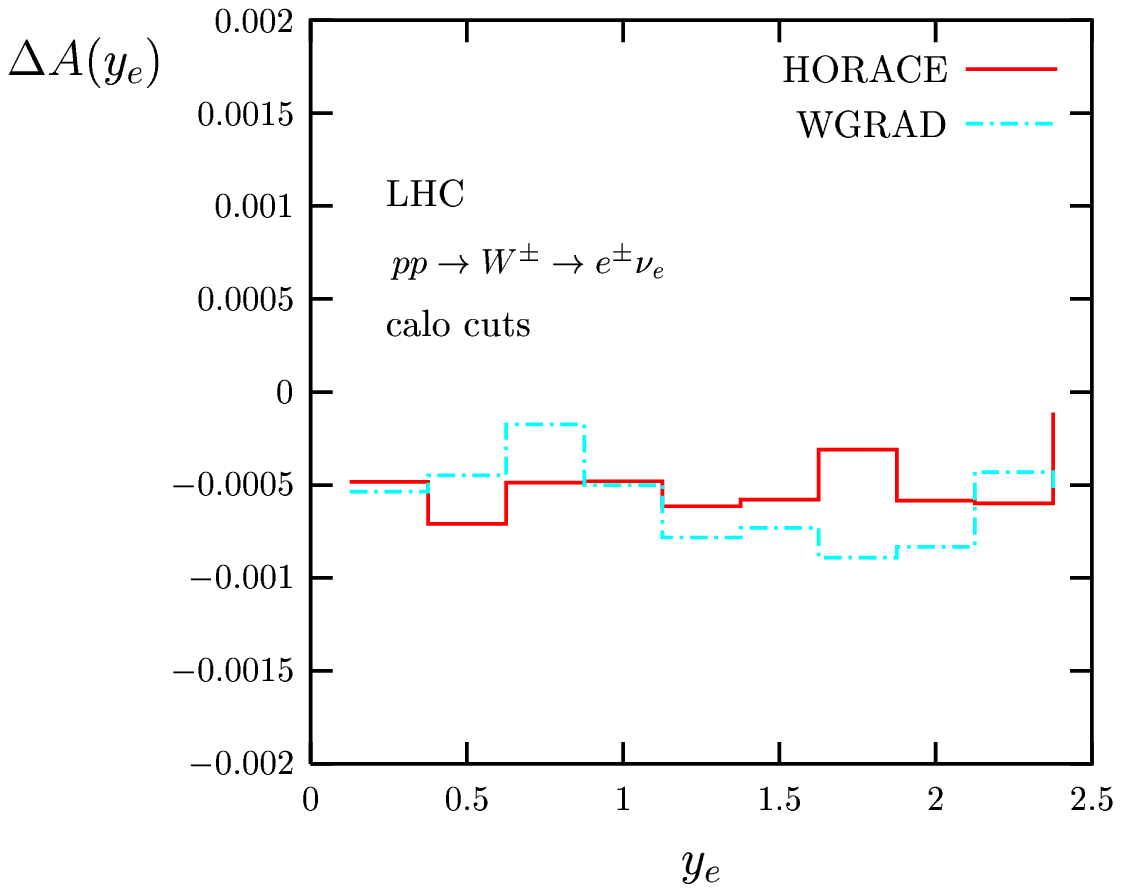}
\hspace*{-.0cm}
  \includegraphics[width=7.1cm,
  keepaspectratio=true]{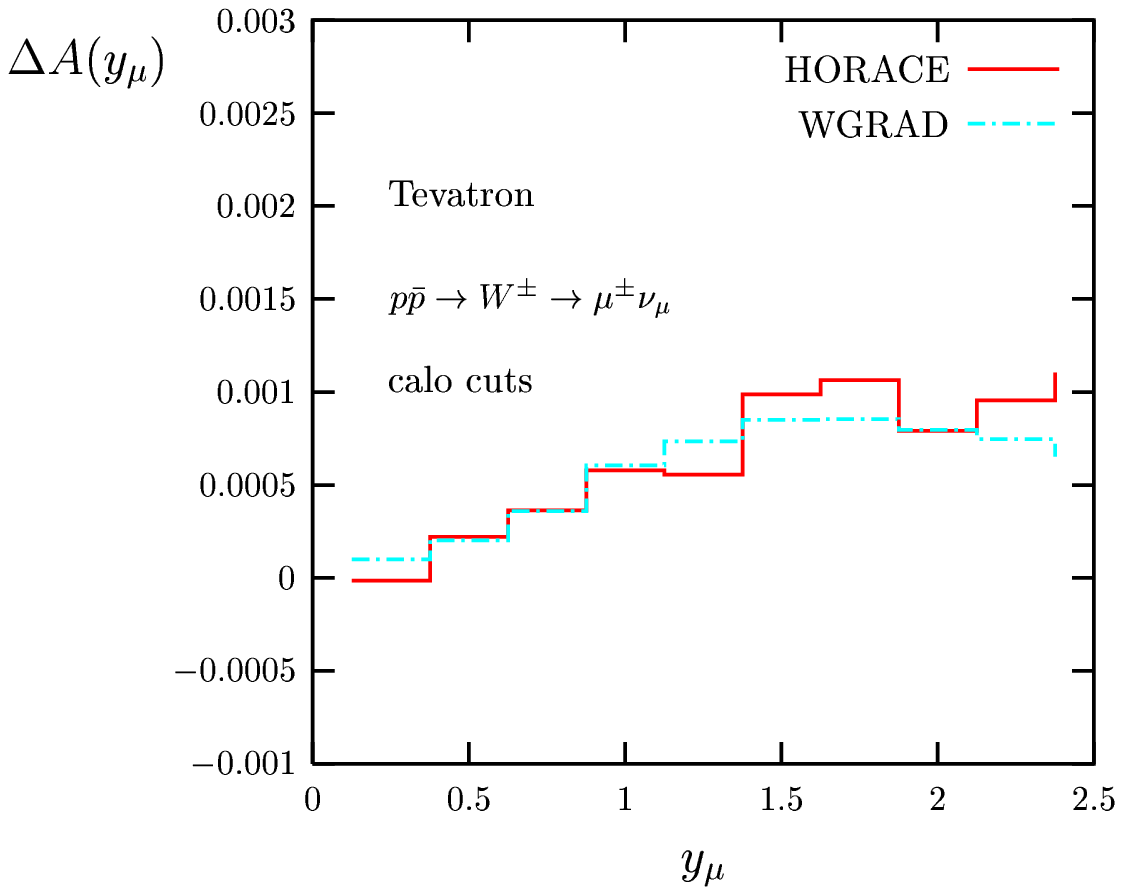}
\hspace*{-.0cm}
  \includegraphics[width=7.1cm,
  keepaspectratio=true]{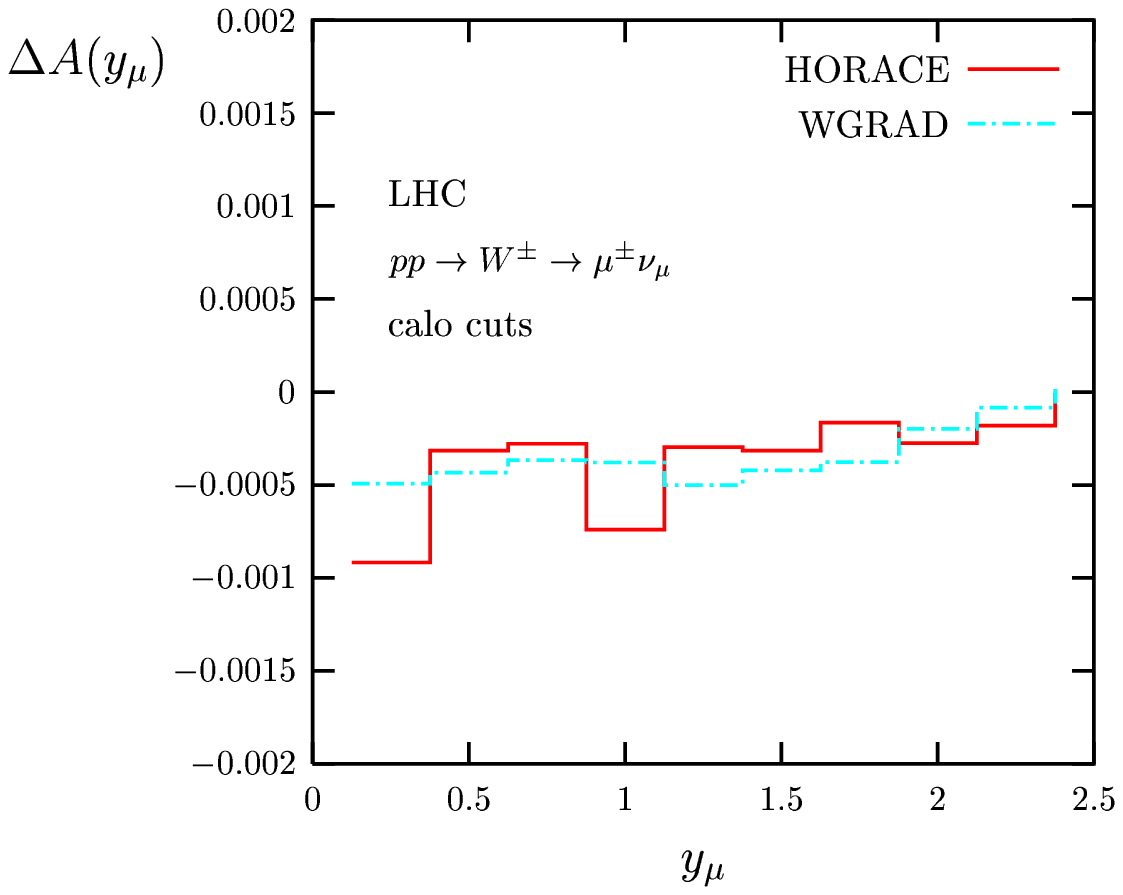}
\end{center}
\caption{The difference between the NLO and LO predictions for $A(y_l)$ 
due to electroweak ${\cal O}(\alpha)$ corrections for single $W^\pm$ production with calo cuts at the Tevatron and the LHC.}\label{fig:th_ewk_yl2}
\end{figure}

\begin{center} {\it $W$ boson observables} \end{center}
\begin{itemize}
\item
$\sigma_W$: total inclusive cross section of $W$ boson production.\\
\noindent
The results for $\sigma_W$ at LO and EW NLO and the corresponding
relative corrections $\Delta$ are provided in Table~\ref{tab:th_ewk_d}.
\item
$\frac{d\sigma}{dM_T(l\nu)}$: transverse mass distribution of the lepton lepton-neutrino pair.\\
\noindent
The transverse mass is defined as
\begin{equation}
M_T(l\nu)=\sqrt{2p_T(\ell)p_T(\nu)(1-\cos\phi^{\ell\nu})}~,
\label{eq:mt}
\end{equation}
where $p_T(\nu)$ is the transverse momentum of the neutrino, and
$\phi^{\ell\nu}$ is the angle between the charged lepton and the
neutrino in the transverse plane. The neutrino transverse momentum is
identified with the missing transverse momentum, $p\llap/_T$, in the
event.

The relative corrections $\Delta$ for different $M_T$ ranges 
are shown in Figs.~\ref{fig:th_ewk_mt11},\ref{fig:th_ewk_mt21}
for bare cuts and in Figs.~\ref{fig:th_ewk_mt12},\ref{fig:th_ewk_mt22} for calo cuts. 
\item
$\frac{d\sigma}{dp_T^l}$: transverse lepton momentum distribution. \\
\noindent
The relative corrections $\Delta$ are shown in Fig.~\ref{fig:th_ewk_pt1}
for bare cuts and in Fig.~\ref{fig:th_ewk_pt2} for calo cuts. 
\item
$A(y_l)$: $W$ charge asymmetry for leptons. \\
\noindent
The charge asymmetry of leptons in $W$ decays~\cite{Baur:1998kt} is defined as
\begin{equation}
A(y_l)={d\sigma^+/dy_l-d\sigma^-/dy_l\over
d\sigma^+/dy_l+d\sigma^-/dy_l}~, 
\end{equation}
where $y_l$ is the lepton rapidity and 
\begin{equation}
d\sigma^\pm=d\sigma(pp,p\bar p\to l^\pm\nu X),
\end{equation}
In Fig.~\ref{fig:th_ewk_yl1} (with bare cuts) and Fig.~\ref{fig:th_ewk_yl2} (with calo cuts) we show the difference $\Delta A(y_l)$ between the NLO EW and LO predictions
for the charge asymmetries at the Tevatron and the LHC. 
\end{itemize}
\begin{table}
\begin{center}
\begin{tabular}{|c|c|c|c|c|c|c|} \hline
\multicolumn{7}{|c|}{\bf Tevatron, $p \bar p \to W^+ \to e^+ \nu_e$} \\ \hline
& \multicolumn{3}{|c|}{bare cuts} & \multicolumn{3}{|c|}{calo cuts} \\ \hline
           & LO [pb]& NLO [pb]& $\Delta$ [\%] & LO [pb]& NLO [pb]& $\Delta$ [\%]  \\ \hline
{\sc HORACE} & 773.509(5)  &  791.14(2) & 2.279(3) & 733.012(5)  &  762.21(3) & 3.983(4) \\
{\sc SANC}   & 773.510(2)  &  791.04(8) & 2.27(1) & 733.024(2)  &  762.03(9) & 3.96(1) \\ 
{\sc WGRAD2} & 773.516(5)   &  791.01(5) & 2.268(7) & 733.004(6)   &  762.00(5) & 3.956(6) \\
\hline
\multicolumn{7}{|c|}{\bf Tevatron, $p \bar p \to W^+ \to \mu^+ \nu_\mu$} \\ \hline
& \multicolumn{3}{|c|}{bare cuts} & \multicolumn{3}{|c|}{calo cuts} \\ \hline
           & LO [pb]& NLO [pb]& $\Delta$ [\%]& LO [pb]& NLO [pb]& $\Delta$ [\%] \\ \hline
{\sc HORACE} & 773.509(5)  &  804.18(2) & 3.965(3) & 732.913(6)  &  738.16(3) & 0.716(4) \\
{\sc SANC}   & 773.510(2)  &  804.07(6) & 3.951(7) & 732.908(2)  &  738.01(5) & 0.696(7) \\ 
{\sc WGRAD2} & 773.516(5)   &  804.11(1) & 3.955(2) & 732.917(6)   &  738.00(1) & 0.693(2) \\
\hline
\multicolumn{7}{|c|}{\bf LHC, $p p \to W^+ \to e^+ \nu_e$} \\ \hline
& \multicolumn{3}{|c|}{bare cuts} & \multicolumn{3}{|c|}{calo cuts} \\ \hline
           & LO [pb]& NLO [pb]& $\Delta$ [\%] & LO [pb]& NLO [pb]& $\Delta$ [\%] \\ \hline
{\sc HORACE} & 5039.11(4)  &  5140.6(1) & 2.014(2) & 4924.17(4)  &  5115.5(2)  & 3.886(4) \\
{\sc SANC}   & 5039.21(1)  &  5139.5(5) & 1.99(1)  & 4925.31(1)  &  5113.5(4) & 3.821(9) \\ 
{\sc WGRAD2} & 5039.16(7)  &  5139.6(6) & 1.99(1)  & 4924.15(5)   &  5114.1(6) & 3.86(1) \\
\hline
\multicolumn{7}{|c|}{\bf LHC, $p p \to W^+ \to \mu^+ \nu_\mu$} \\ \hline
& \multicolumn{3}{|c|}{bare cuts} & \multicolumn{3}{|c|}{calo cuts} \\ \hline
           & LO [pb]& NLO [pb]& $\Delta$ [\%] & LO [pb]& NLO [pb] & $\Delta$ [\%]  \\ \hline
{\sc HORACE} &  5039.11(4) &  5230.5(2) & 3.798(4) & 4925.16(5)  &  4944.5(2) & 0.393(4) \\
{\sc SANC}   &  5039.21(1) & 5229.4(3) & 3.775(7) & 4925.31(1)  &  4942.5(5) & 0.349(9) \\ 
{\sc WGRAD2} &  5039.16(7) &  5229.9(1) & 3.786(3) & 4925.30(7)   &  4943.0(1) & 0.360(3) \\
\hline
\end{tabular}
\caption{Tuned comparison of LO and EW NLO predictions for $\sigma_W$ from {\sc HORACE}, {\sc SANC}, and {\sc WGRAD2}. The statistical error of the Monte Carlo integration is given in parentheses.} 
\label{tab:th_ewk_d}
\end{center}
\end{table}

\begin{figure}
\begin{center}
  \includegraphics[width=7.1cm,
  keepaspectratio=true]{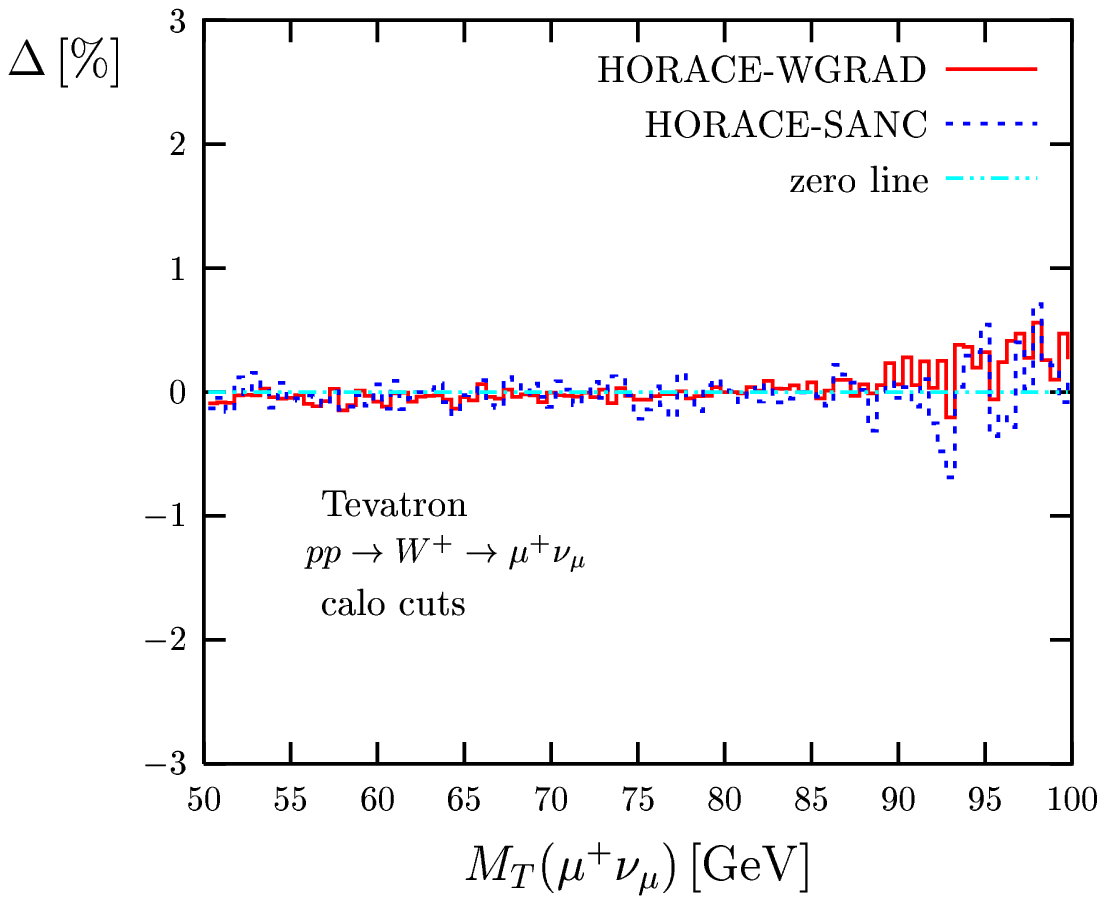}
\hspace*{-.0cm}
  \includegraphics[width=7.1cm,
  keepaspectratio=true]{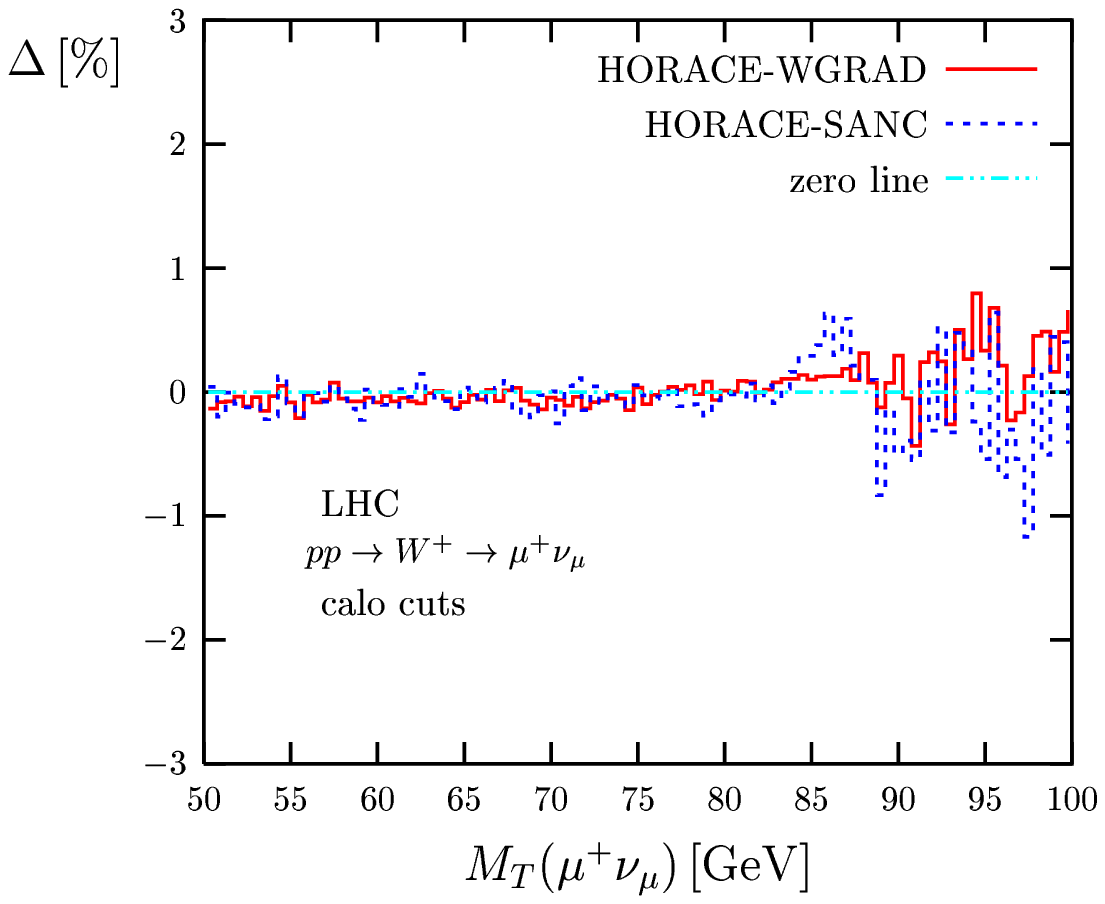}
\hspace*{-.0cm}
  \includegraphics[width=7.1cm,
  keepaspectratio=true]{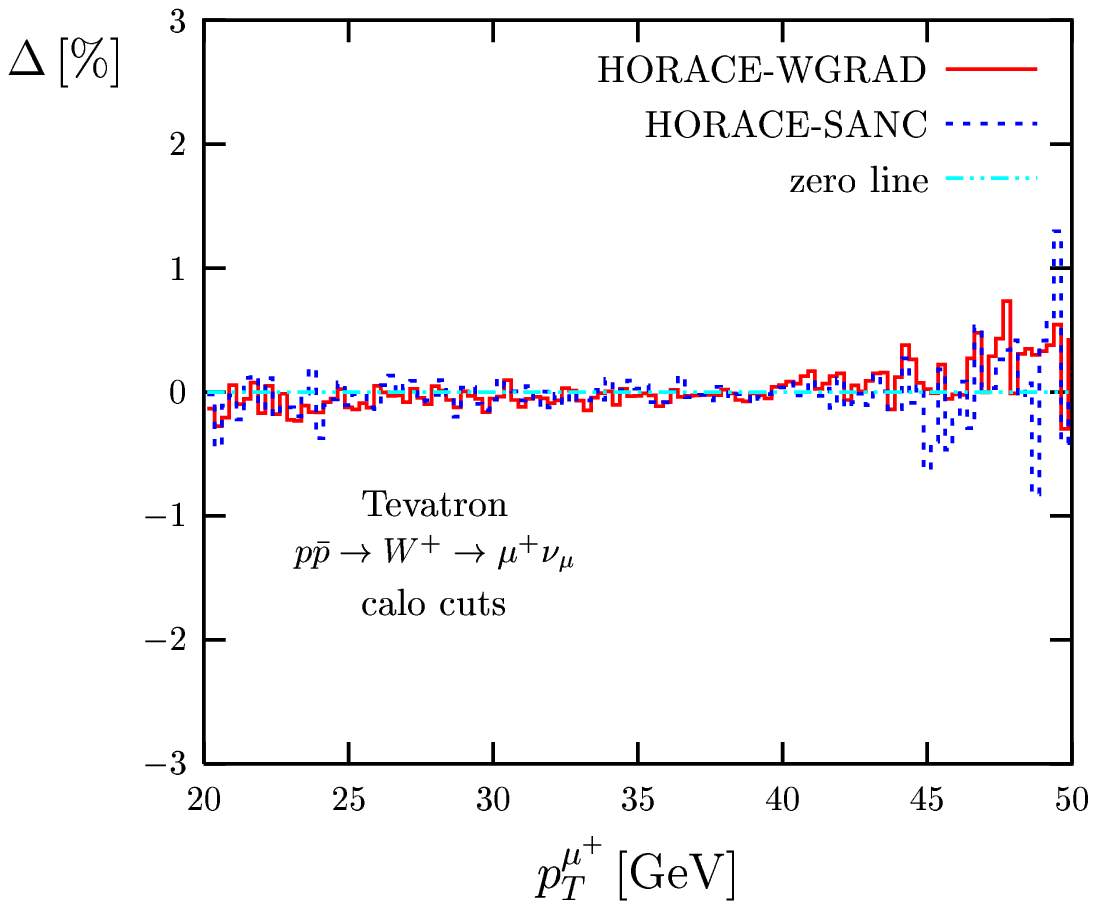}
\hspace*{-.0cm}
  \includegraphics[width=7.1cm,
  keepaspectratio=true]{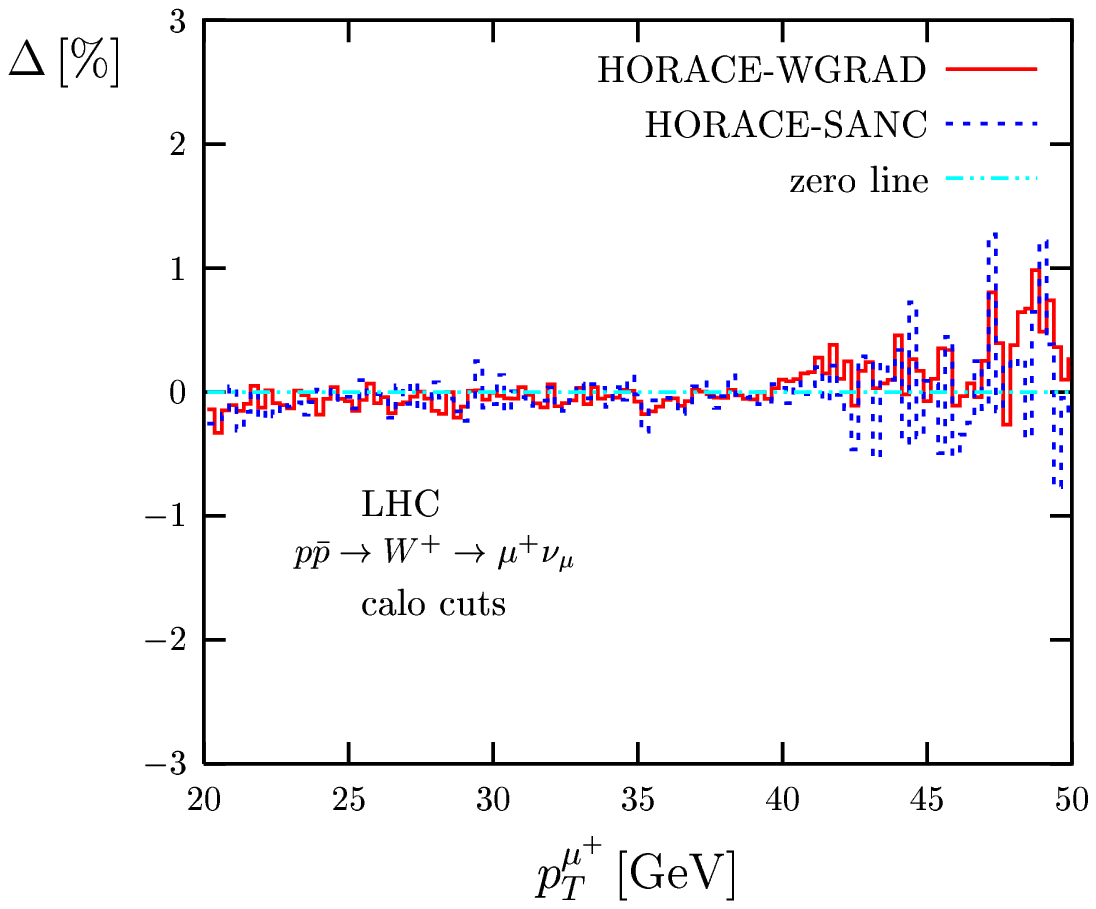}
\end{center}
\caption{Relative difference $\Delta$ between {\sc HORACE} and {\sc SANC} NLO predictions as well as {\sc Horace} and {\sc WGRAD} NLO predictions for the $M_T(\mu^+\nu_\mu)$ and $p_T^{\mu^+}$ distributions
for single $W^+$ production with calo cuts at the Tevatron and the LHC.}\label{fig:th_ewk_mtptdiff}
\end{figure}

\noindent
We find numerical agreement within the statistical uncertainties of the Monte
Carlo integration. In Figs.~\ref{fig:th_ewk_mt12},
\ref{fig:th_ewk_mt22} (upper right figures), we observe a discrepancy
between {\sc SANC} and {\sc WGRAD}/{\sc HORACE} predictions for the
$M_T(e\nu_e)$ distributions at the LHC (with calo cuts).  This
difference is presently under study. We do not expect that it will
persist and, thus, do not consider it in the estimate of the residual
theoretical uncertainties in Section~\ref{subsec:th_ewkerror}.  The
good numerical agreement is also illustrated in detail in
Fig.~\ref{fig:th_ewk_mtptdiff}, where we show the relative differences
$\Delta=$({\sc HORACE}-X)/{\sc HORACE}, X={\sc SANC},{\sc WGRAD}, for
the $M_T(\mu^+\nu_\mu)$ and $p_T^{\mu^+}$ distributions at the LHC and the
Tevatron (with calo cuts).

\newpage
\subsection{Effects of multiple photon radiation}
\label{subsec:th_ewkmulti}
{\bf Contributed by: ~C.~M.~Carloni Calame, G.~Montagna, O.~Nicrosini, A.~Vicini, and D.~Wackeroth}

As discussed in Section~\ref{subsec:th_ewkstatus}, photon radiation off
the charged lepton(s) in the final state (FSR) can considerably affect
the predictions for $W$ and $Z$ boson observables. Therefore, 
the effects of multiple photon radiation (mPR), which is
dominated by final-state radiation, need to be
under good theoretical control when extracting for instance the $W$
mass and width from $W$ observables at the Tevatron and LHC.  The MC
programs {\sc HORACE}, {\sc PHOTOS} and {\sc WINHAC} provide
predictions for $W$ production processes that include multiple FSR as
described in Section~\ref{subsec:th_ewkcodes}.  In the following
discussion of the numerical impact of mPR on the total $W$ production
cross section ($\sigma_W$) and the $M_T(l\nu)$ and $p_T^l$ distributions the
results have been obtained with {\sc HORACE}.

\begin{figure}
\begin{center}
  \includegraphics[width=7.1cm,
  keepaspectratio=true]{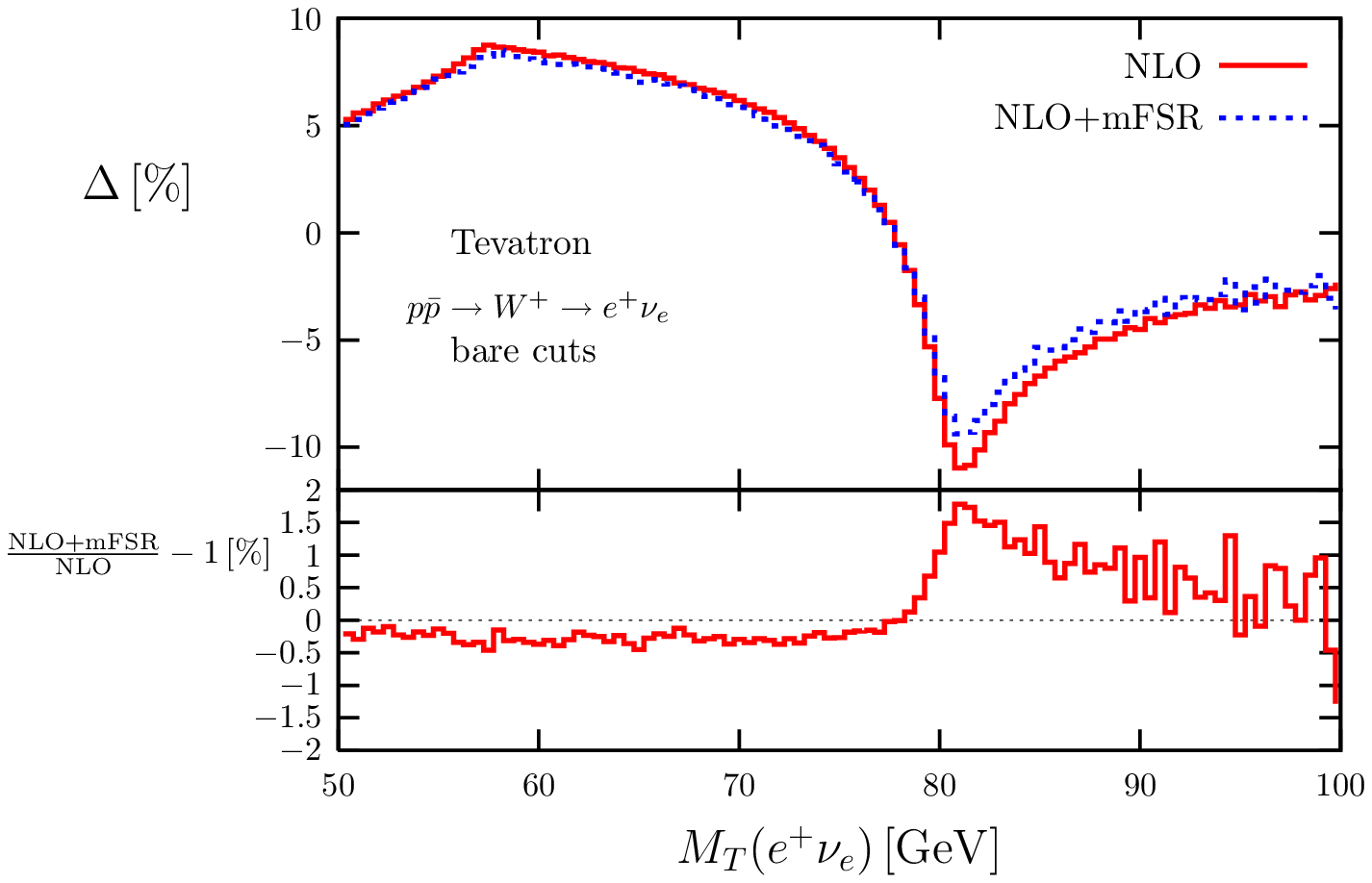}
\hspace*{-.0cm}
  \includegraphics[width=7.1cm,
  keepaspectratio=true]{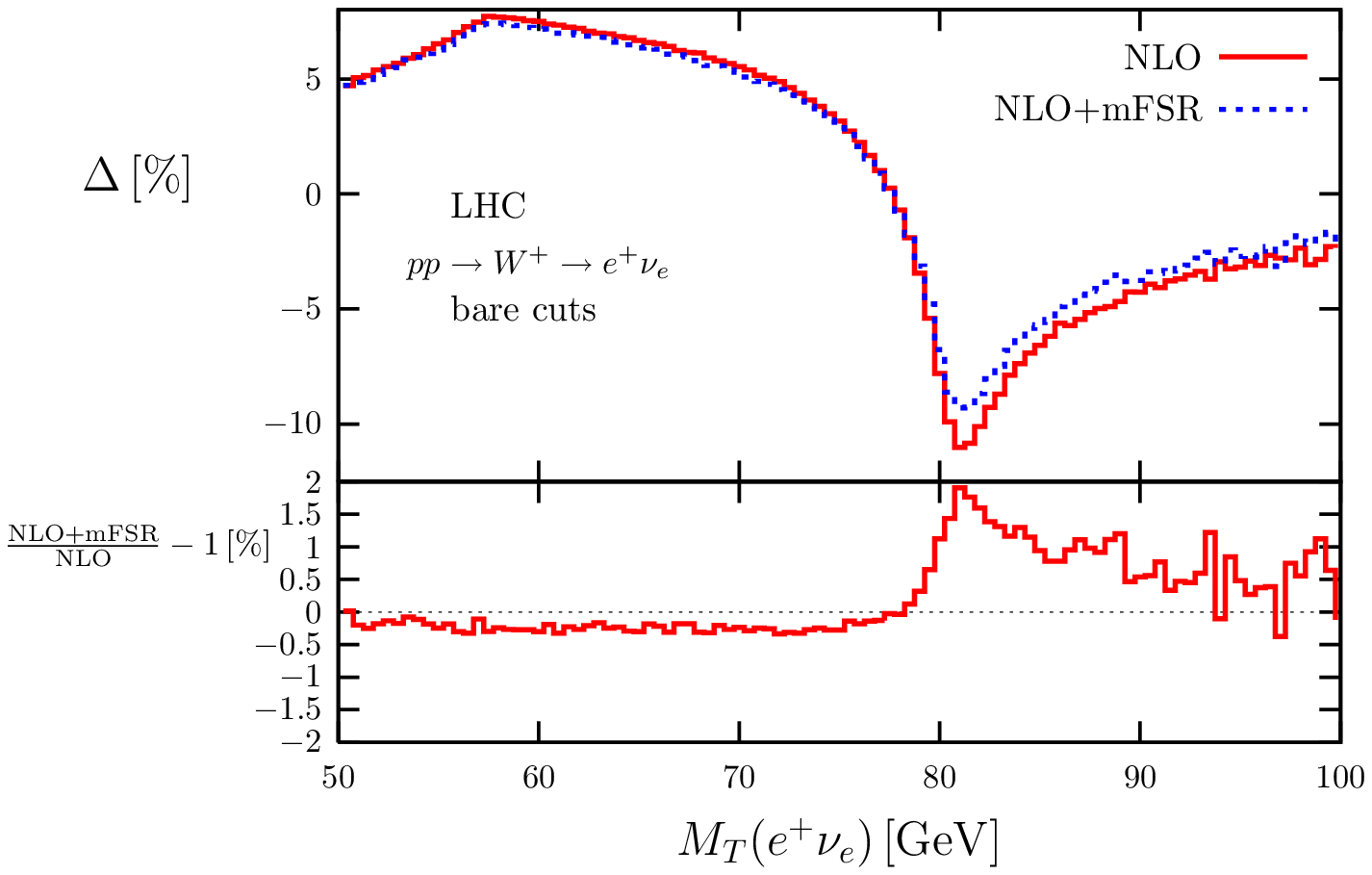}
\hspace*{-.0cm}
  \includegraphics[width=7.1cm,
  keepaspectratio=true]{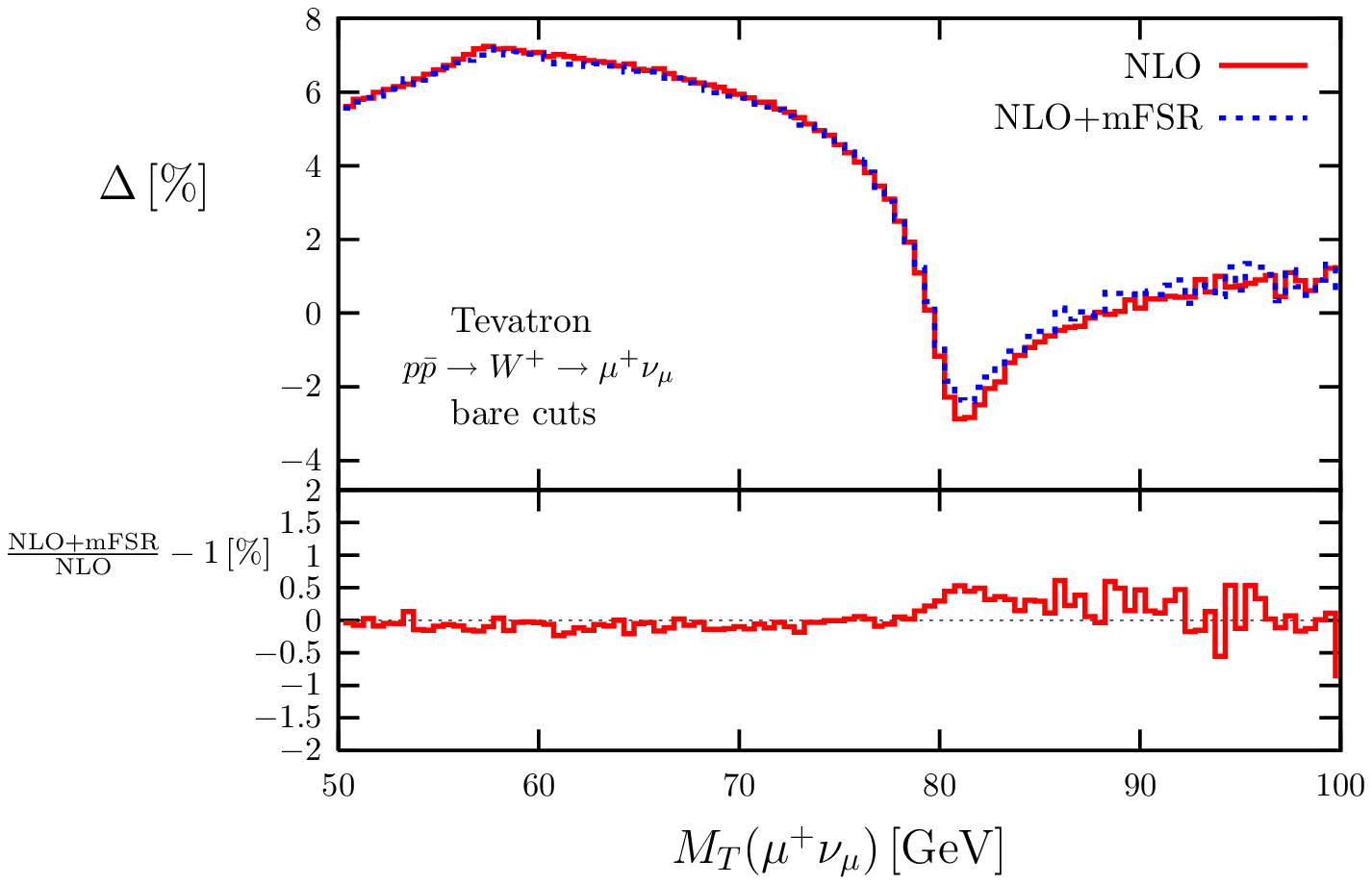}
\hspace*{-.0cm}
  \includegraphics[width=7.1cm,
  keepaspectratio=true]{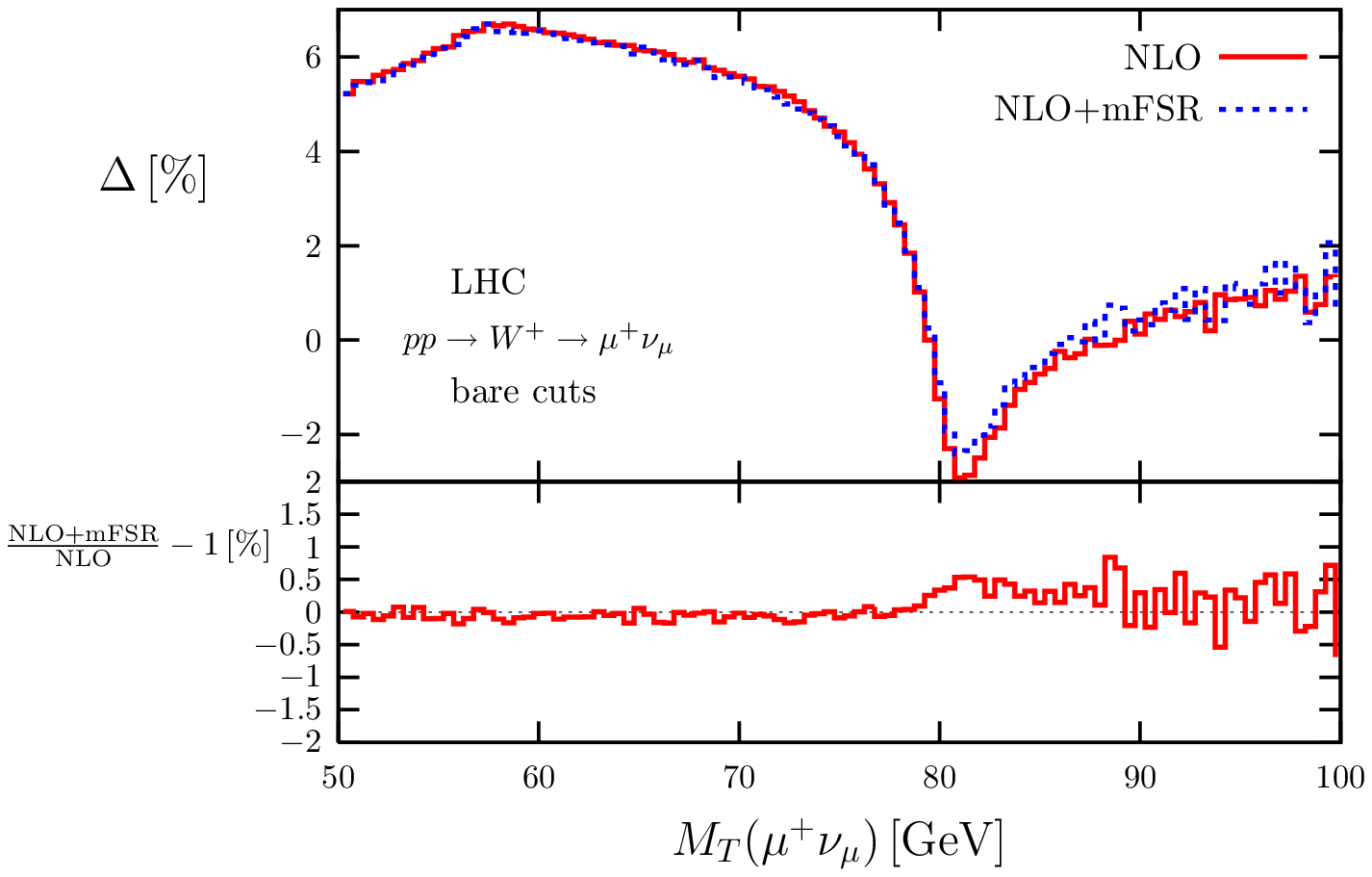}
\end{center}
\caption{The relative correction $\Delta$ due to electroweak ${\cal O}(\alpha)$ corrections ('NLO') and when in addition including multiple final-state photon radiation ('NLO+mPR') to the $M_T(l\nu)$ distribution
  for single $W^+$ production with bare cuts at the Tevatron and the
  LHC.  Also shown in the inset below is the relative difference
  between $M_T(l\nu)$ distributions with and without mPR. The results
  have been obtained with {\sc HORACE}.}\label{fig:th_ewk_mtmfsr11}
\end{figure}

\begin{figure}
\begin{center}
  \includegraphics[width=7.1cm,
  keepaspectratio=true]{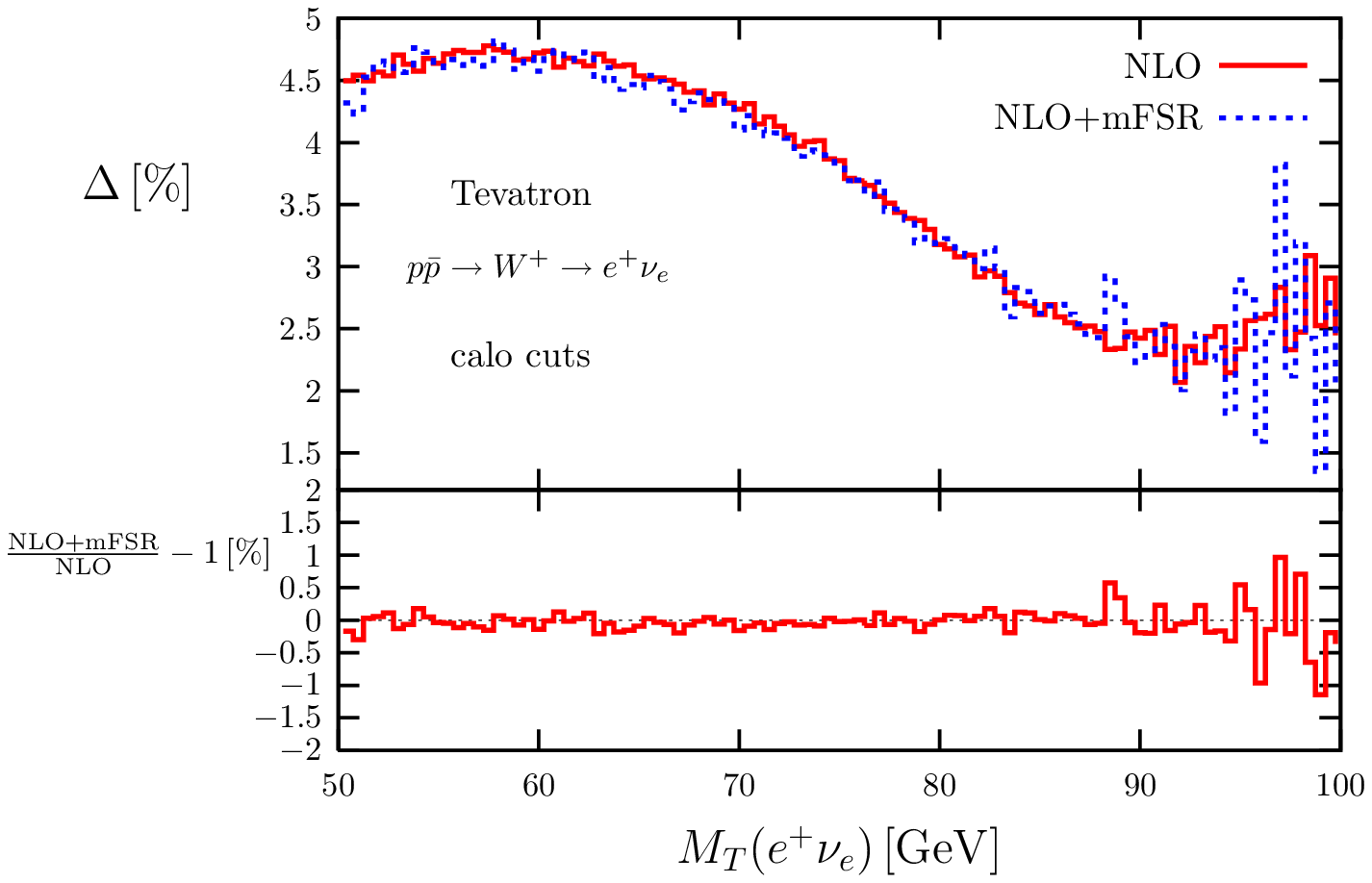}
\hspace*{-.0cm}
  \includegraphics[width=7.1cm,
  keepaspectratio=true]{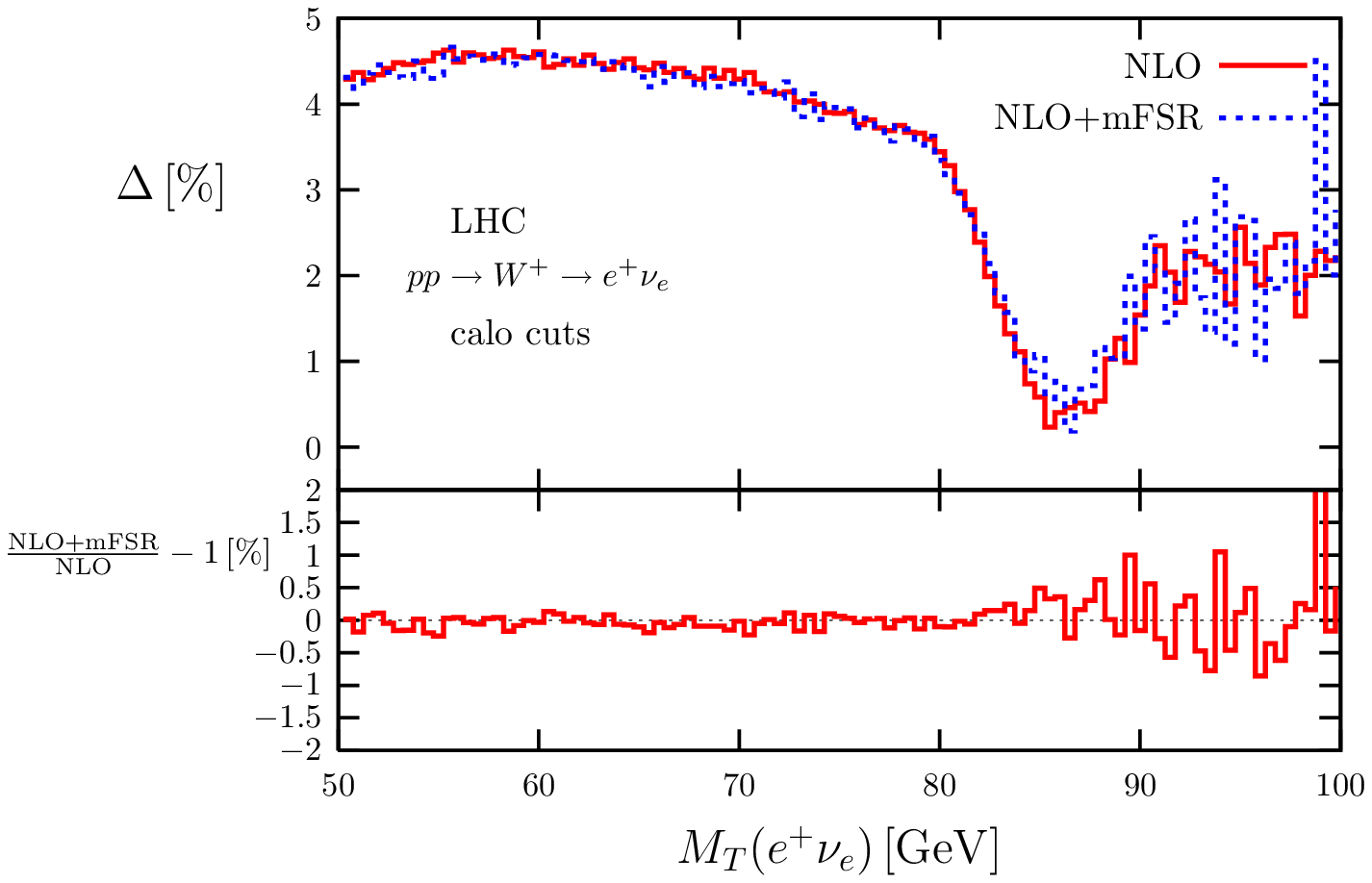}
\hspace*{-.0cm}
  \includegraphics[width=7.1cm,
  keepaspectratio=true]{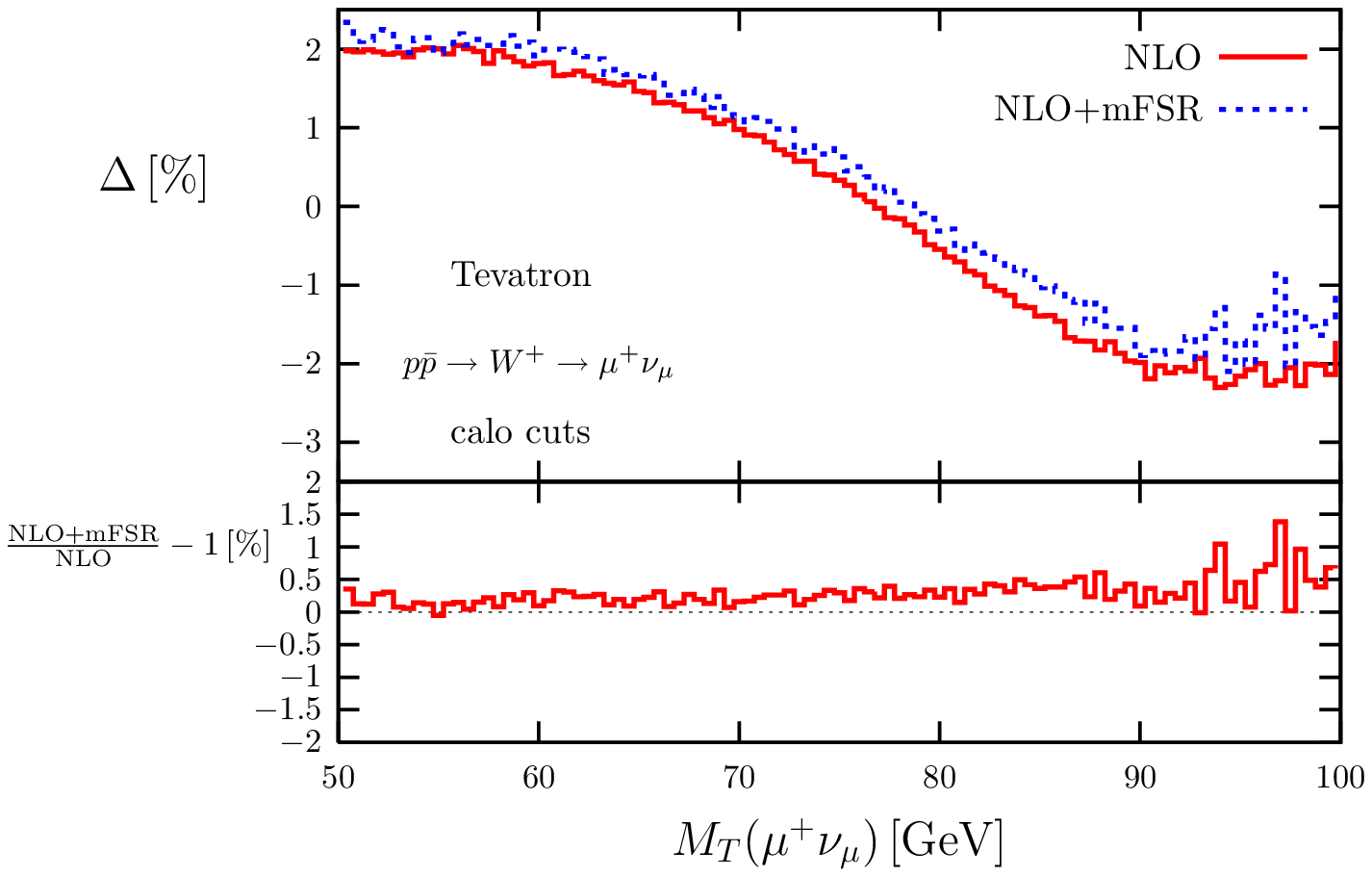}
\hspace*{-.0cm}
  \includegraphics[width=7.1cm,
  keepaspectratio=true]{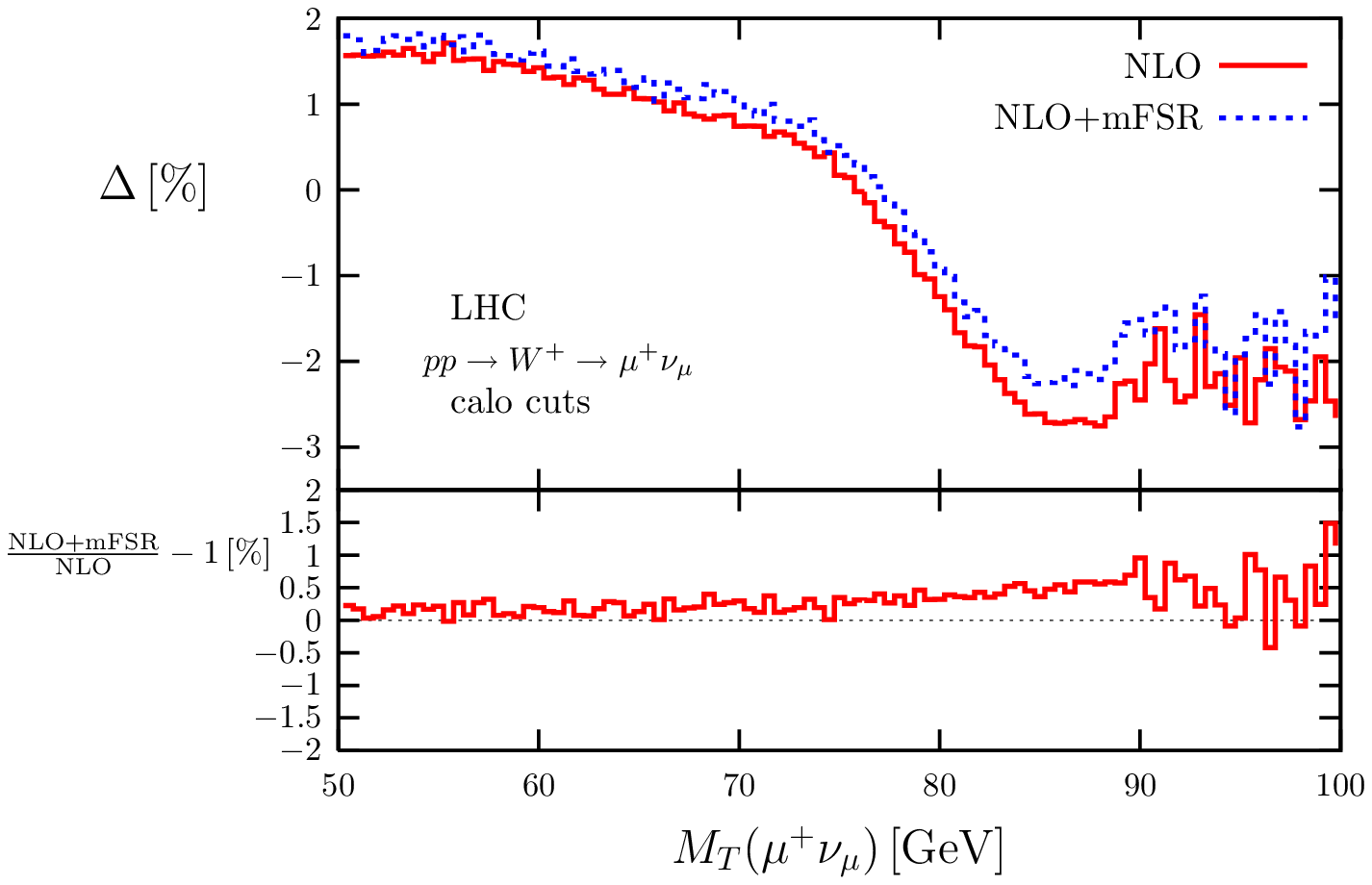}
\end{center}
\caption{The relative correction $\Delta$ due to electroweak ${\cal O}(\alpha)$ corrections ('NLO') and when in addition including multiple final-state photon radiation ('NLO+mPR') to the $M_T(l\nu)$ distribution
for single $W^+$ production with calo cuts at the Tevatron and the LHC. 
Also shown in the inset below is the relative difference between $M_T(l\nu)$ distributions 
with and without mPR. The results have been obtained with {\sc HORACE}.}\label{fig:th_ewk_mtmfsr12}
\end{figure}

\begin{figure}
\begin{center}
  \includegraphics[width=7.1cm,
  keepaspectratio=true]{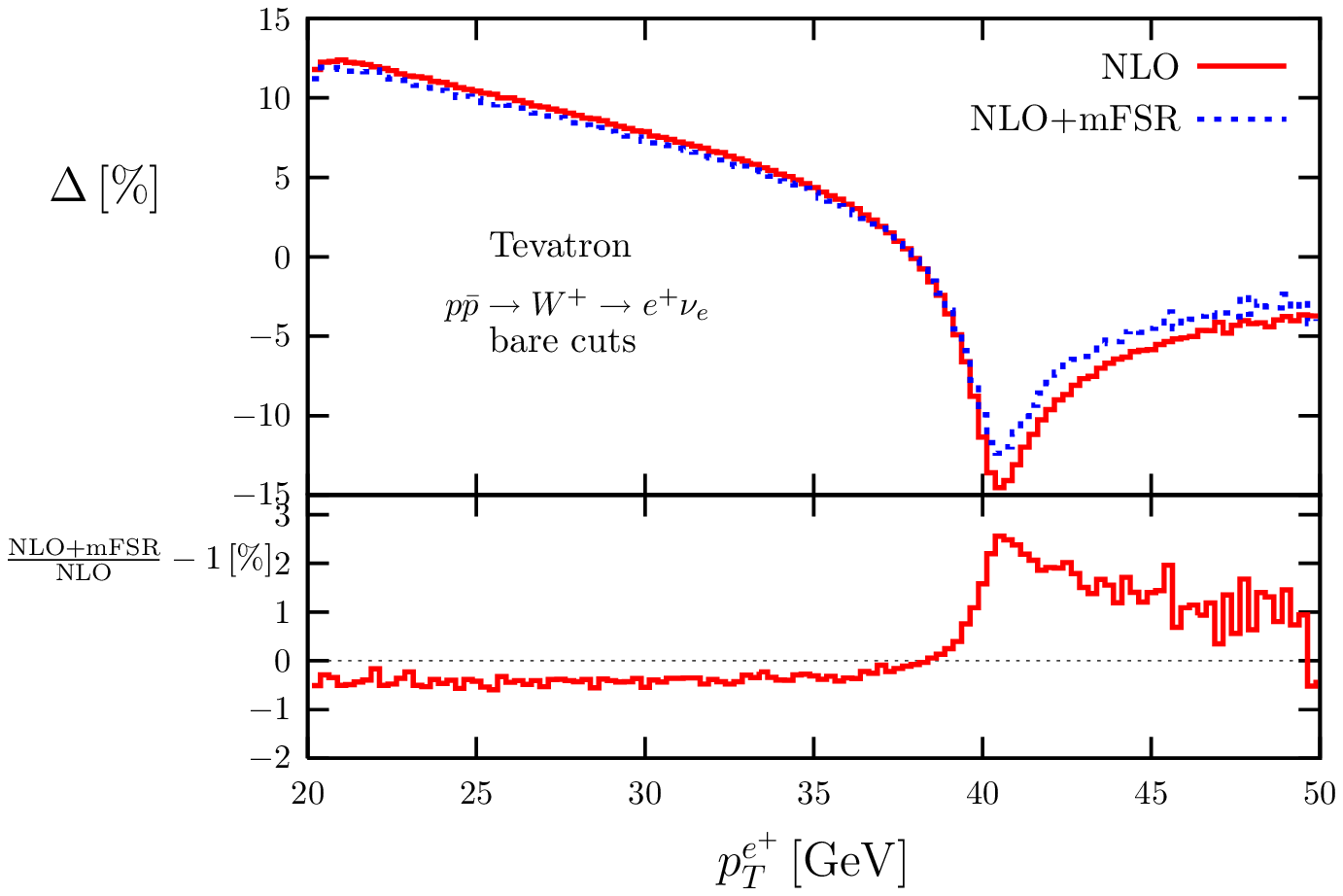}
\hspace*{-.0cm}
  \includegraphics[width=7.1cm,
  keepaspectratio=true]{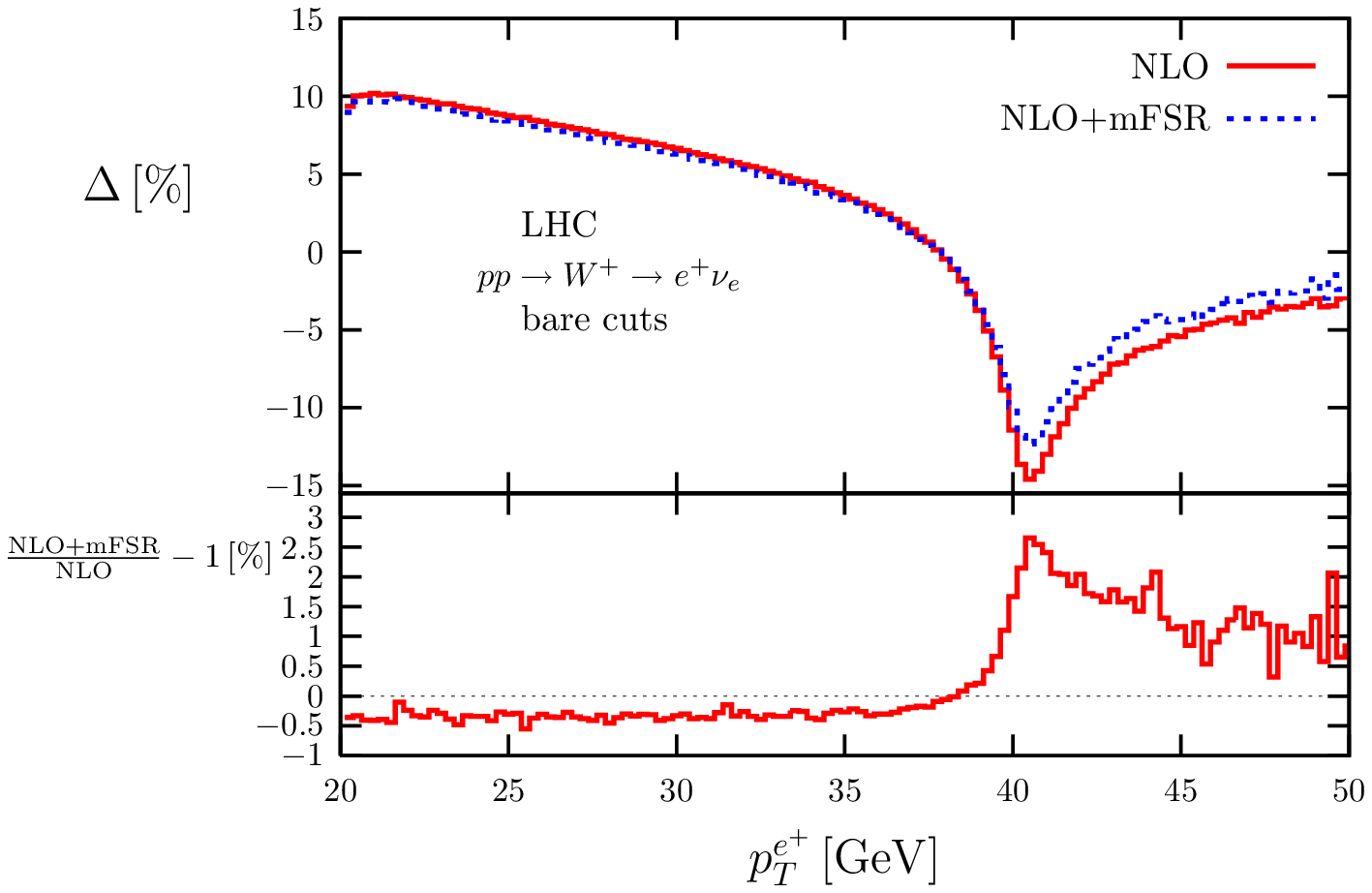}
\hspace*{-.0cm}
  \includegraphics[width=7.1cm,
  keepaspectratio=true]{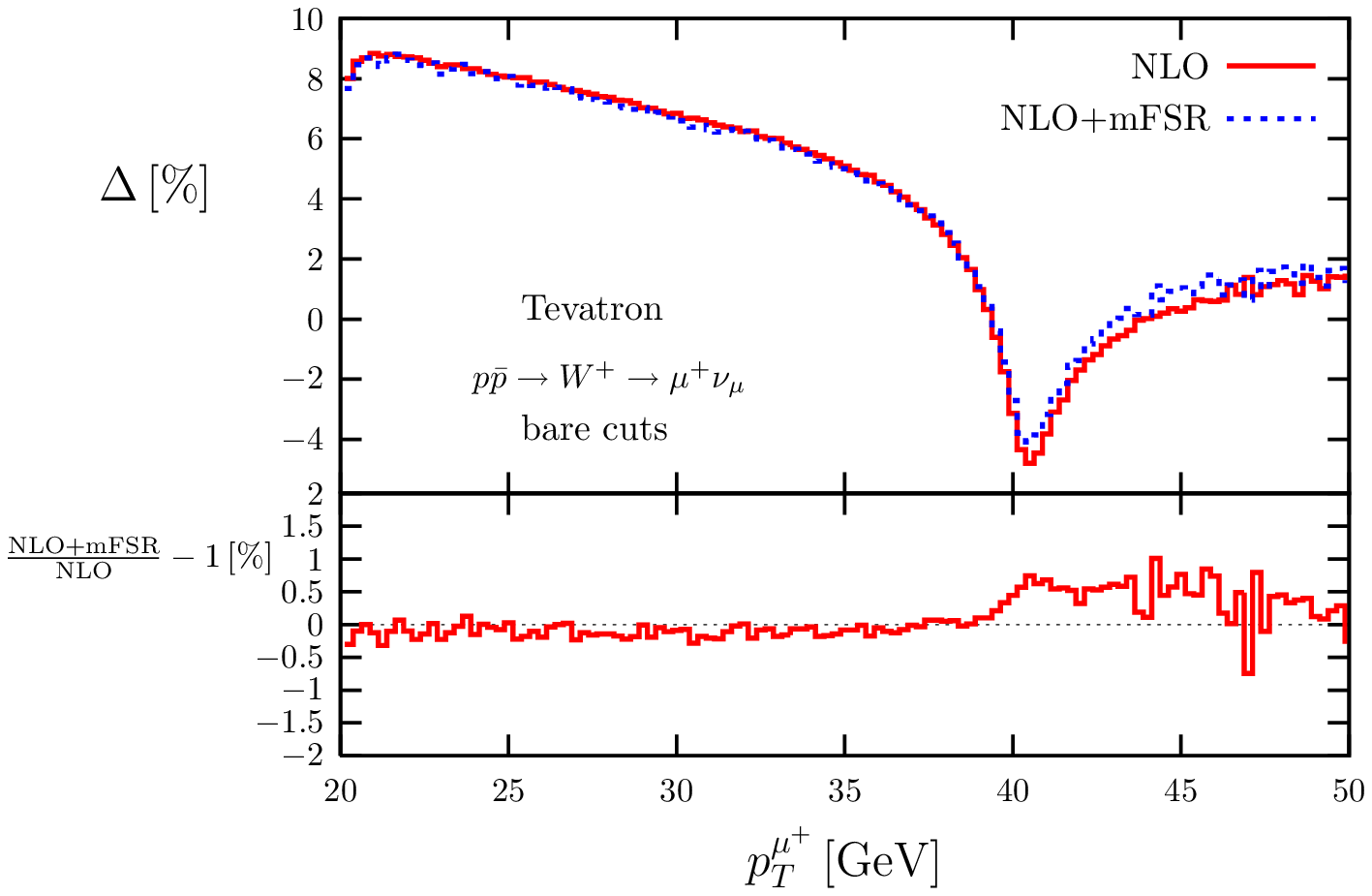}
\hspace*{-.0cm}
  \includegraphics[width=7.1cm,
  keepaspectratio=true]{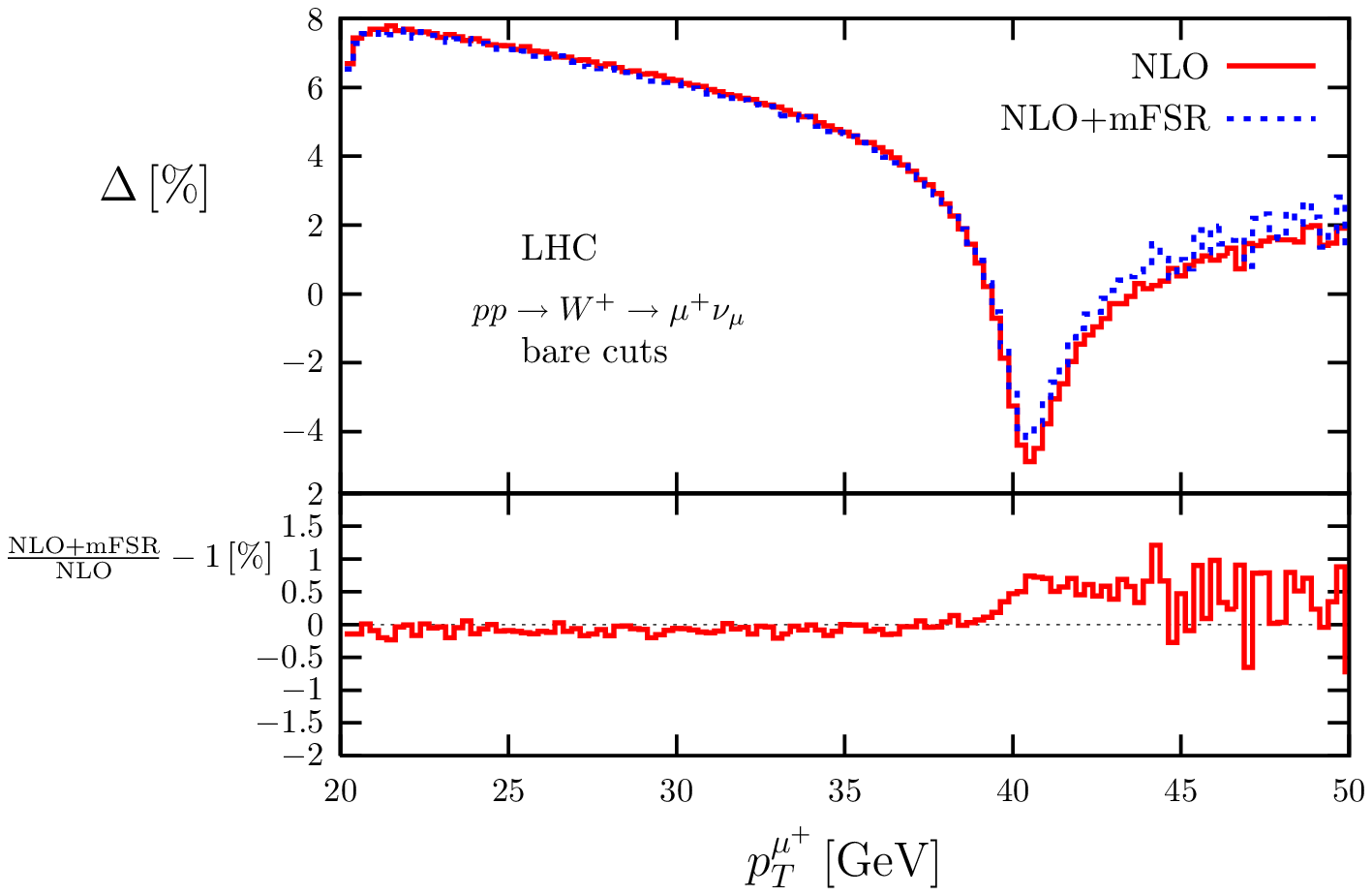}
\end{center}
\caption{The relative correction $\Delta$ due to electroweak ${\cal O}(\alpha)$ corrections ('NLO') and when in addition including multiple final-state photon radiation ('NLO+mPR') to the $p_T^l$ distribution
for single $W^+$ production with bare cuts at the Tevatron and the LHC. 
Also shown in the inset below is the relative difference between $p_T^l$ distributions 
with and without mPR. The results have been obtained with {\sc HORACE}.}\label{fig:th_ewk_ptmfsr1}
\end{figure}

\begin{figure}
\begin{center}
  \includegraphics[width=7.1cm,
  keepaspectratio=true]{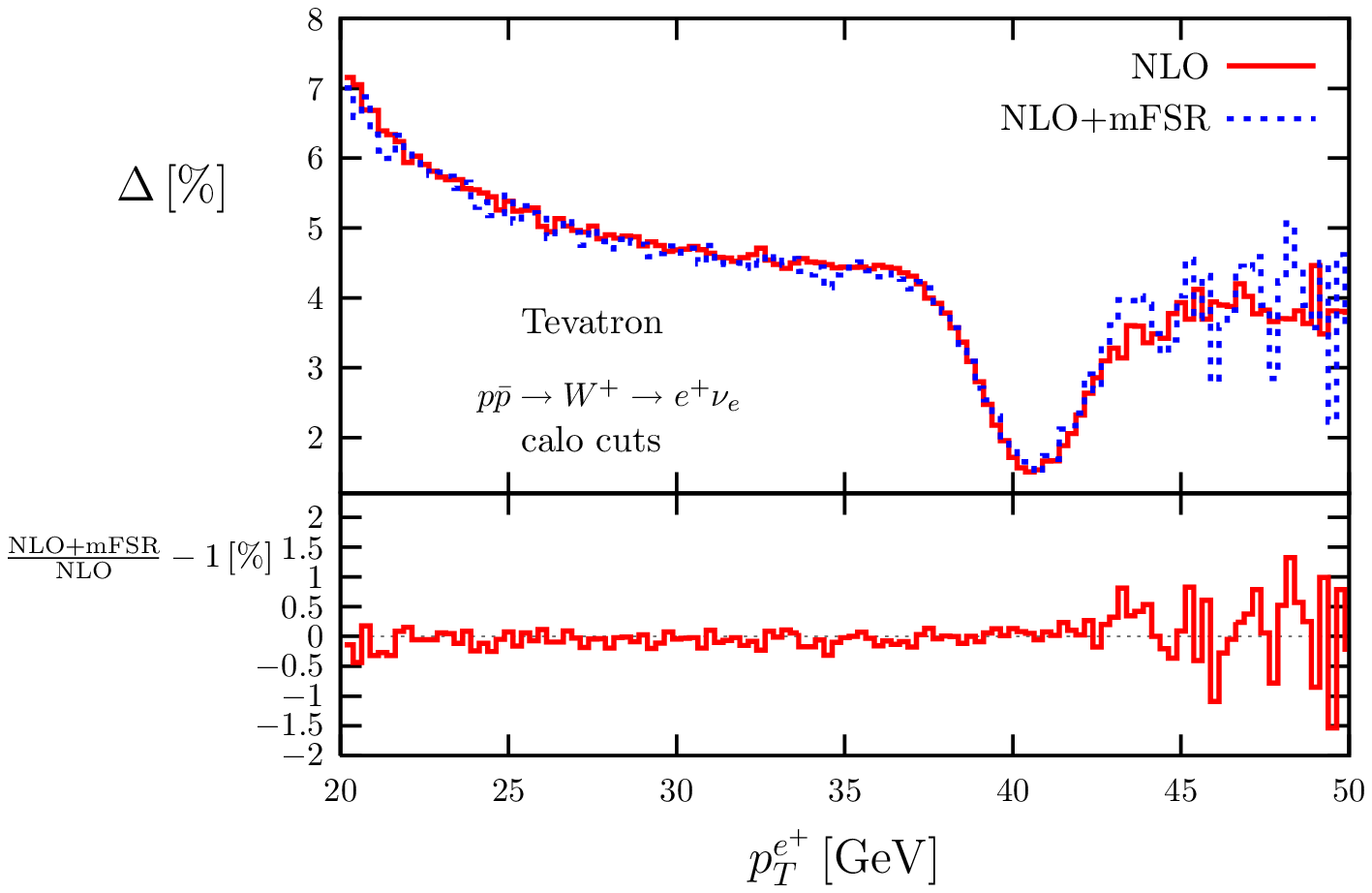}
\hspace*{-.0cm}
  \includegraphics[width=7.1cm,
  keepaspectratio=true]{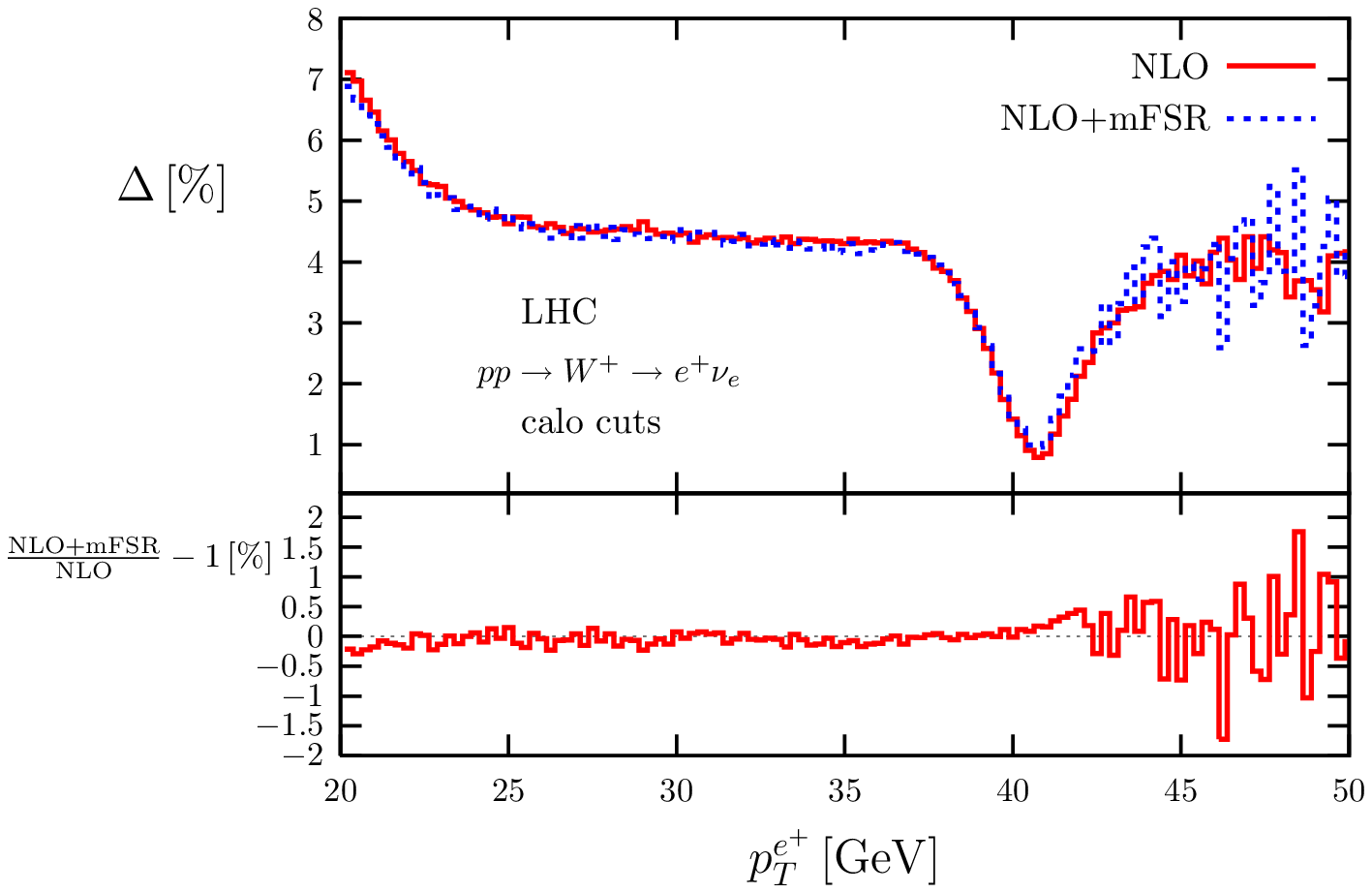}
\hspace*{-.0cm}
  \includegraphics[width=7.1cm,
  keepaspectratio=true]{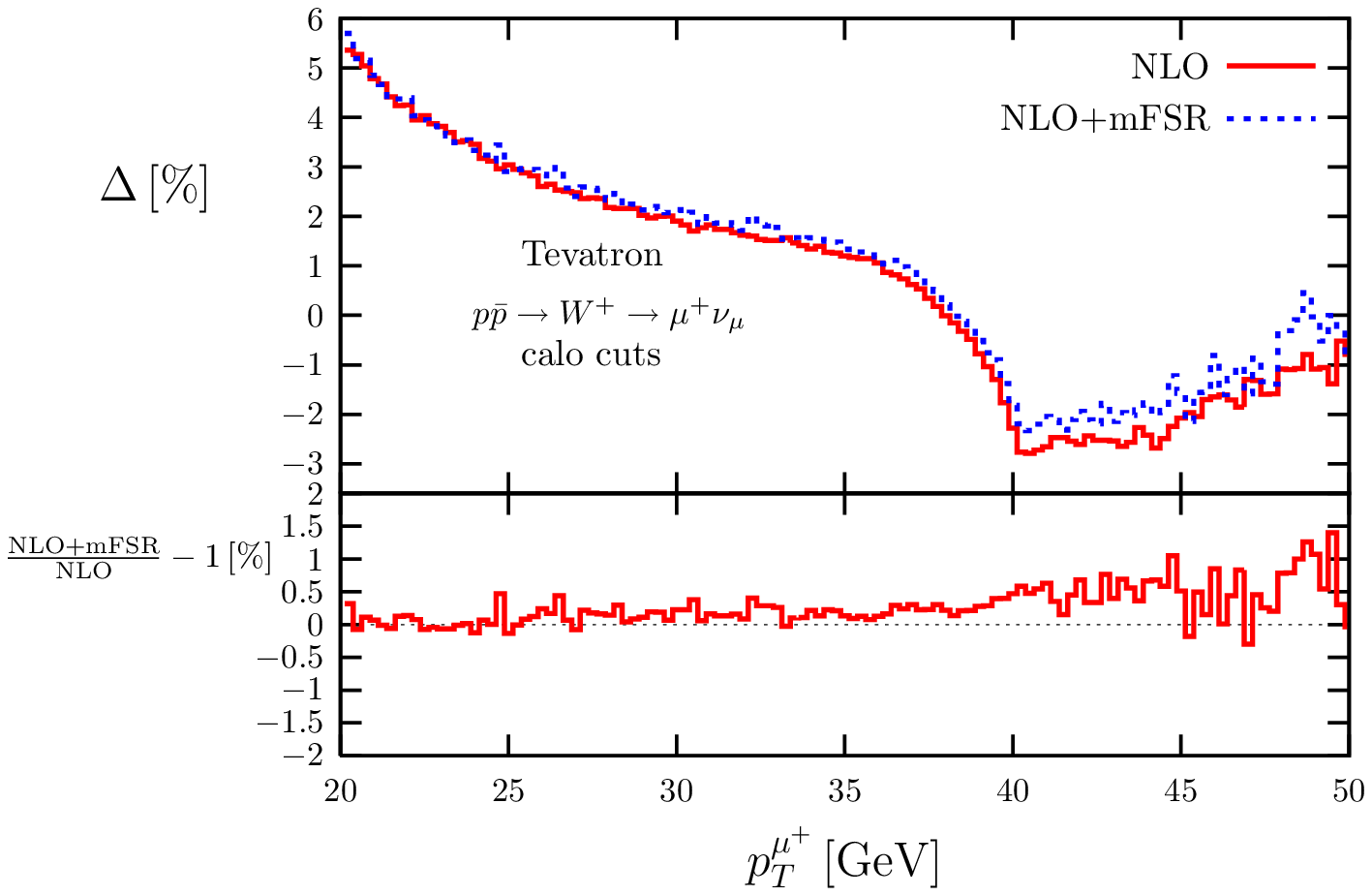}
\hspace*{-.0cm}
  \includegraphics[width=7.1cm,
  keepaspectratio=true]{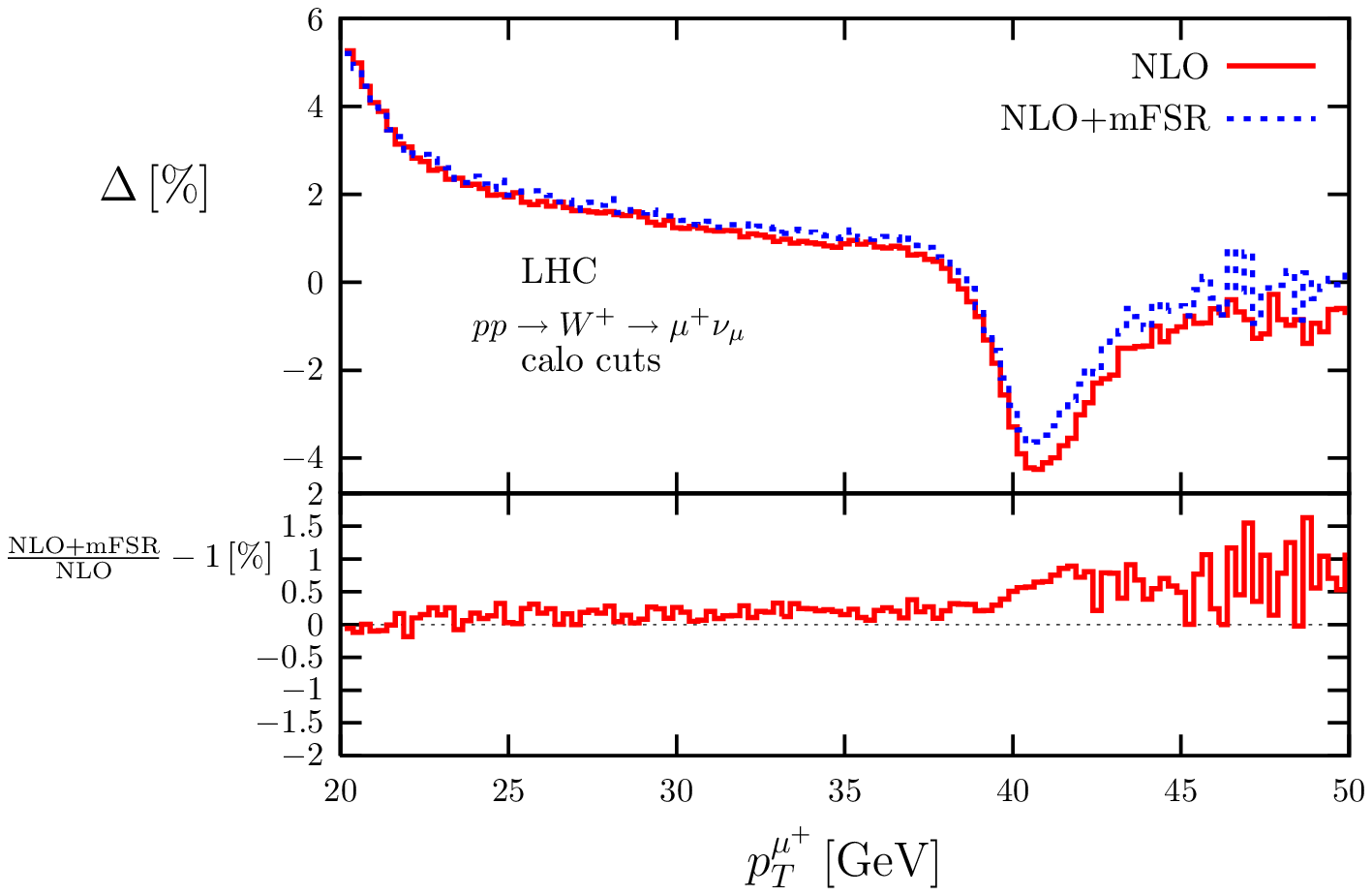}
\end{center}
\caption{The relative correction $\Delta$ due to electroweak ${\cal
    O}(\alpha)$ corrections ('NLO') and when in addition including
  multiple final-state photon radiation ('NLO+mPR') to the $p_T^l$
  distribution for single $W^+$ production with calo cuts at the
  Tevatron and the LHC.  Also shown in the inset below is the relative
  difference between $p_T^l$ distributions with and without mPR. The
  results have been obtained with {\sc
    HORACE}.}\label{fig:th_ewk_ptmfsr2}
\end{figure}

In Table~\ref{tab:th_ewk_f}, NLO EW predictions for $\sigma_W$ are
compared with predictions that include in addition mFSR.  While mPR
does not considerably affect the total cross section, the
$M_T(l\nu_l)$ and $p_T^l$ distributions can be significantly distorted
by mPR, as shown in
Figs~\ref{fig:th_ewk_mtmfsr11},\ref{fig:th_ewk_ptmfsr1}. When only
bare cuts are applied, mPR enhances the NLO EW corrections to the
$M_T(l\nu)(p_T^l)$ distribution in the peak region by up to about
2\%(2.5\%) in the electron case and about 0.5\% in the muon case.
When lepton identification cuts are applied, the effects of mPR are
strongly reduced in the electron case but largely survive in the muon
case, as shown in
Figs~\ref{fig:th_ewk_mtmfsr12},\ref{fig:th_ewk_ptmfsr2}. In
Ref.~\cite{Horace-prd:2004}, the corresponding, additional shift in
$M_W$ due to mPR when extracted from the $M_T(l\nu)$ distribution was
determined to be $10$ MeV in the muon case and negligible in the
electron case, when assuming realistic lepton identification criteria
(similar to the calo cuts used in this report).
  
\begin{table}
\begin{center}
\begin{tabular}{|c|c|c|c|c|} \hline
           & \multicolumn{2}{|c|}{Tevatron} & \multicolumn{2}{|c|}{LHC}  \\ \hline
           & $W^+\to e^+\nu_e$ & $W^+\to \mu^+\nu_\mu$ & $W^+\to e^+\nu_e$ & $W^+\to \mu^+\nu_\mu$ \\ \hline
           & \multicolumn{4}{|c|}{bare cuts}   \\ \hline
NLO [pb]       &  791.14(2) & 804.18(2) &5140.6(1)   &  5230.5(2) \\
NLO+ mPR [pb] & 791.50(5) &  804.39(4) & 5143.4(3)  &  5232.2(3) \\ \hline
           & \multicolumn{4}{|c|}{calo cuts}   \\ \hline
NLO [pb]         &  762.21(3) &  738.16(3)  & 5115.5(2) & 4944.5(2) \\
NLO+ mPR [pb]   & 762.01(6) &  739.86(5) & 5114.5(4)  & 4956.5(3)  \\ \hline
\end{tabular}
\caption{Comparison of EW NLO predictions without and with multiple final-state photon radiation (mPR) for $\sigma_W$ to $pp,p\bar p \to W^+\to e^+\nu_e, \mu^+\nu_\mu$ with bare and calo cuts at the Tevatron and LHC using {\sc HORACE}.} 
\label{tab:th_ewk_f}
\end{center}
\end{table}

\newpage
\subsection{High-precision transverse momentum distributions in $W$ boson production}
\label{subsec:qtresum} 
\textbf{Contributed by: S.~Berge, P. M.~Nadolsky,
and F. I.~Olness}

In this section we discuss theoretical predictions for distributions
in the transverse momentum $q_{T}$ of the $W$ boson, the transverse
momenta $p_{Te}$ and $p_{T\nu}$ of the decay charged lepton and
neutrino, and the transverse mass of the decay lepton pair
$m_{Te\nu}=\sqrt{2(p_{Te}p_{T\nu}-\vec{p}_{Te}\cdot\vec{p}_{T\nu})}$.
Since these distributions are used to extract the $W$ boson mass
$M_{W}$, the associated theoretical uncertainties must be kept under
control.

If the boson's transverse momentum $q_{T}$ is much smaller than the
boson's virtuality $Q$, the calculation of the transverse momentum
distribution must include an all-order sum of large logarithms
$\ln^{n}(q_{T}/Q)$.  The formalism for summation of $q_{T}$ logarithms
in Drell-Yan-like processes is well established at moderate scattering
energies ($Q\sim\sqrt{S}$), when no other large logarithms are present
\cite{Collins:1985kg}.  When formulated in space of the impact
parameter $\vec{b}$ (conjugate to $\vec{q}_{T}$ via a two-dimensional
Fourier transform), it is proven to all orders by a factorization
theorem \cite{Collins:2004nx,Ji:2004wu,Ji:2004xq}.

Resummation in the $b$-space formalism
\cite{Balazs:1997xd,Landry:2002ix,Ladinsky:1993zn} (currently
implemented at NNLL/NLO accuracy) is employed in recent measurements
of $W$ and $Z$ observables at the Tevatron. As precision of the
experimental analysis continues to improve, new effects must be
included in the resummation framework to keep up with modern demands.
The shape of $q_{T}$ spectrum may be appreciably altered by only
partly known NNLO corrections, as well as by variations in parameters
of the PDF's and nonperturbative resummed function. At the LHC, $W$
and $Z$ bosons will be produced by the scattering of partons with
small momentum fractions ($x\sim0.005$) and potentially affected by
radiative contributions associated with $\ln(1/x)$ logarithms
\cite{Berge:2004nt}. A large fraction of the bosons will be produced
in heavy-quark scattering. Heavy-quark masses $m_{Q}$ act as
additional hard momentum scales and suppress multiple parton radiation
at $q_{T}\lesssim m_{Q}$ in charm and bottom scattering, leading to
harder $q_{T}$ distributions than in the dominant process of
quark-antiquark scattering \cite{Berge:2005rv}.  In this report, we
review recent progress in understanding of these factors and quantify
their impact on the measured value of the $W$ boson mass. Further
details pertinent to our discussion can be found in
Refs.~\cite{Landry:2002ix,Berge:2004nt,Berge:2005rv,Konychev:2005iy}.

\begin{figure}
\begin{center}\includegraphics[width=0.65\columnwidth]{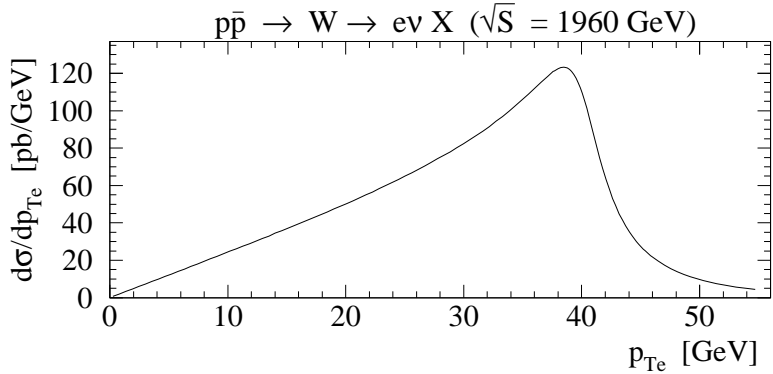}\vspace*{-10pt}\par\end{center}
\caption{\label{fig:th_ewk_W-Tev-pTe} Transverse momentum distribution $d\sigma/dp_{Te}$
of the electrons from the decay of $W$ bosons in the Tevatron Run-2
in the resummation calculation.}
\end{figure}

We concentrate on the $p_{Te}$ distribution of the final-state charged
lepton, since it is more sensitive to the $q_{T}$ of $W$ boson than
the $m_{Te\nu}$ distribution and less affected by experimental
uncertainties than $p_{T\nu}$ distribution. To visualize percent-level
changes in $d\sigma/dp_{Te}$ caused by various effects, we show
several plots of the fractional difference
$\left(d\sigma^{mod}/dp_{Te}\right)/\left(d\sigma^{ref}/dp_{Te}\right)-1$
of the cross sections obtained under {}``reference'' (ref) and
{}``modified'' (\emph{mod}) theoretical assumptions. Our attention
primarily focuses on the cross section near the kinematical (Jacobian)
peak at $p_{Te}\approx M_{W}/2=40$ GeV
(cf.~Fig.~\ref{fig:th_ewk_W-Tev-pTe}), where $d\sigma/dp_{Te}$ is most
sensitive to $M_{W}.$ We compare modifications in $d\sigma/dp_{Te}$
caused by changes in theoretical assumptions with modifications caused
by explicit variations of $M_{W}$.

\subsubsection{Theory overview}

In the $b$-space resummation framework (also called
Collins-Soper-Sterman, or CSS formalism \cite{Collins:1985kg}), the
differential cross section for production of a boson $V$ at small to
moderate $q_{T}$ takes the form\begin{eqnarray}
  \frac{d\sigma}{dQ^{2}dydq_{T}^{2}} & = &
  \int\!\frac{d^{2}b}{(2\pi)^{2}}\,
  e^{-i\vec{q}_{T}\cdot\vec{b}}\,\,\widetilde{W}(b,Q,x_{A},x_{B})+Y(q_{T},Q,x_{A},x_{B}),\end{eqnarray}
where $y$ is the rapidity of the vector boson, and $x_{A,B}\equiv
Qe^{\pm y}/\sqrt{S}$ are the Born-level partonic momentum fractions.
The all-order sum of $\alpha_{s}^{n}\ln^{m}(q_{T}^{2}/Q^{2})$ arising
at $q_{T}\rightarrow0$ is contained in a Fourier-Bessel transform
integral of a $b$-space form factor $\widetilde{W}(b,Q,x_{A},x_{B}$).
It is this integral that has the most impact on the $W$ mass
measurement. The regular NLO contribution $Y(q_{T},Q,x_{A},x_{B})$ is
substantial only at large $q_{T}$ and won't receive much of our
attention.

The form factor $\widetilde{W}_{ab}(b,Q,x_{A},x_{B}$) factorizes at
all $b$ as \begin{eqnarray}
  \widetilde{W}(b,Q,x_{A},x_{B})=\frac{1}{S}\sum_{j,k=u,\bar{u},d,\bar{d},..}{\sigma_{jk}^{(0)}\,
    e^{-{\mathcal{S}}(b,Q)}}\mathcal{P}_{j}(x_{A},b)\mathcal{P}_{k}(x_{B},b).\label{th_ewk_bno_W}\end{eqnarray}
The Sudakov function ${\mathcal{S}}(b,Q)$ and $b$-dependent parton
distributions $\mathcal{P}_{j}(x,b)$ for finding a quark (antiquark)
of flavor $j$ in the proton are universal in Drell-Yan-like processes
and semi-inclusive deep-inelastic scattering (SIDIS)
\cite{Collins:2004nx}.  The coefficient $\sigma_{jk}^{(0)}$ includes
process-specific constant factors from the Born cross section
$q_{j}\bar{q}_{k}\rightarrow V$.  All terms in
Eq.~(\ref{th_ewk_bno_W}) can be computed in perturbative QCD when the
momentum scale $1/b$ is much larger than 1 GeV, i.e., in the dominant
region of $b$ at both colliders.

The contribution of the nonperturbative region at $b\gtrsim1\mbox{
  GeV}^{-1}$ is also tangible and must be properly modeled to describe
the region $q_{T}\lesssim20$ GeV. It is constrained through the global
analysis of $p_{T}$-dependent Drell-Yan and $Z$ boson data
\cite{Landry:2002ix,Konychev:2005iy}.  For this purpose, we separate
the perturbative (small-$b$) and nonperturbative (large-$b$) terms in
$\widetilde{W}(b,Q,x_{A},x_{B})$ by rewriting Eq.~(\ref{th_ewk_bno_W})
as \begin{equation}
  \widetilde{W}(b,Q,x_{A},x_{B})=\widetilde{W}_{LP}(b,Q,x_{A},x_{B})\,
  e^{-\mathcal{\mathcal{F}}_{NP}(b,Q)},\label{th_ewk_bno_W2}\end{equation}
where the leading-power (logarithmic in $b$) term $\widetilde{W}_{LP}$
is given by a model-dependent continuation of the perturbative
contribution to the region $b\gtrsim1\mbox{ GeV}^{-1}$, and the
nonperturbative exponent $e^{-\mathcal{F}_{NP}(b,Q)}$ absorbs
power-suppressed terms proportional to even powers of $b$
\cite{Korchemsky:1994is}. In global fits, the preferred
$\mathcal{\mathcal{F}}_{NP}(b,Q)$ has approximately quadratic
dependence on $b$ (i.e., $\mathcal{\mathcal{F}}_{NP}(b,Q)\propto
b^{2}$).  It may be therefore interpreted as a source of the Gaussian
smearing of the $b$-space form factor (and transverse momentum
distributions) introduced by nonperturbative dynamics.

\subsubsection{Universality of nonperturbative resummed contributions}
\label{subsec:ewkth_nonperturbative}
\begin{figure}
\begin{centering}\includegraphics[width=0.5\columnwidth]{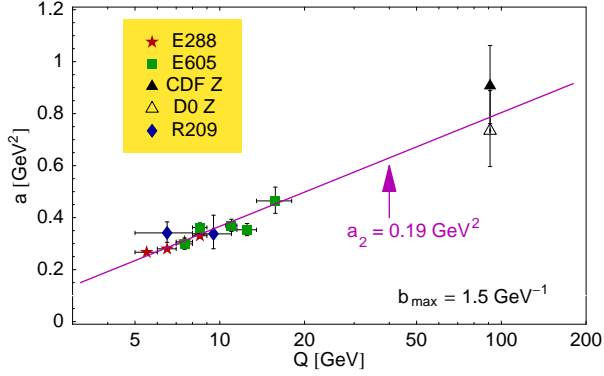}\vspace*{-10pt}\par\end{centering}
\caption{The {}``Gaussian smearing'' parameter $a(Q)$ preferred by the
low-mass Drell-Yan and Tevatron Run-1 $Z$ boson $p_{T}$ data in
the model of Ref.~\cite{Konychev:2005iy}. The derivative of $a(Q)$
with respect to $\ln Q$ observed in the fit (the slope $a_{2}$ of
the solid line) agrees with its independent estimate made in the renormalon
analysis \cite{Tafat:2001in}. \label{fig:th_ewk_aQ}}
\end{figure}

The best-fit form of ${\cal F}_{NP}(b,Q)$ is correlated with the
assumed large-$b$ behavior of $\widetilde{W}_{LP},$ which differs
between the available models
\cite{Collins:1985kg,Tafat:2001in,Qiu:2000hf,Kulesza:2002rh,Guffanti:2000ep}.
We have recently proposed \cite{Konychev:2005iy} a simple revision of
the {}``$b_{*}$ ansatz'' for $\widetilde{W}_{LP}(b,Q,x_{A},x_{B})$
\cite{Collins:1985kg,Landry:2002ix,Ladinsky:1993zn}, which leads to
several improvements over previous studies. The new model extends the
range where $\widetilde{W}_{LP}(b,Q,x_{A},x_{B})$ is approximated by a
known finite-order perturbative prediction to larger values of $b$
and, by doing so, improves agreement with all analyzed $p_{T}$ data
from low-mass Drell-Yan pair and Tevatron $Z$ boson production.  The
best-fit parametrization of ${\cal F}_{NP}(b,Q)=b^{2}a(Q)$ is found to
be in a good agreement with a semi-quantitative estimate in renormalon
analysis and lattice QCD \cite{Tafat:2001in} and has reduced
dependence on the collision energy $\sqrt{S}$. The {}``Gaussian
smearing'' parameter \[
a(Q)=a_{1}+a_{2}\ln\frac{Q}{3.2\,{\textrm{GeV}}}+a_{3}\ln\left(100\,
  x_{A}x_{B}\right)\] grows practically linearly with $\ln Q$ (i.e.,
$a_{1,2}\gg a_{3}$).  The value of the dominant coefficient $a_{2}$
from the fit agrees well with the renormalon analysis estimate, where
it arises from soft gluon subgraphs and does not depend on $\sqrt{S}$
or flavor of initial-state quarks and hadrons. As a result of the
above improvements, the revised $b_{*}$ model leads to more confident
predictions for the nonperturbative contribution at collider energies
by exposing its soft-gluon origin and universality.

\begin{figure}
\includegraphics[width=0.49\columnwidth]{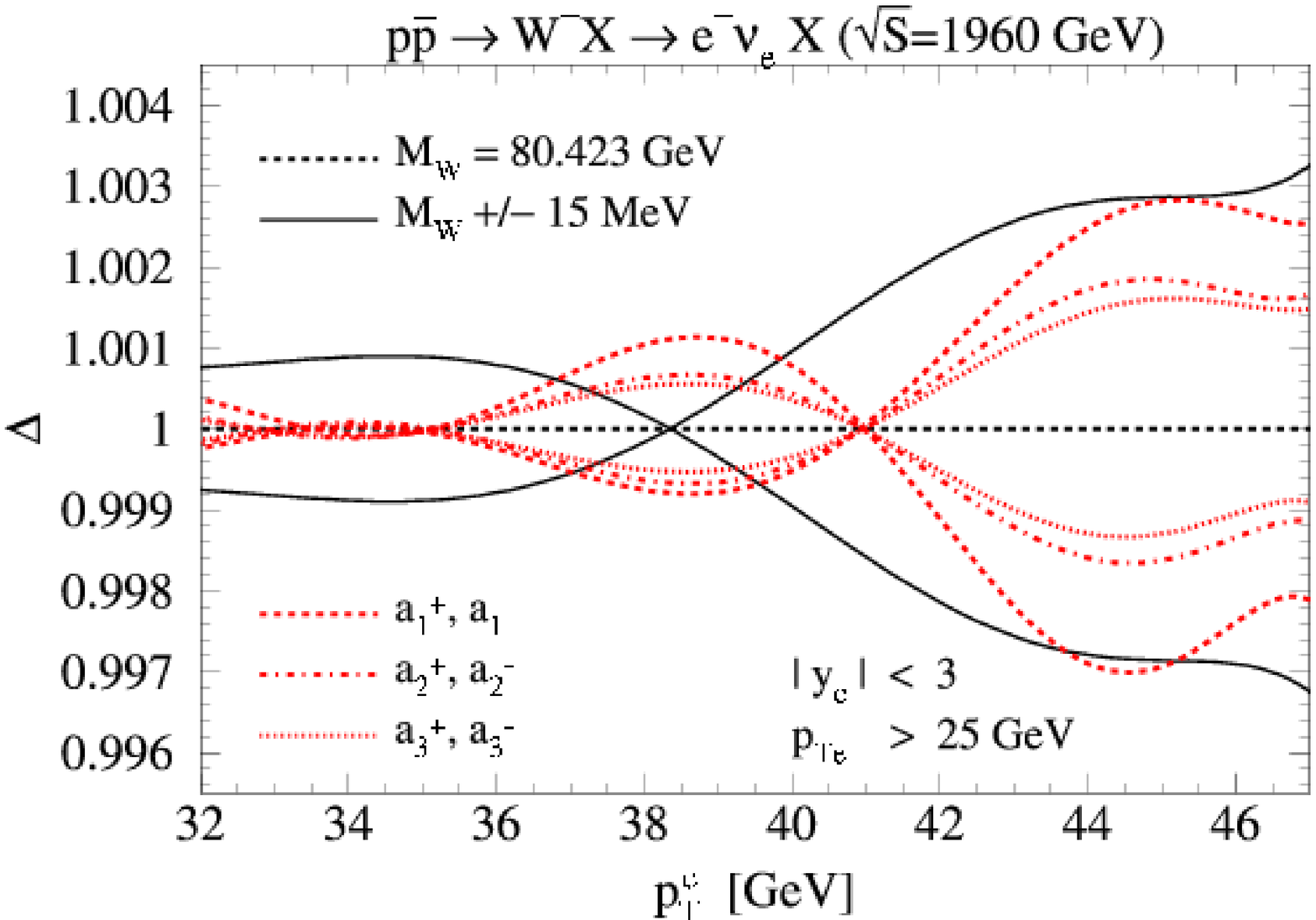}$\,\,\,\,$\includegraphics[width=0.5\columnwidth]{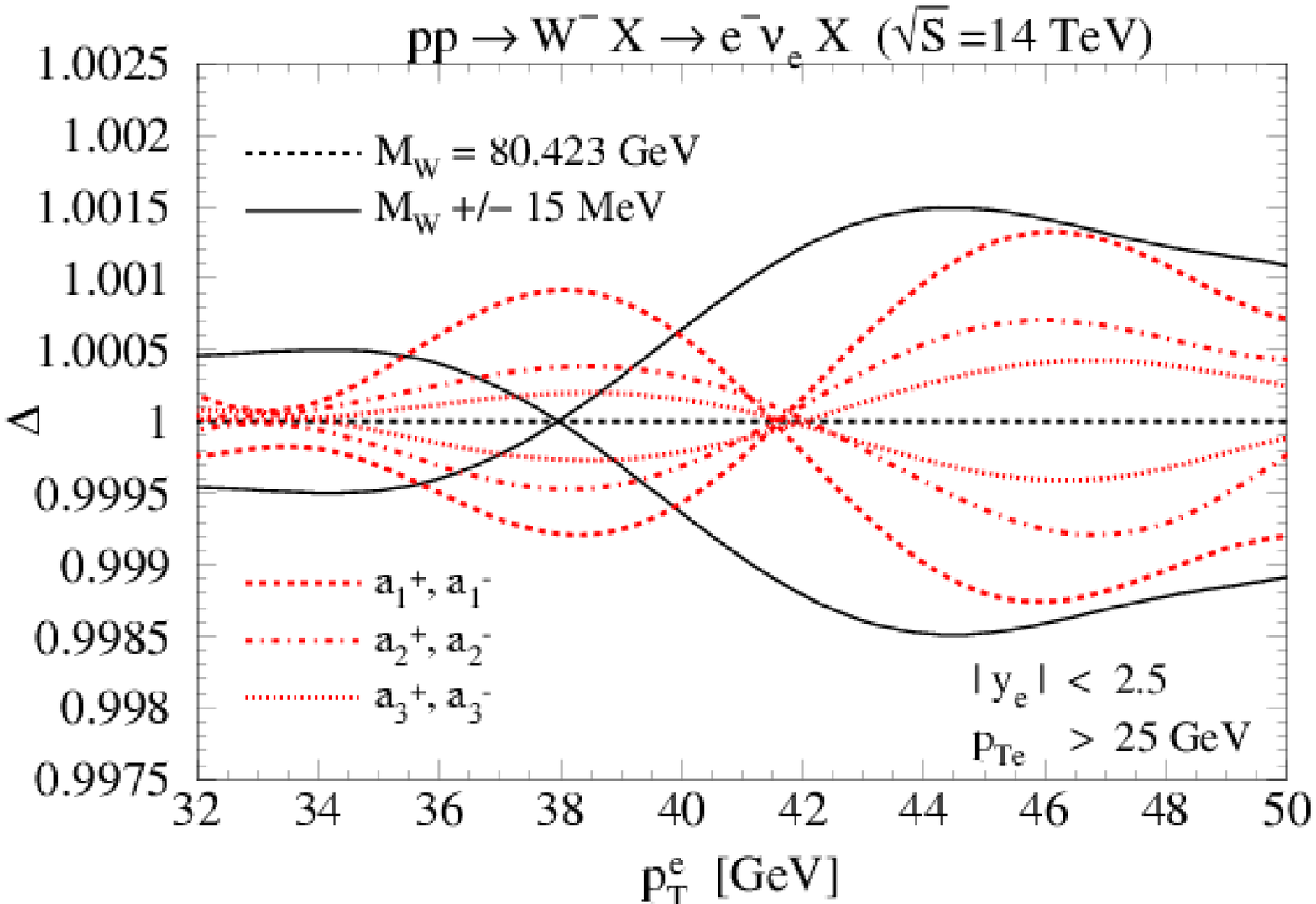}

 \begin{centering}\hspace*{0.5in}(a)\hspace*{3.2in}(b)\par\end{centering}
\caption{Fractional changes in $d\sigma/dp_{Te}$ in $W^{-}$ boson production
at the Tevatron and LHC caused by variations of $M_{W}$ by $\pm15$~MeV
for the central parametrization of ${\cal F}_{NP}(b,Q)$ (solid black
lines), and by six $1\sigma$ variations of ${\cal F}_{NP}(b,Q)$
in the Hessian method for the central value of $M_{W}=80.423$ GeV
(dashed and dotted red lines).\label{fig:th_ewk_S_NP_Tev}}
\end{figure}

Uncertainties in theoretical predictions caused by variations in
${\cal F}_{NP}(b,Q)$ can be estimated with the help of the Hessian
matrix method, developed recently to quantify errors in the global PDF
analysis (see, e.g., Ref.~\cite{Pumplin:2002vw}, and references
therein). In this approach, the central value of the observable $X$ is
computed for the best-fit set ${\bf a}{}_{best}=\left\{ a_{1},\,
  a_{2},\, a_{3}\right\} $ of the nonperturbative function parameters.
In addition, $X$ is computed for six {}``extreme'' parameter sets
${\bf a}_{(i)}^{\pm}$ ($i=1,2,3$) corresponding to the maximal
positive and negative displacements of $\vec{a}$ along three
eigenvectors of the Hessian matrix within the parameter region
satisfying $\chi^{2}-\chi_{best-fit}^{2}\leq1$.  The {}``extreme''
parameter sets are listed in Table~\ref{tab:th_FNP_inp5}.  The
$1\sigma$ error in $X$ is estimated by \[ \delta
X=\frac{1}{2}\sqrt{\sum_{i=1}^{3}\left(X({\bf a}_{(i)}^{+})-X({\bf
      a}_{(i)}^{-})\right)^{2}}.\]
\begin{table}
\begin{centering}\begin{tabular}{|c||c|c|c||c|c|c|}
\hline 
Parametrization:&
\multicolumn{3}{c||}{$C_{3}=b_{0}$}&
\multicolumn{3}{c|}{$C_{3}=2b_{0}$}\tabularnewline
\hline
Set/parameter&
$a_{1}$&
$a_{2}$&
$a_{3}$&
$a_{1}$&
$a_{2}$&
$a_{3}$\tabularnewline
\hline
\hline 
${\bf a}_{(1)}^{+}$&
0.208&
0.198&
-0.034&
0.262&
0.181&
-0.059\tabularnewline
\hline 
${\bf a}_{(1)}^{-}$&
0.192&
0.168&
-0.017&
0.233&
0.135&
-0.039\tabularnewline
\hline 
${\bf a}_{(2)}^{+}$&
0.21&
0.169&
-0.024&
0.240&
0.182&
-0.055\tabularnewline
\hline 
${\bf a}_{(2)}^{-}$&
0.192&
0.199&
-0.029&
0.254&
0.134&
-0.044\tabularnewline
\hline 
${\bf a}_{(3)}^{+}$&
0.208&
0.195&
-0.024&
0.232&
0.153&
-0.057\tabularnewline
\hline 
${\bf a}_{(3)}^{-}$&
0.193&
0.174&
-0.029&
0.262&
0.162&
-0.042\tabularnewline
\hline
\end{tabular}\par\end{centering}
\caption{\label{tab:th_FNP_inp5} Parameters of the six {}``extreme'' sets
${\bf a}_{(i)}^{\pm}$ ($i=1,2,3$) for the nonperturbative functions
${\cal F}_{NP}(b,Q)$ published in Ref.~\cite{Konychev:2005iy}.}
\end{table}

Variations in $d\sigma/dp_{Te}$ for the extreme parametrizations of
${\cal F}_{NP}(b,Q)$ at the Tevatron and LHC are shown in
Fig.~\ref{fig:th_ewk_S_NP_Tev}(a) and
Fig.~\ref{fig:th_ewk_S_NP_Tev}(b). We plot the ratio
$\Delta\equiv(d\sigma^{{a}{}_{(i)}^{\pm}}/dp_{Te})/(d\sigma^{ref}/dp_{Te})$,
where $(d\sigma^{ref}/dp_{Te})$ is the {}``reference'' cross section
evaluated with the central values of $\{ a_{1,2,3}\}$ and a $W$~boson
mass of $M_{W}=80.423$~GeV. $(d\sigma^{a_{(i)}^{\pm}}/dp_{Te})$ are
the cross sections for the extreme parameter sets ${\bf
  a}_{(i)}^{\pm}$ ($i=1,2,3$) and the central $M_{W}$, shown by dashed
and dotted lines. The magnitude of these deviations is comparable to
the effect of a variation of $M_{W}$ by
$\pm15\,{\textrm{{\textrm{MeV}}}}$~(solid black lines), although their
$p_{Te}$ dependence is not exactly the same as the shift in
$d\sigma/dp_{Te}$ caused by the variation of $M_{W}$.
Figure~\ref{fig:th_ewk_S_NP_Tev} indicates that the remaining
uncertainties in ${\cal F}_{NP}(b,Q)$ may introduce an error of up to
10-20 MeV (estimated as in Fig.~\ref{fig:th_ewk_S_NP_Tev}) in the
$M_{W}$ measurement in the $p_{Te}$ channel.

\subsubsection{New features at small $x$ }
\label{subsec:ewkth_newfeatures}

The global $p_{T}$ fits \cite{Landry:2002ix,Konychev:2005iy} analyze
the $p_{T}$-dependent data from low-mass Drell-Yan pair and $Z$ boson
production at $x_{A,B}\gtrsim10^{-2}$. At $x\lesssim10^{-2}$, where no
such data currently exist, $W$ and $Z$ boson production may be subject
to additional transverse momentum broadening, as suggested by fits of
resummed $q_{T}$ distributions to data from semi-inclusive deep
inelastic scattering at $x=10^{-4}\sim10^{-2}$
\cite{Nadolsky:1999kb,Nadolsky:2000ky}.  This broadening may
substantially exceed the range of uncertainties in $d\sigma/dq_{T}$
quoted in the previous subsection.

Using crossing relations, we estimate its magnitude in $W$ and $Z$
boson production based on the SIDIS results~\cite{Berge:2004nt}.  The
BLNY parametrization of ${\cal F}_{NP}(b,Q)$ \cite{Landry:2002ix} is
modified to include an additional term
$\left(\rho(x_{A})+\rho(x_{B})\right)b^{2},$ where the function
$\rho(x)$ parametrizes the cumulative effect of unaccounted
higher-order contributions to the $b$-dependent PDF's ${\cal
  P}_{j}(x,b)$ at nearly nonperturbative impact parameters
($b\sim1\mbox{ GeV}^{-1}$).  Since ${\cal P}_{j}(x,b)$ are included in
the resummed form factors both in Drell-Yan-like processes and SIDIS,
the function $\rho(x)$ can be constrained using the SIDIS data from
HERA. This function satisfies $\rho(x)\propto1/x$ for $x\ll x_{0}$,
and $\rho(x)\sim0$ for $x\gg x_{0}$, where the free parameter $x_{0}$
is chosen in the range $10^{-3}-10^{-2}$.  Since $\rho(x)$ vanishes at
large $x$, this model agrees with the existing Drell-Yan $p_{T}$ data.
At $x<10^{-2}$, the growth of $\rho(x)$ leads to harder $q_{T}$
distributions without affecting the inclusive production rate.

\begin{figure}
\begin{center}\includegraphics[clip,width=0.6\columnwidth]{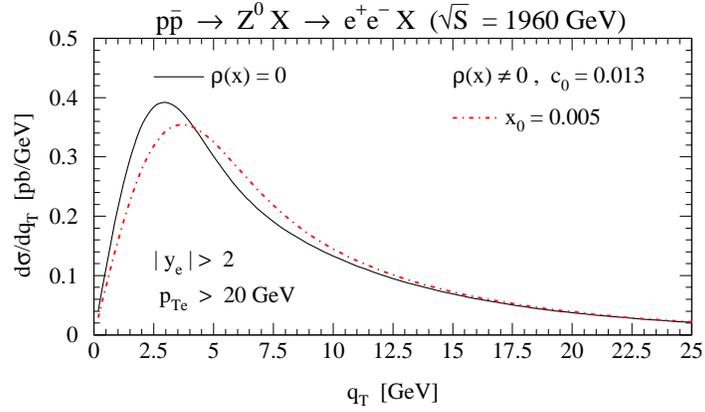}\end{center}\vspace*{-10pt}
\caption{Transverse momentum distributions of $Z$ bosons in the Tevatron
Run-2 for events with both decay electrons registered in the forward
($y_{e^{+}}>2,\,\, y_{e^{-}}>2$) or backward ($y_{e^{+}}<-2,\,\, y_{e^{-}}<-2$)
detector regions. The solid (black) curve is the standard CSS cross
section, calculated using the BLNY nonperturbative function \cite{Landry:2002ix}.
The dashed (red) curve includes the additional term responsible for
the $q_{T}$ broadening in the small-x region. \label{fig:th_ewk_zqt}}
\end{figure}
\begin{figure}
\begin{centering}\includegraphics[width=0.55\columnwidth]{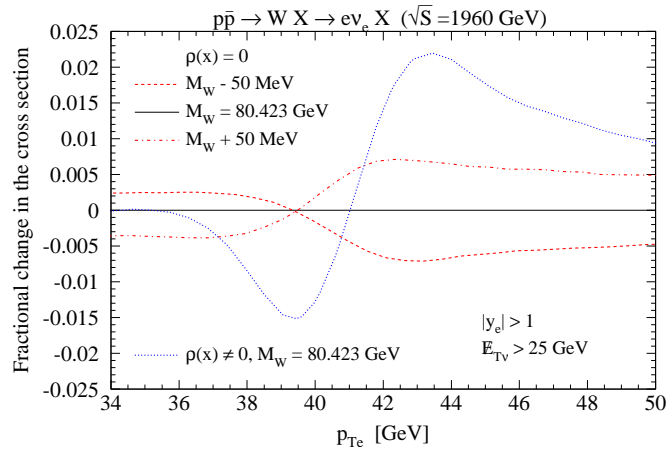}\vspace*{-10pt}\par\end{centering}
\caption{The fractional change in the $d\sigma/dp_{Te}$ distribution in forward
$W$ boson production at the Tevatron ($|y_{e}|>1$, $E_{T}\hspace*{-13pt}/\,\,\,>25$~GeV)
due to the small-$x$ broadening (blue dotted line)~\cite{Berge:2004nt}.
The dashed and dot-dashed lines correspond to ${\delta M}_{W}=\pm50$~MeV
in the case when the broadening is absent. \label{fig:th_ewk_tevMW} }
\end{figure}

At the Tevatron, the small-$x$ broadening may be seen only at large
rapidities, such as forward $Z$ boson production displayed in
Fig.~\ref{fig:th_ewk_zqt}.  It marginally affects the $M_{W}$
measurement, dominated by events with small boson rapidities. The most
pronounced effects may be visible in the $p_{Te}$ distribution (cf.
Fig.~\ref{fig:th_ewk_tevMW}), where variations due to the broadening
are comparable to the effect of a variation of $M_{W}$ by $\sim20$ MeV
($>50$ MeV) at $|y_{e}|<1$ ($|y_{e}|>1$).
\begin{figure}
\begin{centering}\includegraphics[width=0.49\columnwidth,keepaspectratio]{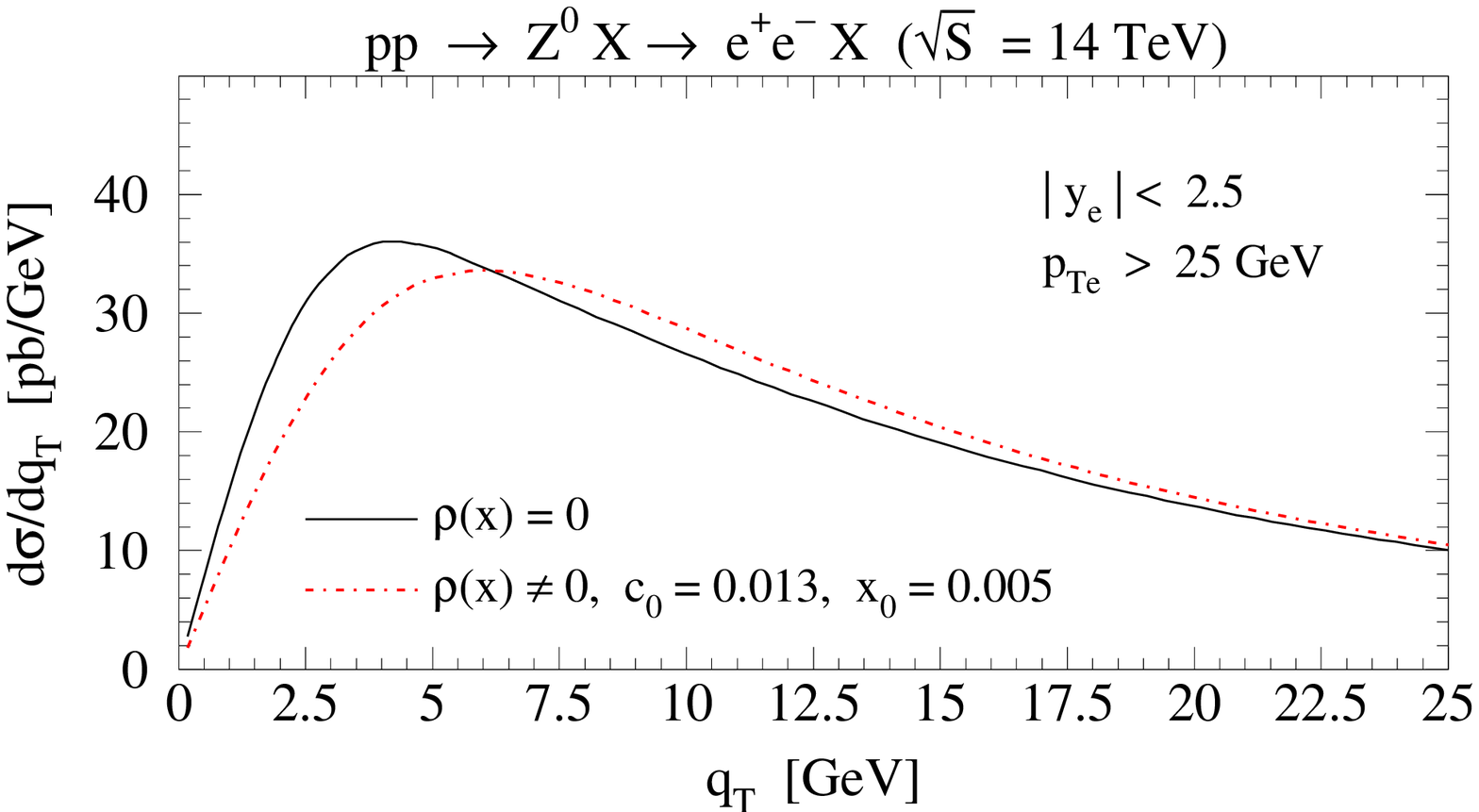}\(\,\,\)\includegraphics[width=0.5\columnwidth,keepaspectratio]{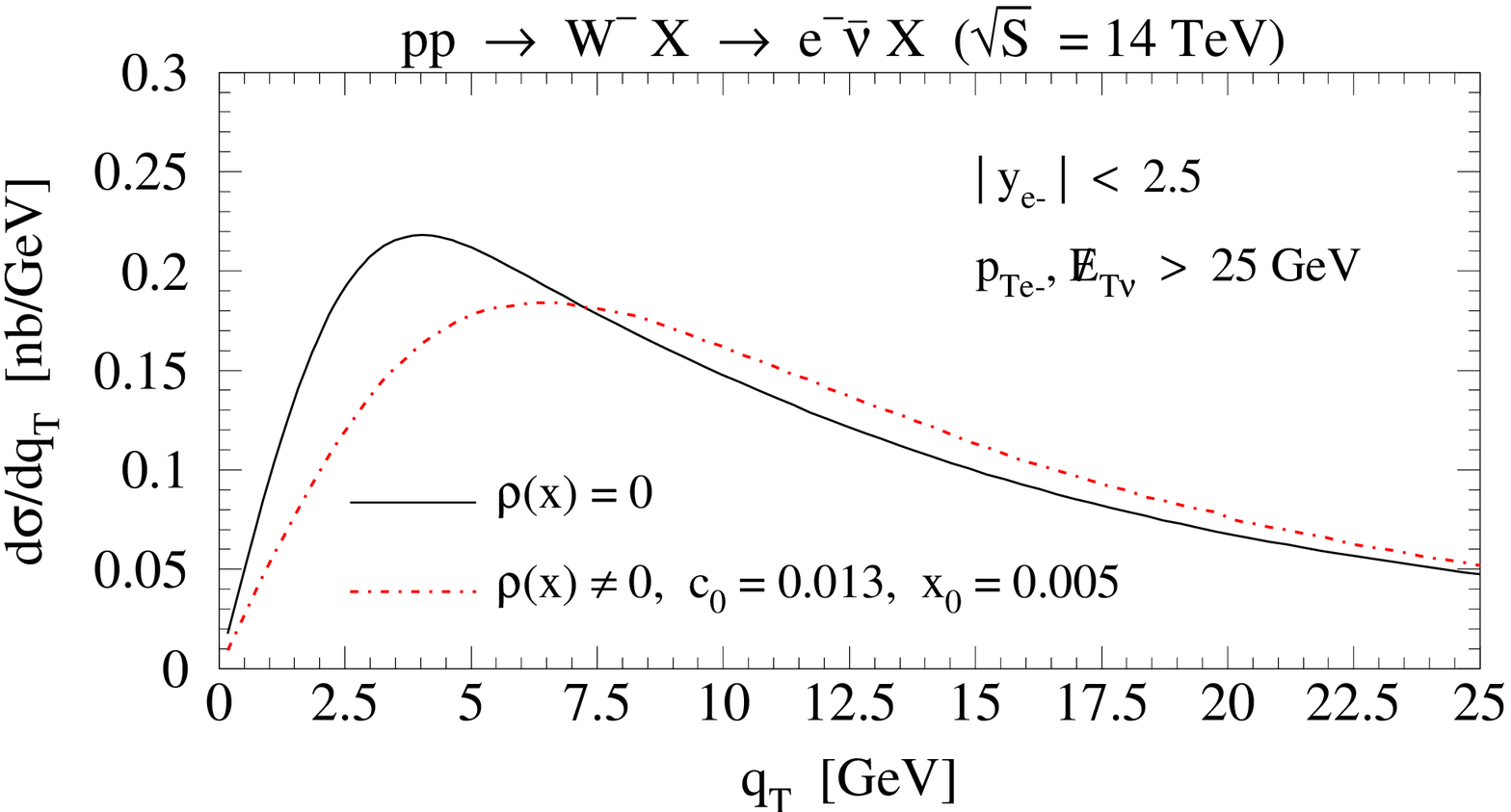}\par\end{centering} \begin{centering}\hspace*{0.5in}(a)\hspace*{3.2in}(b)\par\end{centering}\vspace*{-10pt}
\caption{\label{fig:th_ewk_ZLHC} (a) Transverse momentum distributions of
$Z$ bosons at the Large Hadron Collider with (dashed) and without
(solid) the small-$x$ effects. The events are selected by requiring
$|\, y_{e}\,|<2.5$ and $p_{Te}>25$ GeV for both decay electrons.
(b)~Same for $W^{-}$ bosons. The decay leptons are required to satisfy
$|\, y_{e}\,|<2.5$, $\, p_{Te}>25$ GeV, $\, E_{T}\hspace{-13pt}/\hspace{13pt}>25\,$
GeV. }
\end{figure}

At the LHC, the small-$x$ broadening may be observed at all
rapidities.  Our model can estimate its magnitude for boson rapidities
less than about 2.5, roughly corresponding to the $x$ region covered
by the SIDIS data. In $Z$ boson production
(Fig.~\ref{fig:th_ewk_ZLHC}(a)), the distribution with $\rho(x)\neq0$
is clearly shifted toward higher $q_{T}$. The $q_{T}$ shift is even
larger in the production of $W$ bosons,
cf.~Fig.~\ref{fig:th_ewk_ZLHC}(b), as a result of the smaller boson
mass ($M_{W}<M_{Z}$) and less restrictive leptonic cuts. Furthermore,
the shift is slightly larger in $W^{+}$ boson production than in
$W^{-}$ boson production because of the flatter rapidity distribution
for $W^{+}$ bosons. The shown $q_{T}$ broadening propagates into the
leptonic transverse mass and lepton transverse momentum distributions.
Both the $m_{Te\nu}$ and $p_{T}^{e}$ methods for the measurement of
$M_{W}$ are affected in this case, in contrast to the Tevatron, where
the $m_{Te\nu}$ method is almost not susceptible to the broadening.

\subsubsection{Heavy quark effects}
\label{subsec:ewkth_heavyquark}
\begin{figure}
\begin{centering}\includegraphics[width=0.5\columnwidth]{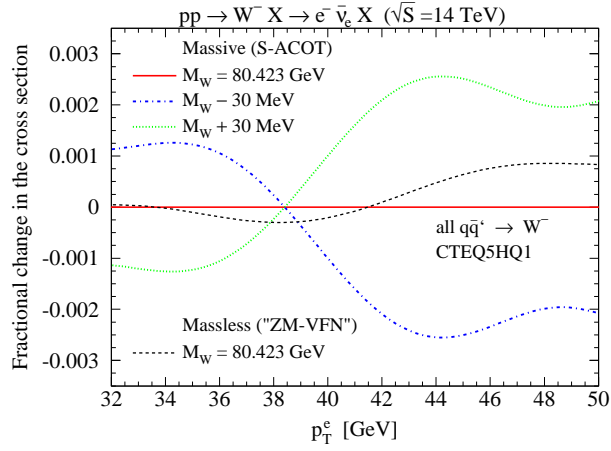}\vspace*{-10pt}\par\end{centering}
\caption{The fractional change in the $d\sigma/dp_{_{T}}^{\, e}$ distribution
for $W^{-}$ boson production at the LHC due to the improved treatment
of heavy-quark mass terms in the CSS resummation \cite{Berge:2005rv}.
The solid red line ($M_{W}=80.423$~GeV), dotted green line ($\delta M_{W}=+30$~MeV)
and dot-dashed blue line ($\delta M_{W}=-30$~MeV) are reference
distributions that include the full $m_{c,b}$ dependence in the general-mass
variable-flavor number (S-ACOT) scheme. The dashed black line is the
distribution in the zero-mass variable-flavor number (ZM-VFN) scheme,
which neglects $m_{c,b}$ dependence in perturbative contributions.
\label{fig:th_ewk_tevHQ}}
\end{figure}

About $20\%,$ $30\%$, and $15\%$ of $W^{+},$ $W^{-},$ and $Z^{0}$
bosons at the LHC will be produced in scattering processes involving
at least one charm or bottom initial-state quark or antiquark. The
tangible rate of heavy-flavor contributions at the LHC contrasts that
at the Tevatron, where only $8\%$ ($3\%$) of $W^{\pm}$ ($Z^{0}$)
bosons are produced in $c$ or $b$ quark scattering. Since the
heavy-quark masses suppress multiple parton radiation at small
transverse momenta, they must be implemented in the resummation
calculation in order to correctly predict $q_{T}$ distributions at the
LHC energy. The improved treatment of heavy-quark masses changes the
$q_{T}$ distribution at the LHC by an amount comparable to other
systematic uncertainties affecting the $W$ boson mass measurement
\cite{Berge:2005rv}.

For this purpose, we formulate the CSS resummation formalism in a
general-mass variable flavor number (S-ACOT) factorization scheme
\cite{Collins:1998rz,Kramer:2000hn}, which preserves correct $m_{c,b}$
dependence at low momentum scales and resums heavy-quark collinear
contributions at large momentum scales. The feasibility of the CSS
resummation in the S-ACOT scheme has been first demonstrated in
Ref.~\cite{Nadolsky:2002jr}.  In $W$ boson production in the
heavy-scattering channels, the S-ACOT scheme predicts harder $q_{T}$
distributions than the zero-mass variable flavor number (ZM-VFN)
scheme used in previous studies. The improved treatment of $m_{c,b}$
in the S-ACOT scheme modifies $p_{Te}$ distributions for $W^{-}$
bosons at the LHC by an amount comparable to the effect of $\delta
M_{W}\sim10$ MeV (see Fig.~\ref{fig:th_ewk_tevHQ}).  The $m_{c,b}$
dependence is somewhat less pronounced in $W^{+}$ and especially
$Z^{0}$ production, as a result of smaller heavy-flavor contents in
these processes. It is negligible at the Tevatron.

\subsubsection{PDF uncertainties}

PDF uncertainties in the $M_{W}$ measurement were estimated in the
Tevatron Run-1 by repeating the analysis for select sets of parton
densities, which did not cover the full span of allowed variations in
the PDF parameters. A more systematical estimate can be realized by
applying the new techniques for the PDF error analysis. The choice of
the PDF set affects $q_{T}$ distributions directly, by changing the
PDF's in the factorized QCD cross section, but also indirectly, by
modifying the nonperturbative function ${\cal F}_{NP}(b,Q)$ in the
resummed form factor. For a chosen form of ${\cal F}_{NP}(b,Q)$, the
PDF errors can be evaluated within the Hessian matrix method, by
repeating the computation of $q_{T}$ distributions for an ensemble of
sample PDF sets.

\begin{figure}
\begin{centering}\includegraphics[width=0.58\columnwidth]{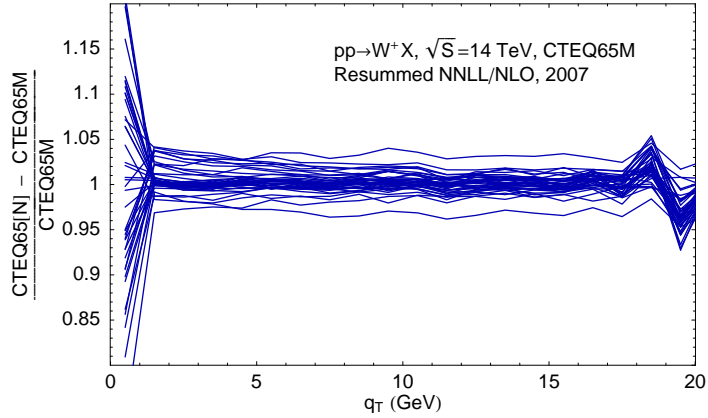}\vspace*{-10pt}\par\end{centering}
\caption{Variations in the $W^{+}$ transverse momentum distribution, $d\sigma/dq_{T}$,
at the LHC for 40~CTEQ6.5 PDF sets \cite{Tung:2006tb} with respect
to the CTEQ6.5M PDF set. \label{fig:th_ewk_qTPDF}}
\end{figure}

Variations in the resummed $q_{T}$ spectrum for $W^{+}$ production at
the LHC are shown in Fig.~\ref{fig:th_ewk_qTPDF} for 41 CTEQ6.5 PDF
sets \cite{Tung:2006tb} and KN1 nonperturbative function
\cite{Konychev:2005iy}.  Depending on the choice of the PDF set,
$d\sigma/dq_{T}$ at small $q_{T}$ changes by up to $\pm4\%$ from its
value for the central PDF set (CTEQ65M), except for very small
$q_{T}$. The variations in the PDF's modify \emph{both} the
normalization and shape of $d\sigma/dq_{T}$.  Although the changes in
the shape are relatively weak at $q_{T}<15$ GeV, they may affect the
measurement of $M_{W}$ in the $p_{Te}$ method. These results do not
reflect possible correlations between the PDF's and ${\cal
  F}_{NP}(b,Q)$ in the global fit to $p_{T}$ data, introduced by the
dependence of ${\cal F}_{NP}(b,Q)$ on the normalizations of the
low-$Q$ Drell-Yan cross sections. The correlation between free
parameters in the PDF's and ${\cal F}_{NP}(b,Q)$ will be explored in
the future by performing a simultaneous global analysis of the
inclusive cross sections and $p_{T}$-dependent data.

\subsubsection{$q_{T}$ spectrum and final-state QED corrections}

As discussed in Sections~\ref{subsec:th_ewkstatus},
\ref{subsec:th_ewkmulti}, electroweak corrections to Drell-Yan $W$ and
$Z$ boson production are dominated by the QED radiation from the
final-state charged lepton, which results in some loss of the charged
lepton's momentum to the surrounding cloud of soft and collinear
photons. The final-state QED (FQED) radiation changes the extracted
value of $M_{W}$ by shifting the Jacobian peak in the $p_{Te}$
distribution in the negative direction.  In contrast, the
initial-state radiation and interference terms mostly change the
overall normalization of the Jacobian peak and have a smaller effect
on the determination of $M_{W}$. The combined effect of the
${\mathcal{O}}(\alpha)$ FQED correction and the resummed QCD
correction was estimated for the Run-2 observables by using a new
computer program \textsc{ResBos-A} (\textsc{ResBos} with FQED
effects)~\cite{Cao:2004yy}.  The FQED and resummed QCD corrections to
the Born-level shape of the Jacobian peak in the $m_{Te\nu}$
distribution were found to be approximately (but not completely)
independent. The reason is that the $m_{Te\nu}$ distribution is almost
invariant with respect to the transverse momentum of $W$ bosons, so
that the QCD correction reduces, to the first approximation, to
rescaling of the Born-level $m_{Te\nu}$ distribution by a constant
factor. The relationship between FQED and QCD corrections is more
involved in the leptonic $p_{T}$ distributions, which depend linearly
on $q_{T}$ of $W$ bosons. In the $p_{T}^{e}$ channel, the combined
effect does not factorize into separate FQED and QCD corrections to
the Born-level cross section.

\subsubsection{Conclusion}

We have reviewed recent advances in the understanding of resummed
$q_{T}$ distributions for electroweak bosons. The $q_{T}$ resummation
formalism is realized at the NNLL/NLO level of accuracy and includes
such new ingredients as the dominant NLO electroweak contributions,
correct dependence on heavy-quark mass terms, and an improved model
for the nonperturbative recoil at $x\gtrsim10^{-2}$. Other important
aspects of $q_{T}$ resummation, such as the behavior of higher-order
radiative contributions at $x<10^{-2}$ and correlations between the
PDF's and nonperturbative resummed function must be assessed to ensure
that the systematic uncertainties in the $M_{W}$ measurements are
under full control. The dynamics of these factors can be tested by
measuring fully differential distributions of lepton pairs in a wide
range of $Q$ and $y$ in the Tevatron Run-2.

\section*{Acknowledgments}
This work was partially supported by the U.S. Department of Energy
under grant DE-FG03-95ER40908, contract DE-AC02-06CH11357, and the 
Lightner-Sams Foundation.

\subsection{Estimate of theoretical uncertainties due to missing higher-order
corrections}
\label{subsec:th_ewkerror}
{\bf Contributed by:~A.~Arbuzov, D.~Bardin, U.~Baur, S.~Bondarenko,
C.~M.~ Carloni Calame, P.~Christova, L.~Kalinovskaya, G.~Montagna, 
O.~Nicrosini, R.~Sadykov, A.~Vicini, and D.~Wackeroth}

\begin{figure}
\begin{center}
  \includegraphics[width=7.1cm,
  keepaspectratio=true]{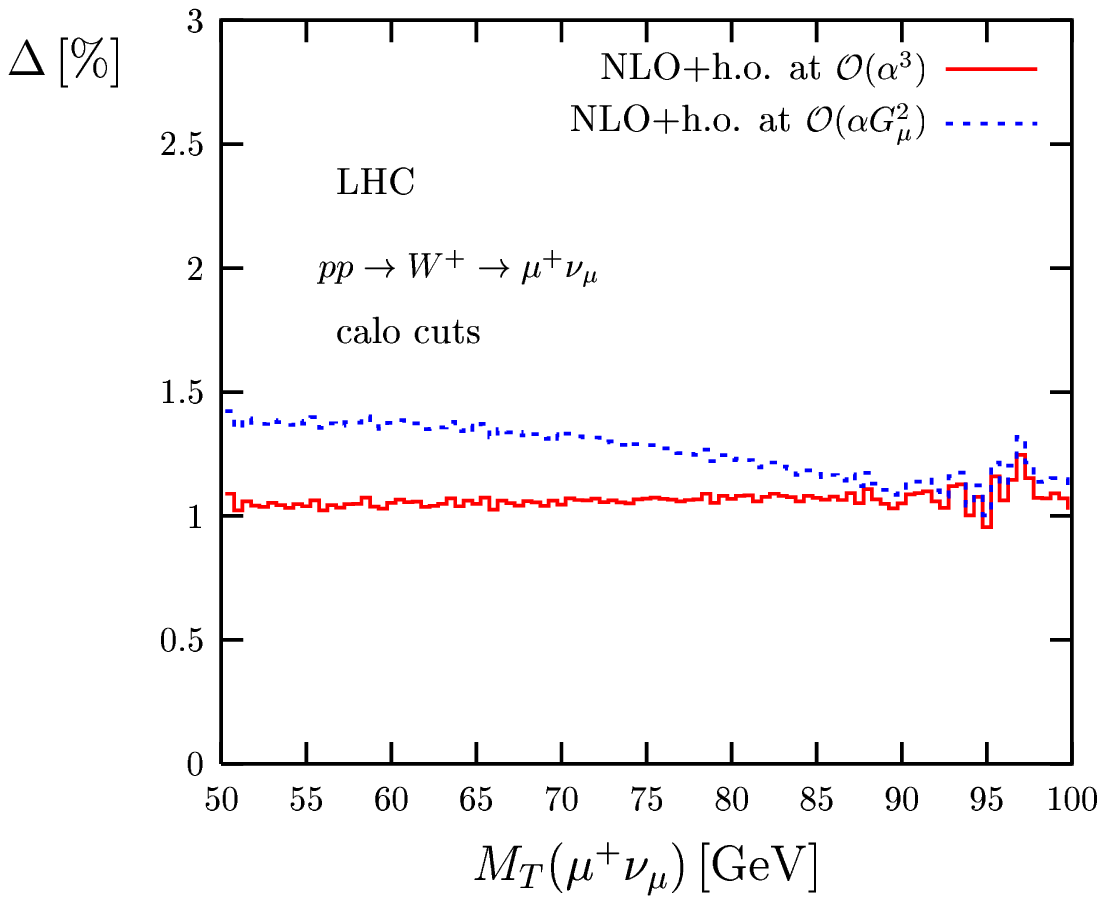}
\hspace*{-.0cm}
  \includegraphics[width=7.1cm,
  keepaspectratio=true]{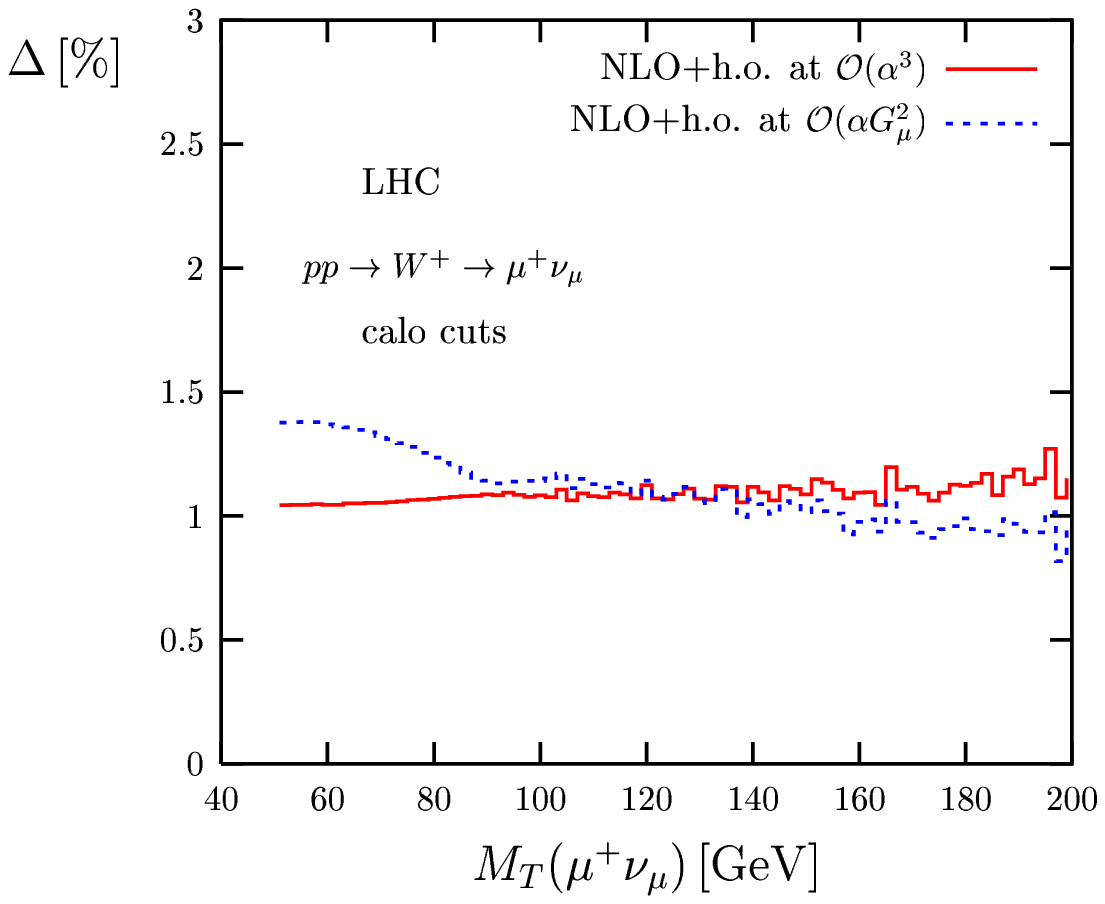}
\hspace*{-.0cm}
  \includegraphics[width=7.1cm,
  keepaspectratio=true]{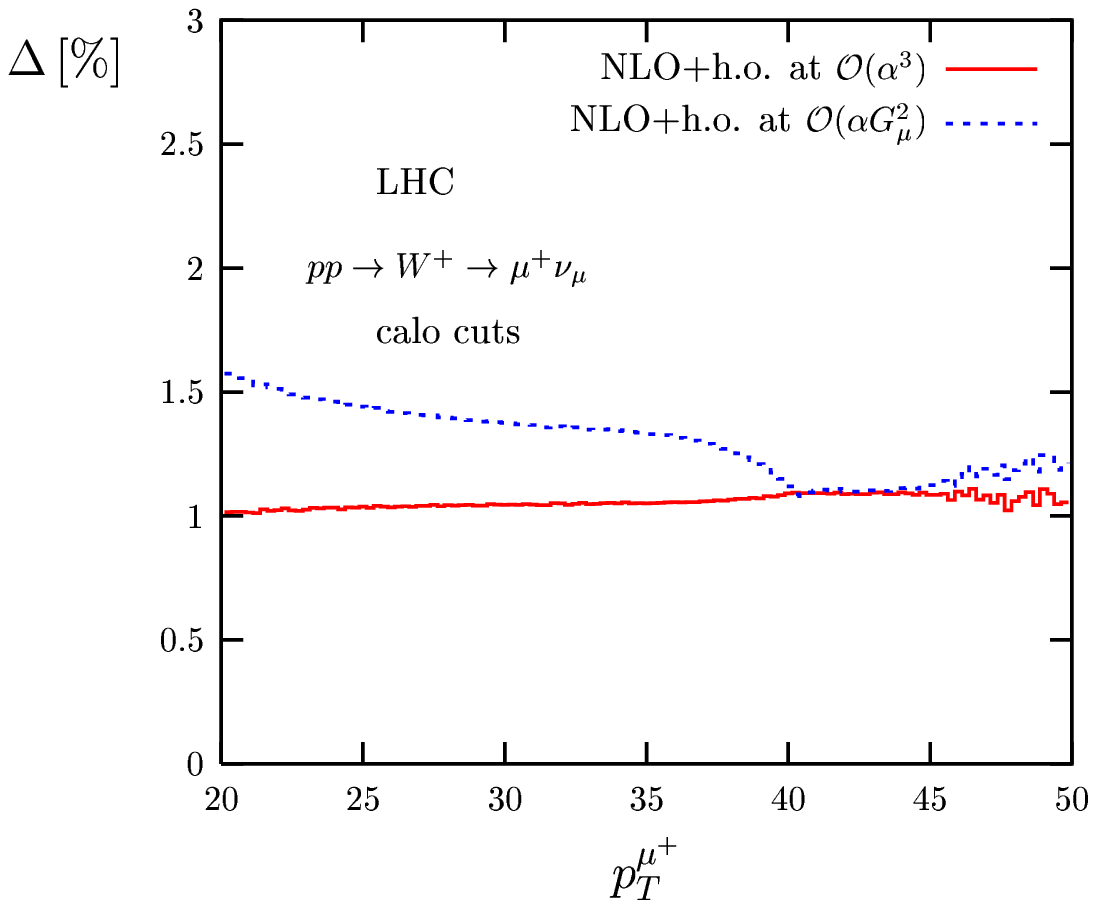}
\hspace*{-.0cm}
  \includegraphics[width=7.1cm,
  keepaspectratio=true]{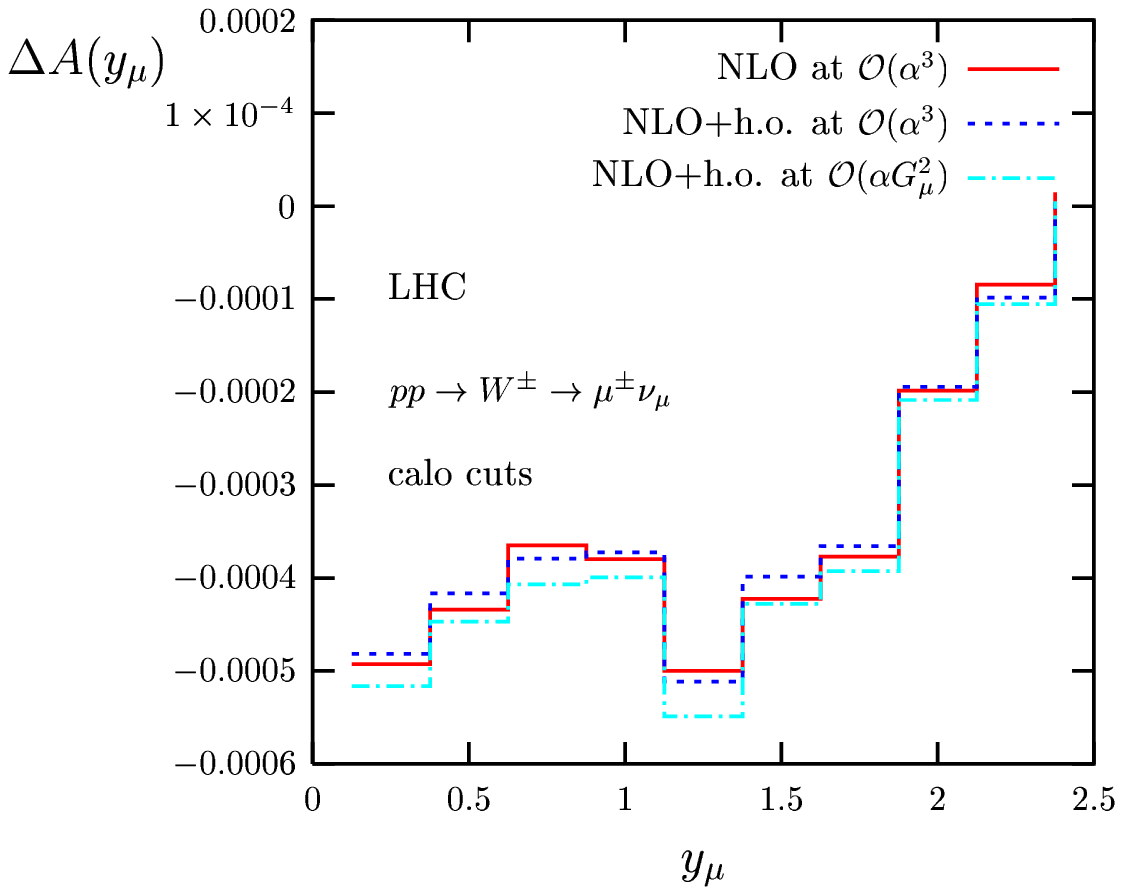}
\end{center}
\caption{
  Relative differences $\Delta$ between NLO and higher-order
  predictions for the $M_T(\mu^+\nu_\mu)$, $p_T^{\mu^+}$ and
  $A(y_\mu)$ distributions for single $W^+$ production with calo cuts
  at the LHC.  The higher-order predictions include corrections beyond
  ${\cal O}(\alpha)$ other than mPR, in addition to the complete set
  of electroweak ${\cal O}(\alpha)$ corrections (see text for more
  details).  For this comparison, we use {\sc WGRAD2} results for
  definiteness.}
\label{fig:th_ewk_lhc_mtptbest}
\end{figure}

In order to estimate the residual theoretical uncertainties due to
missing higher-order corrections of predictions obtained by
electroweak precision tools such as {\sc HORACE}, {\sc SANC}, and {\sc
  WGRAD2}, we study in the following the effects of different choices
for the EW input parameter scheme and of leading higher-order
(irreducible) QCD and EW corrections connected to the $\rho$
parameter. For definiteness we use {\sc WGRAD2}.  Similar results can
be easily obtained with {\sc HORACE} and {\sc SANC} as well.  In
Table~\ref{tab:th_ewk_e} and Fig.~\ref{fig:th_ewk_lhc_mtptbest} we
compare the predictions of the tuned comparison using the setup
described in Section~\ref{subsec:th_ewkcomp} (labeled as 'NLO at
${\cal O}(\alpha^3)$') with predictions that are obtained as follows:
\begin{itemize}
\item 'NLO at ${\cal O}(\alpha^3)$ incl. h.o.':\\ The EW input
  parameter scheme of the tuned comparison is used as described in
  Section~\ref{subsec:th_ewkcomp}.  But we replace the $Z$ mass
  renormalization constant $\delta M_Z^2 = {\cal R}e
  \Bigl(\Sigma^{Z}(M_Z^2)\Bigr)$ by
\begin{equation}
\delta M_Z^2 = {\cal R}e \Bigl(\Sigma^{Z}(M_Z^2)-\frac{(\hat \Sigma^{\gamma Z}(M_Z^2))^2}{M_Z^2+\hat \Sigma^{\gamma}(M_Z^2)} \Bigr), \quad
\delta M_W^2 = {\cal R}e \Sigma^{W}(M_W^2)
\end{equation}
where $\Sigma^V(\hat\Sigma^V)$ denote the transverse parts of
unrenormalized(renormalized) vector boson self energies, and include
higher-order (irreducible) corrections connected to the $\rho$
parameter, $\Delta\rho^{HO}$, by performing the replacement
\begin{equation}
\frac{\delta M_Z^2}{M_Z^2}-\frac{\delta M_W^2}{M_W^2} 
\to \frac{\delta M_Z^2}{M_Z^2}-\frac{\delta M_W^2}{M_W^2}-\Delta\rho^{HO}
\end{equation}
as described in detail in Ref.~\cite{Baur:2001ze} (Appendix A).
\item 'NLO at ${\cal O}(\alpha G_\mu^2)$ incl. h.o.':\\ In addition to
  the modifications described above, we change the EW input parameter
  scheme ($\alpha(0)$ scheme $\to$ $G_\mu$ scheme) by replacing
\[\alpha(0) \to {\sqrt{2}G_\mu M_W^2\over\pi}\left (1-{M_W^2\over
M_Z^2}\right ),\]
so that 
\[{\rm d}\sigma(\alpha G_\mu^2)=[{\rm d}\sigma_{NLO}(\alpha^3)-2 \, \Delta r \, {\rm d}\sigma_{LO}(\alpha^2)] \; \left[\frac{\sqrt{2}G_\mu M_W^2}{\pi \alpha(0)}\left (1-{M_W^2\over M_Z^2}\right )\right]^2 \; ,\]
where $\Delta r$ parametrizes the radiative corrections to muon decay
(see also Ref.~\cite{Buttar:2006zd}).
\end{itemize}
As illustrated in Table~\ref{tab:th_ewk_e} and
Fig.~\ref{fig:th_ewk_lhc_mtptbest} for the LHC the relative
differences between the different predictions, $\Delta={\rm
  d}\sigma_{NLO}(\alpha^3)/({\rm d}\sigma_{\rm h.o.})-1$ and $\Delta
A(y_l)=A(y_l)_{NLO}(\alpha^3)-A(y_l)_{\rm h.o.}$, are at most about
1.5\% for $\sigma_W$, and the $M_T(l\nu)$, $p_T^l$ distributions, and
up to about $4 \cdot 10^{-5}$ for the charge asymmetry of leptons in
$W$ decay.  We find the same relative differences at the Tevatron.
Since switching to the $G_\mu$ scheme changes the shape of the
$M_T(l\nu)$ distribution, a more detail study of how these effects
translate into a shift in $M_W$ is warranted.  Moreover, other sources
of residual theoretical uncertainties, for instance missing
higher-order EW Sudakov logarithms and the QED scale dependence, need
to be under control as well.

\begin{table}
\begin{center}
\begin{tabular}{|c|c|c|} \hline
           & \multicolumn{1}{|c|}{Tevatron, $\sigma_W$ [pb]} & \multicolumn{1}{|c|}{LHC, $\sigma_W$  [pb]}  \\ \hline
           & $p\bar p \to W^+\to \mu^+\nu_\mu$ &  $pp \to W^+\to \mu^+\nu_\mu$ \\ \hline
NLO at ${\cal O}(\alpha^3)$ &  738.00(1)  & 4943.0(1)  \\
NLO at ${\cal O}(\alpha^3)$ incl. h.o.  &  745.80(1)  & 4995.5(1) \\ 
NLO at ${\cal O}(\alpha G_{\mu}^2)$ incl. h.o. &  747.62(1)  & 5006.5(1) \\
\hline
\end{tabular}
\caption{Comparison of predictions for $\sigma_W$ to $pp,p\bar p \to W^+\to \mu^+\nu_\mu$ with calo cuts at the Tevatron and the LHC. The higher-order predictions include corrections beyond ${\cal O}(\alpha)$ other than mPR, in addition to the complete set of electroweak ${\cal O}(\alpha)$ corrections (see text for more details). For this comparison, we use {\sc WGRAD2} results for definiteness.} 
\label{tab:th_ewk_e}
\end{center}
\end{table}

\subsection{Experimental Uncertainties}
\label{subsec:th_ewkexp}
\textbf{Contributed by: C.~Hays and D.~Wackeroth}

The $W$ boson observables studied in this report -- the total $W$
boson production cross section ($\sigma_W$), the $M_T(l\nu)$ and
$p_T^l$ distributions, and the $W$ boson charge asymmetry for leptons
($A(y_l)$) -- have been measured by the CDF and D\O\
collaborations\footnote{A collection of the most recent EW results can
  be found at the CDF and D\O\ physics results websites,
  www-cdf.fnal.gov/physics/ewk and
  www-d0.fnal.gov/Run2Physics/WWW/results/ew.htm}. The $W$ boson mass
is dominantly extracted from the $M_T(l\nu)$ distribution, as
described in Section~5.  Possible improvements in the $W$ boson mass
measurement at the LHC by using the transverse momentum distribution
of the charged lepton have been studied in Ref.~\cite{CMS-PTDR-Vol2}.
In the following we briefly summarize present and anticipated
experimental uncertainties in the measurements of $\sigma_W$, $A(y_W)$, $A(y_l)$
and $M_W$ and discuss their implications on further improvements of
theoretical predictions.

\subsubsection{Total $W$ and $Z$ boson production cross section} 
As pointed out earlier, given the large $W$ and $Z$ boson production
rates at the LHC, the total $W$ and $Z$ boson production cross
sections are expected to be used for detector calibration and as
luminosity monitors~\cite{Dittmar:1997md}.  The $72~{\rm pb}^{-1}$ CDF
combined $e$ and $\mu$ result for the total $W$ boson production cross
section is~\cite{Abulencia:2005ix}:
\[\sigma(W) \times Br(W \to l \nu) =  2775 \pm 10 ({\rm stat}) \pm 53 ({\rm sys})  \; {\rm pb} \; ,\]
which corresponds to a relative precision of $2\%$.
The $96~{\rm pb}^{-1}$ D\O\ $\mu$ result is~\cite{bib:D0wcrosssection}
\[\sigma(W) \times Br(W \to \mu \nu) =  2989 \pm 15 ({\rm stat}) \pm 81 (sys)  \; {\rm pb} \; .\]
The $72~{\rm pb}^{-1}$ CDF combined $e$ and $\mu$ result for the total $Z/\gamma^*$ production cross section is~\cite{Abulencia:2005ix}:
\[\sigma(Z/\gamma^*) \times Br(Z/\gamma^* \to ll) = 254.9 \pm 3.3 ({\rm stat}) \pm 4.6 ({\rm sys})\; {\rm pb} \; .\]
The D\O\ $148~{\rm pb}^{-1}$ result is~\cite{bib:D0zcrosssection}:
\[\sigma(Z/\gamma^*) \times Br(Z/\gamma^* \to \mu\mu) = 329.2 \pm 3.4 ({\rm stat}) \pm 7.8 ({\rm sys})\; {\rm pb} \; .\]
These results exclude the Tevatron luminosity uncertainty of about
$6\%$, of which $\approx 4\%$ is correlated between experiments.  The
$W$ and $Z$ boson measurements have a few systematic uncertainty
components that are different, so combining them should give the most
accurate luminosity measurement.  The measurements of $\sigma_W$ and
$\sigma_Z$ at the LHC are expected to reach a relative precision of
$3.3\%$ ($W\to \mu \nu_\mu$) and $2.3\%$ ($Z\to \mu \mu$) for ${\cal
  L}=1 \, {\rm fb}^{-1}$ and to be limited again by the luminosity
uncertainty of about $5\%$~\cite{CMS-PTDR-Vol2}. As long as the
luminosity uncertainty cannot be drastically improved, a theoretical
uncertainty of $1.5\%$ due to missing higher order EW corrections (see
Section~\ref{subsec:th_ewkerror}) is not worrisome. However, the
impact of these uncertainties on precise electroweak measurements at
the LHC based on $W/Z$ ratios should be studied in more detail.

\subsubsection{$W$ boson and lepton charge asymmetry} 
\begin{figure}
\begin{center}
\includegraphics[height=.28\textheight]{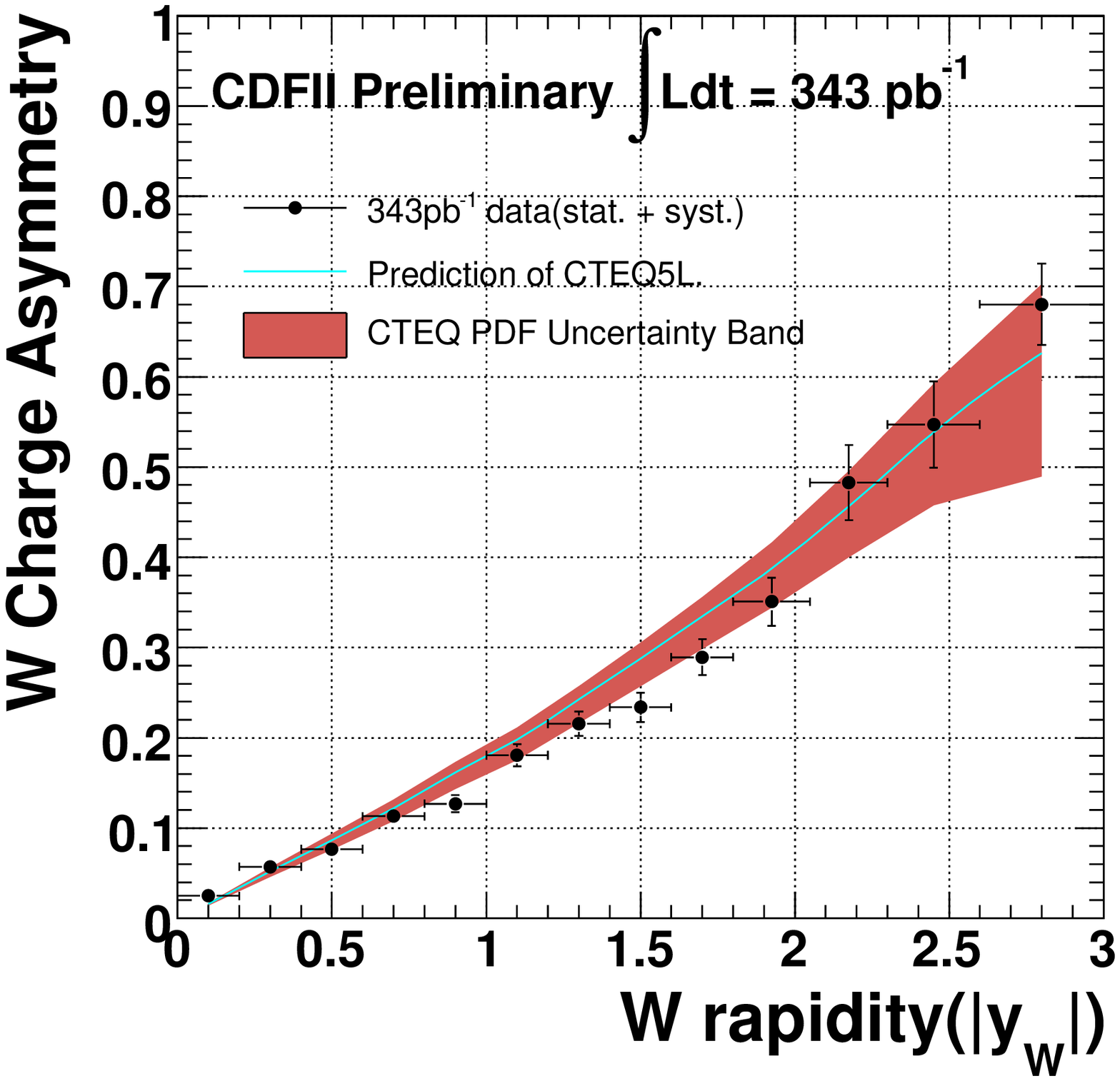}
\includegraphics[height=.25\textheight]{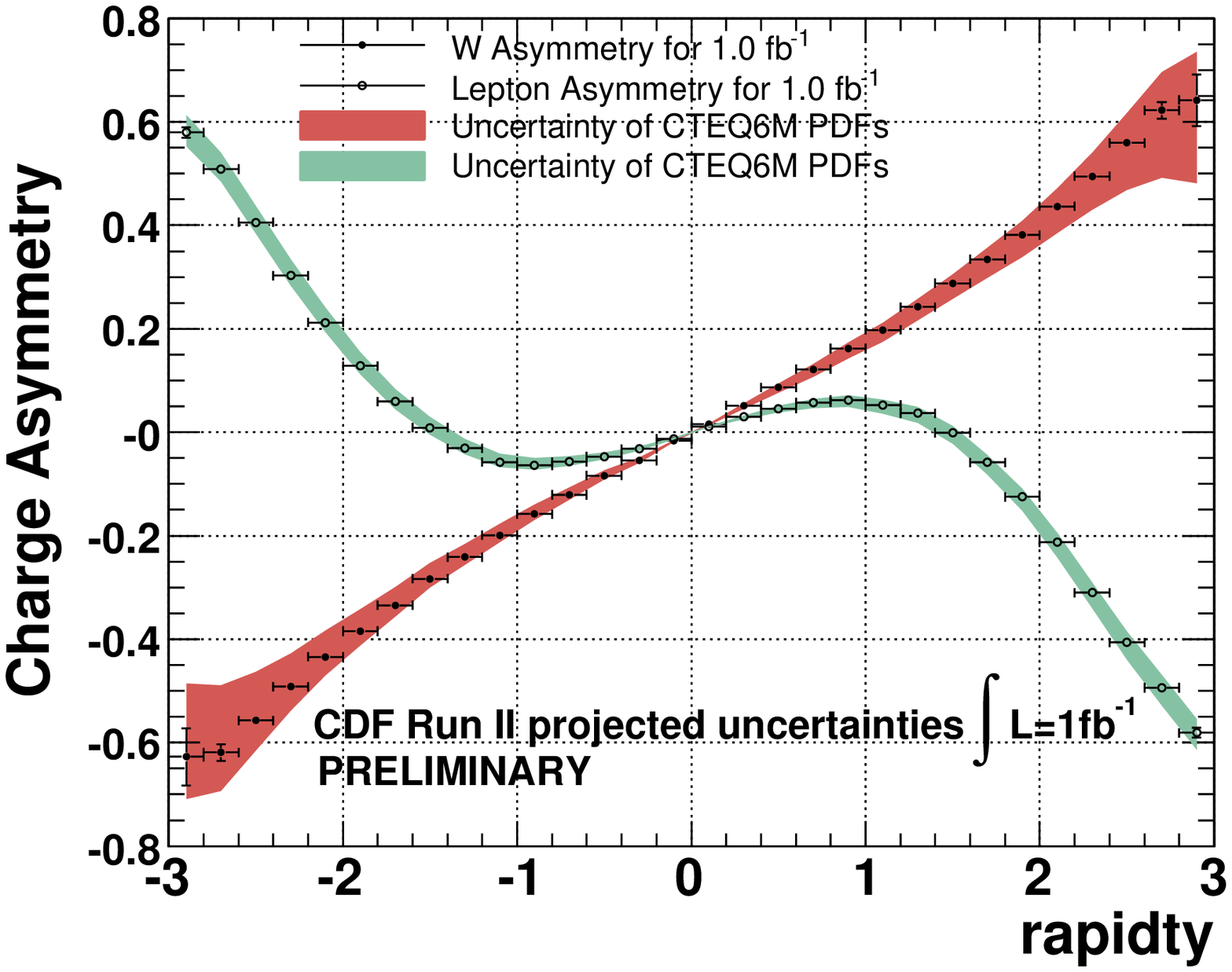}
\end{center}
\caption{ 
  The measured (with ${\cal L}=343 \, {\rm
    pb}^{-1}$)~\cite{James:2007ks,bib:CDF} and projected (${\cal L}=1
  \, {\rm fb}^{-1}$)~\cite{James:2007ks,bib:CDF} CDF $W$ charge
  asymmetry $A(y_W)$ and lepton asymmetry in $p\bar p\to W\to l
  \nu_l$.}\label{fig:th_ewk_asymexp}
\end{figure}

The $W$ boson charge asymmetry ($A(y_W)$) is a sensitive probe of
valence quark PDFs.  Recent theoretical advances include the
calculation of the fully differential cross section at NNLO QCD to
$W/Z$ boson production~\cite{Melnikov:2006kv}, which will help to
further constrain quark PDFs.  In Fig.~\ref{fig:th_ewk_asymexp} we
show the $W$ boson charge asymmetry as measured by CDF with ${\cal
  L}=343 \,{\rm pb}^{-1}$~\cite{James:2007ks,bib:CDF} and a projection
to $1 \,{\rm fb}^{-1}$~\cite{James:2007ks,bib:CDF}.  In
Table~\ref{tab:th_ewk_summary} we provide the combined statistical and
systematic uncertainties for three representative rapidities,
$y_W,y_l=1,1.8,2.6$, of the present D\O\ measurement of the muon
asymmetry with $230\,{\rm pb}^{-1}$~\cite{bib:D0wasymmetry} and the
projected CDF measurement of the $W$ boson and lepton asymmetry with
$1 \,{\rm fb}^{-1}$~\cite{James:2007ks,bib:CDF}. In
Ref.~\cite{Tricoli:2005nx} the PDF uncertainty in a measurement of
$A(y_W)$ at the LHC has been estimated to be $4\%$.  As shown in
Table~\ref{tab:th_ewk_summary}, the impact of different choices of EW
input schemes on $A(y_l)$ is negligible. We expect to observe similar
effects in $A(y_W)$ which, however, needs to be studied in more
detail.

\subsubsection{$W$ boson mass} 
The most precise single $W$ boson mass measurement is presently
provided by CDF~\cite{bib:CDFwmass} (see also Section~5), yielding a
combined CDF and D\O\ measurement of~\cite{lepewwg}
\[M_W= 80.429 \pm 0.039  \, {\rm GeV.}\] 
A Tevatron precision of about 20 MeV is anticipated with $2~{\rm
  fb}^{-1}$.  The extraction of the $W$ boson mass from the
$M_T(l\nu)$ distribution is sensitive to effects that distort the
shape of the distribution around the Jacobian peak.  In
Fig.~\ref{fig:th_ewk_lhc_mtptbest} we observed a distortion of the
$M_T(l\nu)$ and $p_T^l$ distributions when comparing the strictly NLO
results of ${\cal O}(\alpha(0)^3)$ with the result obtained at ${\cal
  O}(\alpha(0) G_\mu^2)$.  Therefore, a more detailed study is
warranted to determine the shift in $M_W$ due to these effects when
using the $M_T(l\nu)$ distribution and ratios of $W$ and $Z$ boson
distributions. In the latter case, they may largely cancel, but this
has to be determined by a careful study.

\begin{table}
\begin{center}
\begin{tabular}{|l|l|l|l|l|l|} \hline
Observable & $\sigma_{W}$ & charge asymmetry & $M_W$ \\ 
& & $y_W,y_l=1; 1.8; 2.6$ &  \\ \hline
\multicolumn{4}{|l|}{{\it experimental precision:} (Section~\ref{subsec:th_ewkexp})} \\ \hline 
Tevatron (now)& 2\% & $A(y_l)$: 0.0078; 0.0484; -  (D\O) & 39 MeV\\ \hline
Tevatron ($1-2~{\rm fb}^{-1}$)& - & $A(y_W)$: 0.0043; 0.0073; 0.030 (CDF) & 20 MeV\\ 
& & $A(y_l)$: 0.0056, 0.0078, 0.076 (CDF) & \\ \hline 
LHC & 3.3\% & - &  15 MeV\\ \hline 
\multicolumn{4}{|l|}{{\it impact of h.o. corrections and theoretical uncertainties:}}\\ \hline
Observable & $\sigma_{W}$ & $A(y_l)$ & $M_W(\mu(e))$ \\ \hline
mPR $(l=\mu)$ (Section~\ref{subsec:th_ewkmulti})& $\lesssim 0.2\%$  & -& 10(2) MeV\\ \hline
EW input scheme/missing h.o. & 1.5\% & $\lesssim 4 \cdot 10^{-5}$ & tbd\\ 
(Sect.~\ref{subsec:th_ewkerror}) & &  &  \\ \hline
$q_T$ broad. (Section~\ref{subsec:ewkth_newfeatures}) &- &  - & 20-50 MeV\\ \hline  
heavy $q$ mass (Section~\ref{subsec:ewkth_heavyquark})&- &  -& $\lesssim 10$ MeV\\ \hline  
nonperturb. (Section~\ref{subsec:ewkth_nonperturbative}) & - & - & $\lesssim 17$ MeV\\ \hline  
\end{tabular}
\caption{Present and anticipated experimental uncertainties of
  $W$ boson observables are compared to effects of higher-order
  corrections, i.e.~beyond ${\cal O}(\alpha)$, as well as theoretical
  uncertainties studied in this report. Details are provided in the
  respective sections.  Experimental uncertainties on $\sigma_W$ do
  not include the $\approx 6$\% luminosity uncertainty.}
\label{tab:th_ewk_summary}
\end{center}
\end{table}

\subsection{Conclusion}
\label{subsec:th_ewkconcl}

In this report we gave an overview of the state-of-the art of
precision calculations for single $W$ production at the Tevatron and
the LHC. We performed a tuned comparison of the Monte Carlo programs
{\sc HORACE, SANC} and {\sc WGRAD2}, taking into account realistic
lepton identification requirements.  As a result of this comparison we
found good numerical agreement of the predictions for the total $W$
production cross section, the $M_T(l\nu)$, $p_T^l$ distributions and
the lepton charge asymmetry.  The effects of higher-order QED
corrections have been studied as well using {\sc HORACE}. To estimate
the residual theoretical uncertainty due to missing higher-order
corrections and different choices of the EW input parameter scheme we
compared the strictly NLO results with predictions that in addition
include leading QCD and EW two-loop corrections and predictions that
use the $G_\mu$ scheme instead of the $\alpha(0)$ scheme.  Moreover,
we discussed important aspects of $q_{T}$ resummation that may affect
significantly the systematic uncertainties in the $M_{W}$ measurement.
Some of our results of these studies of higher-order corrections and
theoretical uncertainties are summarized in
Table~\ref{tab:th_ewk_summary}. When comparing with the anticipated
experimental uncertainties, we conclude that further theoretical
improvements are needed to fully exploit the potential of the LHC for
performing high-precision studies of the electroweak gauge bosons.
Moreover, more detailed studies of the residual uncertainties of
predictions obtained with the available tools are needed, in
particular the impact of these effects on the $W$ mass. For instance,
our study does not include PDF uncertainties, combined QCD and EW
effects, QED/QCD scale uncertainties, and the impact of higher-order
EW Sudakov logarithms.


\clearpage

\newpage
\section{Measurement of the W Mass}
{\bf Contributed by:~C.~Hays}\\

The $m_W$ measurement in $p\bar{p}$ data uses $s$-channel resonant
$W$ bosons with leptonic decays.  The transverse momentum of the decay
$e$ or $\mu$ ($p_T^l$) can be measured with high precision and thus
provides the bulk of the mass information.  Additional information 
comes from the decay $\nu$ transverse momentum ($p_T^{\nu}$), which 
is inferred from the measured energy imbalance in the event.  Since the
lepton energy is well measured, the dominant uncertainty on $p_T^{\nu}$
comes from measuring the hadrons recoiling against the produced
$W$ boson.  Because the $Z$ boson has a similar mass and production mechanism
to the $W$ boson, events with $Z$ bosons can be used to calibrate and model
the detector response to hadronic activity.
\par
The best statistical power for measuring $m_W$ is obtained by combining
$p_T^l$ and $p_T^{\nu}$ into the transverse mass, defined as:

\begin{equation}
m_T = \sqrt{2 p_T^l p_T^{\nu}(1-\cos(\Delta\phi))}.
\label{wmass_eq:mt}
\end{equation}

\noindent
With precise detector calibration, the lepton momentum can be measured to 
a few parts in 10,000.  However, the hadrons resulting from initial-state
radiation are typically measured to a precision of 1\%, degrading the
resolution of the inferred neutrino momentum.  To suppress this
degradation, the transverse hadronic momentum (known as the ``recoil'') is 
required to be less than 15 or 20 GeV.  Alternatively, the lepton
transverse momentum ($p_T$) distribution can be used to measure the $W$ boson 
mass, though this suffers from uncertainties in the theoretical prediction of the
$W$ boson $p_T$, which has not been modelled from first-principles QCD.  In a 
final analysis, the two fits can be combined to utilize the strengths of each.

\subsection{CDF Run 2 Measurement}

The Run 2 $W$ mass measurement proceeds by sequentially calibrating the detector 
response to:
\begin{enumerate}
\item Muon momentum
\item Electron energy
\item Hadronic recoil energy
\end{enumerate}

The muon momentum calibration uses low-mass quarkonia decays to dimuons; the 
electron energy calibration uses the calibrated tracks from $W$ decay electrons;
and the hadronic recoil energy calibration uses the measured recoil in $Z\rightarrow ll$
events.

\subsubsection*{Track Momentum Calibration}

\begin{figure}
  \includegraphics[height=.3\textheight]{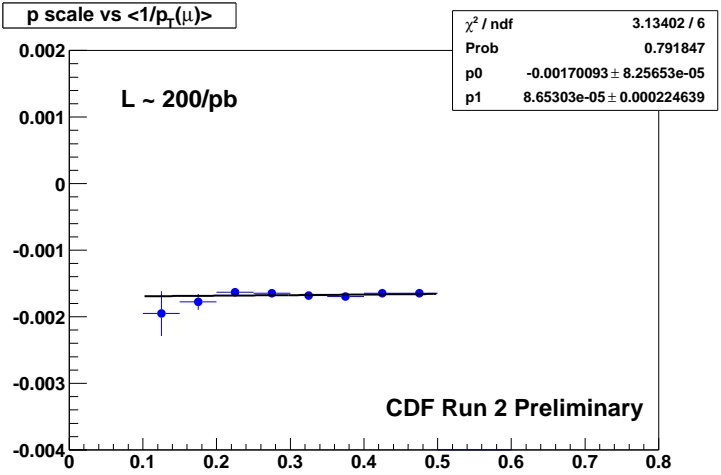}
  \caption{The momentum scale correction as a function of mean inverse muon 
momentum.  The correction is determined by comparing the measured $J/\psi$
mass to that of the PDG. }
  \label{jpsi}
\end{figure}

A charged particle's momentum is measured through its observed curvature
in the tracker.  Since the momentum is inversely proportional to curvature,
the momentum scale is measured as a function of the mean inverse momentum of
$J/\psi$ muons and fit to a line (Fig. \ref{jpsi}).  Improper modelling of 
the muon energy loss in the tracker can lead to a non-zero slope of this line, 
since high-momentum muons lose a smaller fraction of their energy than 
low-momentum muons.  The amount of material contributing to ionization energy 
loss is tuned to make the slope equal to zero.  At CDF, the tuning is a 6\% 
correction to the known material used in {\sc geant} simulation.  To speed up 
event simulation, a material map based on the tracker material is produced and 
used in place of {\sc geant}.  The map contains the material properties 
necessary for electron and muon simulation:  ionization energy loss constants 
from the Bethe-Bloch equation; and radiation lengths.

To improve momentum resolution, muon tracks from $W$ and $Z$ decays use the 
transverse beam position as a point in the track fit.  This constraint is not 
applied to $J/\psi$ decays since they can be separated from the beam line.  Instead, 
$\Upsilon$ decays are used to verify that the beam constraint produces no bias 
on the momentum calibration.  The $\Upsilon$ momentum scale is combined with
that of the $J/\psi$'s to reduce the total uncertainty on the momentum scale.
As a cross-check, the scale is applied to the $Z\rightarrow\mu\mu$ sample and
the extracted $Z$ mass is compared to the LEP measurement of 91.187 GeV.
\par
Aside from the material calibration to model muon energy loss, the simulation 
of multiple Coulomb scattering is necessary to accurately model the 
resolution of low-momentum muons ($< 10$ GeV).  The multiple scattering 
is simulated to have a Gaussian width of:
\begin{equation}
\label{wmass_eq:ms}
\Delta \theta = 13.6~MeV \sqrt{x_0}/p,
\end{equation}

\noindent
where $x_0$ is the fraction of radiation lengths of the detector.  Additional
resolution arises from hard scatters in the tail of the distribution; about
2\% of the scatters have a Gaussian width $\approx$4 times larger than
that of Equation \ref{wmass_eq:ms}.
\par
At high momentum, additional resolution can result from misalignments in the 
drift chamber used for track measurement (the central outer tracker, or COT).  
A detailed alignment procedure based on cosmic rays sets the positions of the 
wires in the COT.  Final curvature corrections, determined using electron 
calorimeter energy and separating electrons from positrons, are applied
to all tracks.  The resulting simulation of the resolution is tested using the
observed width of the $Z\rightarrow \mu\mu$ resonance.  The known hit resolutions
and the transverse beam spot size completely determine the resolution of 
beam-constrained tracks.  Any difference between the observed and simulated $Z$ 
width is removed by tuning the beam spot size in the simulation.
\par
The uncertainties of the momentum scale calibration come from the statistics
and systematics of the $J/\psi$ and $\Upsilon$ samples ($\delta m_W = 16$ MeV), 
and from possible residual misalignments ($\delta m_W = 6$ MeV).

\subsubsection*{Calorimeter Energy Calibration}

\begin{figure}
  \includegraphics[height=.3\textheight]{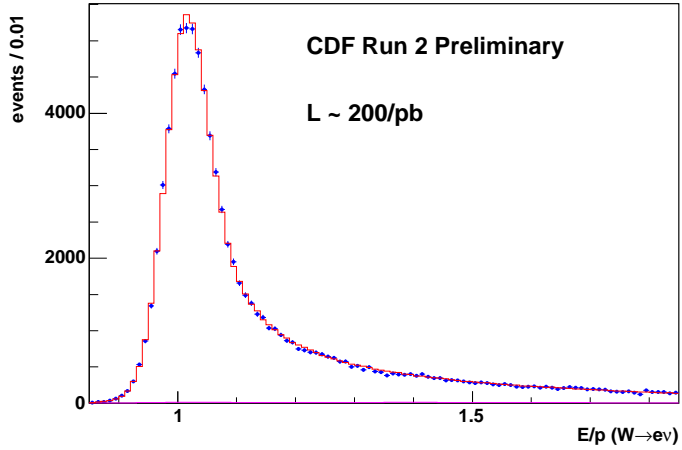}
  \caption{The ratio of calorimeter energy to track momentum for electrons
from $W\rightarrow e\nu$ decays.  The simulated calorimeter energy is scaled
to match the data distribution in the peak. }
  \label{eop}
\end{figure}

Given the momentum calibration, electron tracks from $W$ decays are
used to calibrate the electromagnetic calorimeter.  The simulated 
calorimeter energy is scaled such that the ratio of energy to track
momentum ($E/p$) matches that of the data near the peak (Fig. \ref{eop}).  
This calibration requires a detailed simulation of processes affecting
the shape and position of the peak.  These processes include:  
electron bremsstrahlung and photon conversion in the tracker; 
electron and photon energy loss in the solenoid, which sits inside
of the calorimeter; and electron and photon energy leakage into 
the hadronic section of the calorimeter, which is not used in
the cluster energy measurement. 

\begin{figure}
  \includegraphics[height=.3\textheight]{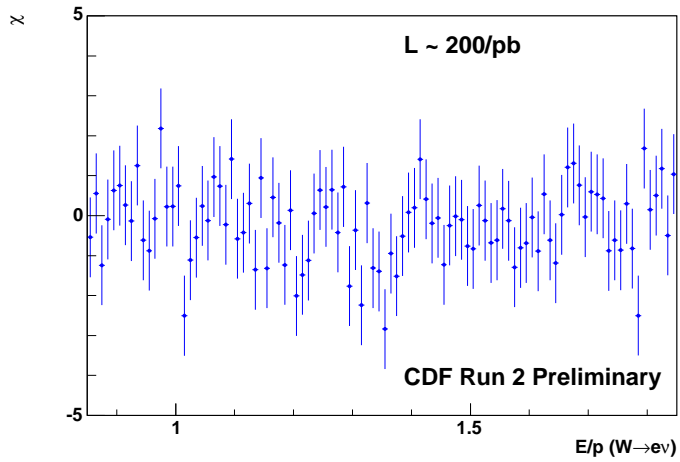}
  \caption{The signed $\chi$ difference between data and simulation for each
bin in the $E/p$ distribution used to extract a calorimeter scale for electrons.}
  \label{eopchi}
\end{figure}

The significant amount of material in the silicon tracker moves the peak
to larger $E/p$ values, since radiated photons enter the calorimeter
cluster but reduce the track momentum.  The material model is tested by the
shape of the $E/p$ distribution at high values, where harder bremsstrahlung
occurs.  Figure \ref{eopchi} shows the difference between simulation and 
data for each 0.01 bin of $E/p$, measured in terms of sigma.  The events in 
the region $1.19 < E/p < 1.85$ are divided into two bins and used to tune
the amount of material contributing to radiation lengths. This tuning can
result in a different correction from the $J/\psi$ material tuning, since
ionization energy loss and radiation lengths scale differently with nuclear 
charge ($Z$).  Thus, to correctly describe both processes \it a priori\rm , 
one would need to know both the amount \it and \rm type of material in the 
tracker.

\begin{figure}
  \includegraphics[height=.3\textheight]{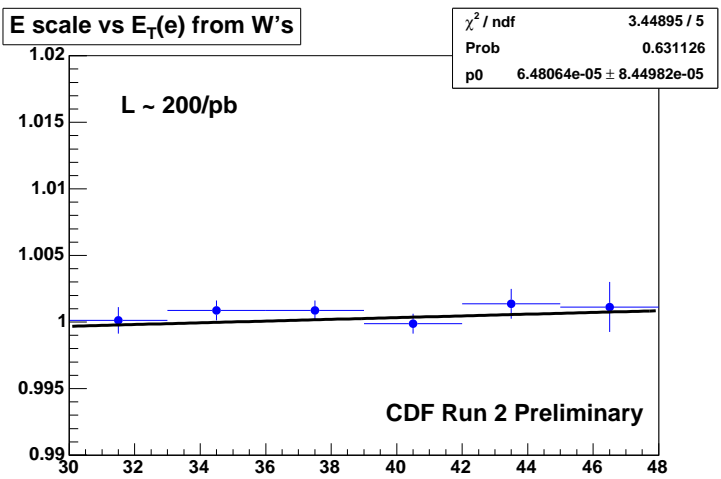}
  \caption{The $E/p$ distribution as a function of $E_T$ for electrons 
from $W\rightarrow e\nu$ decays.  The simulated calorimeter response is
tuned as a function of $E_T$ to produce zero slope for the combined $W$ 
and $Z$ sample.}
  \label{eopet}
\end{figure}

The CDF calorimeter has a non-linear response as a function of particle 
energy.  A non-linearity correction is taken from the $E/p$ distribution
from $W\rightarrow e\nu$ and $Z\rightarrow ee$ decays, separated in bins of 
$E_T$ (Fig. \ref{eopet}).  This correction is applied to each simulated 
electron and photon entering the calorimeter.

The total uncertainty on the calorimeter energy scale arises from 
uncertainties on the material tuning ($\delta m_W = 9$ MeV), 
on the non-linearity correction ($\delta m_W = 23$ MeV), 
on the statistics of the $E/p$ peak ($\delta m_W = 20$ MeV), 
and on the tracker momentum scale ($\delta m_W = 17$ MeV).
This uncertainty is reduced to a total of 30 MeV by incorporating 
the $Z$ boson mass measurement into the calibration.

\subsubsection*{Hadronic Recoil Measurement and Simulation}

The hadronic recoil energy is measured by vectorially summing all the
energy in the calorimeter, excluding that contributed by the lepton.  
Removing the lepton also removes underlying event energy parallel to
the lepton.  The amount of removed energy is estimated using calorimeter
towers separated in $\phi$ from the lepton, and a correction is applied
to the simulation.  
\par
The detector response to the hadronic energy is defined as
$R = u_{meas}/u_{true}$, where $u_{true}$ is the recoil momentum of the
$W$ boson.  The response is measured using $Z\rightarrow ll$ events, 
since leptons are measured more precisely than the hadronic energy.
\par
\begin{figure}
  \includegraphics[height=.3\textheight]{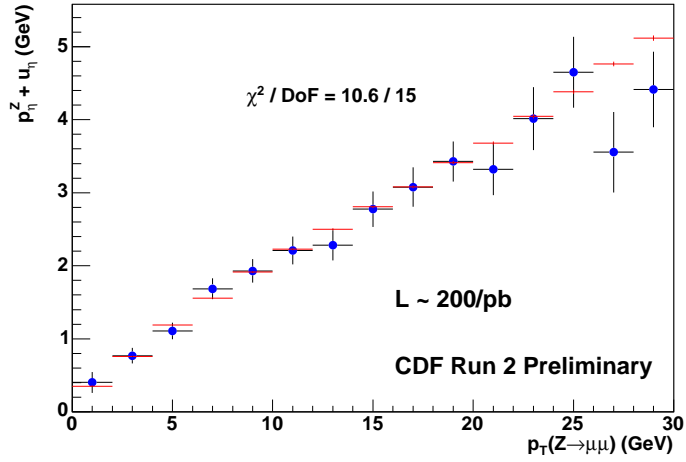}
  \caption{The net momentum along the bisecting axis of the muons in 
$Z\rightarrow \mu\mu$ events, as a function of $Z~p_T$.  The calorimeter 
response to hadronic energy is tuned in the simulation to match the data. }
  \label{response}
\end{figure}
\begin{figure}
  \includegraphics[height=.3\textheight]{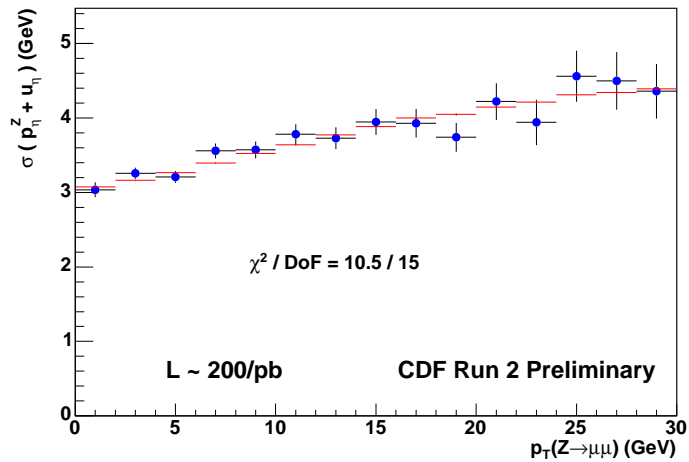}
  \caption{The spread of the net momentum along the bisecting axis of the 
muons in $Z\rightarrow \mu\mu$ events, as a function of $Z~p_T$.  The $\sum E_T$
of the hard interaction is tuned to match the simulation spread to that of
the data. }
  \label{resolution}
\end{figure}

The hadronic energy resolution is modelled as having a component from
the underlying event (independent of recoil) and a component from the
recoiling hadrons.  The model parameters are tuned using the resolution
of $Z\rightarrow ll$ along the axis bisecting the leptons.  This
axis is the least susceptible to fluctuations in lepton energy.
Figures \ref{response} and \ref{resolution} show the response and 
resolution in $Z\rightarrow \mu\mu$ events after tuning the model parameters.
\par
The underlying event resolution component is parametrized in terms of 
$\sum E_T$ in the calorimeter, and incorporated by applying the measured
calorimeter resolution as a function of $\sum E_T$.  The simulated $\sum E_T$ 
distribution contains the hard interaction producing the $W$ or $Z$ plus 
additional interactions at a rate that depends on the instantaneous luminosity.  
The $\sum E_T$ distribution of the additional interactions is taken from an 
inelastic scattering sample.  The hard interaction distribution is extracted 
as a deconvolution of the inelastic scattering $\sum E_T$ distribution.  Since
generic inelastic scatters have a different $Q^2$ momentum transfer than
$W$ and $Z$ events, a tunable scale factor is applied to the $\sum E_T$ of
the hard interaction.  This factor is adjusted to produce the best agreement
between simulation and data of the recoil resolution of $Z$ events.
\par
The uncertainties from the recoil simulation arise from the lepton 
removal ($\delta m_W = 5-8$ MeV), response ($\delta m_W = 9$ MeV), 
and resolution ($\delta m_W = 7$ MeV).

\newpage
\section{Measurement of the $W$ Width}
\label{sec:$W$ Width}
{\bf Contributed by:~J.~Zhu}\\

Width of the $W$ boson is a fundamental parameter in the Standard Model.  
The leptonic partial width for the lepton $l$ can be expressed in terms of
the muon decay constant $G_{\mu}$, the $W$ mass and a small ($<0.5\%$)
radiative correction ($\delta_{SM}$) to the Born-level expression as
$\Gamma(W \rightarrow l~\nu) = G_{\mu}M^{3}_{W}/6\sqrt{2}\pi \times
[1+\delta_{SM}]$ \cite{Rosner:1994ww}. Dividing the partial width
by the leptonic branching ratio $Br(W \rightarrow l~\nu)=3+6 \left [
  1+\alpha_{s}(M_{W})/\pi + O(\alpha^{2}_{s}) \right ]$, gives the SM
prediction for the full decay width $\Gamma_{W}=2.090 \pm 0.008$ GeV
\cite{Hagiwara:2002ww}, where the uncertainty is dominated by the
experimental $M_{W}$ precision. Thus a precise measurement of the W
width can be used to test the SM calculation and probe the physics
beyond SM model since additional particles beyond the SM would
increase the $W$ width. \newline

\begin{figure}
\begin{center}
\includegraphics[scale=0.45]{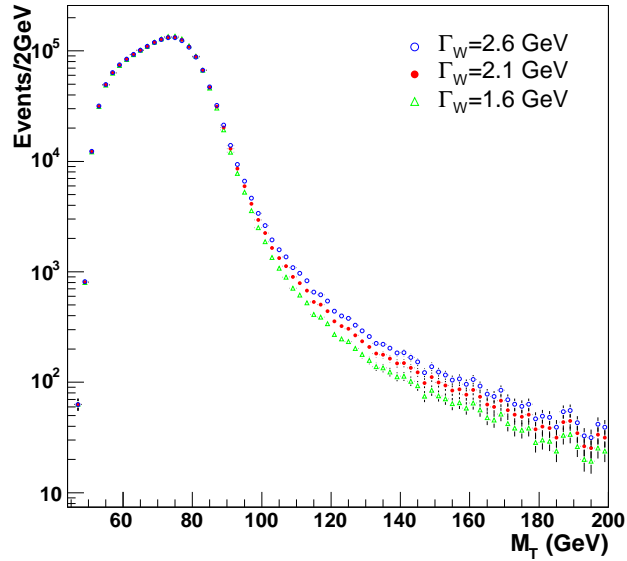}
\end{center}
\caption{\small Transverse mass spectra from Monte Carlo simulation with 
different $W$ widths, normalized to some arbitrary number.
The green triangles are for $\Gamma_{W}=1.6$ GeV, the red dots are 
for $\Gamma_{W}=2.1$ GeV and the blue circles are for $\Gamma_{W}=2.6$ GeV.
}
\label{fig:different_width}
\end{figure}

The $W$ width can be measured indirectly using the ratio of the $W
\rightarrow l\nu$ and $Z \rightarrow ll$ cross sections. 
$\Gamma_{W}$ can also be obtained
directly from a precise determination of the $W$ transverse mass
($M_{T}$) lineshape. Figure \ref{fig:different_width} shows the Monte
Carlo simulated $M_{T}$ spectra for different input $W$ widths. The
$M_{T}$ spectrum has a kinematic upper limit at the value of $M_{W}$,
and events with $M_{T} > M_{W}$ arise due to the combination of the
intrinsic $W$ width and the detector resolution.  In the region
$M_{T}>100$ GeV, the $W$ width component dominates the detector
resolution component. Thus, the transverse mass tail region is
sensitive to $\Gamma_{W}$, and the width can be directly extracted
from a fit to the region $100<M_{T}<200$ GeV. Using this technique,
both CDF and $\dzero$ experiments have published their results using
Run I data \cite{cdf:2000ww} \cite{d0:2002ww},
preliminary Run II result from $\dzero$ has been reported in
\cite{d0:2004ww_prelim}, and the combined result from all Tevatron
direct measurements is $\Gamma_{W} = 2.078 \pm 0.087$ GeV
\cite{width_combined}. \newline

Due to the rapid falling of the Jacobian peak, only a small fraction
of the $W$ events is used in the fitting, and so all previous
measurements are limited by the available statistics. At the LHC,
after all selection cuts about 60 million $W$'s are expected in one
year of data taking at low luminosity (10 fb$^{-1}$)
\cite{gianotti:1999ewk}, the fraction of events in the
fitting region (100 - 200 GeV) is roughly $1\%$, therefore 0.6 million
$W$'s can be used to extract $\Gamma_{W}$. If we scale the statistical
uncertainty with $1/\sqrt{N_{W}}$, the final statistical uncertainty
on the width measurement should be smaller than 5 MeV. $\Gamma_{W}$
measurements from LHC experiments will all be limited by the
systematic uncertainty. \newline

The $W$ width analysis shares most of the issues of $W$ production and
decay modelling and the detector response simulation with the $W$ mass
analysis, the sources of the systematic uncertainty are therefore
similar.  Every input parameter in the MC simulation could alter the
transverse mass lineshape and cause systematic uncertainty on
$\Gamma_{W}$ measurement, these parameters are in most cases
determined by the $Z \rightarrow ll$ data. Although the uncertainties
on these smearing parameters are considered as systematic
uncertainties for the width measurement, they are really statistical
uncertainties which depend on the number of $Z$ events. At LHC, a
large collected $Z \rightarrow ll$ sample ($\sim 6$ million $Z$ events
per channel per experiment) will definitely help to redue the overall
systematic uncertainty.  The systematic uncertainty also depends on
the fitting region, fitting only the high-end region will have a
smaller systematic error since the uncertainties from detector
resolution and SM backgrounds will be smaller. With enough $W$
candidates in the tail region at the LHC, using a smaller fitting
region like $110<M_{T}<200$ GeV or $120<M_{T}<200$ GeV will reduce the
final systematic uncertainty. \newline

The modelling of the $W$ recoil provided the largest uncertaintiy in
all previous width measurements from Tevatron.  The recoil system is
mainly composed of soft hadrons from the underlying event and the
contribution from the pile-up.  $Z \rightarrow ll$ data is used to
measure the detector response and resolution to the underlying event.
For the pile-up contribution, fortunately, the mean number of
interactions per bunch crossing is about 2 at the low luminosity,
which is actually lower than the mean number of interactions per
crossing at the Tevatron Run II. This relatively quiet environment,
together with the large-size $Z$ samples, will reduce the dominant
source of systematic uncertainty. Extrapolating to the LHC data
sample, an error of smaller than 15 MeV per channel should be
achieved. \newline

At the Tevatron Run I, the absolute lepton scale is known with a
precision of about $0.1\%$ and the uncertainty on $\Gamma_{W}$ is
around 20 MeV. If the lepton scale is known to $0.02\%$ at LHC in
order to measure $M_{W}$ with a precision of better than 20 MeV, the
uncertainty due to lepton scale on $\Gamma_{W}$ should be less than 5
MeV. \newline

The leptons from the fitting region tend to have higher transverse
momenta than leptons near the Jacobian edge, thus the lepton scale
non-linearity plays a significant role in the width measurement. The
ability to place bounds on the non-linearity using collider data is a
limiting source of $W$ mass measurement in Run II; 
this is also true for the Run II width measurement. In
$\dzero$ Run Ib measurement, the test beam results were used and the
effect on the width measurement was found to be negligible. At the
LHC, with the help of test beam results, this uncertainty should be on
the order of 5 MeV. \newline

At the Tevatron, the main sources of backgrounds come from QCD
processes, $W \rightarrow \tau \nu$ decays and $Z \rightarrow ll$
decays where one lepton is mismeasured, no new physics processes will
contribute to the tail region of $M_{T}$ spectrum. At the LHC, this
may not be the case, non-SM processes may have large contributions to
the fitting region. It is very difficult to estimate this uncertainty
right now. \newline

For almost all Tevatron measurements, the lepton and recoil
resolutions are parameterized as gaussian functions, the effect on the
non-gaussian part of the detector resolutions was not carefully
estimated. At the LHC, with the extensive studies of the test-beam
results and large collected $Z$ samples, the effect on $\Gamma_{W}$
should be less than 5 MeV. \newline

The theoretical uncertainties on the width measurement mostly come
from $p_{T}(W)$ spectrum (due to QCD corrections), PDF and radiative
corrections. Currently, the estimated uncertainty on $\Gamma_{W}$
associated with modelling the $W$ boson $p_{T}$ spectrum is of the
order of 30 MeV at the Tevatron. In \cite{keller:1998wmass}, 
the authors show
that larger QCD corrections are expected at the LHC, and hence the
uncertainty will also be larger. On the other hand, as mentioned
before, the $W$ boson $p_{T}$ spectrum can be constrained by $Z
\rightarrow ll$ data. With 0.6 million $Z$ events, the uncertainty due
to QCD corrections should be controlled to 10 MeV level. Since the
boson $p_{T}$ form is constrained from $Z$ events, it is imperative
that all effects that are different for $Z$ and $W$ are included in
the generator prescription.  The uncertainties from PDF and radiative
corrections seem under control for all Tevatron measurements ($\sim$
10 MeV for each), but will need improvements to avoid becoming
dominant at the LHC experiments. \newline

The $W$ mass will be measured with a precision of about 30 MeV from
the LEP and Tevatron measurements before this measurement
\cite{keller:1998wmass}, the uncertainty of $M_{W}$ on $\Gamma_{W}$ should be
less than 5 MeV. In this high-precision measurement, assuming SM
$M_{W}$-$\Gamma_{W}$ relation may not be enough. \newline
 
With an integrated luminosity of 10 fb$^{-1}$, which should be
collected in the first year of LHC's low luminosity run and by
considering only one lepton decay channel, a total uncertainty of
smaller than 30 MeV should be achieved by each LHC experiment.

\section{Summary}

This report includes detailed descriptions of experimental methods
used to measure the $W$ boson mass, search for single top production,
and precision electroweak measurements at hadron colliders.  In
addition, it includes numerous new theoretical developments in the
areas of single top production and precision electroweak measurements.
The main conclusions are summarized below. Details of the studies are
found in the respective sections of the report and references cited.

Impressive advances have been made to control the systematic
uncertainties arising from jet energy calibration in the measurement
of the top quark mass, and work is in progress to control the
systematic uncertainties arising from b-jets. Tevatron experience has
shown that the measurement can be significantly improved by combining
the results from the two experiments, D0 and CDF. It is therefore
important to agree on how to classify and apply the uncertainties to
allow for a more straightforward combination. A quantitative study of
the effects of Color reconnections and other final state interactions
is needed to reduce the uncertainties arising from Monte Carlo
generation.

Tevatron experiments are using elaborated multi-variate analysis
techniques to extract the single top quark signal from the
overwhelming $W$+ jets backgrounds.  Recently, D0 announced that it
observes evidence for single top production when it analyzes about
twice the amount of data compared to the one used in the analyses
described in this report. We expect the LHC single top samples to have
much larger event statistics, especially in the $t$-channel, which
should allow the signal to be extracted using a cut-based analysis.
The advanced analysis techniques developed at the Tevatron for the
single top searches will be particularly useful for Higgs and beyond
the Standard Model searches at the LHC.
  
The $W$ mass measurements rely on a detailed calibration of the
detector that will be more difficult to achieve at the LHC compared to
the Tevatron.  Recently, the CDF collaboration completed the most
precise single measurement of the $W$ mass available to-date, to a
stunning precision of 0.06\%, following analysis techniques described in
this report. Together, the precise measurements of the $W$ boson mass
and the top quark mass are constraining the mass of the Higgs boson.






\bibliographystyle{tev4lhc}
\bibliography{topew-main}

\hyphenation{Post-Script Sprin-ger}
\begin{thebibliography}{100}

\bibitem{radiative_corr}
A.~Sirlin,
\newblock Phys. Rev. {\bf D22}, 971 (1980).

\bibitem{tevatron}
{The Tevatron Electroweak Working Group},
\newblock hep-ex/0507091.

\bibitem{Heinemeyer:2004gx}
S.~Heinemeyer, W.~Hollik and G.~Weiglein,
\newblock Phys. Rept. {\bf 425}, 265 (2006), [hep-ph/0412214].

\bibitem{Kondo:1988yd}
K.~Kondo,
\newblock J. Phys. Soc. Jap. {\bf 57}, 4126 (1988).

\bibitem{Abazov:2004cs}
D\O\ Collaboration, V.~M. Abazov {\em et~al.},
\newblock Nature {\bf 429}, 638 (2004), [hep-ex/0406031].

\bibitem{bib:CDFme}
CDF Collaboration, F.~Canelli, B.~Mohr, R.~Wallny and J.~Hauser,
\newblock (2006),
\newblock CDF Note 8151.

\bibitem{bib:D0me}
D\O\ Collaboration, F.~Fiedler {\em et~al.},
\newblock (2006),
\newblock { D\O{} Note 5053}.

\bibitem{Abulencia:2005uq}
CDF Collaboration, A.~Abulencia {\em et~al.},
\newblock Phys. Rev. Lett. {\bf 96}, 152002 (2006), [hep-ex/0512070].

\bibitem{bib:CDFmedil}
{CDF Collaboration},
\newblock (2006),
\newblock {CDF Note 8090}.

\bibitem{Abulencia:2005ak}
CDF Collaboration, A.~Abulencia {\em et~al.},
\newblock Phys. Rev. Lett. {\bf 96}, 022004 (2006), [hep-ex/0510049].

\bibitem{Eur.Phys.J.C2.581}
DELPHI Collaboration, P.~Abreu {\em et~al.},
\newblock Eur. Phys. J. {\bf C2}, 581 (1998).

\bibitem{Phys.Lett.B462.410}
DELPHI Collaboration, P.~Abreu {\em et~al.},
\newblock Phys. Lett. {\bf B462}, 410 (1999).

\bibitem{Phys.Lett.B511.159}
DELPHI Collaboration, P.~Abreu {\em et~al.},
\newblock Phys. Lett. {\bf B511}, 159 (2001), [hep-ex/0104047].

\bibitem{Mulders:2001aa}
DELPHI Collaboration, M.~Mulders,
\newblock Int. J. Mod. Phys. {\bf A16S1A}, 284 (2001).

\bibitem{Mulders:2001mp}
M.~P. Mulders,
\newblock {\em Direct measurement of the W boson mass in e+ e- collisions at
  LEP},
\newblock PhD thesis, University of Amsterdam, 2001.

\bibitem{run1prdmass}
D\O\ Collaboration, B.~Abbott {\em et~al.},
\newblock Phys. Rev. {\bf D58}, 052001 (1998), [hep-ex/9801025].

\bibitem{Pumplin:2002vw}
J.~Pumplin {\em et~al.},
\newblock {JHEP} {\bf 0207}, 012 (2002), [hep-ph/0201195].

\bibitem{bib:Zbible}
{LEP Electroweak Working Group, SLD Electroweak Group, and SLD Heavy Flavour
  Group},
\newblock hep-ex/0509008.

\bibitem{Abulencia:2005aj}
CDF Collaboration, A.~Abulencia {\em et~al.},
\newblock Phys. Rev. {\bf D73}, 032003 (2006), [hep-ex/0510048].

\bibitem{Bhatti:2005ai}
A.~Bhatti {\em et~al.},
\newblock hep-ex/0510047.

\bibitem{Abbott:1998xw}
D\O\ Collaboration, B.~Abbott {\em et~al.},
\newblock Nucl. Instrum. Meth. {\bf A424}, 352 (1999), [hep-ex/9805009].

\bibitem{MtSum05}
{CDF and D\O{} collaborations and TEVEWWG},
\newblock hep-ex/0507006.

\bibitem{Lyons:1988rp}
L.~Lyons, D.~Gibaut and P.~Clifford,
\newblock Nucl. Instrum. Meth. {\bf A270}, 110 (1988).

\bibitem{Valassi:2003mu}
A.~Valassi,
\newblock Nucl. Instrum. Meth. {\bf A500}, 391 (2003).

\bibitem{CdfLjR2}
CDF Collaboration, A.~Abulencia {\em et~al.},
\newblock Phys. Rev. Lett. {\bf 96}, 022004 (2006), [hep-ex/0510049].

\bibitem{CdfDlR2}
CDF Collaboration, A.~Abulencia {\em et~al.},
\newblock Phys. Rev. Lett. {\bf 96}, 152002 (2006), [hep-ex/0512070].

\bibitem{CDFTDR}
R.\ Blair {\it{et al.}}, (CDF Collaboration), {\it{The CDF-II Detector
  Technical Design Report}}, Fermilab- Pub-96/390-E (1996).

\bibitem{Snowmass}
{Snowmass Working Group on Precision Electroweak measurements},
\newblock hep-ph/0202001.

\bibitem{PhDRoy}
P.~Roy,
\newblock PCCFT0202,
\newblock PhD Thesis, Blaise Pascal University.

\bibitem{Beneke:2000hk}
M.~Beneke {\em et~al.},
\newblock (2000), [hep-ph/0003033].

\bibitem{etienvre}
J.-P. Etienvre, A.-I.~Meyer and J.~Schwindling,
\newblock ATLAS Internal note: ATL-PHYS-INT-2005-002.

\bibitem{top_atlas}
I.~Borjanovic {\em et~al.},
\newblock Eur.Phys.J {\bf C39S2}, 63 (2005), [hep-ex/0403021].

\bibitem{top_cms}
L.~Sonnenschein,
\newblock CMS Internal note~: CMS-2001-001.

\bibitem{top_cms2}
J.~D'Hondt, J.~Heyninck and S.~Lowette,
\newblock CMS Note 2006/066.

\bibitem{top_pt_atlas}
I.~Efthymiopoulos,
\newblock ATLAS Internal note~: ATL-COM-PHYS-1999-050.

\bibitem{top_jpsi}
A.~Kharchilava,
\newblock Phys. Lett. {\bf B476}, 73 (2000).

\bibitem{top_jpsi2}
R.~Chierici and A.~Dierlamm,
\newblock CMS Note 2006/058.

\bibitem{top_cms3}
M.~Davids {\em et~al.},
\newblock CMS Note 2006/077.

\bibitem{Abe:1995hr}
CDF Collaboration, F.~Abe {\em et~al.},
\newblock Phys. Rev. Lett. {\bf 74}, 2626 (1995), [hep-ex/9503002].

\bibitem{Abachi:1995iq}
D\O\ Collaboration, S.~Abachi {\em et~al.},
\newblock Phys. Rev. Lett. {\bf 74}, 2632 (1995), [hep-ex/9503003].

\bibitem{Willenbrock:1986cr}
S.~S.~D. Willenbrock and D.~A. Dicus,
\newblock Phys. Rev. {\bf D34}, 155 (1986).

\bibitem{Yuan:1989tc}
C.-P. Yuan,
\newblock Phys. Rev. {\bf D41}, 42 (1990).

\bibitem{Ellis:1992yw}
R.~K. Ellis and S.~J. Parke,
\newblock Phys. Rev. {\bf D46}, 3785 (1992).

\bibitem{Cortese:1991fw}
S.~Cortese and R.~Petronzio,
\newblock Phys. Lett. {\bf B253}, 494 (1991).

\bibitem{Stelzer:1995mi}
T.~Stelzer and S.~Willenbrock,
\newblock Phys. Lett. {\bf B357}, 125 (1995), [hep-ph/9505433].

\bibitem{Heinson:1996zm}
A.~P. Heinson, A.~S. Belyaev and E.~E. Boos,
\newblock Phys. Rev. {\bf D56}, 3114 (1997), [hep-ph/9612424].

\bibitem{Tait:1999cf}
T.~M.~P. Tait,
\newblock Phys. Rev. {\bf D61}, 034001 (2000), [hep-ph/9909352].

\bibitem{Smith:1996ij}
M.~C. Smith and S.~Willenbrock,
\newblock Phys. Rev. {\bf D54}, 6696 (1996), [hep-ph/9604223].

\bibitem{Harris:2002md}
B.~W. Harris, E.~Laenen, L.~Phaf, Z.~Sullivan and S.~Weinzierl,
\newblock Phys. Rev. {\bf D66}, 054024 (2002), [hep-ph/0207055].

\bibitem{Cao:2004ky}
Q.-H. Cao and C.~P. Yuan,
\newblock Phys. Rev. {\bf D71}, 054022 (2005), [hep-ph/0408180].

\bibitem{Cao:2004ap}
Q.-H. Cao, R.~Schwienhorst and C.~P. Yuan,
\newblock Phys. Rev. {\bf D71}, 054023 (2005), [hep-ph/0409040].

\bibitem{Sullivan:2004ie}
Z.~Sullivan,
\newblock Phys. Rev. {\bf D70}, 114012 (2004), [hep-ph/0408049].

\bibitem{Campbell:2004ch}
J.~Campbell, R.~K. Ellis and F.~Tramontano,
\newblock Phys. Rev. {\bf D70}, 094012 (2004), [hep-ph/0408158].

\bibitem{Frixione:2005vw}
S.~Frixione, E.~Laenen, P.~Motylinski and B.~R. Webber,
\newblock hep-ph/0512250.

\bibitem{Chetyrkin:2000mq}
K.~G. Chetyrkin and M.~Steinhauser,
\newblock Phys. Lett. {\bf B502}, 104 (2001), [hep-ph/0012002].

\bibitem{Bordes:1994ki}
G.~Bordes and B.~van Eijk,
\newblock Nucl. Phys. {\bf B435}, 23 (1995).

\bibitem{Stelzer:1997ns}
T.~Stelzer, Z.~Sullivan and S.~Willenbrock,
\newblock Phys. Rev. {\bf D56}, 5919 (1997), [hep-ph/9705398].

\bibitem{Cao:2005pq}
Q.-H. Cao, R.~Schwienhorst, J.~A. Benitez, R.~Brock and C.-P. Yuan,
\newblock Phys. Rev. {\bf D72}, 094027 (2005), [hep-ph/0504230].

\bibitem{Zhu:2002uj}
S.~Zhu,
\newblock Phys. Lett. {\bf B524}, 283 (2002).

\bibitem{Campbell:2005bb}
J.~Campbell and F.~Tramontano,
\newblock Nucl. Phys. {\bf B726}, 109 (2005), [hep-ph/0506289].

\bibitem{Belyaev:2000me}
A.~Belyaev and E.~Boos,
\newblock Phys. Rev. {\bf D63}, 034012 (2001), [hep-ph/0003260].

\bibitem{Stelzer:1998ni}
T.~Stelzer, Z.~Sullivan and S.~Willenbrock,
\newblock Phys. Rev. {\bf D58}, 094021 (1998), [hep-ph/9807340].

\bibitem{Belyaev:1998dn}
A.~S. Belyaev, E.~E. Boos and L.~V. Dudko,
\newblock Phys. Rev. {\bf D59}, 075001 (1999), [hep-ph/9806332].

\bibitem{Sullivan:2005ar}
Z.~Sullivan,
\newblock Phys. Rev. {\bf D72}, 094034 (2005), [hep-ph/0510224].

\bibitem{Bowen:2004my}
M.~T. Bowen, S.~D. Ellis and M.~J. Strassler,
\newblock Phys. Rev. {\bf D72}, 074016 (2005), [hep-ph/0412223].

\bibitem{Bowen:2005xq}
M.~T. Bowen,
\newblock hep-ph/0503110.

\bibitem{Cacciari:2003fi}
M.~Cacciari, S.~Frixione, M.~L. Mangano, P.~Nason and G.~Ridolfi,
\newblock JHEP {\bf 04}, 068 (2004), [hep-ph/0303085].

\bibitem{Acosta:2005hr}
CDF Collaboration, D.~Acosta {\em et~al.},
\newblock Phys. Rev. Lett. {\bf 95}, 102002 (2005), [hep-ex/0505091].

\bibitem{Eidelman:2004wy}
Particle Data Group, S.~Eidelman {\em et~al.},
\newblock Phys. Lett. {\bf B592}, 1 (2004).

\bibitem{Ladinsky:1990ut}
G.~A. Ladinsky and C.~P. Yuan,
\newblock Phys. Rev. {\bf D43}, 789 (1991).

\bibitem{Stange:1994bb}
A.~Stange, W.~J. Marciano and S.~Willenbrock,
\newblock Phys. Rev. {\bf D50}, 4491 (1994), [hep-ph/9404247].

\bibitem{Moretti:1997ng}
S.~Moretti,
\newblock Phys. Rev. {\bf D56}, 7427 (1997), [hep-ph/9705388].

\bibitem{Carlson:1993dt}
D.~O. Carlson and C.~P. Yuan,
\newblock Phys. Lett. {\bf B306}, 386 (1993).

\bibitem{Mahlon:1996pn}
G.~Mahlon and S.~J. Parke,
\newblock Phys. Rev. {\bf D55}, 7249 (1997), [hep-ph/9611367].

\bibitem{Mahlon:1999gz}
G.~Mahlon and S.~J. Parke,
\newblock Phys. Lett. {\bf B476}, 323 (2000), [hep-ph/9912458].

\bibitem{Carlson:1994bg}
D.~O. Carlson, E.~Malkawi and C.~P. Yuan,
\newblock Phys. Lett. {\bf B337}, 145 (1994), [hep-ph/9405277].

\bibitem{Datta:1996gg}
A.~Datta and X.~Zhang,
\newblock Phys. Rev. {\bf D55}, 2530 (1997), [hep-ph/9611247].

\bibitem{Tait:1997fe}
T.~Tait and C.~P. Yuan,
\newblock hep-ph/9710372.

\bibitem{Hikasa:1998wx}
K.-I. Hikasa, K.~Whisnant, J.~M. Yang and B.-L. Young,
\newblock Phys. Rev. {\bf D58}, 114003 (1998), [hep-ph/9806401].

\bibitem{Boos:1999dd}
E.~Boos, L.~Dudko and T.~Ohl,
\newblock Eur. Phys. J. {\bf C11}, 473 (1999), [hep-ph/9903215].

\bibitem{Tait:2000sh}
T.~Tait and C.-P. Yuan,
\newblock Phys. Rev. {\bf D 63}, 014018 (2001).

\bibitem{Chen:2005vr}
C.-R. Chen, F.~Larios and C.~P. Yuan,
\newblock Phys. Lett. {\bf B631}, 126 (2005), [hep-ph/0503040].

\bibitem{Atwood:1996pd}
D.~Atwood, S.~Bar-Shalom, G.~Eilam and A.~Soni,
\newblock Phys. Rev. {\bf D54}, 5412 (1996), [hep-ph/9605345].

\bibitem{Simmons:1996ws}
E.~H. Simmons,
\newblock Phys. Rev. {\bf D55}, 5494 (1997), [hep-ph/9612402].

\bibitem{Li:1996bh}
C.~S. Li, R.~J. Oakes and J.~M. Yang,
\newblock Phys. Rev. {\bf D55}, 1672 (1997), [hep-ph/9608460].

\bibitem{Li:1996ir}
C.~S. Li, R.~J. Oakes and J.~M. Yang,
\newblock Phys. Rev. {\bf D55}, 5780 (1997), [hep-ph/9611455].

\bibitem{Li:1997qf}
C.-S. Li, R.~J. Oakes, J.-M. Yang and H.-Y. Zhou,
\newblock Phys. Rev. {\bf D57}, 2009 (1998), [hep-ph/9706412].

\bibitem{Bar-Shalom:1997si}
S.~Bar-Shalom, D.~Atwood and A.~Soni,
\newblock Phys. Rev. {\bf D57}, 1495 (1998), [hep-ph/9708357].

\bibitem{Malkawi:1995dm}
E.~Malkawi and T.~Tait,
\newblock Phys. Rev. {\bf D54}, 5758 (1996), [hep-ph/9511337].

\bibitem{Datta:1997us}
A.~Datta, J.~M. Yang, B.-L. Young and X.~Zhang,
\newblock Phys. Rev. {\bf D56}, 3107 (1997), [hep-ph/9704257].

\bibitem{Oakes:1997zg}
R.~J. Oakes, K.~Whisnant, J.~M. Yang, B.-L. Young and X.~Zhang,
\newblock Phys. Rev. {\bf D57}, 534 (1998), [hep-ph/9707477].

\bibitem{Han:1998tp}
T.~Han, M.~Hosch, K.~Whisnant, B.-L. Young and X.~Zhang,
\newblock Phys. Rev. {\bf D58}, 073008 (1998), [hep-ph/9806486].

\bibitem{Datta:2000gm}
A.~Datta, P.~J. O'Donnell, Z.~H. Lin, X.~Zhang and T.~Huang,
\newblock Phys. Lett. {\bf B483}, 203 (2000), [hep-ph/0001059].

\bibitem{Chivukula:1995gu}
R.~S. Chivukula, E.~H. Simmons and J.~Terning,
\newblock Phys. Rev. {\bf D53}, 5258 (1996), [hep-ph/9506427].

\bibitem{Muller:1996dj}
D.~J. Muller and S.~Nandi,
\newblock Phys. Lett. {\bf B383}, 345 (1996), [hep-ph/9602390].

\bibitem{Malkawi:1996fs}
E.~Malkawi, T.~Tait and C.~P. Yuan,
\newblock Phys. Lett. {\bf B385}, 304 (1996), [hep-ph/9603349].

\bibitem{He:1999vp}
H.-J. He, T.~Tait and C.~P. Yuan,
\newblock Phys. Rev. {\bf D62}, 011702 (2000), [hep-ph/9911266].

\bibitem{Batra:2003nj}
P.~Batra, A.~Delgado, D.~E. Kaplan and T.~M.~P. Tait,
\newblock JHEP {\bf 02}, 043 (2004), [hep-ph/0309149].

\bibitem{Batra:2004vc}
P.~Batra, A.~Delgado, D.~E. Kaplan and T.~M.~P. Tait,
\newblock JHEP {\bf 06}, 032 (2004), [hep-ph/0404251].

\bibitem{Sullivan:2003xy}
Z.~Sullivan,
\newblock hep-ph/0306266.

\bibitem{Sullivan:2002jt}
Z.~Sullivan,
\newblock Phys. Rev. {\bf D66}, 075011 (2002), [hep-ph/0207290].

\bibitem{Giele:1991vf}
W.~T. Giele and E.~W.~N. Glover,
\newblock Phys. Rev. {\bf D46}, 1980 (1992).

\bibitem{Giele:1993dj}
W.~T. Giele, E.~W.~N. Glover and D.~A. Kosower,
\newblock Nucl. Phys. {\bf B403}, 633 (1993), [hep-ph/9302225].

\bibitem{Keller:1998tf}
S.~Keller and E.~Laenen,
\newblock Phys. Rev. {\bf D59}, 114004 (1999), [hep-ph/9812415].

\bibitem{Alitti:1990aa}
UA2, J.~Alitti {\em et~al.},
\newblock Phys. Lett. {\bf B257}, 232 (1991).

\bibitem{Mahlon:1995zn}
G.~Mahlon and S.~Parke,
\newblock Phys. Rev. {\bf D 53}, 4886 (1996).

\bibitem{Parke:1996pr}
S.~J. Parke and Y.~Shadmi,
\newblock Phys. Lett. {\bf B387}, 199 (1996), [hep-ph/9606419].

\bibitem{Giele:1995kr}
W.~T. Giele, S.~Keller and E.~Laenen,
\newblock Phys. Lett. {\bf B372}, 141 (1996), [hep-ph/9511449].

\bibitem{Fadin:1993kt}
V.~S. Fadin, V.~A. Khoze and A.~D. Martin,
\newblock Phys. Lett. {\bf B320}, 141 (1994), [hep-ph/9309234].

\bibitem{Fadin:1993dz}
V.~S. Fadin, V.~A. Khoze and A.~D. Martin,
\newblock Phys. Rev. {\bf D49}, 2247 (1994).

\bibitem{Melnikov:1993np}
K.~Melnikov and O.~I. Yakovlev,
\newblock Phys. Lett. {\bf B324}, 217 (1994), [hep-ph/9302311].

\bibitem{Pittau:1996rp}
R.~Pittau,
\newblock Phys. Lett. {\bf B386}, 397 (1996), [hep-ph/9603265].

\bibitem{Macesanu:2001bj}
C.~Macesanu,
\newblock Phys. Rev. {\bf D65}, 074036 (2002), [hep-ph/0112142].

\bibitem{Ellis:1980wv}
R.~K. Ellis, D.~A. Ross and A.~E. Terrano,
\newblock Nucl. Phys. {\bf B178}, 421 (1981).

\bibitem{Catani:1996jh}
S.~Catani and M.~H. Seymour,
\newblock Phys. Lett. {\bf B378}, 287 (1996), [hep-ph/9602277].

\bibitem{Catani:1996vz}
S.~Catani and M.~H. Seymour,
\newblock Nucl. Phys. {\bf B485}, 291 (1997), [hep-ph/9605323].

\bibitem{Catani:2002hc}
S.~Catani, S.~Dittmaier, M.~H. Seymour and Z.~Trocsanyi,
\newblock Nucl. Phys. {\bf B627}, 189 (2002), [hep-ph/0201036].

\bibitem{Nagy:1998bb}
Z.~Nagy and Z.~Trocsanyi,
\newblock Phys. Rev. {\bf D59}, 014020 (1999), [hep-ph/9806317].

\bibitem{Nagy:2003tz}
Z.~Nagy,
\newblock Phys. Rev. {\bf D68}, 094002 (2003), [hep-ph/0307268].

\bibitem{Boos:2003yi}
E.~Boos and T.~Plehn,
\newblock Phys. Rev. {\bf D69}, 094005 (2004), [hep-ph/0304034].

\bibitem{Dittmar:1996ss}
M.~Dittmar and H.~K. Dreiner,
\newblock Phys. Rev. {\bf D55}, 167 (1997), [hep-ph/9608317].

\bibitem{Acosta:2001un}
CDF Collaboration, D.~Acosta {\em et~al.},
\newblock Phys. Rev. {\bf D 65}, 091102 (2002).

\bibitem{Acosta:2004un}
CDF Collaboration, D.~Acosta {\em et~al.},
\newblock Phys. Rev. {\bf D 69}, 052003 (2004).

\bibitem{boos2000}
E.~Boos, L.~Dudko and V.~Savrin,
\newblock (2000).

\bibitem{RunII:cdf_result}
CDF Collaboration, D.~Acosta {\em et~al.},
\newblock Phys. Rev. {\bf D 71}, 012005 (2005).

\bibitem{Abazov:2005zz}
D\O\ Collaboration, V.~M. Abazov {\em et~al.},
\newblock Phys. Lett. {\bf B622}, 265 (2005), [hep-ex/0505063].

\bibitem{Stelzer:1994ta}
T.~Stelzer and W.~F. Long,
\newblock Comput. Phys. Commun. {\bf 81}, 357 (1994), [hep-ph/9401258].

\bibitem{Maltoni:2002qb}
F.~Maltoni and T.~Stelzer,
\newblock JHEP {\bf 02}, 027 (2003), [hep-ph/0208156].

\bibitem{Boos:2004kh}
CompHEP, E.~Boos {\em et~al.},
\newblock Nucl. Instrum. Meth. {\bf A534}, 250 (2004), [hep-ph/0403113].

\bibitem{Lueck:2006hz}
J.~Lueck,
\newblock FERMILAB-MASTERS-2006-01; available as Universit{\"a}t Karlsruhe,
  Institut fuer Experimentelle Kernphysik Report No. IEKP-KA/2006-2, Diploma
  Thesis.

\bibitem{Sjostrand:2003wg}
T.~Sjostrand, L.~Lonnblad, S.~Mrenna and P.~Skands,
\newblock hep-ph/0308153  (2003), [hep-ph/0308153].

\bibitem{Boos:cmsnote}
E.~Boos, L.~Dudko and V.~Savrin,
\newblock CMS Note 2000/065  (2000), [CMS Note 2000/065].

\bibitem{Boos:SingleTop}
E.~E. Boos, V.~E. Bunichev, L.~V. Dudko, V.~I. Savrin and A.~V. Sherstnev,
\newblock Phys. Atom. Nucl. {\bf 69}, 1317 (2006).

\bibitem{Tung:2004md}
W.-K. Tung,
\newblock AIP Conf. Proc. {\bf 753}, 15 (2005), [hep-ph/0410139].

\bibitem{Sjostrand:2000wi}
T.~Sjostrand {\em et~al.},
\newblock Comput. Phys. Commun. {\bf 135}, 238 (2001), [hep-ph/0010017].

\bibitem{Marchesini:1991ch}
G.~Marchesini {\em et~al.},
\newblock Comput. Phys. Commun. {\bf 67}, 465 (1992).

\bibitem{Campbell:2002tg}
J.~Campbell and R.~K. Ellis,
\newblock Phys. Rev. {\bf D65}, 113007 (2002), [hep-ph/0202176].

\bibitem{Lai:1999wy}
CTEQ, H.~L. Lai {\em et~al.},
\newblock Eur. Phys. J. {\bf C12}, 375 (2000), [hep-ph/9903282].

\bibitem{Baur:1993zd}
U.~Baur, F.~Halzen, S.~Keller, M.~L. Mangano and K.~Riesselmann,
\newblock Phys. Lett. {\bf B318}, 544 (1993), [hep-ph/9308370].

\bibitem{Acosta:2005zd}
CDF Collaboration, D.~Acosta {\em et~al.},
\newblock Phys. Rev. {\bf D72}, 032002 (2005), [hep-ex/0506001].

\bibitem{Abazov:2004zd}
D\O\ Collaboration, V.~M. Abazov {\em et~al.},
\newblock Phys. Rev. Lett. {\bf 94}, 161801 (2005), [hep-ex/0410078].

\bibitem{Campbell:2003dd}
J.~Campbell, R.~K. Ellis, F.~Maltoni and S.~Willenbrock,
\newblock Phys. Rev. {\bf D69}, 074021 (2004), [hep-ph/0312024].

\bibitem{Aktas:2005iw}
H1 Collaboration, A.~Aktas {\em et~al.},
\newblock Eur. Phys. J. {\bf C45}, 23 (2006), [hep-ex/0507081].

\bibitem{Aktas:2004az}
H1 Collaboration, A.~Aktas {\em et~al.},
\newblock Eur. Phys. J. {\bf C40}, 349 (2005), [hep-ex/0411046].

\bibitem{Breitweg:1999ad}
ZEUS Collaboration, J.~Breitweg {\em et~al.},
\newblock Eur. Phys. J. {\bf C12}, 35 (2000), [hep-ex/9908012].

\bibitem{Chekanov:2003rb}
ZEUS Collaboration, S.~Chekanov {\em et~al.},
\newblock Phys. Rev. {\bf D69}, 012004 (2004), [hep-ex/0308068].

\bibitem{Acosta:2004bs}
CDF Collaboration, D.~Acosta {\em et~al.},
\newblock Phys. Rev. {\bf D71}, 012005 (2005), [hep-ex/0410058].

\bibitem{Bowen:2005rs}
M.~T. Bowen, S.~D. Ellis and M.~J. Strassler,
\newblock Acta Phys. Polon. {\bf B36}, 271 (2005), [hep-ph/0504186].

\bibitem{O'Neil:2002ks}
ATLAS Collaboration, D.~O'Neil, B.~Gonzalez-Pineiro and M.~Lefebvre,
\newblock J. Phys. {\bf G28}, 2657 (2002).

\bibitem{Barisonzi}
M.~Barisonzi,
\newblock private communication, 2005.

\bibitem{Dudko:aihenp}
D\O\ Collaboration, L.~Dudko,
\newblock (1999),
\newblock Prepared for VI International Workshop On Artificial Intelligence in
  High Energy and Nuclear Physics (AIHENP 99), Crete, April, 1999.

\bibitem{Dudko:2000wx}
D\O\ Collaboration, L.~Dudko,
\newblock (2000),
\newblock Prepared for 7th International Workshop on Advanced Computing and
  Analysis Techniques in Physics Research (ACAT 2000), Batavia, Illinois, 16-20
  Oct 2000.

\bibitem{Boos:2003gv}
E.~Boos and L.~Dudko,
\newblock Nucl. Instrum. Meth. {\bf A502}, 486 (2003), [hep-ph/0302088].

\bibitem{Abazov:2001ns}
D\O\ Collaboration, V.~M. Abazov {\em et~al.},
\newblock Phys. Lett. {\bf B517}, 282 (2001), [hep-ex/0106059].

\bibitem{Abbott:2000pa}
D\O\ Collaboration, B.~Abbott {\em et~al.},
\newblock Phys. Rev. {\bf D63}, 031101 (2001), [hep-ex/0008024].

\bibitem{Dudko:d03612}
D\O\ Collaboration, E.~Boos and L.~Dudko,
\newblock (1999),
\newblock D\O\ Note 3612.

\bibitem{Dudko:d03856}
D\O\ Collaboration, E.~Boos, L.~Dudko, A.~Heinson and N.~Sotnikova,
\newblock (2000),
\newblock D\O\ Note 3856.

\bibitem{Dudko:shw}
D\O\ Collaboration, J.~Conway,
\newblock (2000),
\newblock \mbox{http://www.physics.ucdavis.edu/
  conway/research/software/pgs/pgs.html}.

\bibitem{Peterson:1993nk}
C.~Peterson, T.~Rognvaldsson and L.~Lonnblad,
\newblock Comput. Phys. Commun. {\bf 81}, 185 (1994).

\bibitem{Boos:2002xw}
E.~E. Boos and A.~V. Sherstnev,
\newblock Phys. Lett. {\bf B534}, 97 (2002), [hep-ph/0201271].

\bibitem{D0detector}
D0, V.~M. Abazov {\em et~al.},
\newblock Nucl. Instrum. Meth. {\bf A565}, 463 (2006), [physics/0507191].

\bibitem{boos-dudko}
D\O\ Collaboration, E.~Boos and L.~Dudko,
\newblock Nucl. Instrum. Methods A {\bf A 502}, 486 (2003).

\bibitem{rgs}
N.~Amos {\em et~al.},
\newblock prepared for {\it Computing in high energy physics '95}, Rio de
  Janeiro 1995.

\bibitem{mlpfit}
J.~Schwindling,
\newblock MLPfit: A Tool For Designing and Using Multi-Layer Perceptrons, see
  \url{http://schwind.home.cern.ch/schwind/MLPfit.html}.

\bibitem{IainTM2000}
I.~Bertram {\em et~al.},
\newblock FERMILAB-TM-2104  (2000).

\bibitem{SGTOP-LHC-PHENO-TOPREX}
S.~Slabospitsky {\em et~al.},
\newblock Comput. Phys. Commun. {\bf 148}, 87 (2002), [hep-ph/0201292].

\bibitem{SGTOP-LHC-PHENO-MCNLO}
S.~Frixione, B.~Weber and P.~Nason,
\newblock JHEP {\bf 0308}, 007 (2003), [hep-ph/0305252].

\bibitem{SGTOP-LHC-PHENO-HUBAUT}
F.~Hubaut {\em et~al.},
\newblock Eur.Phys.J. {\bf C44S2}, 13 (2005), [hep-ex/0508061].

\bibitem{SGTOP-LHC-PHENO-MCFM-WQQ}
J.~Campbell {\em et~al.},
\newblock Phys.Rev. {\bf D68}, 094021 (2003), [hep-ph/0308195].

\bibitem{SGTOP-LHC-PHENO-HERWIG}
G.~Corcella {\em et~al.},
\newblock JHEP {\bf 0101}, 010 (2001), [hep-ph/0011363].

\bibitem{ATL-PHYS-98-131}
E.~Richter-Was {\em et~al.},
\newblock ATLAS Note PHYS-98-131  (1998).

\bibitem{SGTOP-LHC-NOTE-TOPMASS}
I.~Bojanovic {\em et~al.},
\newblock Eur. Phys.J {\bf C39s2}, 63 (2005), [hep-ex/0403021].

\bibitem{Gunion:1989we}
J.~F. Gunion, H.~E. Haber, G.~L. Kane and S.~Dawson,
\newblock {\em The Higgs Hunter's Guide} (Addison-Wesely, 1989),
\newblock SCIPP-89/13.

\bibitem{SGTOP-LHC-LUCOTTE}
A.~Lucotte and F.~Chevallier,
\newblock ATLAS Note ATL-PHYS-COM-2006-0003 , 1 (2005), [published in {\it
  Hadron Collider Physics 2005}, Eds. M Campanelli, A. Clark, X. Wu, Springer,
  ISBN 3-540-32840-8, p.300].

\bibitem{SGTOP-LHC-ATL-PHYS-ALPHAS}
H.~Stenzel,
\newblock ATLAS Note ATL-PHYS-2001-0003  (2001).

\bibitem{CMS-singletop-T}
V.~Abramov {\em et~al.},
\newblock CMS Note 2006/084  (2006).

\bibitem{CMS-singletop-W}
S.~Blyth {\em et~al.},
\newblock CMS Note 2006/086  (2006).

\bibitem{PTDRv1}
{CMS Collaboration},
\newblock CERN-LHCC-2006-01.

\bibitem{singletop}
V.~S. E.~Boos, L.~Dudko,
\newblock CMS Note 2000/065  (2000).

\bibitem{Mangano:2002ea}
M.~L. Mangano, M.~Moretti, F.~Piccinini, R.~Pittau and A.~D. Polosa,
\newblock JHEP {\bf 07}, 001 (2003), [hep-ph/0206293].

\bibitem{Sjostrand:2001yu}
T.~Sjostrand, L.~Lonnblad and S.~Mrenna,
\newblock hep-ph/0108264.

\bibitem{GARCONMAIN}
S.~Abdullin {\em et~al.},
\newblock [hep-ph/0605143].

\bibitem{FISHER}
R.~Fisher,
\newblock Annals of Eugenics {\bf 7}, 179 (1936).

\bibitem{JHEP_0207_012}
J.~Pumplin {\em et~al.},
\newblock JHEP {\bf 07}, 012 (2002), [hep-ph/0201195].

\bibitem{lepewwg2}
ALEPH, DELPHI, L3, OPAL and LEPEWWG,
\newblock hep-ex/0511027.

\bibitem{Baur:2002gp}
The Snowmass Working Group on Precision Electroweak Measurements, U.~Baur {\em
  et~al.},
\newblock hep-ph/0202001.

\bibitem{Dittmar:1997md}
M.~Dittmar, F.~Pauss and D.~Zurcher,
\newblock Phys. Rev. {\bf D56}, 7284 (1997), [hep-ex/9705004].

\bibitem{Dittmar:2003ir}
M.~Dittmar, A.-S. Nicollerat and A.~Djouadi,
\newblock Phys. Lett. {\bf B583}, 111 (2004), [hep-ph/0307020].

\bibitem{Baur:2004ig}
U.~Baur and D.~Wackeroth,
\newblock Phys. Rev. {\bf D70}, 073015 (2004), [hep-ph/0405191].

\bibitem{Accomando:2005xp}
E.~Accomando and A.~Kaiser,
\newblock Phys. Rev. {\bf D73}, 093006 (2006), [hep-ph/0511088].

\bibitem{Berends:1984xv}
F.~A. Berends, R.~Kleiss, J.~P. Revol and J.~P. Vialle,
\newblock Z. Phys. {\bf C27}, 155 (1985).

\bibitem{Wackeroth:1996hz}
D.~Wackeroth and W.~Hollik,
\newblock Phys. Rev. {\bf D55}, 6788 (1997), [hep-ph/9606398].

\bibitem{Baur:1998kt}
U.~Baur, S.~Keller and D.~Wackeroth,
\newblock Phys. Rev. {\bf D59}, 013002 (1999), [hep-ph/9807417].

\bibitem{CarloniCalame:2003ux}
C.~M. Carloni~Calame, G.~Montagna, O.~Nicrosini and M.~Treccani,
\newblock Phys. Rev. {\bf D69}, 037301 (2004), [hep-ph/0303102].

\bibitem{Abe:1995np}
CDF Collaboration, F.~Abe {\em et~al.},
\newblock Phys. Rev. Lett. {\bf 75}, 11 (1995), [hep-ex/9503007].

\bibitem{Abachi:1996ey}
D\O\ Collaboration, S.~Abachi {\em et~al.},
\newblock Phys. Rev. Lett. {\bf 77}, 3309 (1996), [hep-ex/9607011].

\bibitem{tevewwg}
{Tevatron Electroweak Working Group},
\newblock hep-ex/0510077.

\bibitem{Abazov:2003sv}
CDF Collaboration, V.~M. Abazov {\em et~al.},
\newblock Phys. Rev. {\bf D70}, 092008 (2004).

\bibitem{Buttar:2006zd}
C.~Buttar {\em et~al.},
\newblock hep-ph/0604120.

\bibitem{Anastasiou:2003ds}
C.~Anastasiou, L.~J. Dixon, K.~Melnikov and F.~Petriello,
\newblock Phys. Rev. {\bf D69}, 094008 (2004), [hep-ph/0312266].

\bibitem{Anastasiou:2003yy}
C.~Anastasiou, L.~J. Dixon, K.~Melnikov and F.~Petriello,
\newblock Phys. Rev. Lett. {\bf 91}, 182002 (2003), [hep-ph/0306192].

\bibitem{Melnikov:2006di}
K.~Melnikov and F.~Petriello,
\newblock Phys. Rev. Lett. {\bf 96}, 231803 (2006), [hep-ph/0603182].

\bibitem{Melnikov:2006kv}
K.~Melnikov and F.~Petriello,
\newblock hep-ph/0609070.

\bibitem{Balazs:1997xd}
C.~Balazs and C.~P. Yuan,
\newblock Phys. Rev. {\bf D56}, 5558 (1997), [hep-ph/9704258].

\bibitem{Ellis:1997ii}
R.~K. Ellis and S.~Veseli,
\newblock Nucl. Phys. {\bf B511}, 649 (1998), [hep-ph/9706526].

\bibitem{Dittmaier:2001ay}
S.~Dittmaier and M.~Kr{\"a}mer,
\newblock Phys. Rev. {\bf D65}, 073007 (2002), [hep-ph/0109062].

\bibitem{Arbuzov:2005dd}
A.~Arbuzov {\em et~al.},
\newblock Eur. Phys. J. {\bf C46}, 407 (2006), [hep-ph/0506110].

\bibitem{CarloniCalame:2006zq}
C.~M. Carloni~Calame, G.~Montagna, O.~Nicrosini and A.~Vicini,
\newblock JHEP {\bf 12}, 016 (2006), [hep-ph/0609170].

\bibitem{Baur:2001ze}
U.~Baur, O.~Brein, W.~Hollik, C.~Schappacher and D.~Wackeroth,
\newblock Phys. Rev. {\bf D65}, 033007 (2002), [hep-ph/0108274].

\bibitem{Placzek:2003zg}
W.~Placzek and S.~Jadach,
\newblock Eur. Phys. J. {\bf C29}, 325 (2003), [hep-ph/0302065].

\bibitem{CarloniCalame:2005vc}
C.~M. Carloni~Calame, G.~Montagna, O.~Nicrosini and M.~Treccani,
\newblock JHEP {\bf 05}, 019 (2005), [hep-ph/0502218].

\bibitem{Cao:2004yy}
Q.-H. Cao and C.~P. Yuan,
\newblock Phys. Rev. Lett. {\bf 93}, 042001 (2004), [hep-ph/0401026].

\bibitem{Andonov:2004hi}
A.~Andonov {\em et~al.},
\newblock Comput. Phys. Commun. {\bf 174}, 481 (2006), [hep-ph/0411186].

\bibitem{DeRujula:1979jj}
A.~De~Rujula, R.~Petronzio and A.~Savoy-Navarro,
\newblock Nucl. Phys. {\bf B154}, 394 (1979).

\bibitem{Martin:2004dh}
A.~D. Martin, R.~G. Roberts, W.~J. Stirling and R.~S. Thorne,
\newblock Eur. Phys. J. {\bf C39}, 155 (2005), [hep-ph/0411040].

\bibitem{Ciafaloni:2000df}
M.~Ciafaloni, P.~Ciafaloni and D.~Comelli,
\newblock Phys. Rev. Lett. {\bf 84}, 4810 (2000), [hep-ph/0001142].

\bibitem{Melles:2001ye}
M.~Melles,
\newblock Phys. Rept. {\bf 375}, 219 (2003), [hep-ph/0104232].

\bibitem{Jantzen:2005az}
B.~Jantzen, J.~H. Kuhn, A.~A. Penin and V.~A. Smirnov,
\newblock Nucl. Phys. {\bf B731}, 188 (2005), [hep-ph/0509157].

\bibitem{Denner:2006jr}
A.~Denner, B.~Jantzen and S.~Pozzorini,
\newblock hep-ph/0608326.

\bibitem{Baur:2006sn}
U.~Baur,
\newblock Phys. Rev. {\bf D75}, 013005 (2007), [hep-ph/0611241].

\bibitem{Konychev:2005iy}
A.~V. Konychev and P.~M. Nadolsky,
\newblock Phys. Lett. {\bf B633}, 710 (2006), [hep-ph/0506225].

\bibitem{Berge:2004nt}
S.~Berge, P.~Nadolsky, F.~Olness and C.~P. Yuan,
\newblock Phys. Rev. {\bf D72}, 033015 (2005), [hep-ph/0410375].

\bibitem{Berge:2005rv}
S.~Berge, P.~M. Nadolsky and F.~I. Olness,
\newblock Phys. Rev. {\bf D73}, 013002 (2006), [hep-ph/0509023].

\bibitem{Horace-prd:2004}
C.~Carloni~Calame, G.~Montagna, O.~Nicrosini and M.~Treccani,
\newblock Phys. Rev. {\bf D69}, 037301 (2004), [hep-ph/0303102].

\bibitem{Horace-jhep:2005}
C.~Carloni~Calame, G.~Montagna, O.~Nicrosini and M.~Treccani,
\newblock JHEP {\bf 05}, 019 (2005), [hep-ph/0502218].

\bibitem{Babayaga-npb:2000}
C.~Carloni~Calame, C.~Lunardini, G.~Montagna, O.~Nicrosini and F.~Piccinini,
\newblock Nucl. Phys. {\bf B584}, 459 (2000), [hep-ph/0003268].

\bibitem{Calame-plb:2001}
C.~Carloni~Calame,
\newblock Phys. Lett. {\bf B520}, 16 (2001), [hep-ph/0103117].

\bibitem{CarloniCalame:2004qw}
C.~M. Carloni~Calame, S.~Jadach, G.~Montagna, O.~Nicrosini and W.~Placzek,
\newblock Acta Phys. Polon. {\bf B35}, 1643 (2004), [hep-ph/0402235].

\bibitem{Barberio:1990ms}
E.~Barberio, B.~van Eijk and Z.~Was,
\newblock Comput. Phys. Commun. {\bf 66}, 115 (1991).

\bibitem{Barberio:1994qi}
E.~Barberio and Z.~Was,
\newblock Comput. Phys. Commun. {\bf 79}, 291 (1994).

\bibitem{Golonka:2005pn}
P.~Golonka and Z.~Was,
\newblock Eur. Phys. J. {\bf C45}, 97 (2006), [hep-ph/0506026].

\bibitem{Golonka:2003xt}
P.~Golonka {\em et~al.},
\newblock Comput. Phys. Commun. {\bf 174}, 818 (2006), [hep-ph/0312240].

\bibitem{Nanava:2003cg}
G.~Nanava and Z.~Was,
\newblock Acta Phys. Polon. {\bf B34}, 4561 (2003), [hep-ph/0303260].

\bibitem{Golonka:2002rz}
P.~Golonka, T.~Pierzchala and Z.~Was,
\newblock Comput. Phys. Commun. {\bf 157}, 39 (2004), [hep-ph/0210252].

\bibitem{PhDGolonka}
P.~Golonka,
\newblock {\em Computer simulations in high energy physics: a case for PHOTOS,
  MC-TESTER, TAUOLA, and at2sim},
\newblock PhD thesis, Institute of Nuclear Physics, Krakow, 2006,
\newblock Written under the supervision of Prof. Z. Was.

\bibitem{Golonka:2006tw}
P.~Golonka and Z.~Was,
\newblock Eur. Phys. J. {\bf C50}, 53 (2007), [hep-ph/0604232].

\bibitem{Nanava:2006vv}
G.~Nanava and Z.~Was,
\newblock hep-ph/0607019.

\bibitem{MScGolonka}
P.~Golonka,
\newblock Photos+ : a c++ implementation of a universal monte carlo algorithm
  for qed radiative corrections in particle's decays,
\newblock Master's thesis, Faculty of Nuclear Physics and Techniques, AGH
  University of Science and Technology, 1999,
\newblock Written under the supervision of Z. Was, available at {\tt
  http://cern.ch/Piotr.Golonka/MC/photos}.

\bibitem{Lepage:1977sw}
G.~P. Lepage,
\newblock J. Comput. Phys. {\bf 27}, 192 (1978).

\bibitem{Baer:1990ra}
H.~Baer, J.~Ohnemus and J.~F. Owens,
\newblock Phys. Rev. {\bf D42}, 61 (1990).

\bibitem{Harris:2001sx}
B.~W. Harris and J.~F. Owens,
\newblock Phys. Rev. {\bf D65}, 094032 (2002), [hep-ph/0102128].

\bibitem{Bardeen:1978yd}
W.~A. Bardeen, A.~J. Buras, D.~W. Duke and T.~Muta,
\newblock Phys. Rev. {\bf D18}, 3998 (1978).

\bibitem{Owens:1992hd}
J.~F. Owens and W.-K. Tung,
\newblock Ann. Rev. Nucl. Part. Sci. {\bf 42}, 291 (1992).

\bibitem{Krasny:2005cb}
M.~W. Krasny, S.~Jadach and W.~Placzek,
\newblock Eur. Phys. J. {\bf C44}, 333 (2005), [hep-ph/0503215].

\bibitem{Jadach:2005rd}
S.~Jadach and M.~Skrzypek,
\newblock hep-ph/0509178.

\bibitem{Diener:2005me}
K.~P.~O. Diener, S.~Dittmaier and W.~Hollik,
\newblock Phys. Rev. {\bf D72}, 093002 (2005), [hep-ph/0509084].

\bibitem{Dittmaier:1999mb}
S.~Dittmaier,
\newblock Nucl. Phys. {\bf B565}, 69 (2000), [hep-ph/9904440].

\bibitem{Bredenstein:2005zk}
A.~Bredenstein, S.~Dittmaier and M.~Roth,
\newblock Eur. Phys. J. {\bf C44}, 27 (2005), [hep-ph/0506005].

\bibitem{Awramik:2003rn}
M.~Awramik, M.~Czakon, A.~Freitas and G.~Weiglein,
\newblock Phys. Rev. {\bf D69}, 053006 (2004), [hep-ph/0311148].

\bibitem{Degrassi:1997iy}
G.~Degrassi, P.~Gambino, M.~Passera and A.~Sirlin,
\newblock Phys. Lett. {\bf B418}, 209 (1998), [hep-ph/9708311].

\bibitem{Ferroglia:2002rg}
A.~Ferroglia, G.~Ossola, M.~Passera and A.~Sirlin,
\newblock Phys. Rev. {\bf D65}, 113002 (2002), [hep-ph/0203224].

\bibitem{Jegerlehner:2001wq}
F.~Jegerlehner,
\newblock J. Phys. {\bf G29}, 101 (2003), [hep-ph/0104304].

\bibitem{Denner:1990yz}
A.~Denner and T.~Sack,
\newblock Nucl. Phys. {\bf B347}, 203 (1990).

\bibitem{Collins:1985kg}
J.~C. Collins, D.~E. Soper and G.~Sterman,
\newblock Nucl. Phys. {\bf B250}, 199 (1985).

\bibitem{Collins:2004nx}
J.~C. Collins and A.~Metz,
\newblock Phys. Rev. Lett. {\bf 93}, 252001 (2004), [hep-ph/0408249].

\bibitem{Ji:2004wu}
X.~Ji, J.-P. Ma and F.~Yuan,
\newblock Phys. Rev. {\bf D71}, 034005 (2005).

\bibitem{Ji:2004xq}
X.~Ji, J.-P. Ma and F.~Yuan,
\newblock Phys. Lett. {\bf B597}, 299 (2004).

\bibitem{Landry:2002ix}
F.~Landry, R.~Brock, P.~M. Nadolsky and C.~P. Yuan,
\newblock Phys. Rev. {\bf D67}, 073016 (2003), [hep-ph/0212159].

\bibitem{Ladinsky:1993zn}
G.~A. Ladinsky and C.-P. Yuan,
\newblock Phys. Rev. {\bf D50}, 4239 (1994).

\bibitem{Korchemsky:1994is}
G.~P. Korchemsky and G.~Sterman,
\newblock Nucl. Phys. {\bf B437}, 415 (1995).

\bibitem{Tafat:2001in}
S.~Tafat,
\newblock JHEP {\bf 05}, 004 (2001), [hep-ph/0102237].

\bibitem{Qiu:2000hf}
J.-w. Qiu and X.-f. Zhang,
\newblock Phys. Rev. {\bf D63}, 114011 (2001), [hep-ph/0012348].

\bibitem{Kulesza:2002rh}
A.~Kulesza, G.~Sterman and W.~Vogelsang,
\newblock Phys. Rev. {\bf D66}, 014011 (2002).

\bibitem{Guffanti:2000ep}
A.~Guffanti and G.~E. Smye,
\newblock JHEP {\bf 10}, 025 (2000).

\bibitem{Nadolsky:1999kb}
P.~Nadolsky, D.~R. Stump and C.~P. Yuan,
\newblock Phys. Rev. {\bf D61}, 014003 (2000), [hep-ph/9906280].

\bibitem{Nadolsky:2000ky}
P.~M. Nadolsky, D.~R. Stump and C.~P. Yuan,
\newblock Phys. Rev. {\bf D64}, 114011 (2001), [hep-ph/0012261].

\bibitem{Collins:1998rz}
J.~C. Collins,
\newblock Phys. Rev. {\bf D58}, 094002 (1998), [hep-ph/9806259].

\bibitem{Kramer:2000hn}
M.~Kr{\"a}mer, F.~I. Olness and D.~E. Soper,
\newblock Phys. Rev. {\bf D62}, 096007 (2000), [hep-ph/0003035].

\bibitem{Nadolsky:2002jr}
P.~M. Nadolsky, N.~Kidonakis, F.~I. Olness and C.~P. Yuan,
\newblock Phys. Rev. {\bf D67}, 074015 (2003), [hep-ph/0210082].

\bibitem{Tung:2006tb}
W.~K. Tung {\em et~al.},
\newblock hep-ph/0611254.

\bibitem{CMS-PTDR-Vol2}
{CMS Collaboration},
\newblock CERN/LHCC 2006-021,
\newblock Volume 2 of the CMS Physics TDR.

\bibitem{Abulencia:2005ix}
CDF Collaboration, A.~Abulencia {\em et~al.},
\newblock hep-ex/0508029.

\bibitem{bib:D0wcrosssection}
{D\O\ Collaboration},
\newblock Measurement of the cross section for inclusive $W$ production in the
  muon channel at $\sqrt{s}=1.96$ TeV using the D0 detector, D0note 4750
  (2006).

\bibitem{bib:D0zcrosssection}
{D\O\ Collaboration},
\newblock Measurement of the cross section for inclusive $Z$ production in the
  di-muon final states at $\sqrt{s}=1.96$ TeV, D0note 4573 (2006).

\bibitem{James:2007ks}
CDF and D\O\ Collaborations, E.~B. James,
\newblock hep-ex/0701003.

\bibitem{bib:CDF}
{CDF Collaboration},
\newblock CDF Run 2 Electroweak Public Results at
  \url{http://www-cdf.fnal.gov/physics/ewk}.

\bibitem{bib:D0wasymmetry}
{D\O{} Collaboration},
\newblock A measurement of the $W\to \mu \nu_\mu$ asymmetry with the D0
  detector at $\sqrt{s}=1.96$ TeV, D0note 5061-CONF (2006).

\bibitem{Tricoli:2005nx}
A.~Tricoli, A.~M. Cooper-Sarkar and C.~Gwenlan,
\newblock hep-ex/0509002.

\bibitem{bib:CDFwmass}
{CDF Collaboration},
\newblock First measurement of the $W$ boson mass with CDF in RUN II, CDF Note
  8665 (2006).

\bibitem{lepewwg}
{LEP Electroweak Working Group}, 2005,
\newblock {\tt lepewwg.web.cern.ch/LEPEWWG/Welcome.html}.

\bibitem{Rosner:1994ww}
J.~Rosner, M.~Worah and T.~Takeuchi,
\newblock Phys. Rev. {\bf D49}, 1363 (1994).

\bibitem{Hagiwara:2002ww}
K.~Hagiwara {\em et~al.},
\newblock Phys. Rev. {\bf D66}, 010001 (2002).

\bibitem{cdf:2000ww}
CDF Collaboration, T.~Affolder {\em et~al.},
\newblock Phys. Rev. Lett. {\bf 85}, 3347 (2000).

\bibitem{d0:2002ww}
D\O\ Collaboration, V.~Abazov {\em et~al.},
\newblock Phys. Rev. {\bf D66}, 032008 (2002).

\bibitem{d0:2004ww_prelim}
{D\O\ Collaboration},
\newblock {Direct Measurement of the $W$ Boson Width in $p\bar{p}$ Collisions
  at $\sqrt{s}=1.96$ TeV},
\newblock {D0 Note 7983-CONF}, 2004.

\bibitem{width_combined}
{The Tevatron Electroweak Working Group},
\newblock hep-ex/0510077.

\bibitem{gianotti:1999ewk}
F.~Gianotti,
\newblock (2006),
\newblock in {\it Introduction to LHC physics}, Journal of Physics: conference
  series 53.

\bibitem{keller:1998wmass}
S.~Keller and J.~Womersley,
\newblock Eur. Phys. J. {\bf C5}, 249 (1998).

\end{thebibliography}

\end{document}